\documentclass[structabstract]{aa}  
\usepackage{graphicx}
\usepackage{float}
\usepackage[caption = false]{subfig}
\usepackage{lscape}
\usepackage{txfonts}
\newcommand{\teff}{\ifmmode T_{\rm eff} \else $T_{\mathrm{eff}}$\fi}
\newcommand{\logg}{\ifmmode \log g \else $\log g$\fi}
\newcommand{\lL}{\ifmmode \log \frac{L}{L_{\odot}} \else $\log \frac{L}{L_{\odot}}$\fi}

\newcommand{\vsini}{\ifmmode v \sin\,i \else $v \sin\,i$\fi}
\newcommand{\vinf}{\ifmmode v_{\infty} \else $v_{\infty}$\fi}

\newcommand{\vmac}{\ifmmode v_{\rm mac} \else $v_{\rm mac}$\fi}
\newcommand{\vesc}{\ifmmode  v_{\rm esc} \else $v_{\rm esc}$ \fi}
\newcommand{\kms}{km~s$^{-1}$}
\newcommand{\msun}{\ifmmode M_{\odot} \else $M_{\odot}$\fi}
\newcommand{\zsun}{\ifmmode Z_{\odot} \else $Z_{\odot}$\fi}
\newcommand{\lsun}{\ifmmode L_{\odot} \else $L_{\odot}$\fi}
\newcommand{\rsun}{\ifmmode R_{\odot} \else $R_{\odot}$\fi}
\newcommand{\qh}{\ifmmode Q_{\rm H} \else $Q_{\rm H}$\fi}
\newcommand{\qhei}{\ifmmode Q_{\ion{He}{i}} \else $Q_{\ion{He}{i}}$\fi}

\begin{document}
   \title{The Tarantula Massive Binary Monitoring:}

   \subtitle{III. Atmosphere analysis of double-lined spectroscopic systems \thanks{Based on observations collected at the European Southern Observatory (Paranal and La Silla, Chile) under program IDs 090.D-0323 and 092.D-0136 (PI: Sana)}}

   \author{L. Mahy\inst{1}
          \and
          H. Sana\inst{1}
          \and
          M. Abdul-Masih\inst{1}
          \and
          L. A. Almeida\inst{2,3}
          \and
          N. Langer\inst{4}
          \and
          T. Shenar\inst{1}
          \and
          A. de Koter\inst{5,1}
          \and
          S. E. de Mink\inst{6,5}
          \and
          S. de Wit\inst{5,16}
          \and
          N. J. Grin\inst{4}
          \and
          C. J. Evans\inst{7}
          \and
          A. F. J. Moffat\inst{8}
          \and
          F. R. N. Schneider\inst{9,10}
          \and
          R. Barb{\'a}\inst{11}
          \and
          J. S. Clark\inst{12}
          \and 
          P. Crowther\inst{13}
          \and
          G. Gr{\"a}fener\inst{4}
          \and
          D. J. Lennon\inst{14,15}
          \and
          F. Tramper\inst{16}
          \and
          J. S. Vink\inst{17}
          }

   \offprints{L. Mahy}

   \institute{Instituut voor Sterrenkunde, KU Leuven, Celestijnlaan 200D, Bus 2401, B-3001 Leuven, Belgium\\
   \email{laurent.mahy@kuleuven.be}
   \and
   Departamento de F\'isica, Universidade do Estado do Rio Grande do Norte, Mossor\'o, RN, Brazil
   \and 
   Departamento de F\'isica, Universidade Federal do Rio Grande do Norte, UFRN, CP 1641, Natal, RN, 59072-970, Brazil
   \and
   Argelander-Institut f{\"u}r Astronomie der Universit{\"a}t Bonn, Auf dem H{\"u}gel 71, 53121 Bonn, Germany
   \and
   Astronomical Institute Anton Pannekoek, Amsterdam University, Science Park 904, 1098 XH Amsterdam, The Netherlands
   \and 
   Center for Astrophysics, Harvard \& Smithsonian, 60 Garden Street, Cambridge, MA 02138, USA
   \and
   UK Astronomy Technology Centre, Royal Observatory Edinburgh, Blackford Hill, Edinburgh EH9 3HJ, United Kingdom
   \and
   D{\'e}partement de physique, Universit{\'e} de Montr{\'e}al, and Centre de Recherche en Astrophysique de Qu{\'e}bec, CP 6128, Succ. C-V, Montr{\'e}al, QC, H3C 3J7, Canada
   \and
   Zentrum f\"{u}r Astronomie der Universit\"{a}t Heidelberg, Astronomisches Rechen-Institut, M\"{o}nchhofstr. 12-14, 69120 Heidelberg, Germany
   \and
   Heidelberger Institut f\"{u}r Theoretische Studien, Schloss-Wolfsbrunnenweg 35, 69118 Heidelberg, Germany
   \and
   Departamento de F{\'i}sica y Astronom{\'i}a, Universidad de La Serena, Av. Cisternas 1200 Norte, La Serena, Chile
   \and
   School of Physical Sciences, The Open University, Walton Hall, Milton Keynes, MK7 6AA, UK
   \and
   Dept of Physics and Astronomy, University of Sheffield, Hounsfield Road, Sheffield, S3 7RH, UK
   \and
   Instituto de Astrof\'isica de Canarias, E-38205 La Laguna, Tenerife, Spain
   \and
   ESA, European Space Astronomy Centre, Apdo. de Correos 78, E-28691 Villanueva de la Ca{\~n}ada, Madrid, Spain
   \and
   IAASARS, National Observatory of Athens, 15326 Penteli, Greece
   \and
   Armagh Observatory, College Hill, Armagh, BT61 9DG, UK
   }

   \date{Received ...; accepted ...}

% \abstract{}{}{}{}{} 
% 5 {} token are mandatory
 
  \abstract
   {Accurate stellar parameters of individual objects in binary systems are essential to constrain the effects of binarity on stellar evolution. These parameters serve as a prerequisite to probing existing and future theoretical evolutionary models.  }
   {We aim to derive the atmospheric parameters of the 31 double-lined spectroscopic binaries in the Tarantula Massive Binary Monitoring sample. This sample, composed of detached, semi-detached and contact systems with at least one of the components classified as an O-type star, is an excellent test-bed to study how binarity can impact our knowledge of the evolution of massive stars.  }
   {In the present paper, 32 epochs of FLAMES/GIRAFFE spectra are analysed by using spectral disentangling to construct the individual spectra of 62 components. We then apply the CMFGEN atmosphere code to determine their stellar parameters and their helium, carbon, and nitrogen surface abundances.  }
   {Among the 31 systems that we study in the present paper, we identify between 48 and 77\% of them as detached, likely pre-interacting systems,  16\% as semi-detached systems, and between 5 and 35\% as systems in or close to contact phase. Based on the properties of their components, we show that the effects of tides on chemical mixing are limited. Components on longer-period orbits show higher nitrogen enrichment at their surface than those on shorter-period orbits, in contrast to expectations of rotational or tidal mixing, implying that other mechanisms play a role in this process. For semi-detached systems, components that fill their Roche lobe are mass donors. They exhibit higher nitrogen content at their surface and rotate more slowly than their companions. By accreting new material, their companions spin faster and are likely rejuvenated. Their locations in the $N-\vsini$ diagram tend to show that binary products are good candidates to populate the two groups of stars (slowly rotating, nitrogen-enriched objects and rapidly rotating non-enriched objects) that cannot be reproduced through single-star population synthesis. Finally, we find no peculiar surface abundances for the components in (over-)contact systems, as has been suggested by evolutionary models for tidal mixing.  }
   {This sample, consisting of 31 massive binary systems, is the largest sample of binaries composed of at least one O-type star to be studied in such a homogeneous way by applying spectral disentangling and atmosphere modelling. The study of these objects gives us strong observational constraints to test theoretical binary evolutionary tracks.  }


   \maketitle

%%#####################################################################
%%-------------------------------   Introduction  --------------------- 
\section{Introduction}
\label{intro}

Massive stars are among the most important cosmic engines, playing a considerable role in the ecology of galaxies via their intense winds, ultraviolet radiation fields, and their explosions as supernovae. A high fraction of massive OB stars are found in multiple systems. In young open clusters or OB associations in the Milky Way, such as IC\,1805 \citep{debecker06,hillwig06}, IC\,1848 \citep{hillwig06}, NGC\,6231 \citep{sana08}, NGC\,2244 \citep{mahy09}, NGC\,6611 \citep{sana09}, Tr\,16 \citep{rauw09}, IC\,2944 \citep{sana11}, Cyg\,OB1, 3, 8, 9 \citep{mahy13}, and Cyg\,OB2 \citep{kobulnicky14}, the detected  spectroscopic binary fraction ranges between 30 and 60\%, and makes up over 90\% once observational biases and orbital periods beyond the spectroscopic binary regime are taken into account \citep{sana12,sana14}. The binary fraction among the massive star population also remains high outside our Galaxy. In the Tarantula region in the Large Magellanic Cloud (LMC), the spectroscopic binary fraction of massive stars was estimated to be 51\% \citep{sana13, dunstall15} after correction for the observational biases. 

This abundance of multiple systems challenges the usual view of the predominance of the single-star evolutionary channel. Even though pre-interacting binaries are probably ideal laboratories to test {\it single} star evolution, the presence of a nearby companion modifies the evolution of a star through tides and mass transfer. However, large uncertainties remain in the physics of binary evolution, which is mostly related to internal mixing and the efficiency of mass and angular momentum transfer. To answer these questions, observational constraints from large homogeneous samples of well-characterised binaries are needed. 

The Tarantula Nebula (30 Doradus) in the LMC offers an ideal laboratory to investigate the above-listed physical processes, thanks to its large massive star population and young age. Located at a distance of 50\,kpc, and with a metallicity of one half of solar (0.5 $Z_{\odot}$), 30\,Dor is the brightest \ion{H}{ii} region in the Local Group and the closest massive starburst region known so far. The VLT- FLAMES Tarantula Survey (VFTS, \citealt{evans11}) obtained multi-epoch spectra of over 800 massive stars located in this region. These data lead to better constraints on  the parameters of the single O- and B-type star populations, and to the identification of binary systems \citep{sana13, dunstall15}. Among this population of binaries, we highlight VFTS\,527, the most massive binary system known to contain two O supergiants \citep{taylor11}, VFTS\,450 and VFTS\,652, two high-mass analogues of classical Algol systems \citep{howarth15},  VFTS\,352, one of the most massive contact binaries \citep{almeida15,abdul-masih19}, VFTS\,399, a potential X-ray binary \citep{clark15} and R\,145, which is the most evolved object in the sample \citep{shenar17}.

While \citet{sana13} were able to detect spectroscopic binary systems, the secured number of epochs was nevertheless insufficient to characterise their individual orbital properties. In this context, the Tarantula Massive Binary Monitoring (TMBM) project was designed to measure the orbital properties of systems with orbital periods from about one day up to slightly over one year. The measurements of the radial velocities (RVs) and the determinations of the orbital properties for 82 systems (51 single-lined - SB1 - and 31 double-lined - SB2 - spectroscopic binaries) were performed by \citet[][hereafter Paper\,I]{almeida17}. These authors also studied the period, eccentricity, and mass-ratio distributions and compared them with those derived in the Galaxy. They pointed out a universality of the incidence rate of massive binaries and their orbital properties in the metallicity range from solar ($Z_{\odot}$) to about half of solar. We are therefore curious as to whether the effects on the properties of the individual components are also similar through different metallicity regimes.

To study the effects of the presence of a companion on stellar evolution, we must derive the individual properties of each object. The technique of spectral disentangling can be used to generate individual spectra of the component stars, notably in double-lined spectroscopic binaries. These spectra, reconstructed from observations covering the range of orbital phases as uniformly as possible, can then be analysed with atmosphere models. Since the resulting disentangled spectra have a much higher signal-to-noise ratio than the observed spectra, the spectral disentangling is very useful for chemical abundance analysis and the determination of the physical properties of the stars. So far, very few massive galactic systems have in fact been studied with this technique \citep[see e.g.,][among others]{mahy13, mahy17, martins17, pavlovski18, raucq16, raucq17}. These were mostly individual studies and a more global view is currently lacking to make the bridge between observations and theory.

The present paper aims to study the physical properties of each component of the 31 SB2 systems in the TMBM. This is the first time that such a study has been homogeneously performed on a sample of this size. In addition to the stellar parameters, the surface abundances are also discussed and compared with respect to stellar evolution. This paper is organised as follows: Section\,\ref{obs} presents the data and Sect.\,\ref{method} summarises the methodology of our analysis. Section\,\ref{discussion} discusses the effects of the rotational and tidal mixings, and of the mass transfer on the evolution of these objects as well as the age and rotation distributions between the single and binary populations in 30 Dor. Our conclusions are given in Sect.\,\ref{conc}. 

%%#####################################################################
%%-------------------------------   Observations  --------------------- 

\section{Observations and data reduction}
\label{obs}
\subsection{Spectroscopic observations}
\label{obs:spec}

\begin{figure*}[t!]
  \centering
pdf    \includegraphics[trim=40 35 70 70,clip,width=16cm]{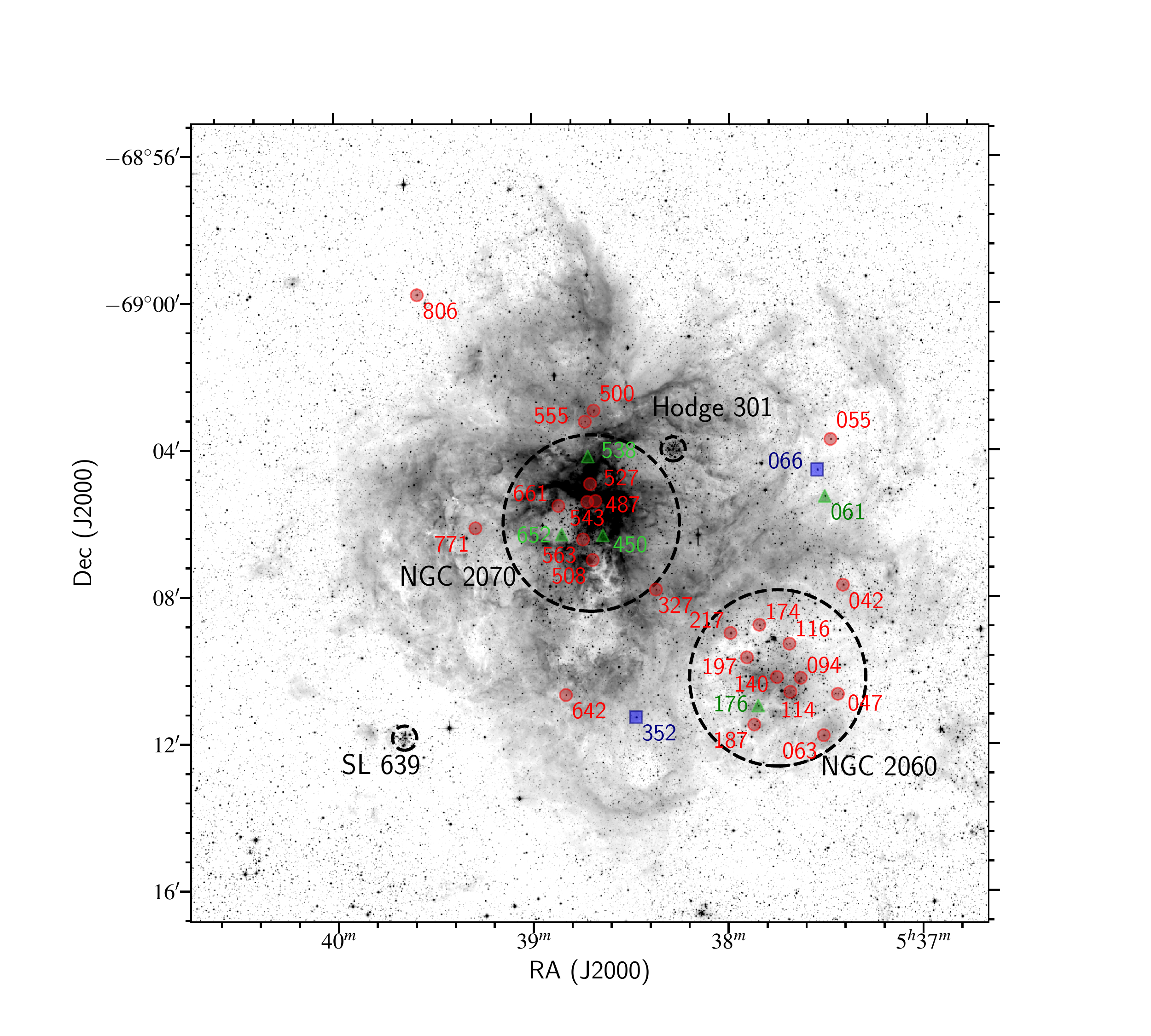}
    \caption{Location of the 31 SB2 massive systems in the Tarantula nebula. Red circles indicate the detached systems, the green triangles the semi-detached systems and the blue squares the contact systems. Dashed circles indicate the NGC\,2070 and NGC\,2060 OB associations (radius of 2.4') as well as the older B-type clusters SL\,639 and Hodge\,301 (radius of 0.33').}\label{fig:FOV} 
\end{figure*}

The initial monitoring targeted 102 massive stars in the Tarantula region. Observations were carried out with the FLAMES/GIRAFFE spectrograph mounted on the VLT/UT2 operated in its MEDUSA$+$UVES mode. We obtained 32 unevenly sampled epochs of each object using the L427.2 (LR02) grating which provides continuous spectral coverage of the 3964--4567\AA\ wavelength range at a spectral resolving power of $R= 6400$. We refer to Paper\,I for further information on this monitoring and the data reduction procedure. In the present paper we only focus on the 31 SB2 systems. Figure\,\ref{fig:FOV} shows their location in the Tarantula nebula.

\subsection{Photometric observations}
\label{obs:phot}

Thirteen objects in our sample were observed by the Optical Gravitational Lensing Experiment (OGLE) project. From the photometric light curves (Mahy et al. 2019, Paper IV), we use the relative brightness of the different components of these systems determined from the light-curve analysis as input to our spectral disentangling code (see below). The orbital configurations derived photometrically from the inclinations and the Roche lobe filling factors are discussed in detail in Paper IV.

%%#####################################################################
%%-------------------------   Methodology    --------------------- 

\section{Methodology}
\label{method}

\subsection{Spectral disentangling}
\label{method:disentangling}
The spectral disentangling code that we used is based on the Fourier method \citep{simon94, hadrava95, ilijic04} and allows one to obtain the separation of the spectral contributions of both components in a system as well as the orbital parameters of the system through a minimisation procedure based on the Nelder \& Mead simplex \citep{nelder65}. The resulting spectra of each component can then be analysed as if the stars were single.

The RVs and the orbital solutions of the different systems were derived in Paper I. They are used as initial input for the spectral disentangling code. The orbital solutions are then refined to minimise the $\chi^2$ between the observed spectra and the reconstruction of the disentangled spectra. A comparison between the disentangled and the observed spectra is shown in the Appendix for each system. For these figures, we shift the disentangled spectra by the RV of each component, multiply them by the corresponding brightness factor and sum up the two spectra. These figures allow us to see the impact of the nebular contamination in each different system. Table\,\ref{tab:orbit} summarises the new determinations of the orbital parameters for the 31 SB2 systems under the definition that the primaries correspond to the (current) most massive components of the systems. We note that, in Table\,\ref{tab:orbit}, the spectral types are from \citet{walborn14} and were established by visual inspection of the strengths of the lines and comparison with spectral standards. The latter definition assumes that the primary stars are the most luminous objects of the systems which is different from our definition (based on the most massive stars as primaries). We also emphasise that HJD$_0$ refers to the time when the primary is in front of the secondary for circular systems and to the time of periastron passage for eccentric orbits.

\begin{sidewaystable*}
\tiny
\caption{\label{tab:orbit}Spectral types and orbital parameters of the SB2 binaries refined from Paper I.}
\centering
\addtolength{\tabcolsep}{-3.5pt}   
\begin{tabular}{llrrrrrrrrrrr}
\hline\hline
Star& Previous spectral type& $P_{\mathrm{orb}}$ & $e$ & $\omega$ & $M_P/M_S$ & $K_P$ & $K_S$ & HJD$_0 -$ & $a_P \sin i$ & $a_S \sin i $ & $m_P \sin^3 i$ & $m_S \sin^3 i$ \\
        &  \citet{walborn14}   &  [days] &     &  [\degr]    &       &   [\kms]  &  [\kms]   &  2456000 & [\rsun] & [\rsun] & [\msun] & [\msun] \\
\hline
\hline
042 &  O9.5\,III((n))  & $29.300249 \pm 0.015689$ & $0.184 \pm 0.024$ & $145.23 \pm   8.68$ & $1.23 \pm 0.06$ & $ 52.19 \pm   1.67$ & $ 64.34 \pm   2.06$ & $219.32 \pm 0.67$ & $ 29.68 \pm   0.96$ & $ 36.59 \pm   1.18$ & $  2.52 \pm   0.22$ & $2.04 \pm 0.18$ \\ [2pt]
047 &  O9\,V $+$ O9.5\,V  & $5.930379 \pm 0.000704$ & $0.133 \pm 0.023$ & $125.39 \pm  10.68$ & $1.10 \pm 0.05$ & $167.41 \pm   5.38$ & $184.89 \pm   5.94$ & $206.68 \pm 0.17$ & $ 19.43 \pm   0.63$ & $ 21.46 \pm   0.69$ & $ 13.72 \pm   1.00$ & $12.42 \pm 0.89$ \\ [2pt]
055 & O8.5\,V $+$ O9.5\,IV   & $6.444405 \pm 0.000562$ & $0.101 \pm 0.018$ & $255.49 \pm   9.85$ & $1.05 \pm 0.03$ & $141.03 \pm   3.20$ & $147.46 \pm   3.34$ & $210.67 \pm 0.17$ & $ 17.86 \pm   0.41$ & $ 18.67 \pm   0.42$ & $  8.06 \pm   0.48$ & $7.71 \pm 0.46$ \\ [2pt]
061 &  ON8.5\,III: $+$ O9.7:\,V:  & $2.333440 \pm 0.000007$ & $0.000$ & $270.00$ & $1.88 \pm 0.11$ & $151.84 \pm   6.28$ & $284.76 \pm  11.85$ & $209.59 \pm 0.01$ & $  7.00 \pm   0.29$ & $ 13.22 \pm   0.55$ & $ 13.33 \pm   1.15$ & $7.06 \pm 0.47$ \\ [2pt]
063 &  O5\,III(n)(fc) $+$ sec  & $85.767054 \pm 0.066141$ & $0.647 \pm 0.039$ & $359.87 \pm   2.14$ & $1.86 \pm 0.29$ & $ 85.89 \pm   9.62$ & $159.79 \pm  17.89$ & $213.63 \pm 0.26$ & $110.91 \pm  13.31$ & $206.35 \pm  24.76$ & $ 37.96 \pm  13.55$ & $20.40 \pm 7.26$ \\ [2pt]
066 &  O9.5\,III(n)  & $1.141161 \pm 0.000004$ & $0.000$ & $270.00$ & $1.91 \pm 0.05$ & $ 56.62 \pm   1.07$ & $108.02 \pm   2.04$ & $209.62 \pm 0.01$ & $  1.28 \pm   0.02$ & $  2.43 \pm   0.05$ & $  0.35 \pm   0.01$ & $0.18 \pm 0.01$ \\ [2pt]
094 & O3.5\,Inf$^{*}$p $+$ sec?   & $2.256327 \pm 0.000006$ & $0.000$ & $270.00$ & $1.08 \pm 0.11$ & $142.74 \pm  10.40$ & $154.46 \pm  11.25$ & $208.67 \pm 0.04$ & $  6.36 \pm   0.46$ & $  6.88 \pm   0.50$ & $  3.19 \pm   0.59$ & $2.95 \pm 0.54$ \\ [2pt]
114 &  O8.5\,IV $+$ sec  & $27.782612 \pm 0.003570$ & $0.514 \pm 0.035$ & $166.14 \pm   4.14$ & $1.10 \pm 0.12$ & $106.78 \pm   7.95$ & $117.84 \pm   8.77$ & $215.20 \pm 0.22$ & $ 50.27 \pm   3.94$ & $ 55.48 \pm   4.35$ & $ 10.81 \pm   2.27$ & $9.79 \pm 2.05$ \\ [2pt]
116 &  O9.7:\,V: $+$ B0:\,V:  & $23.921162 \pm 0.001600$ & $0.257 \pm 0.040$ & $357.17 \pm   8.04$ & $1.25 \pm 0.09$ & $ 85.56 \pm   4.42$ & $107.02 \pm   5.53$ & $207.02 \pm 0.41$ & $ 39.06 \pm   2.06$ & $ 48.86 \pm   2.58$ & $  8.87 \pm   1.17$ & $7.09 \pm 0.91$ \\ [2pt]
140 &  O8.5\,Vz  & $1.611768 \pm 0.000006$ & $0.089 \pm 0.019$ & $186.15 \pm  13.37$ & $1.02 \pm 0.04$ & $100.25 \pm   2.79$ & $102.33 \pm   2.85$ & $210.51 \pm 0.06$ & $  3.18 \pm   0.09$ & $  3.24 \pm   0.09$ & $  0.69 \pm   0.05$ & $0.68 \pm 0.04$ \\ [2pt]
174 &  O8\,V $+$ B0:\,V:  & $4.756063 \pm 0.000128$ & $0.271 \pm 0.006$ & $ 28.06 \pm   1.61$ & $1.47 \pm 0.02$ & $159.64 \pm   1.48$ & $234.44 \pm   2.17$ & $211.79 \pm 0.02$ & $ 14.43 \pm   0.14$ & $ 21.20 \pm   0.20$ & $ 15.99 \pm   0.37$ & $10.89 \pm 0.24$ \\ [2pt]
176 &  O6\,V:((f)) $+$ O9.5:\,V:  & $1.777593 \pm 0.000003$ & $0.000$ & $270.00$ & $1.62 \pm 0.05$ & $240.56 \pm   5.08$ & $388.77 \pm   8.20$ & $208.70 \pm 0.01$ & $  8.45 \pm   0.18$ & $ 13.65 \pm   0.29$ & $ 28.34 \pm   1.48$ & $17.54 \pm 0.85$ \\ [2pt]
187 & O9\,IV $+$ B0:\,V:   & $3.542258 \pm 0.000025$ & $0.000$ & $270.00$ & $1.29 \pm 0.06$ & $123.77 \pm   3.79$ & $159.87 \pm   4.89$ & $210.39 \pm 0.02$ & $  8.66 \pm   0.27$ & $ 11.19 \pm   0.34$ & $  4.72 \pm   0.35$ & $3.65 \pm 0.26$ \\ [2pt]
197 &  O9\,III  & $69.925180 \pm 0.004200$ & $0.157 \pm 0.021$ & $5.75 \pm  8.02$ & $1.57 \pm 0.06$ & $ 56.00 \pm   2.19$ & $87.93 \pm   3.19$ & $150.03 \pm 1.38$ & $76.46 \pm   3.00$ & $120.05 \pm   4.38$ & $ 12.74 \pm   2.14$ & $8.11 \pm 1.39$ \\ [2pt]
217 &  O4\,V((fc)): $+$ O5\,V((fc)):  & $1.855341 \pm 0.000002$ & $0.000$ & $270.00$ & $1.20 \pm 0.01$ & $222.88 \pm   1.92$ & $268.29 \pm   2.31$ & $208.92 \pm 0.01$ & $  8.17 \pm   0.07$ & $  9.83 \pm   0.08$ & $ 12.44 \pm   0.22$ & $10.33 \pm 0.18$ \\ [2pt]
327 &  O8.5\,V(n) $+$ sec  & $2.955204 \pm 0.000022$ & $0.176 \pm 0.022$ & $286.08 \pm   6.51$ & $1.50 \pm 0.06$ & $107.98 \pm   2.96$ & $161.77 \pm   4.43$ & $209.03 \pm 0.05$ & $  6.20 \pm   0.17$ & $  9.29 \pm   0.26$ & $  3.44 \pm   0.25$ & $2.29 \pm 0.16$ \\ [2pt]
352 &  O4.5\,V(n)((fc))):z: $+$ & $1.124142 \pm 0.000001$ & $0.000$ & $270.00$ & $1.02 \pm 0.02$ & $301.97 \pm   5.05$ & $308.28 \pm   5.16$ & $209.34 \pm 0.01$ & $  6.70 \pm   0.11$ & $  6.84 \pm   0.11$ & $ 13.36 \pm   0.56$ & $13.09 \pm 0.55$ \\ 
 &  O5.5\,V(n)((fc))):z:  & &  & &   & &  & &  & &   &  \\ [2pt]
450 &  O9.7\,III: $+$ O7::   & $6.892336 \pm 0.000027$ & $0.000$ & $270.00$ & $1.04 \pm 0.08$ & $188.47 \pm   9.81$ & $196.46 \pm  10.23$ & $211.05 \pm 0.06$ & $ 25.65 \pm   1.34$ & $ 26.74 \pm   1.39$ & $ 20.78 \pm   2.78$ & $19.93 \pm 2.65$ \\ [2pt]
487 &  O6.5:\,IV:((f)): $+$ O6.5:\,IV:((f)):   & $4.121518 \pm 0.000412$ & $0.064 \pm 0.028$ & $231.19 \pm  23.88$ & $1.21 \pm 0.05$ & $158.71 \pm   4.94$ & $191.53 \pm   5.97$ & $208.87 \pm 0.28$ & $ 12.89 \pm   0.40$ & $ 15.56 \pm   0.49$ & $  9.96 \pm   0.89$ & $8.26 \pm 0.73$ \\ [2pt]
500 & O6.5\,IV((fc)) $+$ O6.5\,V((fc))   & $2.875371 \pm 0.000004$ & $0.000$ & $270.00$ & $1.05 \pm 0.01$ & $233.55 \pm   2.30$ & $246.09 \pm   2.42$ & $208.39 \pm 0.01$ & $ 13.26 \pm   0.13$ & $ 13.97 \pm   0.14$ & $ 16.86 \pm   0.36$ & $16.00 \pm 0.33$ \\ [2pt]
508 &  O9.5\,V  & $128.586000 \pm 0.025000$ & $0.436 \pm 0.030$ & $256.67 \pm   5.02$ & $1.35 \pm 0.09$ & $ 85.18 \pm   3.80$ & $115.20 \pm   5.14$ & $227.07 \pm 1.32$ & $194.65 \pm   9.24$ & $263.25 \pm  12.49$ & $ 44.88 \pm   5.55$ & $33.18 \pm 3.99$ \\ [2pt]
527 & O6.5\,Iafc $+$ O6\,Iaf   & $153.941600 \pm 0.009200$ & $0.382 \pm 0.016$ & $134.91 \pm   2.32$ & $1.29 \pm 0.04$ & $ 93.49 \pm   2.08$ & $120.44 \pm   2.69$ & $195.46 \pm 0.79$ & $262.71 \pm   6.14$ & $338.46 \pm   7.91$ & $ 69.38 \pm   4.10$ & $53.86 \pm 3.09$ \\ [2pt]
538 & ON9\,Ia: $+$ O7.5:\,I:(f):  & $4.159758 \pm 0.000023$ & $0.000$ & $270.00$ & $1.35 \pm 0.11$ & $161.62 \pm   9.46$ & $218.73 \pm  12.80$ & $210.30 \pm 0.02$ & $ 13.28 \pm   0.78$ & $ 17.97 \pm   1.05$ & $ 13.63 \pm   1.71$ & $10.07 \pm 1.16$ \\ [2pt]
543 &  O9\,IV $+$ O9.7:\,V  & $1.383989 \pm 0.000003$ & $0.000$ & $270.00$ & $1.13 \pm 0.04$ & $189.52 \pm   4.18$ & $213.58 \pm   4.71$ & $209.92 \pm 0.01$ & $  5.18 \pm   0.11$ & $  5.84 \pm   0.13$ & $  4.97 \pm   0.28$ & $4.41 \pm 0.25$ \\ [2pt]
555 &  O9.5\,Vz  & $66.099500 \pm 0.001500$ & $0.821 \pm 0.029$ & $ 98.46 \pm   4.49$ & $1.50 \pm 0.22$ & $ 78.69 \pm   8.19$ & $117.76 \pm  12.26$ & $158.34 \pm 0.09$ & $ 58.68 \pm   7.45$ & $ 87.82 \pm  11.15$ & $  5.80 \pm   2.15$ & $3.88 \pm 1.43$ \\ [2pt]
563 &  O9.7\,III: $+$B0:\,V:  & $1.217342 \pm 0.000006$ & $0.000$ & $270.00$ & $1.31 \pm 0.07$ & $153.71 \pm   5.53$ & $200.72 \pm   7.22$ & $210.15 \pm 0.01$ & $  3.70 \pm   0.13$ & $  4.83 \pm   0.17$ & $  3.18 \pm   0.24$ & $2.43 \pm 0.17$ \\ [2pt]
642 & O5:\,Vz: $+$ O8:\,Vz:   & $1.726824 \pm 0.000008$ & $0.000$ & $270.00$ & $1.55 \pm 0.02$ & $141.31 \pm   1.38$ & $219.08 \pm   2.14$ & $210.59 \pm 0.01$ &  $  4.82 \pm   0.05$ & $  7.47 \pm   0.07$ & $5.09 \pm 0.12$ & $  3.28 \pm   0.07$  \\ [2pt]
652 &  B2\,Ip $+$O9\,III:  & $8.589090 \pm 0.000150$ & $0.000$ & $270.00$ & $2.77 \pm 0.36$ & $ 71.87 \pm   6.51$ & $199.35 \pm  18.06$ & $205.34 \pm 0.05$ & $ 12.19 \pm   1.10$ & $ 33.82 \pm   3.06$ & $ 13.04 \pm   2.55$ & $4.70 \pm 0.53$ \\ [2pt]
661 &  O6.5\,V(n) $+$ O9.7:\,V:   & $1.266430 \pm 0.000004$ & $0.000$ & $270.00$ & $1.41 \pm 0.02$ & $266.02 \pm   3.51$ & $374.01 \pm   4.09$ & $209.41 \pm 0.01$ & $  6.65 \pm   0.09$ & $  9.35 \pm   0.12$ & $ 20.09 \pm   0.64$ & $14.29 \pm 0.43$ \\ [2pt]
771 &  O9.7\,III:(n)  & $29.842733 \pm 0.021269$ & $0.545 \pm 0.039$ & $346.43 \pm   4.25$ & $1.01 \pm 0.11$ & $ 70.49 \pm   5.38$ & $ 70.97 \pm   5.42$ & $211.61 \pm 0.27$ & $ 34.83 \pm   2.86$ & $ 35.07 \pm   2.88$ & $  2.59 \pm   0.60$ & $2.57 \pm 0.60$ \\ [2pt]
806 &  O5.5\,V((fc)):z $+$ O7\,Vz:  & $2.584883 \pm 0.000004$ & $0.000$ & $270.00$ & $1.26 \pm 0.04$ & $156.50 \pm   3.53$ & $197.68 \pm   4.46$ & $210.21 \pm 0.01$ & $  7.99 \pm   0.18$ & $ 10.09 \pm   0.23$ & $  6.64 \pm   0.30$ & $5.25 \pm 0.22$ \\ [2pt]
%\tablefootmark{b}\\
\hline
\end{tabular}
\tablefoot{$M_P/M_S$ corresponds to the mass ratio between the primary ($P$) and the secondary ($S$). $K$ is the semi-amplitude of the RV curve. HJD$_0$ refers to the time of the primary inferior conjunction for circular systems and to the periastron passage for the eccentric ones. $a \sin i$ and $m \sin^3 i$ are the projected semi-major axis and the minimum mass, respectively.}
\end{sidewaystable*}

Almost all objects in our sample have spectra affected by nebular contamination. The strengths of these emission lines vary from one system to another and from one epoch to another. To eliminate the nebular contamination from the disentangled spectra, we first consider these nebular features as a third component with a static RV but with a variable intensity.
We proceed by running the spectral disentangling code a first time, fixing the brightness factor ($l$) of the nebular emissions to 1.0 even though the strengths of these lines vary from epoch to epoch due to small variations in the sky and nebula sampled around each target, and fibre positions. The result consists of a mean spectrum of the nebular contamination. We then scale this to fit the observed nebular contamination at each epoch. We use this scaling value as brightness factor and run the spectral disentangling code again to obtain the final disentangled spectra of the primary and secondary components as well as for the nebular contamination. 

As mentioned above, the only parameter useful for the spectral disentangling that cannot be determined through spectroscopy is the brightness factor of each component. When photometry is available, we adopt the value of this parameter from the light-curve fit with the assumption that $l_P + l_S = 1$. Otherwise, if no light curve exists for the system, we need to estimate the brightness ratio in another way. Since the spectral disentangling yields the strengths of the lines in the primary and secondary stars relative to the combined spectrum, we estimate the brightness ratio by computing the equivalent widths (EWs) of the \ion{He}{i} and \ion{He}{ii} lines within the wavelength range of our spectra and by comparing these values to those calculated from synthetic spectra corresponding to stars with the same spectral types as the components of our systems. The ratio between the EWs of the disentangled spectra and the synthetic ones gives us the theoretical brightness factor for each component. The final condition before adopting these values is that $l_P + l_S = 1$ as previously mentioned.

%we scale the resulting spectra in a way that the strengths of the helium lines of each component fit a synthetic spectrum with appropriate effective temperature and surface gravity. The latter method is thus based on the determinations of the equivalent widths of the helium lines.

Once the resulting spectra, scaled with the brightness factors, are obtained, a new normalisation procedure is required. Indeed, during the spectral disentangling procedure and when one component is not totally eclipsed, an ambiguity exists in the determination of the continuum level. This ambiguity creates low-frequency oscillations in the continuum of the disentangled spectrum that need to be removed before analysing the spectra. Further discussions about the spectral disentangling can be found in \citet{pavlovski10} and in \citet{pavlovski12}. We stress that the uncertainties that could arise from the normalisation procedure are not taken into account in the global uncertainties on the presented properties characterising the stars. 

\subsection{Atmosphere modelling}
\label{method:atmosphere}

We use the CMFGEN atmosphere code \citep{hillier98} to determine the fundamental parameters of all the components in our sample. CMFGEN computes non-LTE (Local Thermodynamic Equilibrium) models of massive star atmospheres, including winds and line-blanketing. The hydrodynamical structure is prescribed and consists of an essentially hydrostatic photosphere, where the density structure dictates the velocity structure, and a wind where an adopted velocity structure dictates the density structure.  The trans-sonic velocity profile is assumed to be a $\beta$-velocity law; density and velocity are related through the mass-continuity equation. The level populations are calculated through the rate equations. A super-level approach is used to reduce the size of the problem (and thus the computing time). To ensure good modelling of our observed spectra, we include the following elements in our calculations: H, He, C, N, O, Ne, Mg, Al, Si, S, Ar, Ca, Fe, and Ni. The LMC baseline abundances were estimated from observations of both \ion{H}{ii} regions \citep{kurtDufour98,garnett99} and early-type stars \citep[e.g.,][]{hunter07}. We use the baseline values summarised by \citet[][and references therein]{brott11} for H, He, C, N, O, Mg, Ne, Al, Si, and Fe. However, these values were not given for S, Ar, Ca, and Ni. We therefore extrapolate from the solar abundances provided by \citet{grevesse10} to reach the LMC metallicity of $0.5$\,Z$_{\odot}$. 

Once the atmospheric structure is obtained, a formal solution of the radiative transfer equation is performed. A depth-dependent microturbulent velocity is assumed to compute the emergent spectrum. This microturbulence varies from 10\,\kms\ at the photosphere to 10\% of the terminal wind velocity at the top of the atmosphere. 

CMFGEN assumes spherical symmetry and since we are dealing with binaries, the shape of the components can deviate from sphericity because of rapid rotation and tidal effects. However, these effects are expected to be significant only for contact, or close-to-contact systems even though a detailed quantification of these effects is still lacking. Dedicated tools are being developed (e.g. \citealt{palate12} and Abdul-Masih et al. in prep.) to improve the study of these (over-)contact systems. For the current study, the uncertainties that can result from this drop shape are not taken into account.

As mentioned in Sect.\,\ref{obs:spec}, the TMBM LR02 dataset only covers the 3964--4567\AA\ range. As we detail below, this range provides a sufficient list of diagnostic lines to constrain stellar parameters (effective temperature, surface gravity, projected rotational velocity) and chemical abundances of He, C and N, but lacks diagnostics for stellar winds and O surface abundance. We therefore keep the oxygen surface abundance fixed to the LMC baseline value. To constrain the stellar properties and the surface abundances of all the components in our sample, we therefore perform the spectroscopic analysis as follows (Sect.\,\ref{subsubsec:vsini} - \,\ref{subsubsect:Spclass}).

\subsubsection{Projected rotational velocities and macroturbulence}
\label{subsubsec:vsini}
Projected rotational velocities and macroturbulence velocities are determined in an independent way by applying the {\it iacob\_broad} package from \citet{simondiaz14} on the \ion{He}{i}~4388, \ion{He}{i}~4471 and \ion{He}{ii}~4542 spectral lines. Before validating the two parameters, we compare the different values provided by the Fourier transform method and the goodness-of-fit method and check the agreement. We use as final values for the projected rotational velocities the average values between the Fourier transform and the goodness-of-fit methods. These parameters are then kept fixed in the atmosphere analysis. We moreover emphasise that, given the signal-to-noise of our data and the fact that our analysis relies on the He lines only, the macroturbulence parameter is approximate \citep[see e.g.][]{ramirez15}.

\subsubsection{Effective temperature (\teff) and surface gravity (\logg)}
\label{subsubsec:teff}
To constrain the effective temperature and the surface gravity of our components, we build a grid of synthetic spectra computed with CMFGEN by varying the effective temperature in steps of $\Delta \teff =1000$\,K and the surface gravity in steps of $\Delta \logg = 0.1$ (cgs). Our grid covers $27000\,\mathrm{K} < \teff < 47000 $\,K and $3.1 < \logg < 4.4$. Luminosities are assigned according to \citet{brott11} evolutionary tracks from the combination $(\teff,\logg)$ by assuming an initial rotational velocity of 150\,\kms. For these models, we use for the mass-loss rate prescriptions of \citet{vink00, vink01} for LMC metallicity. As emphasised by \citet{ramirez17}, the mass-loss predictions of \citet{vink01} for LMC metallicity are reasonable proxies assuming a certain wind volume-filling factor. The terminal wind velocities are estimated to be 2.6 times the effective escape velocity from the photosphere ($v_{\rm esc}$, \citealt{lamers95}). Even though this relation is uncertain with a 30\% scatter in the observed values, it is in reasonable concordance with empirically determined terminal velocities of O-stars \citep{prinja91,lamers95}. The exponent $\beta$ of the velocity law was set to $1.0$ \citep[see e.g.][for more details]{repolust04, bouret12} and the clumping filling factor was adopted as $f_{\mathrm{cl}} = 0.1$. For each object, a subset of the model grid was convolved with rotational profiles and radial-tangential profiles corresponding to the measured projected rotational velocities $v \sin i$ and macroturbulence $v_{\rm mac}$ (see Sect\,\ref{subsubsec:vsini}). A third convolution was finally done to take the instrumental broadening into account. The resulting synthetic spectra are then shifted in RV and compared to the disentangled spectrum of each object. 

The effective temperature was constrained from the ionisation balance between the \ion{He}{i} and \ion{He}{ii} lines. We used the \ion{He}{i+ii}~4026, \ion{He}{i}~4143, \ion{He}{i}~4388, \ion{He}{i}~4471, \ion{He}{ii}~4200, \ion{He}{ii}~4542 lines. However, for several components, mostly B-type stars, the effective temperatures are too low to produce \ion{He}{ii} lines. For those objects, we consider the ionisation balance between \ion{Si}{ii},\ion{Si}{iii}, and \ion{Si}{iv} and between the \ion{He}{i}~4471 and \ion{Mg}{ii}~4481 lines as secondary diagnostics. 

The surface gravity was obtained from the wings of the H$\delta$, and H$\gamma$ lines. Because of the low-frequency oscillations resulting from the spectral disentangling code, the normalisation of the Balmer lines was challenging for some objects, potentially introducing systematic errors that are difficult to quantify. \teff\ and \logg\ were derived simultaneously from the grid of synthetic spectra. The quality of the fit was quantified by means of a $\chi^2$ analysis on the H and He lines (mainly sensitive to the surface gravity and effective temperature, respectively). The $\chi^2$ is computed for each model of the grid and linearly interpolated between the grid points in steps of $\Delta \teff =100$\,K and $\Delta \logg=0.01$ (cgs) to produce the $\chi^2$ maps given in the Appendix for each component. The error bars in \teff\ and \logg\ are correlated. The uncertainties at 1-, 2-, and 3-$\sigma$ on \teff\ and \logg\ were estimated from $\Delta \chi^2 = 2.30, 6.18,$ and $11.83$ (two degrees of freedom), respectively \citep{press07}. All systems were treated in this way, except for VFTS\,527; the strong stellar winds observed in this system affect the shapes of the hydrogen and helium lines (some are observed as P-Cygni profiles). The line profiles for the two components require a dedicated modelling, for which the quality is assessed by eye. 

\subsubsection{Luminosity}
The luminosity of the components was estimated in two different ways. 
For the systems that have an OGLE light curve, we determine the luminosities of each component from their radius and their effective temperature (Paper IV). For the other systems, we proceed iteratively:
\begin{enumerate}
\item We first compute the absolute $V$ magnitude using the apparent $V$ magnitude, the extinction ($R_{5495} \times E(4405-5495)$, with $R_{5495} = 4.5$, see \citealt{maiz14}), and the distance modulus of the LMC (DM $= 18.495$; \citealt{pietrzynski13}). We then compute the bolometric magnitudes by correcting the absolute $V$ magnitude for the bolometric correction computed from the effective temperature and Eq. (3) of \citet{martins06}. A first estimation is derived from the bolometric magnitude. 
\item We use this value as input to compute new CMFGEN models for the primary and the secondary, with their \teff\ and \logg\ obtained from the grid of CMFGEN synthetic spectra (see Sect.\,\ref{subsubsec:teff}). 
\item From the new models, we fit the F275W, F336W, F555W, F658N, F775U, F110W and F160W fluxes \citep{sabbi16} and the UBVJHK fluxes to constrain the extinction (listed in Table\,\ref{tab:magnitude}). We proceed by combining the primary and secondary best-fit models scaled by their respective brightness factor (in the B band\footnote{The K band was used in previous VFTS analyses \citep{bestenlehner14,ramirez17} because an average extinction value was assumed for all the objects and because the reddening in this region is less than in the B and V bands. Here, we need the flux ratio in the B- and V-bands to properly scale the disentangled spectra in their wavelength domain.}). The reddening law is taken from \citet{maiz14} with $R_{5495} = 4.5$.
\item Finally, the revised and combined models are again compared to the Spectral Energy Distribution (SED) of the system (see figures in the Appendix). The extinction as well as the magnitudes in the different bands are provided in Table\,\ref{tab:magnitude}. The whole process is iteratively repeated until satisfying the fit of the SED of the systems and the individual disentangled spectra. 
\end{enumerate} 

\subsubsection{Surface abundances}

To help constrain the effects of binary interaction, it is essential to determine the surface abundances of the components. Once the fundamental parameters are constrained, we apply the same method as described in \citet{martins15}. We run several models changing the He, C and N abundances. Since the wavelength range of our spectra is limited, so is the number of diagnostic lines. Here we provide the list of lines that we use:
  \begin{itemize}
  \item {\it helium:} \ion{He}{i+ii}~4026, \ion{He}{i}~4143, \ion{He}{ii}~4200, \ion{He}{i}~4388, \ion{He}{i}~4471, and \ion{He}{ii}~4542,
  \item {\it carbon:} \ion{C}{iii}~4068--70,
  \item {\it nitrogen:} \ion{N}{ii}~4035, \ion{N}{ii}~4041, \ion{N}{iv}~4058, \ion{N}{iii}~4097, \ion{N}{iii}~4197, \ion{N}{iii}~4379, \ion{N}{iii}~4511--15--18.
  \end{itemize}

Because the helium lines are used to scale the disentangled spectra, this creates a bias in the determination of the helium surface abundances, and therefore, this parameter will not be discussed further. 

\subsubsection{Spectral classification}
\label{subsubsect:Spclass}
Finally, we revise the spectral classifications of all the components of our sample from the disentangled spectra (the previous estimations from \citealt{walborn14} are reiterated in Table\,\ref{tab:orbit} whilst the new ones are provided in Table\,\ref{tab:parameter}). For this purpose, we use the classification criteria of \citet{conti71}, \citet{conti73}, and \citet{mathys88,mathys89} as well as more recent criteria from Sana et al. (in prep.). These criteria are based on the measurements of the equivalent widths of specific spectral lines. We emphasise that the luminosity classes were only given for stars with spectral types later than O7. These criteria indeed do not apply to stars with earlier spectral types. We are also not able to use the spectral classification criteria provided by \citet[][and subsequent papers]{walborn71} because we do not have access to the \ion{N}{iii}\,4634--41 and \ion{He}{ii}\,4686 lines in the disentangled spectra. 

All the individual parameters, spectral classifications and $1-\sigma$ uncertainties are provided in Table\,\ref{tab:parameter}. We reiterate that in the present paper we adopt the primary star to be the currently most massive star of the system. This definition does not take the evolution of the components into account (e.g. possible mass-transfer episodes). Moreover, this definition is also different from that given by \citet{walborn14}, which focused on the brightness of the components.

%%###################################################################
%%-------------------------------   Discussion  --------------------- 

\section{Discussion}
\label{discussion}

\begin{figure*}[t!]
  \centering
    \includegraphics[trim=15 0 35 40,clip,width=16cm]{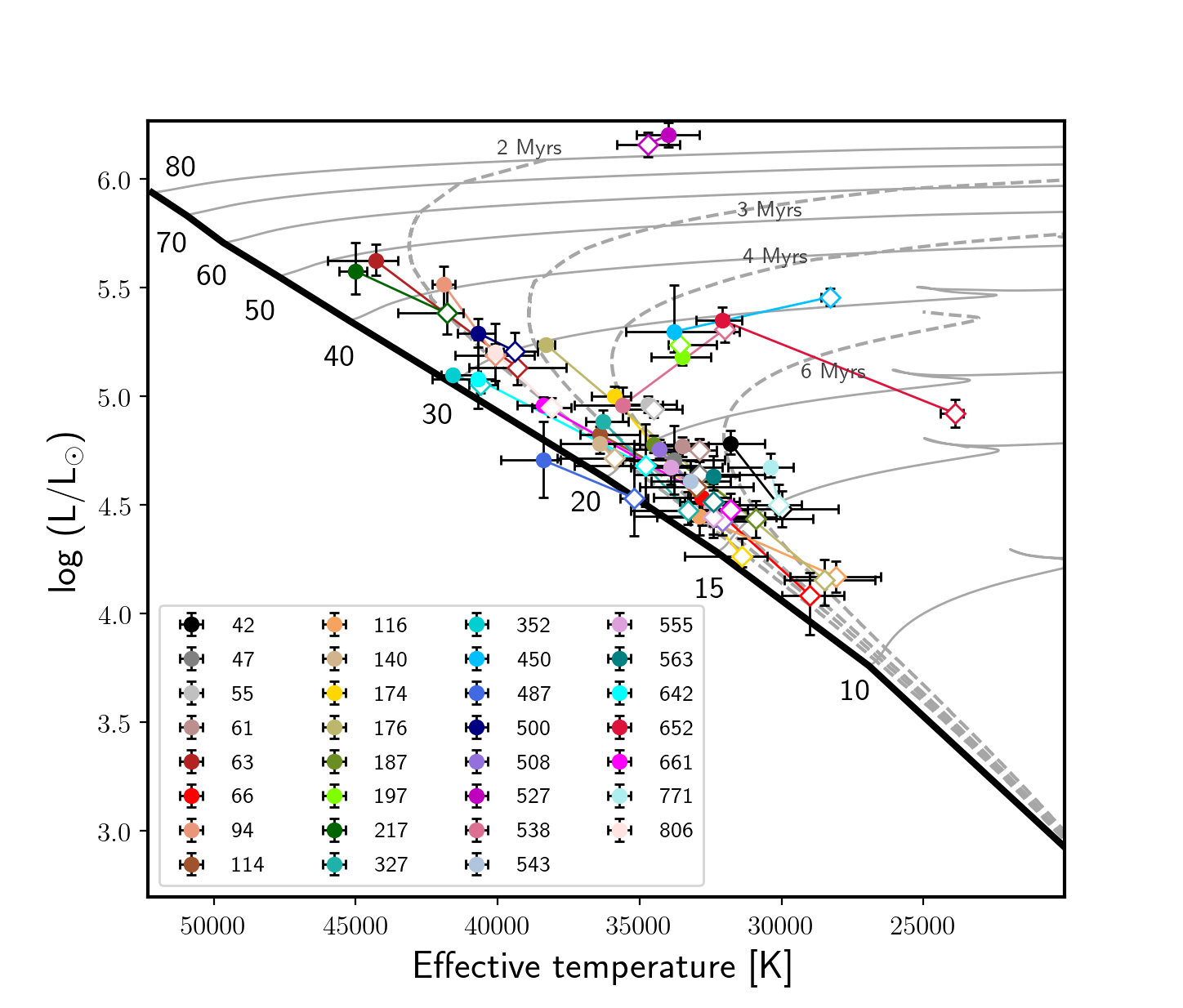}
    \caption{Hertzsprung-Russell diagram. The tracks and the isochrones are from \citet{brott11}, computed with an initial rotational velocity of 150\,\kms. Filled (open) circles (diamonds) refer to the primary (secondary) of each binary system. The names of the systems are color-coded in the legend of the figure. This same color-code is used for the different figures of the paper.}\label{fig:HR} 
\end{figure*}

\begin{figure*}[t!]
  \centering
    \includegraphics[trim=0 0 20 10,clip,width=16cm]{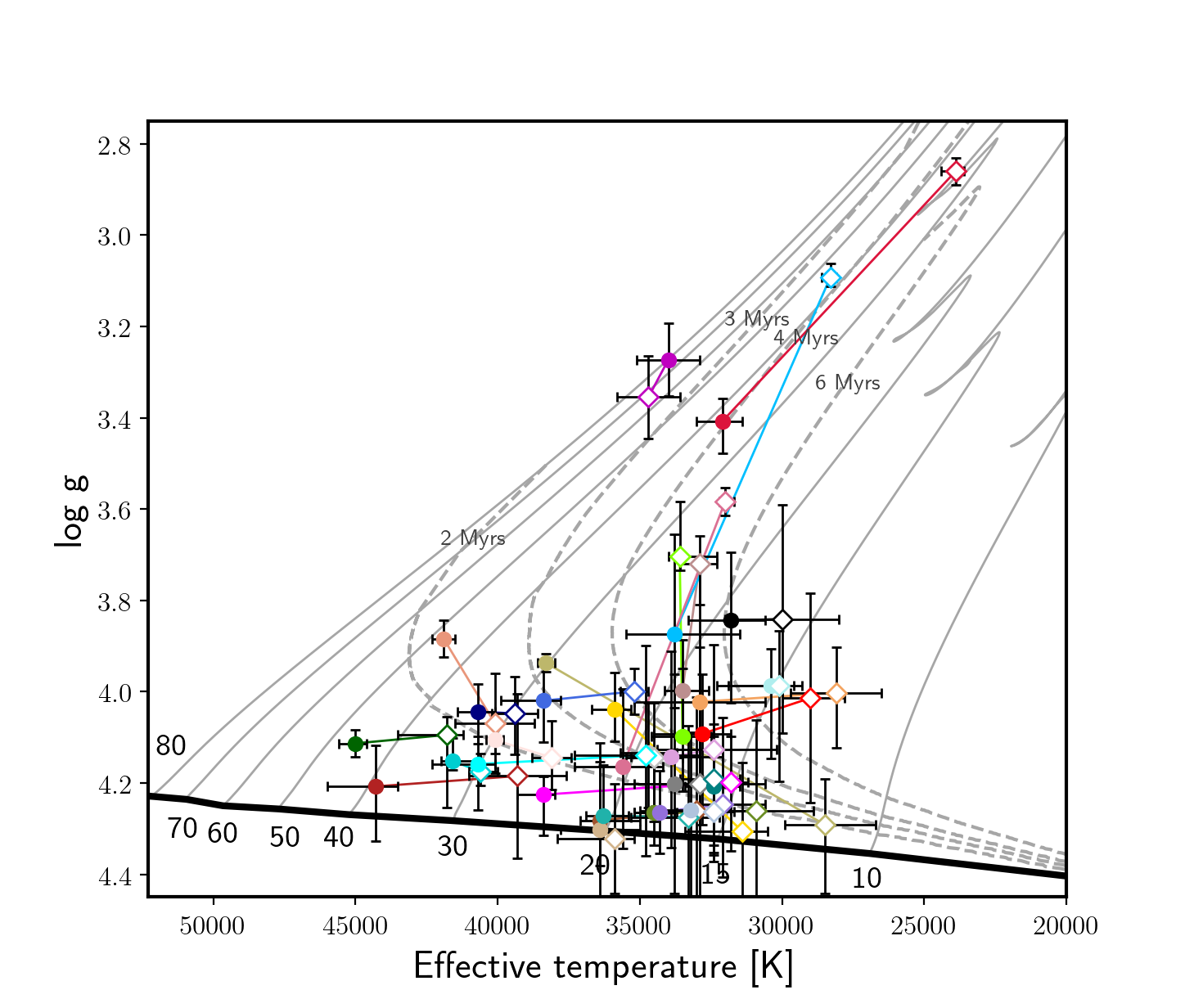}
    \caption{Surface gravity as a function of effective temperature for the sample stars. Evolutionary tracks are from \citet{brott11} computed with an initial rotational velocity of 150\,\kms. The colour code is the same as Fig.\,\ref{fig:HR}.}\label{fig:teff-logg} 
\end{figure*}

From the luminosities, effective temperatures, and surface gravities listed in Table\,\ref{tab:parameter} and computed as described in Sect.\,\ref{method:atmosphere}, we place all components in a Hertzsprung-Russell diagram (HRD, Fig.\,\ref{fig:HR}) and in the Kiel (\logg--\teff) diagram (Fig.\,\ref{fig:teff-logg}). 

We also use the luminosities, effective temperatures, surface gravities, and projected rotational velocities that we determine as inputs of the BONNSAI code (BONN Stellar Astrophysics Interface, \citealt{schneider14} and \citealt{schneider17}). BONNSAI is a Bayesian tool that allows us to compare the properties of the stars with single-star evolutionary models. The ages and the evolutionary masses provided by BONNSAI, are also given in Table\,\ref{tab:parameter} with their $1-\sigma$ uncertainties. The stars of our sample populate the areas corresponding to ages between 0 and 9 Myrs and to evolutionary masses ranging from 10 to 90\,\msun. Individual HRDs are displayed in the Appendix for clarity. The evolutionary tracks in the Appendix are shown for models that are computed with an initial rotational velocity that matches the currently observed $\vsini$ most closely.

By looking at the geometry of the systems and the physical parameters of their components but ignoring the photometry, we classify our systems in five different subsamples:

\begin{enumerate}
\item {\bf Long-period ($P > 20$ days) non-interacting systems:} both components are well detached and move in a long-period orbit. They are expected to evolve through the main sequence as single stars and interact later, when the most massive star leaves the main sequence through Case B mass transfer. We identify VFTS\,042, VFTS\,063, VFTS\,114, VFTS\,116, VFTS\,197, VFTS\,508, VFTS\,527, VFTS\,555, and VFTS\,771 as belonging to this subsample. 
\item {\bf Eccentric short-period ($P < 10$ days) systems:} the components are well within their Roche lobes, and no sign of interaction between the components is observed yet. This subsample contains  VFTS\,047, VFTS\,055, VFTS\,140, VFTS\,174, VFTS\,327, and VFTS\,487. 
\item {\bf Circular short-period ($P < 10$ days) systems:} the components are affected by tides and they have large Roche lobe filling factors.  VFTS\,094, VFTS\,187, VFTS\,217, VFTS\,500, VFTS\,543, VFTS\,563, VFTS\,642, VFTS\,661 and VFTS\,806 belong to this subsample. 
\item {\bf Identified contact systems:} the two components fill their Roche lobe well. We find  VFTS\,066, and VFTS\,352 as belonging to this subsample. 
\item {\bf Semi-detached systems:} these systems show clear indication of recent or ongoing Case A mass transfer (the less massive stars fill their Roche lobe and are in synchronous rotation, while the more massive stars do not fill their Roche lobe, and show super-synchronous rotation as is expected for stars that have accreted mass and angular momentum). Five systems are in this subsample VFTS\,061, VFTS\,176, VFTS\,450, VFTS\,538, and VFTS\,652.
\end{enumerate}

\begin{figure}[t!]
  \centering
    \includegraphics[trim=30 0 30 30,clip,width=9cm]{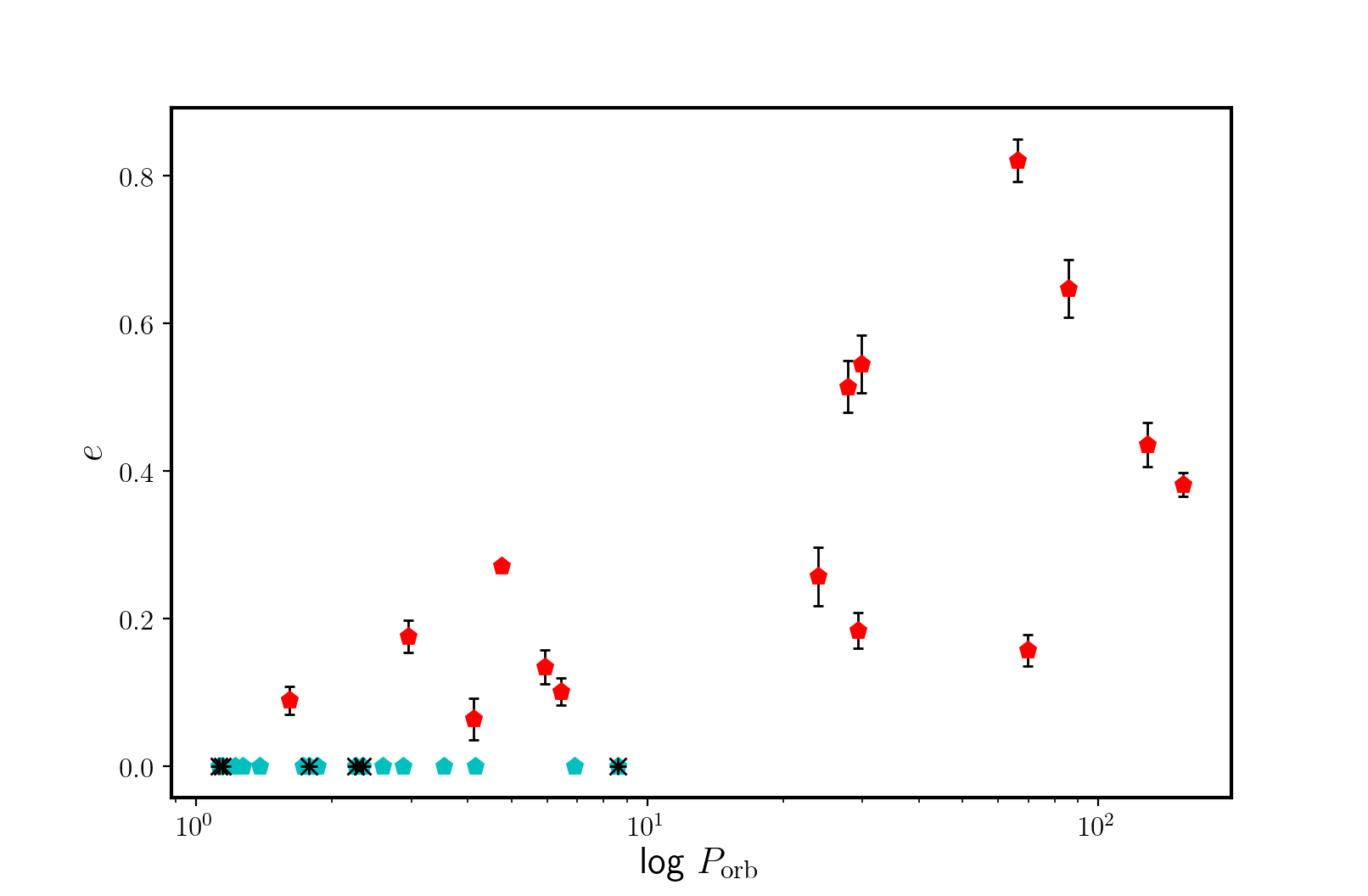}
    \caption{Period-eccentricity diagram for all the SB2 in our sample. The blue dots represent the systems displaying eclipses or ellipsoidal variations and the star symbols mark the systems that have one component filling its Roche lobe.}\label{fig:period-ecc} 
\end{figure}

In Fig.\,\ref{fig:period-ecc}, we show the period-eccentricity diagram for all the SB2 in our sample. The blue dots represent the systems displaying eclipses or ellipsoidal variations and the star symbols mark the systems that have one component filling its Roche lobe. From this figure, we see that all the systems that show mass transfer have an eccentricity of zero and a period shorter than 10 days. There is also a complete absence of systems with low eccentricities and long periods in our sample. 

From the individual properties of the components, we have between 15 and 24 systems that are potentially in a pre-interaction phase (subsamples 1, 2, and possibly 3), which is between 48 and 77\% of our sample. Five systems are in a semi-detached configuration, making up 16\% of the sample, and between 5 and 35\% of our systems are potentially in contact (subsamples 3 and 4). 

Adopting the most recent binary parameter statistics (from Sana et al. 2012) and assuming continuous star formation for a Salpeter initial mass function, \citet{demink14} showed that those systems that show radial velocity variations $\Delta RV > 20$\,km\,s$^{-1}$ are dominated by pre-interaction binary systems (82\,\%). Semi-detached systems make up 7\,\% and post mass-transfer systems, where the secondary is an O star, constitute the final 11\,\%. Given that all the binary systems in our sample have limited periods ($P< 400$\,days), and mass ratios ($M_P/M_S$) between 1 and 3, we argue that the percentages in the different categories mentioned through both studies are fairly consistent. For all systems in our sample with clear evidence for strong interaction (Roche lobe filling), both components are undergoing core hydrogen burning, which is consistent with the expectation from simulated populations based on binary evolution models \citep{demink14}. Our numbers provide only a lower limit on the total fraction of post-interaction binaries, which is expected to grow when considering also the SB1 systems in the TMBM. Amongst those, there could be binaries with evolved components, including compact objects.
In the following, we discuss the data and the trends for objects in these different categories in the context of different physical processes.

\subsection{Detached systems}
\subsubsection{Rotational mixing}

Rotation is a key ingredient that has been suggested to impact the fate of the stars, along with the initial mass and metallicity. It deforms the star to an oblate shape and induces instabilities in the internal layers, producing turbulent mixing. The rotational induced mixing can bring fresh material processed in the core to the surface of the star, with the consequence of higher observed surface abundances, as it is the case for nitrogen or helium. The faster the star rotates, the higher its nitrogen and helium surface abundances should be \citep{maeder00}. Studies of the carbon, nitrogen and oxygen abundances for OB stars in the Galaxy and the Magellanic Clouds often show a correlation between these surface abundances and projected rotational velocities \citep{hunter08, hunter09, mahy15, martins15, grin17}. Even though a general trend is present between the projected rotational velocities and the nitrogen content, two groups remain unexplained. An unexpectedly large number of objects in these historical data were indeed found to be either slow rotators and highly nitrogen enriched, or fast rotators with relatively non-enriched material at their surface. The origins of these two groups are still a matter of debate but they might be products of binary interactions (tides, mass transfer, etc.; see \citealt{demink09,demink13,brott11b}). Looking at the nitrogen surface abundances of the components in detached systems that have a priori not undergone previous binary interactions must in principle help to distinguish the role of binarity in the nitrogen-abnormal stars. 

For our discussion, we decided to focus only on the nitrogen and to discard the helium and carbon contents. This decision was made based on the following: (1) the helium lines were used for 18 systems to correct the disentangled spectra for the brightness ratio in order that they match with the synthetic spectra; (2) the helium abundances can be significantly affected by the microturbulence velocity that we adopt to compute the synthetic emergent spectra \citep{mcerlean98}; and (3) the carbon surface abundance was only determined from the \ion{C}{iii}~4068--70 doublet. The latter might not be representative of the global carbon content of the stars \citep[see][for more details]{martins15,cazorla17a,cazorla17b}. The nitrogen abundance is supposed to gradually increase when the star converts its carbon via the CN-cycle, while the oxygen remains approximately constant. Later during its evolution, the star converts the oxygen into nitrogen via the CNO-cycle, producing more nitrogen. 

%\begin{figure}[t!]
%  \centering
%    \includegraphics[trim=10 0 35 40,clip,width=9cm]{Hunter_group1-2.png}
%    \caption{Projected rotational velocity versus the nitrogen content of the stars in detached systems (i.e., subsamples 1 and 2: see Sect.\,\ref{discussion}). The circles represent the primaries whilst the triangles represent the secondaries of our sample. As a comparison, we also plot in grey the population of single giant and supergiant O-type stars in the 30 Doradus region \citep{grin17}. Tracks are from \citet{brott11}. In the top panel, they correspond to an initial mass of 25\,\msun. The color-bar represents the spectroscopic masses of the components.}\label{fig:rotation} 
%\end{figure}

\begin{figure*}
\subfloat[]{\includegraphics[trim=10 0 35 40,clip,width=9cm]{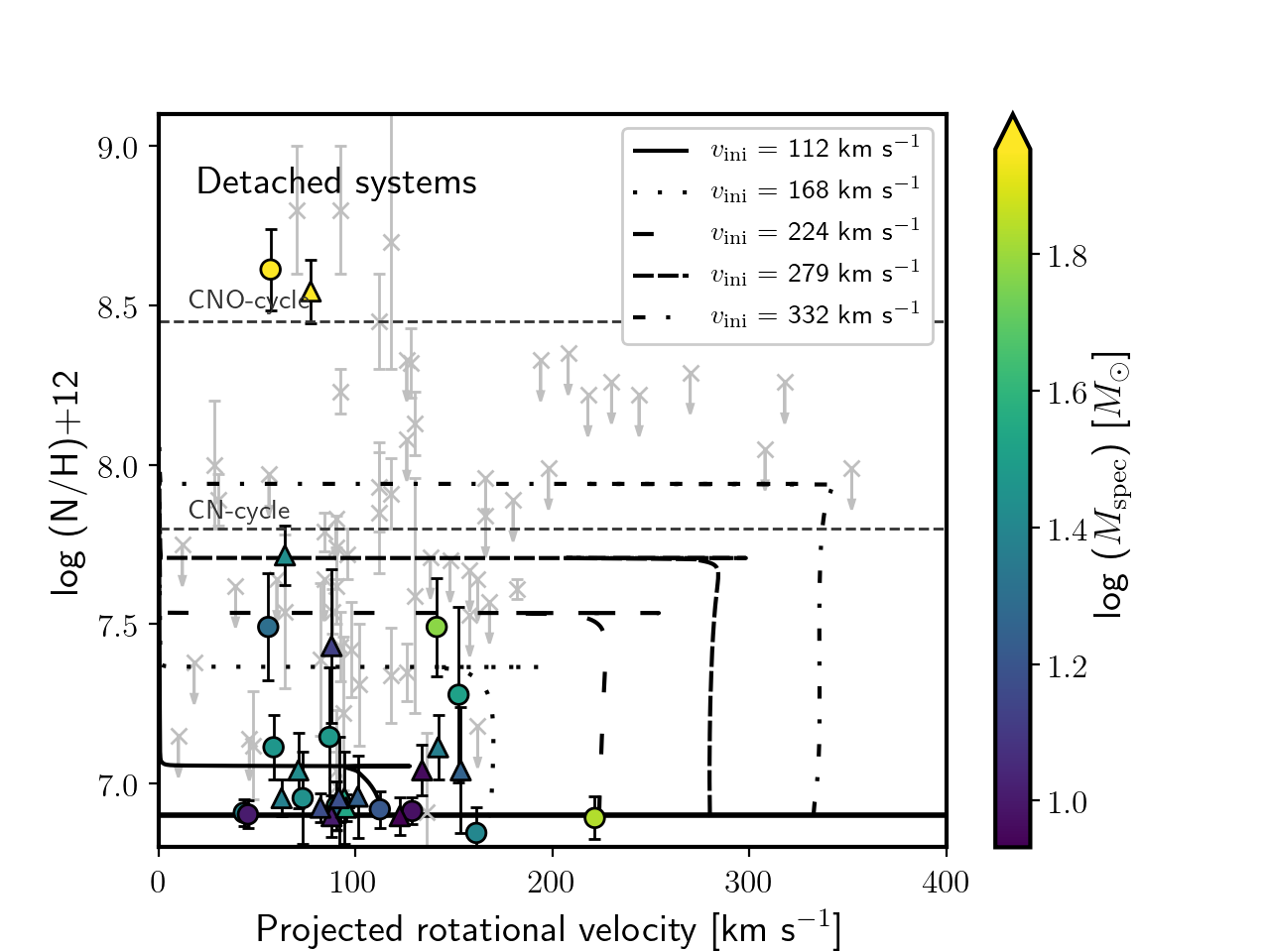}}
\subfloat[]{\includegraphics[trim=10 0 35 40,clip,width=9cm]{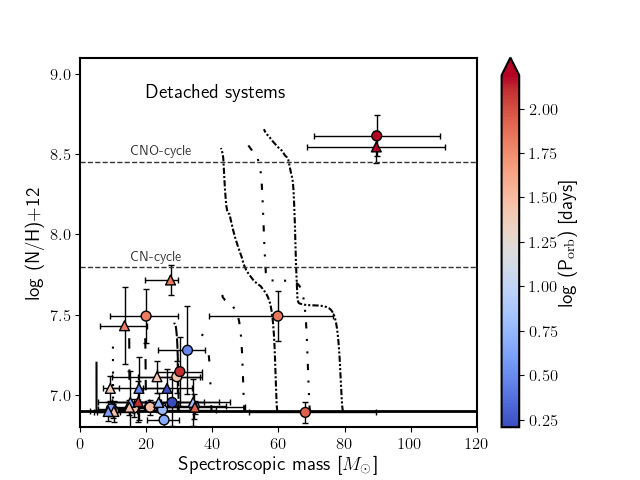}}\\
\subfloat[]{\includegraphics[trim=10 0 35 40,clip,width=9cm]{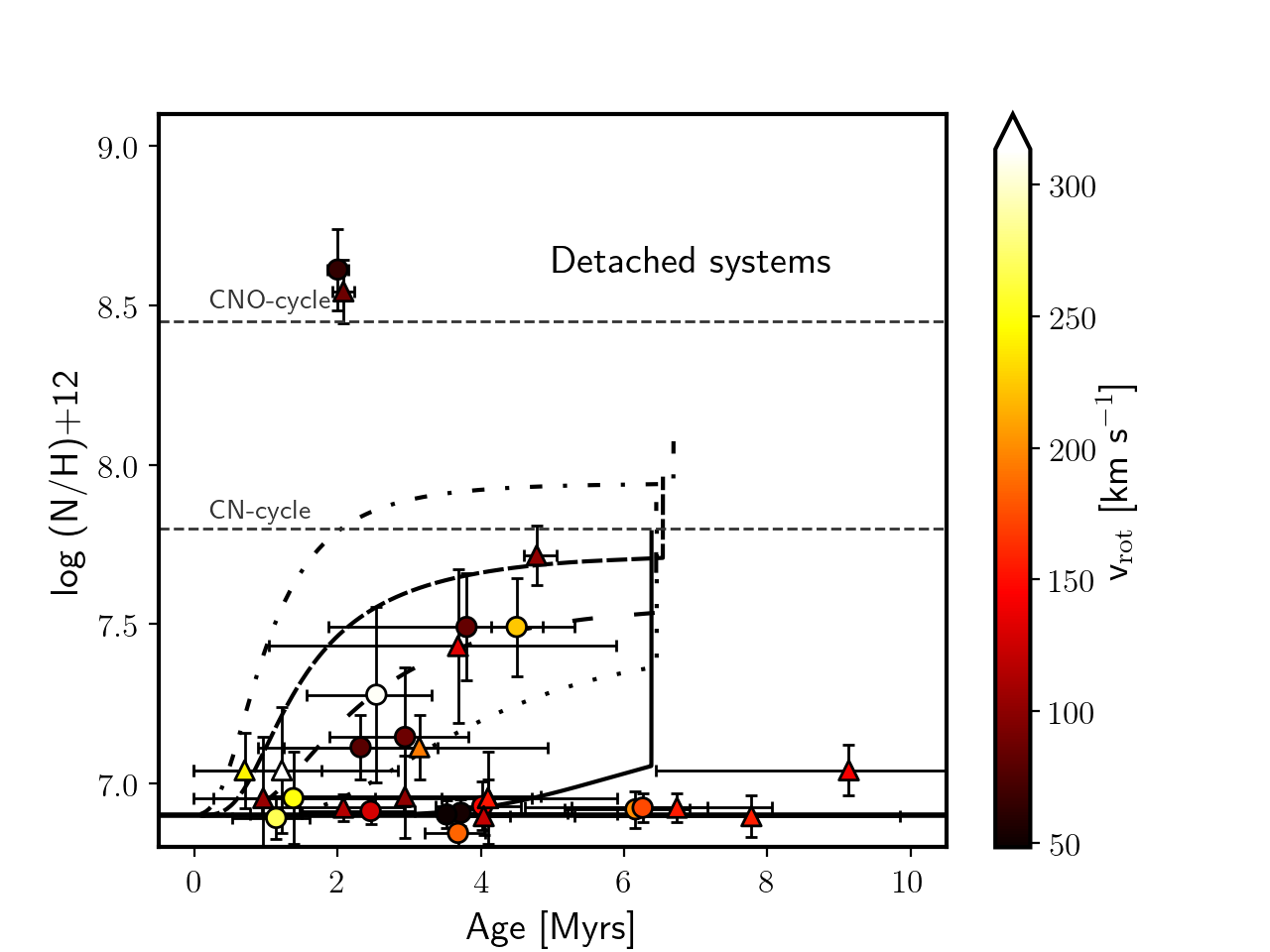}}
\subfloat[]{\includegraphics[trim=10 0 35 40,clip,width=9cm]{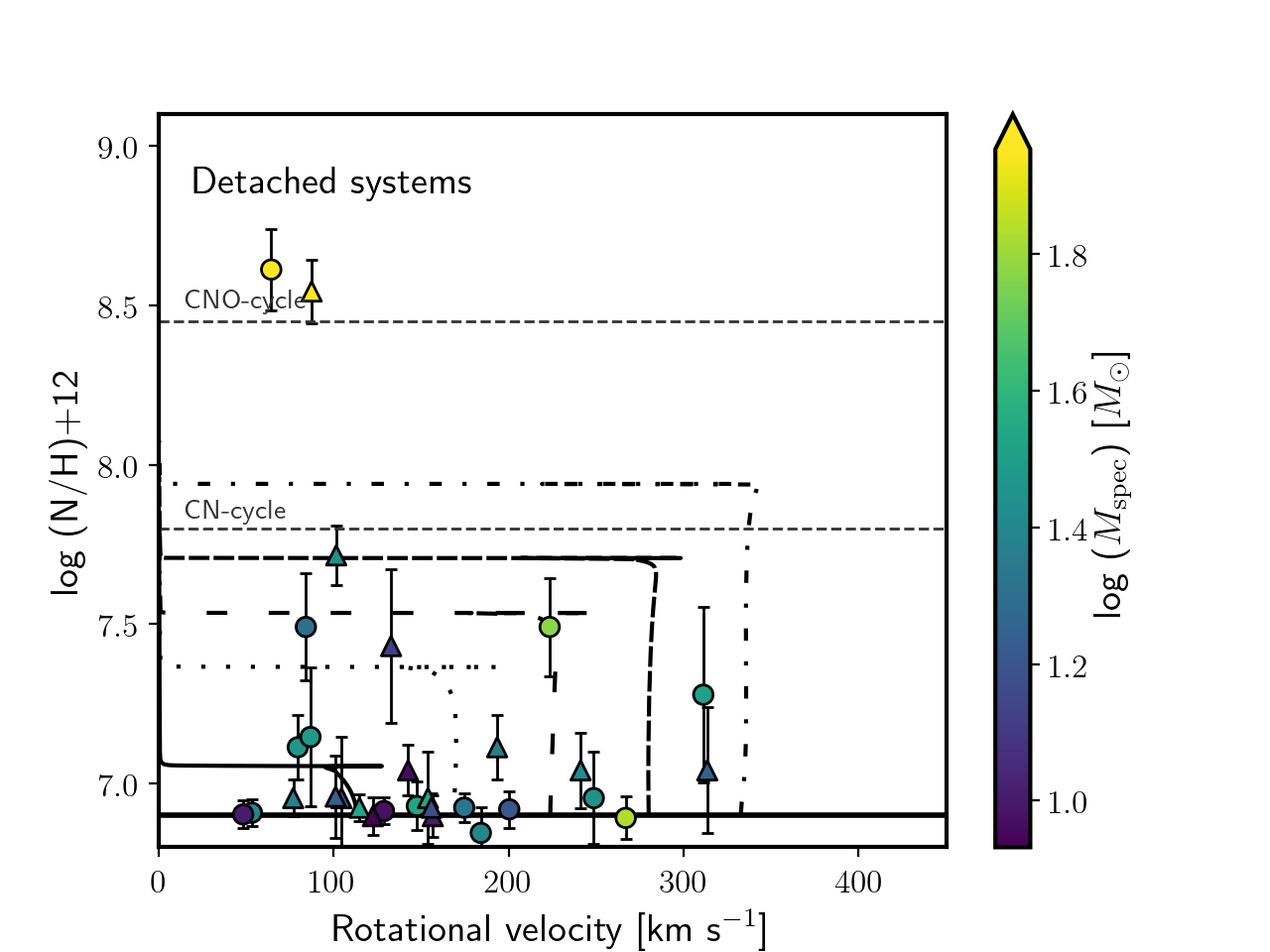}}
\caption{a) Projected rotational velocity vs. the nitrogen content of the stars in detached systems (i.e., subsamples 1 and 2: see Sect.\,\ref{discussion}). The circles represent the primaries whilst the triangles represent the secondaries of our sample. As a comparison, we also plot in grey the population of single giant and supergiant O-type stars in the 30 Doradus region \citep{grin17}. Tracks are from \citet{brott11}; in the top panel, they correspond to an initial mass of 25\,\msun. The colour-bar represents the spectroscopic masses of the components. b) Spectroscopic mass vs. the nitrogen content of the stars in detached systems (i.e. subsamples 1 and 2: see Sect.\,\ref{discussion}). The colour-bar represents the orbital period of the systems. c) Age vs. the nitrogen content of the stars in detached systems (i.e. subsamples 1 and 2: see Sect.\,\ref{discussion}). The colour-bar represents the initial rotational velocity of the stars. d) Rotational velocity vs. nitrogen content of the stars in detached systems (i.e. groups 1 and 2). The colour-bar represents the spectroscopic masses of the components.}
\label{fig:hunter_detached}
\end{figure*}

Figure\,\ref{fig:hunter_detached}a displays the so-called Hunter diagram, which is the projected rotational velocity versus the nitrogen content of the stars. We compare the nitrogen content in short- and long-period systems that do not exhibit any signs of interaction between both components (subsamples 1 and 2) in order to detect the influence of the rotation on the mixing. For systems with long periods, the effects of the tides are more limited and are possibly expected at periastron. We compare these values to those determined by \citet{grin17} on the population of single giant and supergiant O-type stars in the 30 Doradus region (the massive dwarfs not having been analysed yet in a dedicated study). 

In our sample, the comparison between projected rotational velocity and nitrogen surface abundance shows that most stars with $50 < \vsini < 230$\,\kms\ are barely enriched (Fig.\,\ref{fig:hunter_detached}a). However, a distinct group is observed with a higher nitrogen enrichment and a slow projected rotational velocity. These six stars are members of long-period systems: VFTS\,197, VFTS\,555 ($\epsilon_{\rm N} = \log(N/H)+12 \sim 7.5$ in Fig.\,\ref{fig:hunter_detached}a), and VFTS\,527 ($\epsilon_{\rm N} > 8.5$ in Fig.\,\ref{fig:hunter_detached}a) with $65 < P_{\mathrm{orb}} < 160$\,days and eccentricities of 0.11, 0.83, and 0.46, respectively. Tidal effects should play a role in these enrichments, especially at periastron, but their magnitude is expected to be relatively small given the orbital separation between the two components. We indeed compute projected separations of 192, 513 and 48\,\rsun\ for VFTS\,197, VFTS\,527, and VFTS\,555, respectively. 

 The comparison between our sample of detached SB2s and the sample of \citet{grin17} shows a general agreement in the sense that no non-enriched rapid rotator is found and the nitrogen surface enrichment of the stars remains in similar ranges (Fig.\,\ref{fig:hunter_detached}a). We argue that the inclusion of dwarf O-type stars from the Tarantula region will probably strengthen this assessment.
 
 Rotational mixing is also mass dependent. When we compare the spectroscopic masses to the nitrogen surface abundances (Fig.\,\ref{fig:hunter_detached}a and Fig.\,\ref{fig:hunter_detached}b), the majority of the components have a surface enrichment that is compatible with single-star evolutionary tracks. However, this enrichment is clearly too high compared to the error bars for the secondary star of VFTS\,197 and for the two components of VFTS\,555 while it explains the locations of the two objects in VFTS\,527. 
 
 If we compare the evolutionary ages of the components to the observed nitrogen surface abundances (Fig.\,\ref{fig:hunter_detached}c), the discrepancy observed for VFTS\,197, VFTS\,555, and VFTS\,527 remains valid only for the latter system. Therefore, we conclude that the rotational mixing in well-detached binary systems is comparable to that observed for single stars. Both components of VFTS\,527 (at $\epsilon_{\rm N} > 8.5$ in Fig.\,\ref{fig:hunter_detached}a) show the signature of strong stellar winds and these winds are probably responsible for the ejection of a part of the outer layers of the stars, displaying more enriched material.
 
  When we take the inclination (either spectroscopic, estimated from the ratios between minimum and spectroscopic masses, or photometric, computed from the light curves; see Appendix\,\ref{results} for further details) into account, the situation shows more objects with rotational velocities higher than 200\,\kms\ and with almost no enrichment (Fig.\,\ref{fig:hunter_detached}d). Although almost all these systems have orbital periods shorter than 10 days, the general picture seems however consistent with the models. Most of the stars indeed appear to be very young, meaning that the nitrogen has not yet reached the surface.

\subsubsection{Tidal mixing}

In close binaries, both fast rotation and tidal forces due to the proximity of the two components are expected to enhance the internal mixing processes, at least according to contemporary models. The tides increase or decrease the rotational periods of the stars until they synchronise with the orbital period; they may be an important additional effect on the internal and chemical structure of the stars, leading to an increase of the effects of the rotational mixing, allowing the stars in binary systems to be even more enriched than single stars with similar stellar parameters. 

\begin{figure}[t!]
  \centering
    \includegraphics[trim=10 0 35 40,clip,width=9cm]{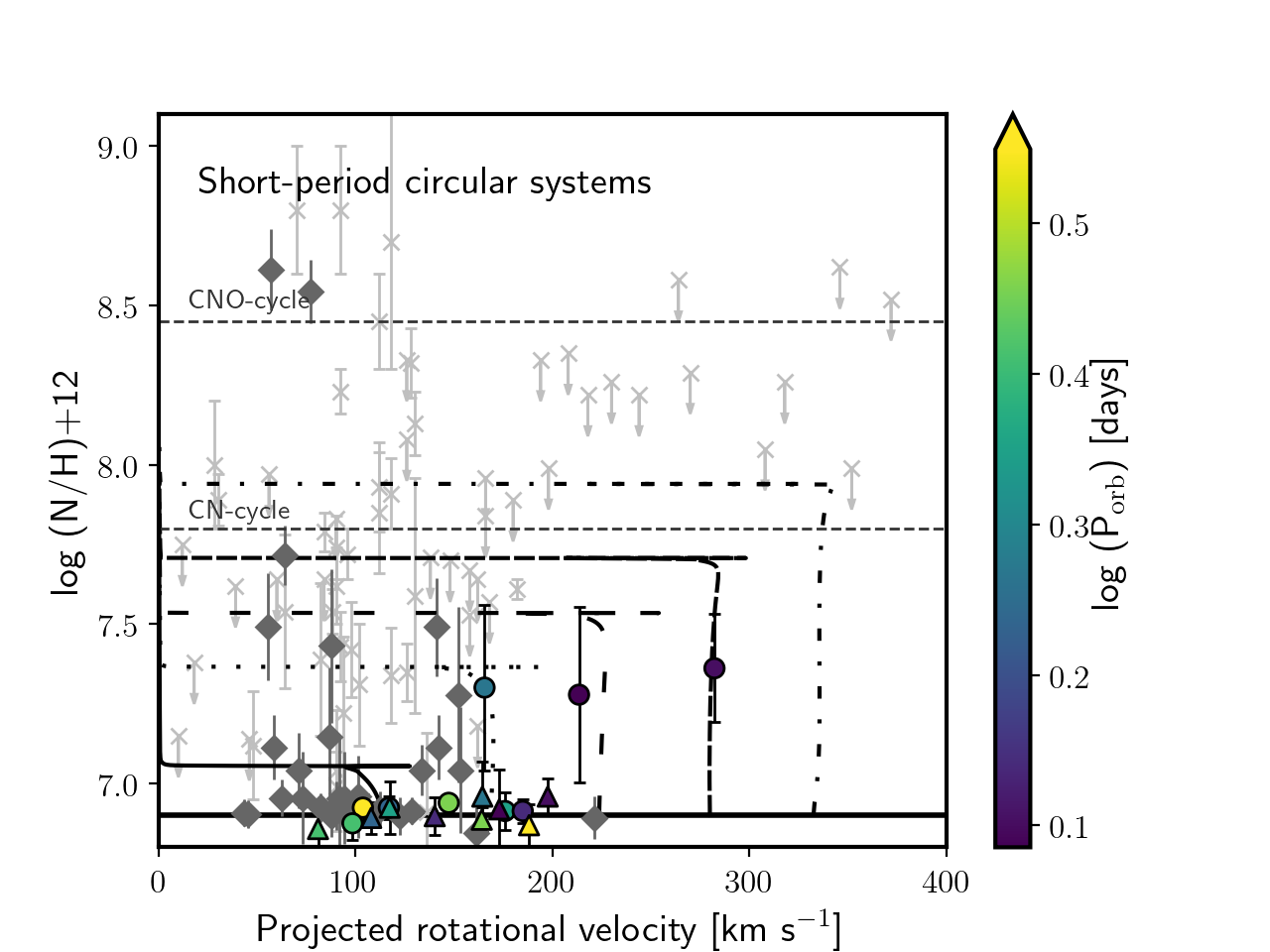}
    \caption{Projected rotational velocity vs. nitrogen content of the stars in our sample belonging to subsample 3 with an orbital period shorter than 10 days. Line styles and symbols are as Fig.\,\ref{fig:hunter_detached}a, diamonds indicate the components belonging to subsamples 1 and 2. The colour-bar represents the orbital period of the systems.}\label{fig:tidal} 
\end{figure}

\citet{demink09} studied the effects of the tides and rotational mixing on detached binary systems. These latter authors found significant dependences of the surface nitrogen abundances on these latter factors for short-period systems ($P_{\mathrm{orb}} < 2$ days) throughout a considerable fraction of their main sequence lifetime. The effects of the tides are stronger on short-period systems. We therefore focus on subsample 3 (the circular short-period systems; see Sect.\,\ref{discussion}). In Fig.\,\ref{fig:tidal}, we display the projected rotational velocity versus nitrogen content. An enrichment is detected for the systems with higher projected rotational velocities. However, the extra surface-enrichment produced by the tides is limited. This conclusion was also drawn by \citet{martins17} based on observations of six short-period systems in the Galaxy and by \citet{abdul-masih19}. 

It therefore appears from the above discussion that components in detached systems, for which no signatures of mass transfer have been detected, have nitrogen surface enrichment comparable to single massive stars. This tends to show that the role of the tides on the nitrogen surface abundance has only a limited effect.  

\subsection{Semi-detached systems}

Five systems in our sample are classified as semi-detached binaries. Among these systems, four have a secondary that fills its Roche lobe (Algol systems, subsample 5 in Sect.\,\ref{discussion}). Initially, the stars born with the highest mass filled their Roche lobe first, but rapidly, they transfer their material to their lower-mass companion. This mass transfer is initially fast (thermal time scale). After so much mass is transferred that the mass ratio reverses, the orbit starts to widen and the mass transfer rate decreases. There is thus a short gap without mass transfer. Models show that this would be followed by a phase of slow case-A mass transfer driven by the nuclear expansion of the now less massive donor star. The second transfer episode is thus slower and it is this one that we observe in four of the five semi-detached systems. 

To clarify the terminology between theory and observations, it is however important to specify that the definition that we take in the current paper for primary or secondary relies on the current (observed) more or less massive object of the system. For Algol systems, the primaries and secondaries in our analysis were the initially less and more massive object, respectively. In this context, the primary (currently the most massive star of the system) is the mass gainer, and shows less enrichment in nitrogen and rotates faster. Such stars are accreting material from their companion in a reverse case-A mass-transfer episode, which explains their high rotational velocity. The accreted material can cause the star to develop a larger convective core that will then mix fresh hydrogen fuel into the core. The star will therefore be rejuvenated, appearing younger than its companion. These stars still show an overabundance in nitrogen at their surface, which can testify that the currently most massive stars manage to accrete a significant amount of mass. 

\begin{figure}[t!]
  \centering
    \includegraphics[trim=10 0 45 35,clip,width=9cm]{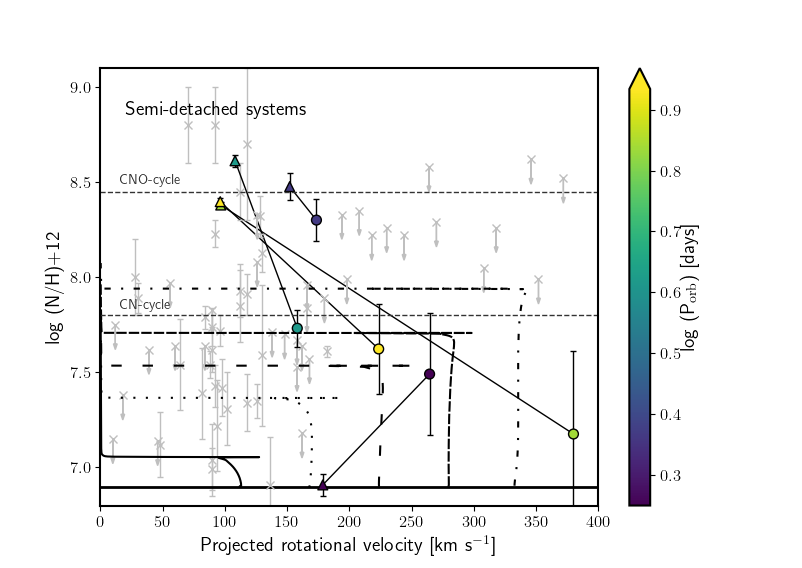}
    \caption{Projected rotational velocity vs. nitrogen content of the stars in semi-detached configuration. Circles and triangles represent the primary and secondary component, respectively. Tracks are from \citet{brott11} and correspond to an initial mass of 25\,\msun. The colour-bar represents the orbital period of the systems.}\label{fig:interact} 
\end{figure}

\begin{figure}[t!]
  \centering
    \includegraphics[trim=10 0 45 35,clip,width=9cm]{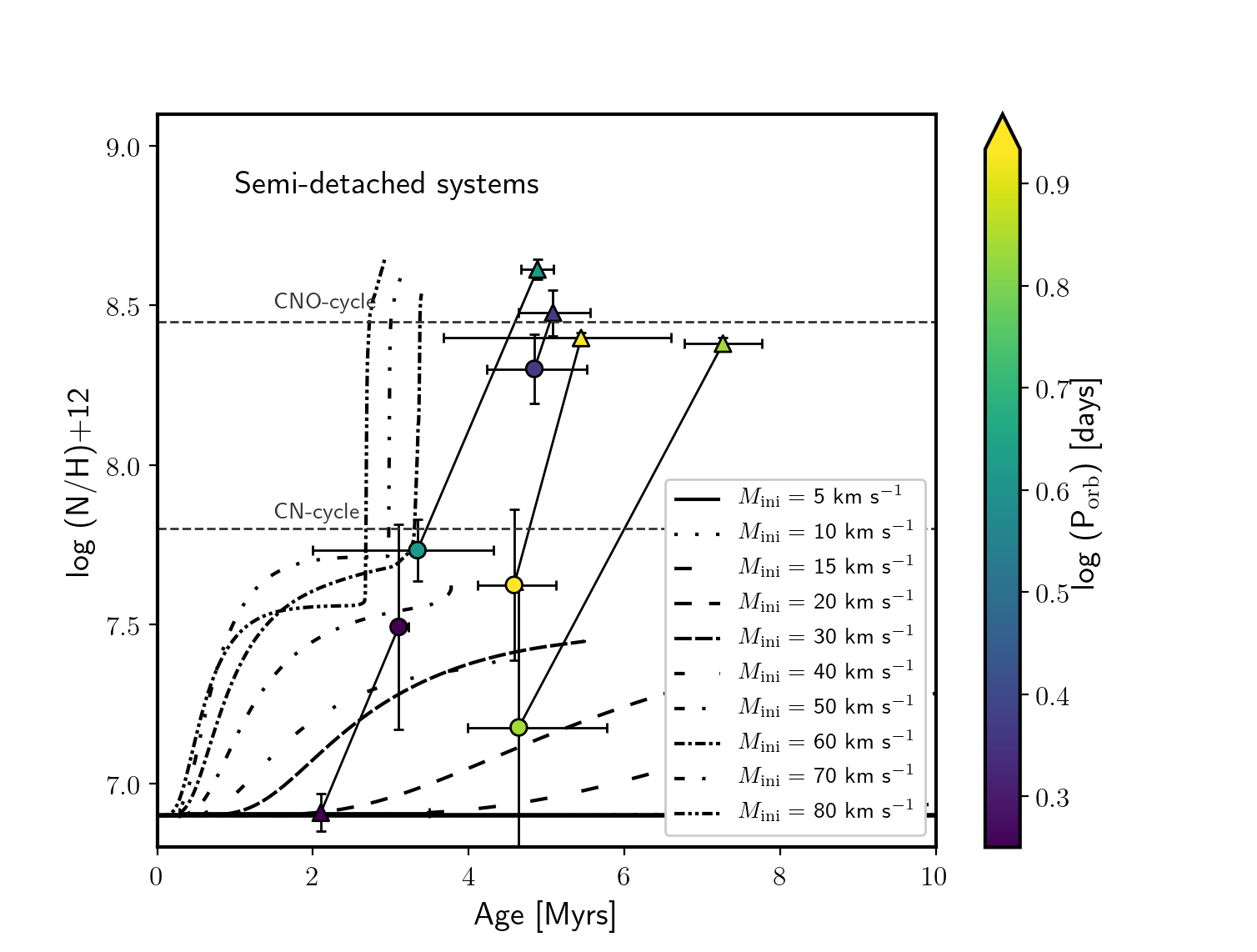}
    \caption{Age vs. nitrogen content of the stars in semi-detached configuration. Circles and triangles represent the primary and secondary component, respectively. Tracks are from \citet{brott11}. Line styles and symbols are as in Fig.\,\ref{fig:interact}.}\label{fig:interact_age} 
\end{figure}

The secondaries (currently the less massive object of the system) are filling their Roche lobe. They are highly enriched in nitrogen, which can be explained by the stripping of their outer layers. Their abundances are consistent with the CNO cycle (Fig.\,\ref{fig:interact}) and they rotate more slowly and appear older than their companion (Fig.\,\ref{fig:interact_age}). Similar conclusions were observed in binary systems where the secondary fills its Roche lobe. This is notably the case for XZ~Cep \citep{martins17} and HD\,149404 \citep{raucq16} in the Galaxy. For these two systems, their secondary is also more enriched and rotates more slowly than the primary. Their surface abundances are also similar to those produced by the CNO cycle. 

From Fig.\,\ref{fig:interact}, mass accretion onto stars can have two outcomes: (i) Only little mass is accreted because the accretor rapidly spins-up which then prevents further mass accretion. In this case, the star is spun-up and is probably not enriched with nitrogen or helium on the surface because those layers that carry nitrogen or helium are so deep in the donor that they are only transferred onto the accretor at a time when the star cannot accrete any more mass. (ii) Much more mass (maybe even all) is accreted because there is a way to get rid of angular momentum such that the accretor never reaches critical rotation. Tides are a prime mechanism here to allow for prolonged mass accretion. In this case, a star can spin up slightly and will be enriched with nitrogen and helium on the surface. From our sample,  we see that the systems with an orbital period longer than 6 days tend to follow channel (i). The donor has lost a lot of mass and is enriched in nitrogen on the surface. The accretor, the faster spinning star, does not show the same nitrogen enrichment which implies that it could not accrete nitrogen-enriched material from deeper within the donor. These stars also spin more rapidly than the donor. The longer period could then be consistent with a weaker influence of tides (or alternatively, the wider systems accrete mass from a disk rather than by direct impact). The wider separation could also affect the amount of angular momentum that is accreted.
In the shorter-period binaries, the spins are more comparable and both stars tend to show more similar nitrogen surface enrichment. The tides tend to synchronise the rotation of the stars with the orbit of the system. For the two systems with the shortest periods, the effects of the tides are stronger so that the discrepancy between the primary (mass gainer) and the secondary (mass donor) is smaller. As mentioned above, the role of the tides has a limited impact on the nitrogen surface enrichment, but seems very efficient at synchronising the rotations of the stars.

The only system where the most massive star fills its Roche lobe is VFTS\,176. This component rotates faster and is also highly enriched in nitrogen whilst its companion does not show any. Unlike the other semi-detached systems, the observed primary appears slightly younger than the secondary even though they are close to having the same age, within the error bars. The orbital period is very short ($P_{\mathrm{orb}} = 1.78$ days), and it is therefore possible that either the system is seen at the beginning of the mass transfer episode or the system faces up to the beginning of the reverse mass transfer episode. The most massive star is synchronised with the system, and the secondary rotates 30\% super-synchronously, which could indicate mass transfer from the primary onto the secondary. 

\subsection{Contact systems}
\label{contact}
As mentioned in Sect.\,\ref{discussion}, we found two systems in contact configuration. Other systems are close to being in contact but given the uncertainties on their inclinations and thus on the radii of their components, they will not be part of the present discussion. 
VFTS\,352 was thoroughly analysed by \citet{abdul-masih19} both from UV and optical spectra. These authors used the FASTWIND atmosphere code \citep{puls05} whilst we used CMFGEN. Comparisons between these two codes were done by \citet{massey13} showing no difference in effective temperatures but a systematic difference in $\log g$ of about 0.12 dex. Taking these systematics into account, the parameters of the two components agree for both components within 1- or 2-$\sigma$ uncertainties. The differences could come from the normalisation process (which was not taken into account for the determination of the uncertainties) but do not change the conclusions of the two papers. Furthermore, we use Vink's mass-loss recipes and only the lines in the optical while \citet{abdul-masih19} focused on both UV and optical domains, which increases their number of diagnostic lines used to constrain the wind parameters and the CNO abundances. We nevertheless confirm that the nitrogen content of the two objects in VFTS\,352 does not show any enrichment in both analyses. Regarding the carbon surface abundances, we only have one diagnostic line to constrain them, which can lead to discrepancies in the real values as demonstrated by \citet{martins15} or \citet{cazorla17a,cazorla17b}. The same conclusions can be drawn for VFTS\,066 since no significant enrichment for the nitrogen surface abundances for the two components is derived. We also do not detect any depletion in carbon for those objects. 

Both systems VFTS\,066 and VFTS\,352 have short periods ($P_{\mathrm{orb}}< 1.15$ days). It is also worth noting that their rotational velocities are large ($v_{\mathrm{rot}} > 330$\,\kms) and that their masses do not correspond to what single stars with similar properties would have. This is particularly true for VFTS\,352 since a difference in mass of about 10\,\msun\ is expected \citep[see also][]{almeida15}. 

One scenario that was discussed by \citet{almeida15} and \citet{abdul-masih19} to explain the properties of VFTS\,352 was the chemically homogeneous evolution. Under this scenario, the internal mixing is extremely efficient. The two components would become hotter and more luminous, but would remain compact, and probably never merge, leading to the creation of binary systems with two black holes \citep{marchant16,demink16}. Although both components have a high rotational velocity, which could infer high internal mixing, the fact that no enrichment is observed in the nitrogen lines seems to reject this assumption. If we assume a more classical evolutionary scheme, this absence of enrichment therefore tends to suggest that the contact phase occurs at the very beginning of the stellar evolution, when significant envelope removal did not occur yet. 

Interestingly, they both appear in relative isolation (Fig.\,\ref{fig:FOV}) which, if their young age is confirmed, would suggest that they have been formed outside the main OB associations in 30 Dor. 

Moreover, VFTS\,066 and VFTS\,352 are not the only contact systems where no enrichment was found so far. The same conclusions were given for the contact systems V382\,Cyg \citep{martins17} and HD\,100213 (Mahy et al. in prep.) in the Galaxy, for which no peculiar abundance was determined. The method of analysis may also be questioned since it is based on the use of 1D atmosphere models with spherical symmetry whilst the components in evolved binary systems are no longer spherical. These systems presumably necessitate more adequate tools that can account for their non-spherical geometry as suggested by \citet{palate12} or Abdul-Masih et al. (in prep.).

\subsection{Age distribution} 

\begin{figure}[htbp]
 \centering
    \includegraphics[trim=5 0 20 10,clip,width=9cm]{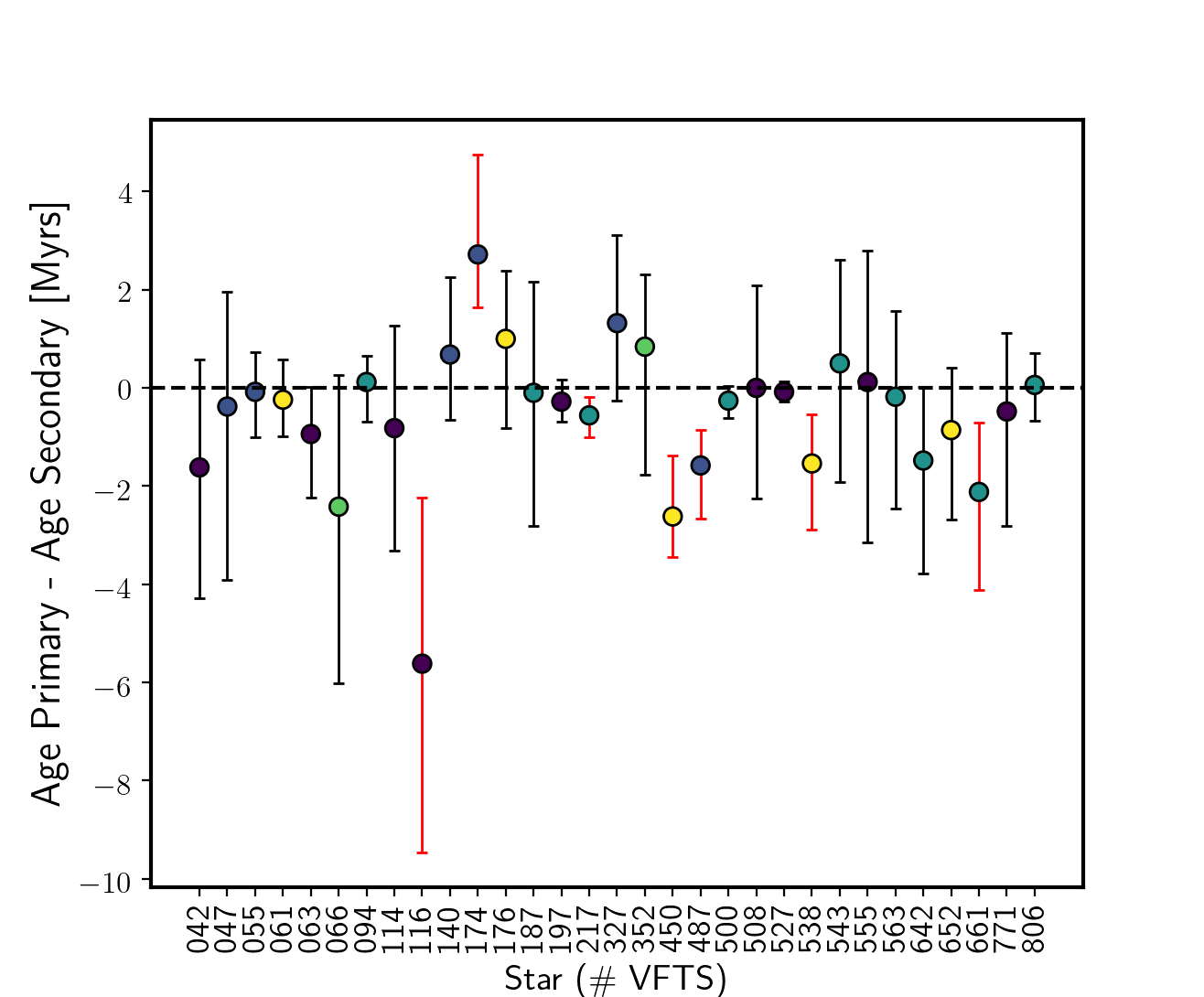}
    \caption{Difference between the ages of the primary and secondary stars as a function of the system identification. The ages are from BONNSAI. The colour-code is given for the different subsamples given in Sect.\,\ref{discussion}, from dark blue for subsample 1 to yellow for group 5. The red error bars indicate that the systems are not coeval. }\label{fig:age_comp} 
\end{figure}

Figure.\,\ref{fig:age_comp} shows the difference between the evolutionary ages of the primary and the secondary stars of the systems. If the binaries are non-interacting, the estimated ages should be the same within the uncertainties because we expect the two stars to be coeval. If the estimated ages are not coeval, this could indicate mass transfers, interactions or any other causes that affect the evolution of the companion. Seven systems (marked with red error bars in Fig.\,\ref{fig:age_comp}) display a significant difference between the estimated ages of their two components. Among them, VFTS\,450, and VFTS\,538 show clear interactions. This figure points out that the primary components in these systems are rejuvenated by the fact that these stars are accreting materials from their companion. Surprisingly, the differences in age in VFTS\,061 and VFTS\,652 do not appear to be significant within $1-\sigma$ error. Four other systems (VFTS\,174, VFTS\,217, VFTS\,487 and VFTS\,661) have short periods ($P_{\rm orb} < 5$ days) that can favour interactions such as tides. Finally the age discrepancy observed for the  last system, VFTS\,116, is difficult to explain given the rather long-period orbit and its eccentricity. With these orbital properties, we indeed expect that the two stars evolve like single stars (i.e. with only very minor influence from their companion). 

\begin{figure}[htbp]
  \centering
    \includegraphics[trim=10 0 0 40,clip,width=9cm]{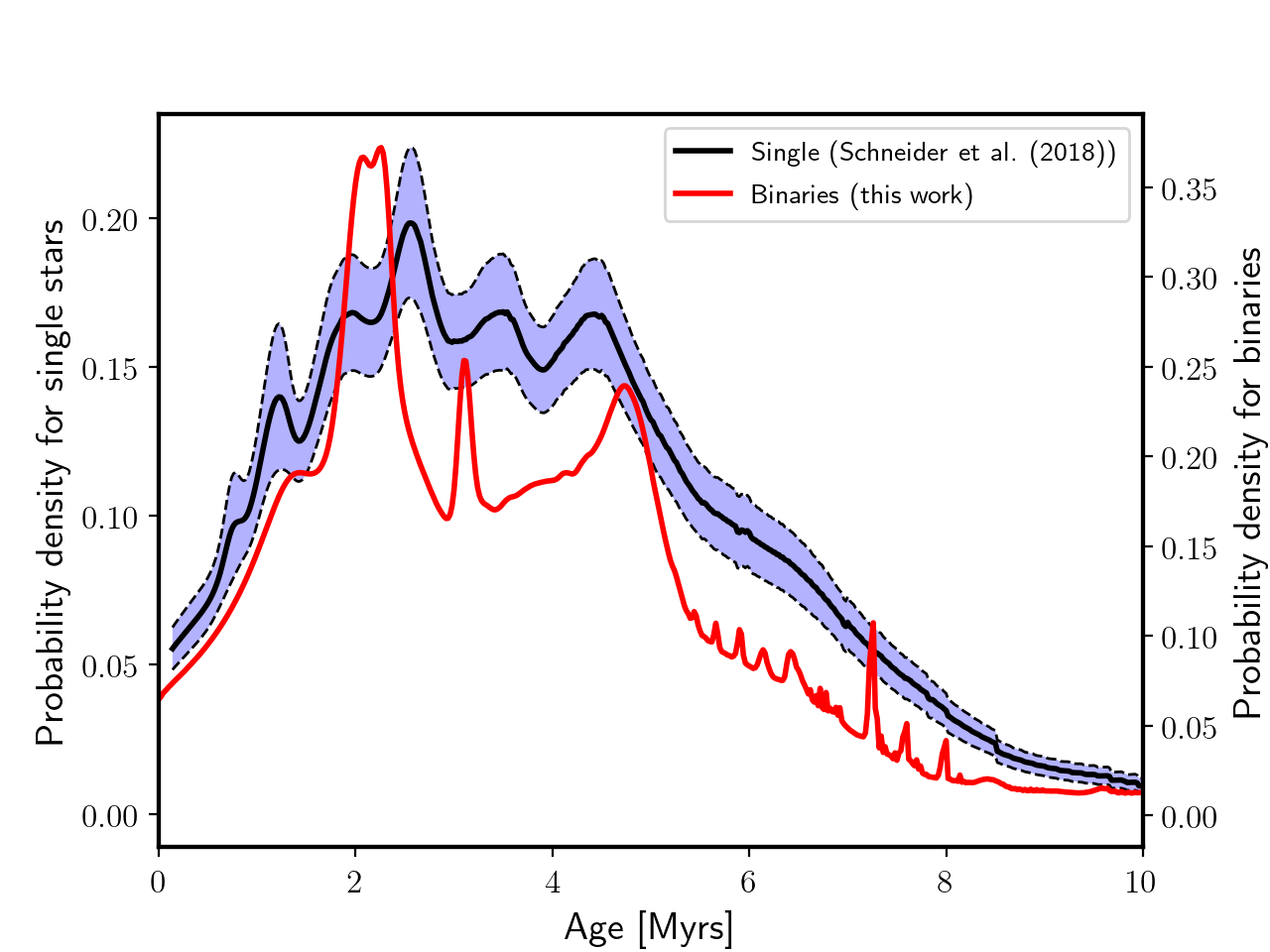}
    \caption{Probability density function of the TMBM SB2 sample (red line) compared to the PDF of the VFTS single stars more massive than 15\,\msun (black line, \citealt{schneider18}). The ages of the binaries are computed as an average value between the ages of the primaries and the secondaries. }\label{fig:age_pdf} 
\end{figure}

We examine the age distribution of the SB2 sample, computed {by summing the probability density functions of each individual star provided by BONNSAI}. We compare this distribution to that published by \citet{schneider18} from the massive single O-type stars ($M > 15\,\msun$) located in the 30 Doradus region. Figure\,\ref{fig:age_pdf} displays the probability density function of the ages of the spectroscopic binaries (in red) and we overplot that of the massive single O-type star population (in black). We identify the highest broad peak at 2 Myrs that is also observed in the single massive star population of 30 Dor. A second broad peak is observed at 4.5 Myrs and a third one is detected at 7 Myrs. The binary population appears to have comparable ages as the single-star population, albeit perhaps slightly younger. This small difference between the single and the binary samples could be explained by two biases: one observational bias linked to the sample and one evolutionary bias. For the former, we indeed observe a relative dearth of systems with orbital periods between 6 and 20 days. The lack of such systems is not fully understood, and we do not know whether this is due to a selection effect or to real phenomena that occur during the evolution of the components. For the latter, in order to be part of our sample, two massive stars must orbit around each other with a significant $\Delta RV$. The most massive systems could have interacted or merged, excluding them from our selection criteria, creating this gap between 2 and 4.5 Myrs. 

When we display the spatial distribution of the mean evolutionary ages (i.e., the mean age between the primary and the secondary of each system), most of the oldest systems are found in the field outside NGC\,2070 and NGC\,2060. Three exceptions are VFTS\,055, VFTS\,642 and VFTS\,806, with ages of about 2.5 Myrs, while the other systems are between 5 and 8 Myrs. This result confirms what was obtained by \citet{schneider18} for the single massive star population in 30 Dor. It also indicates that the youngest systems are found in NGC\,2070 and, more specifically, close to the core region of this cluster, which was also observed by \citet{schneider18}. The two exceptions are VFTS\,450 and VFTS\,652, two systems in case-A mass transfer. These interactions clearly affect the comparison to single-star evolutionary tracks, and thus create a bias. \citet{schneider18} also found ages of 2 to 5 Myrs for the massive single-star population of NGC\,2060. Our analysis of binary SB2 systems yields the same conclusion provided by \citet{schneider18}.

\begin{figure*}[t!]
  \centering
    \includegraphics[width=16cm]{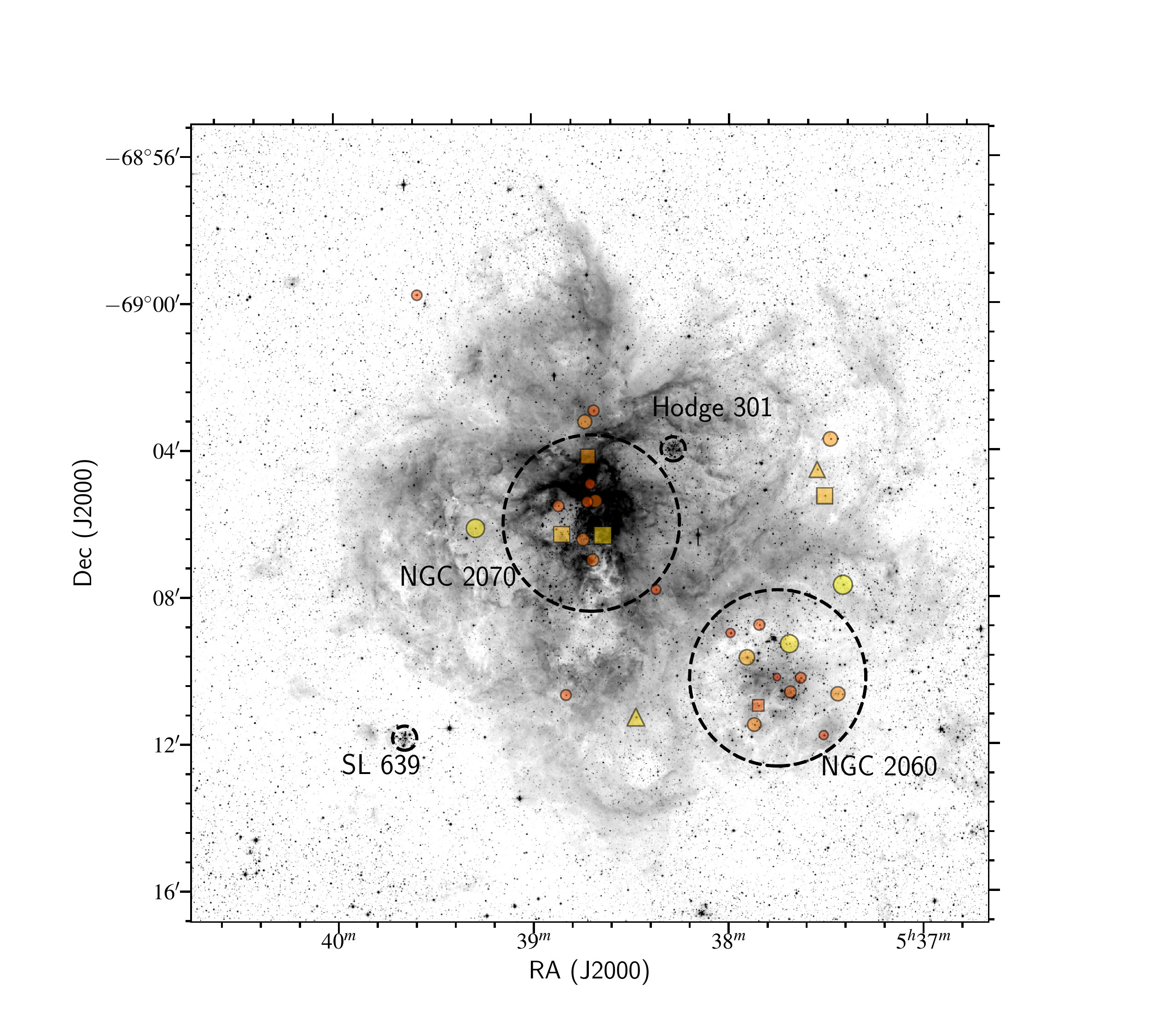}
    \caption{Spatial distribution of the mean ages of the systems. The youngest systems are shown in red while the oldest ones are shown in yellow. The circles indicate the detached systems (subsamples 1, 2, and 3), the triangles the over-contact systems (subsample 4), and the squares the semi-detached systems (subsample 5).}\label{fig:age_tmbm} 
\end{figure*}

\subsection{Rotational rates}

\begin{figure}[htbp]
\centering
\includegraphics[trim=10 0 0 40,clip,width=9cm]{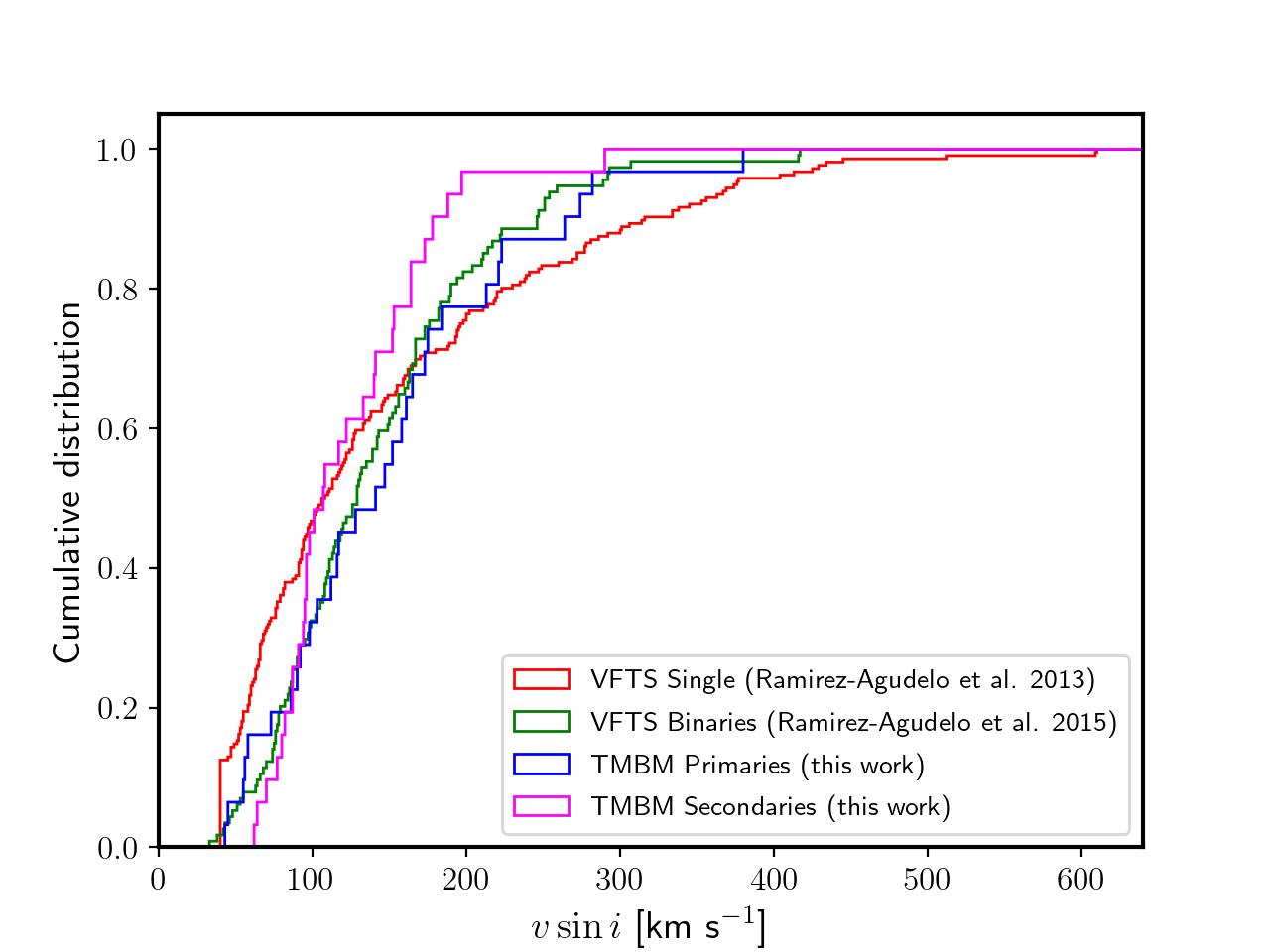}
\caption{Cumulative distributions of the projected rotational velocities of the massive single and binary populations in the 30 Doradus region. Red: VFTS `presumably' single stars \citep{ramirez13}, Green: VFTS primaries in detected binary system (most luminous components, \citealt{ramirez15}); Blue: primaries in the present 31 SB2 TMBM sample (=currently most massive component, this work); Magenta: secondaries in the present 31 SB2 TMBM sample (=currently least massive component, this work). }\label{fig:cumul} 
\end{figure}

\citet{ramirez13,ramirez15} probed the stellar rotational rates of the single-star and binary populations in the 30 Doradus region, respectively. Now that we have undertaken a careful analysis of the stellar properties of the 31 SB2 binary systems (62 objects) through spectral disentangling and atmosphere modelling, we can refine the projected rotational velocities of each component of these systems. Unlike these latter authors, who took the whole binary population into account, our sample only accounts for 62 components and contains a larger fraction of shorter period systems. As in \citet{ramirez13,ramirez15} , we also performed Kuiper tests on the projected rotational velocities of the binary and single-star populations to test the null hypothesis, namely that the samples come from the same parent distributions.

First, Fig.~\ref{fig:cumul} shows that there is no statistical differences between the \vsini\ distributions of our primary star sample and that of \citet{ramirez15} despite the fact that we have used slightly different definitions to identify the primary stars.
 
Compared to the sample of `presumably' single stars in the 30\,Dor region, primary stars in our SB2 sample present a larger average rotation rate but a significantly less pronounced high-velocity tail. These differences are even more pronounced for the secondary stars in our sample (Fig.~\ref{fig:cumul}). This is qualitatively compatible with the following interpretations: the higher average velocity results from tidal interactions while the lack of a strong high-velocity tail results from the fact that our sample is dominated by pre-interaction, detached binaries \citep[see also discussions in ][]{demink13, ramirez15}. 
 
Below 100\,\kms, the two distributions of primary and secondary rotational velocities are rather similar but they deviate at larger projected rotational velocities. These differences are interesting but they may not be significant as they failed to trigger the Kuiper test: ($p-$value=18\%). Indeed, from a physical point of view, the primaries must have radii larger than those of the secondaries. Assuming that the tidal synchronisation is achieved before the first mass-transfer episode, the projected rotational velocities of the primaries should be larger than those of their companions. The fact that the cumulative distribution for the secondary stars is steeper is therefore consistent with what we observe. It is also to be expected that some of the stars in our primary sample (= currently more massive star) are accreting stars that have gained mass and angular momentum from the now less massive -- hence secondary -- star. This is sufficient to explain the more pronounced high-velocity tail in our `primary' star sample. 

%%###################################################################
%%-------------------------------   Conclusion  --------------------- 

\section{Conclusions}
\label{conc}

Here, we present our study of the individual properties of 31 double-lined spectroscopic binaries in the 30 Doradus region. We applied spectral disentangling and atmosphere modelling to determine the stellar parameters and the helium, carbon and nitrogen surface abundances for each component (although the helium and carbon surface abundances were not discussed in the paper). 

In our sample, we detected between 58 and 77\% pre-interacting systems, 16\% semi-detached systems and between 5 and 26\% systems in contact or where the two components fill a large fraction of their Roche lobe. 

For the detached systems, their nitrogen surface abundances are consistent with those of single stars, within the uncertainties, except for two systems, VFTS 197 and VFTS 527. For the latter, the strong stellar winds of the components can explain this high enrichment since under their effects, a part of the external layers of the stars can be ejected. For the detached systems with short periods, we show that the effects of the tides are stronger on the rotation than on the enrichment of nitrogen, since this enrichment is relatively limited in comparison to mass transfer. Our analysis shows that, for the systems displaying evidence of mass transfer, the surface abundances are highly impacted for the mass donors. The removal of the external layers makes chemically processed material appear at the surface; the stars are also more luminous and their rotation spins down. For the mass gainers, the effects of the mass transfer are a higher rotational velocity and rejuvenation. The locations of these components in the Hunter diagram tend to show that binary products are good candidates to populate the two groups of stars (slowly rotating, nitrogen-enriched objects and rapidly rotating non-enriched objects) that cannot be reproduced through single-star population synthesis. Finally, we observe no peculiar abundance for the two systems in (over-)contact, reminiscent of what was observed for several binary systems in the Milky Way.

The comparison between the evolutionary ages of the components in binary systems to those determined for single stars in 30 Dor reveals a general agreement, which confirms the different ages for NGC\,2070 and NGC\,2060. We also find that the older stars are located in the field, having formed ahead of the stellar populations in NGC~2060 and NGC~2070.

%%#####################################################################
\begin{acknowledgements}
The authors thank the anonymous referee for his/her constructive remarks. This work is based on observations collected at the European Southern Observatory under program IDs 090.D-0323 and 092.D-0136. The authors are grateful to the ESO staff for their support in the preparation and execution of the observations. L.M. thanks Prof. D. J. Hillier for making CMFGEN publicly available. H.S. and M.A.M. acknowledge support from the FWO-Odysseus program under project G0F8H6NT. H.S. and T.S. acknowledge support from the European Research Council (ERC) under the European Union’s DLV-772225-MULTIPLES Horizon 2020 research and innovation programme. L.A.A. thanks to Aperfei{\c c}oamento de Pessoal de N{\'i}vel Superior (CAPES) and Funda{\c c}{\~a}o de Amparo {\`a} Pesquisa do Estado de S{\~a}o Paulo (FAPESP – 2011/51680-6, 2012/09716-6, 2013/18245-0) for financial support. S.dM. acknowledges funding by the European Union's Horizon 2020 research and innovation programme from the European Research Council (ERC) (Grant agreement No.\ 715063), and by the Netherlands Organisation for Scientific Research (NWO) as part of the Vidi research program BinWaves with project number 639.042.728. A.F.J.M. acknowledges financial aid from NSERC (Canada). G.G. acknowledges financial support from Deutsche Forschunsgemeinschaft (DFG) under grant GR 1717/5-1.
\end{acknowledgements}

%%#####################################################################
%%#####################################################################
\bibliography{TMBM}

\begin{thebibliography}{85}
\expandafter\ifx\csname natexlab\endcsname\relax\def\natexlab#1{#1}\fi

\bibitem[{{Abdul-Masih} {et~al.}(2019){Abdul-Masih}, {Sana}, {Sundqvist},
  {Mahy}, {Menon}, {Almeida}, {De Koter}, {de Mink}, {Justham}, {Langer},
  {Puls}, {Shenar}, \& {Tramper}}]{abdul-masih19}
{Abdul-Masih}, M., {Sana}, H., {Sundqvist}, J., {et~al.} 2019, \apj, 880, 115

\bibitem[{{Almeida} {et~al.}(2015){Almeida}, {Sana}, {de Mink}, {Tramper},
  {Soszy{\'n}ski}, {Langer}, {Barb{\'a}}, {Cantiello}, {Damineli}, {de Koter},
  {Garcia}, {Gr{\"a}fener}, {Herrero}, {Howarth}, {Ma{\'{\i}}z Apell{\'a}niz},
  {Norman}, {Ram{\'{\i}}rez-Agudelo}, \& {Vink}}]{almeida15}
{Almeida}, L.~A., {Sana}, H., {de Mink}, S.~E., {et~al.} 2015, \apj, 812, 102

\bibitem[{{Almeida} {et~al.}(2017){Almeida}, {Sana}, {Taylor}, {Barb{\'a}},
  {Bonanos}, {Crowther}, {Damineli}, {de Koter}, {de Mink}, {Evans}, {Gieles},
  {Grin}, {H{\'e}nault-Brunet}, {Langer}, {Lennon}, {Lockwood}, {Ma{\'{\i}}z
  Apell{\'a}niz}, {Moffat}, {Neijssel}, {Norman}, {Ram{\'{\i}}rez-Agudelo},
  {Richardson}, {Schootemeijer}, {Shenar}, {Soszy{\'n}ski}, {Tramper}, \&
  {Vink}}]{almeida17}
{Almeida}, L.~A., {Sana}, H., {Taylor}, W., {et~al.} 2017, \aap, 598, A84

\bibitem[{{Bestenlehner} {et~al.}(2014){Bestenlehner}, {Gr{\"a}fener}, {Vink},
  {Najarro}, {de Koter}, {Sana}, {Evans}, {Crowther}, {H{\'e}nault-Brunet},
  {Herrero}, {Langer}, {Schneider}, {Sim{\'o}n-D{\'{\i}}az}, {Taylor}, \&
  {Walborn}}]{bestenlehner14}
{Bestenlehner}, J.~M., {Gr{\"a}fener}, G., {Vink}, J.~S., {et~al.} 2014, \aap,
  570, A38

\bibitem[{{Bouret} {et~al.}(2012){Bouret}, {Hillier}, {Lanz}, \&
  {Fullerton}}]{bouret12}
{Bouret}, J.~C., {Hillier}, D.~J., {Lanz}, T., \& {Fullerton}, A.~W. 2012,
  \aap, 544, A67

\bibitem[{{Brott} {et~al.}(2011{\natexlab{a}}){Brott}, {de Mink}, {Cantiello},
  {Langer}, {de Koter}, {Evans}, {Hunter}, {Trundle}, \& {Vink}}]{brott11}
{Brott}, I., {de Mink}, S.~E., {Cantiello}, M., {et~al.} 2011{\natexlab{a}},
  \aap, 530, A115

\bibitem[{{Brott} {et~al.}(2011{\natexlab{b}}){Brott}, {Evans}, {Hunter}, {de
  Koter}, {Langer}, {Dufton}, {Cantiello}, {Trundle}, {Lennon}, {de Mink},
  {Yoon}, \& {Anders}}]{brott11b}
{Brott}, I., {Evans}, C.~J., {Hunter}, I., {et~al.} 2011{\natexlab{b}}, \aap,
  530, A116

\bibitem[{{Cazorla} {et~al.}(2017{\natexlab{a}}){Cazorla}, {Morel}, {Naz{\'e}},
  {Rauw}, {Semaan}, {Daflon}, \& {Oey}}]{cazorla17a}
{Cazorla}, C., {Morel}, T., {Naz{\'e}}, Y., {et~al.} 2017{\natexlab{a}}, \aap,
  603, A56

\bibitem[{{Cazorla} {et~al.}(2017{\natexlab{b}}){Cazorla}, {Naz{\'e}}, {Morel},
  {Georgy}, {Godart}, \& {Langer}}]{cazorla17b}
{Cazorla}, C., {Naz{\'e}}, Y., {Morel}, T., {et~al.} 2017{\natexlab{b}}, \aap,
  604, A123

\bibitem[{{Clark} {et~al.}(2015){Clark}, {Bartlett}, {Broos}, {Townsley},
  {Taylor}, {Walborn}, {Bird}, {Sana}, {de Mink}, {Dufton}, {Evans}, {Langer},
  {Ma{\'{\i}}z Apell{\'a}niz}, {Schneider}, \& {Soszy{\'n}ski}}]{clark15}
{Clark}, J.~S., {Bartlett}, E.~S., {Broos}, P.~S., {et~al.} 2015, \aap, 579,
  A131

\bibitem[{{Conti}(1973)}]{conti73}
{Conti}, P.~S. 1973, \apj, 179, 161

\bibitem[{{Conti} \& {Alschuler}(1971)}]{conti71}
{Conti}, P.~S. \& {Alschuler}, W.~R. 1971, \apj, 170, 325

\bibitem[{{De Becker} {et~al.}(2006){De Becker}, {Rauw}, {Manfroid}, \&
  {Eenens}}]{debecker06}
{De Becker}, M., {Rauw}, G., {Manfroid}, J., \& {Eenens}, P. 2006, \aap, 456,
  1121

\bibitem[{{de Mink} {et~al.}(2009){de Mink}, {Cantiello}, {Langer}, {Pols},
  {Brott}, \& {Yoon}}]{demink09}
{de Mink}, S.~E., {Cantiello}, M., {Langer}, N., {et~al.} 2009, \aap, 497, 243

\bibitem[{{de Mink} {et~al.}(2013){de Mink}, {Langer}, {Izzard}, {Sana}, \& {de
  Koter}}]{demink13}
{de Mink}, S.~E., {Langer}, N., {Izzard}, R.~G., {Sana}, H., \& {de Koter}, A.
  2013, \apj, 764, 166

\bibitem[{{de Mink} \& {Mandel}(2016)}]{demink16}
{de Mink}, S.~E. \& {Mandel}, I. 2016, \mnras, 460, 3545

\bibitem[{{de Mink} {et~al.}(2014){de Mink}, {Sana}, {Langer}, {Izzard}, \&
  {Schneider}}]{demink14}
{de Mink}, S.~E., {Sana}, H., {Langer}, N., {Izzard}, R.~G., \& {Schneider},
  F.~R.~N. 2014, \apj, 782, 7

\bibitem[{{Dunstall} {et~al.}(2015){Dunstall}, {Dufton}, {Sana}, {Evans},
  {Howarth}, {Sim{\'o}n-D{\'{\i}}az}, {de Mink}, {Langer}, {Ma{\'{\i}}z
  Apell{\'a}niz}, \& {Taylor}}]{dunstall15}
{Dunstall}, P.~R., {Dufton}, P.~L., {Sana}, H., {et~al.} 2015, \aap, 580, A93

\bibitem[{{Evans} {et~al.}(2011){Evans}, {Taylor}, {H{\'e}nault-Brunet},
  {Sana}, {de Koter}, {Sim{\'o}n-D{\'{\i}}az}, {Carraro}, {Bagnoli}, {Bastian},
  {Bestenlehner}, {Bonanos}, {Bressert}, {Brott}, {Campbell}, {Cantiello},
  {Clark}, {Costa}, {Crowther}, {de Mink}, {Doran}, {Dufton}, {Dunstall},
  {Friedrich}, {Garcia}, {Gieles}, {Gr{\"a}fener}, {Herrero}, {Howarth},
  {Izzard}, {Langer}, {Lennon}, {Ma{\'{\i}}z Apell{\'a}niz}, {Markova},
  {Najarro}, {Puls}, {Ramirez}, {Sab{\'{\i}}n-Sanjuli{\'a}n}, {Smartt},
  {Stroud}, {van Loon}, {Vink}, \& {Walborn}}]{evans11}
{Evans}, C.~J., {Taylor}, W.~D., {H{\'e}nault-Brunet}, V., {et~al.} 2011, \aap,
  530, A108

\bibitem[{{Gaia Collaboration} {et~al.}(2018){Gaia Collaboration}, {Brown},
  {Vallenari}, {Prusti}, {de Bruijne}, {Babusiaux}, {Bailer-Jones}, {Biermann},
  {Evans}, {Eyer}, {Jansen}, {Jordi}, {Klioner}, {Lammers}, {Lindegren},
  {Luri}, {Mignard}, {Panem}, {Pourbaix}, {Randich}, {Sartoretti}, {Siddiqui},
  {Soubiran}, {van Leeuwen}, {Walton}, {Arenou}, {Bastian}, {Cropper},
  {Drimmel}, {Katz}, {Lattanzi}, {Bakker}, {Cacciari}, {Casta{\~n}eda},
  {Chaoul}, {Cheek}, {De Angeli}, {Fabricius}, {Guerra}, {Holl}, {Masana},
  {Messineo}, {Mowlavi}, {Nienartowicz}, {Panuzzo}, {Portell}, {Riello},
  {Seabroke}, {Tanga}, {Th{\'e}venin}, {Gracia-Abril}, {Comoretto},
  {Garcia-Reinaldos}, {Teyssier}, {Altmann}, {Andrae}, {Audard},
  {Bellas-Velidis}, {Benson}, {Berthier}, {Blomme}, {Burgess}, {Busso},
  {Carry}, {Cellino}, {Clementini}, {Clotet}, {Creevey}, {Davidson}, {De
  Ridder}, {Delchambre}, {Dell'Oro}, {Ducourant},
  {Fern{\'a}ndez-Hern{\'a}ndez}, {Fouesneau}, {Fr{\'e}mat}, {Galluccio},
  {Garc{\'\i}a-Torres}, {Gonz{\'a}lez-N{\'u}{\~n}ez}, {Gonz{\'a}lez-Vidal},
  {Gosset}, {Guy}, {Halbwachs}, {Hambly}, {Harrison}, {Hern{\'a}ndez},
  {Hestroffer}, {Hodgkin}, {Hutton}, {Jasniewicz}, {Jean-Antoine-Piccolo},
  {Jordan}, {Korn}, {Krone-Martins}, {Lanzafame}, {Lebzelter}, {L{\"o}ffler},
  {Manteiga}, {Marrese}, {Mart{\'\i}n-Fleitas}, {Moitinho}, {Mora}, {Muinonen},
  {Osinde}, {Pancino}, {Pauwels}, {Petit}, {Recio-Blanco}, {Richards},
  {Rimoldini}, {Robin}, {Sarro}, {Siopis}, {Smith}, {Sozzetti}, {S{\"u}veges},
  {Torra}, {van Reeven}, {Abbas}, {Abreu Aramburu}, {Accart}, {Aerts},
  {Altavilla}, {{\'A}lvarez}, {Alvarez}, {Alves}, {Anderson}, {Andrei},
  {Anglada Varela}, {Antiche}, {Antoja}, {Arcay}, {Astraatmadja}, {Bach},
  {Baker}, {Balaguer-N{\'u}{\~n}ez}, {Balm}, {Barache}, {Barata}, {Barbato},
  {Barblan}, {Barklem}, {Barrado}, {Barros}, {Barstow}, {Bartholom{\'e}
  Mu{\~n}oz}, {Bassilana}, {Becciani}, {Bellazzini}, {Berihuete}, {Bertone},
  {Bianchi}, {Bienaym{\'e}}, {Blanco-Cuaresma}, {Boch}, {Boeche}, {Bombrun},
  {Borrachero}, {Bossini}, {Bouquillon}, {Bourda}, {Bragaglia}, {Bramante},
  {Breddels}, {Bressan}, {Brouillet}, {Br{\"u}semeister}, {Brugaletta},
  {Bucciarelli}, {Burlacu}, {Busonero}, {Butkevich}, {Buzzi}, {Caffau},
  {Cancelliere}, {Cannizzaro}, {Cantat-Gaudin}, {Carballo}, {Carlucci},
  {Carrasco}, {Casamiquela}, {Castellani}, {Castro-Ginard}, {Charlot},
  {Chemin}, {Chiavassa}, {Cocozza}, {Costigan}, {Cowell}, {Crifo}, {Crosta},
  {Crowley}, {Cuypers}, {Dafonte}, {Damerdji}, {Dapergolas}, {David}, {David},
  {de Laverny}, {De Luise}, {De March}, {de Martino}, {de Souza}, {de Torres},
  {Debosscher}, {del Pozo}, {Delbo}, {Delgado}, {Delgado}, {Di Matteo},
  {Diakite}, {Diener}, {Distefano}, {Dolding}, {Drazinos}, {Dur{\'a}n},
  {Edvardsson}, {Enke}, {Eriksson}, {Esquej}, {Eynard Bontemps}, {Fabre},
  {Fabrizio}, {Faigler}, {Falc{\~a}o}, {Farr{\`a}s Casas}, {Federici},
  {Fedorets}, {Fernique}, {Figueras}, {Filippi}, {Findeisen}, {Fonti},
  {Fraile}, {Fraser}, {Fr{\'e}zouls}, {Gai}, {Galleti}, {Garabato},
  {Garc{\'\i}a-Sedano}, {Garofalo}, {Garralda}, {Gavel}, {Gavras}, {Gerssen},
  {Geyer}, {Giacobbe}, {Gilmore}, {Girona}, {Giuffrida}, {Glass}, {Gomes},
  {Granvik}, {Gueguen}, {Guerrier}, {Guiraud}, {Guti{\'e}rrez-S{\'a}nchez},
  {Haigron}, {Hatzidimitriou}, {Hauser}, {Haywood}, {Heiter}, {Helmi}, {Heu},
  {Hilger}, {Hobbs}, {Hofmann}, {Holland}, {Huckle}, {Hypki}, {Icardi},
  {Jan{\ss}en}, {Jevardat de Fombelle}, {Jonker}, {Juh{\'a}sz}, {Julbe},
  {Karampelas}, {Kewley}, {Klar}, {Kochoska}, {Kohley}, {Kolenberg},
  {Kontizas}, {Kontizas}, {Koposov}, {Kordopatis}, {Kostrzewa-Rutkowska},
  {Koubsky}, {Lambert}, {Lanza}, {Lasne}, {Lavigne}, {Le Fustec}, {Le
  Poncin-Lafitte}, {Lebreton}, {Leccia}, {Leclerc}, {Lecoeur-Taibi},
  {Lenhardt}, {Leroux}, {Liao}, {Licata}, {Lindstr{\o}m}, {Lister}, {Livanou},
  {Lobel}, {L{\'o}pez}, {Managau}, {Mann}, {Mantelet}, {Marchal}, {Marchant},
  {Marconi}, {Marinoni}, {Marschalk{\'o}}, {Marshall}, {Martino}, {Marton},
  {Mary}, {Massari}, {Matijevi{\v{c}}}, {Mazeh}, {McMillan}, {Messina},
  {Michalik}, {Millar}, {Molina}, {Molinaro}, {Moln{\'a}r}, {Montegriffo},
  {Mor}, {Morbidelli}, {Morel}, {Morris}, {Mulone}, {Muraveva}, {Musella},
  {Nelemans}, {Nicastro}, {Noval}, {O'Mullane}, {Ord{\'e}novic},
  {Ord{\'o}{\~n}ez-Blanco}, {Osborne}, {Pagani}, {Pagano}, {Pailler},
  {Palacin}, {Palaversa}, {Panahi}, {Pawlak}, {Piersimoni}, {Pineau}, {Plachy},
  {Plum}, {Poggio}, {Poujoulet}, {Pr{\v{s}}a}, {Pulone}, {Racero}, {Ragaini},
  {Rambaux}, {Ramos-Lerate}, {Regibo}, {Reyl{\'e}}, {Riclet}, {Ripepi}, {Riva},
  {Rivard}, {Rixon}, {Roegiers}, {Roelens}, {Romero-G{\'o}mez}, {Rowell},
  {Royer}, {Ruiz-Dern}, {Sadowski}, {Sagrist{\`a} Sell{\'e}s}, {Sahlmann},
  {Salgado}, {Salguero}, {Sanna}, {Santana-Ros}, {Sarasso}, {Savietto},
  {Schultheis}, {Sciacca}, {Segol}, {Segovia}, {S{\'e}gransan}, {Shih},
  {Siltala}, {Silva}, {Smart}, {Smith}, {Solano}, {Solitro}, {Sordo}, {Soria
  Nieto}, {Souchay}, {Spagna}, {Spoto}, {Stampa}, {Steele},
  {Steidelm{\"u}ller}, {Stephenson}, {Stoev}, {Suess}, {Surdej}, {Szabados},
  {Szegedi-Elek}, {Tapiador}, {Taris}, {Tauran}, {Taylor}, {Teixeira},
  {Terrett}, {Teyssand ier}, {Thuillot}, {Titarenko}, {Torra Clotet}, {Turon},
  {Ulla}, {Utrilla}, {Uzzi}, {Vaillant}, {Valentini}, {Valette}, {van Elteren},
  {Van Hemelryck}, {van Leeuwen}, {Vaschetto}, {Vecchiato}, {Veljanoski},
  {Viala}, {Vicente}, {Vogt}, {von Essen}, {Voss}, {Votruba}, {Voutsinas},
  {Walmsley}, {Weiler}, {Wertz}, {Wevers}, {Wyrzykowski}, {Yoldas},
  {{\v{Z}}erjal}, {Ziaeepour}, {Zorec}, {Zschocke}, {Zucker}, {Zurbach}, \&
  {Zwitter}}]{gaia18}
{Gaia Collaboration}, {Brown}, A.~G.~A., {Vallenari}, A., {et~al.} 2018, \aap,
  616, A1

\bibitem[{{Garnett}(1999)}]{garnett99}
{Garnett}, D.~R. 1999, in IAU Symposium, Vol. 190, New Views of the Magellanic
  Clouds, ed. Y.-H. {Chu}, N.~{Suntzeff}, J.~{Hesser}, \& D.~{Bohlender}, 266

\bibitem[{{Grevesse} {et~al.}(2010){Grevesse}, {Asplund}, {Sauval}, \&
  {Scott}}]{grevesse10}
{Grevesse}, N., {Asplund}, M., {Sauval}, A.~J., \& {Scott}, P. 2010, \apss,
  328, 179

\bibitem[{{Grin} {et~al.}(2017){Grin}, {Ram{\'{\i}}rez-Agudelo}, {de Koter},
  {Sana}, {Puls}, {Brott}, {Crowther}, {Dufton}, {Evans}, {Gr{\"a}fener},
  {Herrero}, {Langer}, {Lennon}, {van Loon}, {Markova}, {de Mink}, {Najarro},
  {Schneider}, {Taylor}, {Tramper}, {Vink}, \& {Walborn}}]{grin17}
{Grin}, N.~J., {Ram{\'{\i}}rez-Agudelo}, O.~H., {de Koter}, A., {et~al.} 2017,
  \aap, 600, A82

\bibitem[{{Hadrava}(1995)}]{hadrava95}
{Hadrava}, P. 1995, \aaps, 114, 393

\bibitem[{{Hillier} \& {Miller}(1998)}]{hillier98}
{Hillier}, D.~J. \& {Miller}, D.~L. 1998, \apj, 496, 407

\bibitem[{{Hillwig} {et~al.}(2006){Hillwig}, {Gies}, {Bagnuolo}, {Huang},
  {McSwain}, \& {Wingert}}]{hillwig06}
{Hillwig}, T.~C., {Gies}, D.~R., {Bagnuolo}, Jr., W.~G., {et~al.} 2006, \apj,
  639, 1069

\bibitem[{{Howarth} {et~al.}(2015){Howarth}, {Dufton}, {Dunstall}, {Evans},
  {Almeida}, {Bonanos}, {Clark}, {Langer}, {Sana}, {Sim{\'o}n-D{\'{\i}}az},
  {Soszy{\'n}ski}, \& {Taylor}}]{howarth15}
{Howarth}, I.~D., {Dufton}, P.~L., {Dunstall}, P.~R., {et~al.} 2015, \aap, 582,
  A73

\bibitem[{{Hunter} {et~al.}(2009){Hunter}, {Brott}, {Langer}, {Lennon},
  {Dufton}, {Howarth}, {Ryans}, {Trundle}, {Evans}, {de Koter}, \&
  {Smartt}}]{hunter09}
{Hunter}, I., {Brott}, I., {Langer}, N., {et~al.} 2009, \aap, 496, 841

\bibitem[{{Hunter} {et~al.}(2008){Hunter}, {Brott}, {Lennon}, {Langer},
  {Dufton}, {Trundle}, {Smartt}, {de Koter}, {Evans}, \& {Ryans}}]{hunter08}
{Hunter}, I., {Brott}, I., {Lennon}, D.~J., {et~al.} 2008, \apjl, 676, L29

\bibitem[{{Hunter} {et~al.}(2007){Hunter}, {Dufton}, {Smartt}, {Ryans},
  {Evans}, {Lennon}, {Trundle}, {Hubeny}, \& {Lanz}}]{hunter07}
{Hunter}, I., {Dufton}, P.~L., {Smartt}, S.~J., {et~al.} 2007, \aap, 466, 277

\bibitem[{{Ilijic}(2004)}]{ilijic04}
{Ilijic}, S. 2004, in Astronomical Society of the Pacific Conference Series,
  Vol. 318, Spectroscopically and Spatially Resolving the Components of the
  Close Binary Stars, ed. R.~W. {Hilditch}, H.~{Hensberge}, \& K.~{Pavlovski},
  107--110

\bibitem[{{Kato} {et~al.}(2007){Kato}, {Nagashima}, {Nagayama}, {Kurita},
  {Koerwer}, {Kawai}, {Yamamuro}, {Zenno}, {Nishiyama}, {Baba}, {Kadowaki},
  {Haba}, {Hatano}, {Shimizu}, {Nishimura}, {Nagata}, {Sato}, {Murai},
  {Kawazu}, {Nakajima}, {Nakaya}, {Kandori}, {Kusakabe}, {Ishihara},
  {Kaneyasu}, {Hashimoto}, {Tamura}, {Tanab{\'e}}, {Ita}, {Matsunaga},
  {Nakada}, {Sugitani}, {Wakamatsu}, {Glass}, {Feast}, {Menzies}, {Whitelock},
  {Fourie}, {Stoffels}, {Evans}, \& {Hasegawa}}]{kato07}
{Kato}, D., {Nagashima}, C., {Nagayama}, T., {et~al.} 2007, \pasj, 59, 615

\bibitem[{{Kobulnicky} {et~al.}(2014){Kobulnicky}, {Kiminki}, {Lundquist},
  {Burke}, {Chapman}, {Keller}, {Lester}, {Rolen}, {Topel}, {Bhattacharjee},
  {Smullen}, {Vargas {\'A}lvarez}, {Runnoe}, {Dale}, \&
  {Brotherton}}]{kobulnicky14}
{Kobulnicky}, H.~A., {Kiminki}, D.~C., {Lundquist}, M.~J., {et~al.} 2014,
  \apjs, 213, 34

\bibitem[{{Kurt} \& {Dufour}(1998)}]{kurtDufour98}
{Kurt}, C.~M. \& {Dufour}, R.~J. 1998, in Revista Mexicana de Astronomia y
  Astrofisica, vol.~27, Vol.~7, Revista Mexicana de Astronomia y Astrofisica
  Conference Series, ed. R.~J. {Dufour} \& S.~{Torres-Peimbert}, 202

\bibitem[{{Lamers} {et~al.}(1995){Lamers}, {Snow}, \& {Lindholm}}]{lamers95}
{Lamers}, H.~J.~G.~L.~M., {Snow}, T.~P., \& {Lindholm}, D.~M. 1995, \apj, 455,
  269

\bibitem[{{Maeder} \& {Meynet}(2000)}]{maeder00}
{Maeder}, A. \& {Meynet}, G. 2000, \araa, 38, 143

\bibitem[{{Mahy} {et~al.}(2017){Mahy}, {Damerdji}, {Gosset}, {Nitschelm},
  {Eenens}, {Sana}, \& {Klotz}}]{mahy17}
{Mahy}, L., {Damerdji}, Y., {Gosset}, E., {et~al.} 2017, \aap, 607, A96

\bibitem[{{Mahy} {et~al.}(2009){Mahy}, {Naz{\'e}}, {Rauw}, {Gosset}, {De
  Becker}, {Sana}, \& {Eenens}}]{mahy09}
{Mahy}, L., {Naz{\'e}}, Y., {Rauw}, G., {et~al.} 2009, \aap, 502, 937

\bibitem[{{Mahy} {et~al.}(2013){Mahy}, {Rauw}, {De Becker}, {Eenens}, \&
  {Flores}}]{mahy13}
{Mahy}, L., {Rauw}, G., {De Becker}, M., {Eenens}, P., \& {Flores}, C.~A. 2013,
  \aap, 550, A27

\bibitem[{{Mahy} {et~al.}(2015){Mahy}, {Rauw}, {De Becker}, {Eenens}, \&
  {Flores}}]{mahy15}
{Mahy}, L., {Rauw}, G., {De Becker}, M., {Eenens}, P., \& {Flores}, C.~A. 2015,
  \aap, 577, A23

\bibitem[{{Ma{\'\i}z Apell{\'a}niz} {et~al.}(2014){Ma{\'\i}z Apell{\'a}niz},
  {Evans}, {Barb{\'a}}, {Gr{\"a}fener}, {Bestenlehner}, {Crowther},
  {Garc{\'\i}a}, {Herrero}, {Sana}, {Sim{\'o}n-D{\'\i}az}, {Taylor}, {van
  Loon}, {Vink}, \& {Walborn}}]{maiz14}
{Ma{\'\i}z Apell{\'a}niz}, J., {Evans}, C.~J., {Barb{\'a}}, R.~H., {et~al.}
  2014, \aap, 564, A63

\bibitem[{{Marchant} {et~al.}(2016){Marchant}, {Langer}, {Podsiadlowski},
  {Tauris}, \& {Moriya}}]{marchant16}
{Marchant}, P., {Langer}, N., {Podsiadlowski}, P., {Tauris}, T.~M., \&
  {Moriya}, T.~J. 2016, \aap, 588, A50

\bibitem[{{Martins} {et~al.}(2015){Martins}, {Herv{\'e}}, {Bouret},
  {Marcolino}, {Wade}, {Neiner}, {Alecian}, {Grunhut}, \& {Petit}}]{martins15}
{Martins}, F., {Herv{\'e}}, A., {Bouret}, J.-C., {et~al.} 2015, \aap, 575, A34

\bibitem[{{Martins} {et~al.}(2017){Martins}, {Mahy}, \&
  {Herv{\'e}}}]{martins17}
{Martins}, F., {Mahy}, L., \& {Herv{\'e}}, A. 2017, \aap, 607, A82

\bibitem[{{Martins} \& {Plez}(2006)}]{martins06}
{Martins}, F. \& {Plez}, B. 2006, \aap, 457, 637

\bibitem[{{Massey} {et~al.}(2013){Massey}, {Neugent}, {Hillier}, \&
  {Puls}}]{massey13}
{Massey}, P., {Neugent}, K.~F., {Hillier}, D.~J., \& {Puls}, J. 2013, \apj,
  768, 6

\bibitem[{{Mathys}(1988)}]{mathys88}
{Mathys}, G. 1988, \aaps, 76, 427

\bibitem[{{Mathys}(1989)}]{mathys89}
{Mathys}, G. 1989, \aaps, 81, 237

\bibitem[{{McErlean} {et~al.}(1998){McErlean}, {Lennon}, \&
  {Dufton}}]{mcerlean98}
{McErlean}, N.~D., {Lennon}, D.~J., \& {Dufton}, P.~L. 1998, \aap, 329, 613

\bibitem[{{Morrell} {et~al.}(2014){Morrell}, {Massey}, {Neugent}, {Penny}, \&
  {Gies}}]{morrell14}
{Morrell}, N.~I., {Massey}, P., {Neugent}, K.~F., {Penny}, L.~R., \& {Gies},
  D.~R. 2014, \apj, 789, 139

\bibitem[{Nelder \& Mead(1965)}]{nelder65}
Nelder, J.~A. \& Mead, R. 1965, Computer Journal, 7, 308

\bibitem[{{Palate} \& {Rauw}(2012)}]{palate12}
{Palate}, M. \& {Rauw}, G. 2012, \aap, 537, A119

\bibitem[{{Pavlovski} \& {Hensberge}(2010)}]{pavlovski10}
{Pavlovski}, K. \& {Hensberge}, H. 2010, in Astronomical Society of the Pacific
  Conference Series, Vol. 435, Binaries - Key to Comprehension of the Universe,
  ed. A.~{Pr{\v s}a} \& M.~{Zejda}, 207

\bibitem[{{Pavlovski} \& {Southworth}(2012)}]{pavlovski12}
{Pavlovski}, K. \& {Southworth}, J. 2012, in IAU Symposium, Vol. 282, From
  Interacting Binaries to Exoplanets: Essential Modeling Tools, ed. M.~T.
  {Richards} \& I.~{Hubeny}, 359--364

\bibitem[{{Pavlovski} {et~al.}(2018){Pavlovski}, {Southworth}, \&
  {Tamajo}}]{pavlovski18}
{Pavlovski}, K., {Southworth}, J., \& {Tamajo}, E. 2018, \mnras, 481, 3129

\bibitem[{{Pietrzy{\'n}ski} {et~al.}(2013){Pietrzy{\'n}ski}, {Graczyk},
  {Gieren}, {Thompson}, {Pilecki}, {Udalski}, {Soszy{\'n}ski}, {Koz{\l}owski},
  {Konorski}, {Suchomska}, {Bono}, {Moroni}, {Villanova}, {Nardetto},
  {Bresolin}, {Kudritzki}, {Storm}, {Gallenne}, {Smolec}, {Minniti}, {Kubiak},
  {Szyma{\'n}ski}, {Poleski}, {Wyrzykowski}, {Ulaczyk}, {Pietrukowicz},
  {G{\'o}rski}, \& {Karczmarek}}]{pietrzynski13}
{Pietrzy{\'n}ski}, G., {Graczyk}, D., {Gieren}, W., {et~al.} 2013, \nat, 495,
  76

\bibitem[{Press {et~al.}(2007)Press, Teukolsky, Vetterling, \&
  Flannery}]{press07}
Press, W.~H., Teukolsky, S.~A., Vetterling, W.~T., \& Flannery, B.~P. 2007,
  Numerical Recipes 3rd Edition: The Art of Scientific Computing, 3rd edn. (New
  York, NY, USA: Cambridge University Press)

\bibitem[{{Prinja} {et~al.}(1991){Prinja}, {Barlow}, \& {Howarth}}]{prinja91}
{Prinja}, R.~K., {Barlow}, M.~J., \& {Howarth}, I.~D. 1991, \apj, 383, 466

\bibitem[{{Puls} {et~al.}(2005){Puls}, {Urbaneja}, {Venero}, {Repolust},
  {Springmann}, {Jokuthy}, \& {Mokiem}}]{puls05}
{Puls}, J., {Urbaneja}, M.~A., {Venero}, R., {et~al.} 2005, \aap, 435, 669

\bibitem[{{Ram{\'{\i}}rez-Agudelo} {et~al.}(2017){Ram{\'{\i}}rez-Agudelo},
  {Sana}, {de Koter}, {Tramper}, {Grin}, {Schneider}, {Langer}, {Puls},
  {Markova}, {Bestenlehner}, {Castro}, {Crowther}, {Evans}, {Garc{\'{\i}}a},
  {Gr{\"a}fener}, {Herrero}, {van Kempen}, {Lennon}, {Ma{\'{\i}}z
  Apell{\'a}niz}, {Najarro}, {Sab{\'{\i}}n-Sanjuli{\'a}n},
  {Sim{\'o}n-D{\'{\i}}az}, {Taylor}, \& {Vink}}]{ramirez17}
{Ram{\'{\i}}rez-Agudelo}, O.~H., {Sana}, H., {de Koter}, A., {et~al.} 2017,
  \aap, 600, A81

\bibitem[{{Ram{\'{\i}}rez-Agudelo} {et~al.}(2015){Ram{\'{\i}}rez-Agudelo},
  {Sana}, {de Mink}, {H{\'e}nault-Brunet}, {de Koter}, {Langer}, {Tramper},
  {Gr{\"a}fener}, {Evans}, {Vink}, {Dufton}, \& {Taylor}}]{ramirez15}
{Ram{\'{\i}}rez-Agudelo}, O.~H., {Sana}, H., {de Mink}, S.~E., {et~al.} 2015,
  \aap, 580, A92

\bibitem[{{Ram{\'{\i}}rez-Agudelo} {et~al.}(2013){Ram{\'{\i}}rez-Agudelo},
  {Sim{\'o}n-D{\'{\i}}az}, {Sana}, {de Koter}, {Sab{\'{\i}}n-Sanjul{\'{\i}}an},
  {de Mink}, {Dufton}, {Gr{\"a}fener}, {Evans}, {Herrero}, {Langer}, {Lennon},
  {Ma{\'{\i}}z Apell{\'a}niz}, {Markova}, {Najarro}, {Puls}, {Taylor}, \&
  {Vink}}]{ramirez13}
{Ram{\'{\i}}rez-Agudelo}, O.~H., {Sim{\'o}n-D{\'{\i}}az}, S., {Sana}, H.,
  {et~al.} 2013, \aap, 560, A29

\bibitem[{{Raucq} {et~al.}(2017){Raucq}, {Gosset}, {Rauw}, {Manfroid}, {Mahy},
  {Mennekens}, \& {Vanbeveren}}]{raucq17}
{Raucq}, F., {Gosset}, E., {Rauw}, G., {et~al.} 2017, \aap, 601, A133

\bibitem[{{Raucq} {et~al.}(2016){Raucq}, {Rauw}, {Gosset}, {Naz{\'e}}, {Mahy},
  {Herv{\'e}}, \& {Martins}}]{raucq16}
{Raucq}, F., {Rauw}, G., {Gosset}, E., {et~al.} 2016, \aap, 588, A10

\bibitem[{{Rauw} {et~al.}(2009){Rauw}, {Naz{\'e}}, {Fern{\'a}ndez Laj{\'u}s},
  {Lanotte}, {Solivella}, {Sana}, \& {Gosset}}]{rauw09}
{Rauw}, G., {Naz{\'e}}, Y., {Fern{\'a}ndez Laj{\'u}s}, E., {et~al.} 2009,
  \mnras, 398, 1582

\bibitem[{{Repolust} {et~al.}(2004){Repolust}, {Puls}, \&
  {Herrero}}]{repolust04}
{Repolust}, T., {Puls}, J., \& {Herrero}, A. 2004, \aap, 415, 349

\bibitem[{{Sabbi} {et~al.}(2016){Sabbi}, {Lennon}, {Anderson}, {Cignoni}, {van
  der Marel}, {Zaritsky}, {De Marchi}, {Panagia}, {Gouliermis}, {Grebel},
  {Gallagher}, {Smith}, {Sana}, {Aloisi}, {Tosi}, {Evans}, {Arab}, {Boyer}, {de
  Mink}, {Gordon}, {Koekemoer}, {Larsen}, {Ryon}, \& {Zeidler}}]{sabbi16}
{Sabbi}, E., {Lennon}, D.~J., {Anderson}, J., {et~al.} 2016, \apjs, 222, 11

\bibitem[{{Sana} {et~al.}(2013){Sana}, {de Koter}, {de Mink}, {Dunstall},
  {Evans}, {H{\'e}nault-Brunet}, {Ma{\'{\i}}z Apell{\'a}niz},
  {Ram{\'{\i}}rez-Agudelo}, {Taylor}, {Walborn}, {Clark}, {Crowther},
  {Herrero}, {Gieles}, {Langer}, {Lennon}, \& {Vink}}]{sana13}
{Sana}, H., {de Koter}, A., {de Mink}, S.~E., {et~al.} 2013, \aap, 550, A107

\bibitem[{{Sana} {et~al.}(2012){Sana}, {de Mink}, {de Koter}, {Langer},
  {Evans}, {Gieles}, {Gosset}, {Izzard}, {Le Bouquin}, \& {Schneider}}]{sana12}
{Sana}, H., {de Mink}, S.~E., {de Koter}, A., {et~al.} 2012, Science, 337, 444

\bibitem[{{Sana} {et~al.}(2009){Sana}, {Gosset}, \& {Evans}}]{sana09}
{Sana}, H., {Gosset}, E., \& {Evans}, C.~J. 2009, \mnras, 400, 1479

\bibitem[{{Sana} {et~al.}(2008){Sana}, {Gosset}, {Naz{\'e}}, {Rauw}, \&
  {Linder}}]{sana08}
{Sana}, H., {Gosset}, E., {Naz{\'e}}, Y., {Rauw}, G., \& {Linder}, N. 2008,
  \mnras, 386, 447

\bibitem[{{Sana} {et~al.}(2011){Sana}, {James}, \& {Gosset}}]{sana11}
{Sana}, H., {James}, G., \& {Gosset}, E. 2011, \mnras, 416, 817

\bibitem[{{Sana} {et~al.}(2014){Sana}, {Le Bouquin}, {Lacour}, {Berger},
  {Duvert}, {Gauchet}, {Norris}, {Olofsson}, {Pickel}, {Zins}, {Absil}, {de
  Koter}, {Kratter}, {Schnurr}, \& {Zinnecker}}]{sana14}
{Sana}, H., {Le Bouquin}, J.-B., {Lacour}, S., {et~al.} 2014, \apjs, 215, 15

\bibitem[{{Schneider} {et~al.}(2017){Schneider}, {Castro}, {Fossati}, {Langer},
  \& {de Koter}}]{schneider17}
{Schneider}, F.~R.~N., {Castro}, N., {Fossati}, L., {Langer}, N., \& {de
  Koter}, A. 2017, \aap, 598, A60

\bibitem[{{Schneider} {et~al.}(2014){Schneider}, {Langer}, {de Koter}, {Brott},
  {Izzard}, \& {Lau}}]{schneider14}
{Schneider}, F.~R.~N., {Langer}, N., {de Koter}, A., {et~al.} 2014, \aap, 570,
  A66

\bibitem[{{Schneider} {et~al.}(2018){Schneider}, {Sana}, {Evans},
  {Bestenlehner}, {Castro}, {Fossati}, {Gr{\"a}fener}, {Langer},
  {Ram{\'{\i}}rez-Agudelo}, {Sab{\'{\i}}n-Sanjuli{\'a}n},
  {Sim{\'o}n-D{\'{\i}}az}, {Tramper}, {Crowther}, {de Koter}, {de Mink},
  {Dufton}, {Garcia}, {Gieles}, {H{\'e}nault-Brunet}, {Herrero}, {Izzard},
  {Kalari}, {Lennon}, {Ma{\'{\i}}z Apell{\'a}niz}, {Markova}, {Najarro},
  {Podsiadlowski}, {Puls}, {Taylor}, {van Loon}, {Vink}, \&
  {Norman}}]{schneider18}
{Schneider}, F.~R.~N., {Sana}, H., {Evans}, C.~J., {et~al.} 2018, Science, 359,
  69

\bibitem[{{Shenar} {et~al.}(2017){Shenar}, {Richardson}, {Sablowski},
  {Hainich}, {Sana}, {Moffat}, {Todt}, {Hamann}, {Oskinova}, {Sander},
  {Tramper}, {Langer}, {Bonanos}, {de Mink}, {Gr{\"a}fener}, {Crowther},
  {Vink}, {Almeida}, {de Koter}, {Barb{\'a}}, {Herrero}, \&
  {Ulaczyk}}]{shenar17}
{Shenar}, T., {Richardson}, N.~D., {Sablowski}, D.~P., {et~al.} 2017, \aap,
  598, A85

\bibitem[{{Simon} \& {Sturm}(1994)}]{simon94}
{Simon}, K.~P. \& {Sturm}, E. 1994, \aap, 281, 286

\bibitem[{{Sim{\'o}n-D{\'{\i}}az} \& {Herrero}(2014)}]{simondiaz14}
{Sim{\'o}n-D{\'{\i}}az}, S. \& {Herrero}, A. 2014, \aap, 562, A135

\bibitem[{{Stroud} {et~al.}(2010){Stroud}, {Clark}, {Negueruela}, {Roche},
  {Norton}, \& {Vilardell}}]{stroud10}
{Stroud}, V.~E., {Clark}, J.~S., {Negueruela}, I., {et~al.} 2010, \aap, 511,
  A84

\bibitem[{{Taylor} {et~al.}(2011){Taylor}, {Evans}, {Sana}, {Walborn}, {de
  Mink}, {Stroud}, {Alvarez-Candal}, {Barb{\'a}}, {Bestenlehner}, {Bonanos},
  {Brott}, {Crowther}, {de Koter}, {Friedrich}, {Gr{\"a}fener},
  {H{\'e}nault-Brunet}, {Herrero}, {Kaper}, {Langer}, {Lennon}, {Ma{\'{\i}}z
  Apell{\'a}niz}, {Markova}, {Morrell}, {Monaco}, \& {Vink}}]{taylor11}
{Taylor}, W.~D., {Evans}, C.~J., {Sana}, H., {et~al.} 2011, \aap, 530, L10

\bibitem[{{Vink} {et~al.}(2000){Vink}, {de Koter}, \& {Lamers}}]{vink00}
{Vink}, J.~S., {de Koter}, A., \& {Lamers}, H.~J.~G.~L.~M. 2000, \aap, 362, 295

\bibitem[{{Vink} {et~al.}(2001){Vink}, {de Koter}, \& {Lamers}}]{vink01}
{Vink}, J.~S., {de Koter}, A., \& {Lamers}, H.~J.~G.~L.~M. 2001, \aap, 369, 574

\bibitem[{{Walborn}(1971)}]{walborn71}
{Walborn}, N.~R. 1971, \apj, 167, 357

\bibitem[{{Walborn} {et~al.}(2014){Walborn}, {Sana}, {Sim{\'o}n-D{\'{\i}}az},
  {Ma{\'{\i}}z Apell{\'a}niz}, {Taylor}, {Evans}, {Markova}, {Lennon}, \& {de
  Koter}}]{walborn14}
{Walborn}, N.~R., {Sana}, H., {Sim{\'o}n-D{\'{\i}}az}, S., {et~al.} 2014, \aap,
  564, A40

\end{thebibliography}

\longtab{
\begin{landscape}
\begin{longtable}{llrccrrrccccrr}
\caption{\label{tab:parameter}  Stellar parameters. Errors represent 1-$\sigma$.} \\            % title of Table
\hline\hline
Star   &  Sp. Type   &   $\log (L/L_{\odot})$ &  $\teff$  &   $\logg$   &   $R$    &   $\vsini $ &  $\vmac$  &  $Y_{\mathrm{He}}\tablefootmark{a}$   &  $\epsilon_{\mathrm{C}}\tablefootmark{b}$            & $\epsilon_{\mathrm{N}}\tablefootmark{c}$               &  Age    &  $M_{spec} $      &    $M_{evol}$  \\
           &               &                   &    [K]   &              & [\rsun]  &   [\kms] &  [\kms]  &   &   &    &  [Myrs] &  [\msun]  & [\msun]  \\
\hline     
\endfirsthead
\caption{continued.}\\
\hline\hline
Star   &  Sp. Type   &   $\log (L/L_{\odot})$ &  $\teff$  &   $\logg$   &   $R$    &   $\vsini $ &  $\vmac$  &  $Y_{\mathrm{He}}\tablefootmark{a}$   &  $\epsilon_{\mathrm{C}}\tablefootmark{b}$            & $\epsilon_{\mathrm{N}}\tablefootmark{c}$               &  Age    &  $M_{spec} $      &    $M_{evol}$  \\
           &               &                   &    [K]   &              & [\rsun]  &   [\kms] &  [\kms]  &   &   &    &  [Myrs] &  [\msun]  & [\msun] \\
\hline
\endhead
\hline
\endfoot
042-P &  O9.2III & $4.78_{-0.05}^{+0.06}$ & $31.8_{- 1.2}^{+ 1.5}$ & $3.83_{-0.15}^{+0.18}$ & $ 8.1_{- 0.3}^{+ 0.3}$ & $112_{- 26}^{+ 13}$ & $ 82_{- 33}^{+ 25}$ & $10.93 \pm 0.03$ & $ 7.72 \pm 0.06$ & $ 6.92 \pm 0.06$ & $6.2_{-1.0}^{+0.8}$ & $ 16.2_{-  5.7}^{+  6.8}$ & $ 18.2_{-  0.8}^{+  1.0}$ \\ [2pt]

042-S &  B0V     & $4.48_{-0.08}^{+0.08}$ & $30.0_{- 2.0}^{+ 2.0}$ & $3.83_{-0.25}^{+0.25}$ & $ 6.4_{- 0.3}^{+ 0.3}$ & $ 88_{- 25}^{+ 16}$ & $ 79_{- 16}^{+ 22}$ & $10.88 \pm 0.07$ & $ 7.73 \pm 0.07$ & $ 6.90 \pm 0.07$ & $7.8_{-2.5}^{+2.1}$ & $ 10.2_{-  6.0}^{+  6.0}$ & $ 14.6_{-  1.1}^{+  1.1}$ \\ [2pt]

047-P &  O8V     & $4.71_{-0.16}^{+0.16}$ & $33.8_{- 1.8}^{+ 1.8}$ & $4.20_{-0.24}^{+0.24}$ & $ 6.6_{- 1.1}^{+ 1.1}$ & $ 43_{- 18}^{+ 18}$ & $ 14_{- 14}^{+ 26}$ & $10.92 \pm 0.09$ & $ 7.75 \pm 0.03$ & $ 6.91 \pm 0.04$ & $3.7_{-2.2}^{+1.5}$ & $ 24.9_{- 16.2}^{+ 16.2}$ & $ 18.0_{-  1.7}^{+  1.9}$ \\ [2pt]

047-S &  O8.5V   & $4.64_{-0.17}^{+0.17}$ & $32.9_{- 2.3}^{+ 2.3}$ & $4.20_{-0.30}^{+0.30}$ & $ 6.4_{- 1.1}^{+ 1.1}$ & $ 62_{- 33}^{+ 36}$ & $ 44_{- 31}^{+ 34}$ & $10.94 \pm 0.11$ & $ 7.72 \pm 0.04$ & $ 6.95 \pm 0.06$ & $4.1_{-2.8}^{+1.8}$ & $ 23.8_{- 18.4}^{+ 18.4}$ & $ 17.0_{-  1.9}^{+  1.9}$ \\ [2pt]

055-P &  O8V     & $4.96_{-0.04}^{+0.04}$ & $34.7_{- 1.0}^{+ 1.0}$ & $4.13_{-0.11}^{+0.12}$ & $ 8.4_{- 0.2}^{+ 0.2}$ & $ 90_{- 15}^{+ 17}$ & $ 45_{- 28}^{+ 35}$ & $10.94 \pm 0.04$ & $ 7.79 \pm 0.09$ & $ 6.93 \pm 0.08$ & $4.0_{-0.7}^{+0.5}$ & $ 34.5_{-  8.9}^{+  9.7}$ & $ 22.4_{-  0.9}^{+  0.8}$ \\ [2pt]

055-S &  O9V     & $4.94_{-0.04}^{+0.04}$ & $34.5_{- 1.0}^{+ 1.0}$ & $4.14_{-0.12}^{+0.14}$ & $ 8.2_{- 0.2}^{+ 0.2}$ & $ 94_{- 27}^{+ 24}$ & $ 46_{- 30}^{+ 37}$ & $10.94 \pm 0.02$ & $ 7.81 \pm 0.09$ & $ 6.95 \pm 0.14$ & $4.1_{-0.7}^{+0.6}$ & $ 34.2_{-  9.6}^{+ 11.1}$ & $ 21.8_{-  0.8}^{+  0.9}$ \\ [2pt]

061-P &  O9V     & $4.77_{-0.05}^{+0.04}$ & $33.5_{- 0.9}^{+ 0.6}$ & $3.97_{-0.11}^{+0.09}$ & $ 7.2_{- 0.2}^{+ 0.2}$ & $174_{- 28}^{+ 33}$ & $103_{- 63}^{+ 73}$ & $10.97 \pm 0.04$ & $ 7.64 \pm 0.36$ & $ 8.30 \pm 0.11$ & $4.8_{-0.6}^{+0.7}$ & $ 17.8_{-  4.5}^{+  3.7}$ & $ 19.0_{-  0.7}^{+  0.6}$ \\ [2pt]

061-S &  O9III   & $4.75_{-0.05}^{+0.05}$ & $32.9_{- 0.6}^{+ 0.7}$ & $3.68_{-0.06}^{+0.09}$ & $ 7.3_{- 0.3}^{+ 0.3}$ & $152_{- 40}^{+ 20}$ & $ 44_{- 33}^{+ 37}$ & $11.03 \pm 0.05$ & $ 6.00 \pm 0.22$ & $ 8.48 \pm 0.07$ & $5.1_{-0.4}^{+0.5}$ & $  9.2_{-  2.1}^{+  2.7}$ & $ 22.2_{-  1.6}^{+  3.6}$ \\ [2pt]

063-P &  O4.5    & $5.62_{-0.07}^{+0.08}$ & $44.3_{- 0.8}^{+ 1.7}$ & $4.19_{-0.09}^{+0.12}$ & $11.0_{- 0.8}^{+ 0.8}$ & $222_{- 38}^{+ 37}$ & $ 38_{- 30}^{+ 38}$ & $10.91 \pm 0.04$ & $ 7.71 \pm 0.07$ & $ 6.89 \pm 0.07$ & $1.1_{-0.6}^{+0.5}$ & $ 68.2_{- 17.1}^{+ 21.4}$ & $ 47.8_{-  3.9}^{+  3.8}$ \\ [2pt]

063-S &  O5.5    & $5.13_{-0.08}^{+0.08}$ & $39.3_{- 1.7}^{+ 1.7}$ & $4.18_{-0.18}^{+0.18}$ & $ 7.9_{- 0.6}^{+ 0.6}$ & $ 95_{- 39}^{+ 32}$ & $ 43_{- 38}^{+ 32}$ & $10.91 \pm 0.04$ & $ 7.77 \pm 0.04$ & $ 6.92 \pm 0.04$ & $2.1_{-1.1}^{+0.8}$ & $ 34.7_{- 15.3}^{+ 15.3}$ & $ 28.0_{-  2.0}^{+  2.3}$ \\ [2pt]

066-P &  O9V     & $4.54_{-0.13}^{+0.09}$ & $32.8_{- 1.0}^{+ 1.7}$ & $4.08_{-0.13}^{+0.20}$ & $ 5.8_{- 0.8}^{+ 0.5}$ & $117_{- 21}^{+ 17}$ & $ 18_{- 12}^{+ 26}$ & $10.95 \pm 0.04$ & $ 7.73 \pm 0.04$ & $ 6.86 \pm 0.06$ & $3.8_{-1.8}^{+1.2}$ & $ 14.6_{-  4.3}^{+  5.2}$ & $ 16.6_{-  0.9}^{+  1.1}$ \\ [2pt]

066-S &  B0.2V   & $4.08_{-0.18}^{+0.10}$ & $29.0_{- 1.2}^{+ 1.0}$ & $4.00_{-0.23}^{+0.23}$ & $ 4.4_{- 0.8}^{+ 0.4}$ & $ 99_{- 29}^{+ 18}$ & $ 33_{- 20}^{+ 17}$ & $10.94 \pm 0.03$ & $ 7.75 \pm 0.05$ & $ 6.90 \pm 0.07$ & $6.2_{-3.1}^{+2.4}$ & $  6.9_{-  2.9}^{+  3.0}$ & $ 12.0_{-  0.8}^{+  0.7}$ \\ [2pt]

094-P &  O4      & $5.51_{-0.13}^{+0.09}$ & $41.9_{- 0.4}^{+ 0.4}$ & $3.86_{-0.04}^{+0.04}$ & $10.9_{- 1.7}^{+ 1.1}$ & $176_{- 26}^{+ 30}$ & $ 15_{- 12}^{+ 24}$ & $10.94 \pm 0.04$ & $ 7.79 \pm 0.11$ & $ 6.91 \pm 0.06$ & $2.3_{-0.1}^{+0.1}$ & $ 31.4_{-  9.8}^{+  6.5}$ & $ 43.8_{-  3.1}^{+  3.9}$ \\ [2pt]

094-S &  O6      & $5.19_{-0.12}^{+0.14}$ & $40.1_{- 1.4}^{+ 1.4}$ & $4.06_{-0.11}^{+0.11}$ & $ 8.1_{- 1.1}^{+ 1.2}$ & $117_{- 22}^{+ 24}$ & $107_{- 38}^{+ 48}$ & $10.93 \pm 0.05$ & $ 7.78 \pm 0.09$ & $ 6.92 \pm 0.08$ & $2.2_{-0.8}^{+0.5}$ & $ 27.7_{-  9.7}^{+ 11.0}$ & $ 29.8_{-  2.1}^{+  2.7}$ \\ [2pt]

114-P &  O7.5V   & $4.82_{-0.05}^{+0.04}$ & $36.4_{- 1.4}^{+ 0.7}$ & $4.28_{-0.17}^{+0.10}$ & $ 6.5_{- 0.3}^{+ 0.2}$ & $ 58_{- 28}^{+ 23}$ & $ 50_{- 30}^{+ 32}$ & $10.94 \pm 0.04$ & $ 7.76 \pm 0.08$ & $ 7.11 \pm 0.10$ & $2.3_{-1.1}^{+1.1}$ & $ 29.3_{- 11.7}^{+  7.0}$ & $ 21.4_{-  1.0}^{+  1.0}$ \\ [2pt]

114-S &  O9.5V   & $4.58_{-0.08}^{+0.08}$ & $33.0_{- 2.0}^{+ 2.0}$ & $4.25_{-0.25}^{+0.25}$ & $ 6.0_{- 0.3}^{+ 0.3}$ & $142_{- 21}^{+ 16}$ & $ 51_{- 25}^{+ 30}$ & $10.94 \pm 0.03$ & $ 7.72 \pm 0.08$ & $ 7.11 \pm 0.10$ & $3.1_{-2.2}^{+1.8}$ & $ 23.3_{- 13.6}^{+ 13.6}$ & $ 17.4_{-  1.3}^{+  1.3}$ \\ [2pt]

116-P &  O9.7V   & $4.45_{-0.09}^{+0.09}$ & $32.9_{- 2.3}^{+ 2.3}$ & $4.02_{-0.30}^{+0.30}$ & $ 5.1_{- 0.2}^{+ 0.2}$ & $ 45_{- 14}^{+ 24}$ & $ 44_{- 23}^{+ 22}$ & $10.94 \pm 0.03$ & $ 7.75 \pm 0.02$ & $ 6.90 \pm 0.04$ & $3.5_{-2.7}^{+2.4}$ & $ 10.1_{-  7.0}^{+  7.0}$ & $ 15.4_{-  1.2}^{+  1.4}$ \\ [2pt]

116-S &  B1V     & $4.17_{-0.07}^{+0.07}$ & $28.1_{- 1.6}^{+ 1.6}$ & $3.98_{-0.10}^{+0.12}$ & $ 5.1_{- 0.2}^{+ 0.2}$ & $134_{- 23}^{+ 19}$ & $ 35_{- 19}^{+ 17}$ & $10.94 \pm 0.04$ & $ 7.73 \pm 0.03$ & $ 7.04 \pm 0.08$ & $9.1_{-2.7}^{+2.4}$ & $  9.2_{-  2.2}^{+  2.6}$ & $ 11.8_{-  0.7}^{+  0.8}$ \\ [2pt]

140-P &  O7.5V   & $4.78_{-0.05}^{+0.05}$ & $36.4_{- 1.4}^{+ 1.4}$ & $4.30_{-0.15}^{+0.15}$ & $ 6.2_{- 0.1}^{+ 0.1}$ & $ 73_{- 15}^{+ 13}$ & $ 34_{- 22}^{+ 29}$ & $10.94 \pm 0.04$ & $ 7.71 \pm 0.07$ & $ 6.95 \pm 0.14$ & $1.4_{-1.1}^{+1.1}$ & $ 27.9_{-  9.7}^{+  9.7}$ & $ 21.4_{-  1.1}^{+  1.2}$ \\ [2pt]

140-S &  O7.5V   & $4.71_{-0.02}^{+0.07}$ & $35.9_{- 0.7}^{+ 2.0}$ & $4.32_{-0.12}^{+0.12}$ & $ 5.9_{- 0.1}^{+ 0.2}$ & $ 71_{- 20}^{+ 22}$ & $ 36_{- 16}^{+ 39}$ & $10.96 \pm 0.05$ & $ 7.72 \pm 0.07$ & $ 7.04 \pm 0.12$ & $0.7_{-0.7}^{+1.1}$ & $ 26.3_{-  7.3}^{+  7.5}$ & $ 21.0_{-  0.9}^{+  1.0}$ \\ [2pt]

174-P &  O7.5V   & $5.00_{-0.04}^{+0.04}$ & $35.9_{- 0.6}^{+ 0.8}$ & $4.02_{-0.08}^{+0.07}$ & $ 8.2_{- 0.3}^{+ 0.4}$ & $161_{- 39}^{+ 20}$ & $  8_{-  8}^{+ 19}$ & $10.91 \pm 0.04$ & $ 7.90 \pm 0.11$ & $ 6.85 \pm 0.08$ & $3.7_{-0.5}^{+0.4}$ & $ 25.4_{-  5.2}^{+  4.7}$ & $ 23.4_{-  0.7}^{+  1.0}$ \\ [2pt]

174-S &  O9.7V   & $4.26_{-0.05}^{+0.08}$ & $31.4_{- 0.9}^{+ 2.0}$ & $4.30_{-0.15}^{+0.15}$ & $ 4.6_{- 0.2}^{+ 0.3}$ & $ 92_{- 15}^{+ 19}$ & $ 37_{- 29}^{+ 40}$ & $10.95 \pm 0.03$ & $ 7.71 \pm 0.11$ & $ 6.95 \pm 0.19$ & $1.0_{-1.0}^{+2.0}$ & $ 15.2_{-  5.4}^{+  5.5}$ & $ 14.2_{-  0.6}^{+  0.9}$ \\ [2pt]

176-P &  O6I     & $5.24_{-0.02}^{+0.02}$ & $38.3_{- 0.3}^{+ 0.3}$ & $3.88_{-0.02}^{+0.02}$ & $ 9.4_{- 0.1}^{+ 0.1}$ & $264_{- 21}^{+ 16}$ & $  8_{-  8}^{+ 13}$ & $10.94 \pm 0.04$ & $ 7.95 \pm 0.19$ & $ 7.49 \pm 0.32$ & $3.1_{-0.1}^{+0.1}$ & $ 24.6_{-  1.3}^{+  1.3}$ & $ 32.2_{-  1.4}^{+  1.2}$ \\ [2pt]

176-S &  B0.2V   & $4.15_{-0.12}^{+0.09}$ & $28.5_{- 1.8}^{+ 1.4}$ & $4.27_{-0.10}^{+0.15}$ & $ 4.9_{- 0.2}^{+ 0.2}$ & $179_{- 33}^{+ 40}$ & $ 19_{- 12}^{+ 18}$ & $10.95 \pm 0.05$ & $ 7.71 \pm 0.07$ & $ 6.91 \pm 0.06$ & $2.1_{-1.8}^{+1.4}$ & $ 16.3_{-  2.1}^{+  5.0}$ & $ 12.2_{-  0.8}^{+  1.0}$ \\ [2pt]

187-P &  O8.5V   & $4.78_{-0.04}^{+0.04}$ & $34.5_{- 0.7}^{+ 0.7}$ & $4.26_{-0.13}^{+0.07}$ & $ 6.9_{- 0.2}^{+ 0.2}$ & $104_{- 21}^{+ 31}$ & $ 58_{- 36}^{+ 39}$ & $10.95 \pm 0.04$ & $ 7.79 \pm 0.08$ & $ 6.92 \pm 0.04$ & $3.5_{-0.9}^{+0.8}$ & $ 31.2_{-  9.6}^{+  5.5}$ & $ 20.0_{-  0.7}^{+  0.7}$ \\ [2pt]

187-S &  O9.7V   & $4.43_{-0.08}^{+0.08}$ & $30.9_{- 2.0}^{+ 2.0}$ & $4.24_{-0.20}^{+0.20}$ & $ 5.8_{- 0.3}^{+ 0.3}$ & $188_{- 28}^{+ 32}$ & $  8_{-  8}^{+ 27}$ & $10.94 \pm 0.02$ & $ 7.76 \pm 0.11$ & $ 6.87 \pm 0.06$ & $3.6_{-2.6}^{+2.1}$ & $ 21.0_{-  9.9}^{+  9.9}$ & $ 15.2_{-  1.0}^{+  1.3}$ \\ [2pt]

197-P &  O9.5V   & $5.18_{-0.04}^{+0.04}$ & $33.5_{- 1.0}^{+ 1.1}$ & $4.09_{-0.15}^{+0.12}$ & $11.6_{- 0.3}^{+ 0.3}$ & $141_{- 30}^{+ 15}$ & $ 81_{- 38}^{+ 27}$ & $10.94 \pm 0.04$ & $ 7.76 \pm 0.10$ & $ 7.49 \pm 0.15$ & $4.5_{-0.4}^{+0.4}$ & $ 59.8_{- 20.9}^{+ 16.8}$ & $ 25.4_{-  1.0}^{+  1.0}$ \\ [2pt]

197-S &  O8.5III & $5.24_{-0.05}^{+0.02}$ & $33.6_{- 1.3}^{+ 0.4}$ & $3.70_{-0.12}^{+0.03}$ & $12.2_{- 0.3}^{+ 0.2}$ & $ 64_{- 22}^{+ 20}$ & $ 33_{- 11}^{+ 11}$ & $10.93 \pm 0.04$ & $ 7.65 \pm 0.12$ & $ 7.72 \pm 0.09$ & $4.8_{-0.2}^{+0.3}$ & $ 27.4_{-  7.7}^{+  2.1}$ & $ 25.6_{-  0.8}^{+  0.9}$ \\ [2pt]

217-P &  O4      & $5.57_{-0.10}^{+0.13}$ & $45.0_{- 0.4}^{+ 0.6}$ & $4.10_{-0.03}^{+0.03}$ & $10.1_{- 1.2}^{+ 1.5}$ & $166_{- 30}^{+ 31}$ & $ 58_{- 50}^{+ 59}$ & $10.94 \pm 0.02$ & $ 7.76 \pm 0.04$ & $ 7.30 \pm 0.26$ & $1.3_{-0.3}^{+0.2}$ & $ 46.6_{-  8.8}^{+ 11.2}$ & $ 44.8_{-  2.8}^{+  3.3}$ \\ [2pt]

217-S &  O5.5    & $5.38_{-0.10}^{+0.13}$ & $41.8_{- 0.6}^{+ 1.7}$ & $4.08_{-0.04}^{+0.16}$ & $ 9.4_{- 1.0}^{+ 1.4}$ & $164_{- 31}^{+ 14}$ & $ 59_{- 32}^{+ 26}$ & $10.94 \pm 0.04$ & $ 7.85 \pm 0.07$ & $ 6.96 \pm 0.11$ & $1.9_{-0.3}^{+0.3}$ & $ 38.7_{-  8.0}^{+  8.6}$ & $ 35.4_{-  2.3}^{+  2.7}$ \\ [2pt]

327-P &  O7.5V   & $4.88_{-0.06}^{+0.05}$ & $36.3_{- 0.9}^{+ 0.6}$ & $4.26_{-0.11}^{+0.05}$ & $ 7.0_{- 0.4}^{+ 0.4}$ & $152_{- 24}^{+ 17}$ & $ 56_{- 32}^{+ 27}$ & $10.92 \pm 0.03$ & $ 7.77 \pm 0.05$ & $ 7.28 \pm 0.27$ & $2.5_{-1.0}^{+0.8}$ & $ 32.5_{-  9.1}^{+  5.2}$ & $ 22.2_{-  0.9}^{+  1.0}$ \\ [2pt]

327-S &  O9.5V   & $4.47_{-0.06}^{+0.09}$ & $33.3_{- 1.1}^{+ 2.0}$ & $4.26_{-0.20}^{+0.20}$ & $ 5.2_{- 0.3}^{+ 0.3}$ & $153_{- 29}^{+ 33}$ & $ 54_{- 46}^{+ 41}$ & $10.95 \pm 0.02$ & $ 7.72 \pm 0.07$ & $ 7.04 \pm 0.20$ & $1.2_{-1.2}^{+1.6}$ & $ 17.8_{-  8.5}^{+  8.6}$ & $ 16.8_{-  0.9}^{+  1.1}$ \\ [2pt]

352-P &  O5.5V   & $5.10_{-0.04}^{+0.02}$ & $41.6_{- 1.0}^{+ 0.4}$ & $4.10_{-0.05}^{+0.02}$ & $ 6.8_{- 0.2}^{+ 0.1}$ & $275_{- 40}^{+ 38}$ & $ 46_{- 46}^{+ 51}$ & $10.95 \pm 0.07$ & $ 7.85 \pm 0.09$ & $ 6.91 \pm 0.05$ & $6.6_{-1.0}^{+0.8}$ & $ 21.5_{-  1.9}^{+  1.1}$ & $ 22.8_{-  1.0}^{+  1.0}$ \\ [2pt]

352-S &  O6.5V   & $5.05_{-0.04}^{+0.02}$ & $40.6_{- 0.6}^{+ 0.4}$ & $4.12_{-0.04}^{+0.03}$ & $ 6.8_{- 0.2}^{+ 0.1}$ & $290_{- 36}^{+ 34}$ & $ 54_{- 48}^{+ 52}$ & $10.95 \pm 0.06$ & $ 7.73 \pm 0.07$ & $ 6.92 \pm 0.04$ & $5.7_{-2.4}^{+1.2}$ & $ 22.2_{-  1.9}^{+  1.1}$ & $ 22.6_{-  1.0}^{+  2.2}$ \\ [2pt]

450-P &  O8V     & $5.30_{-0.10}^{+0.21}$ & $33.8_{- 2.3}^{+ 1.7}$ & $3.77_{-0.22}^{+0.16}$ & $13.0_{- 0.6}^{+ 3.0}$ & $380_{- 40}^{+ 42}$ & $  8_{-  8}^{+ 12}$ & $10.94 \pm 0.15$ & $ 7.77 \pm 0.07$ & $ 7.18 \pm 0.43$ & $4.6_{-0.7}^{+1.1}$ & $ 36.1_{- 11.7}^{+ 10.4}$ & $ 25.0_{-  1.3}^{+  4.0}$ \\ [2pt]

450-S &  O9.7I   & $5.46_{-0.04}^{+0.04}$ & $28.3_{- 0.2}^{+ 0.3}$ & $3.07_{-0.03}^{+0.02}$ & $22.2_{- 0.4}^{+ 0.3}$ & $ 97_{- 12}^{+ 12}$ & $100_{- 23}^{+ 25}$ & $11.08 \pm 0.05$ & $ 7.51 \pm 0.16$ & $ 8.38 \pm 0.02$ & $7.3_{-0.5}^{+0.5}$ & $ 21.1_{-  2.1}^{+  2.4}$ & $ 24.6_{-  2.0}^{+  2.0}$ \\ [2pt]

487-P &  O6V     & $4.71_{-0.17}^{+0.18}$ & $38.4_{- 0.6}^{+ 1.5}$ & $4.00_{-0.08}^{+0.09}$ & $ 5.1_{- 1.0}^{+ 1.0}$ & $129_{- 28}^{+ 15}$ & $ 76_{- 10}^{+ 26}$ & $10.92 \pm 0.05$ & $ 7.76 \pm 0.10$ & $ 6.91 \pm 0.04$ & $2.5_{-1.0}^{+0.6}$ & $  9.4_{-  4.1}^{+  4.2}$ & $ 24.8_{-  1.4}^{+  2.2}$ \\ [2pt]

487-S &  O6V     & $4.53_{-0.17}^{+0.17}$ & $35.2_{- 0.5}^{+ 0.5}$ & $3.98_{-0.05}^{+0.05}$ & $ 5.0_{- 1.0}^{+ 1.0}$ & $122_{- 37}^{+ 34}$ & $108_{- 48}^{+ 45}$ & $10.96 \pm 0.03$ & $ 7.79 \pm 0.11$ & $ 6.90 \pm 0.06$ & $4.0_{-0.4}^{+0.4}$ & $  8.6_{-  3.5}^{+  3.5}$ & $ 20.6_{-  0.8}^{+  1.3}$ \\ [2pt]

500-P &  O6V     & $5.29_{-0.06}^{+0.07}$ & $40.7_{- 0.6}^{+ 0.7}$ & $4.03_{-0.06}^{+0.07}$ & $ 8.9_{- 0.6}^{+ 0.6}$ & $147_{- 26}^{+ 27}$ & $ 51_{- 28}^{+ 26}$ & $10.95 \pm 0.04$ & $ 7.71 \pm 0.06$ & $ 6.94 \pm 0.02$ & $2.3_{-0.2}^{+0.2}$ & $ 31.1_{-  5.8}^{+  4.7}$ & $ 32.8_{-  1.3}^{+  1.5}$ \\ [2pt]

500-S &  O6.5V   & $5.21_{-0.07}^{+0.08}$ & $39.4_{- 0.8}^{+ 0.8}$ & $4.03_{-0.08}^{+0.09}$ & $ 8.7_{- 0.6}^{+ 0.8}$ & $164_{- 19}^{+ 11}$ & $ 60_{- 33}^{+ 27}$ & $10.98 \pm 0.05$ & $ 7.77 \pm 0.07$ & $ 6.89 \pm 0.04$ & $2.5_{-0.3}^{+0.2}$ & $ 29.3_{-  5.0}^{+  5.0}$ & $ 29.8_{-  1.2}^{+  1.6}$ \\ [2pt]

508-P &  O9.7V   & $4.75_{-0.04}^{+0.05}$ & $34.3_{- 0.6}^{+ 0.9}$ & $4.26_{-0.09}^{+0.09}$ & $ 6.8_{- 0.3}^{+ 0.3}$ & $ 87_{- 23}^{+ 20}$ & $ 67_{- 14}^{+ 15}$ & $10.94 \pm 0.03$ & $ 7.90 \pm 0.14$ & $ 7.15 \pm 0.22$ & $2.9_{-1.1}^{+0.9}$ & $ 30.2_{-  6.8}^{+  6.8}$ & $ 19.8_{-  0.7}^{+  1.0}$ \\ [2pt]

508-S &  O9.7V   & $4.42_{-0.07}^{+0.05}$ & $32.1_{- 1.5}^{+ 1.1}$ & $4.24_{-0.19}^{+0.16}$ & $ 5.3_{- 0.3}^{+ 0.2}$ & $101_{- 23}^{+ 31}$ & $ 58_{- 32}^{+ 36}$ & $10.93 \pm 0.04$ & $ 7.72 \pm 0.17$ & $ 6.96 \pm 0.13$ & $2.9_{-2.0}^{+1.9}$ & $ 17.6_{-  7.9}^{+  6.7}$ & $ 15.4_{-  0.8}^{+  0.9}$ \\ [2pt]

527-P &  O6.5I   & $6.20_{-0.06}^{+0.06}$ & $34.0_{- 1.1}^{+ 1.1}$ & $3.27_{-0.08}^{+0.08}$ & $36.4_{- 1.9}^{+ 1.9}$ & $ 57_{- 13}^{+  8}$ & $ 92_{- 10}^{+ 10}$ & $11.00 \pm 0.05$ & $ 7.60 \pm 0.22$ & $ 8.61 \pm 0.13$ & $2.0_{-0.1}^{+0.1}$ & $ 89.7_{- 19.1}^{+ 19.1}$ & $ 81.6_{-  7.2}^{+  7.5}$ \\ [2pt]

527-S &  O7I     & $6.16_{-0.06}^{+0.06}$ & $34.7_{- 1.1}^{+ 1.1}$ & $3.35_{-0.09}^{+0.09}$ & $33.1_{- 1.8}^{+ 1.8}$ & $ 77_{- 43}^{+ 25}$ & $ 92_{- 30}^{+ 23}$ & $11.00 \pm 0.05$ & $ 7.60 \pm 0.22$ & $ 8.54 \pm 0.10$ & $2.1_{-0.1}^{+0.1}$ & $ 89.6_{- 20.8}^{+ 20.8}$ & $ 76.4_{-  6.7}^{+  7.1}$ \\ [2pt]

538-P &  O8.5III & $4.96_{-0.07}^{+0.08}$ & $35.6_{- 1.4}^{+ 1.7}$ & $4.15_{-0.17}^{+0.18}$ & $ 7.9_{- 0.6}^{+ 0.6}$ & $158_{- 33}^{+ 28}$ & $110_{- 59}^{+ 47}$ & $10.92 \pm 0.03$ & $ 7.00 \pm 0.26$ & $ 7.73 \pm 0.10$ & $3.3_{-1.3}^{+1.0}$ & $ 32.4_{- 13.5}^{+ 14.2}$ & $ 23.0_{-  1.6}^{+  1.6}$ \\ [2pt]

538-S &  O9.5III & $5.31_{-0.06}^{+0.06}$ & $32.0_{- 0.3}^{+ 0.2}$ & $3.57_{-0.03}^{+0.03}$ & $14.7_{- 1.0}^{+ 1.0}$ & $108_{- 21}^{+ 14}$ & $ 84_{- 12}^{+ 14}$ & $10.97 \pm 0.03$ & $ 5.00 \pm 0.43$ & $ 8.61 \pm 0.03$ & $4.9_{-0.2}^{+0.2}$ & $ 29.1_{-  4.4}^{+  4.4}$ & $ 27.0_{-  1.5}^{+  0.9}$ \\ [2pt]

543-P &  O9.5V   & $4.61_{-0.14}^{+0.09}$ & $33.2_{- 1.4}^{+ 1.4}$ & $4.24_{-0.17}^{+0.20}$ & $ 6.1_{- 0.8}^{+ 0.2}$ & $185_{- 35}^{+ 38}$ & $102_{- 54}^{+ 43}$ & $10.91 \pm 0.03$ & $ 7.77 \pm 0.08$ & $ 6.91 \pm 0.04$ & $2.6_{-2.0}^{+1.6}$ & $ 23.4_{- 11.6}^{+ 10.8}$ & $ 17.4_{-  1.4}^{+  1.5}$ \\ [2pt]

543-S &  O9.7V   & $4.51_{-0.15}^{+0.06}$ & $32.4_{- 0.7}^{+ 1.0}$ & $4.25_{-0.14}^{+0.09}$ & $ 5.7_{- 0.9}^{+ 0.2}$ & $140_{- 30}^{+ 36}$ & $131_{- 38}^{+ 42}$ & $10.95 \pm 0.03$ & $ 7.79 \pm 0.11$ & $ 6.90 \pm 0.06$ & $2.1_{-1.4}^{+1.4}$ & $ 21.2_{-  7.2}^{+  4.5}$ & $ 16.0_{-  1.0}^{+  1.0}$ \\ [2pt]

555-P &  O8.5V   & $4.67_{-0.08}^{+0.07}$ & $33.9_{- 1.8}^{+ 1.4}$ & $4.14_{-0.23}^{+0.20}$ & $ 6.3_{- 0.4}^{+ 0.4}$ & $ 56_{-  9}^{+ 16}$ & $ 29_{-  9}^{+  8}$ & $10.93 \pm 0.03$ & $ 7.78 \pm 0.06$ & $ 7.49 \pm 0.17$ & $3.8_{-1.9}^{+1.5}$ & $ 20.0_{- 10.9}^{+  9.5}$ & $ 18.2_{-  1.2}^{+  1.3}$ \\ [2pt]

555-S &  O9V     & $4.44_{-0.10}^{+0.09}$ & $32.4_{- 2.2}^{+ 2.0}$ & $4.12_{-0.23}^{+0.21}$ & $ 5.3_{- 0.4}^{+ 0.4}$ & $ 88_{- 24}^{+ 34}$ & $ 38_{- 36}^{+ 35}$ & $10.95 \pm 0.03$ & $ 7.64 \pm 0.12$ & $ 7.43 \pm 0.24$ & $3.7_{-2.6}^{+2.2}$ & $ 13.5_{-  7.4}^{+  6.8}$ & $ 15.4_{-  1.2}^{+  1.3}$ \\ [2pt]

563-P &  O9.5V   & $4.63_{-0.10}^{+0.09}$ & $32.4_{- 0.9}^{+ 1.0}$ & $4.18_{-0.11}^{+0.15}$ & $ 6.6_{- 0.7}^{+ 0.4}$ & $214_{- 28}^{+ 22}$ & $ 49_{- 42}^{+ 36}$ & $10.93 \pm 0.03$ & $ 7.73 \pm 0.05$ & $ 7.28 \pm 0.27$ & $3.1_{-1.6}^{+1.1}$ & $ 23.8_{-  5.2}^{+  8.0}$ & $ 17.6_{-  1.1}^{+  1.2}$ \\ [2pt]

563-S &  O9.5V   & $4.52_{-0.12}^{+0.08}$ & $32.4_{- 0.9}^{+ 1.2}$ & $4.17_{-0.12}^{+0.18}$ & $ 5.8_{- 0.6}^{+ 0.4}$ & $173_{- 28}^{+ 31}$ & $ 77_{- 61}^{+ 61}$ & $10.94 \pm 0.02$ & $ 7.72 \pm 0.07$ & $ 6.92 \pm 0.13$ & $3.2_{-1.6}^{+1.3}$ & $ 17.8_{-  4.6}^{+  4.0}$ & $ 16.4_{-  1.1}^{+  1.1}$ \\ [2pt]

642-P &  O5.5V   & $5.08_{-0.13}^{+0.18}$ & $40.7_{- 0.6}^{+ 1.6}$ & $4.15_{-0.06}^{+0.10}$ & $ 7.0_{- 1.0}^{+ 1.4}$ & $117_{- 22}^{+ 33}$ & $ 83_{- 41}^{+ 49}$ & $10.93 \pm 0.05$ & $ 7.77 \pm 0.07$ & $ 6.92 \pm 0.04$ & $1.5_{-0.8}^{+0.5}$ & $ 24.9_{-  6.6}^{+  9.2}$ & $ 29.0_{-  1.9}^{+  2.0}$ \\ [2pt]

642-S &  O9V     & $4.68_{-0.19}^{+0.19}$ & $34.8_{- 1.9}^{+ 2.5}$ & $4.13_{-0.24}^{+0.22}$ & $ 6.0_{- 1.1}^{+ 1.1}$ & $108_{- 18}^{+ 25}$ & $ 62_{- 28}^{+ 24}$ & $10.94 \pm 0.04$ & $ 7.72 \pm 0.08$ & $ 6.89 \pm 0.05$ & $3.0_{-2.2}^{+1.4}$ & $ 17.8_{- 10.9}^{+ 11.4}$ & $ 18.6_{-  2.0}^{+  2.2}$ \\ [2pt]

652-P &  O8V     & $5.35_{-0.04}^{+0.06}$ & $32.1_{- 0.7}^{+ 0.9}$ & $3.32_{-0.05}^{+0.07}$ & $15.4_{- 0.3}^{+ 0.7}$ & $224_{- 45}^{+ 38}$ & $ 56_{- 50}^{+ 42}$ & $10.95 \pm 0.03$ & $ 7.63 \pm 0.13$ & $ 7.62 \pm 0.24$ & $4.6_{-0.5}^{+0.5}$ & $ 18.0_{-  1.8}^{+  3.2}$ & $ 29.6_{-  2.5}^{+  2.3}$ \\ [2pt]

652-S &  B1I     & $4.92_{-0.07}^{+0.06}$ & $23.9_{- 0.3}^{+ 0.5}$ & $2.81_{-0.03}^{+0.03}$ & $16.8_{- 0.2}^{+ 0.7}$ & $ 96_{- 13}^{+  8}$ & $ 56_{-  9}^{+  9}$ & $11.05 \pm 0.07$ & $ 7.65 \pm 0.14$ & $ 8.40 \pm 0.02$ & $5.4_{-1.8}^{+1.1}$ & $  6.6_{-  2.0}^{+  2.0}$ & $ 24.8_{-  3.9}^{+ 11.5}$ \\ [2pt]

661-P &  O6.5V   & $4.96_{-0.02}^{+0.04}$ & $38.4_{- 0.4}^{+ 0.9}$ & $4.18_{-0.04}^{+0.09}$ & $ 6.8_{- 0.0}^{+ 0.0}$ & $282_{- 38}^{+ 36}$ & $ 47_{- 40}^{+ 45}$ & $10.92 \pm 0.04$ & $ 7.79 \pm 0.08$ & $ 7.36 \pm 0.17$ & $1.4_{-1.0}^{+0.6}$ & $ 25.8_{-  2.4}^{+  4.2}$ & $ 26.0_{-  1.0}^{+  1.3}$ \\ [2pt]

661-S &  O9.7V   & $4.48_{-0.03}^{+0.08}$ & $31.8_{- 0.6}^{+ 1.4}$ & $4.17_{-0.10}^{+0.15}$ & $ 5.7_{- 0.0}^{+ 0.0}$ & $198_{- 33}^{+ 27}$ & $ 87_{- 28}^{+ 33}$ & $10.94 \pm 0.05$ & $ 7.51 \pm 0.28$ & $ 6.96 \pm 0.06$ & $3.6_{-1.7}^{+1.3}$ & $ 17.5_{-  3.6}^{+  6.0}$ & $ 16.6_{-  0.8}^{+  0.9}$ \\ [2pt]

771-P &  O9.7V   & $4.67_{-0.04}^{+0.07}$ & $30.4_{- 0.8}^{+ 1.5}$ & $3.98_{-0.08}^{+0.16}$ & $ 7.8_{- 0.3}^{+ 0.3}$ & $ 92_{- 20}^{+ 19}$ & $ 40_{- 21}^{+ 20}$ & $10.94 \pm 0.03$ & $ 7.76 \pm 0.18$ & $ 6.92 \pm 0.05$ & $6.3_{-1.0}^{+0.9}$ & $ 21.2_{-  4.2}^{+  8.0}$ & $ 17.0_{-  0.8}^{+  0.8}$ \\ [2pt]

771-S &  O9.7V   & $4.50_{-0.04}^{+0.09}$ & $30.1_{- 0.8}^{+ 2.2}$ & $3.98_{-0.12}^{+0.21}$ & $ 6.5_{- 0.2}^{+ 0.4}$ & $ 82_{- 24}^{+  9}$ & $ 33_{- 20}^{+ 18}$ & $10.94 \pm 0.03$ & $ 7.76 \pm 0.18$ & $ 6.92 \pm 0.05$ & $6.7_{-2.1}^{+1.3}$ & $ 14.9_{-  4.3}^{+  7.4}$ & $ 15.4_{-  0.8}^{+  1.1}$ \\ [2pt]

806-P &  O6V     & $5.20_{-0.04}^{+0.04}$ & $40.1_{- 0.3}^{+ 0.3}$ & $4.10_{-0.03}^{+0.03}$ & $ 8.3_{- 0.3}^{+ 0.3}$ & $ 98_{- 21}^{+ 21}$ & $ 40_{- 15}^{+ 18}$ & $10.94 \pm 0.03$ & $ 7.77 \pm 0.03$ & $ 6.88 \pm 0.05$ & $2.1_{-0.2}^{+0.2}$ & $ 31.4_{-  3.4}^{+  3.4}$ & $ 29.8_{-  0.6}^{+  0.8}$ \\ [2pt]

806-S &  O7V     & $4.95_{-0.04}^{+0.04}$ & $38.1_{- 0.7}^{+ 0.7}$ & $4.14_{-0.08}^{+0.07}$ & $ 6.8_{- 0.3}^{+ 0.3}$ & $ 81_{- 19}^{+ 10}$ & $ 45_{-  9}^{+ 22}$ & $10.94 \pm 0.03$ & $ 7.76 \pm 0.06$ & $ 6.86 \pm 0.07$ & $2.1_{-0.7}^{+0.6}$ & $ 23.4_{-  4.8}^{+  4.3}$ & $ 24.4_{-  1.0}^{+  0.9}$ \\ [2pt]

\end{longtable}
\tablefoot{ 
\tablefoottext{a}{Helium baseline in the LMC is $Y_{\mathrm{He}} = 10.93 $.}
\tablefoottext{b}{Carbon baseline in the LMC is $\epsilon_{\mathrm{C}} = 7.75$. }
\tablefoottext{c}{Nitrogen baseline in the LMC is $\epsilon_{\mathrm{C}} = 6.90$.}
}
\end{landscape}
}

\begin{appendix}
\section{Results for individual binary systems}
\label{results}

\subsection{VFTS\,042}
\label{result:object:VFTS042}

VFTS\,042 has a period of $29.3$ days and an eccentricity of $0.18$. Given that the spectral separation between the two components is of the order of the rotation rate, the spectral lines are never fully separated during the orbital cycle. This renders the disentangled spectra somewhat uncertain, which may affect our results. We determine spectral types of O9.2\,III for the primary and B0\,V for the secondary. The effective temperature and the \logg\ of the primary are estimated to 31800\,K and 3.83, respectively, whilst those of the secondary are 30000\,K and 3.83. The primary and secondary rotational periods are similar (projected rotational periods of 3.3 days for each object). We determine spectroscopic masses of 12.72\,\msun\ for the primary and 8.03\,\msun\ for the secondary, with errors of about 4.5\,\msun. These masses are in reasonable agreement with their evolutionary masses given the rather large errors. If we compare these masses to the minimum masses, we can compute an inclination of $34 \pm 5$\degr\ for the system. No clear enrichment in nitrogen or depletion in carbon is derived for either objects. Given the large eccentricity, the two components of this system do not seem to interact with each other. 

\subsection{VFTS\,047}
\label{result:object:VFTS047}

The orbital solution of VFTS\,047 was revised by the application of the spectral disentangling. The orbital period of VFTS\,047 is $P =5.93$ days and its eccentricity $e = 0.13$ compared to the previous period $P =5.93$ days and eccentricity $e=0.05$  given in Paper I. From the disentangled spectra, we derive spectral types of O8\,V for the primary and O8.5\,V for the secondary. We estimate $\teff = 33800$\,K and $\logg=4.20$ for the primary and $\teff=32900$\,K and $\logg=4.20$ for the secondary. The projected rotational velocities are measured to $\vsini= 43$\,\kms\ for the primary and $\vsini=62$\,\kms\ for the secondary. Given the signal-to-noise ratio of the disentangled spectra, the uncertainties on the \vsini\ and \vmac\ are large. The two components are not yet synchronised (projected rotational period of 5.9 days for the primary and 4.0 days for the secondary, with mean errors of 1.5 days). No significant modification of their surface abundances is found for either object. If we compare the minimum masses with the spectroscopic masses for both components, we estimate an inclination $i = 54 \degr \pm 14\degr$. We emphasise that choosing the evolutionary rather than the spectroscopic mass does not change our estimation within the error bars. Both components are thus well within their Roche lobes and there is no sign of mass transfer. 

\subsection{VFTS\,055}
\label{result:object:VFTS055}

The system has a short orbital period ($P_{\mathrm{orb}}=6.45$ days) with an eccentricity of 0.1. VFTS\,055 is an O8\,V $+$ O9\,V binary system. The primary has $\teff=34700$\,K and $\logg=4.13$ and the secondary $\teff=34500$\,K and $\logg=4.14$. Both components have a similar rotational period (projected rotational period of 3.3 days for the primary and 3.1 days for the secondary, with mean errors of 1.3 days) and show surface abundances similar to the LMC baseline. By comparing minimum masses with spectroscopic masses, we roughly estimate an inclination of $37\degr \pm 3\degr$. For this inclination, the two stars do not fill their Roche lobes, and there is no sign of mass transfer. 

\subsection{VFTS\,061}
\label{result:object:VFTS061}

VFTS\,061 has a period of 2.3 days with a circular orbit. The primary is classified as O8.5\,V whilst the secondary is O9\,III. We determine $\teff=33500$K and $\logg=3.97$ for the primary and $\teff=32700$K and $\logg=3.68$ for the secondary. The surface gravity of the secondary is much lower than that of the primary. The secondary has a rotational period similar to the orbital period whilst the primary is super synchronous. The light curve of this system shows eclipses, which allows us to derive an inclination of $i =69.1\degr \pm 0.9\degr$ (see Paper IV). The secondary clearly fills its Roche lobe. We derive depletions in carbon and overabundances in nitrogen for both components. The abundance changes relative to baseline are larger for the secondary. This system is classified as an Algol-type binary in a phase of slow case-A mass transfer. We do not observe an age discrepancy but there is an important mass discrepancy for the secondary. Whilst the ratio between the evolutionary masses is close to unity, the observed mass ratio (from the orbital solution, see Table\,\ref{tab:orbit}) indicates that the primary is twice more massive than the secondary.  

\subsection{VFTS\,063}
\label{result:object:VFTS063}

VFTS\,063 is a system with an 85.8-day orbit and an eccentricity of 0.65. It is composed of two objects, classified as O4.5 for the primary and O5.5 for the secondary. We estimated $\teff = 44300$\,K for the primary and $39300$\,K for the secondary and very similar \logg\ of about 4.20. The primary is a fast rotator with $\vsini \sim 220$\,\kms\ while the secondary has a \vsini\ close to 100\,\kms. The two components are not synchronised (projected rotational period of $2.0 \pm 0.5$ days for the primary and $3.4 \pm 0.9$ days for the secondary). Their surface abundances are similar to the baseline for both objects. From the spectroscopic masses and minimum masses, we infer an inclination of $56\degr \pm 11\degr$. There are no age and mass discrepancies observed for this system. Given that both stars reside well within their Roche lobes, we expect that the system is currently not interacting.

\subsection{VFTS\,066}
\label{result:object:VFTS066}

VFTS\,066 is a short-period ($P_{\mathrm{orb}} = 1.14$ days) system with a circular orbit. The two components are classified as O9\,V for the primary and as B0.2\,V for the secondary. The \teff\ of the primary is 32800\,K and that of the secondary is 29000\,K. Both objects have similar surface gravities reminiscent of dwarf stars ($\sim 4.05$), and projected rotational velocities of about $\sim 100$\,\kms. Their rotational periods indicate two stars in super-synchronous rotation (rotational period of 0.7 days for the two components). Based on the photometry, this system is an over-contact system seen under a low inclination of $17.5_{-2.5}^{+3.2}$\degr\ (Paper IV). We detect no significant change in the surface abundances of the two stars. There is also a large mass discrepancy for the secondary and a large age discrepancy between the two objects. 

\subsection{VFTS\,094}
\label{result:object:VFTS094}

VFTS\,094 has a period of 2.26 days and a circular orbit. We classified this object as an O4 $+$ O6 system. Given that the \ion{He}{ii}~4686 and H$\alpha$ lines are absent from our disentangled spectra and that the spectral types of both components are too early to use Conti's and Mathys' criteria, we cannot infer a luminosity class for the either object. We determined $\teff= 41900$\,K and $\logg=3.86$ for the primary and $\teff = 40100$\,K and $\logg=4.06$ for the secondary. The projected rotational velocity of the primary is estimated to be 175\,\kms\ whilst that of the secondary is 120\,\kms. The two components rotate faster than the orbital period of the system (rotational period of 1.5 and 1.7 days for the primary and the secondary, respectively). From the photometry, the system seems to have an inclination of $28.1_{-3.0}^{+3.4}$\degr\ and the primary is at the limit of filling its Roche lobe (Paper IV).  However, we do not find any evidence of depletion in carbon or enrichment in nitrogen for either object. The primary shows a mass discrepancy (but no age discrepancy for this system). We do not exclude mass transfer even though no clear evidence is found. 

\subsection{VFTS\,114}
\label{result:object:VFTS114}

VFTS\,114 is a system with a period of $27.8$ days and an eccentricity of $0.5$. It is composed of an O7.5\,V primary and of an O9.5\,V secondary. We determine $\teff = 36400$\,K for the primary and $\teff=33000$\,K for the secondary. Both components have a similar surface gravity ($\logg \sim 4.25$). We measure $\vsini = 58$\,\kms\ for the primary and $142$\,\kms\ for the secondary. The two stars are young and their synchronisation has not yet occurred (projected rotational period of $4.9 \pm 1.3$ days for the primary and $1.9 \pm 0.5$ days for the secondary). The comparison between the minimum masses and the spectroscopic masses provides a rough inclination of $47\degr \pm 8\degr$ for the system. From our analysis, both objects present an enrichment in nitrogen (1.25--2.0 times the baseline value) at their surface but their surface carbon content is close to baseline. This system is probably non-interacting given its relatively wide orbit. 

\subsection{VFTS\,116}
\label{result:object:VFTS116}

As for VFTS\,114, this system has a long period (23.9 days) and high eccentricity ($e =0.3$). The spectral types of the primary and the secondary are estimated to be O9.7\,V and B1\,V respectively. We measure an $\teff = 32900$\,K for the primary and $\teff=28100$\,K for the secondary. Both components have a surface gravity of about 4.0. The difference between the projected rotational velocities of the two components is relatively large (45\,\kms\ for the primary and 134\,\kms\ for the secondary), meaning that the components are not synchronised (projected rotational period of $7.5 \pm 1.4$ days for the primary and $2.6 \pm 0.7$ days for the secondary). The comparison between the minimum and spectroscopic masses yields an inclination of $70 \degr \pm 23\degr$. As for VFTS\,114, we derive carbon and nitrogen surface abundance similar to the baseline, within the error bars. This system is also probably non-interacting at the moment given its large eccentricity and the fact that no peculiar surface abundances are detected. However, the fact that we do observe an age discrepancy between the two objects and that the secondary star rotates much faster than the primary is curious. 

\subsection{VFTS\,140}
\label{result:object:VFTS140}

VFTS\,140 is a circular system with a period of 1.6 days. We classify VFTS\,140 as a system composed of two O7.5\,V stars. The stellar parameters (\teff, \logg, \vsini) of both objects are similar to each other with $\teff \sim 36000$\,K, $\logg \sim 4.3,$ and $\vsini \sim 70$\,\kms. {The two stars have projected rotational periods of about 4.0 days, showing that they are synchronised.} This system is probably seen under a small inclination. The comparison between the spectroscopic masses and the minimum masses indeed indicates an inclination of $17 \degr \pm 2\degr$. Given the probable low inclination of this system, we cannot exclude that this system is in contact. Given the S/N of the disentangled spectra, the error bars on the surface abundances are large, but the carbon and nitrogen values are estimated close to baseline. 

\subsection{VFTS\,174}
\label{result:object:VFTS174}

VFTS\,174 is a system with a period of $4.76$ days and an eccentricity of $0.27$. The eccentricity is relatively high with respect to its short orbital period. This system is composed of an O7.5\,V primary and an O9.7\,V secondary. We measure $\teff=35900$\,K for the primary and $31400$\,K for the secondary. The primary has a lower \logg\ ($4.02 \pm 0.10$) than the secondary ($4.30 \pm 0.12$). The two components have a similar rotational period (projected rotational period of $2.7 \pm 1.1$ days for the two objects). From the ratio between minimum masses and spectroscopic masses, we estimate an inclination of $61 \degr \pm 8\degr$. The carbon and nitrogen contents of both components are similar to the LMC abundances. We also observe that the primary is older than the secondary.

\subsection{VFTS\,176}
\label{result:object:VFTS176}

VFTS\,176 is an O6 $+$ B0.2\,V eclipsing system with an orbital period of 1.78 days and a circular orbit. The surface gravity of the primary is smaller than that of the secondary and its $\teff$ is estimated to be 38300\,K whilst that of the secondary is 28500\,K. We determine $\vsini = 265$\,\kms\ for the primary and $\vsini = 180$\,\kms\ for the secondary. The system has an inclination close to 90\degr (Paper IV). The analysis of the light curve shows that the primary fills its Roche lobe. It shows a nitrogen enrichment and a carbon surface abundance close to the LMC baseline whilst these abundances for the secondary do not show any variations. In this semi-detached system, the most massive star fills its Roche lobe. The primary is synchronised with the system whilst the secondary is super-synchronous. We suspect that the less massive star is likely accreting. We also observe a mass discrepancy for both components. 

\subsection{VFTS\,187}
\label{result:object:VFTS187}

The spectral disentangling suggests a circular orbit for the system rather than an eccentric one ($e \sim 0.2$), as reported in Paper I. The period of this system is 3.5 days. The disentangled spectra give us an O8.5\,V primary and an O9.7\,V secondary. The primary has $\teff= 34500$\,K and the secondary $\teff=30900$\,K. Both components have similar surface gravities ($\logg \sim 4.25$). The two stars rotate faster than the orbital period but we cannot exclude that the two objects are synchronised within the error bars (the rotational period of the primary is about $1.7 \pm 0.8$ days and the secondary has a rotational period of about $0.8 \pm 0.3$ days). From the ratio between minimum mass and spectroscopic mass, the system should have an inclination of $33 \degr \pm 4\degr$. Both stars have surface abundances similar to the LMC baseline.  

\subsection{VFTS\,197}
\label{result:object:VFTS197}

VFTS\,197 has an orbital period of 69.7 days and an eccentricity of 0.1. It is composed of an O8.5\,V primary and an O9\,V secondary. We estimate $\teff = 33500$\,K for the primary and $\teff = 33600$\,K for the secondary. We compute \logg\ of 4.09 for the primary and of 3.70 for the secondary, as well as $\vsini$ of 140\,\kms\ for the primary and of 64\,\kms\ for the secondary. The rotational periods are not synchronised. Both components are enriched in nitrogen but the secondary appears to be more enriched (4 times higher than baseline for the primary and 6.5 times for the secondary). Their carbon surface abundances are similar to the LMC baseline. The comparison between the minimum masses and the spectroscopic masses gives an inclination of $39\degr \pm 4\degr$. No age discrepancy is detected for these objects, but the primary presents a mass discrepancy with the evolutionary mass larger than the spectroscopic one.

\subsection{VFTS\,217}
\label{result:object:VFTS217}

VFTS\,217 is a short-period ($P_{\mathrm{orb}}=1.86$ days) system with a circular orbit. The components are classified as O4 and O5.5. We determine $\teff = 45000$\,K and $\logg = 4.10$ for the primary and $\teff = 41800$\,K and $\logg = 4.08$ for the secondary. This system was observed by OGLE and its light curve displays ellipsoidal variations. Paper IV determines an inclination of $40 \degr \pm 4.0$\degr. The two components almost fill their Roche lobe, but no evidence of mass transfer has been observed. We however observe a small age discrepancy between the two components.
 %According to Fig\,2 of \citet{mandel16}, the relation between the companion mass and the orbital period of the system makes VFTS\,217 a good candidate for chemically homogeneous evolution. The projected rotational  velocities are computed for both object to 165\,\kms, giving equatorial velocities of about 260\,\kms. However, we do not detect any significant variations in the surface abundances for both components. 

\subsection{VFTS\,327}
\label{result:object:VFTS327}

VFTS\,327 is a short-period binary ($P_{\mathrm{orb}}=2.96$ days) with an eccentricity of $e \sim 0.2$. We classified the primary as an O7.5\,V and the secondary as an O9.5\,V. The primary has $\teff=36300$\,K whilst the secondary has $\teff=33300$\,K. The surface gravities and the projected rotational velocities of both components are similar ($\logg \sim 4.30$ and $\vsini = 150$\,\kms). Within the error bars, the two components are almost synchronised (projected rotational period of $2.0 \pm 0.5$ days for the primary and $1.5 \pm 0.4$ days for the secondary). The ratio between minimum mass and spectroscopic mass points to an inclination of $29 \degr \pm 3\degr$. The analysis of their surface abundances shows that both objects are slightly enriched in nitrogen but their carbon abundance remains close to the LMC baseline. We observe an age discrepancy between the two components.

\subsection{VFTS\,352}
\label{result:object:VFTS352}

VFTS\,352 is an overcontact system with a period of 1.12 days, composed by an O4.5\,V primary and an O5.5\,V secondary. We determine $\teff = 41600$\,K for the primary and $\teff=40600$\,K for the secondary. Both components have $\logg \sim 4.10$ and their projected rotational velocities are computed to be 274\,\kms\ and 290\,\kms\ for the primary and secondary, respectively. The two components have rotational periods shorter than the orbital period, and are synchronous. The light curve was analysed by \citet{almeida15} and an independent analysis was performed in Paper IV. The inclination of this system is estimated to be $i = 53.6_{-0.9}^{+1.3}$\degr. A complete analysis of UV and optical data is performed by \citet{abdul-masih19}. We obtain a lower effective temperature for the primary and a higher surface gravity for the secondary. Several arguments can be given to explain these differences. Our analysis relies on different assumptions for the wind parameters whilst the analysis of Abdul-Masih et al. used the UV spectra to determine these parameters. It is not clear whether or not these parameters can have such an impact on the stellar parameters of both components. We also emphasise that for the current analysis we used the CMFGEN atmosphere code whilst Abdul-Masih et al. used FASTWIND \citep{puls05}. Systematic differences between the two codes were reported by \citet{massey13}, with surface gravities determined using FASTWIND being systematically lower by 0.12~dex compared to CMFGEN. Finally, differences in the re-normalisation process (after the spectral disentangling) can also impact the final set of stellar parameters. In any case,   from optical spectra we determine C and N surface abundances consistent with the baseline values, which is in agreement with the results from Abdul-Masih et al. 

\subsection{VFTS\,450}
\label{result:object:VFTS450}
VFTS\,450 has a period of 6.9 days and a circular orbit. This system was analysed by \citet{howarth15}. These authors showed that it is a high-mass analogue of the classical Algol systems. They estimated $\teff = 34000$\,K and $\logg = 3.6$ for the primary and $\teff = 27000$\,K and $\logg = 2.9$ for the secondary. We determine $\teff = 33800$\,K and $\logg = 3.77$ for the primary and $\teff = 28300$\,K and $\logg = 3.07$ for the secondary. The secondary is thus slightly hotter in our analysis. For both components we find significant surface nitrogen enrichment, the more so for the secondary. The disentangled spectrum of the primary remains contaminated by some features of an unknown origin (removal of the nebular emission, accretion disks, and so on). These features prevent us from determining accurate stellar parameters, which explains the large error bars in Table\,\ref{tab:parameter}. The primary rotates extremely fast with $\vsini \sim 380$\,\kms\ while the secondary reaches only about 100\,\kms. We estimate the inclination of this system to be about 65\degr (Paper IV). Together with a strong surface nitrogen enrichment and strong surface carbon depletion, this indicates that the system is probably transferring material to the primary. We observe age and mass discrepancies in this system. 

\subsection{VFTS\,487}
\label{result:object:VFTS487}

VFTS\,487 is composed of two O6\,V stars orbiting with a short period ($P_{\mathrm{orb}} = 4.12$ days) and a small eccentricity ($e = 0.06$) around each other. The two components have surface gravities close to 4.0, and effective temperatures of 38400\,K for the primary and 35200\,K for the secondary. The projected rotational velocities are of about 125\,\kms\ for both components. The two stars are synchronised within the error bars (projected rotational period of $2.0 \pm 0.8$ days for the primary and $2.1 \pm 0.8$ days for the secondary). The comparison between the minimum masses listed in Table\,\ref{tab:orbit} and the spectroscopic masses does not provide an inclination for the system. We do not detect any peculiar surface abundances for these two stars, with surface abundances similar to the LMC baselines but the stars do present age and mass discrepancies. 

\subsection{VFTS\,500}
\label{result:object:VFTS500}

VFTS\,500 is a system with a period of $2.88$ days and a circular orbit. It was analysed by \citet{morrell14}. These latter authors classified the primary as an O5\,V star and the secondary as an O6.5\,V((f)) star. They obtained effective temperatures of $40500 \pm 1000$\,K for the primary and of $40000 \pm 1000$\,K for the secondary as well as $\logg = 4.1\pm 0.1$ and $4.0\pm0.1$, respectively. Our analysis shows that the primary has an $\teff = 40700$\,K and a $\logg = 4.03$ whilst the secondary has an $\teff = 39400$\,K and a $\logg = 4.03$. Both studies thus agree with each other within the error bars. We compute $\vsini = 147$\,\kms for the primary and $\vsini=164$\,\kms\ for the secondary. The rotational periods of the two components are synchronised (rotational period of $2.7 \pm 1.1$ days for the primary and $2.3 \pm 0.9$ days for the secondary). The inclination of the system is estimated to about $61.2 \degr \pm 2.0$\degr\ (Paper IV). We derive surface abundances close to the LMC values, indicating no significant enrichment for both components. No evidence of interaction between both components is detected. 

\subsection{VFTS\,508}
\label{result:object:VFTS508}

The period of VFTS\,508 is 128.59 days and its eccentricity 0.4. This system is composed of an O8.5\,V primary and an O9.7\,V secondary. We determined $\teff = 34300$\,K and $\logg = 4.26$ for the primary and $\teff = 32100$\,K and $\logg = 4.24$ for the secondary. The projected rotational velocity of the primary is  87\,\kms\ and that of the secondary 100\,\kms. The projected rotational period of the primary is $3.3 \pm 1.1$ days and of the secondary $2.2 \pm 1.1$ days. We derive evolutionary and spectroscopic masses between two and three times smaller than the minimum masses provided in Table\,\ref{tab:orbit} (which are indeed high for their spectral types). This could be due to a poor estimation of the orbital period of the system due to the time-scale of the different observations. Finally, within the error bars, both components have surface abundances close to baselines. Given the orbital period and the eccentricity of the orbit, and that the two stars do not fill their Roche lobes, it is unlikely that they interact with each other.

\subsection{VFTS\,527}
\label{result:object:VFTS527}
The orbital parameters were first determined by \citet{taylor11} and then revised in Paper I. The period of this system is estimated to be 153.96 days and its eccentricity is 0.46. VFTS\,527 is composed of two supergiant O-type stars. It is overluminous in X-rays (L. Townsley priv. comm.) probably due to wind-wind collision. We determine $\teff =  34000$\,K, $\logg = 3.27$ and $\vsini = 57$\,\kms\ for the primary and $\teff =  34700$\,K, $\logg = 3.35$ and $\vsini=77$\,\kms\ for the secondary. The spectroscopic masses of the two objects are estimated to about 67.5\,\msun, making it so far one of the most massive binary systems known to contain two evolved Of supergiants. By comparing the spectroscopic masses with the minimum masses, we estimate that the inclination of the system is $62\degr \pm 6\degr$. We also derive an overabundance of nitrogen and a depletion of carbon for both objects. The nitrogen content is estimated to be close to what is expected for CNO equilibrium. VFTS\,527 is a very interesting system, notably because the two components evolve on a wide orbit, and thus without any interactions between them (except at periastron). The system offers an excellent test-case to constrain accurate properties of supergiant O-type stars, similar to e.g. HD\,166734 \citep{mahy17} and V1827\,Cyg \citep{stroud10} in the Galaxy. 

\subsection{VFTS\,538}
\label{result:object:VFTS538}

VFTS\,538 is a circular system with a period of 4.15 days, composed of an O8.5\,III primary and an O9.5\,III secondary. With its surface gravity of about 3.6, the secondary is the brightest component of the system ($\logg \sim 4.2$). We determine $\teff = 35600$\,K for the primary and $\teff = 32000$\,K for the secondary. We measure projected rotational velocities of $158$\,\kms\ for the primary and $108$\,\kms\ for the secondary. The projected rotational period of the primary is $1.6 \pm 1.0$ days and that of the secondary is $4.3 \pm 2.2$ days. The comparison between minimum masses and spectroscopic masses provides an inclination of $47\degr \pm 5\degr$ for the system. The secondary shows strong enrichment in nitrogen and helium whilst the carbon lines have almost vanished. The primary also displays nitrogen enrichment and carbon depletion, but not as extreme as for its companion. The secondary fills its Roche lobe, and is in synchronous rotation with the system, probably transferring its material onto its companion. We observed an age discrepancy between the two objects and a mass discrepancy for the secondary star. We classified VFTS\,538 as a high-mass Algol system.

\subsection{VFTS\,543}
\label{result:object:VFTS543}

VFTS\,543 is a short-period circular system with a period of 1.38 days. The primary has a spectral type of O9.5\,V and the secondary is an O9.7\,V star. We derive $\teff = 33200$\,K and $\logg=4.24$ for the primary and $\teff=32400$\,K and $\logg=4.25$ for the secondary. We also compute $\vsini = 185$\,\kms\ for the primary and $\vsini = 140$\,\kms for the secondary. The two stars are in synchronous rotation and are also synchronised to the system (the rotational period of the primary is $1.0 \pm 0.5$ days and that of the secondary is $1.3 \pm 0.6$ days). The relatively short period of the system implies that both stars occupy a large volume of their Roche lobe, but no evidence of mass transfer is detected yet since their surface abundances are close to the LMC baseline values. The light curve of the system shows ellipsoidal variations that allow us to estimate its inclination, about 37\degr\ (Paper IV). This inclination is consistent with what we get when we compare the minimum masses with their spectroscopic masses. Given the low inclination, we cannot exclude that this system is in contact or close to contact. We do not observe any age or mass discrepancy in this system.

\subsection{VFTS\,555}
\label{result:object:VFTS555}

VFTS\,555 is a system of high eccentricity ($e \sim 0.8$) with a long period ($P_{\mathrm{orb}}=66.1$ days). We attribute to the primary an O8.5\,V spectral type and to the secondary type O9\,V. The effective temperature and surface gravity of the primary are estimated to $33900$\,K and $4.14$, respectively. For the secondary, we find $\teff = 32400$\,K and $\logg = 4.12$. Their projected rotational velocities are computed to be 56\,\kms\ for the primary and 88\,\kms\ for the secondary. This indicates that the components are not synchronised with each other (the projected rotational period is $5.6 \pm 1.5$ days  for the primary and $3.3 \pm 0.9$ days for the secondary). The comparison between minimum and spectroscopic masses roughly provides an inclination of $41 \degr \pm 8\degr$. We derive an overabundance of the nitrogen for both objects (3.5--4 times the baseline value) but the carbon surface abundance is similar to the LMC baseline. Given its relatively wide orbit, this system is probably non-interacting. We do not observe any age or mass discrepancy in this system. 

\subsection{VFTS\,563}
\label{result:object:VFTS563}

This circular system is composed of two O9.5\,V stars orbiting around each other with a period of 1.22 days. We determine effective temperatures of about 32400\,K, surface gravities of about 4.2 for both components and projected rotational velocities of 214\,\kms\ for the primary and 173\,\kms\ for the secondary. The two components are synchronised with each other, but their rotational periods are 1.5 times shorter than the orbital period. The light curve of VFTS\,563 displays ellipsoidal variations that provide an inclination of about $29.7_{-2.0}^{+4.9}$\degr (Paper IV). Nitrogen and carbon surface abundances remains similar to LMC baselines for the two stars, within the error bars. Despite the small separation, there is no evidence for mass transfer. Even though no age or mass discrepancies are detected for this system, we emphasise that the ratio between the evolutionary masses does not match with the mass ratio determined from the orbital solution (Table\,\ref{tab:orbit}). Given the low inclination of the system we cannot however exclude that this system is in contact or close to contact configuration. 

\subsection{VFTS\,642}
\label{result:object:VFTS642}

VFTS\,642 is a circular system with a period of 1.73 days. The primary and secondary are classified as O5.5\,V and O9\,V, respectively. We estimate effective temperatures of $40700$\,K and $34800$\,K and \logg\ of $4.15$ and $4.13$ for the primary and the secondary, respectively. Their projected rotational velocities are quite similar and estimated to be about $110$\,\kms. The two components are synchronised with the system. The light curve displays ellipsoidal variations with grazing eclipses. This allows us to estimate an inclination of $33.7\degr \pm 5.0$\degr (see Paper IV). The surface abundances of the two stars are in line with the LMC baseline. No age and no mass discrepancy is observed in this system.

\subsection{VFTS\,652}
\label{result:object:VFTS652}

Similar to VFTS\,450, VFTS\,652 is composed of a more massive dwarf star (O8\,V) and a cooler more evolved companion (B1\,I). This system with a period of 8.59 days and a circular orbit was also analysed by \citet{howarth15}. These authors estimated the temperature and \logg\ of the primary to be 35000\,K and 3.7 and for the secondary to be 22000\,K and 2.8. We find different values for the effective temperatures and surface gravities, especially for the primary. Indeed, we determine $\teff = 32100$\,K and $\logg = 3.32$ for the primary and $\teff = 23900$\,K and $\logg = 2.81$ for the secondary. We measure a large difference between the projected rotational velocities of the two stars ($\vsini = 224$\,\kms\ for the primary and 96\,\kms\ for the secondary). The analysis of the light curve is presented in Paper IV, where the inclination is estimated to be $63.7_{-4.8}^{+0.9}$\degr (Paper IV). The disentangled spectrum of the primary does not appear to be contaminated by nebular emission as was the case in the analysis of \citet{howarth15}. This contamination may have compromised their analysis. The secondary is highly enriched in helium and nitrogen and strongly depleted in carbon (and in oxygen as can be inferred from the \ion{O}{ii} lines in its spectrum). The primary also shows enrichment in nitrogen and depletion in carbon but to a lesser extent. VFTS\,652 also seems to be a high-mass version of Algol systems such as VFTS\,450. It appears plausible that VFTS\,652 is in a post Roche lobe episode of case-A mass transfer. 

\subsection{VFTS\,661}
\label{result:object:VFTS661}

VFTS\,661 is an eclipsing binary system with an orbital period of 1.27 days in a circular orbit. It is composed of an O6.5\,V primary and an O9.7\,V secondary. We determine $\teff = 38400$\,K and $\logg = 4.18$ for the primary and $\teff = 31800$\,K and $4.17$ for the secondary. The primary has $\vsini \sim 280$\,\kms and the secondary, $\vsini \sim 200$\,\kms. The rotational period of the primary is $1.1 \pm 0.4$ days and that of the secondary is $1.3 \pm 0.6$ days. This indicates that the two components are synchronised with the system. It is not clear whether or not a RLOF episode of mass transfer already occurred in this system. The inclination of the system is $64.5_{-0.6}^{+0.2}$\degr. For both objects, the carbon content is similar to the LMC baseline. Nitrogen is enriched, the more so in the primary (2--3 higher than baseline for the primary). We observe an age discrepancy between the two objects.

\subsection{VFTS\,771}
\label{result:object:VFTS771}

With a period of 29.87 days and an eccentricity of 0.51, it is composed of an O9.7\,V primary and an O9.7\,V secondary. We estimate $\teff = 30400$\,K and $\logg = 3.98$ for the primary and $\teff = 30100$\,K and $\logg = 3.98$ for the secondary. We compute $\vsini = 92$\,\kms\ for the primary and $\vsini = 82$\,\kms\ for the secondary. These values indicate that synchronisation has happened in this system (the projected rotational period of the primary is $4.0 \pm 1.0$ days and that of the secondary is $3.6 \pm 1.2$ days). From the ratio between minimum and spectroscopic masses, we roughly estimate an inclination of $32 \degr \pm 4\degr$. The carbon and nitrogen surface abundances are similar to those of the LMC baseline. We do not observe any age or mass discrepancies in this system. The two components are probably not interacting in this system, given the large eccentricity of the orbit.

\subsection{VFTS\,806}
\label{result:object:VFTS806}

VFTS\,806 is a O6\,V $+$ O7\,V circular system with a period of 2.58 days. This system is reported as a possible runaway. For the primary, we obtain $\teff = 40100$\,K and $\logg = 4.10$ and, for the secondary, $\teff = 38100$\,K and $4.14$. The projected rotational velocities are estimated to be $\vsini = 99$\,\kms\ for the primary and $\vsini =81 $\,\kms\ for the secondary, which indicates that both components are synchronised with the system. From the minimum and the spectroscopic masses, we compute an inclination of $i\sim 37\degr \pm 2\degr$. No peculiarities in surface abundances are detected for both components. There are no age or mass discrepancies in this system.

  \section{Best fits and chi-square maps}
  In this Appendix, we compare the best-fit models to the disentangled spectra and we provide the chi-square maps in effective temperature - $\logg$ for all the components. We also give the spectral energy distribution for each object, the comparison between the disentangled spectra and the observed ones at a particular phase. Finally, we provide an individual HRD and a Kiel (\logg--\teff) diagram of the components in each system.

    \begin{figure*}[t!]
    \centering
        \includegraphics[width=6.cm]{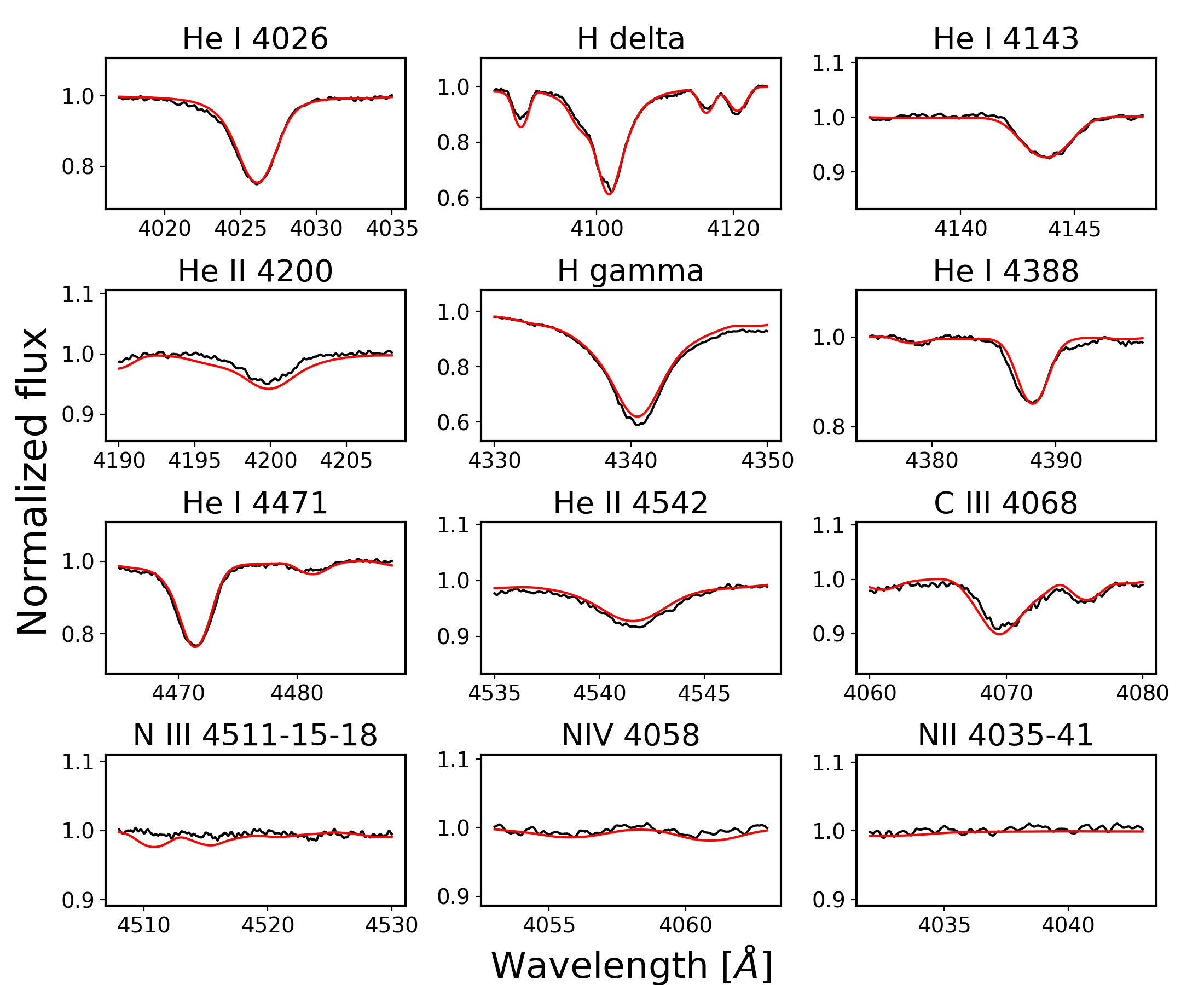}
    \includegraphics[width=7.cm, bb=5 0 453 346,clip]{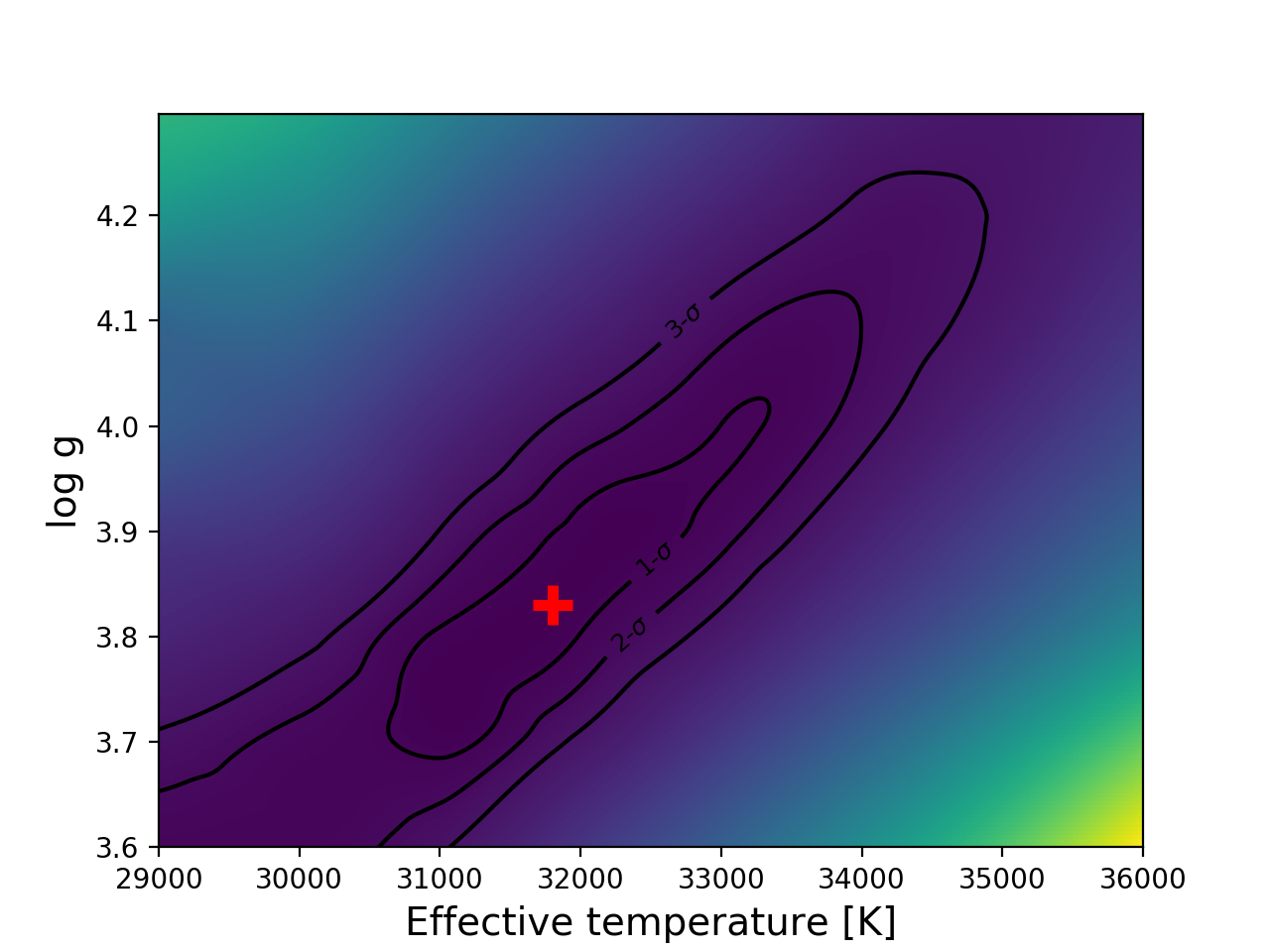}
    \includegraphics[width=6.cm]{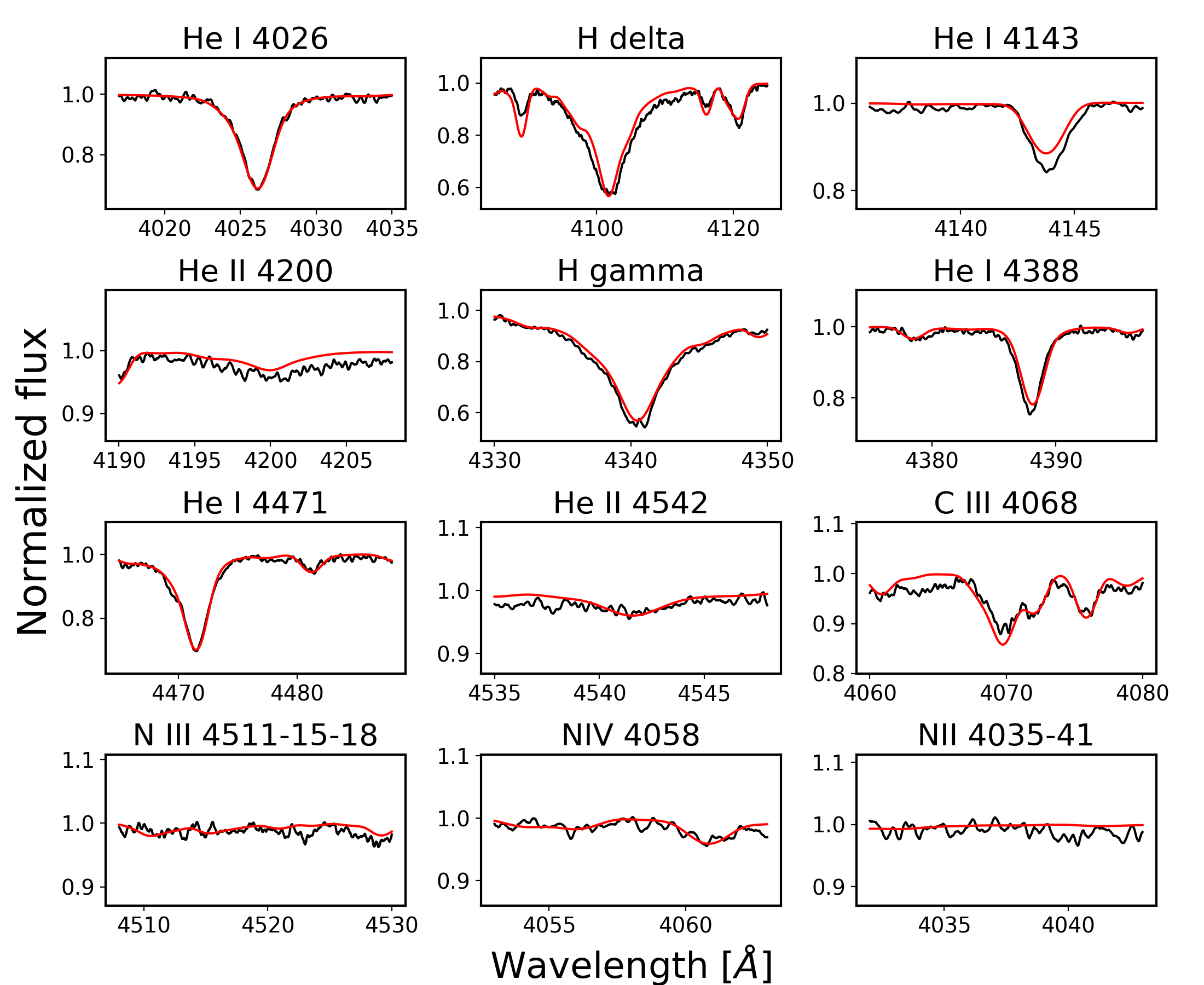}
    \includegraphics[width=7.cm, bb=5 0 453 346,clip]{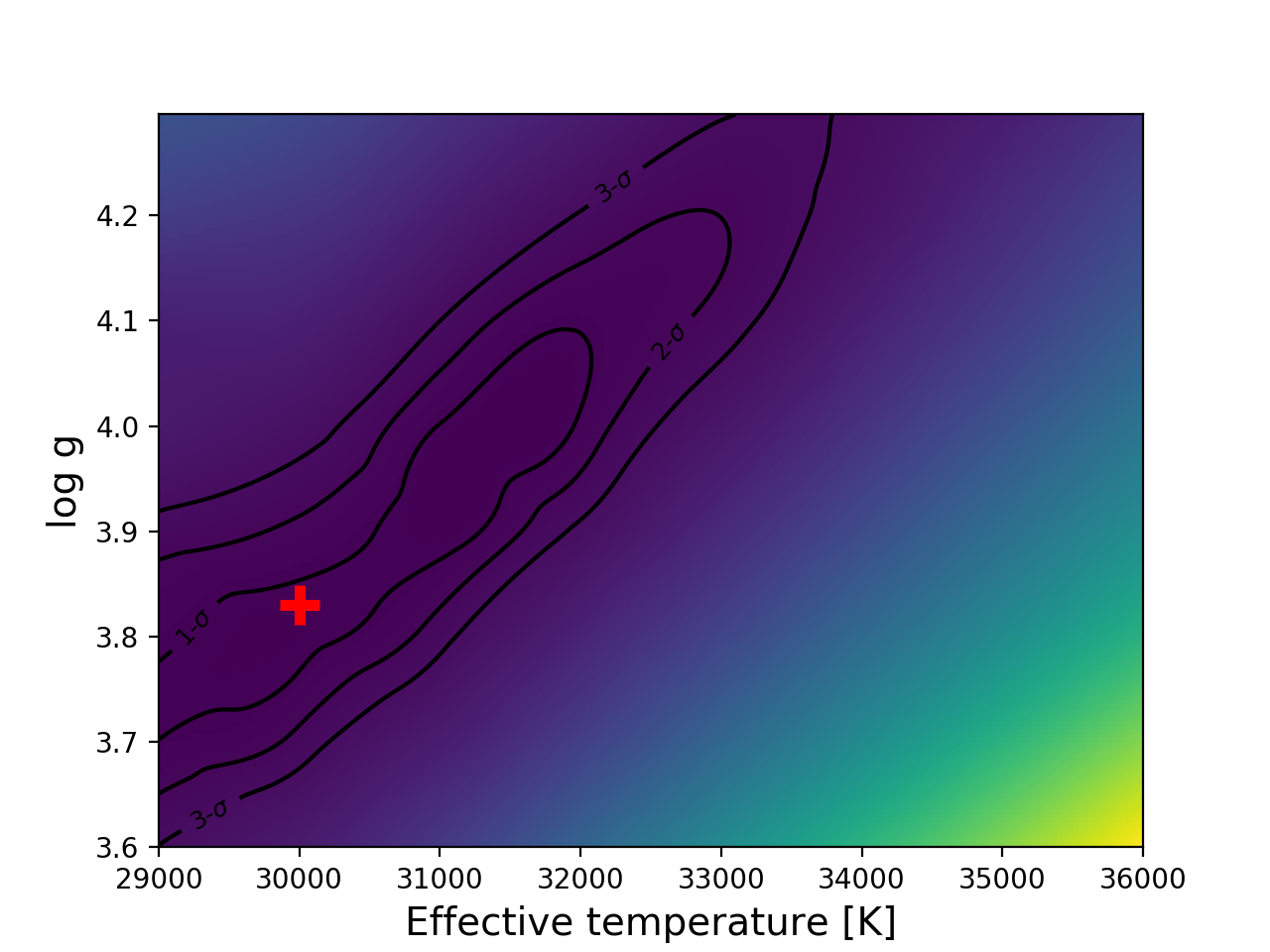}
    \includegraphics[width=7cm]{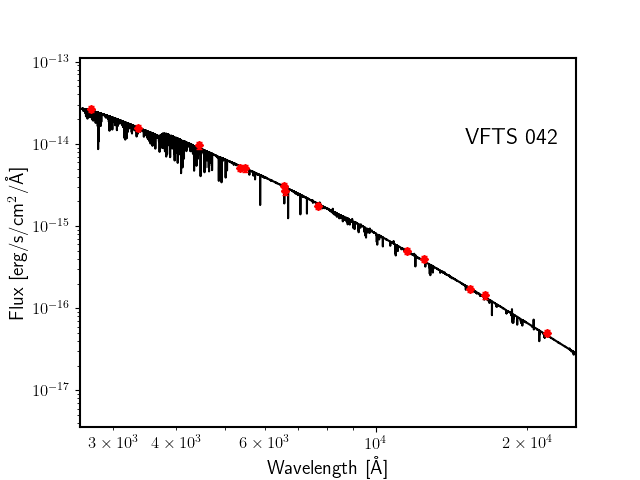}
    \includegraphics[width=6.5cm]{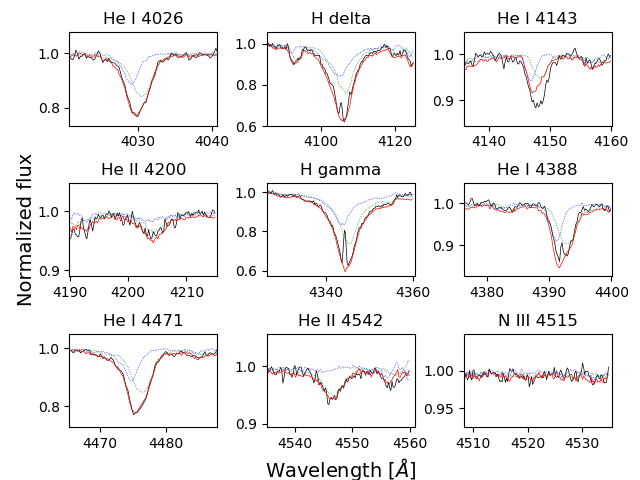}
    \includegraphics[width=7cm, bb=5 0 453 346,clip]{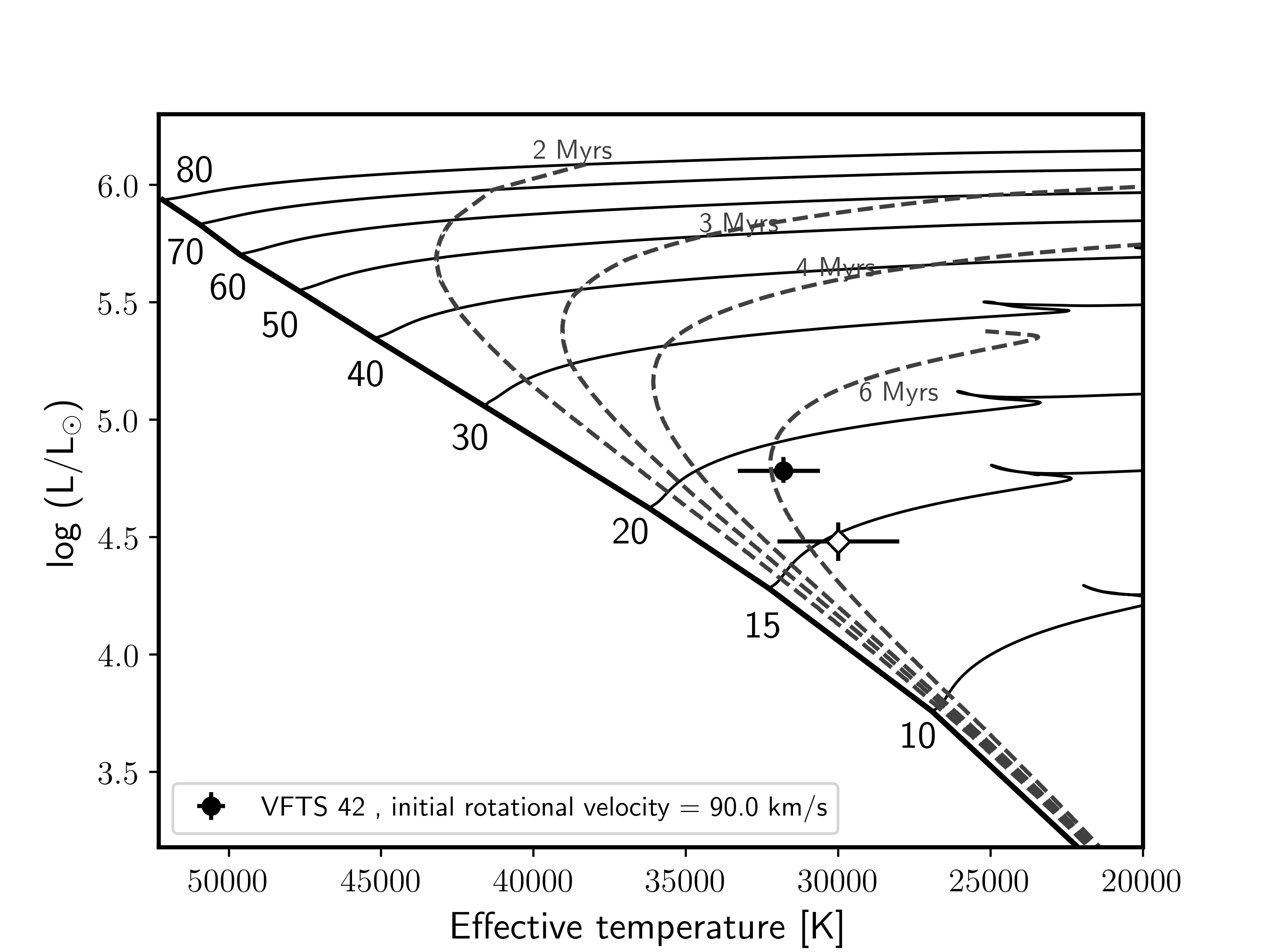}
    \includegraphics[width=7cm, bb=5 0 453 346,clip]{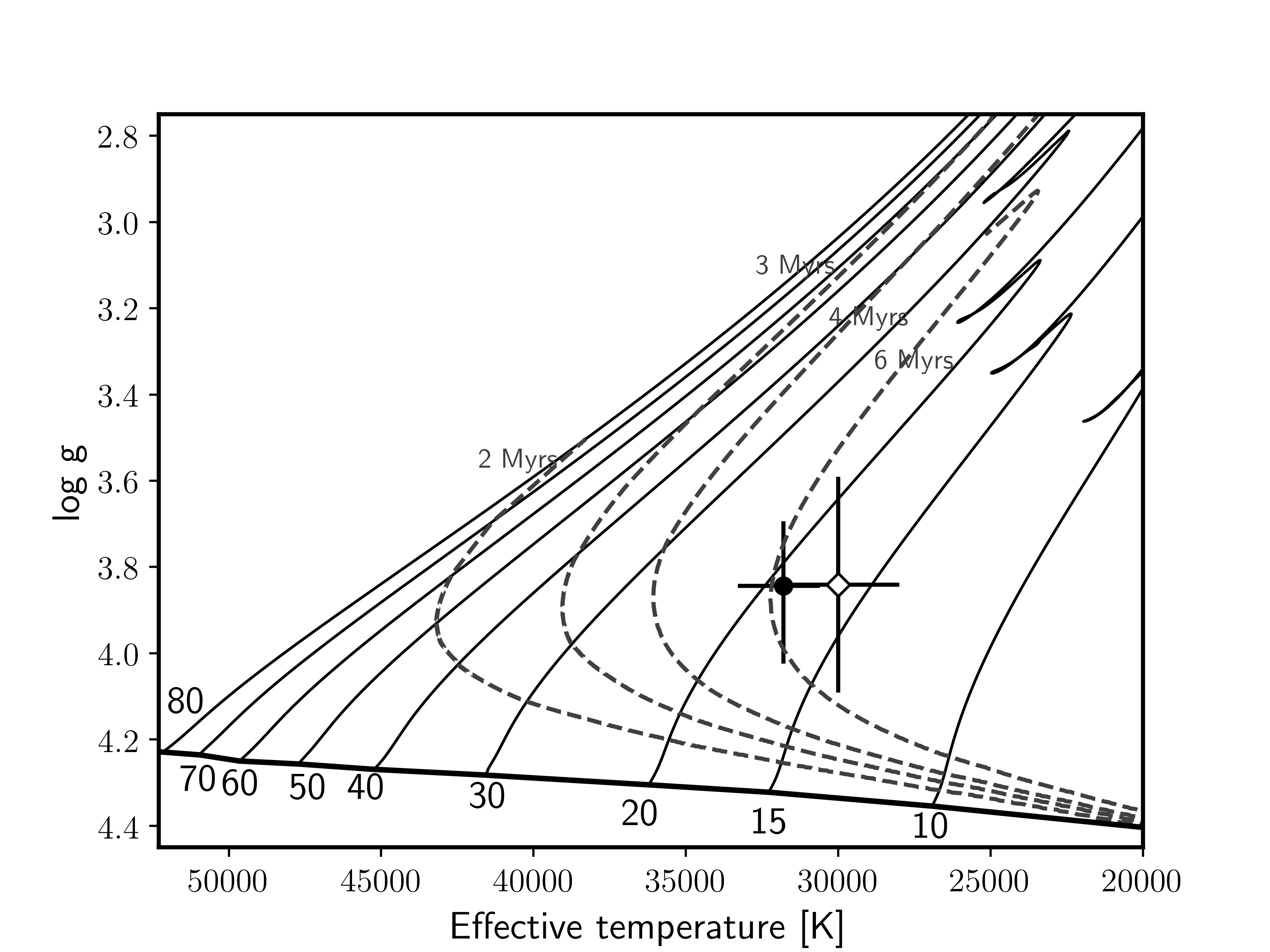}
    \caption{From top to bottom and left to right: (1) Best-fit model of VFTS\,042 primary (red line) compared with the disentangled spectrum (black line), (2) determination of the \teff\ and \logg\ via the chi-square map for this star. The red cross indicates the position of the minimum $\chi^2$, (3) and (4) same as for (1) and (2), but for the secondary. (5) Spectral Energy Distribution of VFTS\,042. (6) Comparison between the disentangled spectra (scaled by the brightness factor of each component and shifted by their radial velocities) and one observed spectrum (7) Individual HRD (Left) and (8) $\logg - \teff$ diagram of the stars of VFTS\,042.} \label{fig:042} 
  \end{figure*}
 \clearpage
    %\subsection{VFTS\,047 primary - O8\,V}

    \begin{figure*}[t!]
    \centering
    \includegraphics[width=6.cm]{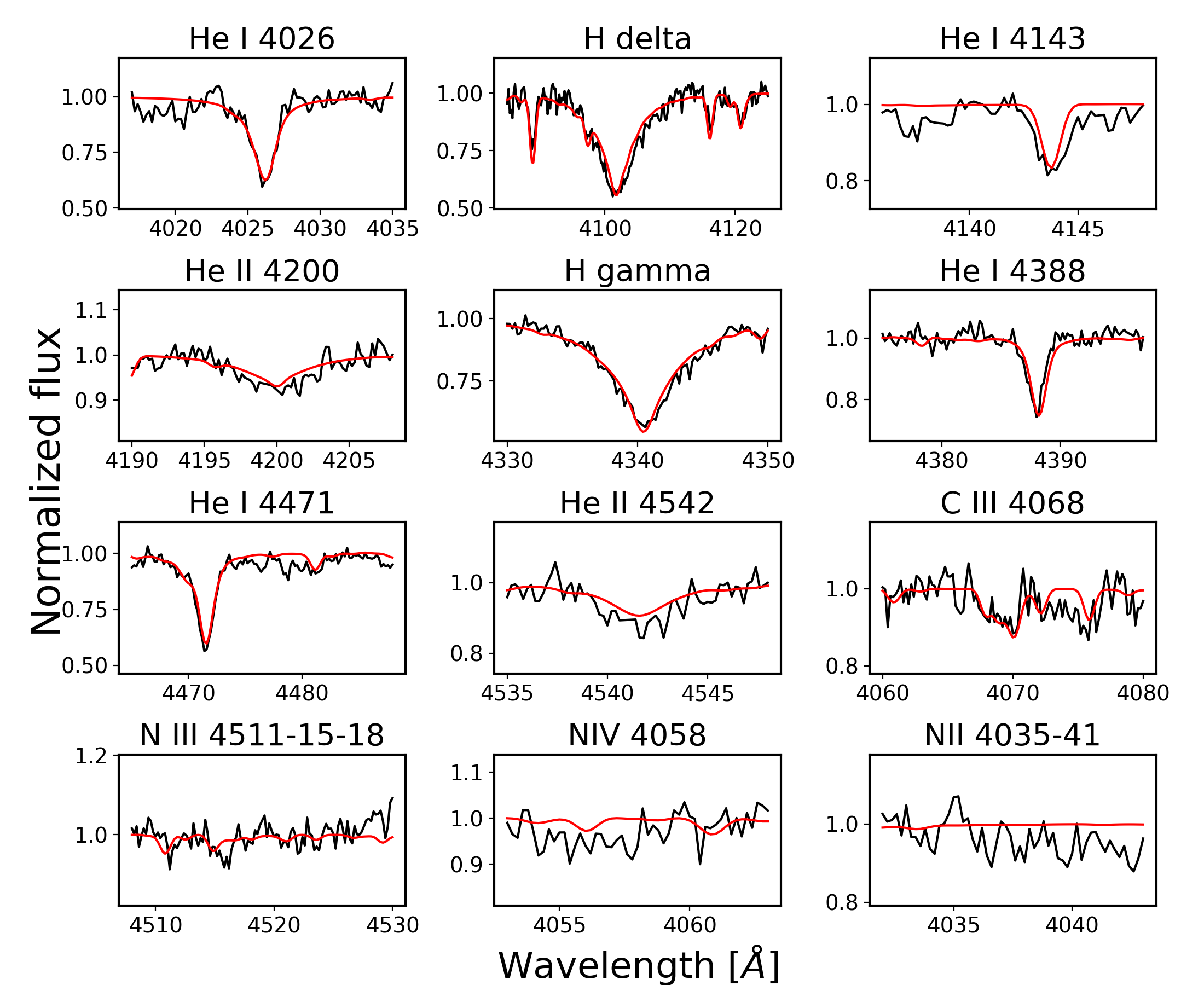}
    \includegraphics[width=7.cm, bb=5 0 453 346,clip]{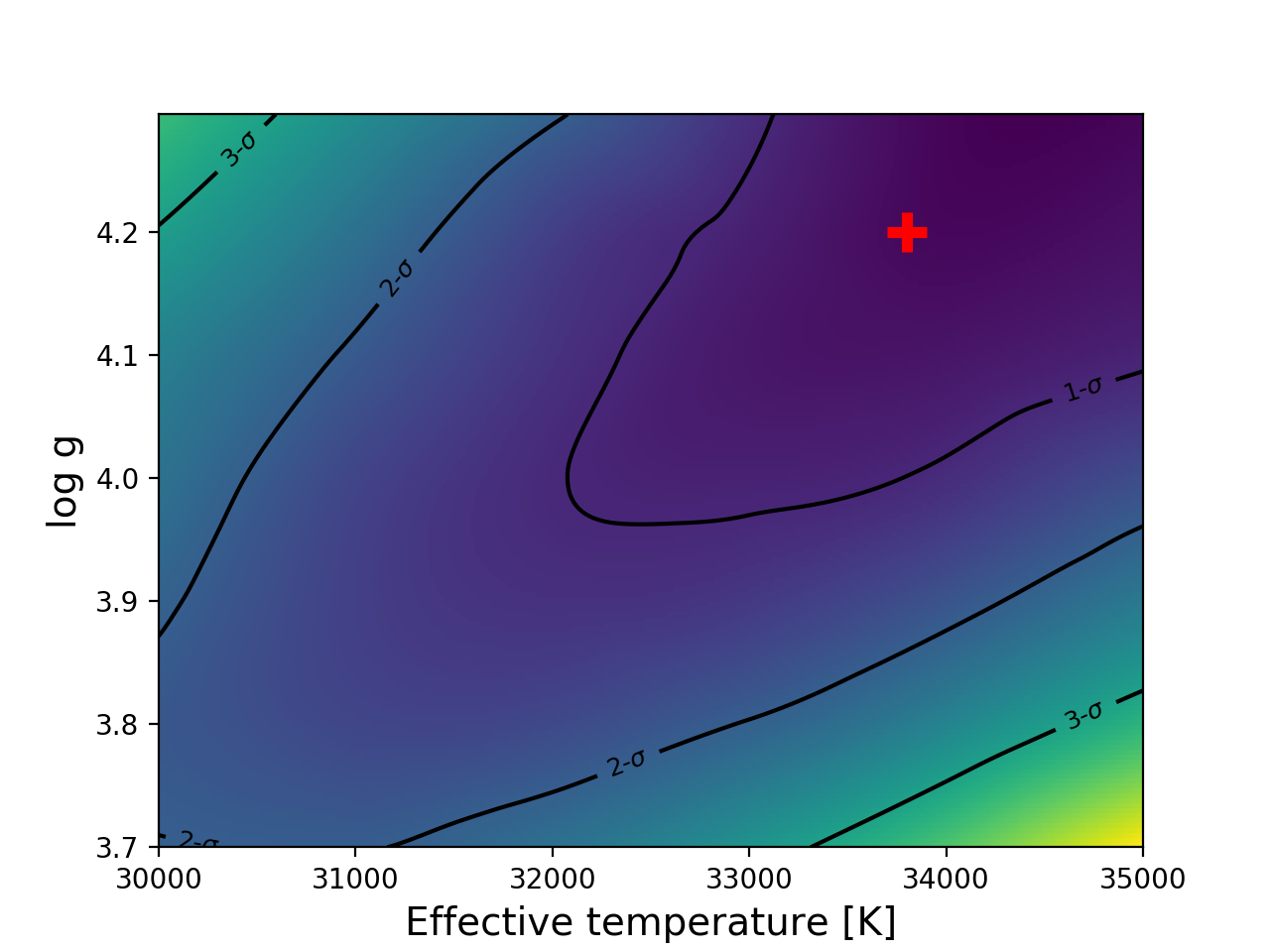}
    \includegraphics[width=6.cm]{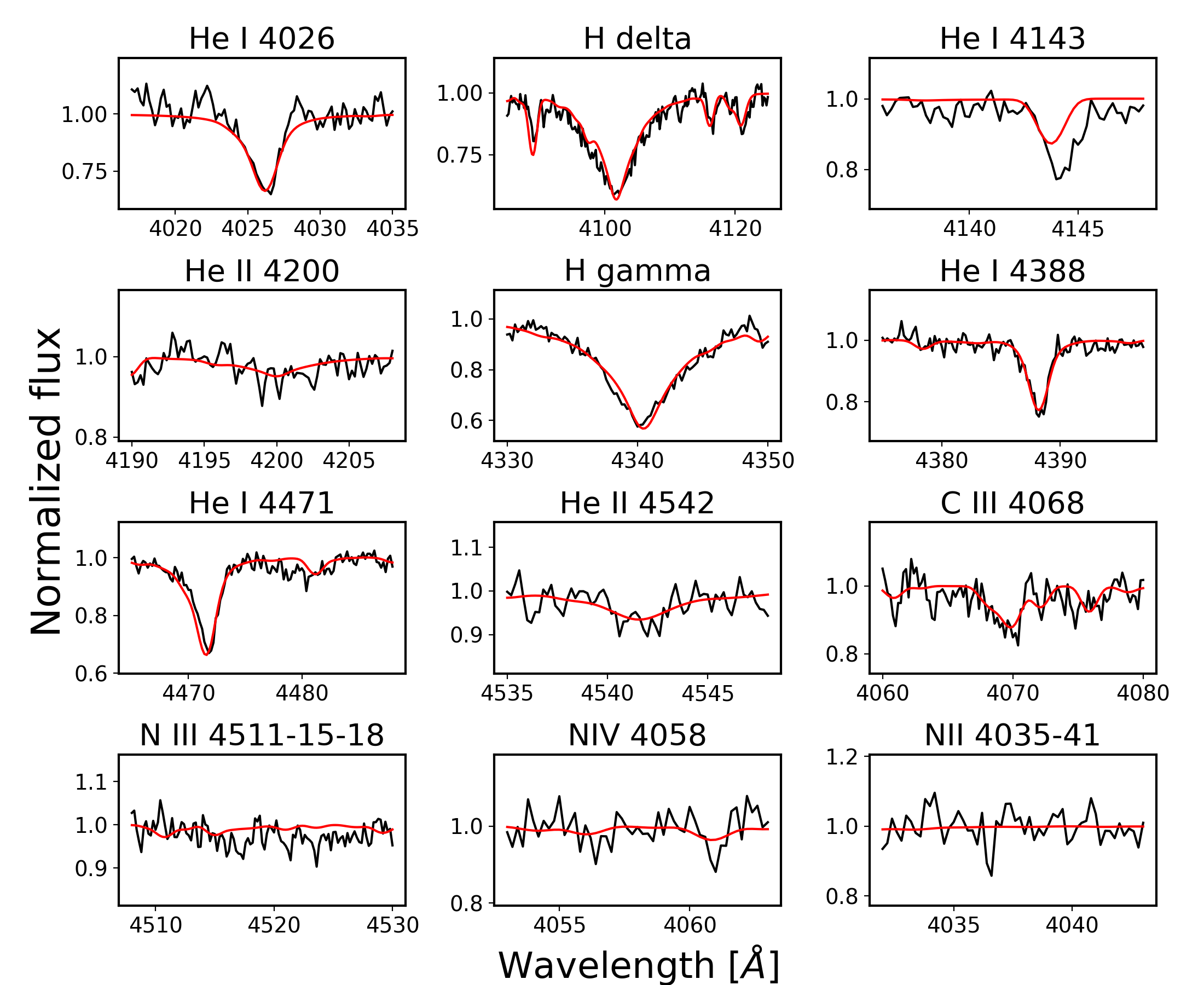}
    \includegraphics[width=7.cm, bb=5 0 453 346,clip]{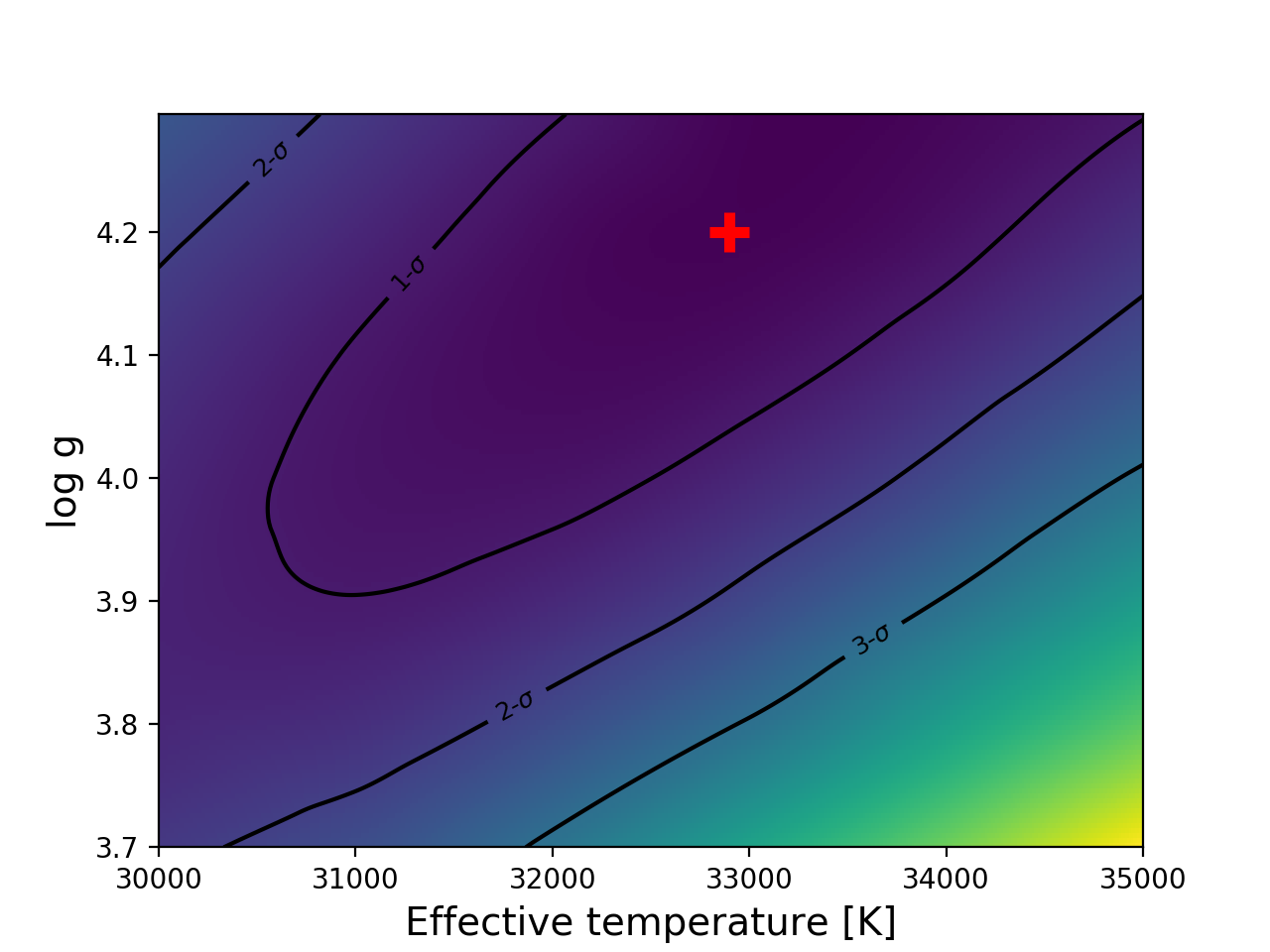}
    \includegraphics[width=7cm]{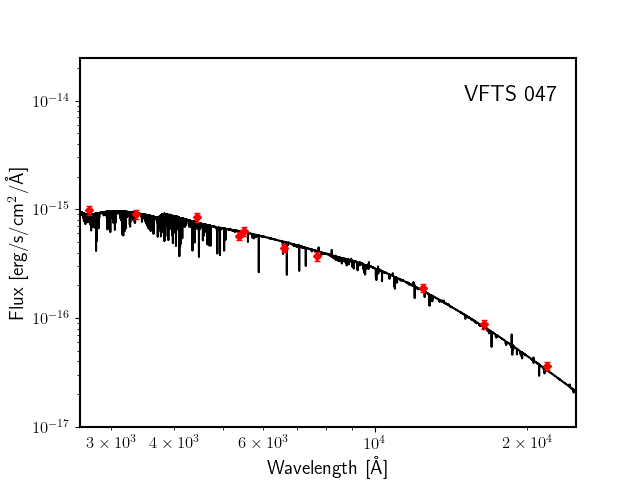}
    \includegraphics[width=6.5cm]{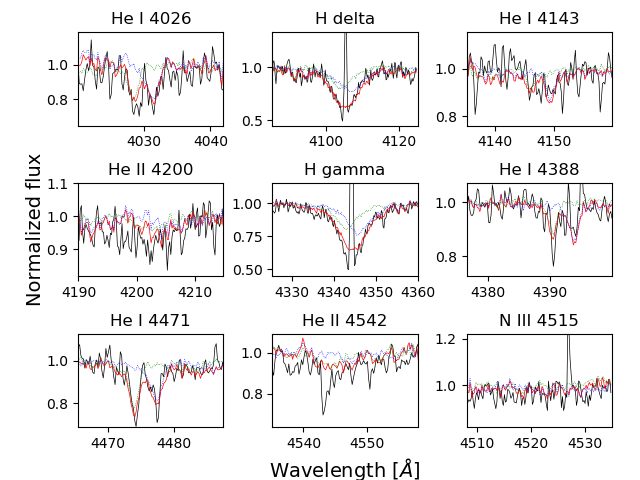}
    \includegraphics[width=7cm, bb=5 0 453 346,clip]{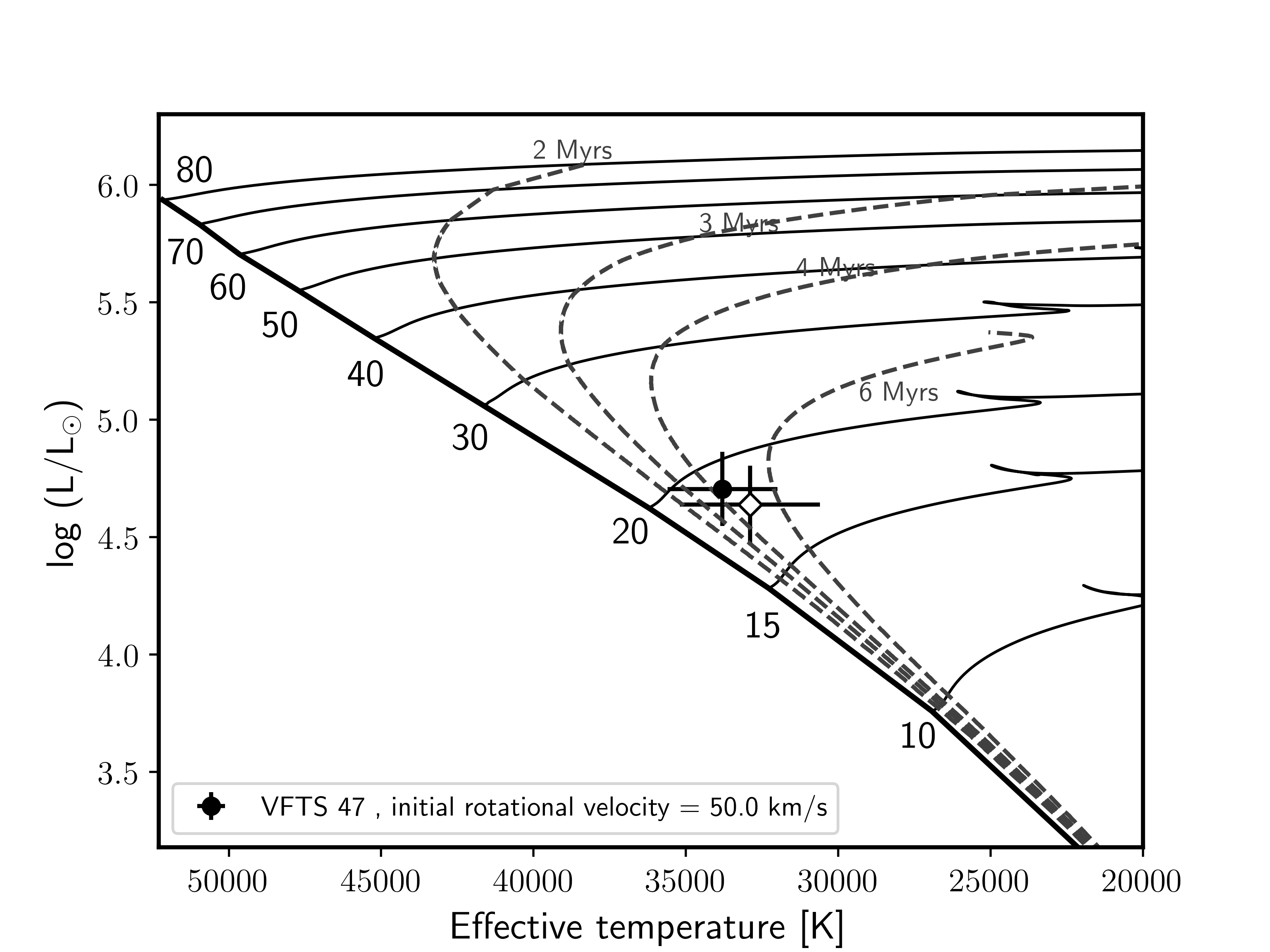}
    \includegraphics[width=7cm, bb=5 0 453 346,clip]{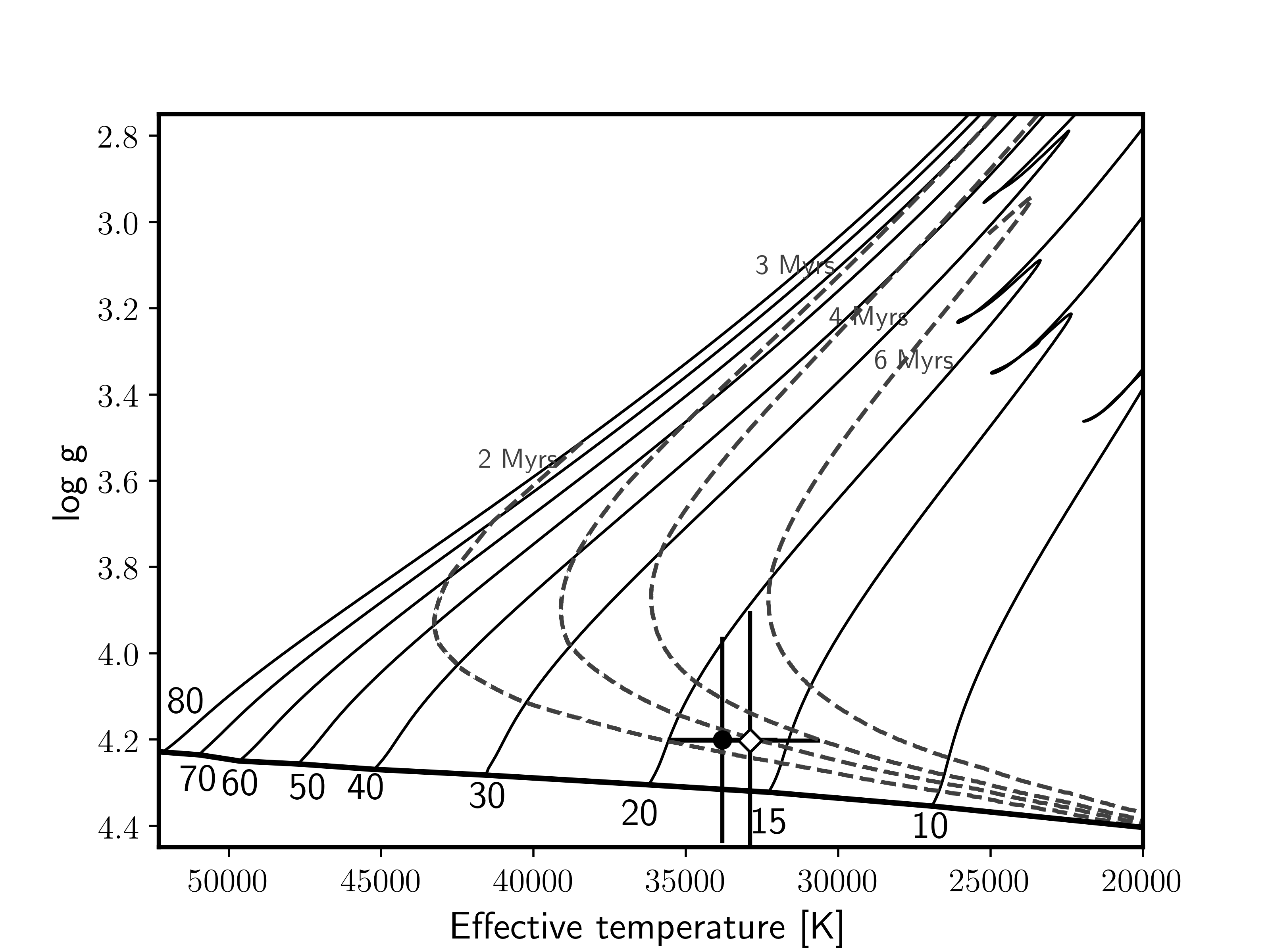}
    \caption{Same as Fig.\,\ref{fig:042} but for VFTS\,047.}\label{fig:047} 
  \end{figure*}
   \clearpage

    \begin{figure*}[t!]
    \centering
    \includegraphics[width=6.cm]{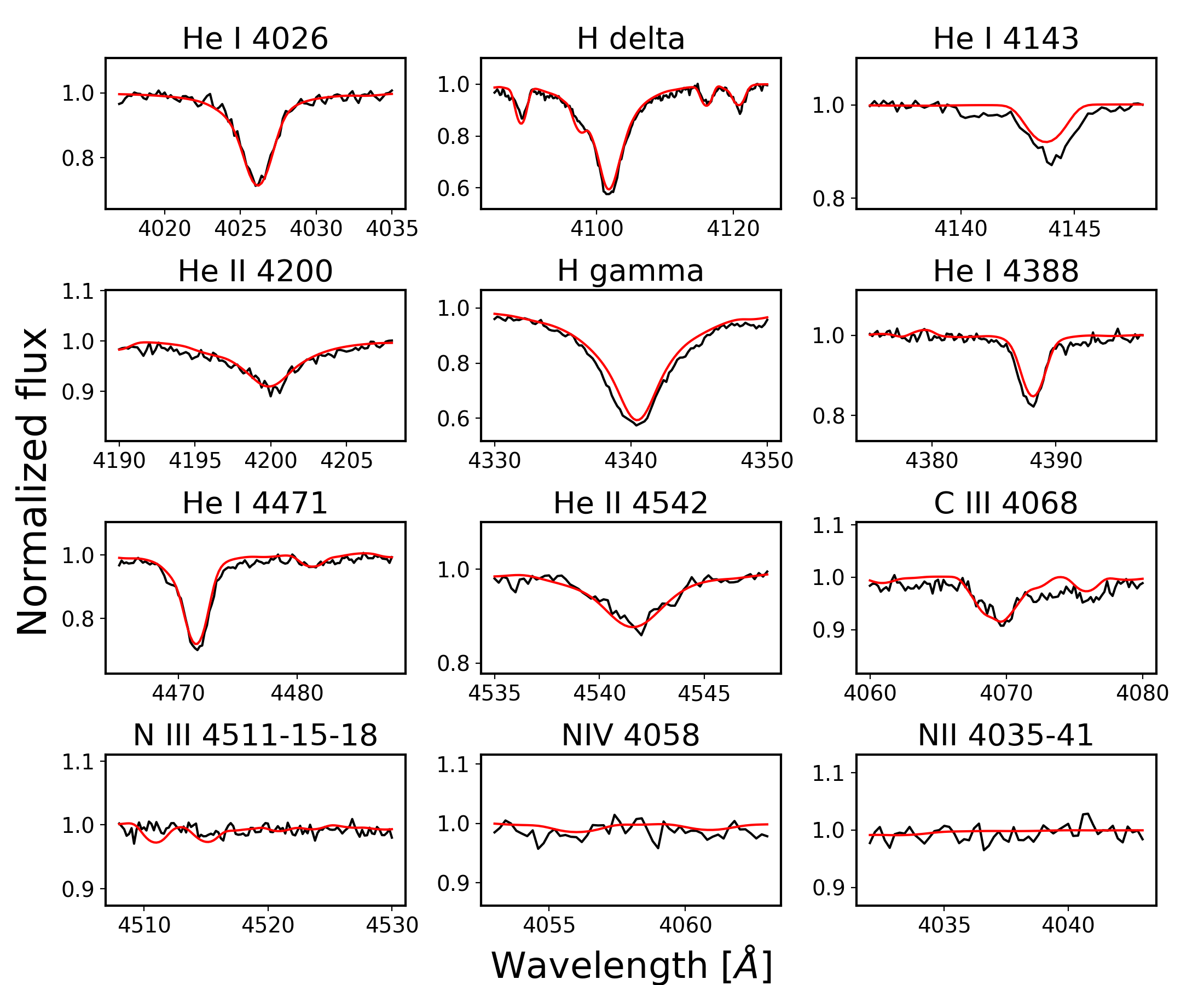}
    \includegraphics[width=7.cm, bb=5 0 453 346,clip]{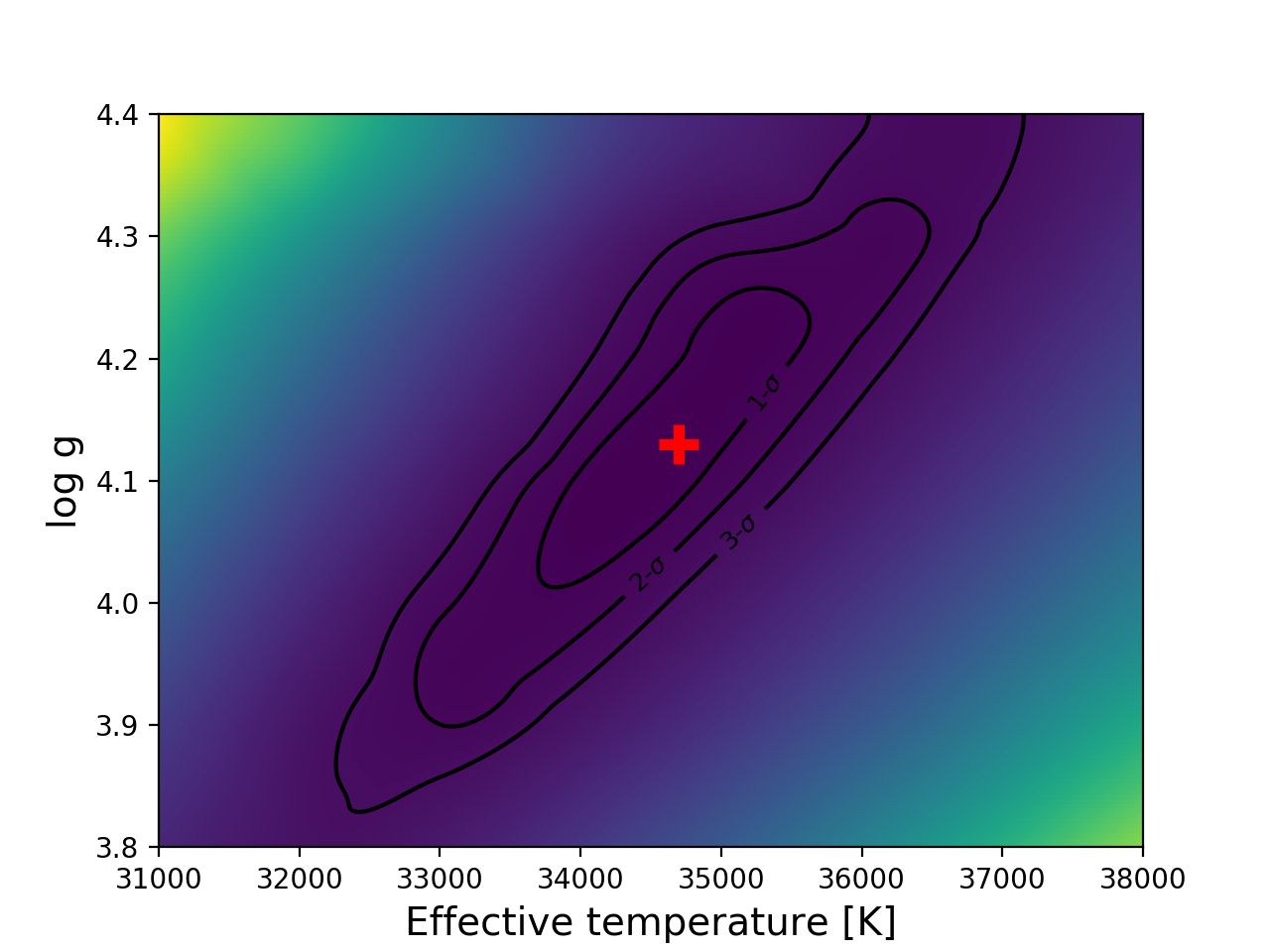}
    \includegraphics[width=6.cm]{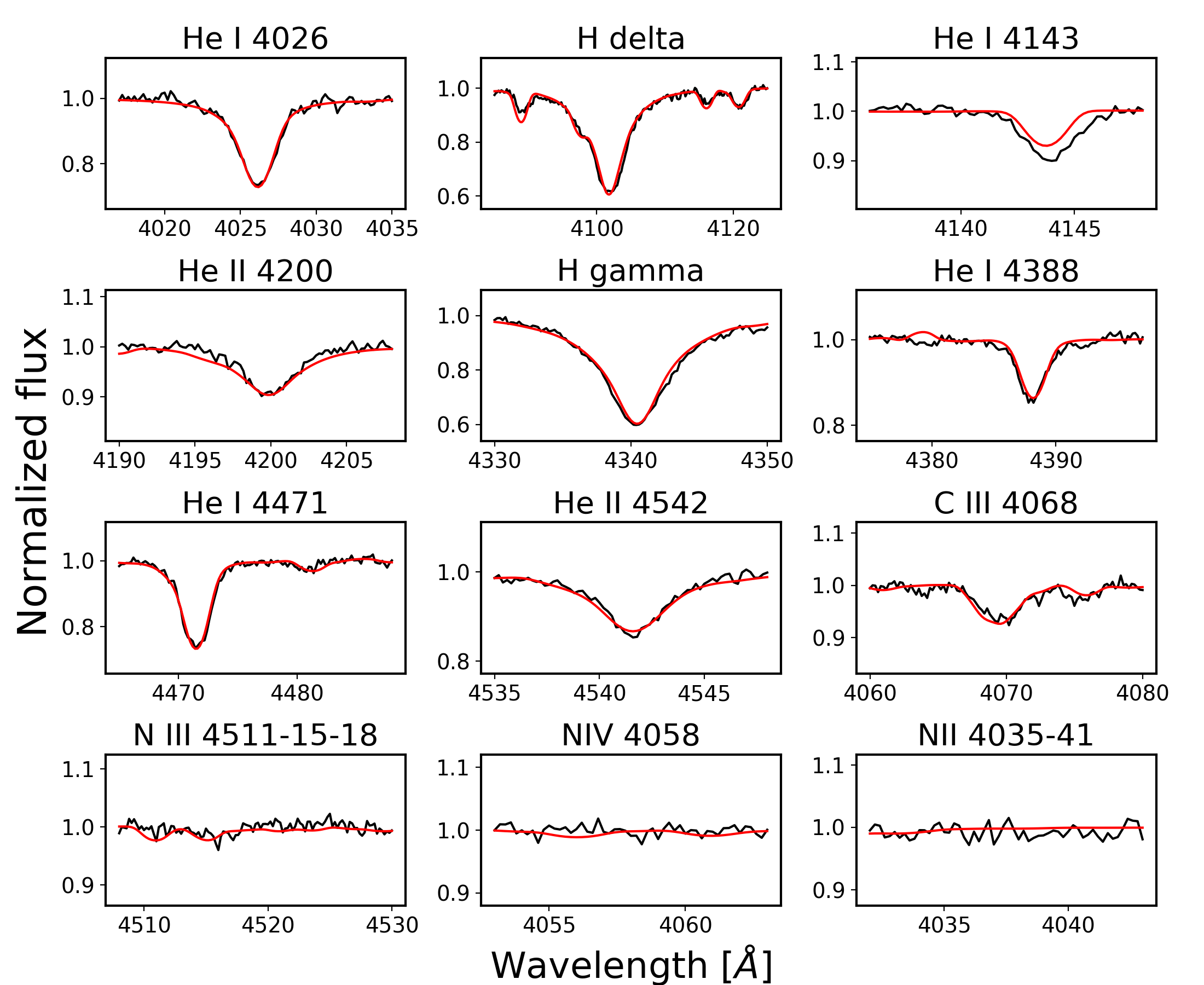}
    \includegraphics[width=7.cm, bb=5 0 453 346,clip]{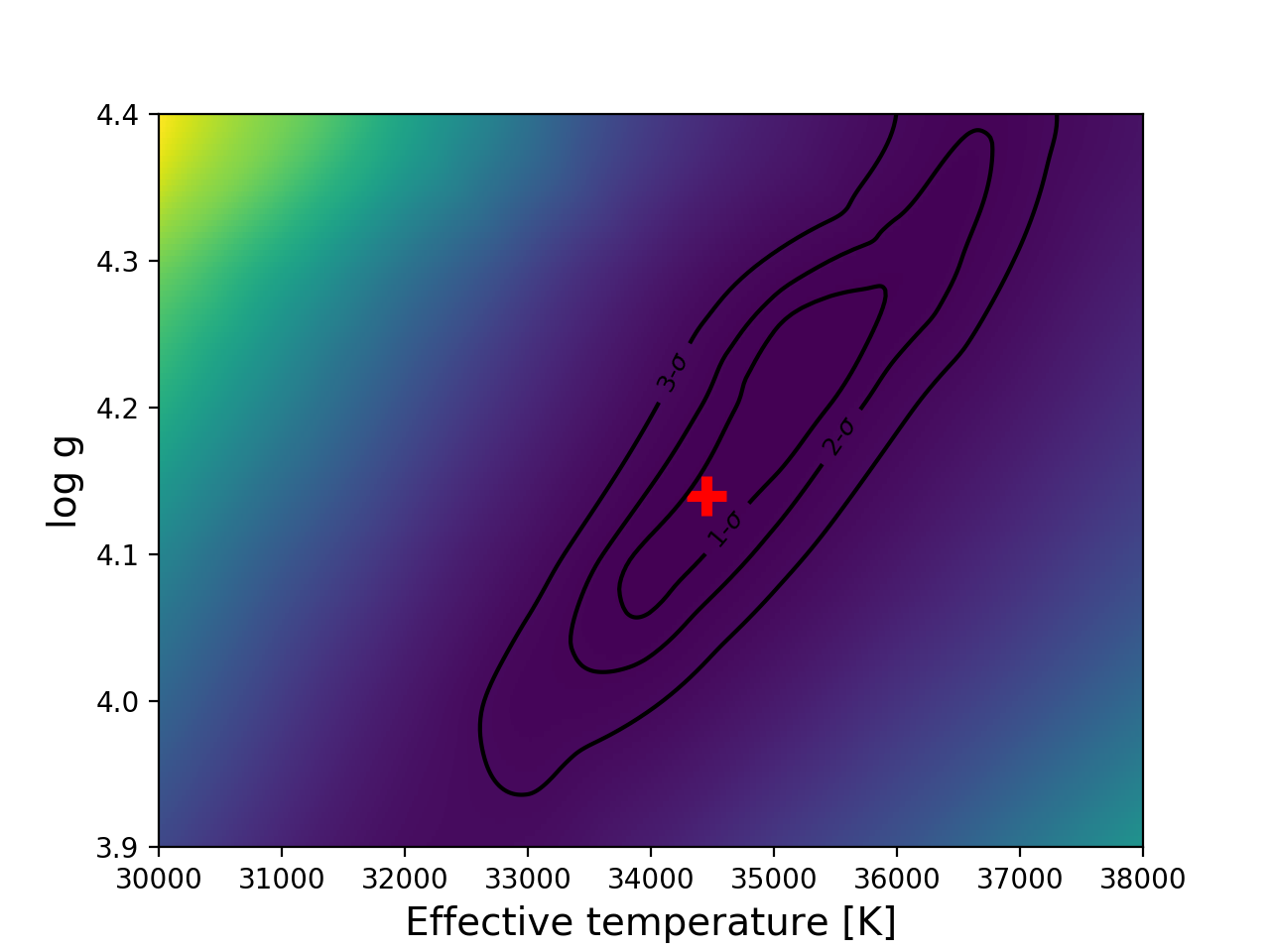}
    \includegraphics[width=7cm]{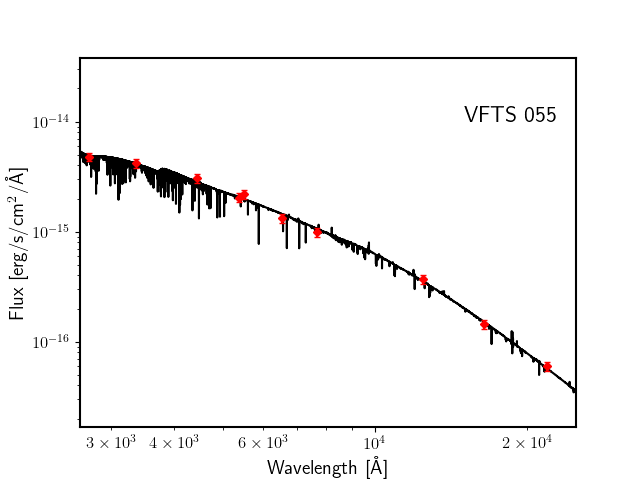}
    \includegraphics[width=6.5cm]{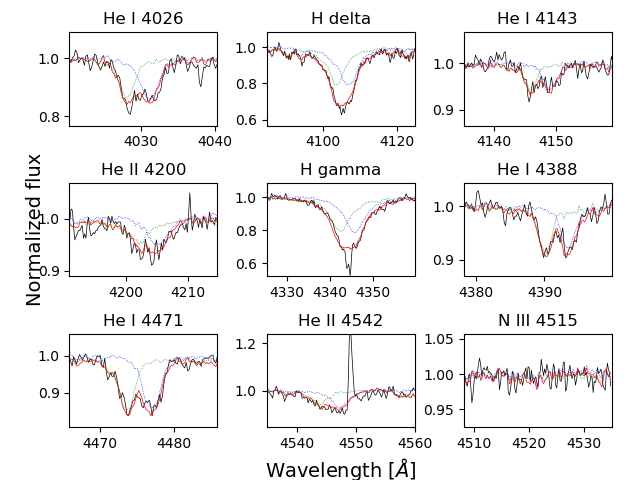}
    \includegraphics[width=7cm, bb=5 0 453 346,clip]{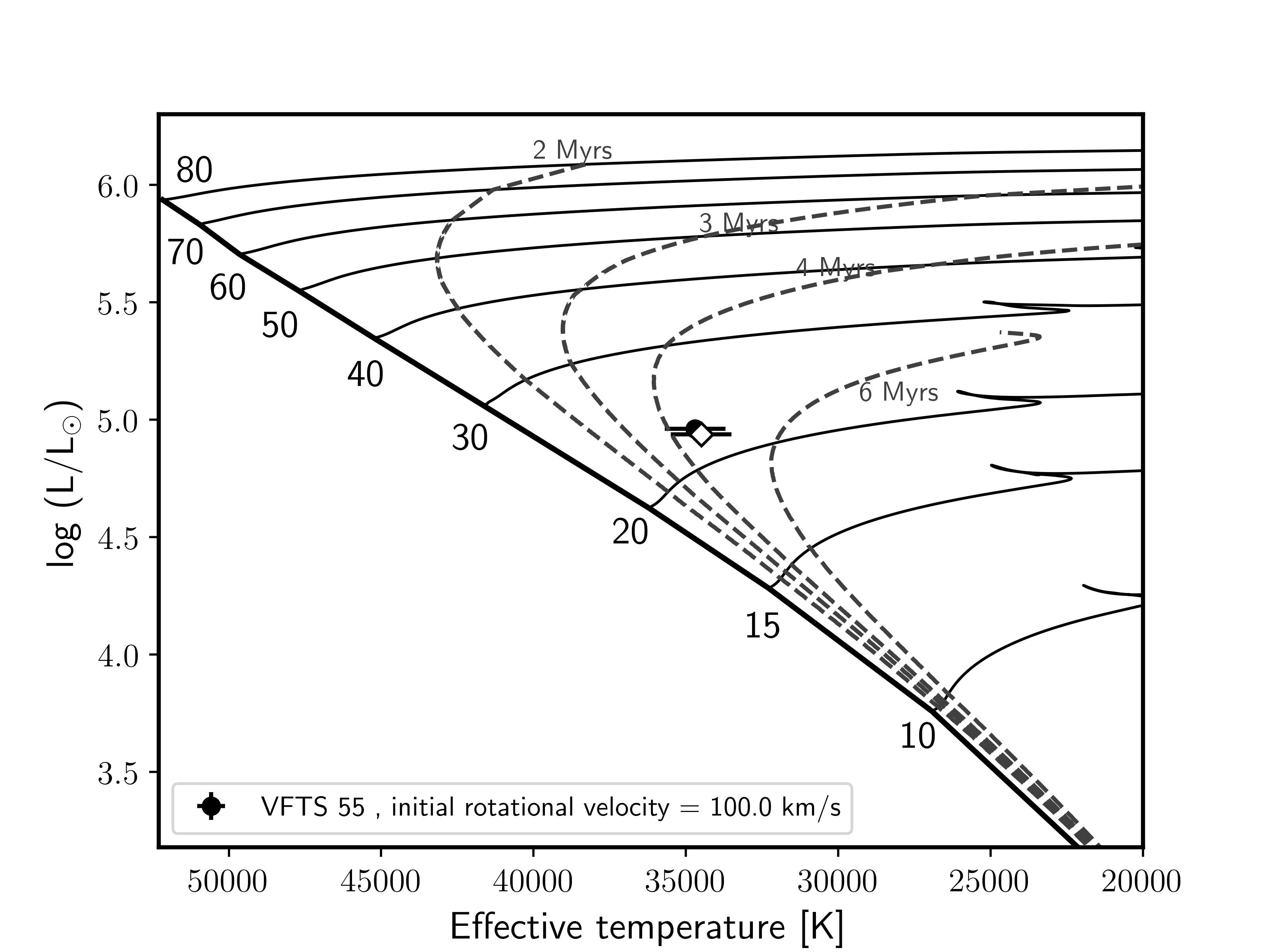}
    \includegraphics[width=7cm, bb=5 0 453 346,clip]{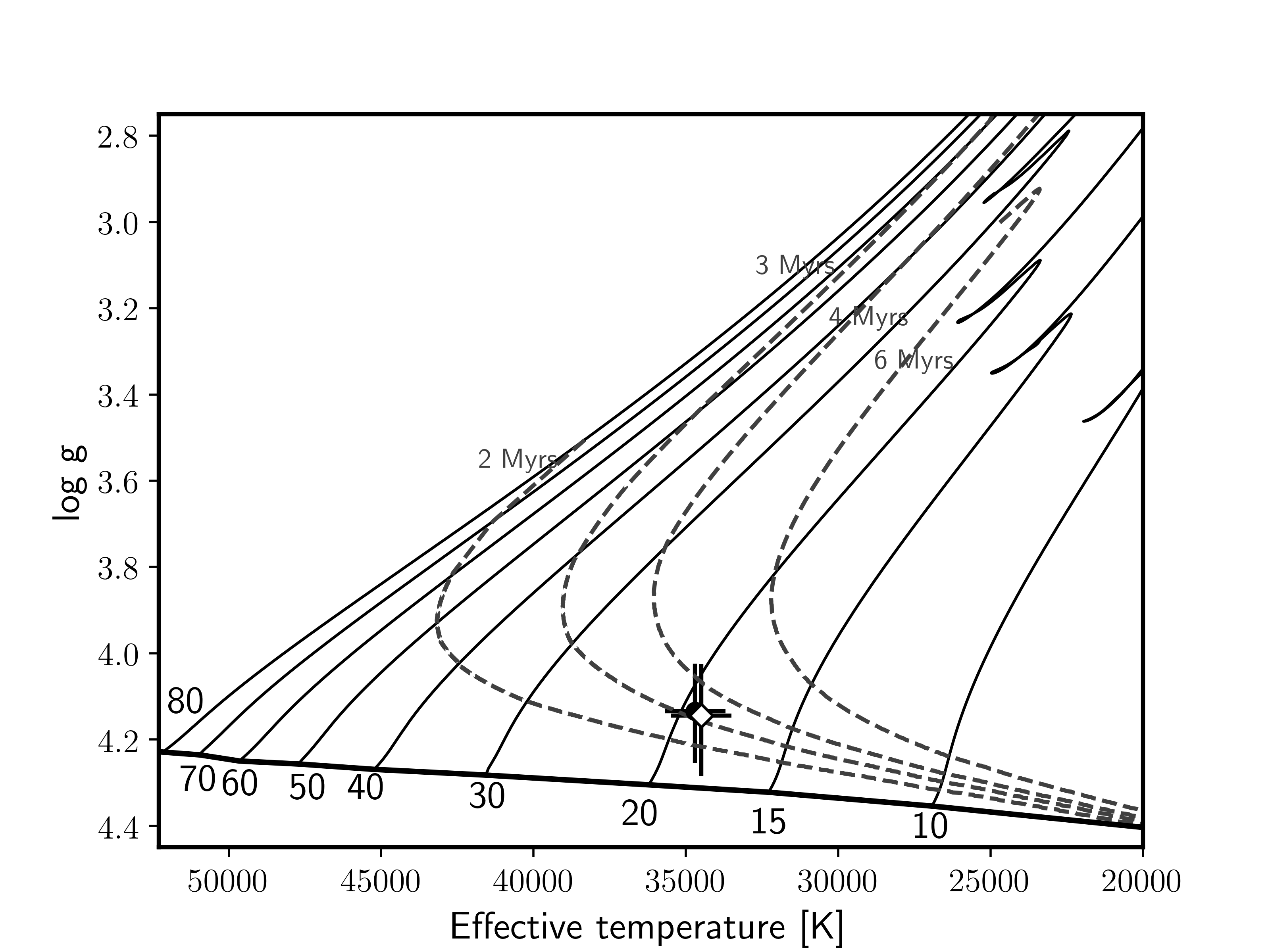}
    \caption{Same as Fig.\,\ref{fig:042} but for VFTS\,055.} \label{fig:055} 
  \end{figure*}
 \clearpage
    
 \begin{figure*}[t!]
    \centering
    \includegraphics[width=6.cm]{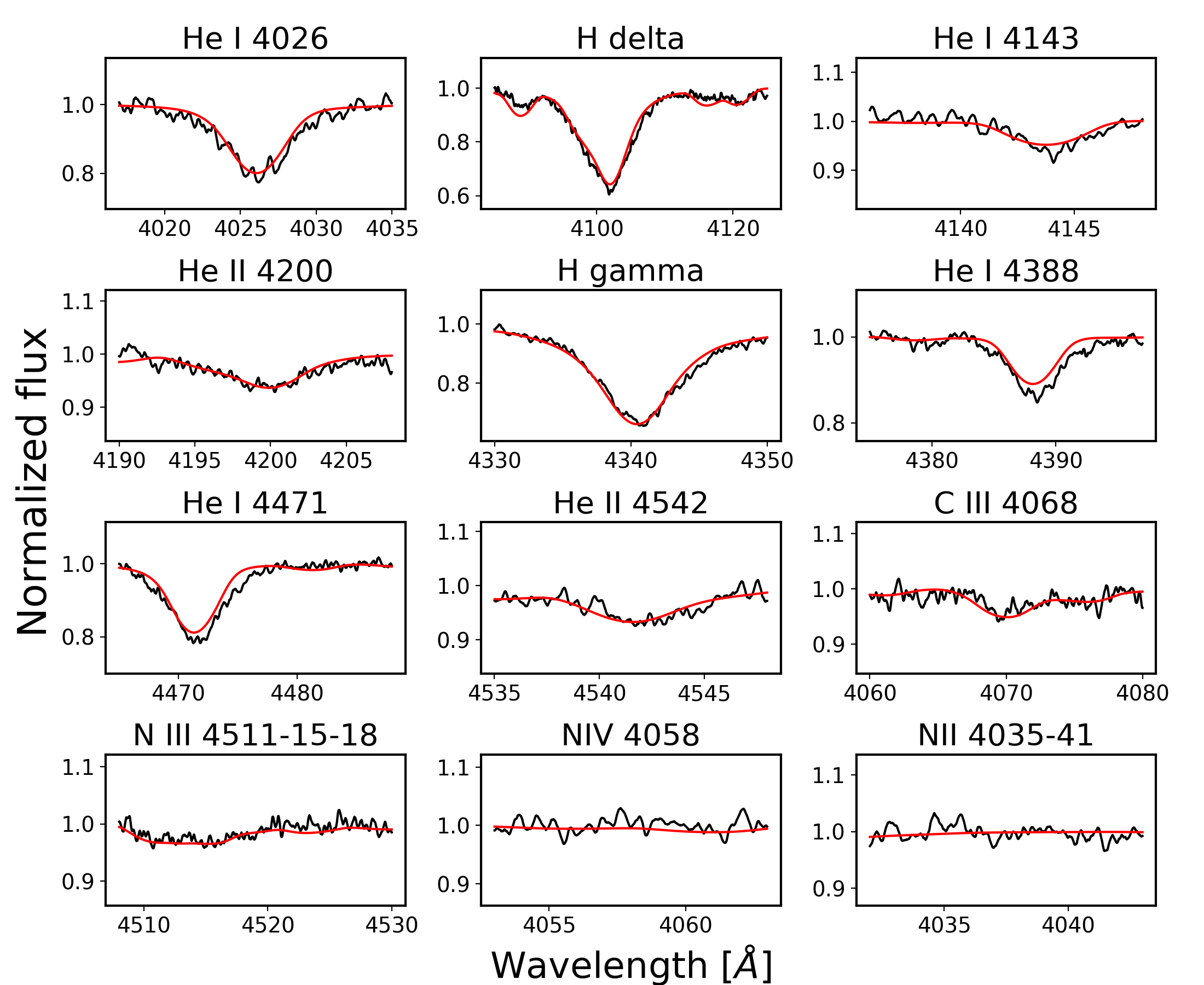}
    \includegraphics[width=7.cm, bb=5 0 453 346,clip]{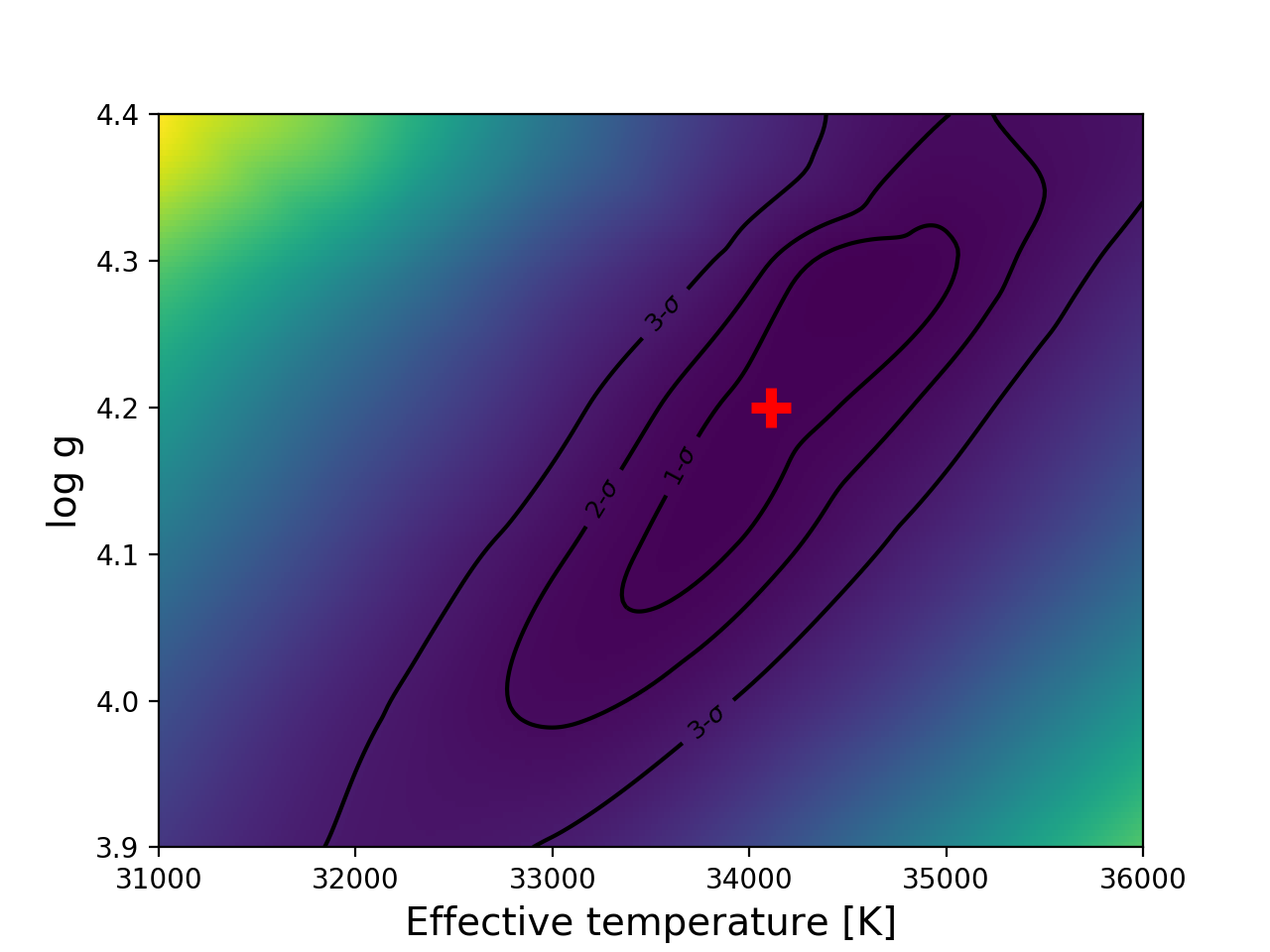}
    \includegraphics[width=6.cm]{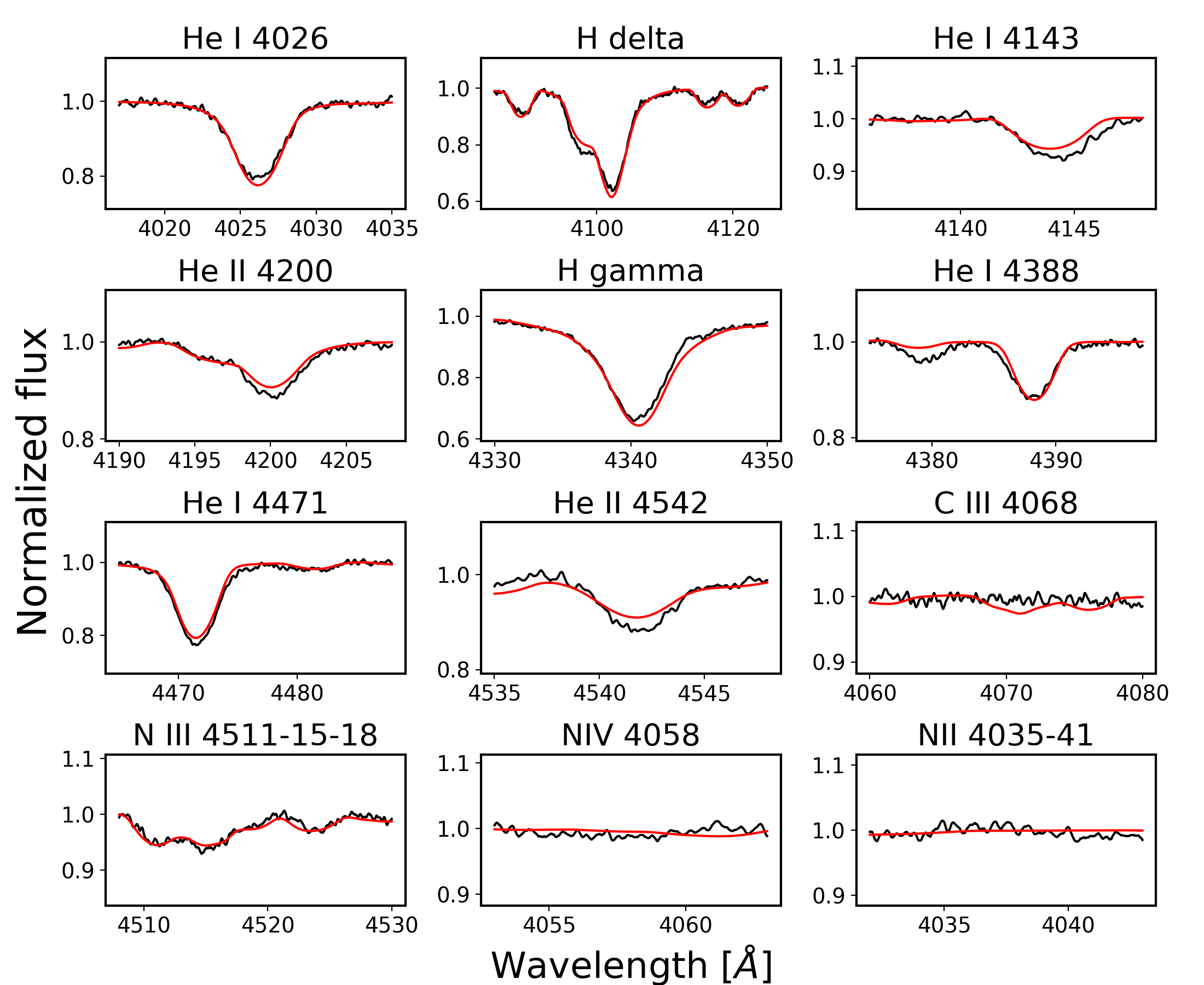}
    \includegraphics[width=7.cm, bb=5 0 453 346,clip]{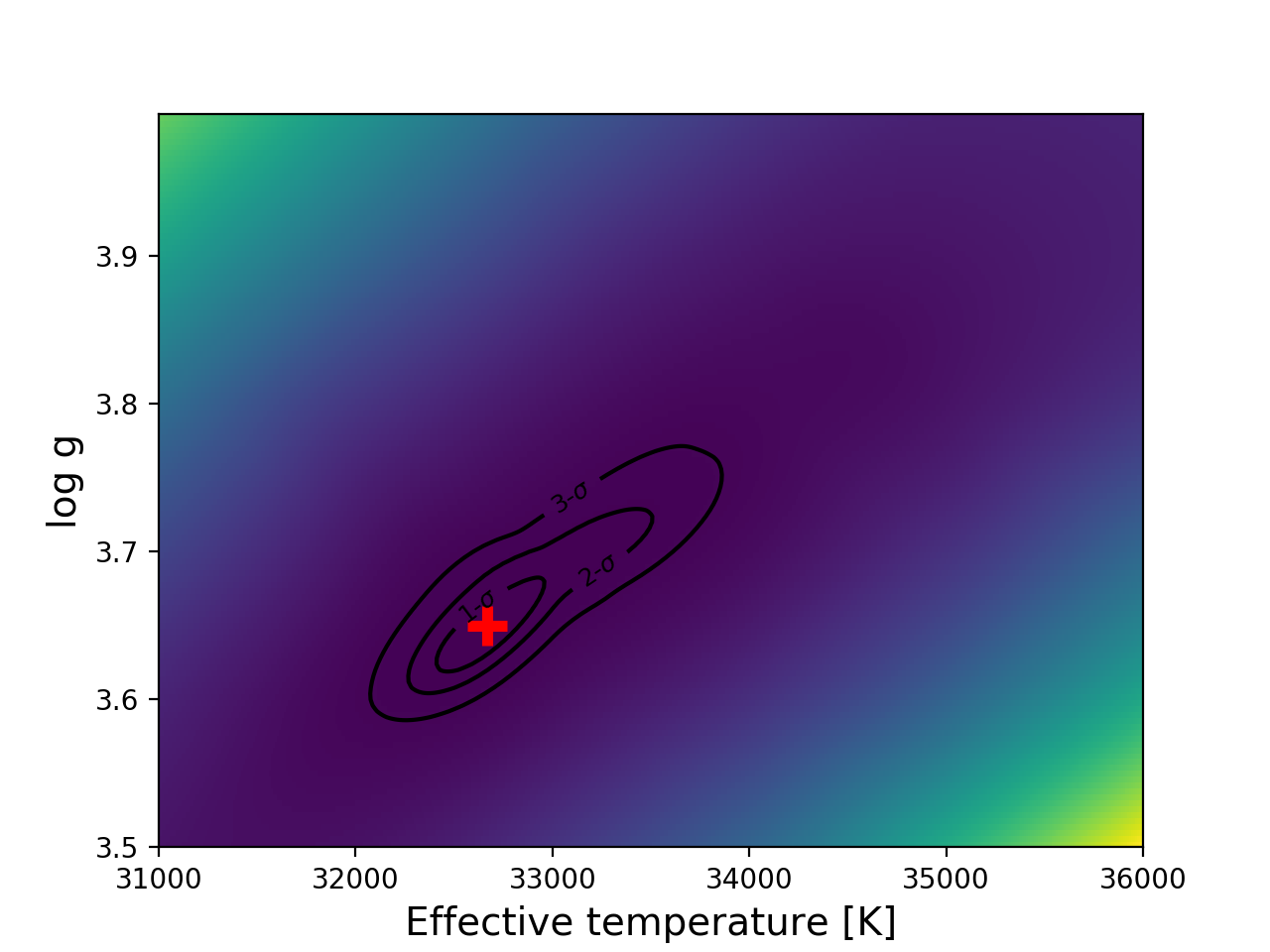}
    \includegraphics[width=7cm]{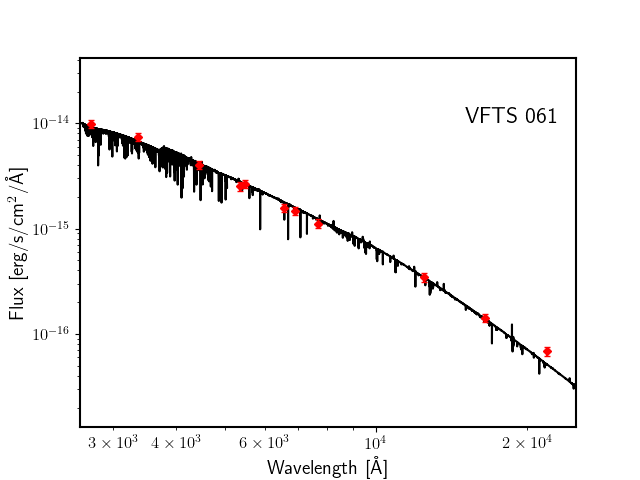}
    \includegraphics[width=6.5cm]{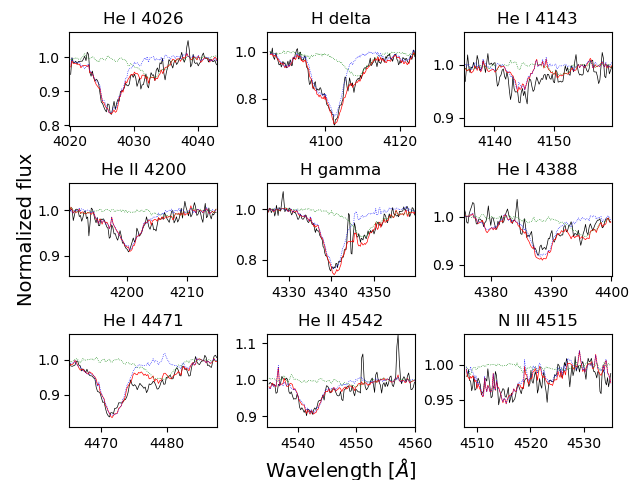}
    \includegraphics[width=7cm, bb=5 0 453 346,clip]{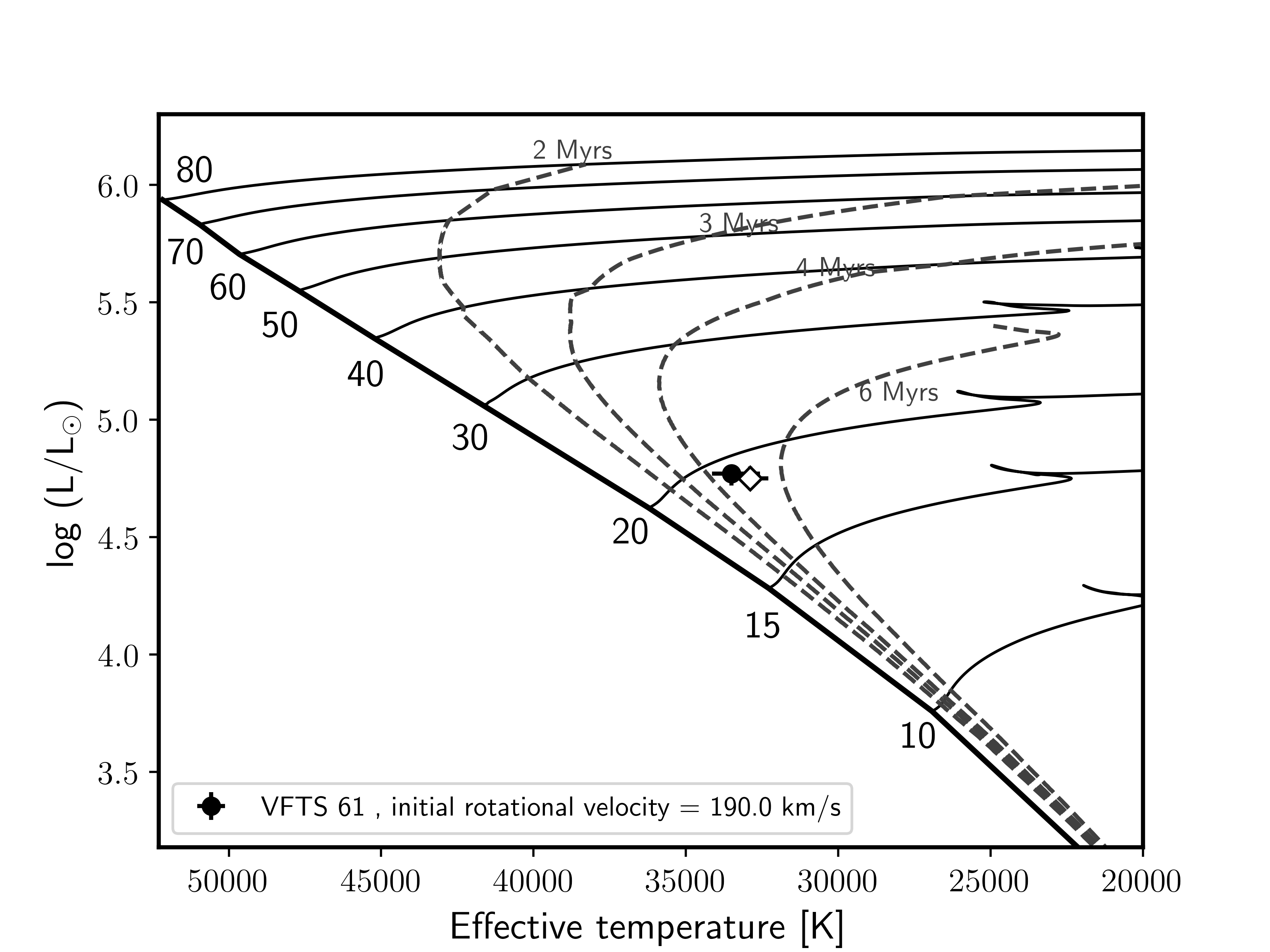}
    \includegraphics[width=7cm, bb=5 0 453 346,clip]{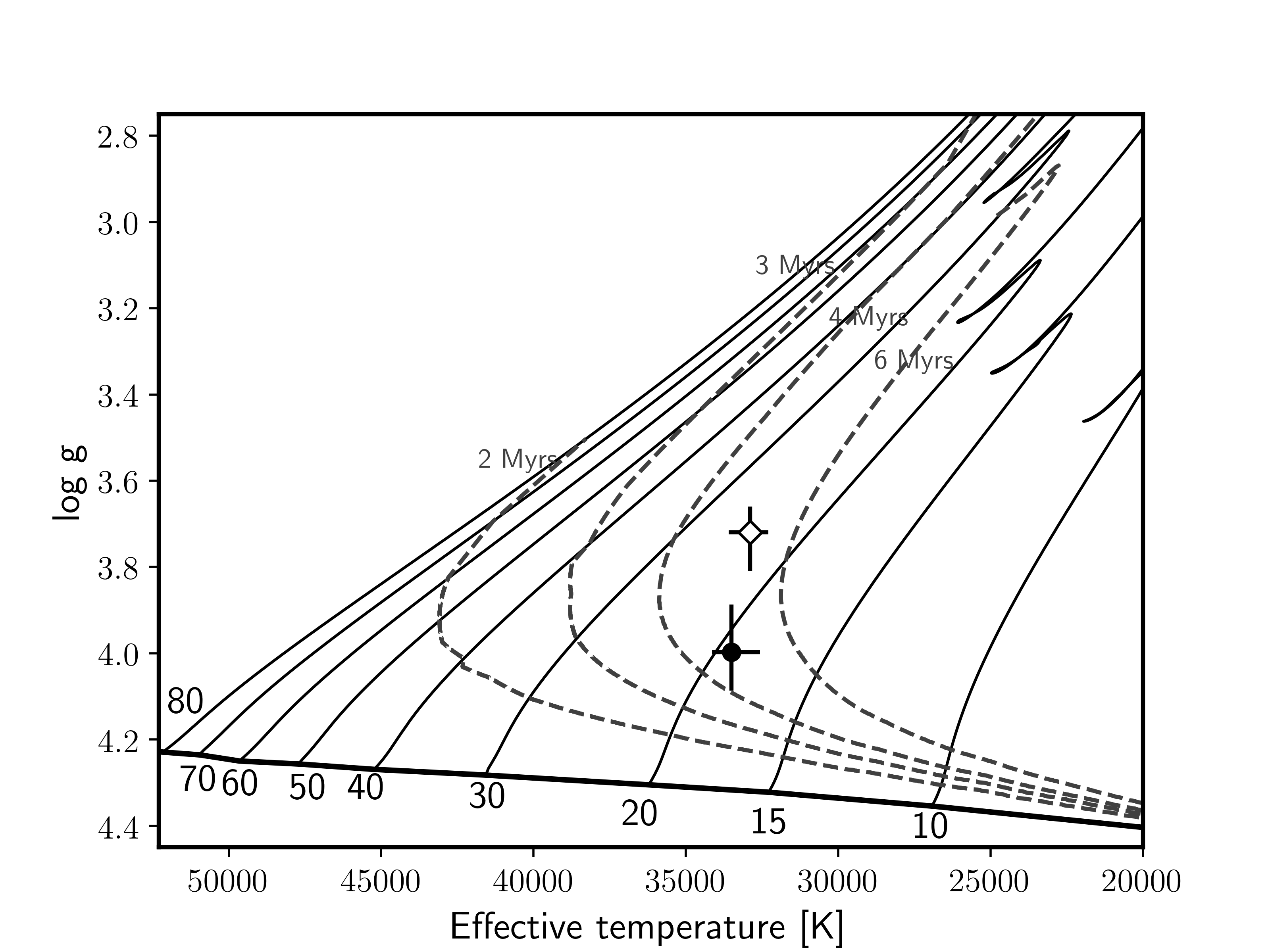}
    \caption{Same as Fig.\,\ref{fig:042} but for VFTS\,061.} \label{fig:061} 
  \end{figure*}
 \clearpage

 \begin{figure*}[t!]
    \centering
    \includegraphics[width=6.cm]{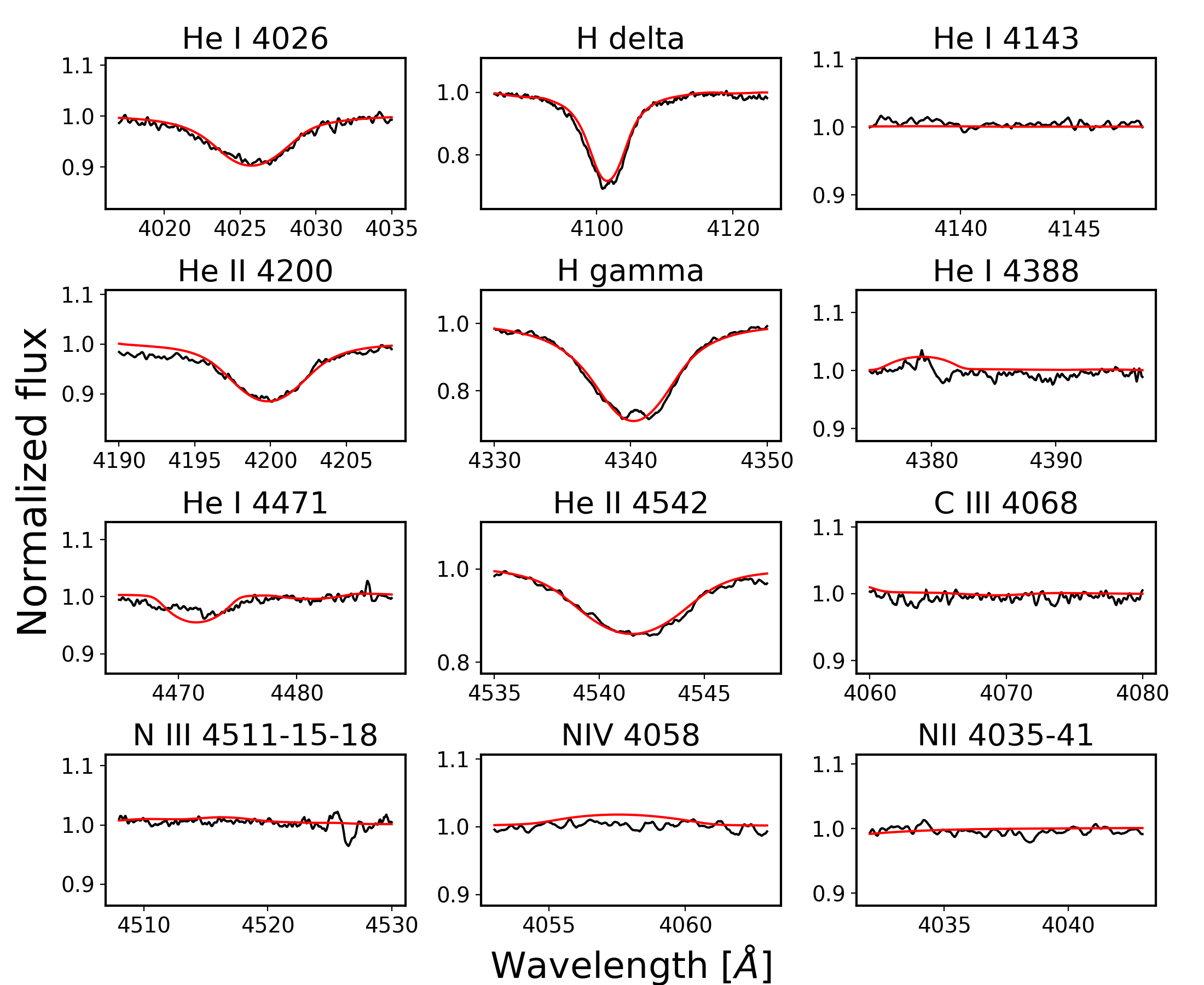}
    \includegraphics[width=7.cm, bb=5 0 453 346,clip]{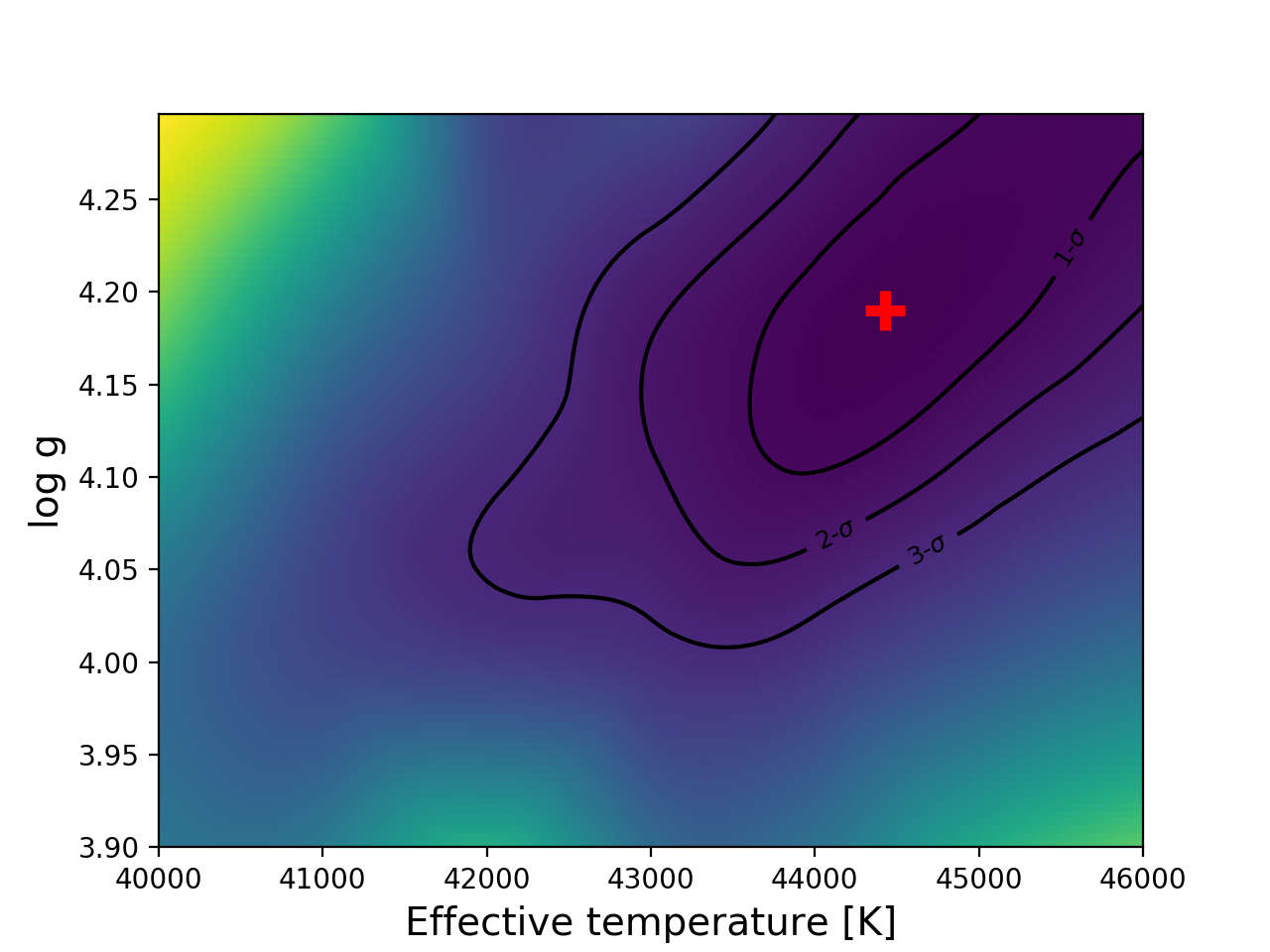}
    \includegraphics[width=6.cm]{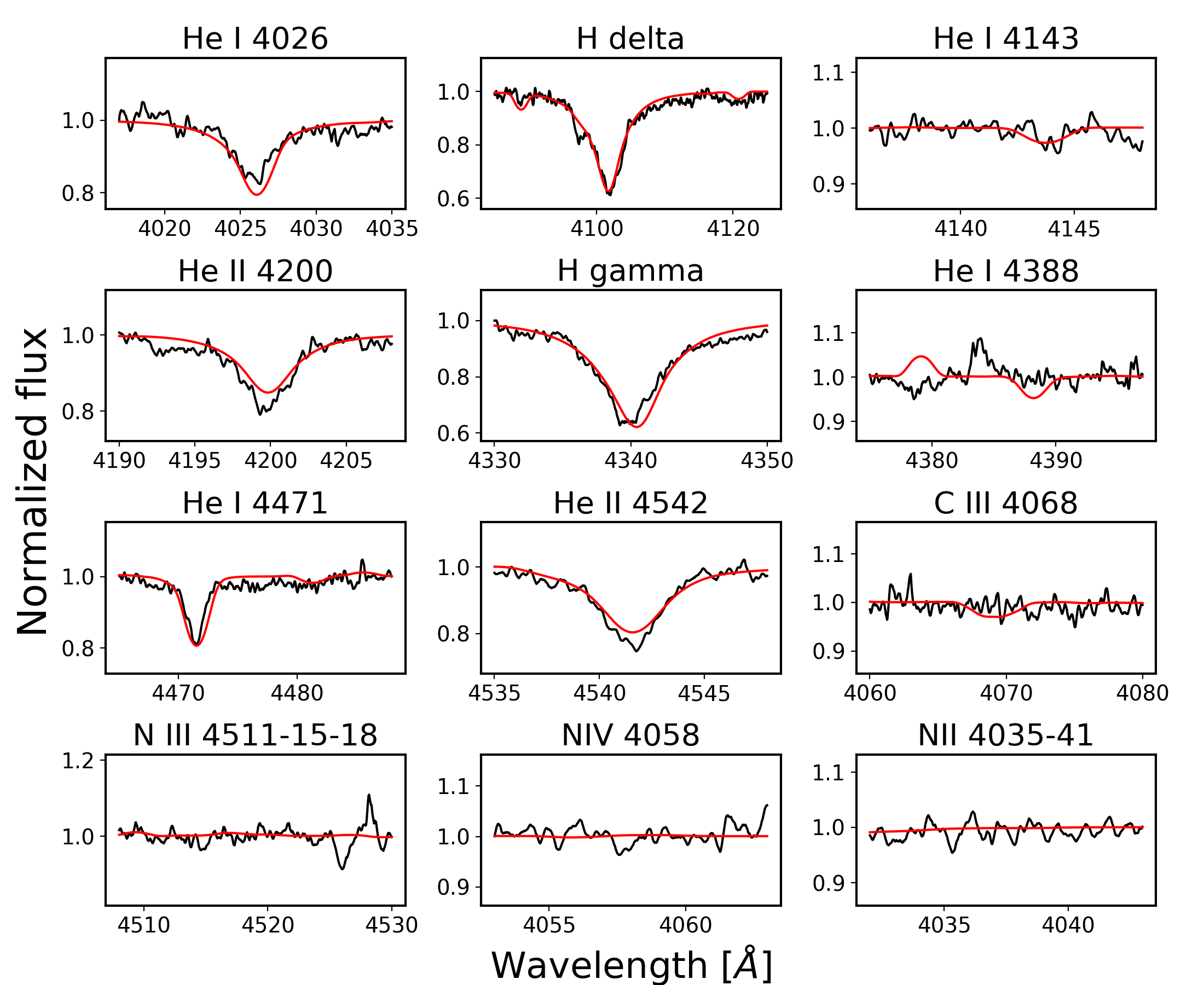}
    \includegraphics[width=7.cm, bb=5 0 453 346,clip]{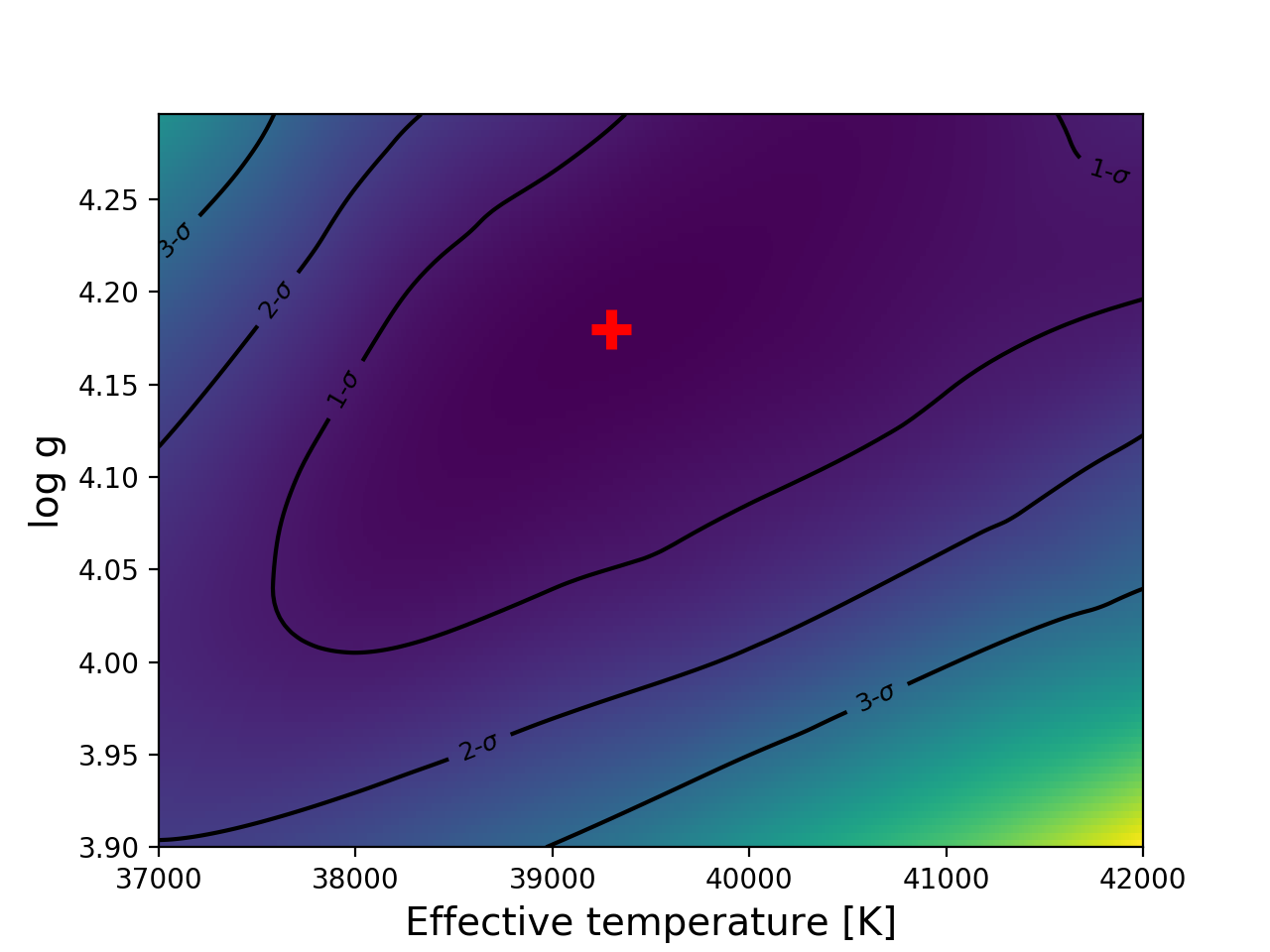}
    \includegraphics[width=7cm]{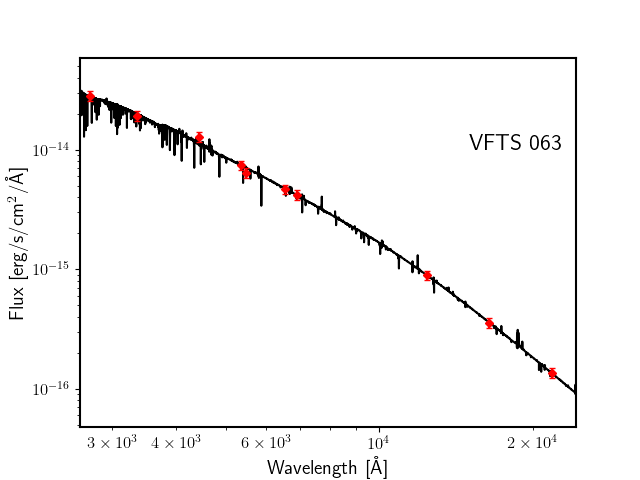}
    \includegraphics[width=6.5cm]{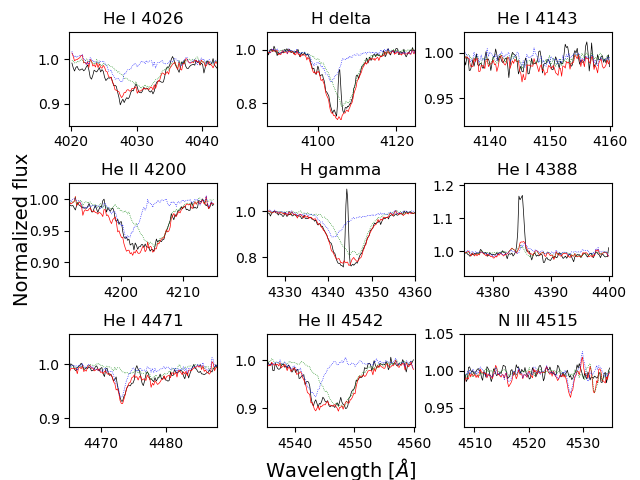}
    \includegraphics[width=7cm, bb=5 0 453 346,clip]{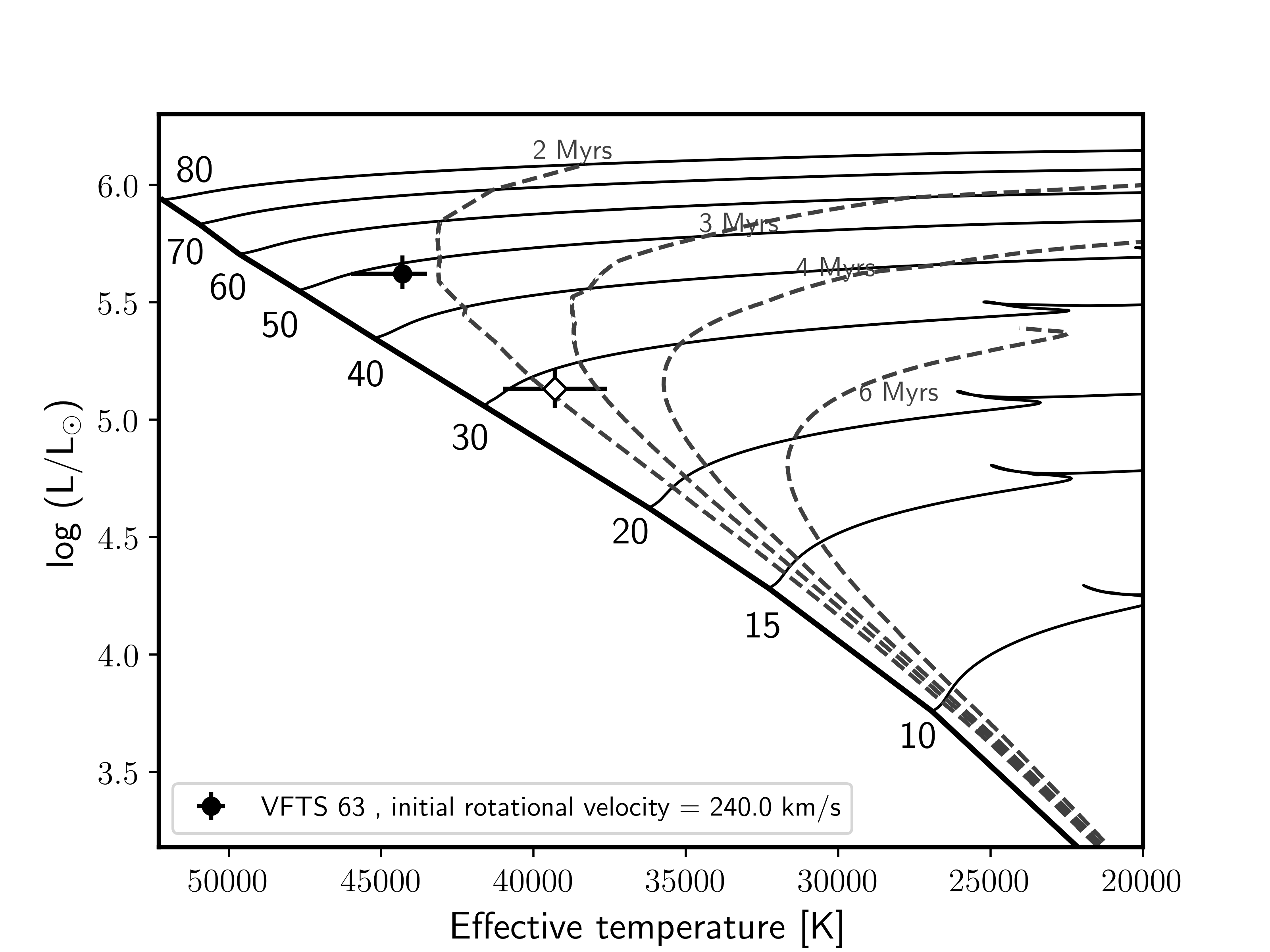}
    \includegraphics[width=7cm, bb=5 0 453 346,clip]{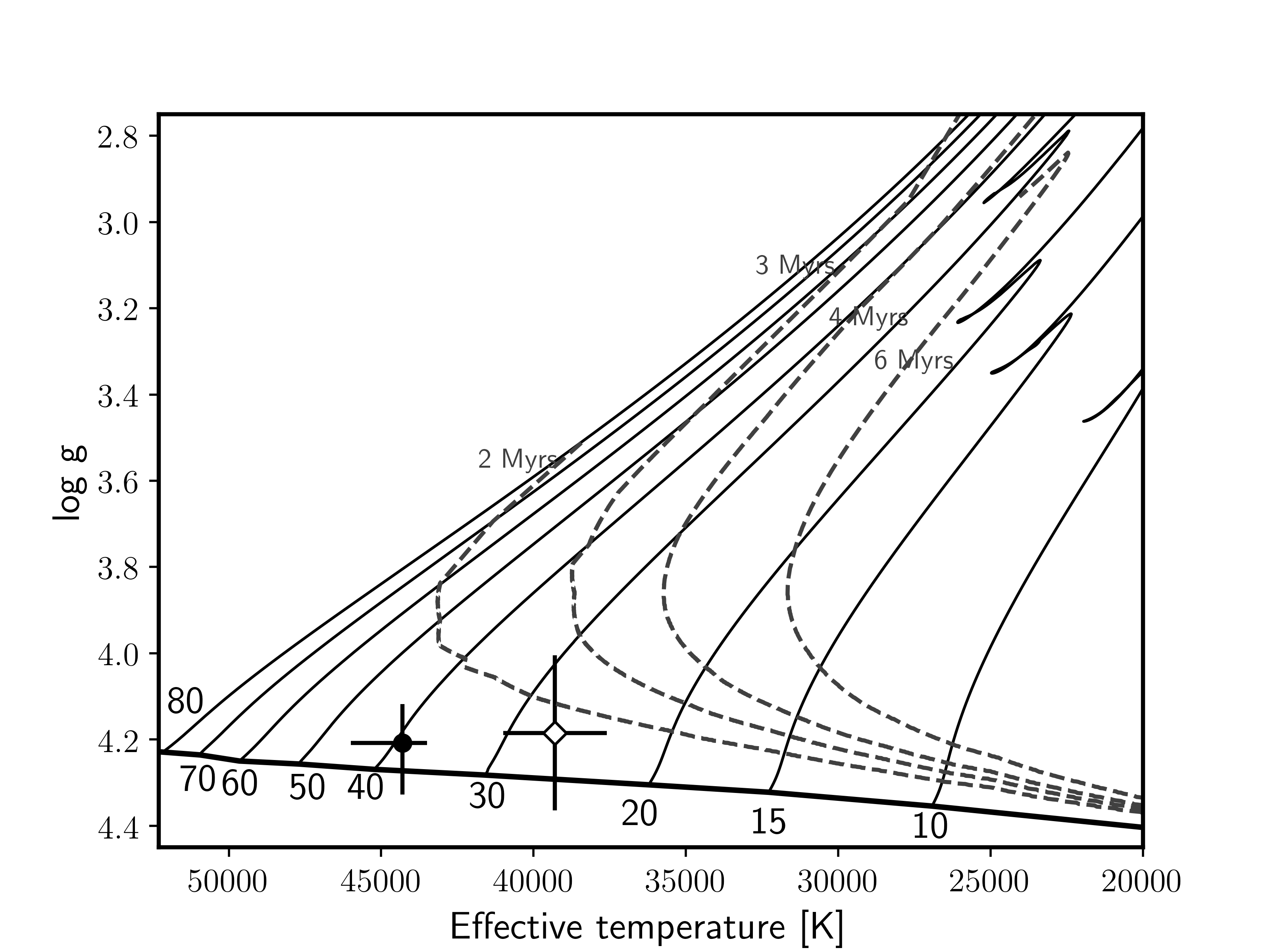}
    \caption{Same as Fig.\,\ref{fig:042} but for VFTS\,063.}\label{fig:063} 
  \end{figure*} \clearpage

 \begin{figure*}[t!]
    \centering
    \includegraphics[width=6.cm]{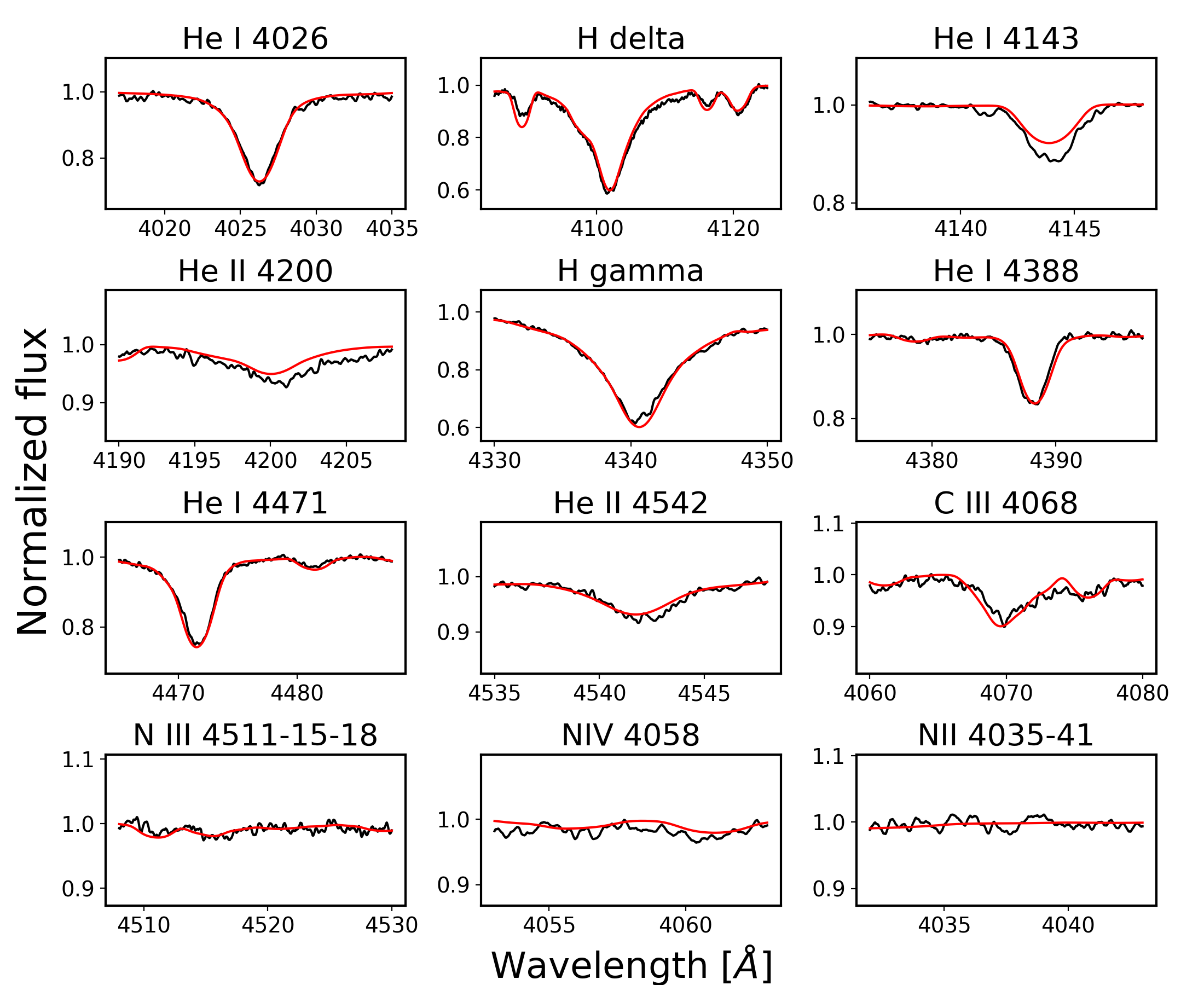}
    \includegraphics[width=7.cm, bb=5 0 453 346,clip]{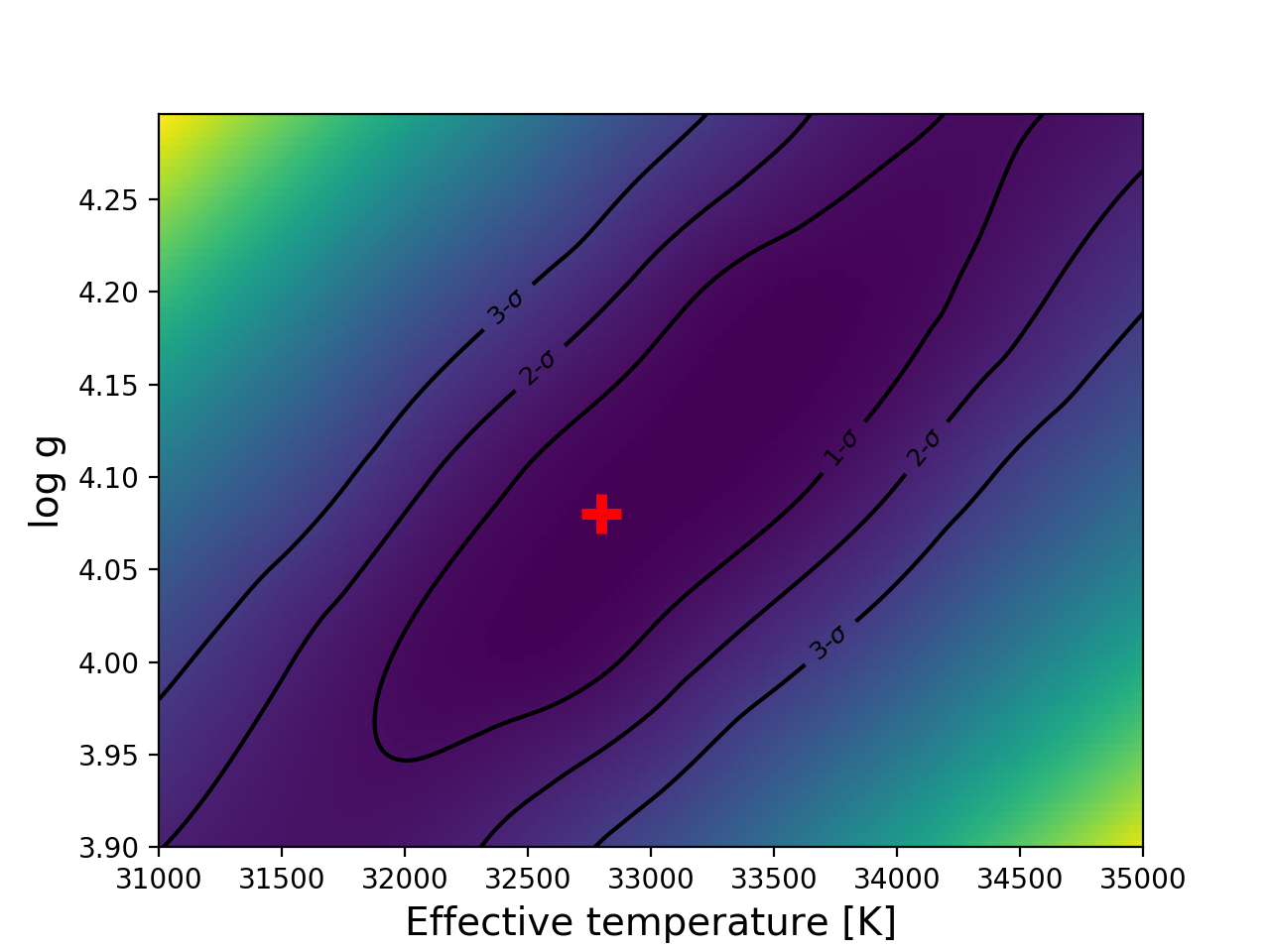}
    \includegraphics[width=6.cm]{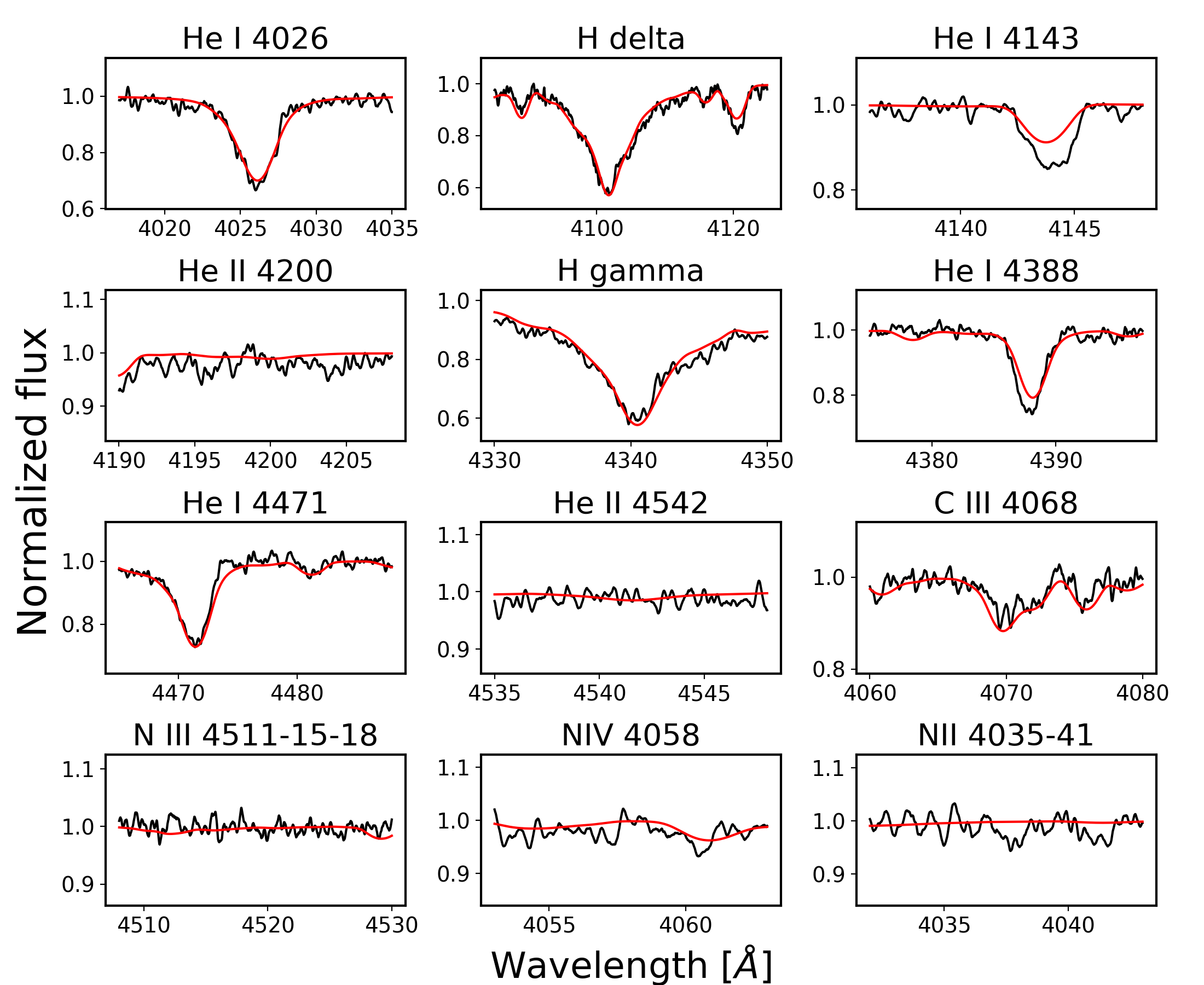}
    \includegraphics[width=7.cm, bb=5 0 453 346,clip]{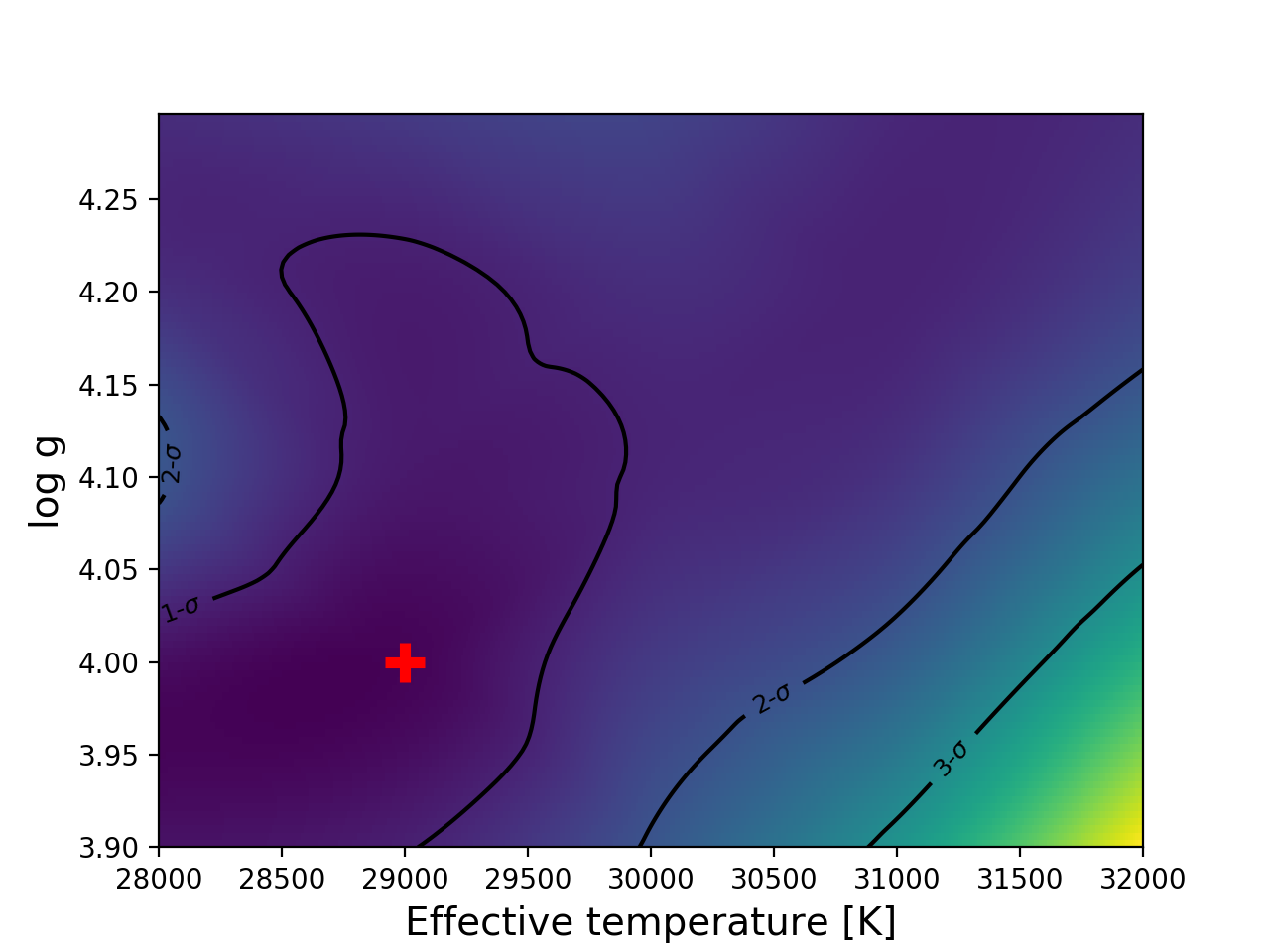}
    \includegraphics[width=7cm]{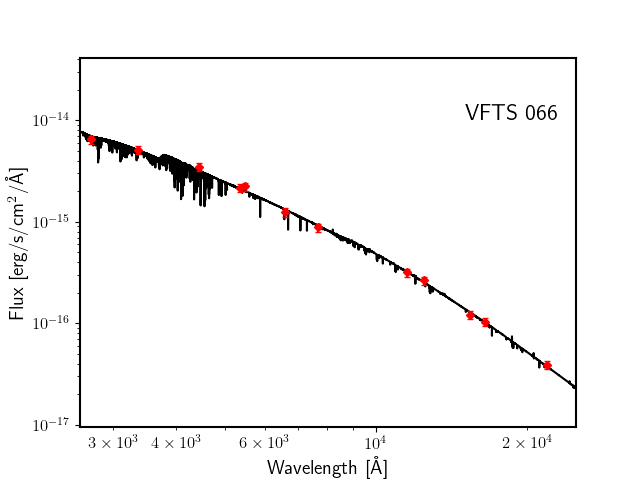}
    \includegraphics[width=6.5cm]{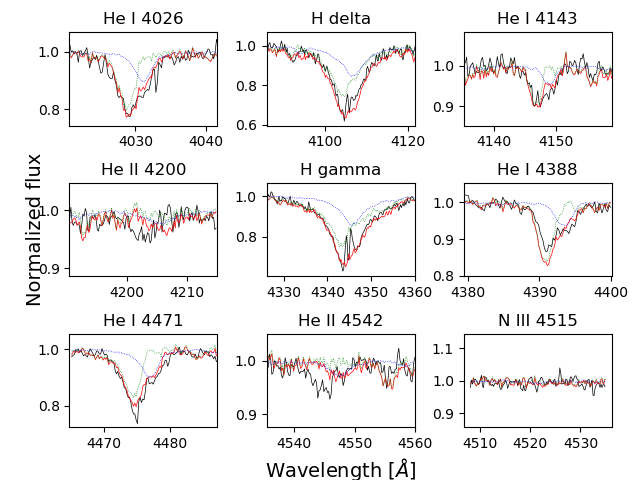}
    \includegraphics[width=7cm, bb=5 0 453 346,clip]{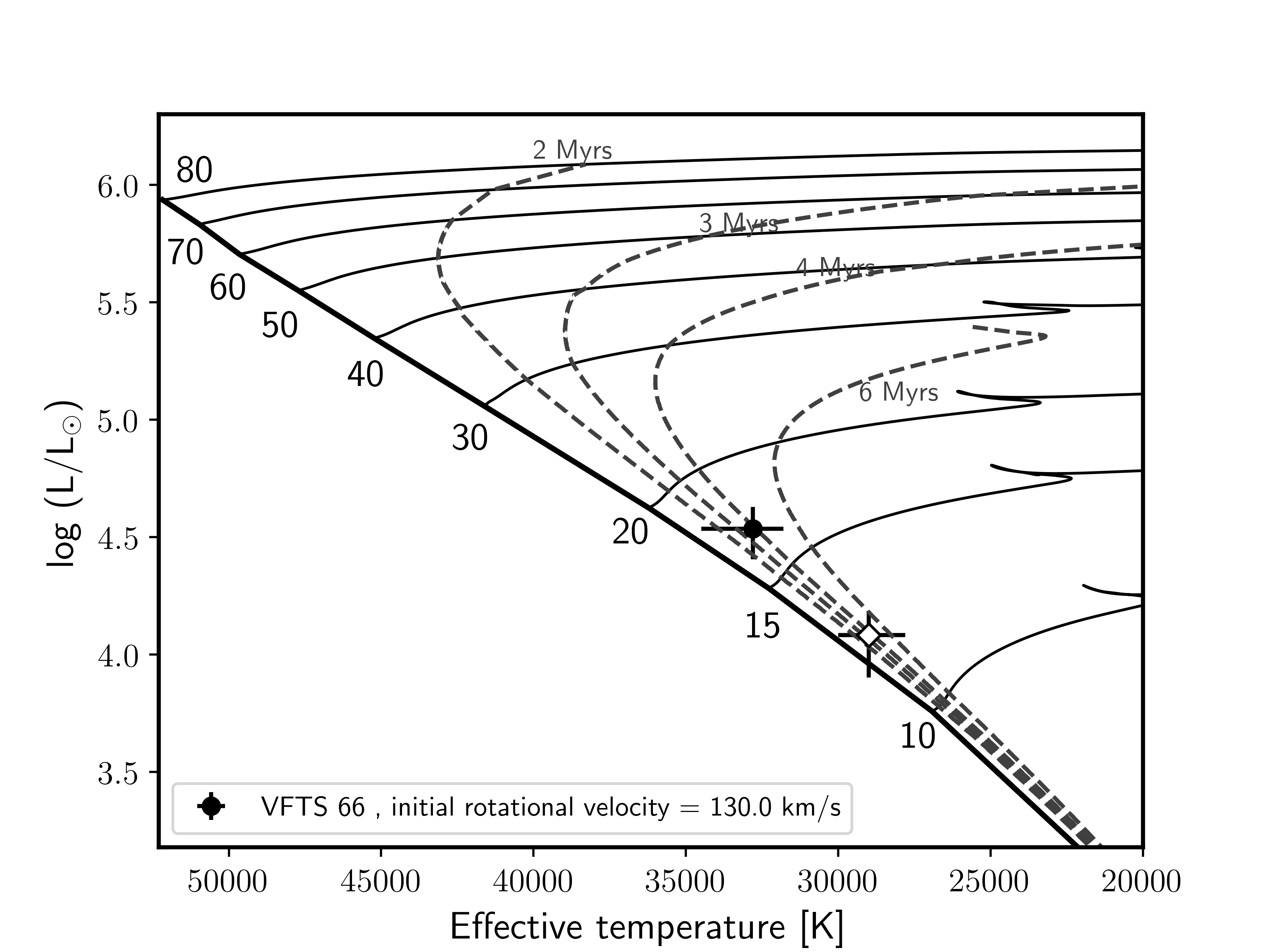}
    \includegraphics[width=7cm, bb=5 0 453 346,clip]{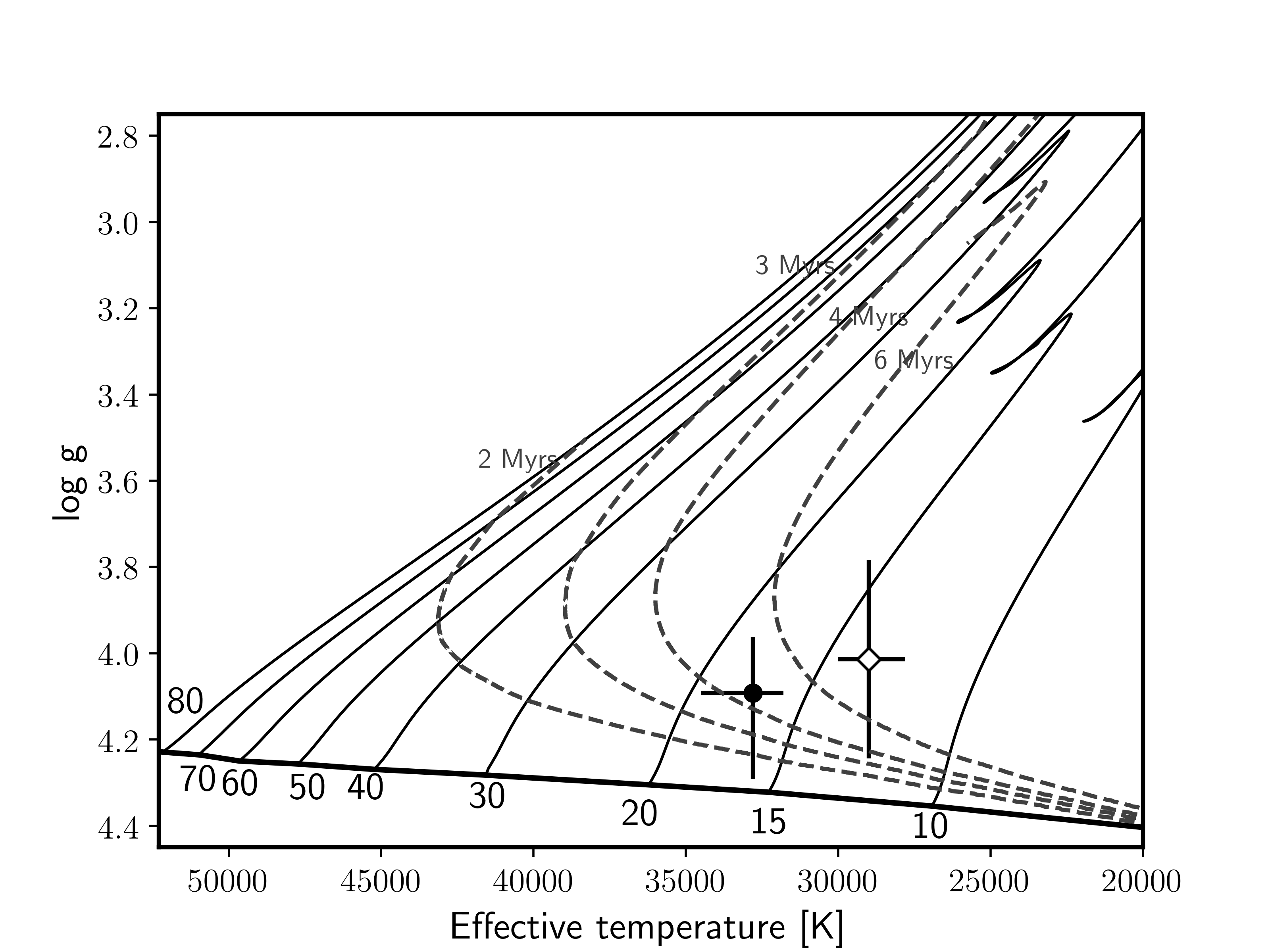}
    \caption{Same as Fig.\,\ref{fig:042} but for VFTS\,066.} \label{fig:066} 
  \end{figure*}
 \clearpage

 \begin{figure*}[t!]
    \centering
    \includegraphics[width=6.cm]{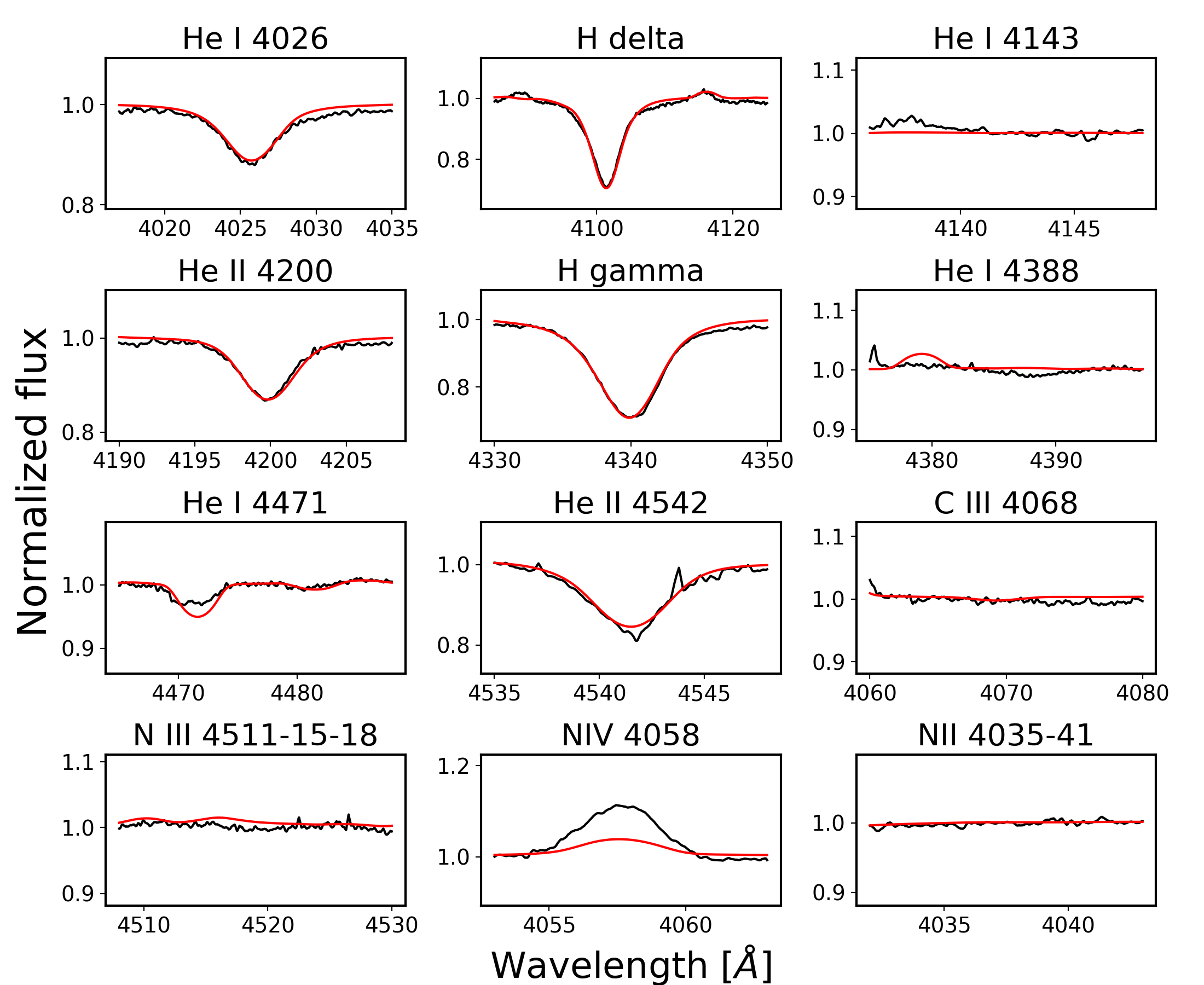}
    \includegraphics[width=7.cm, bb=5 0 453 346,clip]{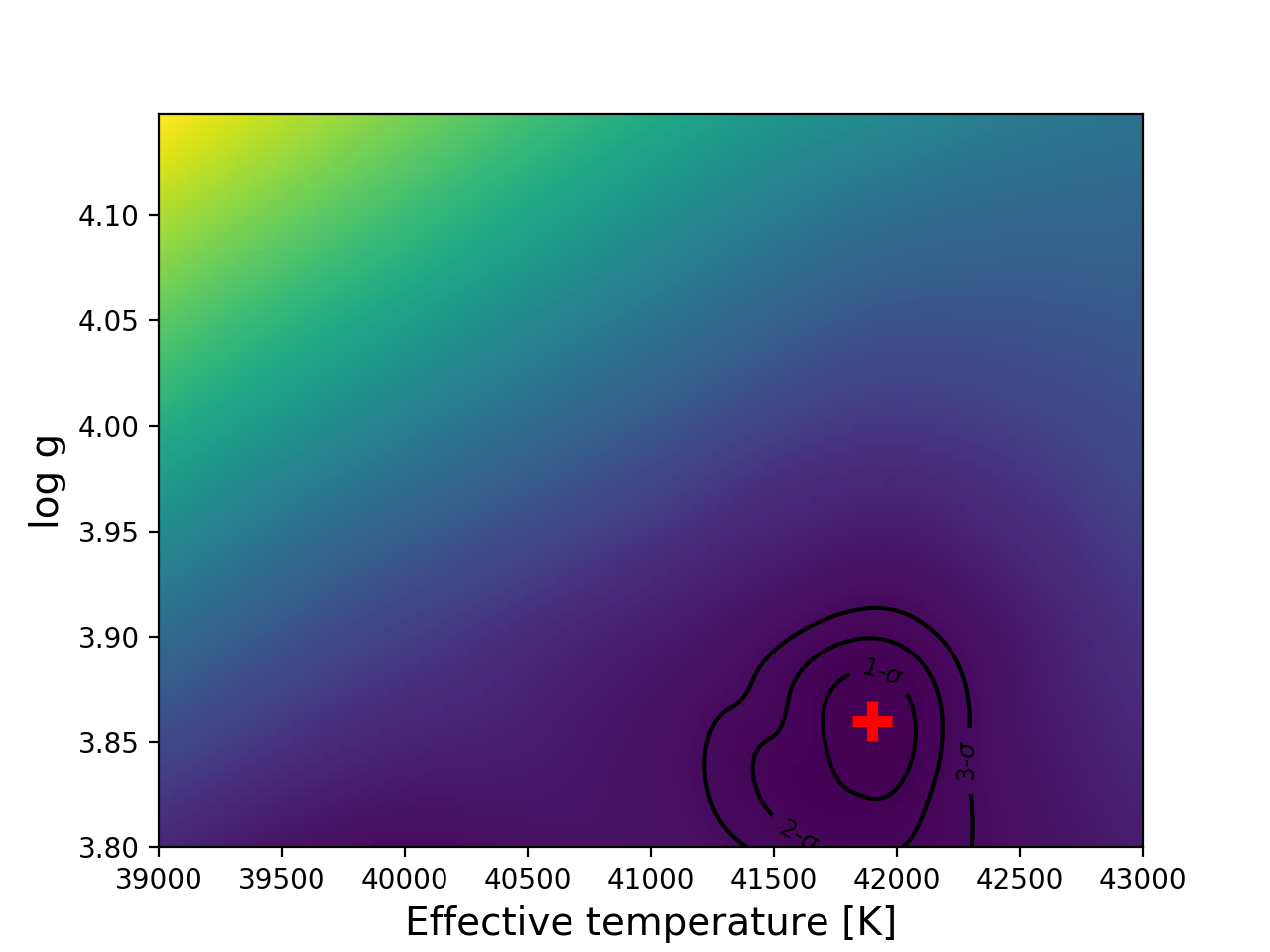}
    \includegraphics[width=6.cm]{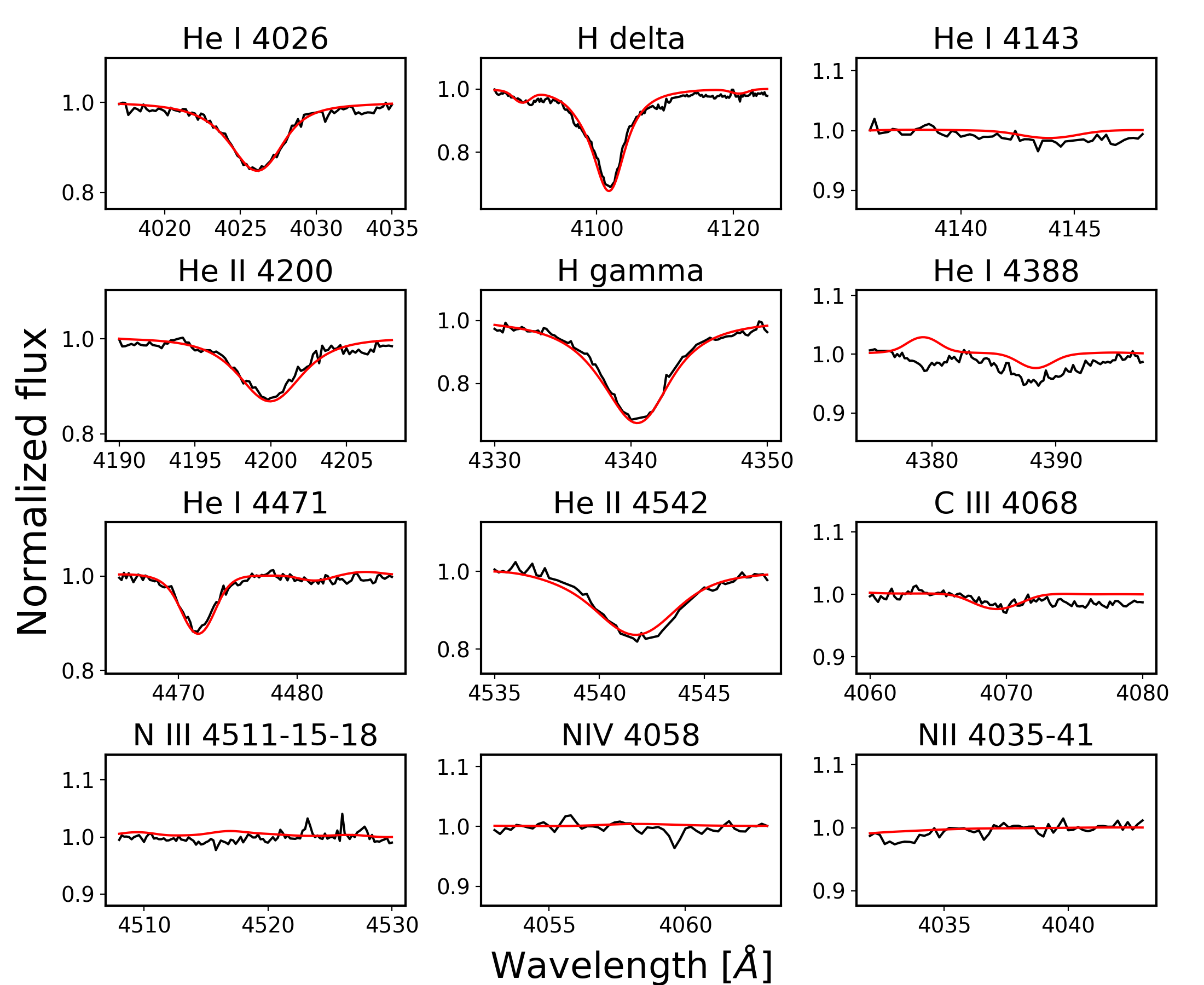}
    \includegraphics[width=7.cm, bb=5 0 453 346,clip]{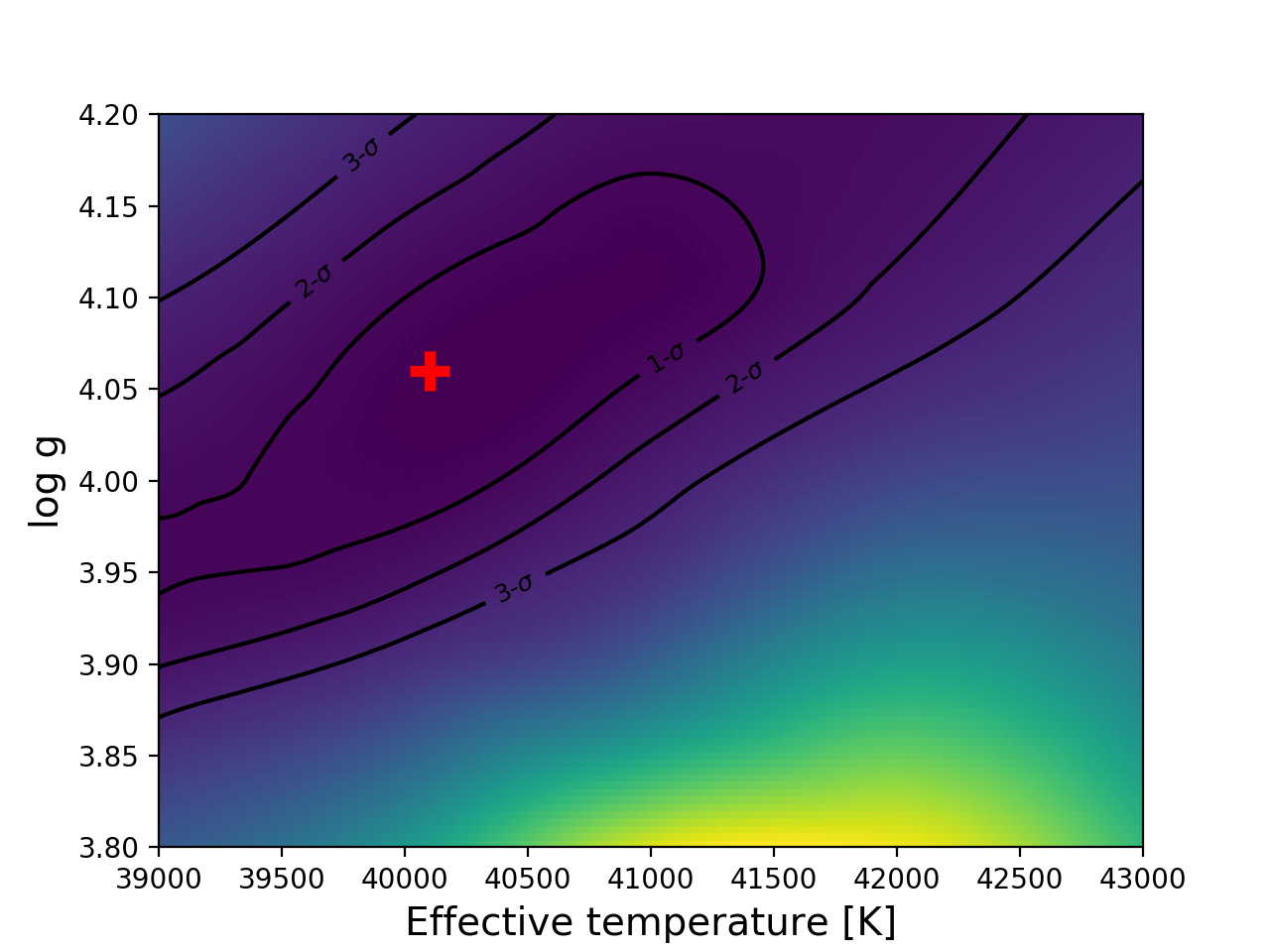}
    \includegraphics[width=7cm]{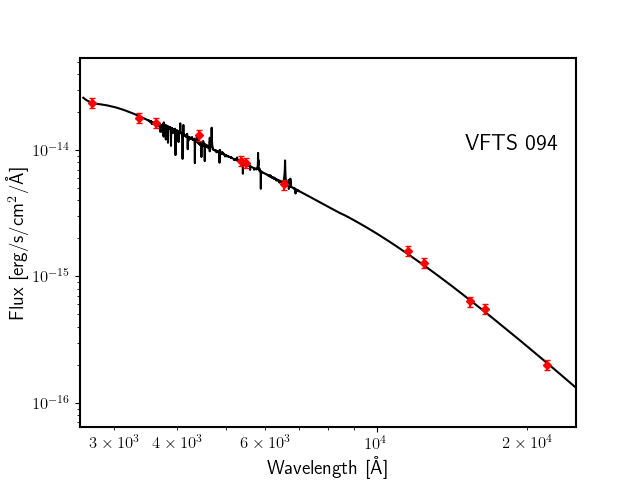}
    \includegraphics[width=6.5cm]{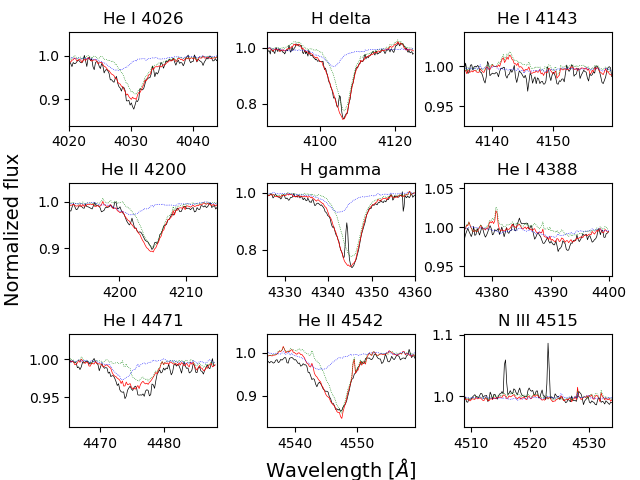}
    \includegraphics[width=7cm, bb=5 0 453 346,clip]{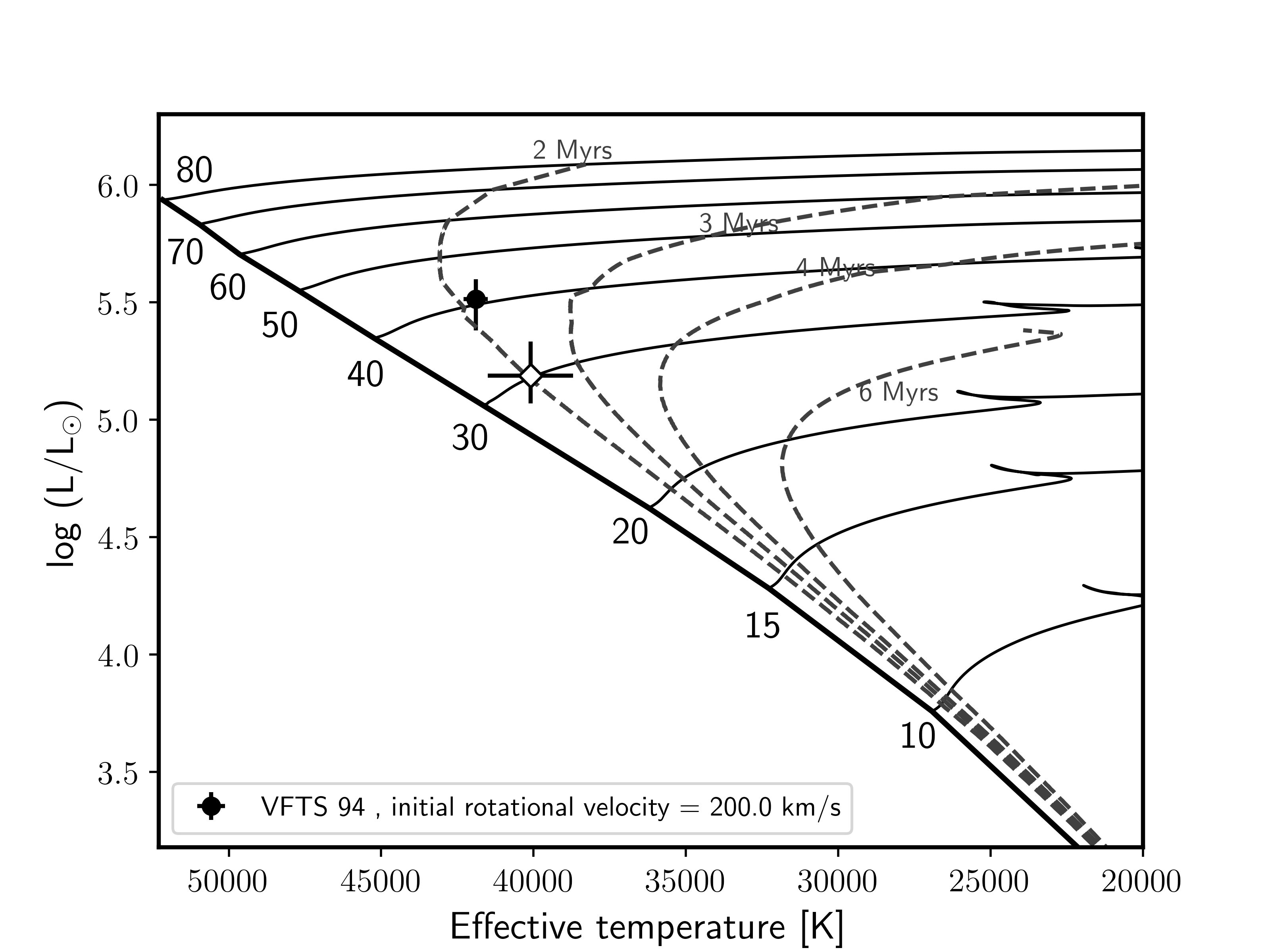}
    \includegraphics[width=7cm, bb=5 0 453 346,clip]{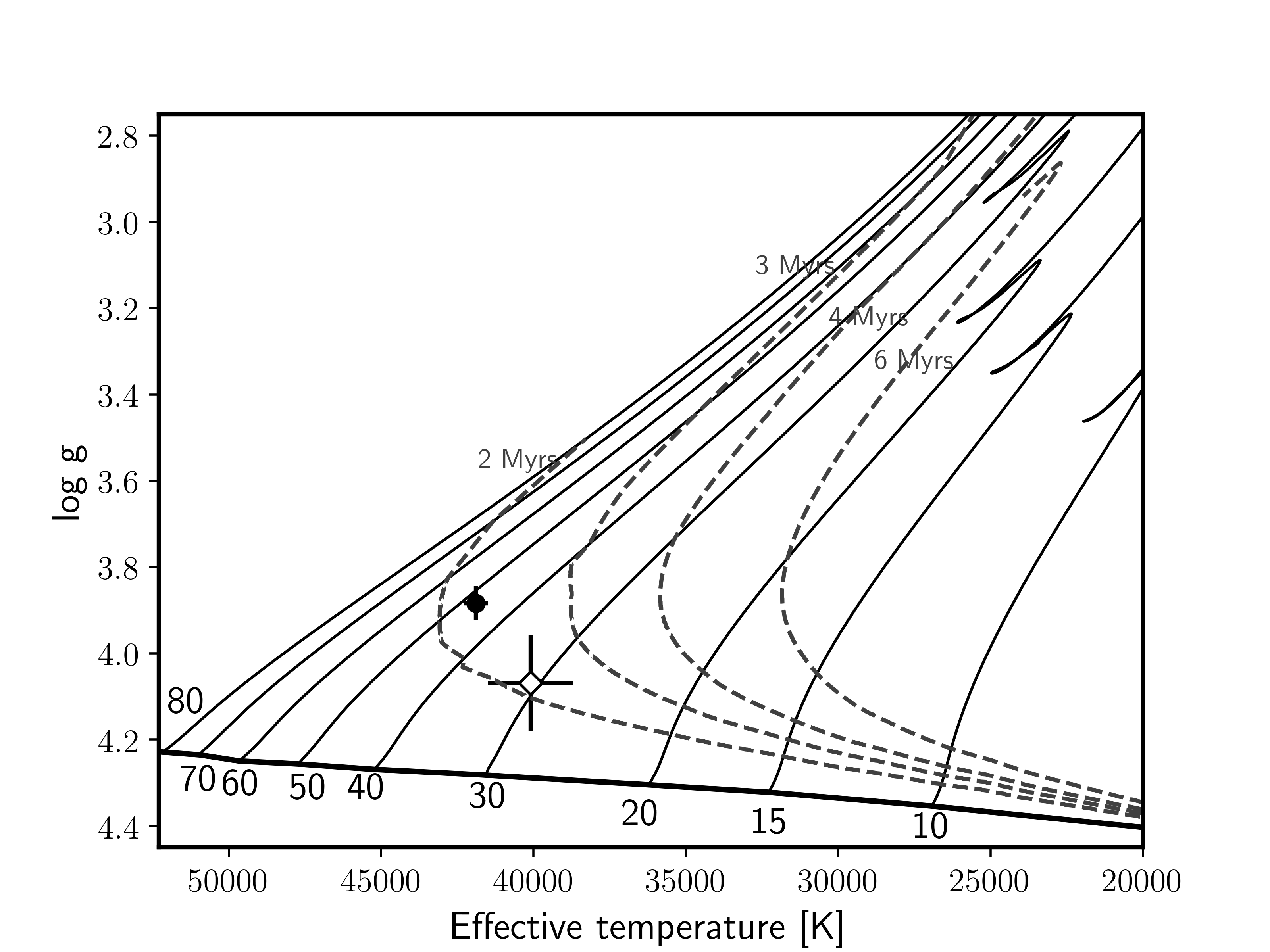}
    \caption{Same as Fig.\,\ref{fig:042} but for VFTS\,094.}\label{fig:094} 
  \end{figure*}
 \clearpage

 \begin{figure*}[t!]
    \centering
    \includegraphics[width=6.cm]{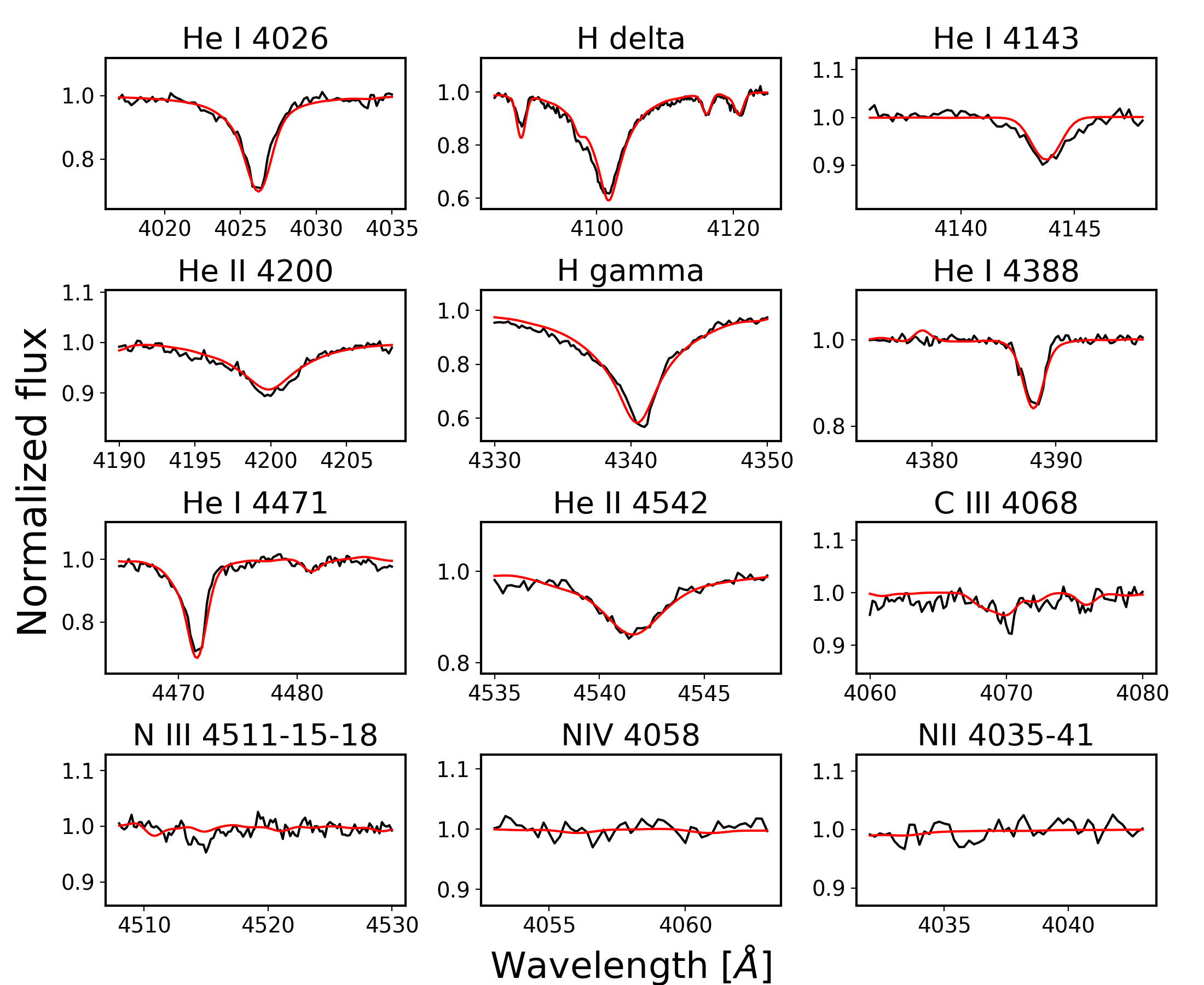}
    \includegraphics[width=7.cm, bb=5 0 453 346,clip]{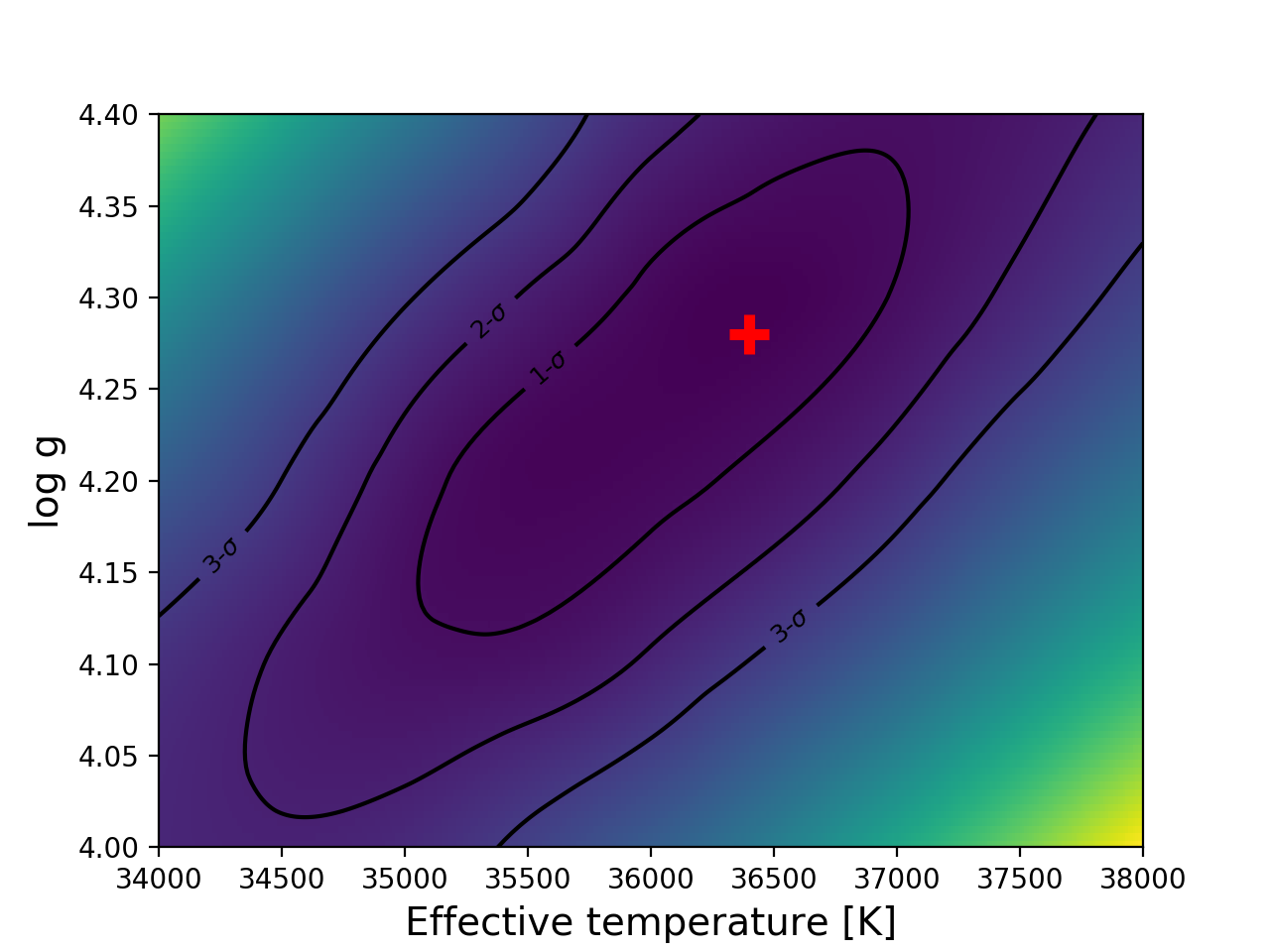}
    \includegraphics[width=6.cm]{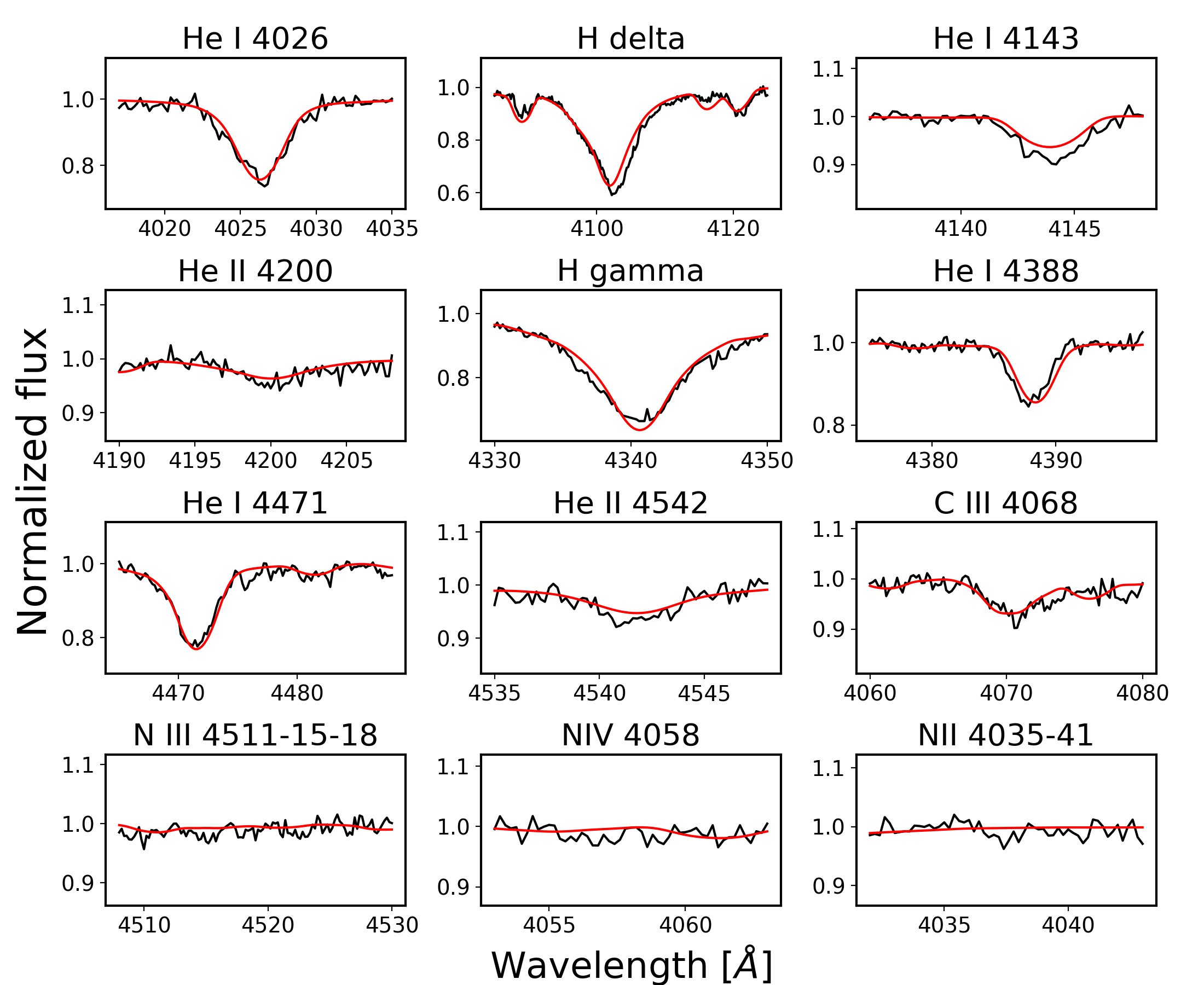}
    \includegraphics[width=7.cm, bb=5 0 453 346,clip]{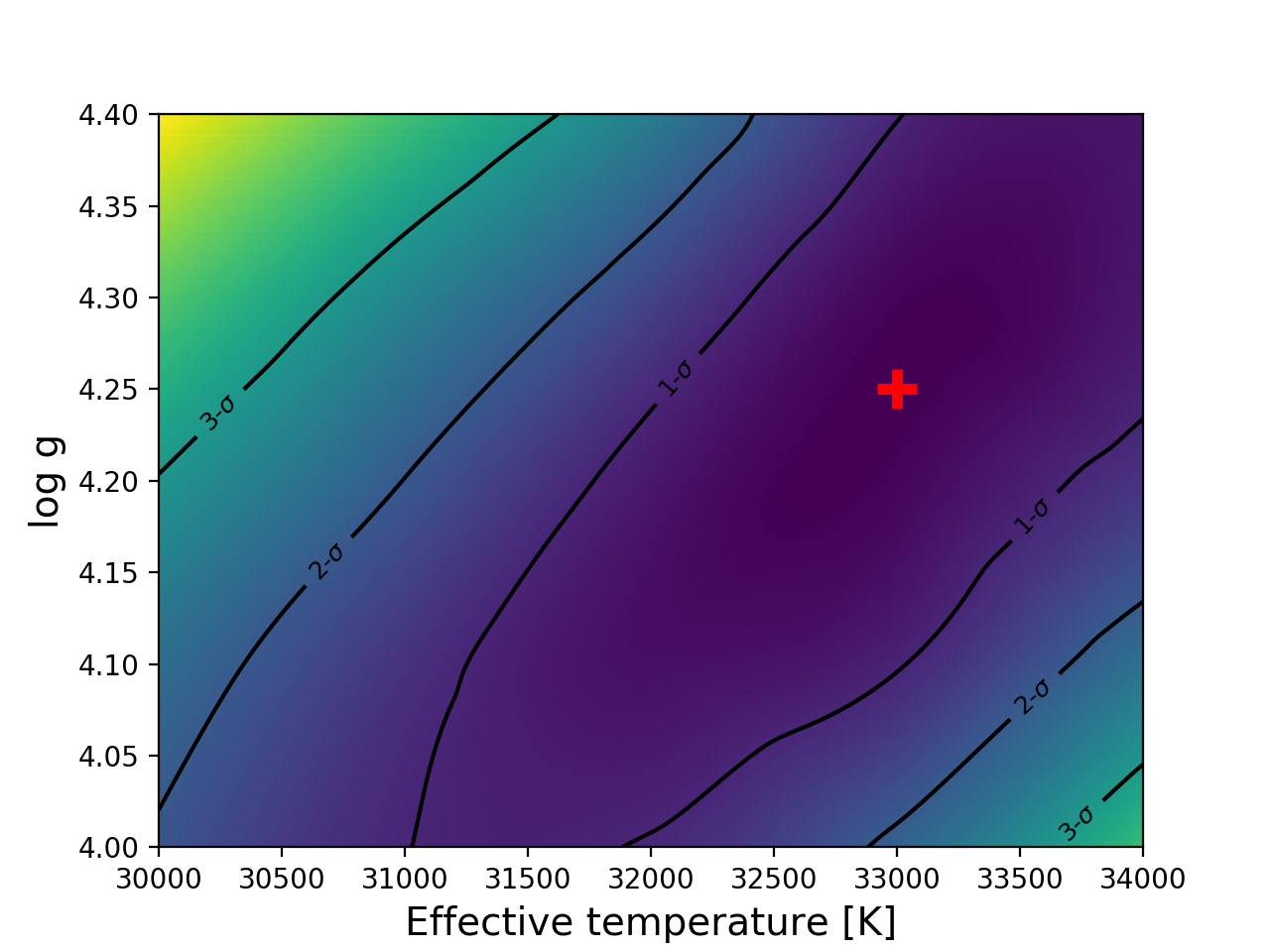}
    \includegraphics[width=7cm]{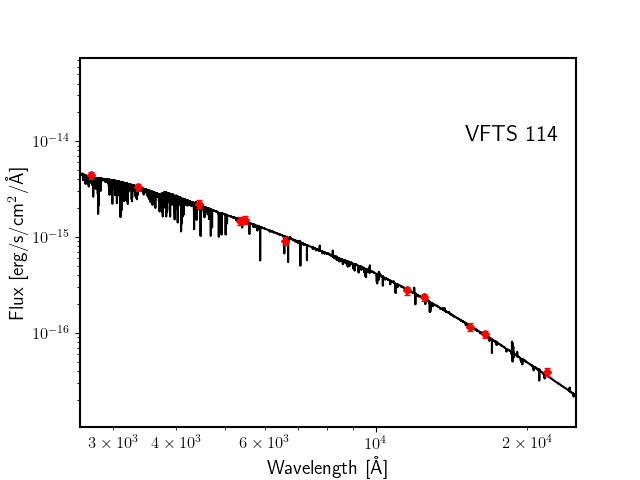}
    \includegraphics[width=6.5cm]{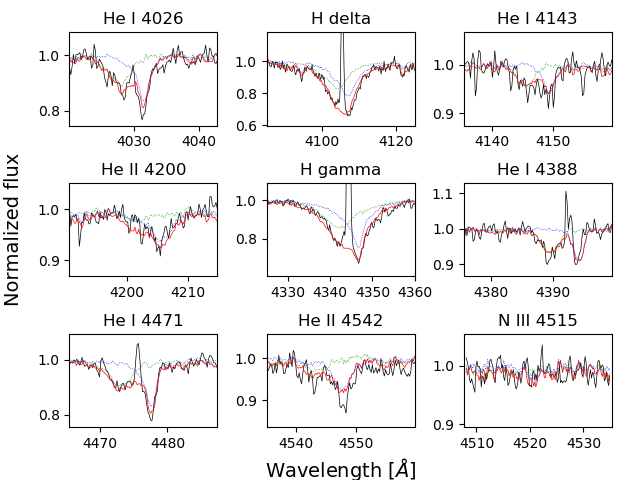}
    \includegraphics[width=7cm, bb=5 0 453 346,clip]{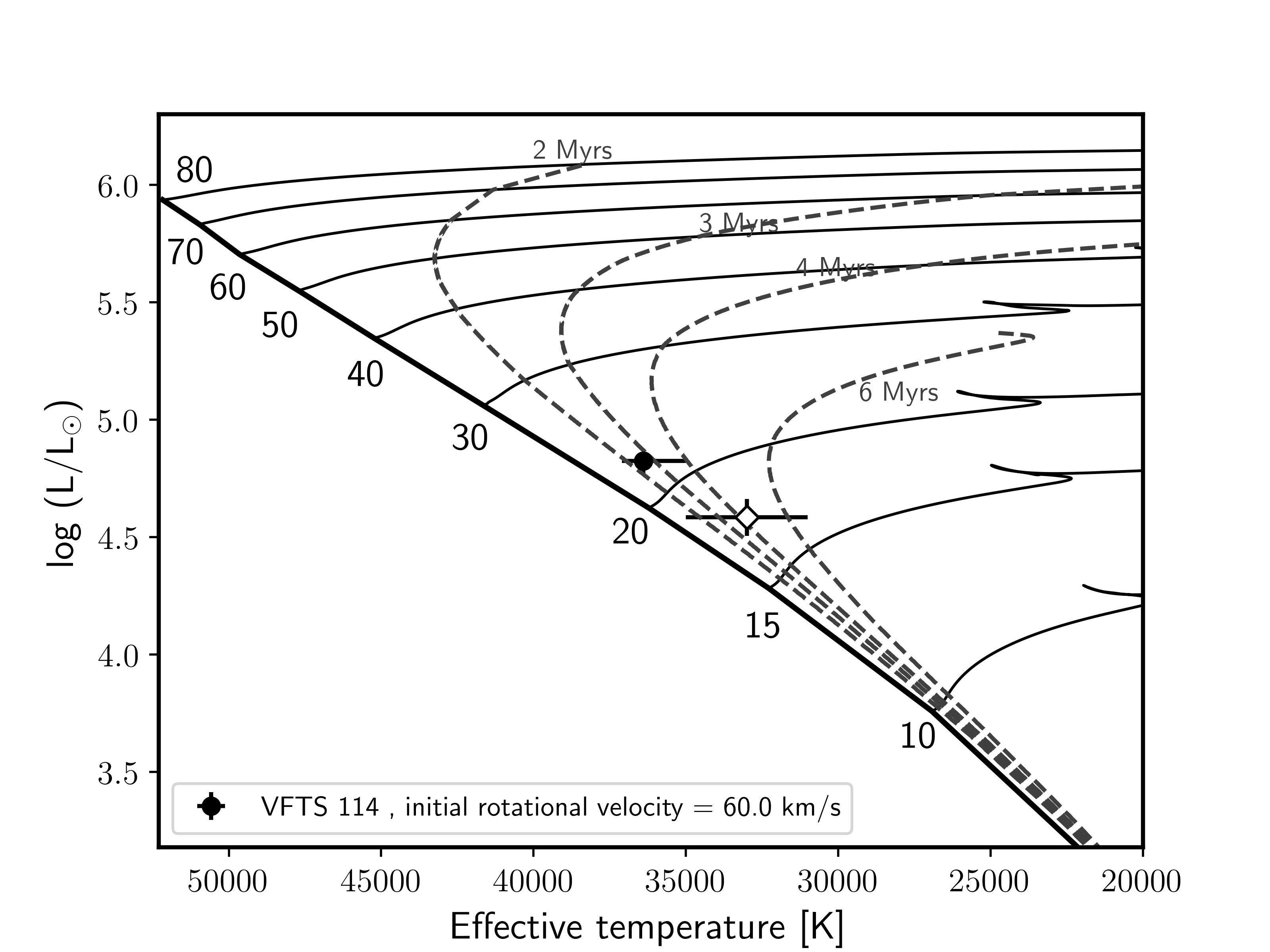}
    \includegraphics[width=7cm, bb=5 0 453 346,clip]{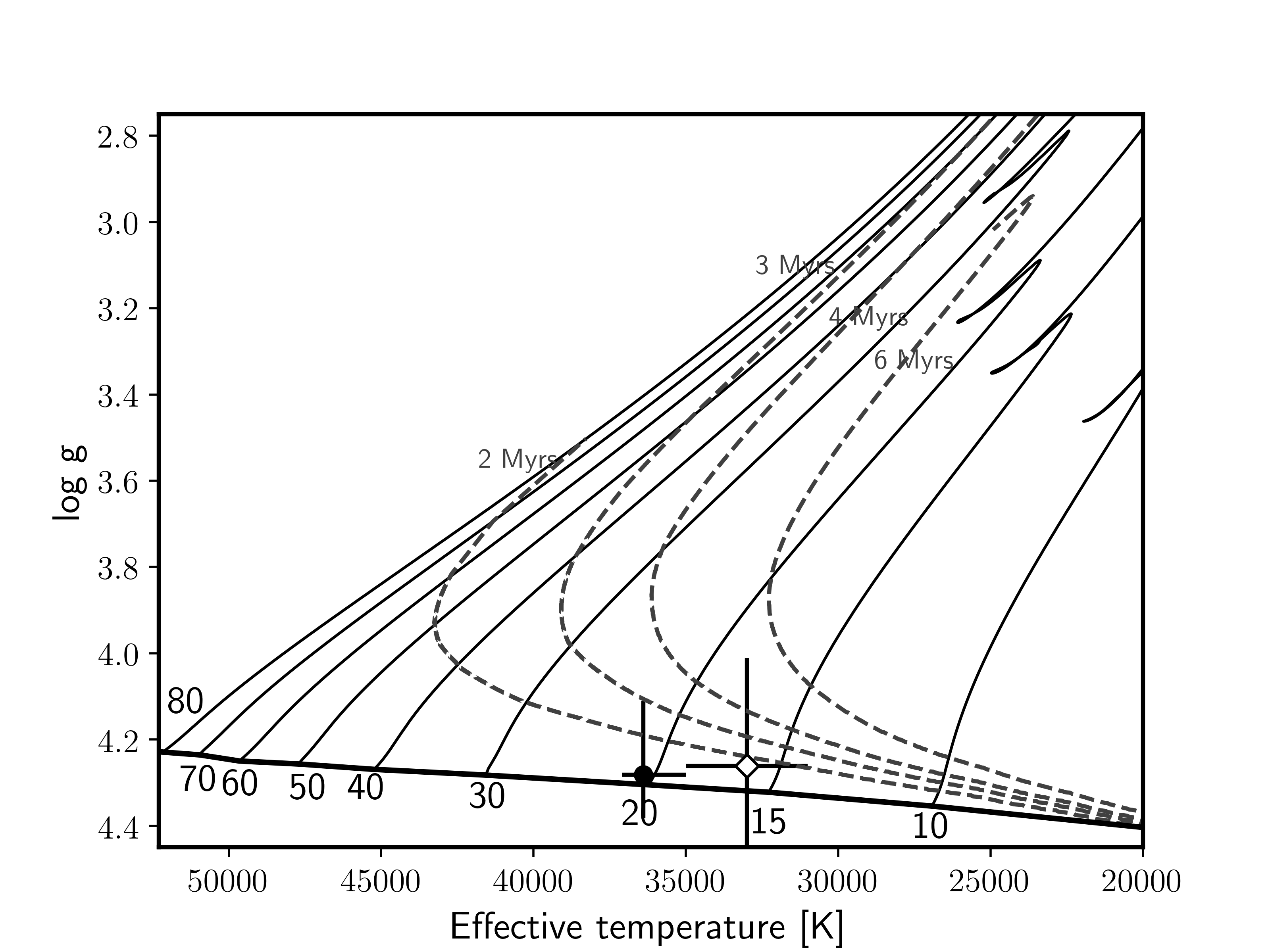}
    \caption{Same as Fig.\,\ref{fig:042} but for VFTS\,114.}\label{fig:114} 
  \end{figure*}
 \clearpage
  
   \begin{figure*}[t!]
    \centering
    \includegraphics[width=6.cm]{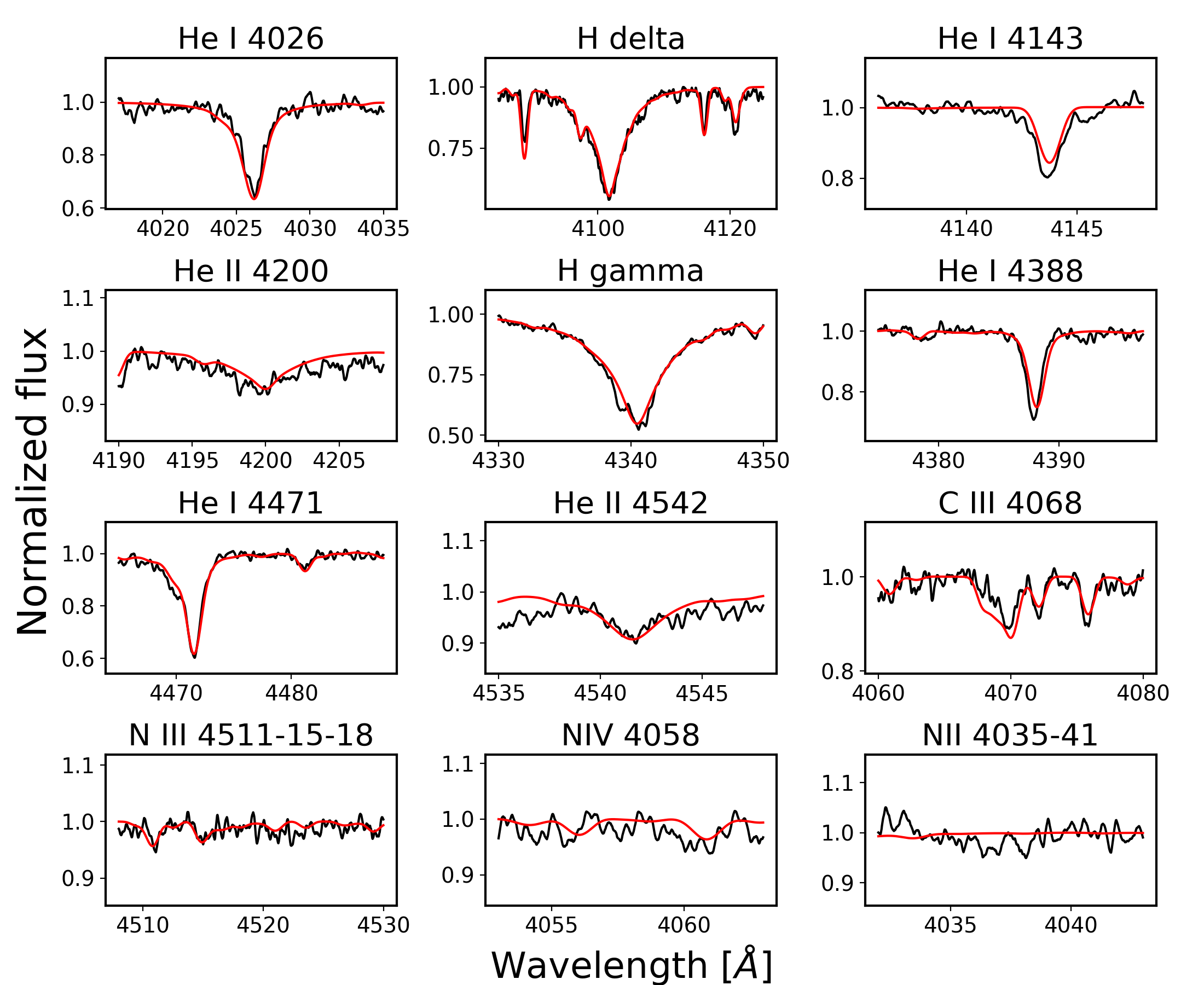}
    \includegraphics[width=7.cm, bb=5 0 453 346,clip]{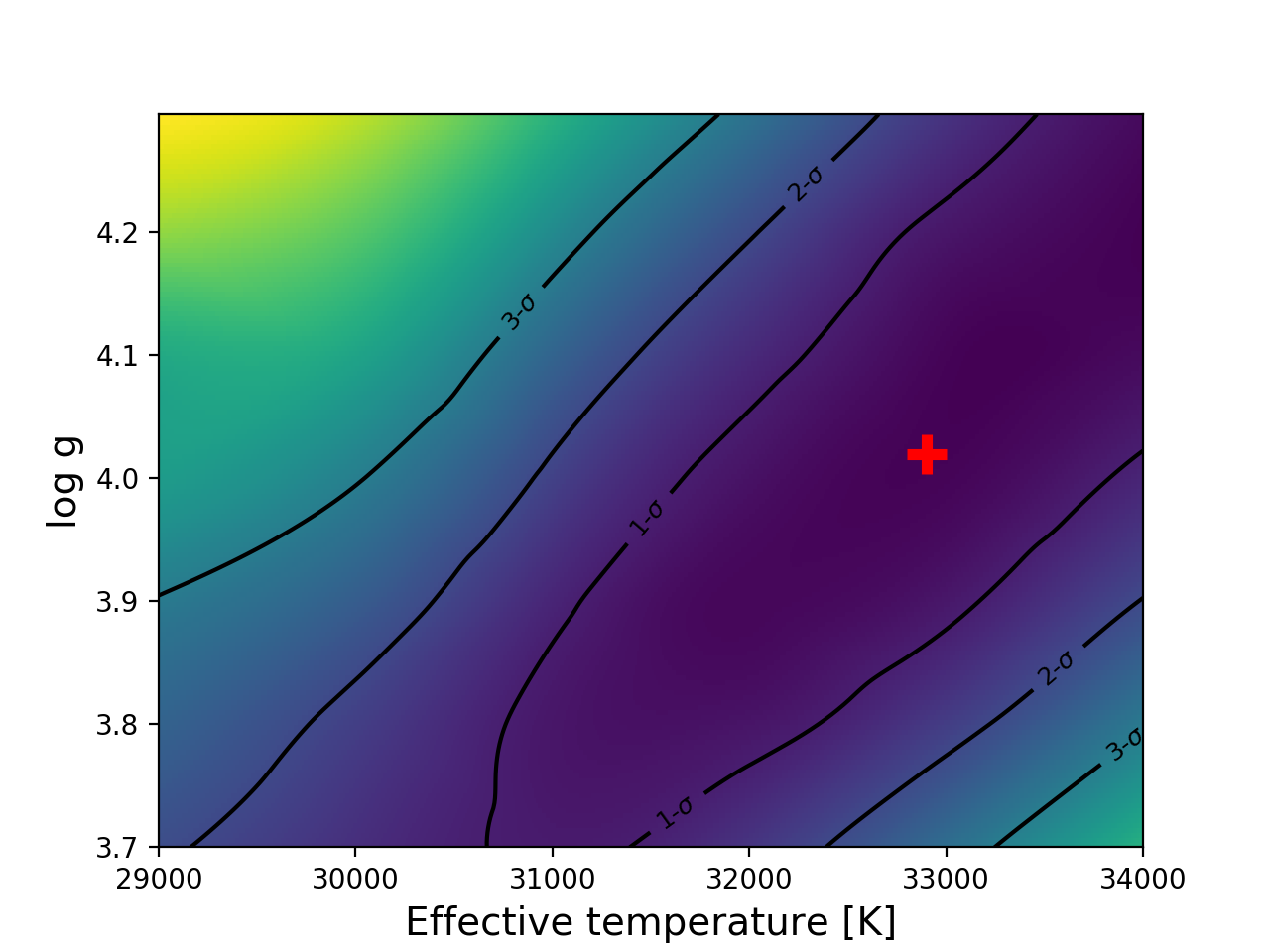}
    \includegraphics[width=6.cm]{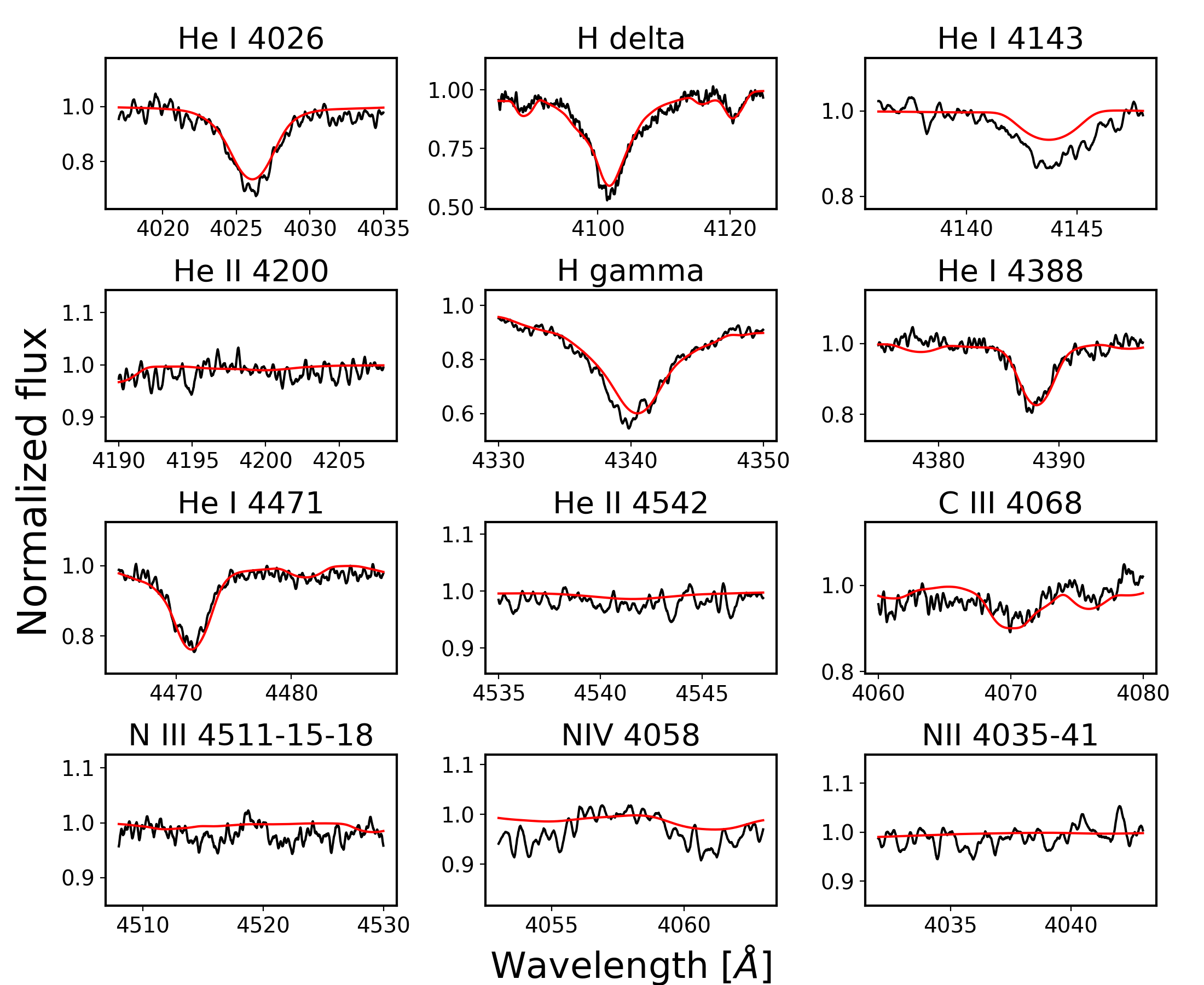}
    \includegraphics[width=7.cm, bb=5 0 453 346,clip]{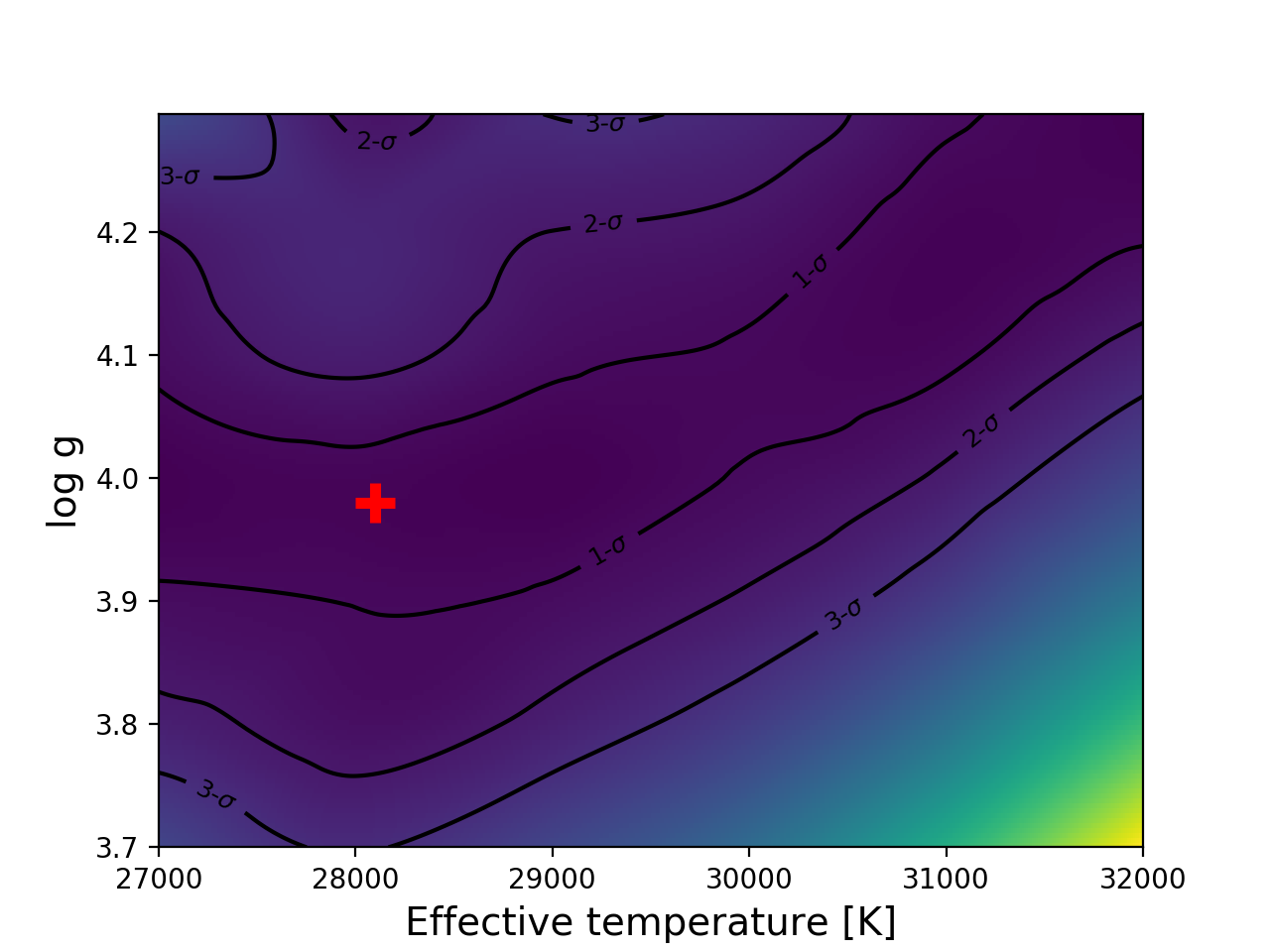}
    \includegraphics[width=7cm]{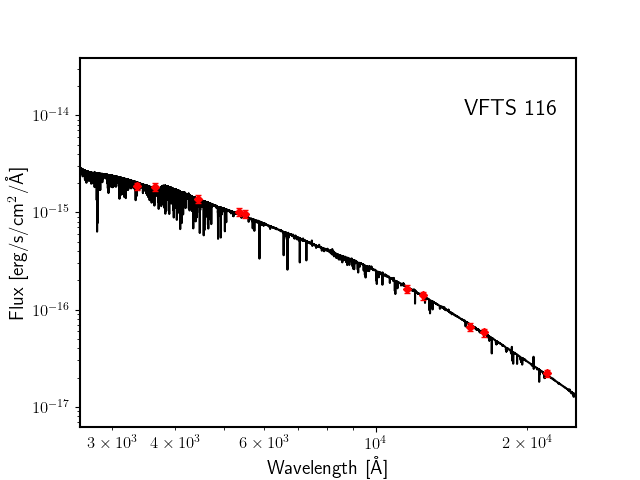}
    \includegraphics[width=6.5cm]{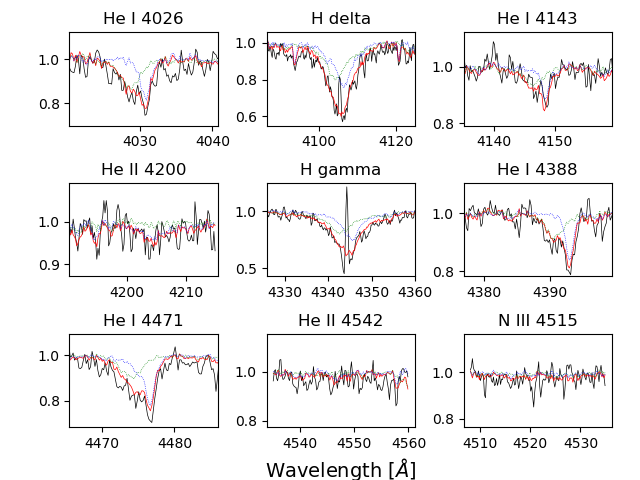}
    \includegraphics[width=7cm, bb=5 0 453 346,clip]{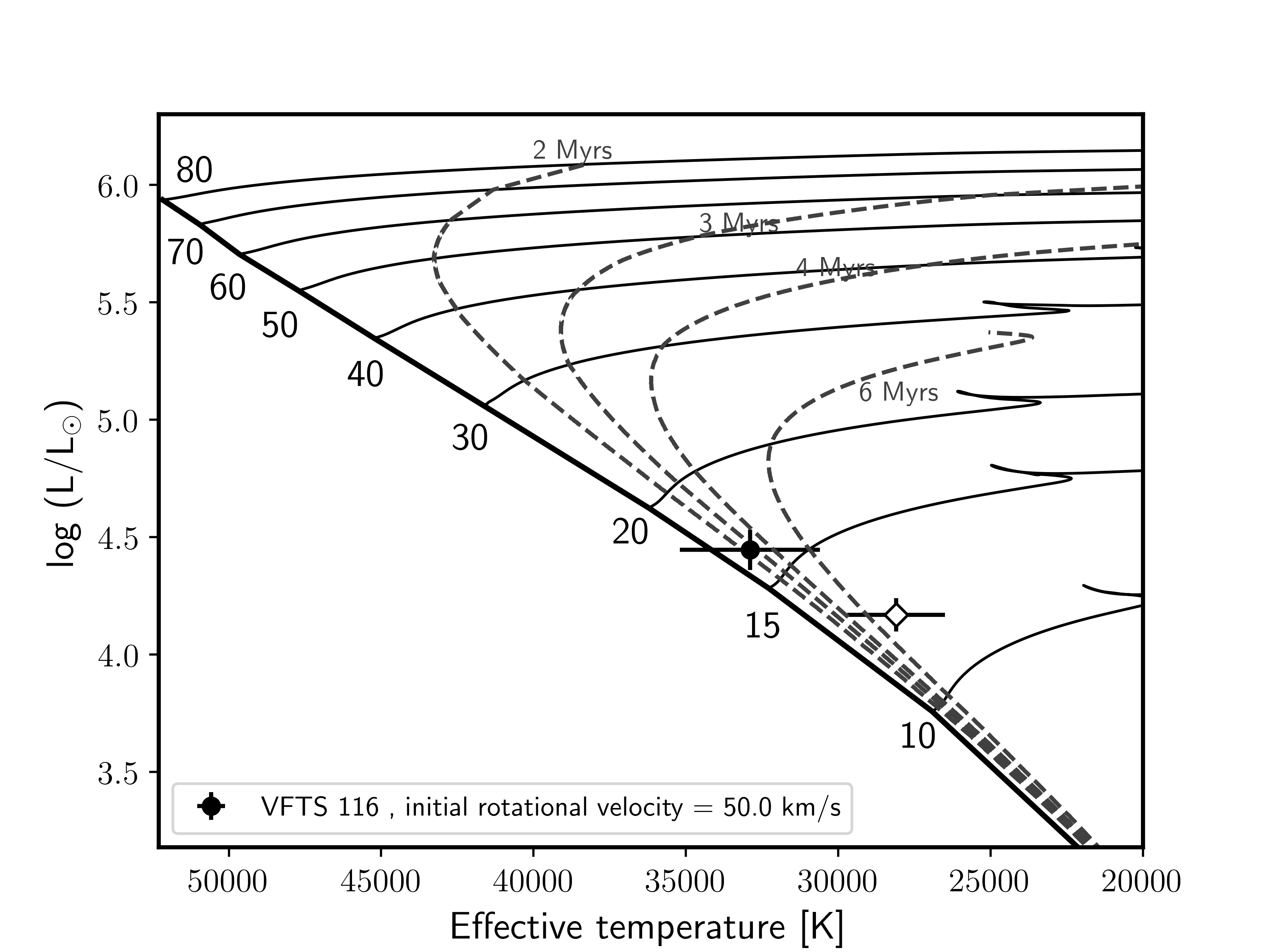}
    \includegraphics[width=7cm, bb=5 0 453 346,clip]{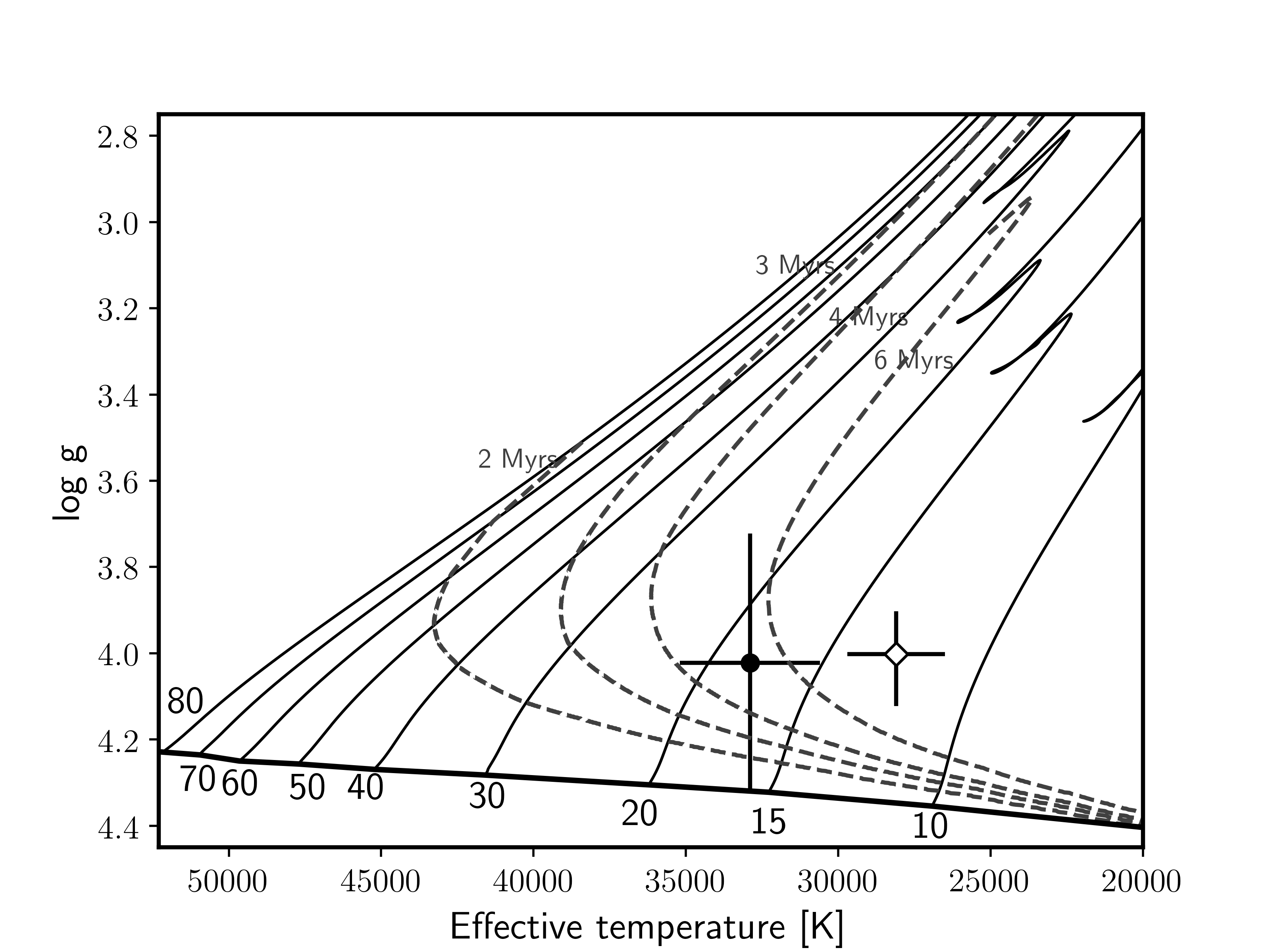}
    \caption{Same as Fig.\,\ref{fig:042} but for VFTS\,116.}\label{fig:116} 
  \end{figure*}
    \clearpage
    
 \begin{figure*}[t!]
    \centering
    \includegraphics[width=6.cm]{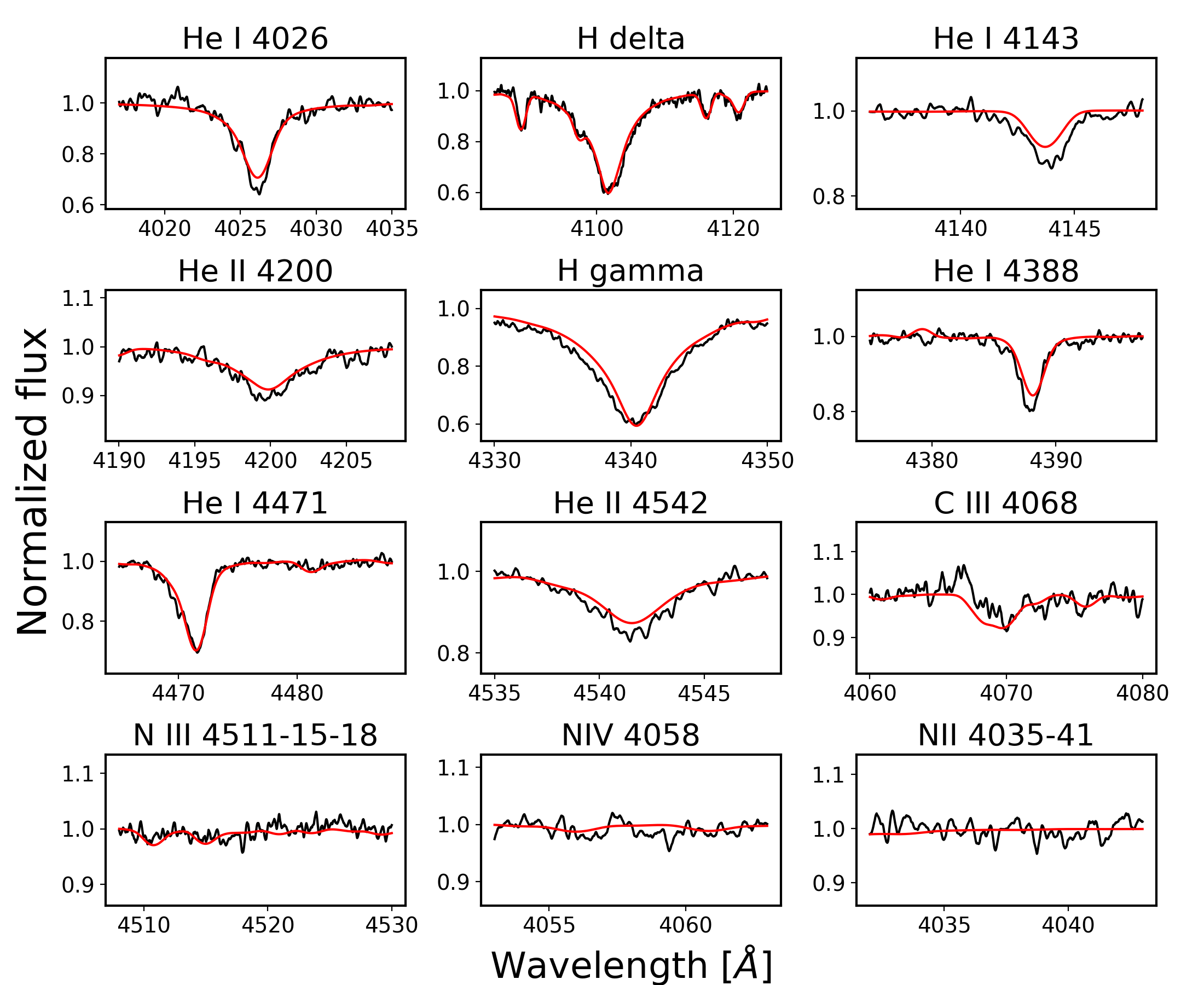}
    \includegraphics[width=7.cm, bb=5 0 453 346,clip]{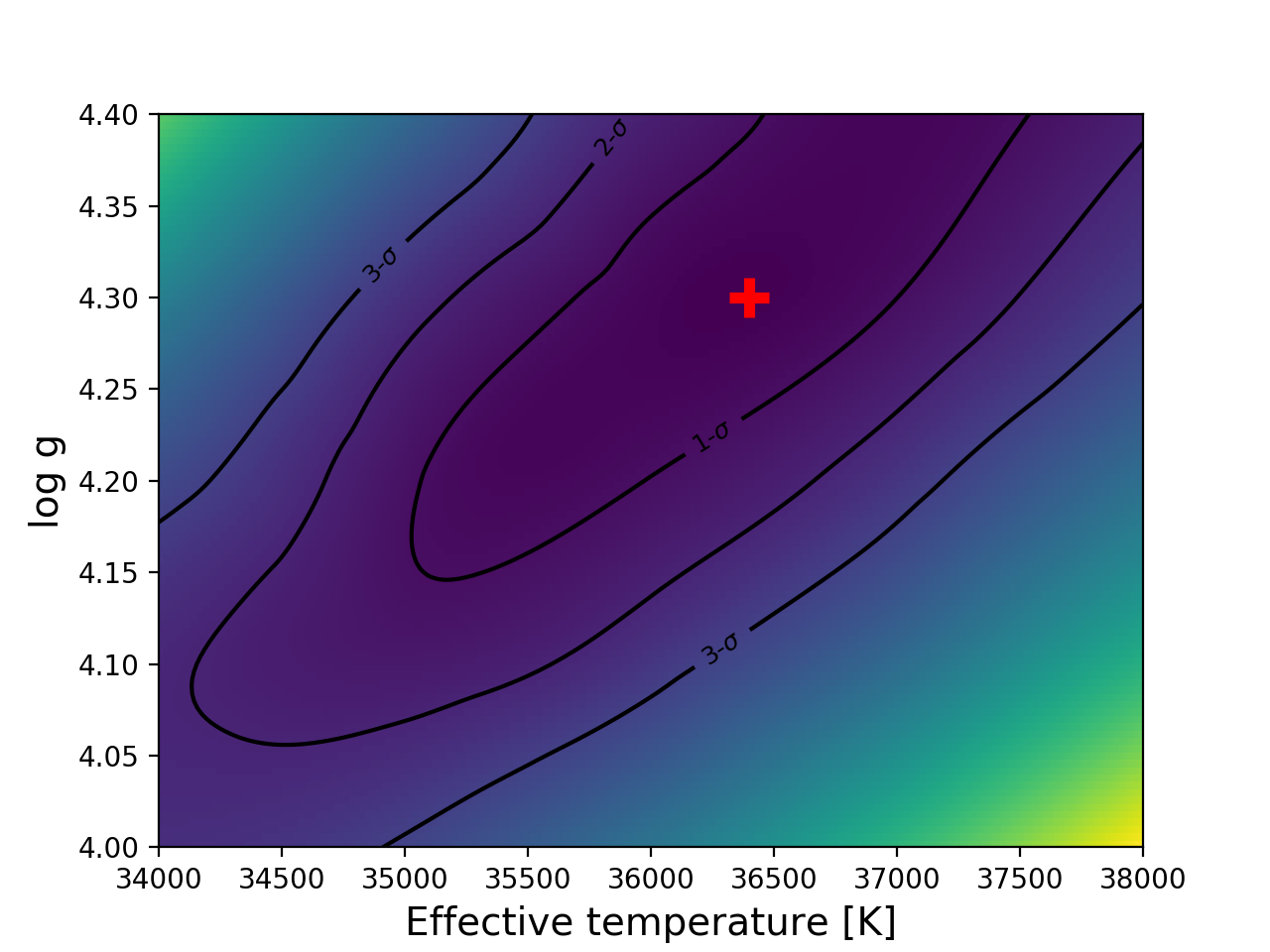}
    \includegraphics[width=6.cm]{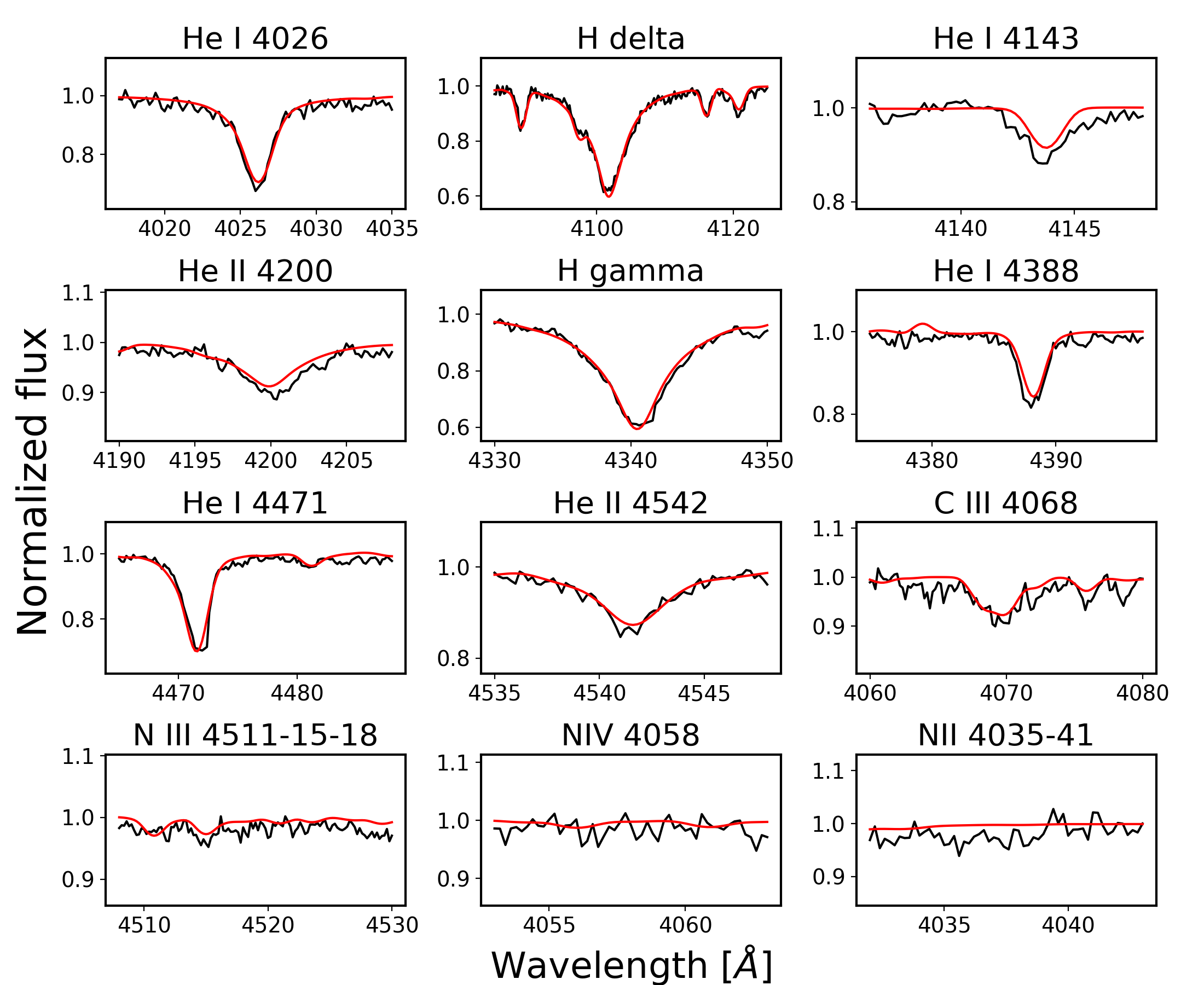}
    \includegraphics[width=7.cm, bb=5 0 453 346,clip]{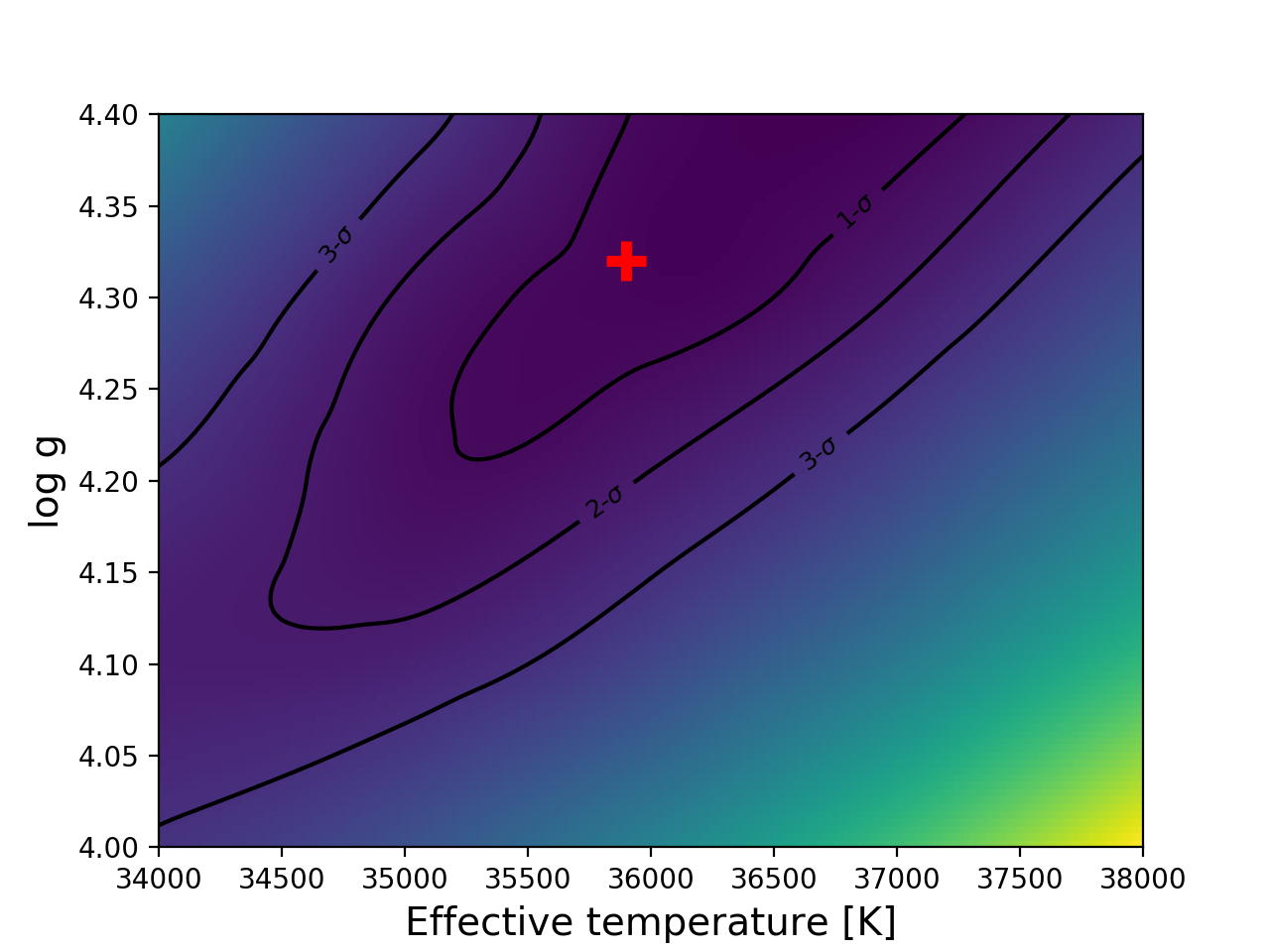}
    \includegraphics[width=7cm]{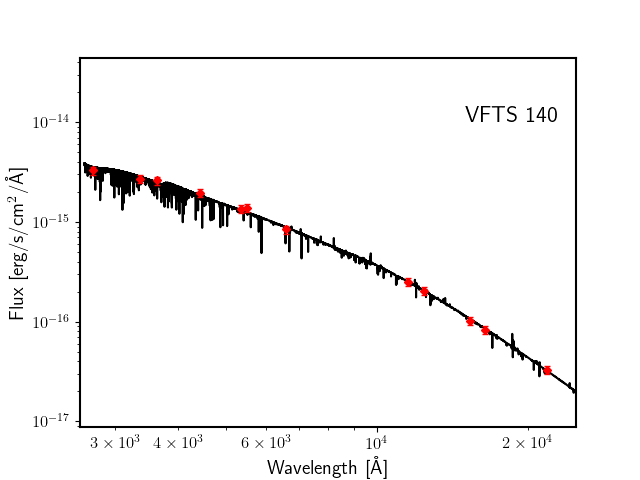}
    \includegraphics[width=6.5cm]{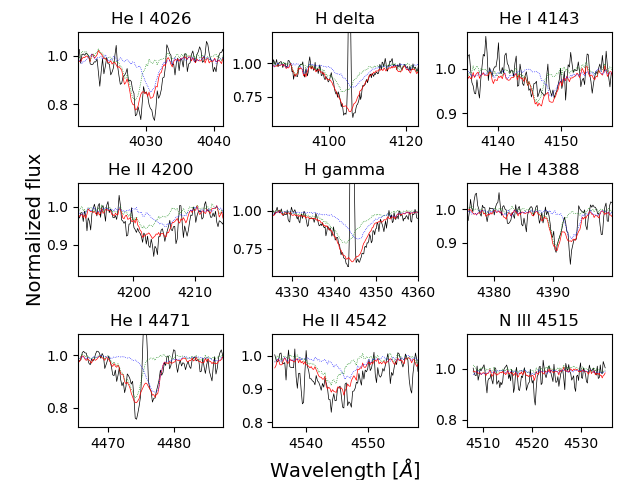}
    \includegraphics[width=7cm, bb=5 0 453 346,clip]{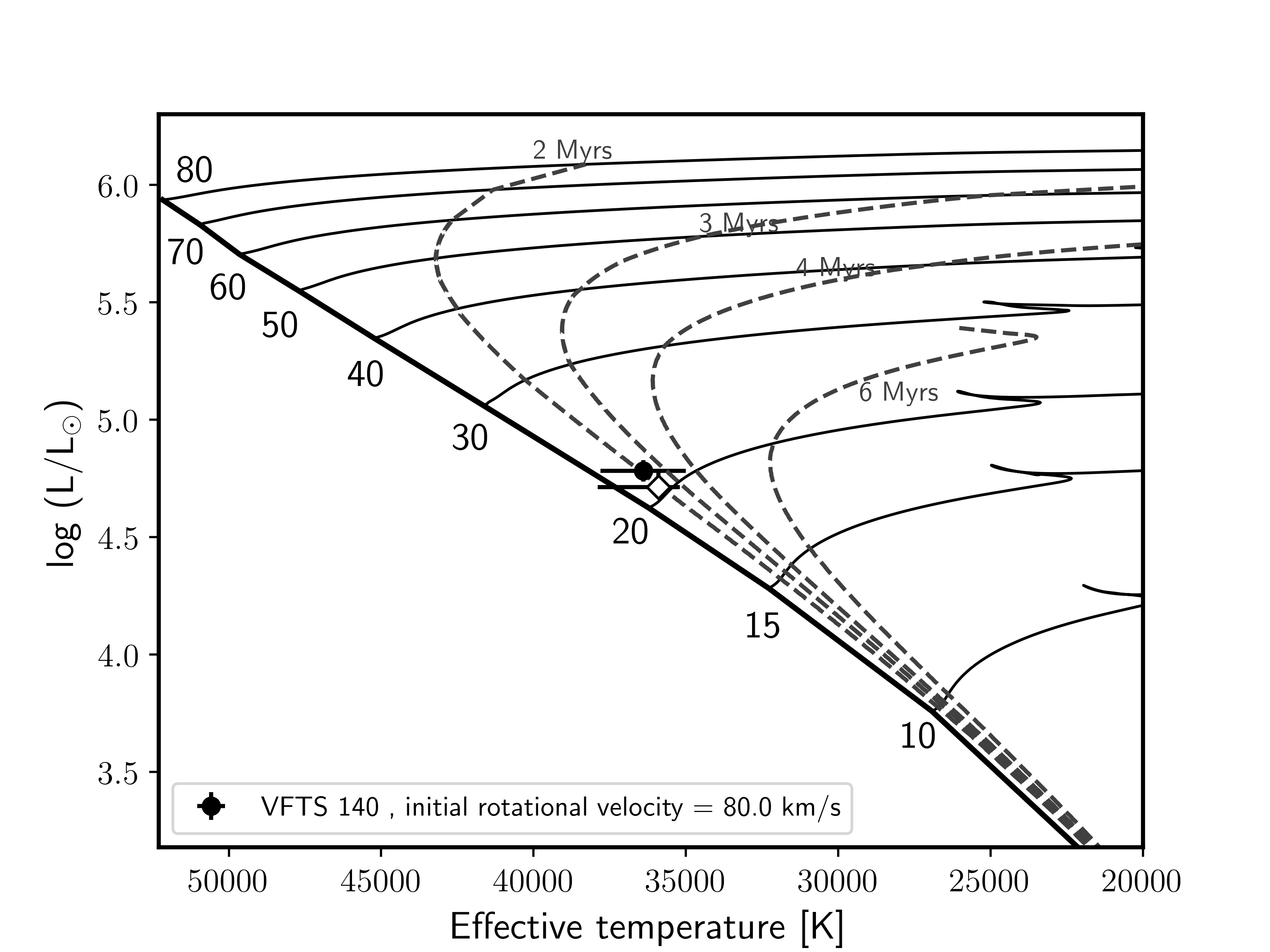}
    \includegraphics[width=7cm, bb=5 0 453 346,clip]{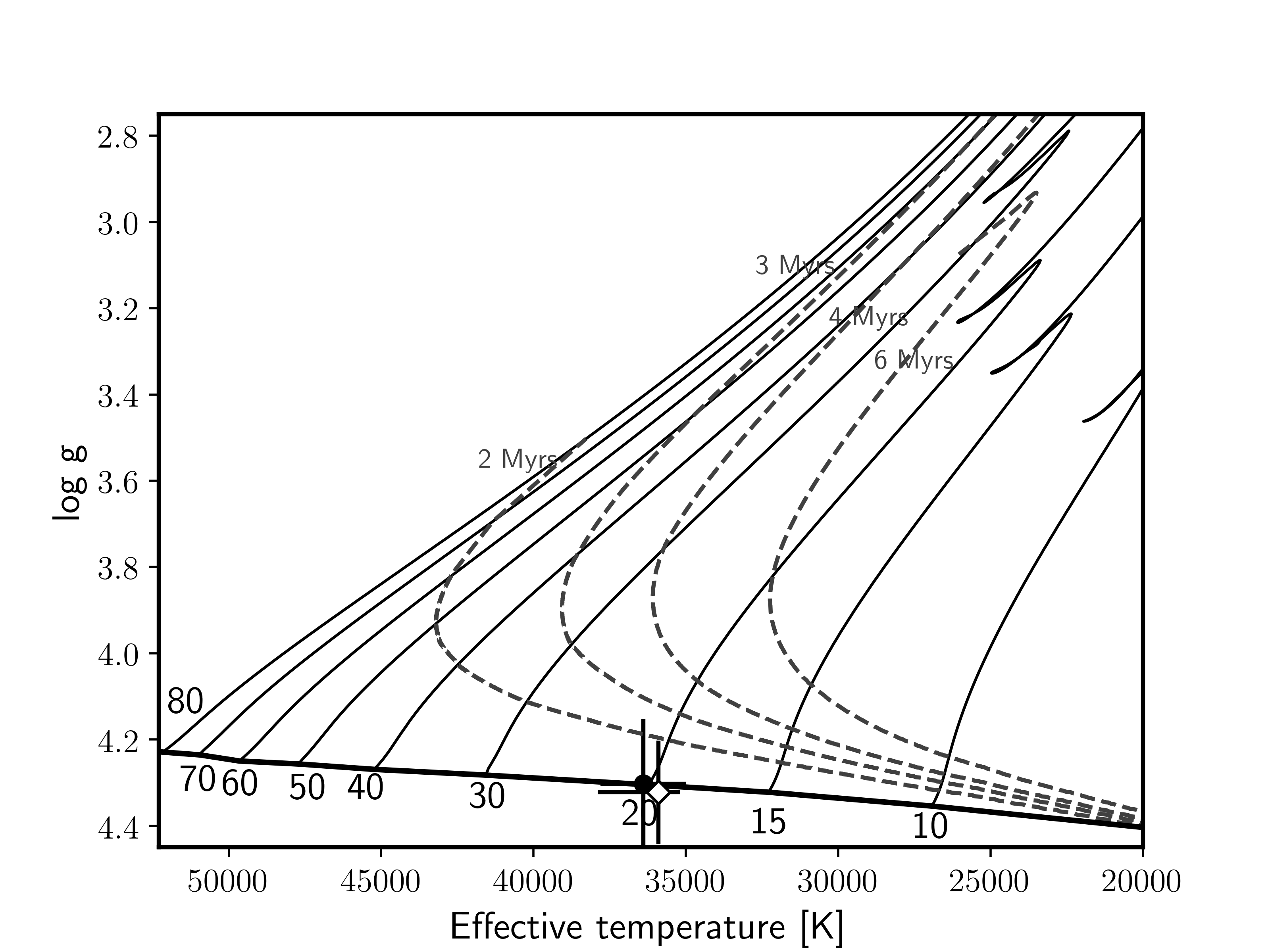}
    \caption{Same as Fig.\,\ref{fig:042} but for VFTS\,140.}\label{fig:140} 
  \end{figure*}
   \clearpage

 \begin{figure*}[t!]
    \centering
    \includegraphics[width=6.cm]{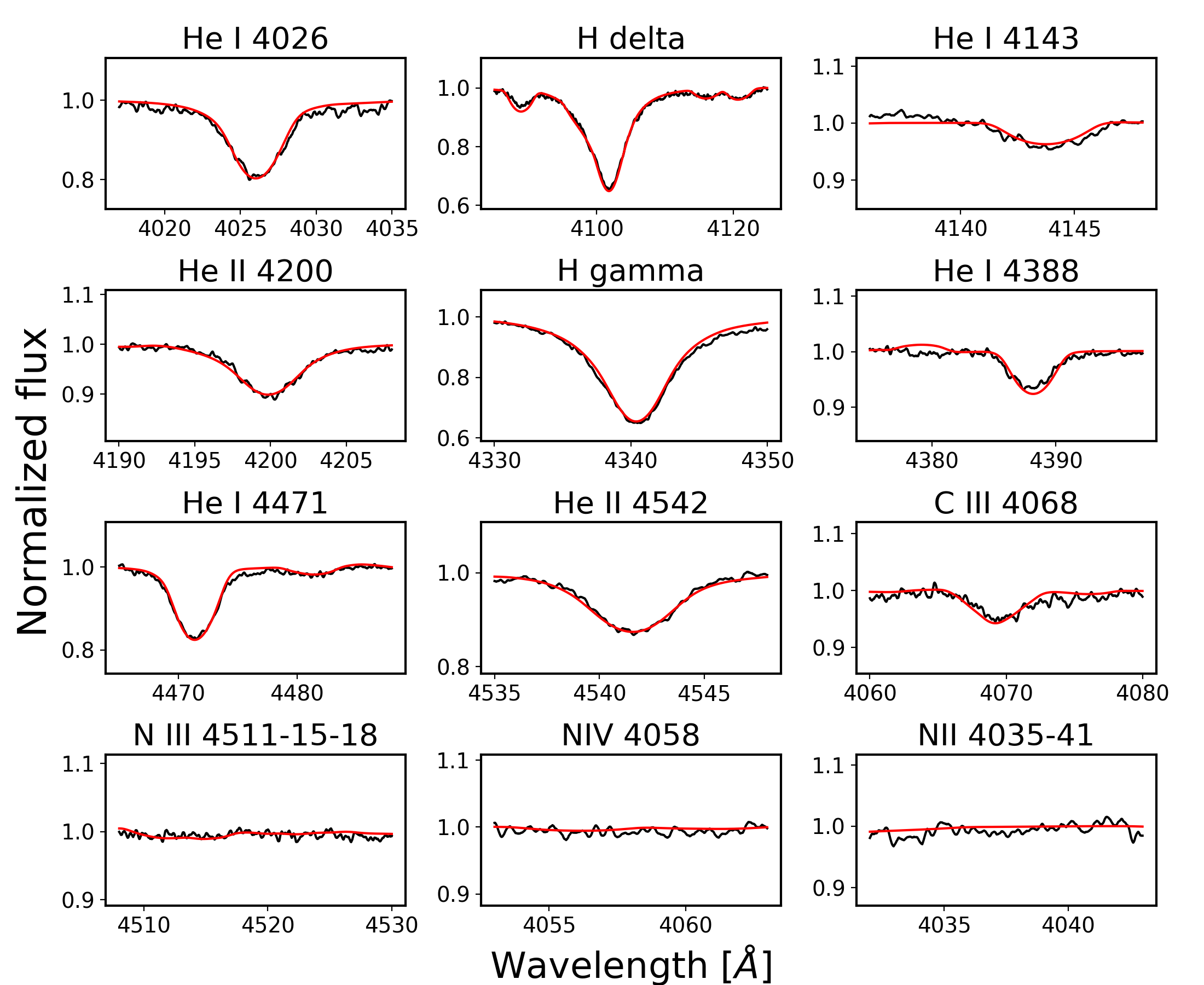}
    \includegraphics[width=7.cm, bb=5 0 453 346,clip]{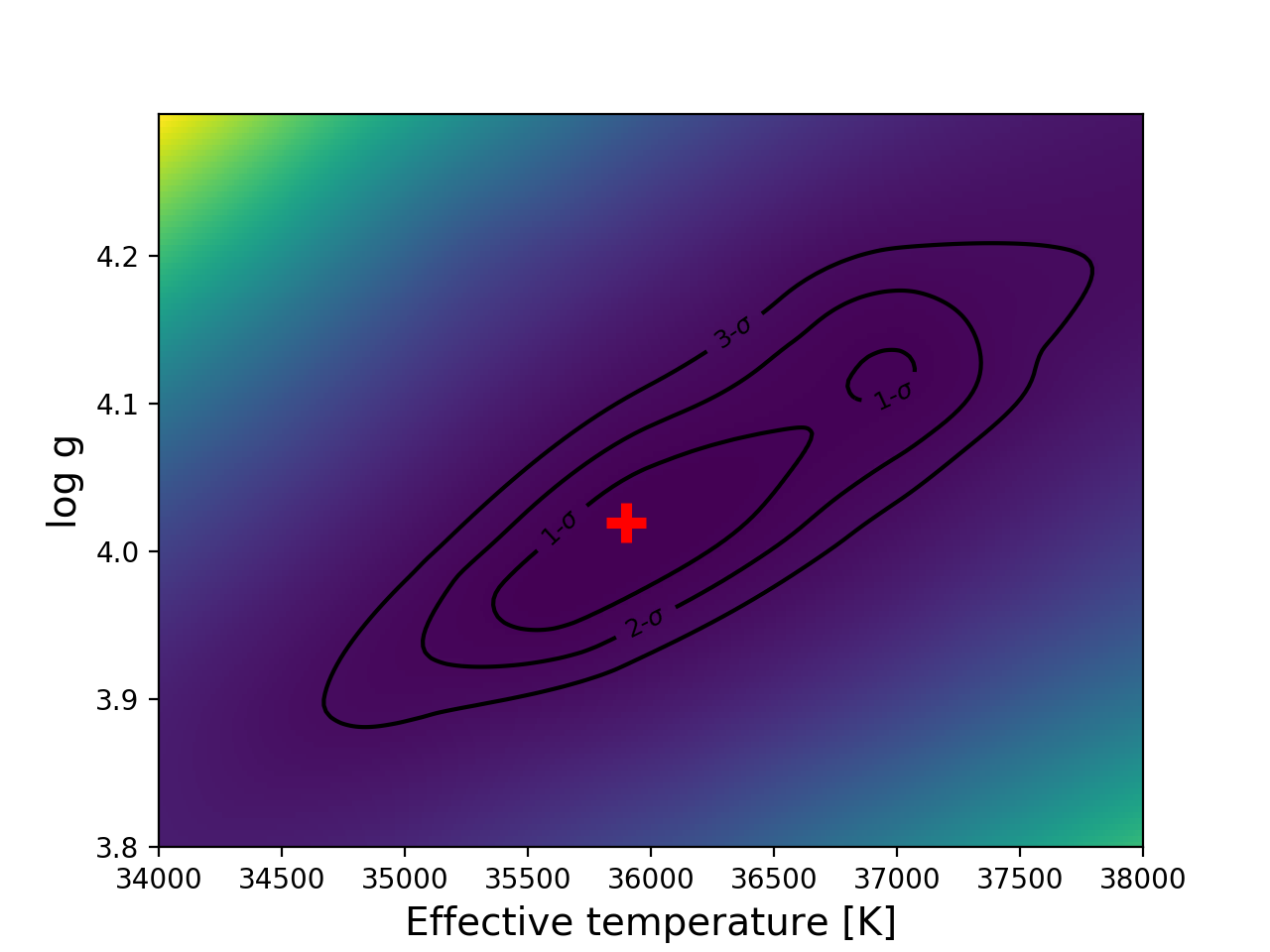}
    \includegraphics[width=6.cm]{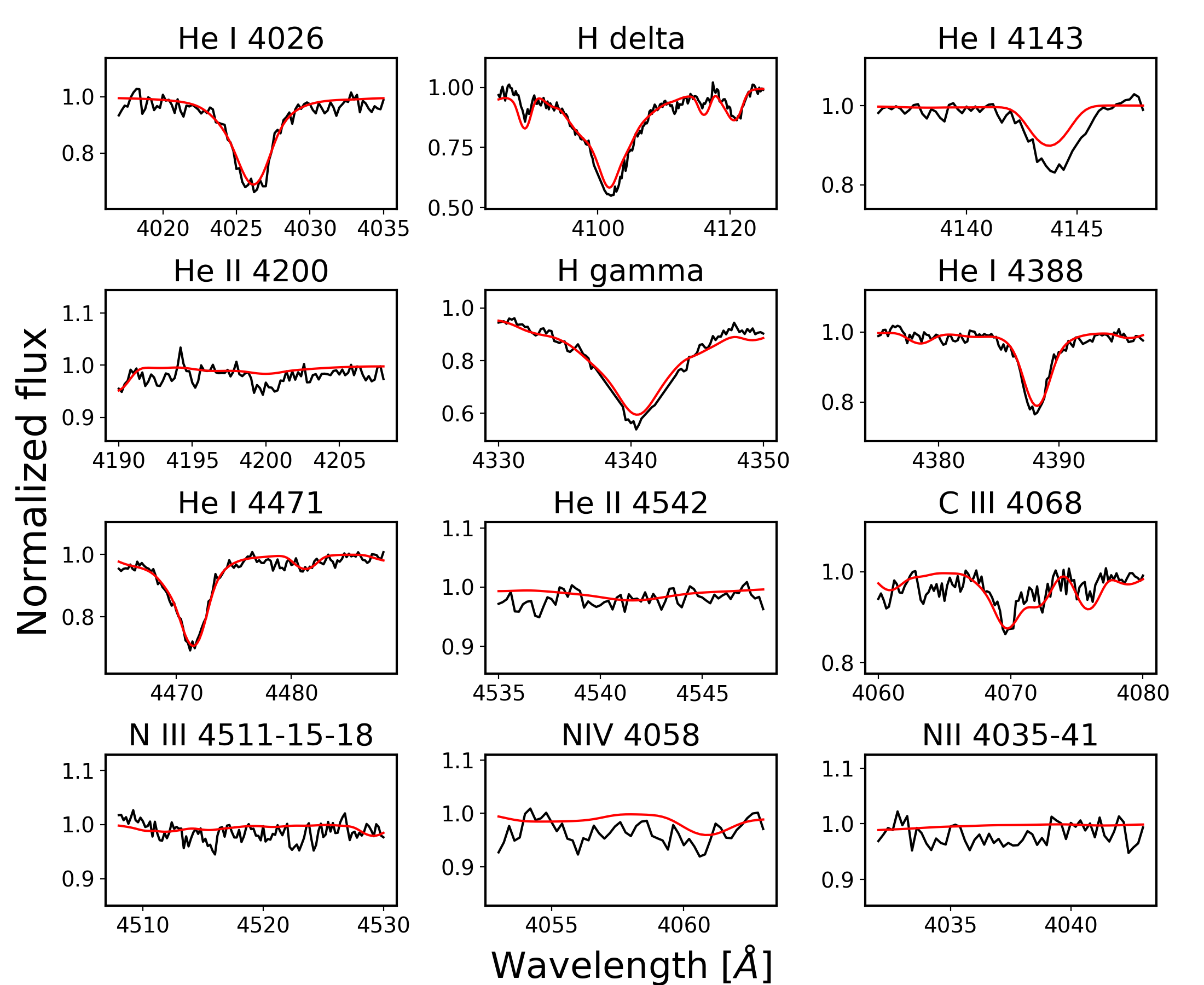}
    \includegraphics[width=7.cm, bb=5 0 453 346,clip]{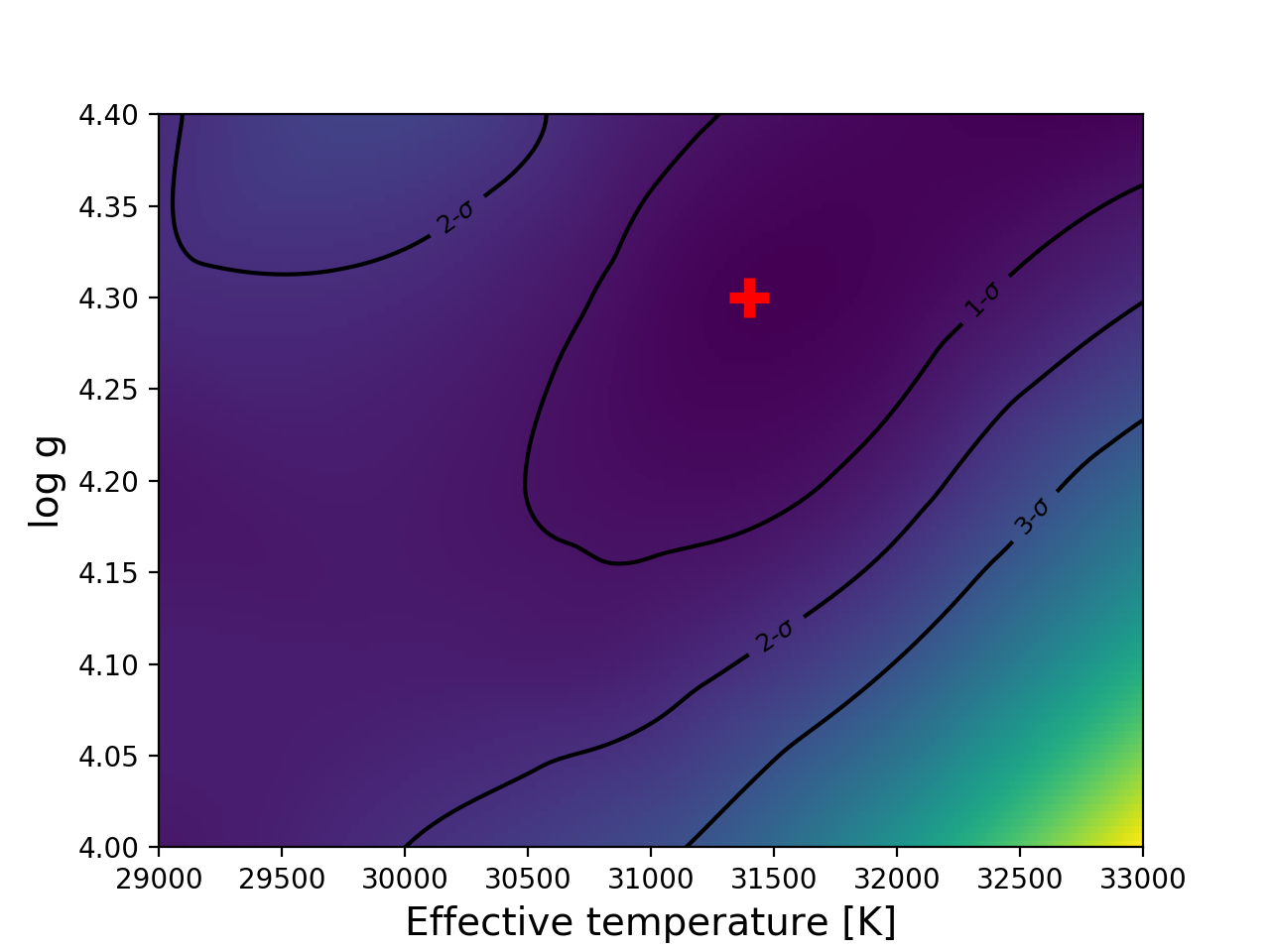}
    \includegraphics[width=7cm]{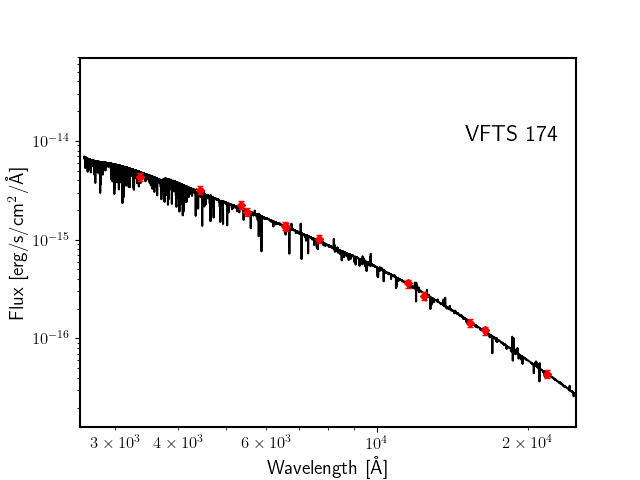}
    \includegraphics[width=6.5cm]{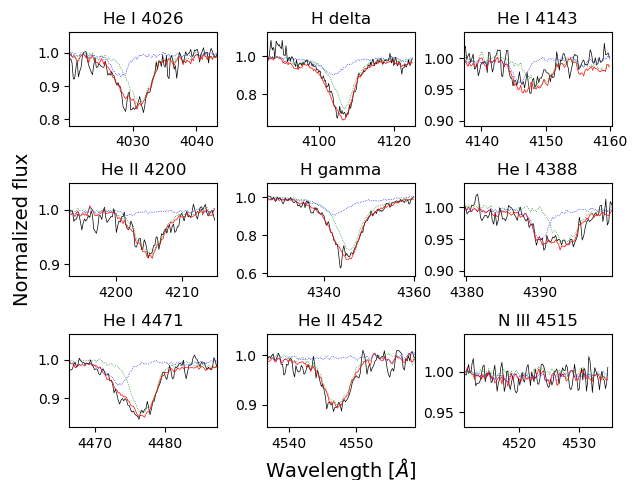}
    \includegraphics[width=7cm, bb=5 0 453 346,clip]{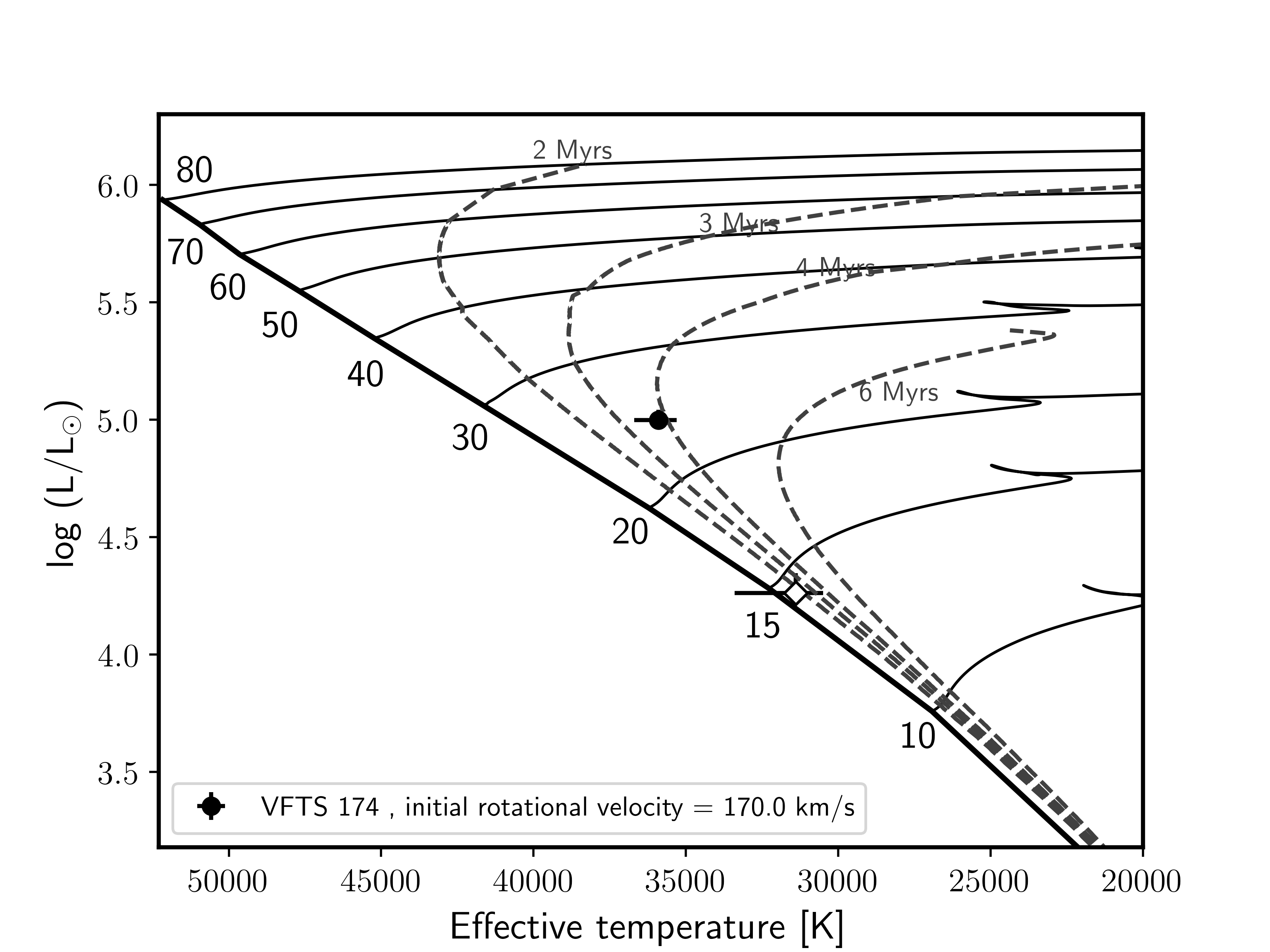}
    \includegraphics[width=7cm, bb=5 0 453 346,clip]{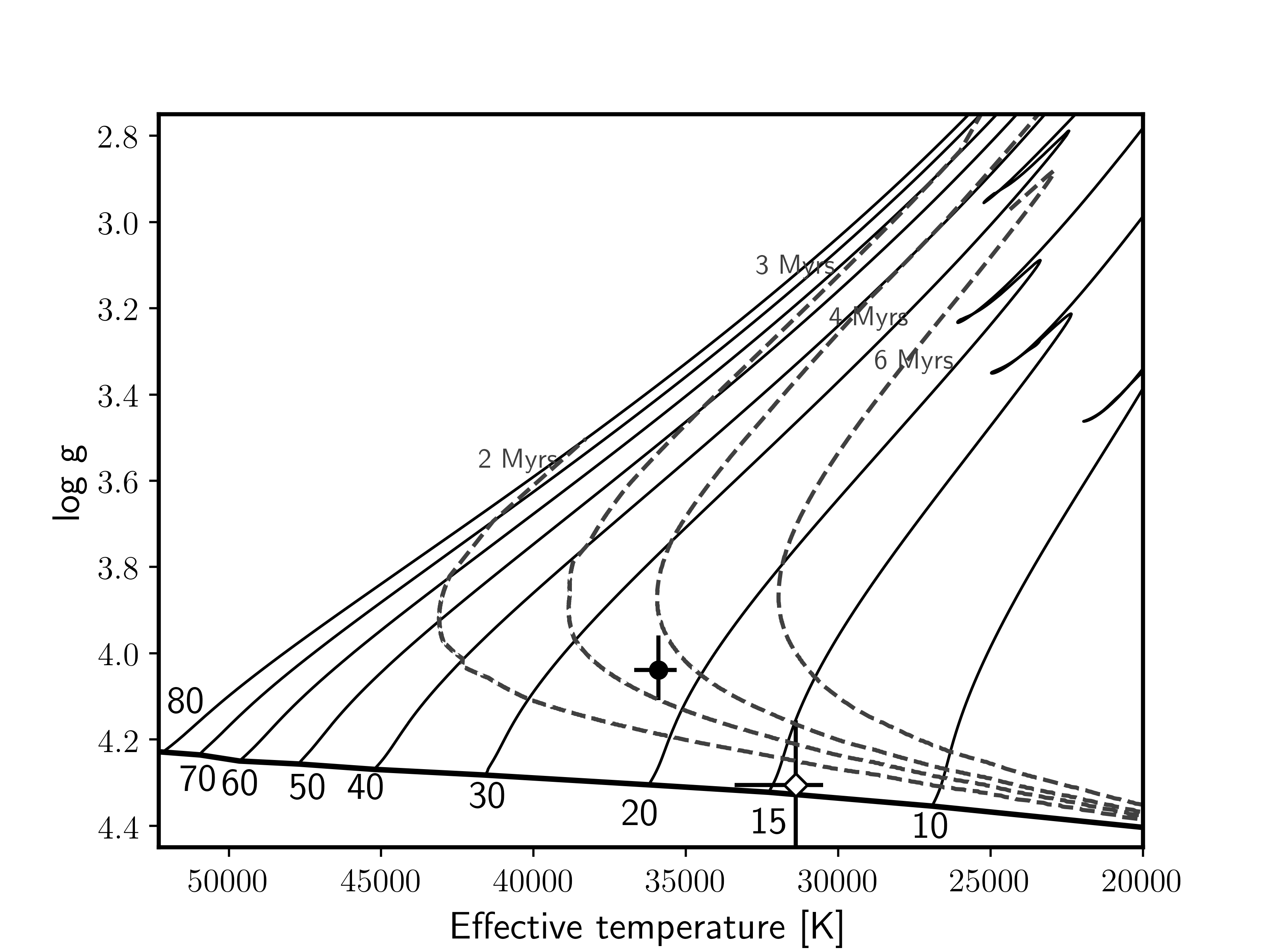}
    \caption{Same as Fig.\,\ref{fig:042} but for VFTS\,174.} \label{fig:174} 
  \end{figure*}
   \clearpage

 \begin{figure*}[t!]
    \centering
     \includegraphics[width=6.cm]{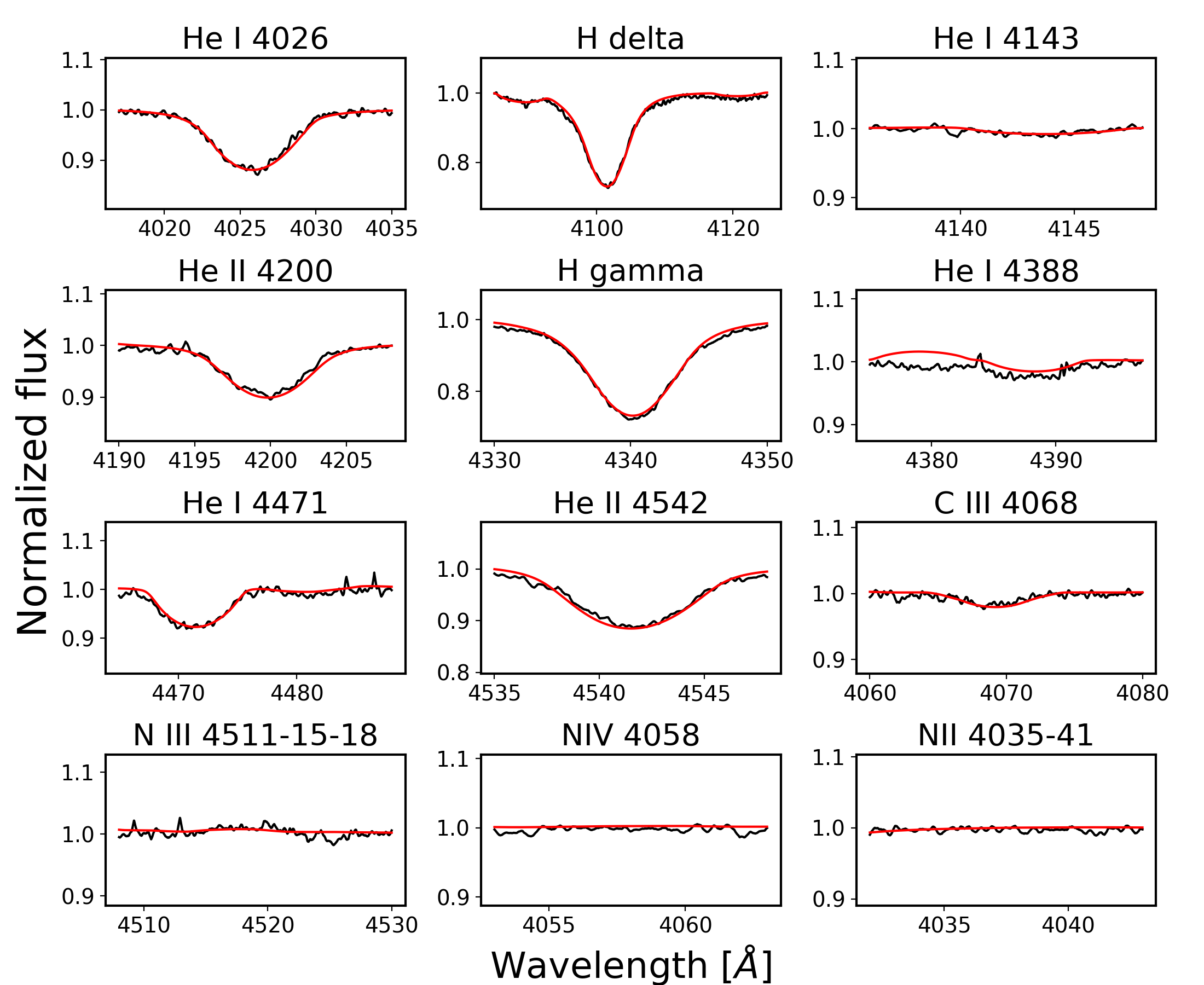}
    \includegraphics[width=7.cm, bb=5 0 453 346,clip]{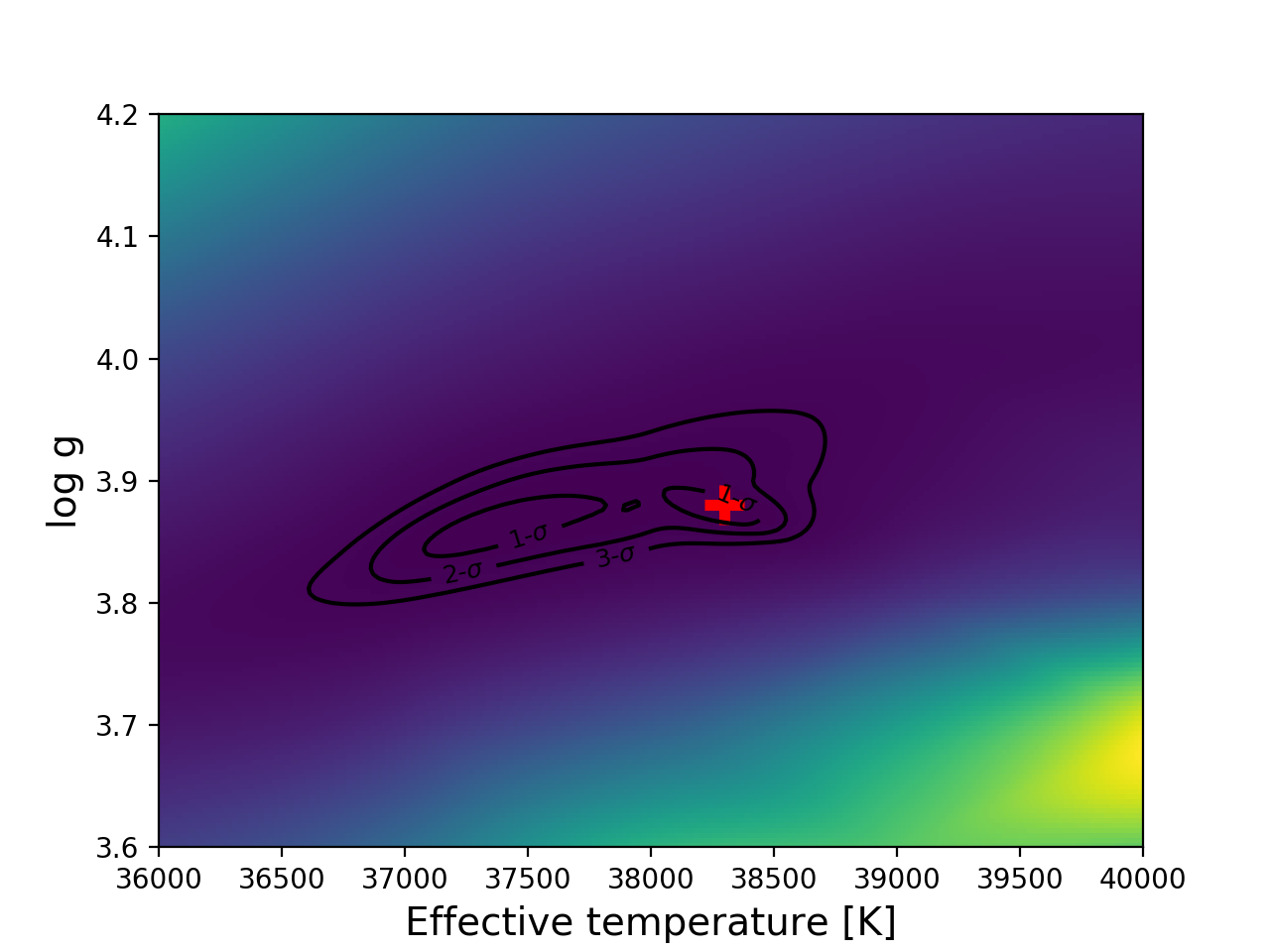}
    \includegraphics[width=6.cm]{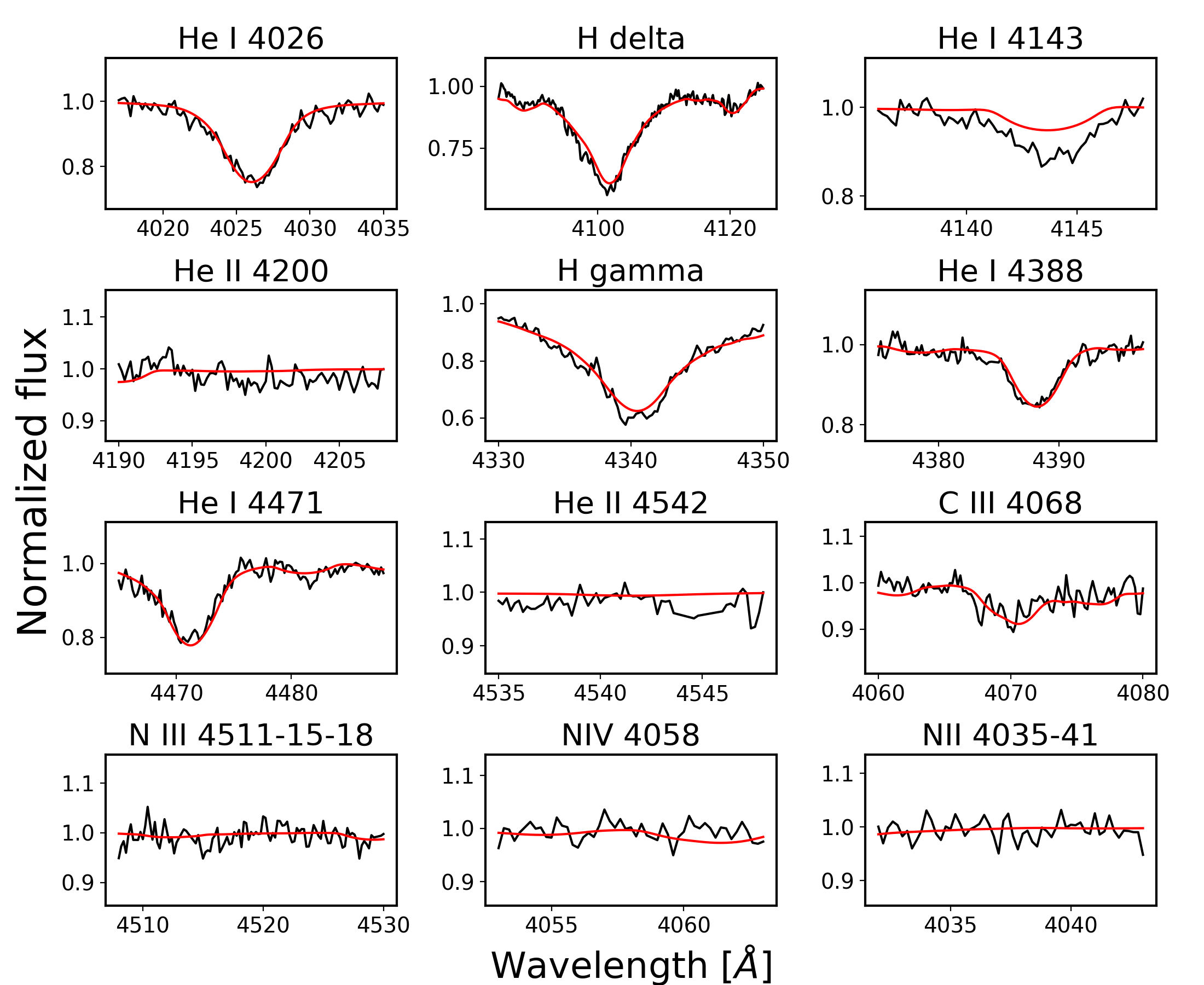}
    \includegraphics[width=7.cm, bb=5 0 453 346,clip]{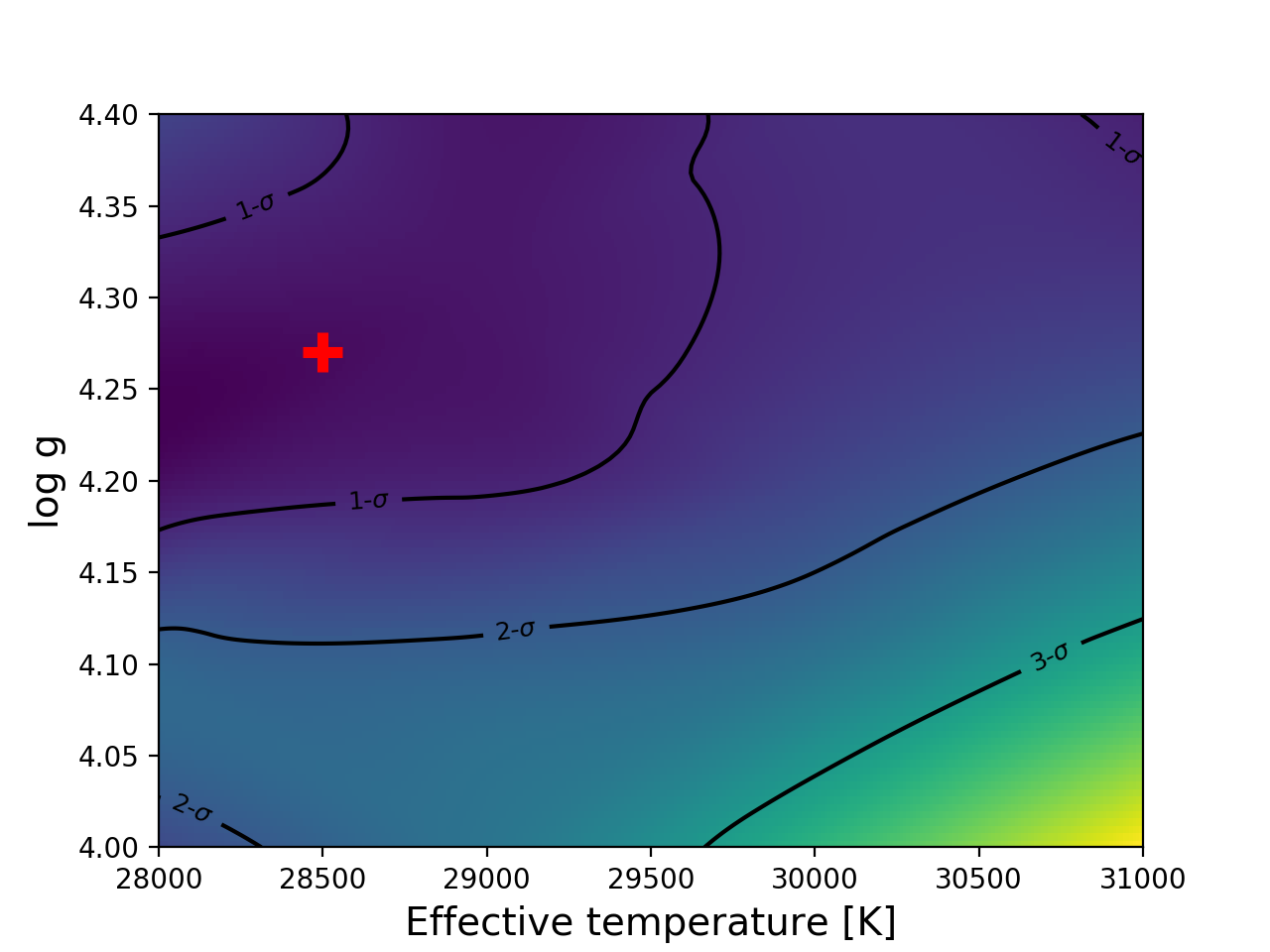}
    \includegraphics[width=7cm]{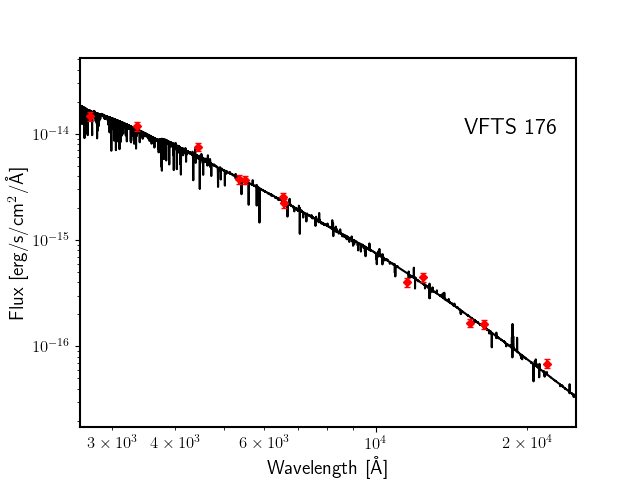}
    \includegraphics[width=6.5cm]{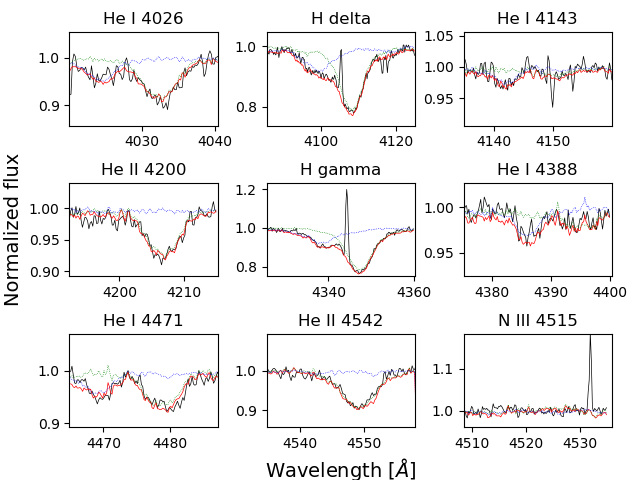}
    \includegraphics[width=7cm, bb=5 0 453 346,clip]{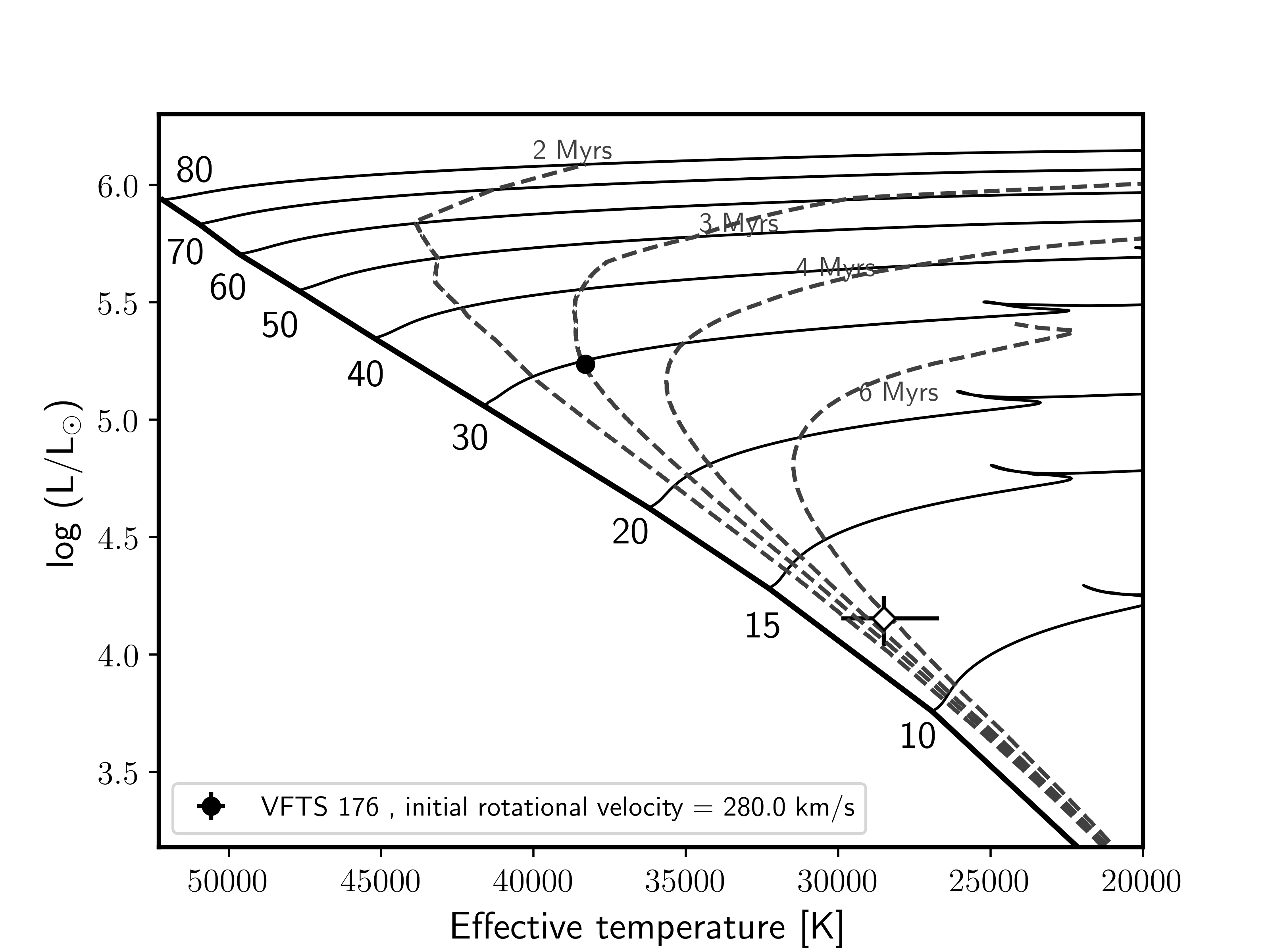}
    \includegraphics[width=7cm, bb=5 0 453 346,clip]{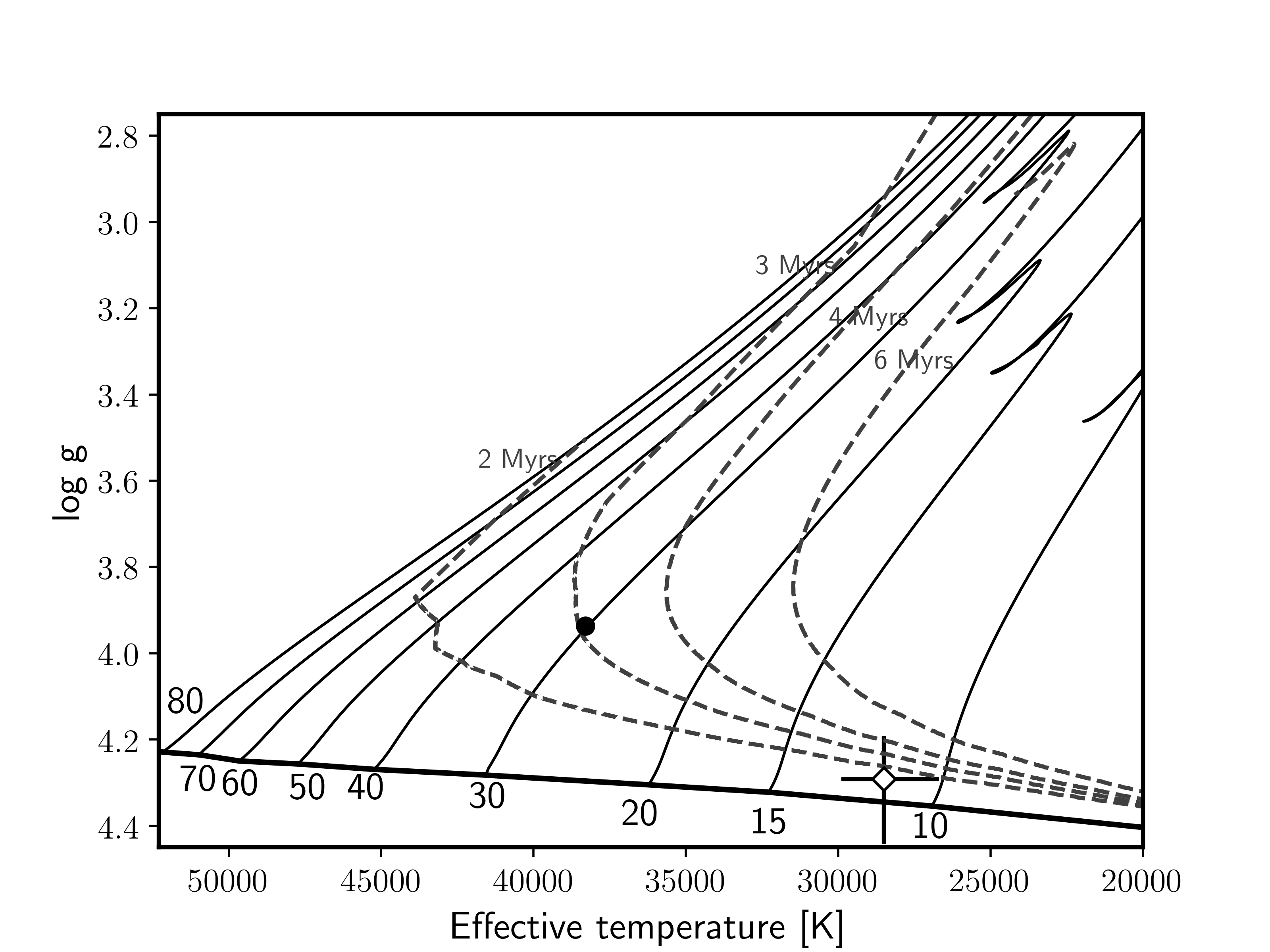}
    \caption{Same as Fig.\,\ref{fig:042} but for VFTS\,176.} \label{fig:176} 
  \end{figure*}
   \clearpage

  \begin{figure*}[t!]
    \centering
    \includegraphics[width=6.cm]{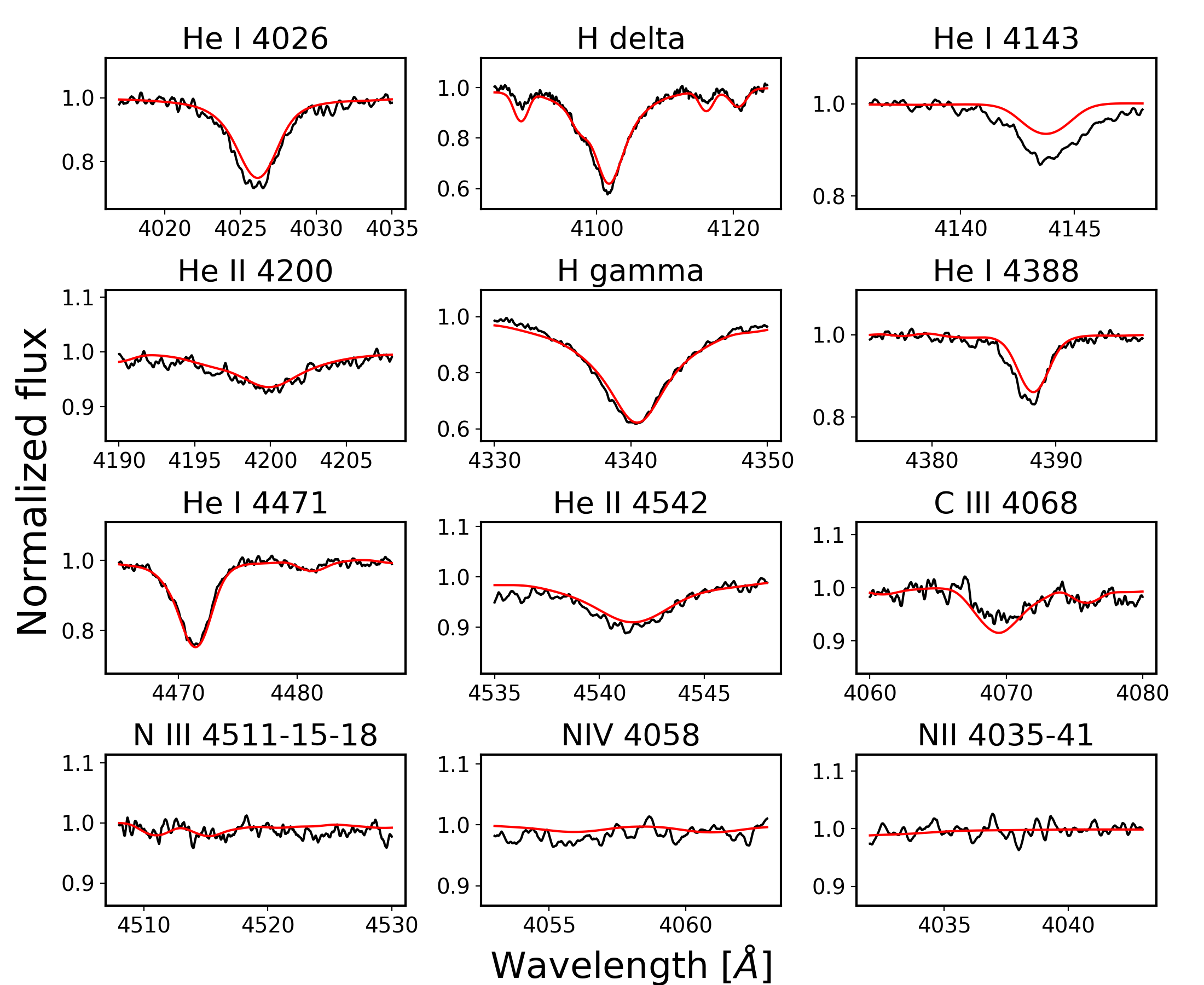}
    \includegraphics[width=7.cm, bb=5 0 453 346,clip]{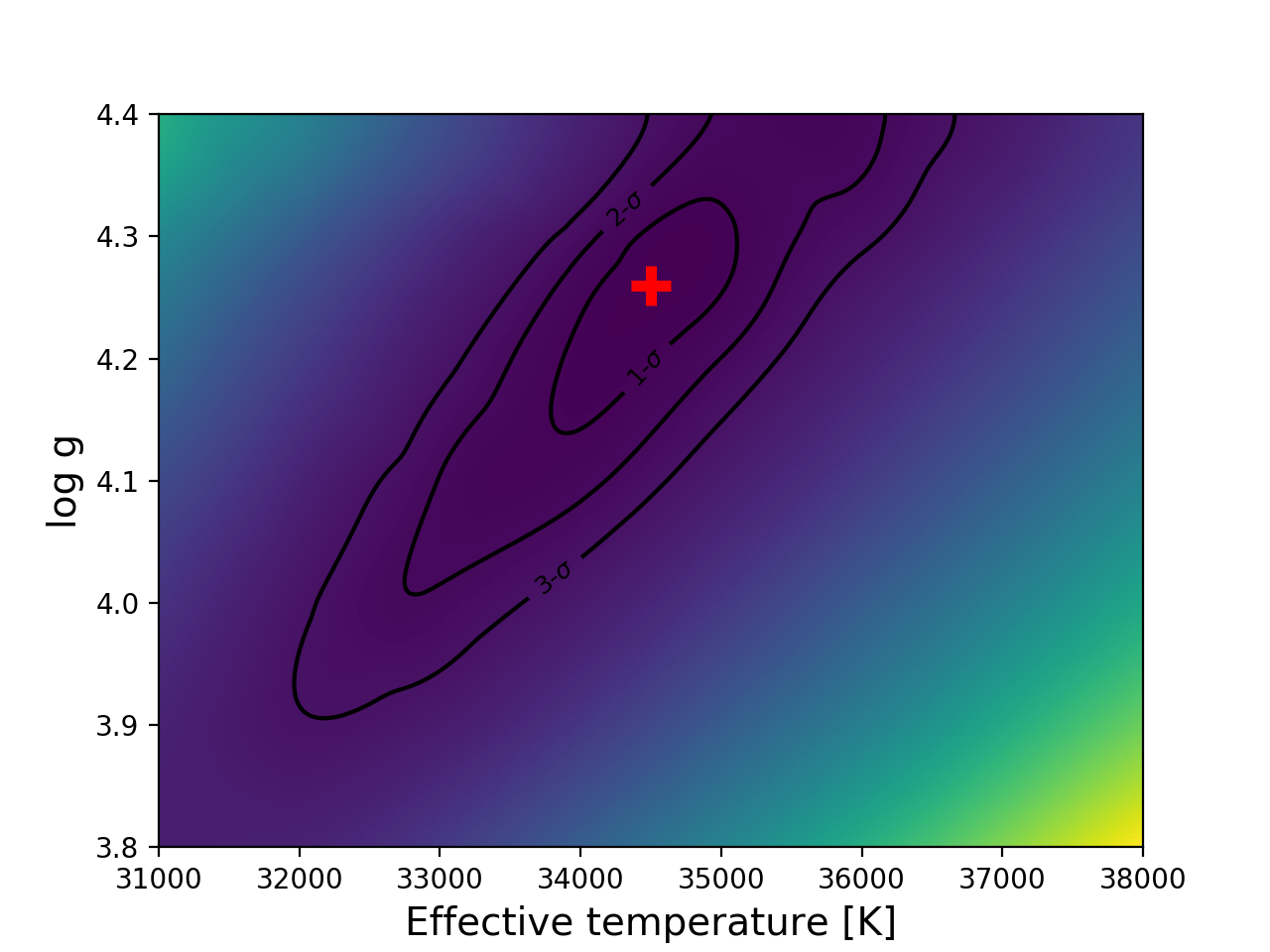}
    \includegraphics[width=6.cm]{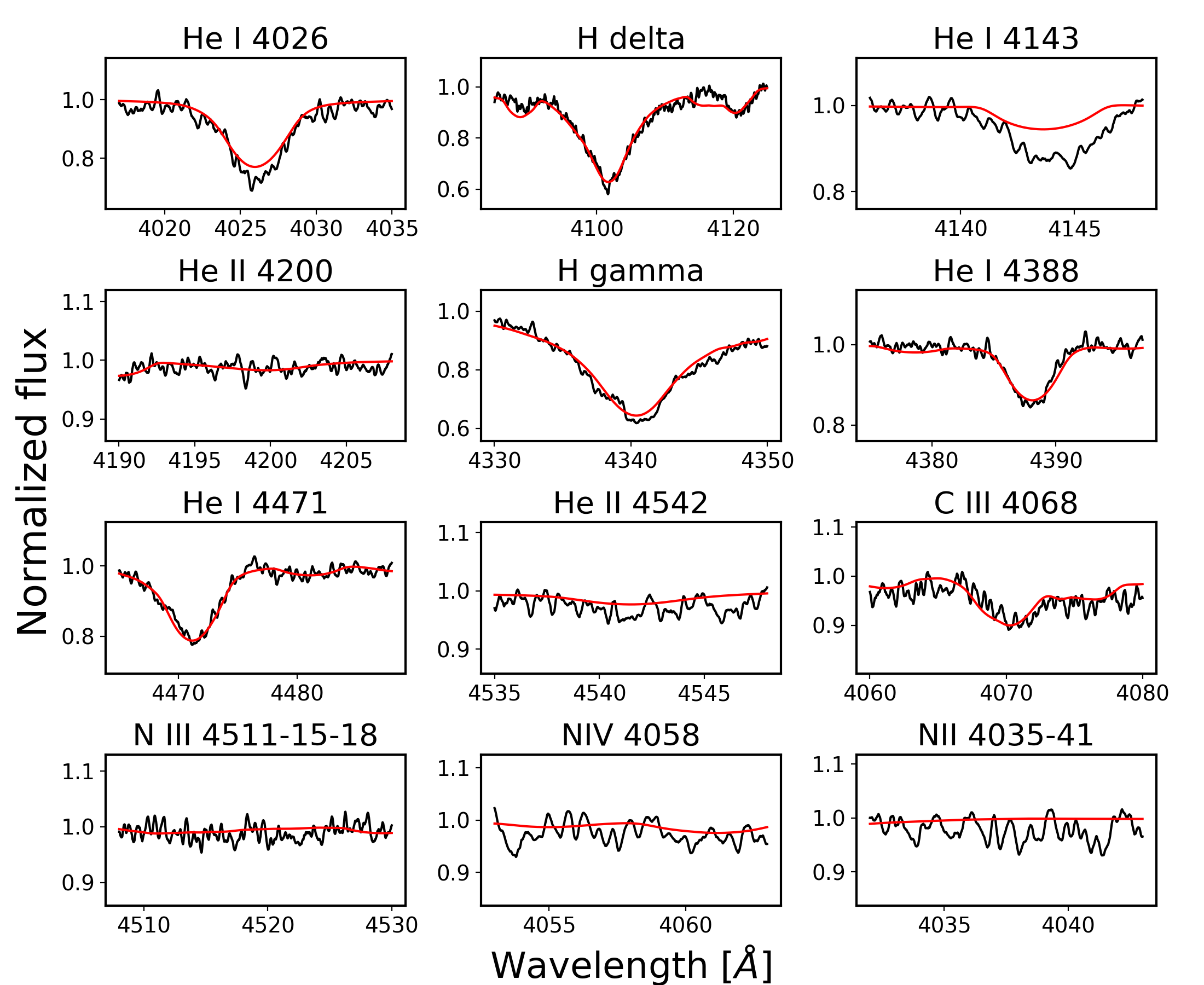}
    \includegraphics[width=7.cm, bb=5 0 453 346,clip]{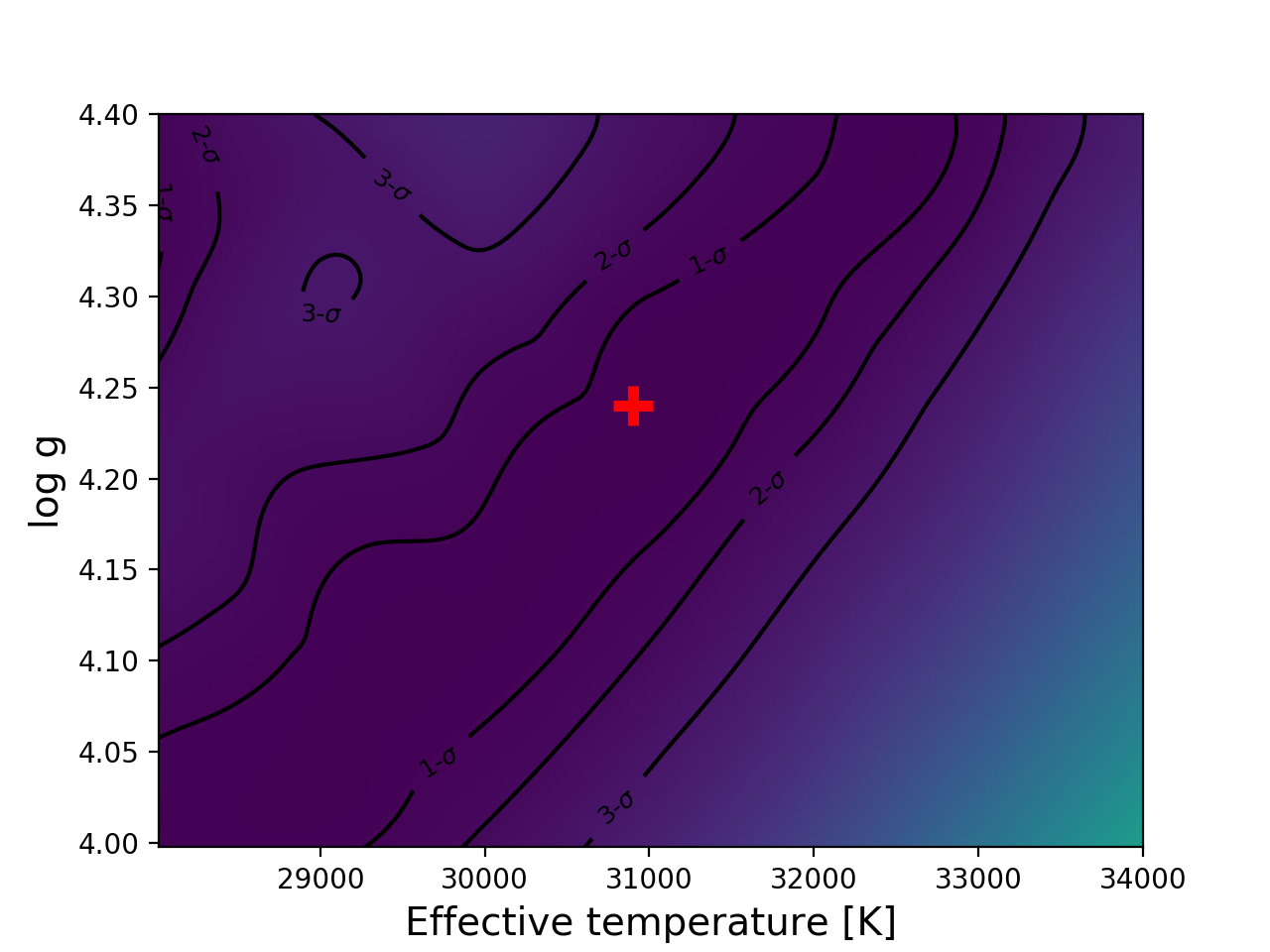}
    \includegraphics[width=7cm]{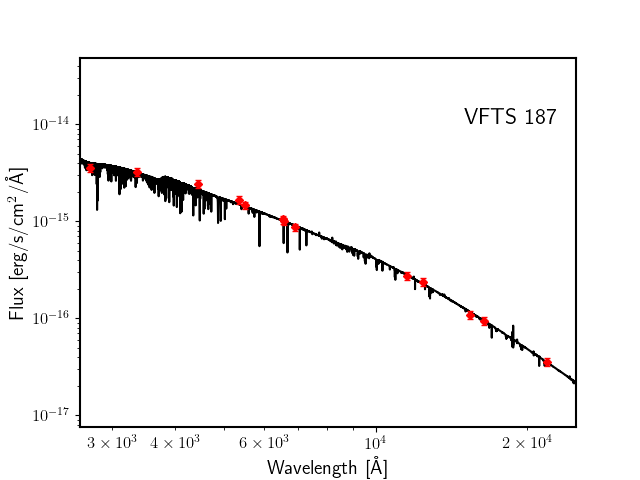}
    \includegraphics[width=6.5cm]{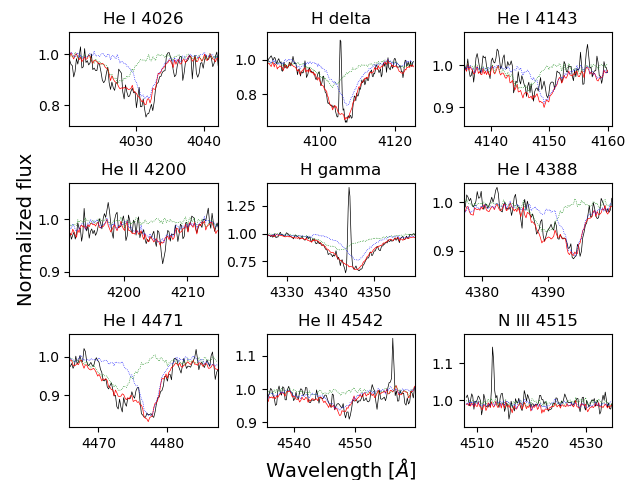}
    \includegraphics[width=7cm, bb=5 0 453 346,clip]{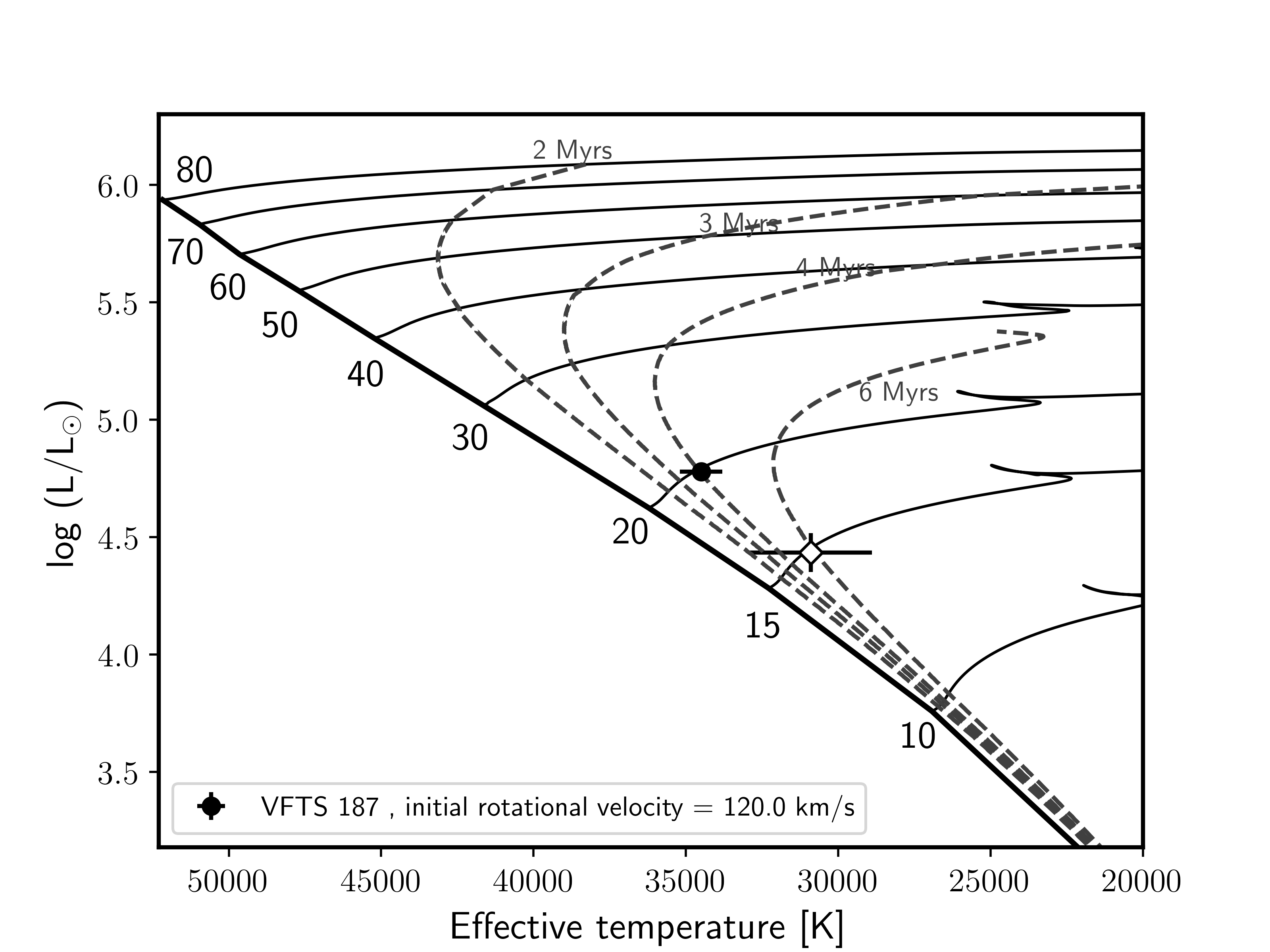}
    \includegraphics[width=7cm, bb=5 0 453 346,clip]{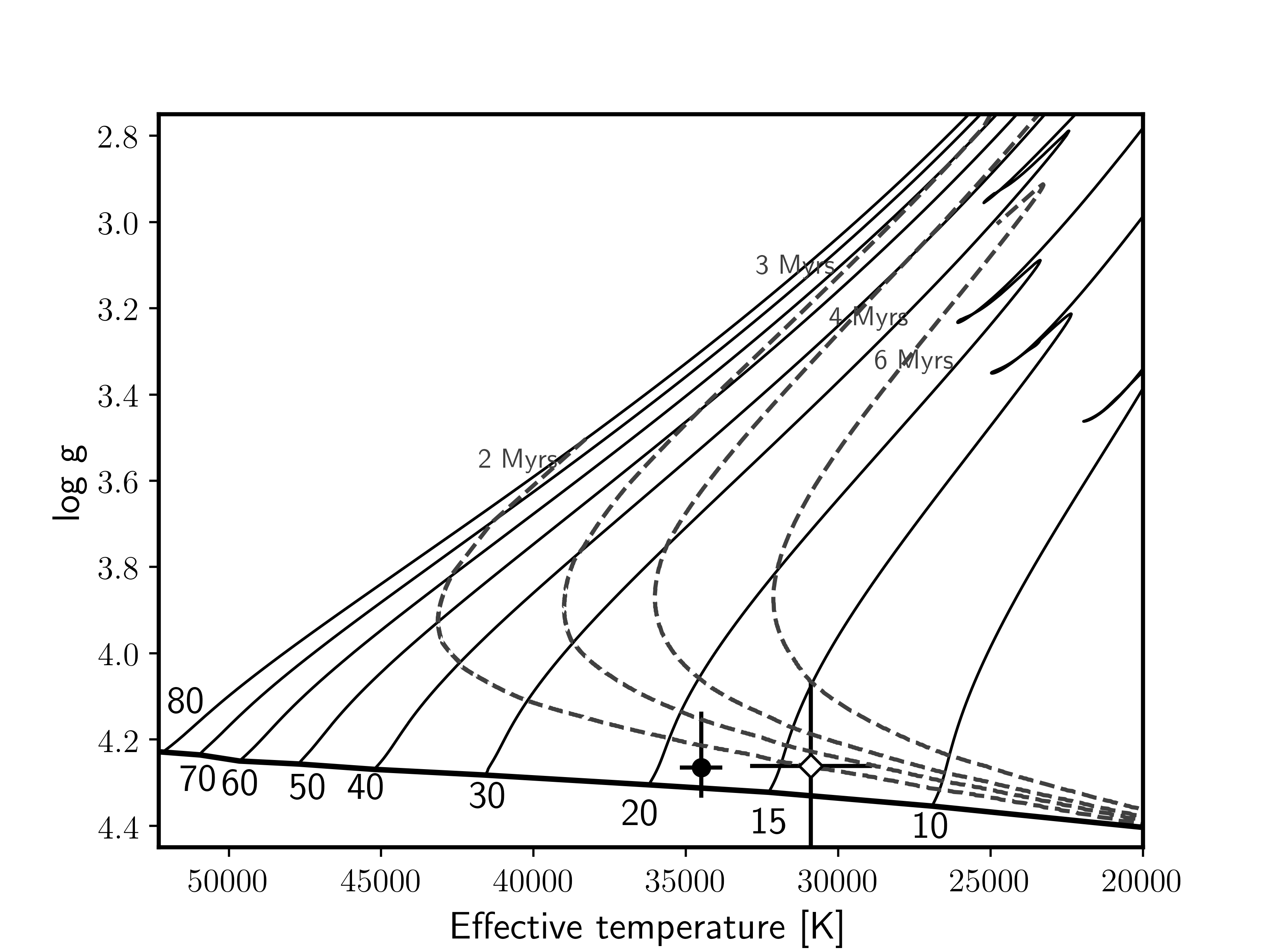}
    \caption{Same as Fig.\,\ref{fig:042} but for VFTS\,187.} \label{fig:187} 
  \end{figure*}
 \clearpage

 \begin{figure*}[t!]
    \centering
    \includegraphics[width=6.cm]{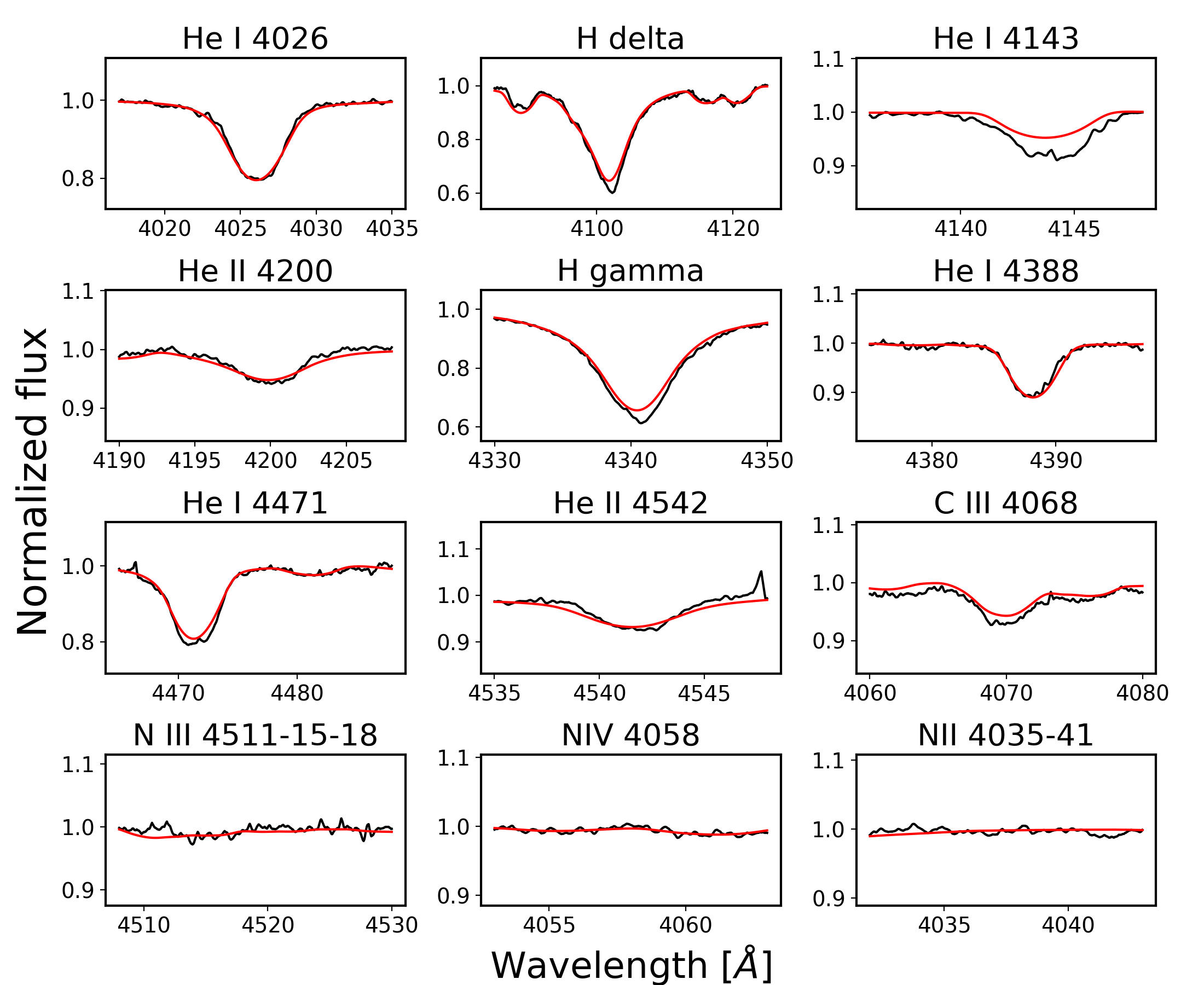}
    \includegraphics[width=7.cm, bb=5 0 453 346,clip]{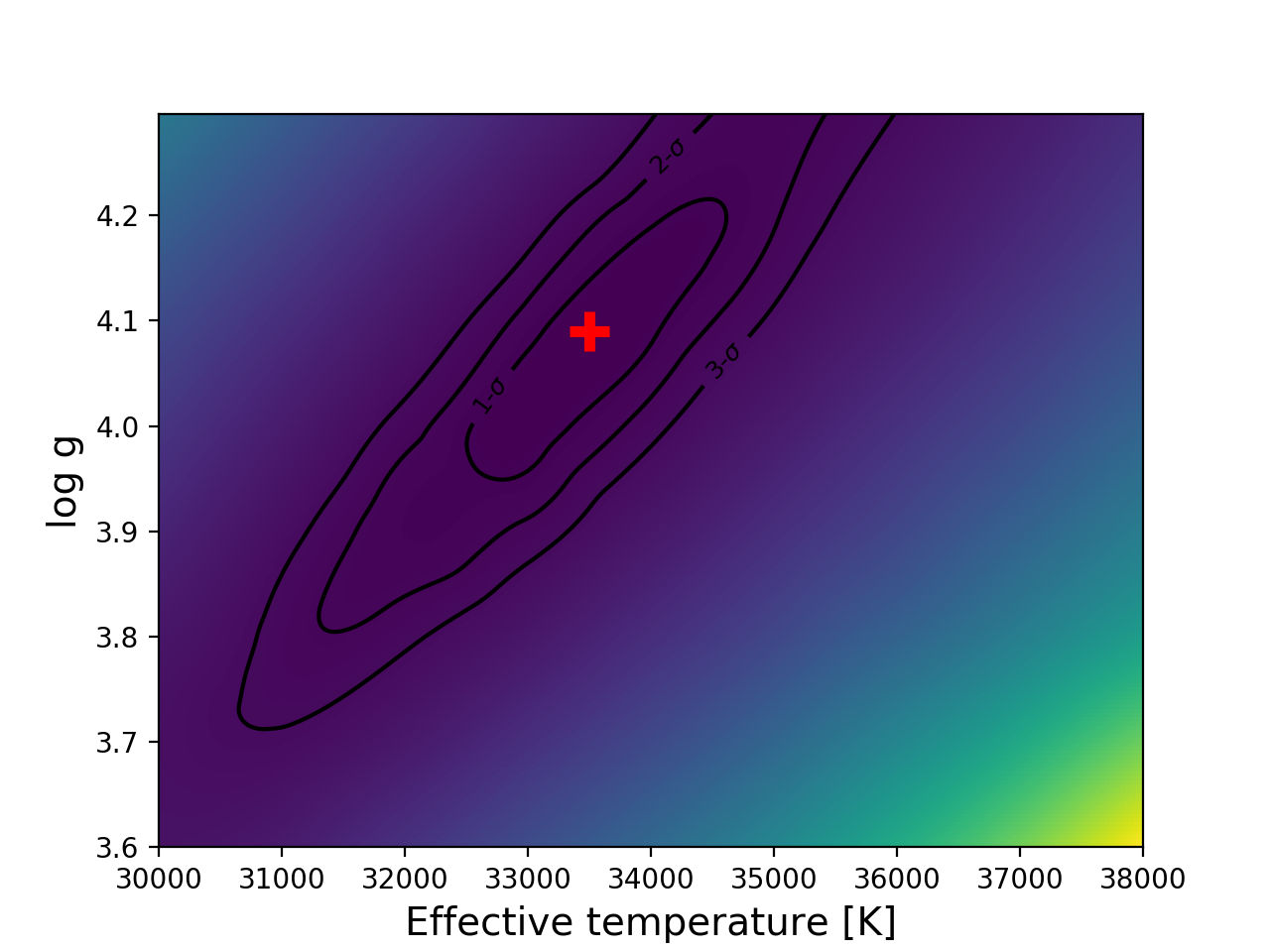}
    \includegraphics[width=6.cm]{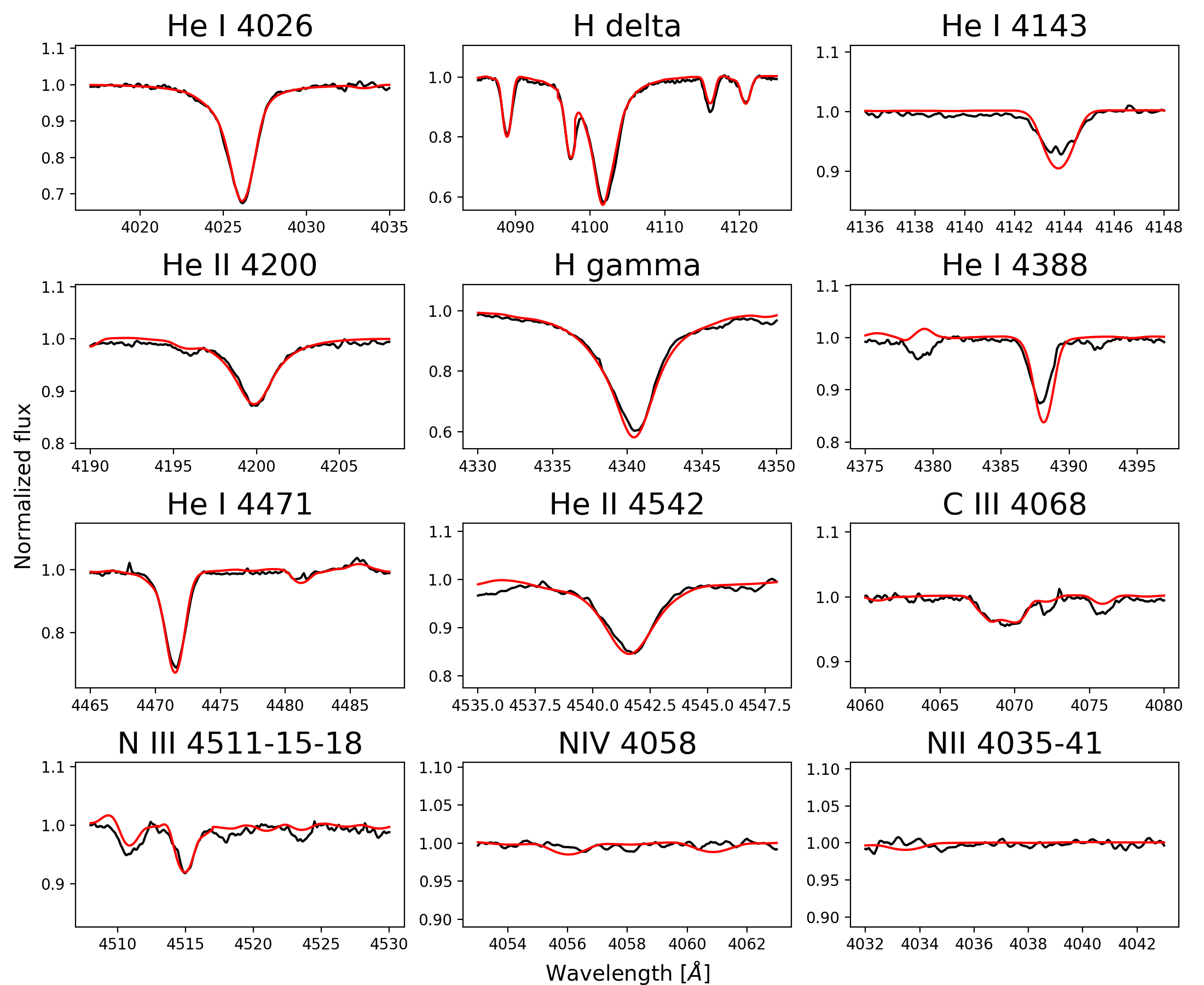}
    \includegraphics[width=7.cm, bb=5 0 453 346,clip]{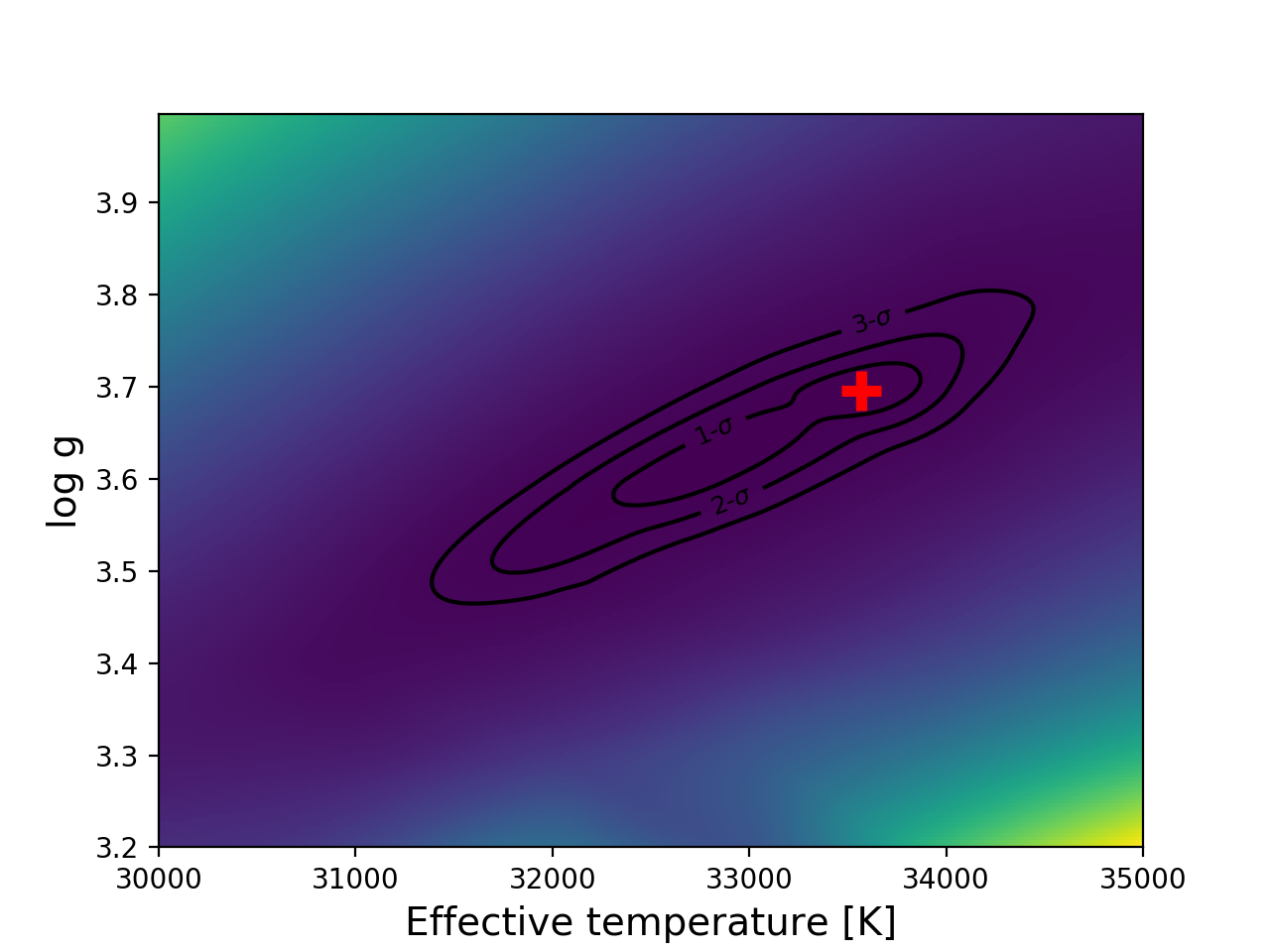}
    \includegraphics[width=7cm]{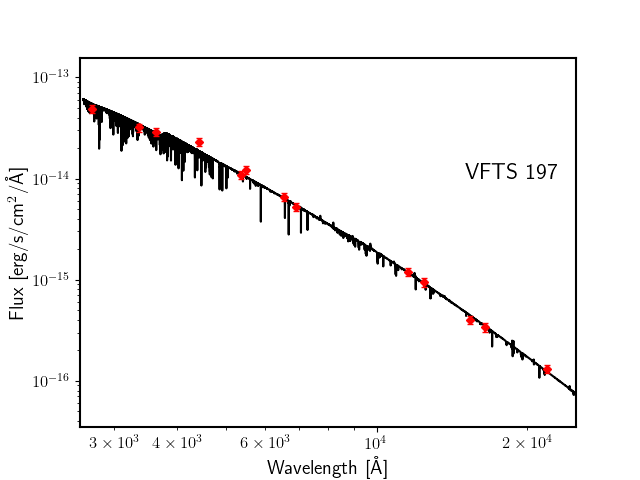}
    \includegraphics[width=6.5cm]{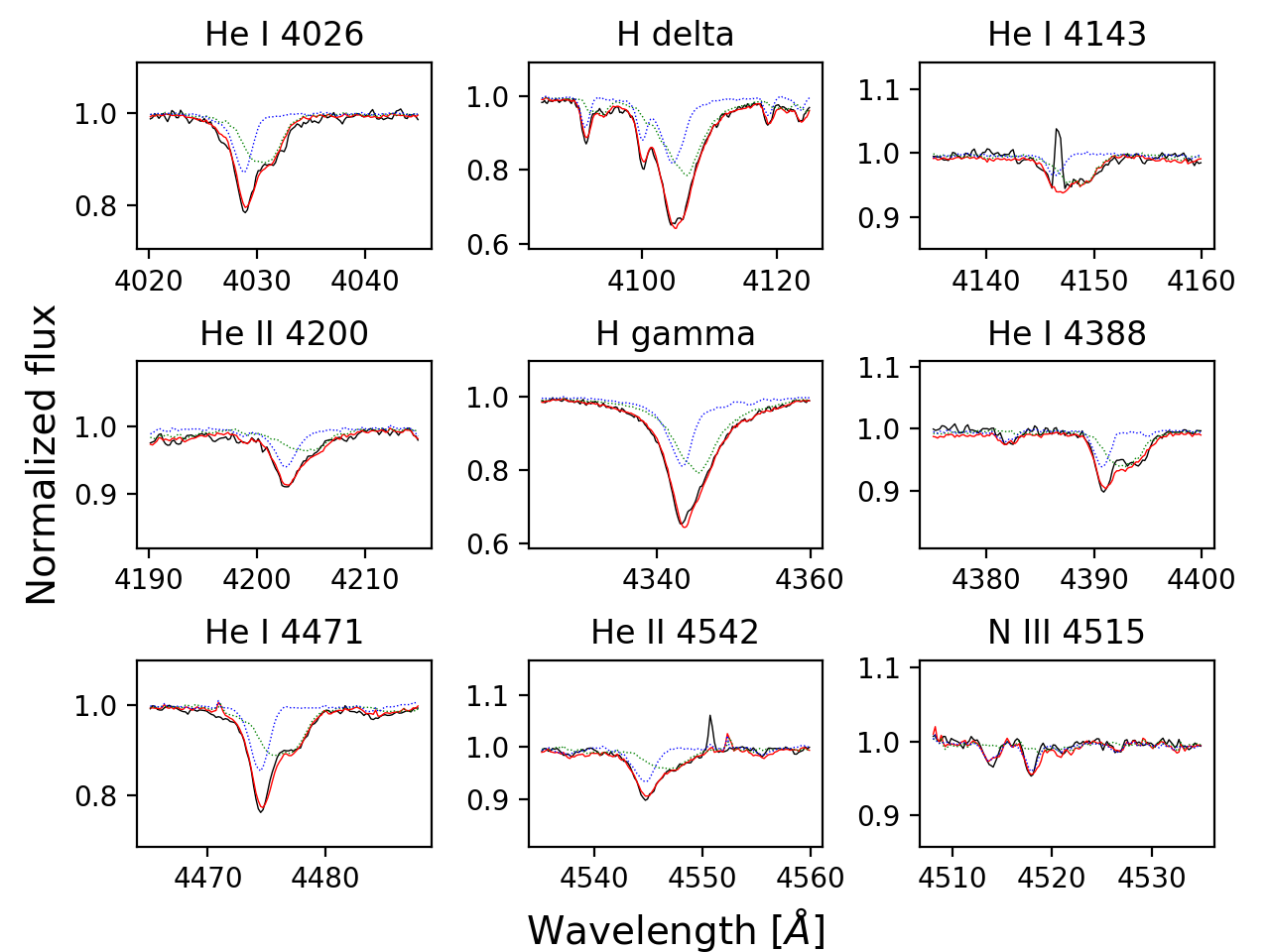}
    \includegraphics[width=7cm, bb=5 0 453 346,clip]{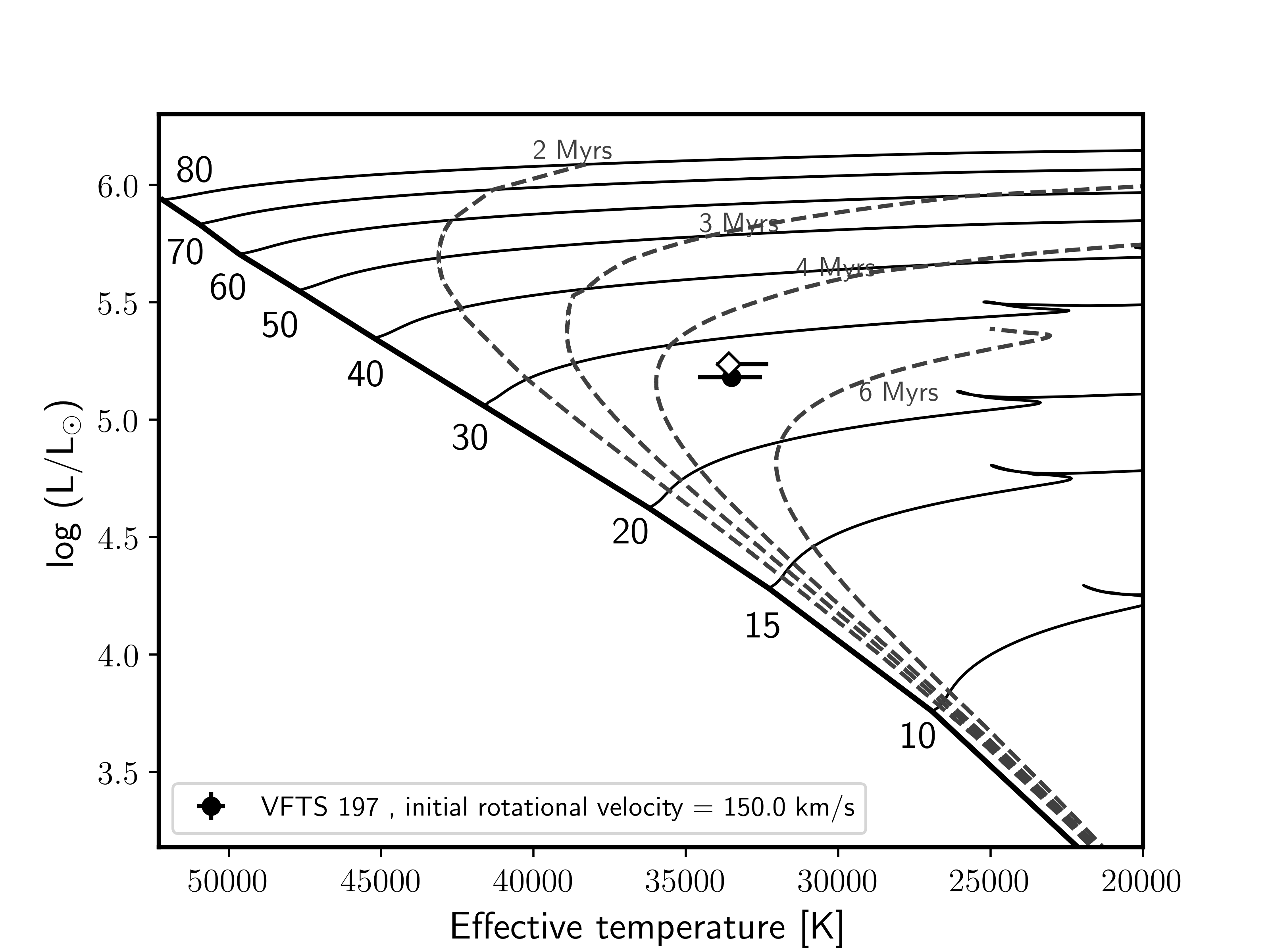}
    \includegraphics[width=7cm, bb=5 0 453 346,clip]{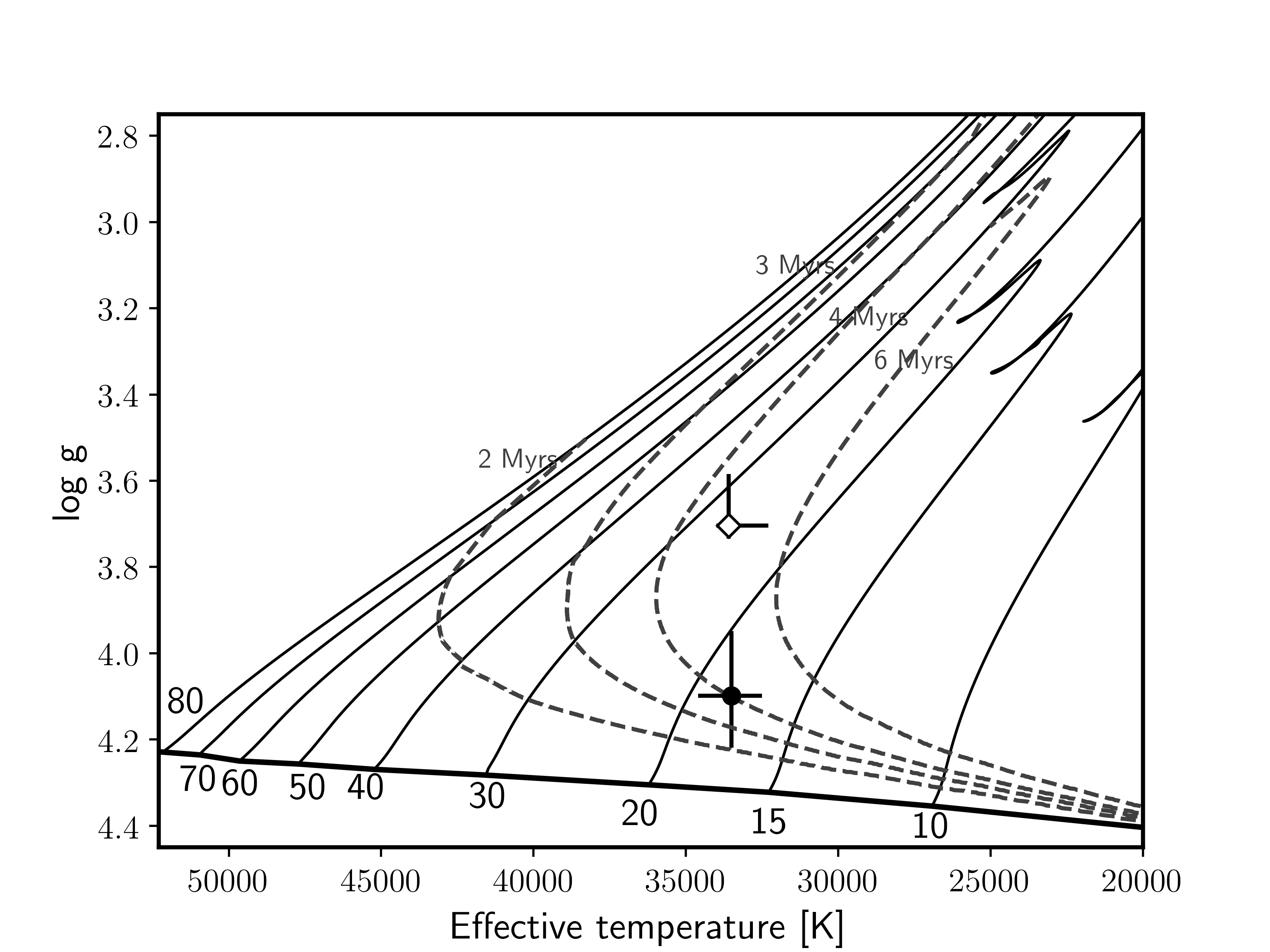}
    \caption{Same as Fig.\,\ref{fig:042} but for VFTS\,197.} \label{fig:197} 
  \end{figure*} 
  \clearpage

 \begin{figure*}[t!]
    \centering
    \includegraphics[width=6.cm]{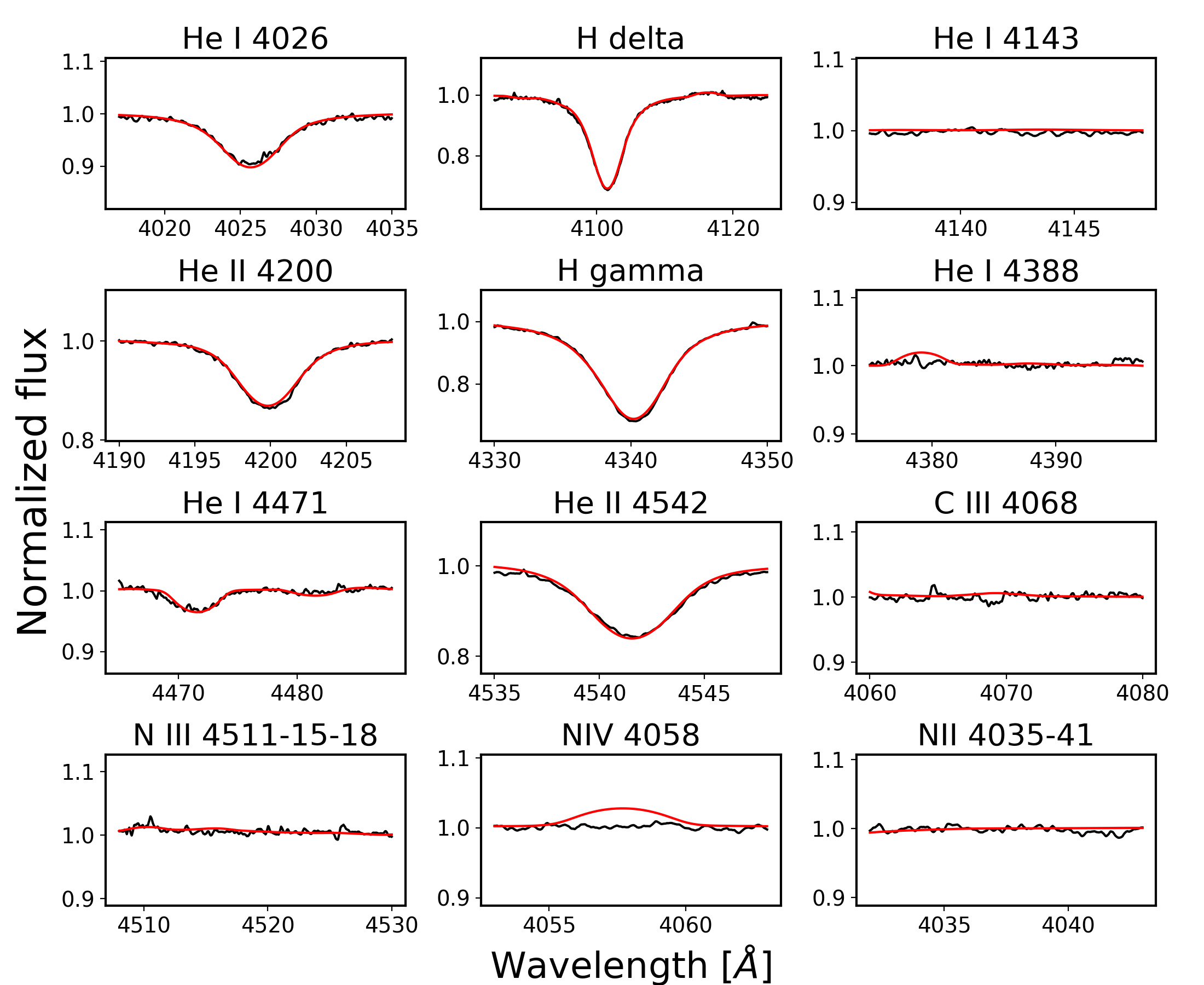}
    \includegraphics[width=7.cm, bb=5 0 453 346,clip]{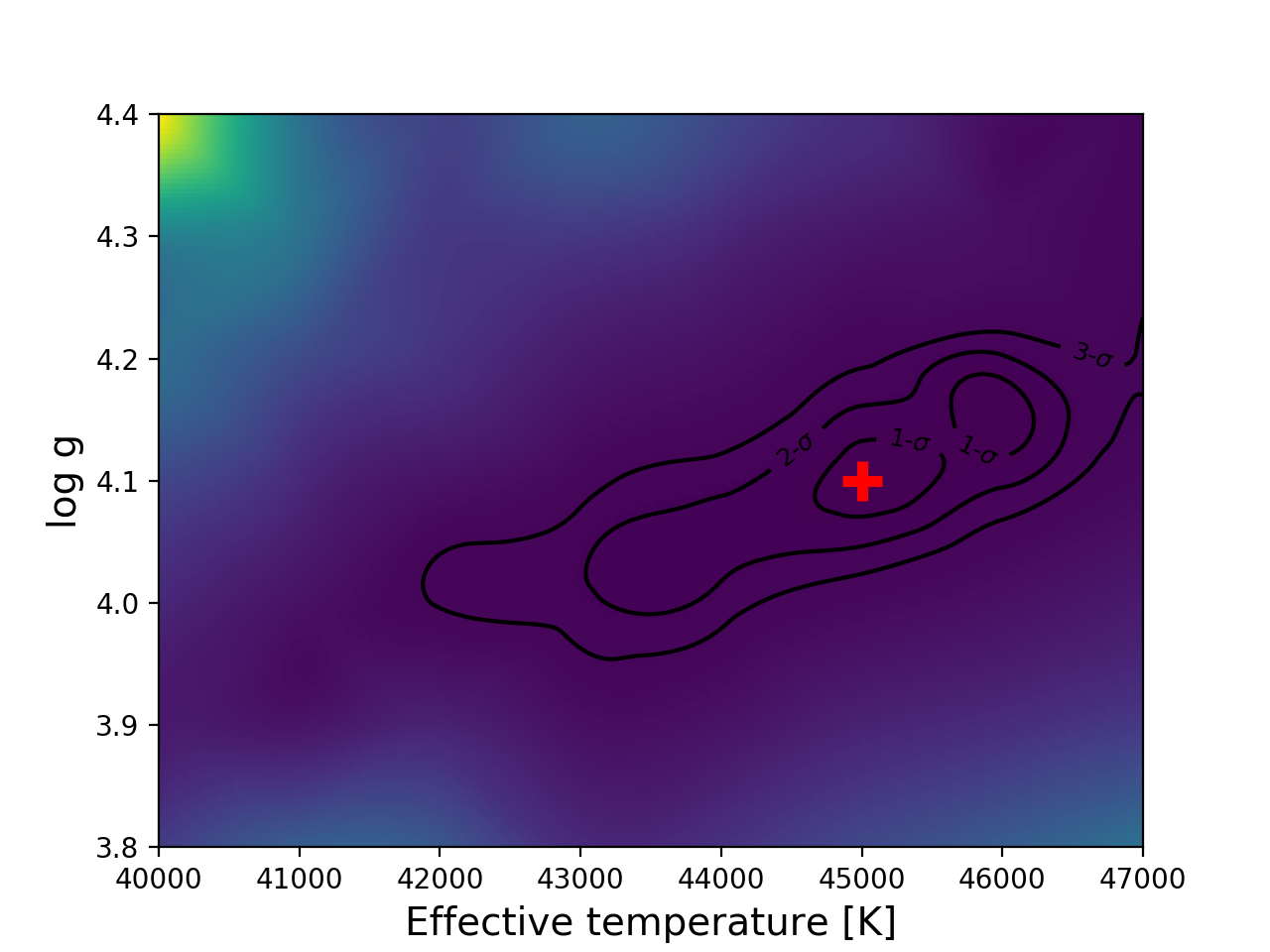}
    \includegraphics[width=6.cm]{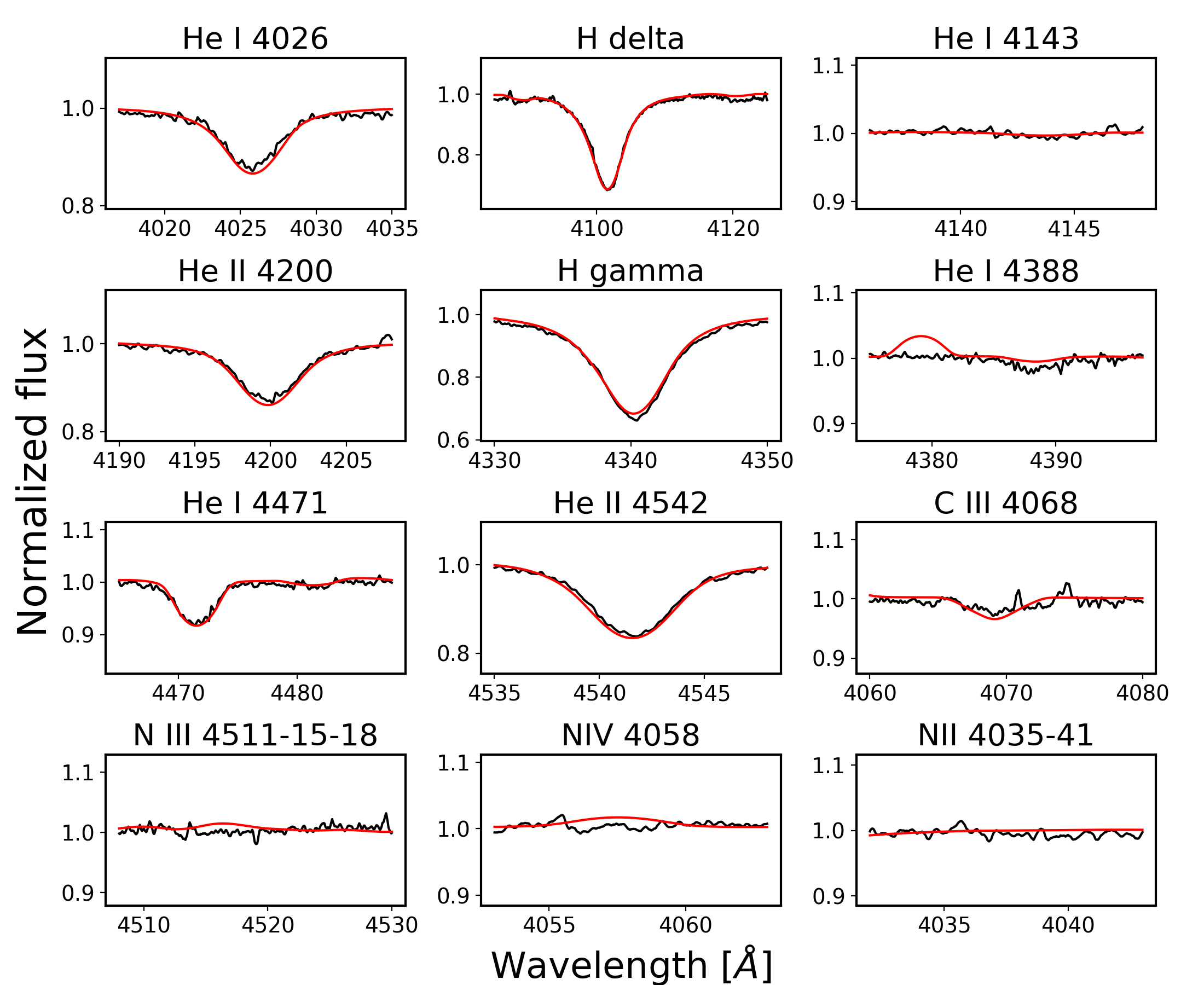}
    \includegraphics[width=7.cm, bb=5 0 453 346,clip]{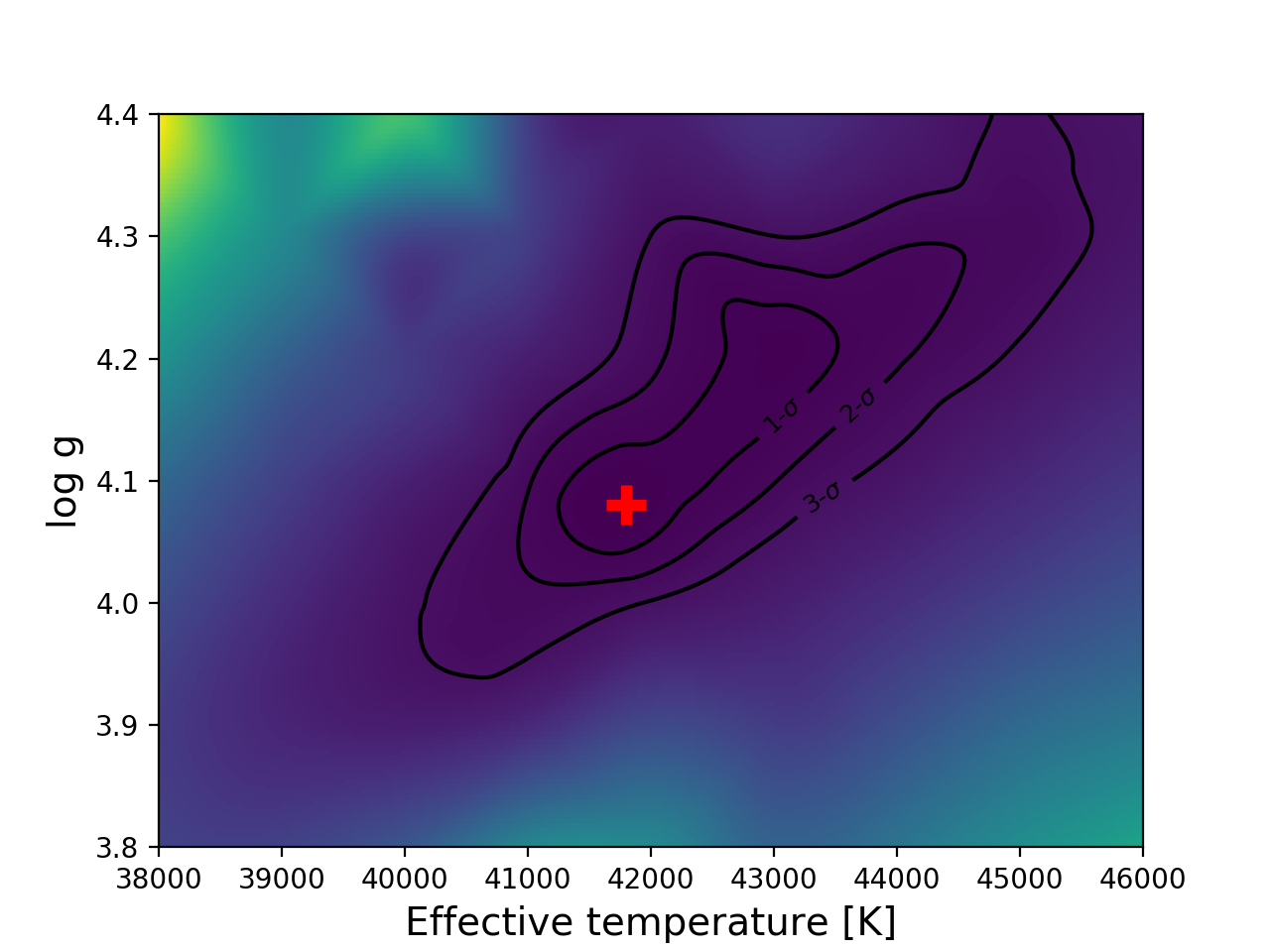}
    \includegraphics[width=7cm]{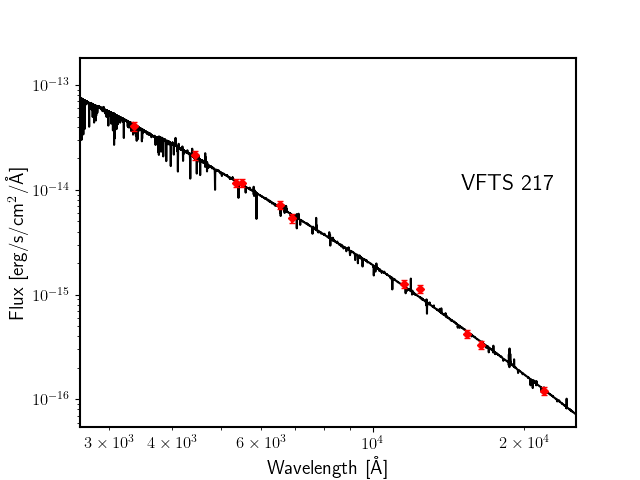}
    \includegraphics[width=6.5cm]{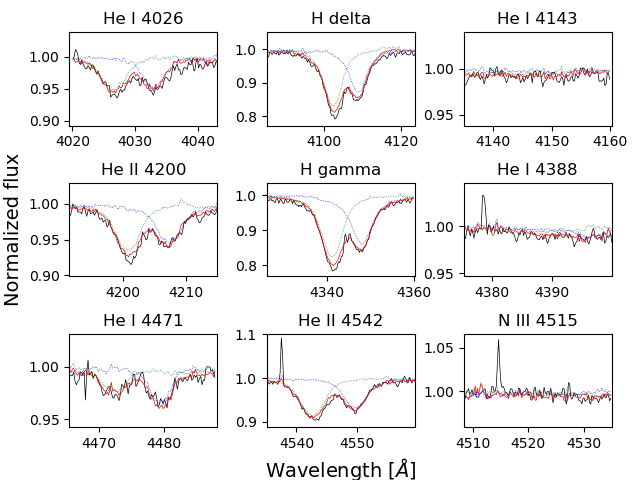}
    \includegraphics[width=7cm, bb=5 0 453 346,clip]{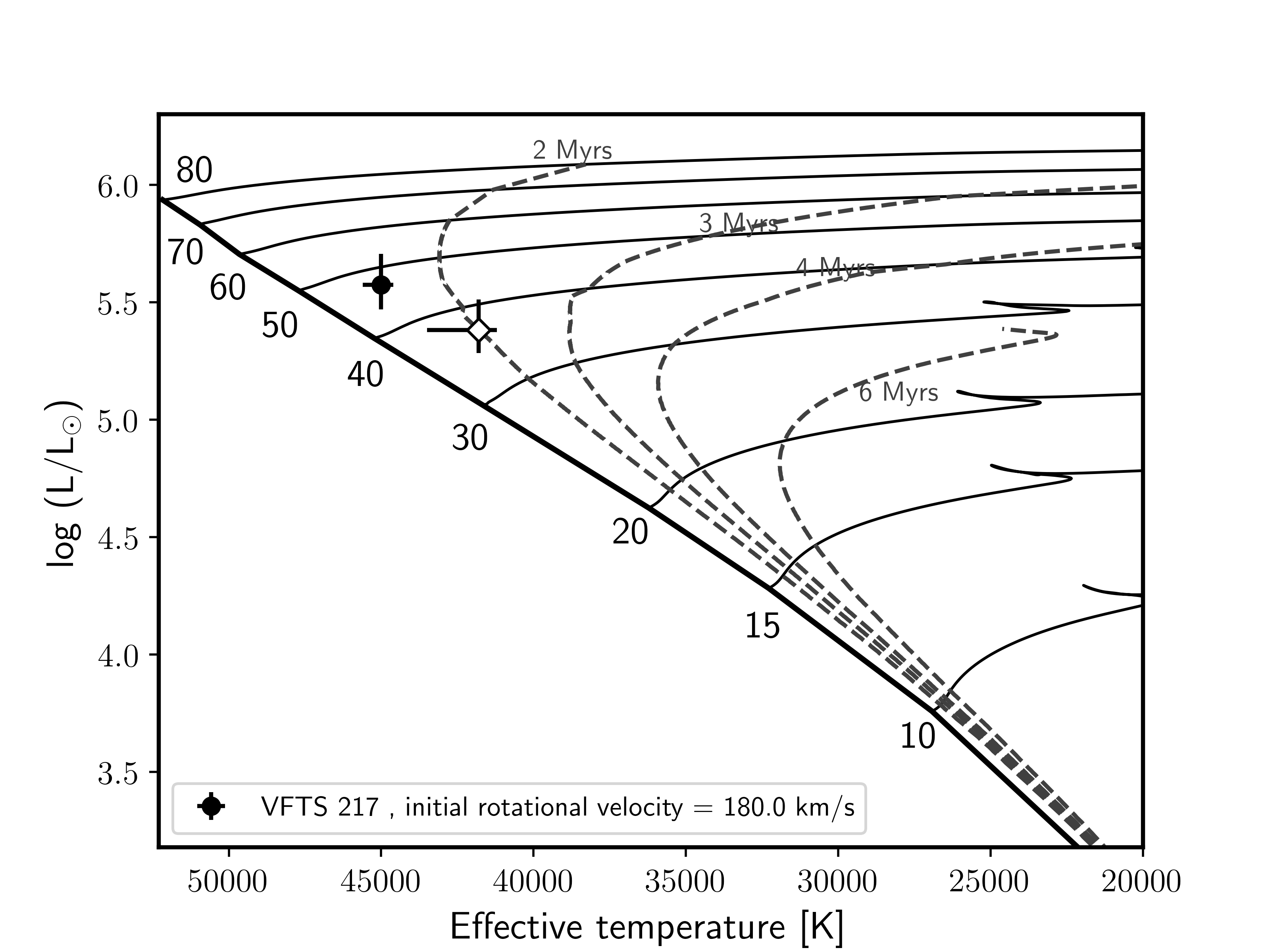}
    \includegraphics[width=7cm, bb=5 0 453 346,clip]{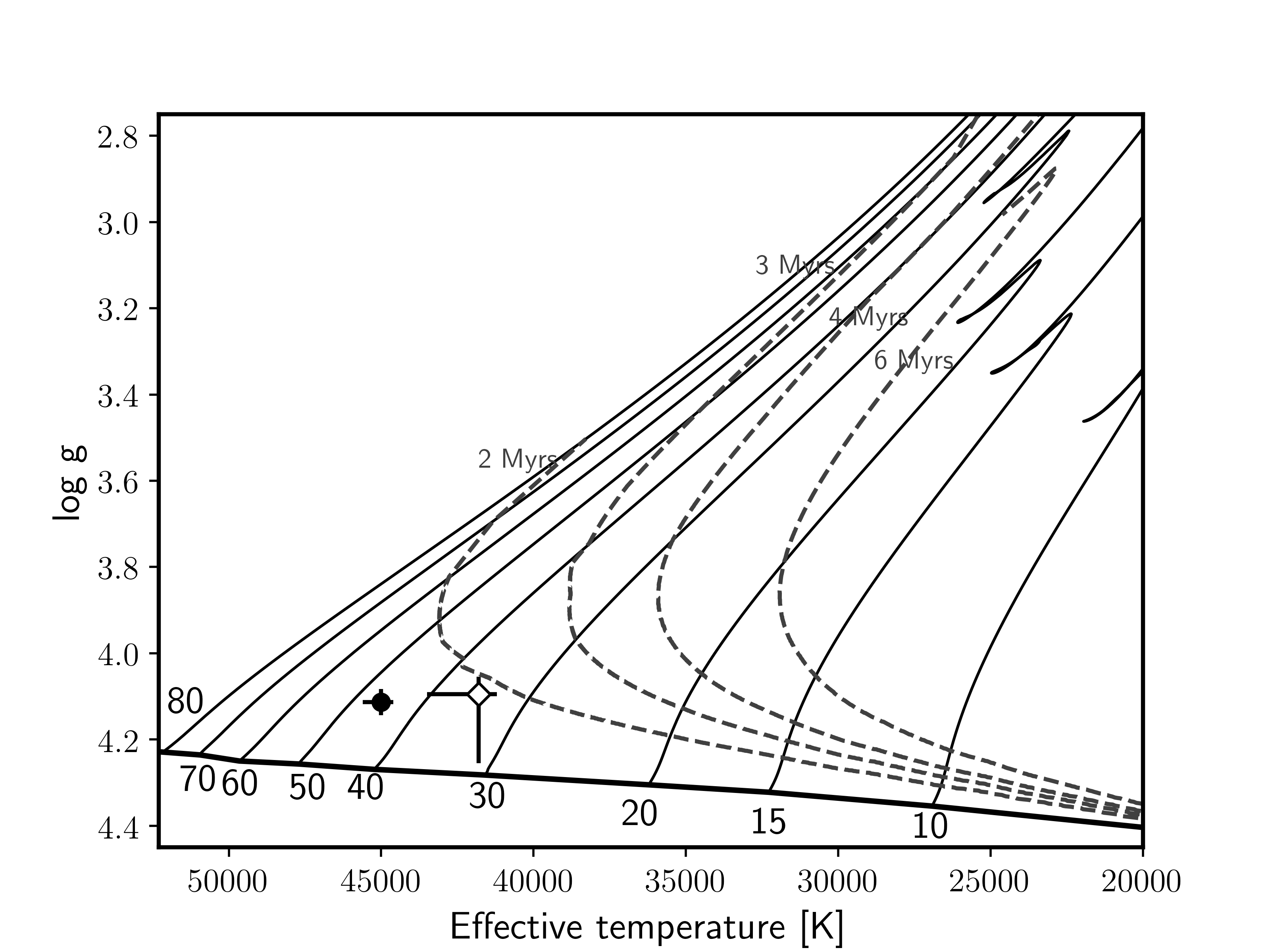}
    \caption{Same as Fig.\,\ref{fig:042} but for VFTS\,217.} \label{fig:217} 
  \end{figure*} 
   \clearpage
           
 \begin{figure*}[t!]
    \centering
    \includegraphics[width=6.cm]{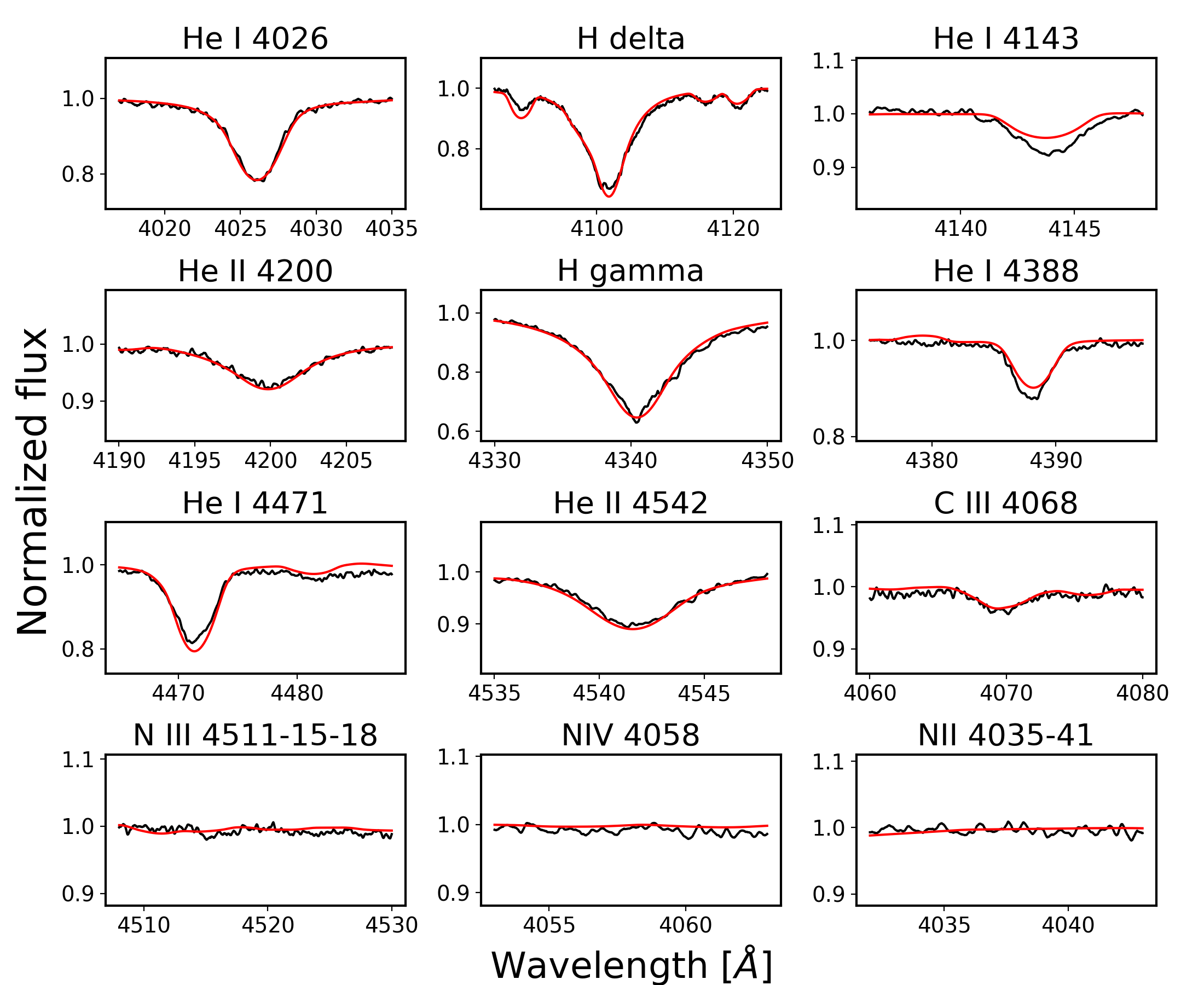}
    \includegraphics[width=7.cm, bb=5 0 453 346,clip]{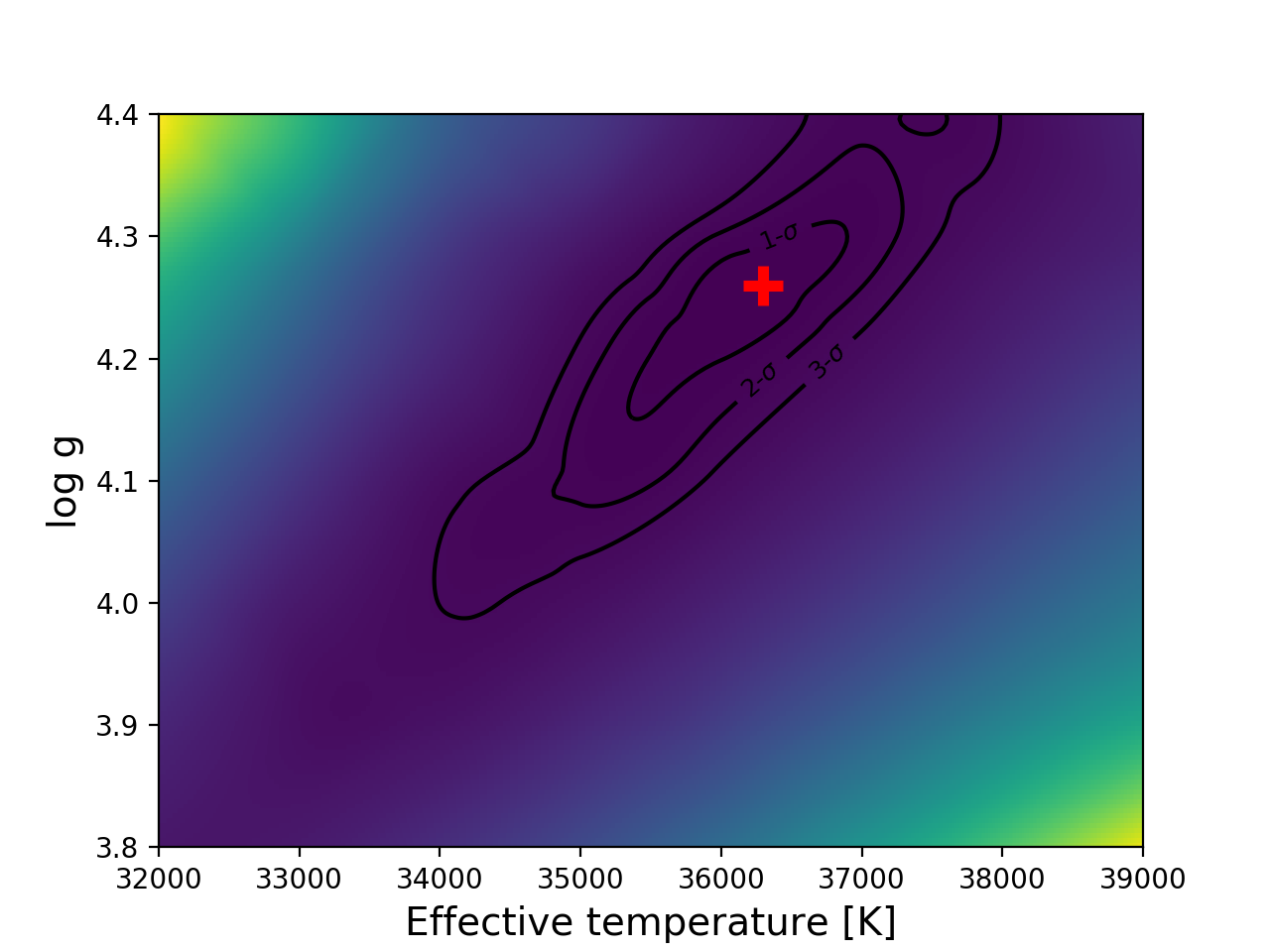}
    \includegraphics[width=6.cm]{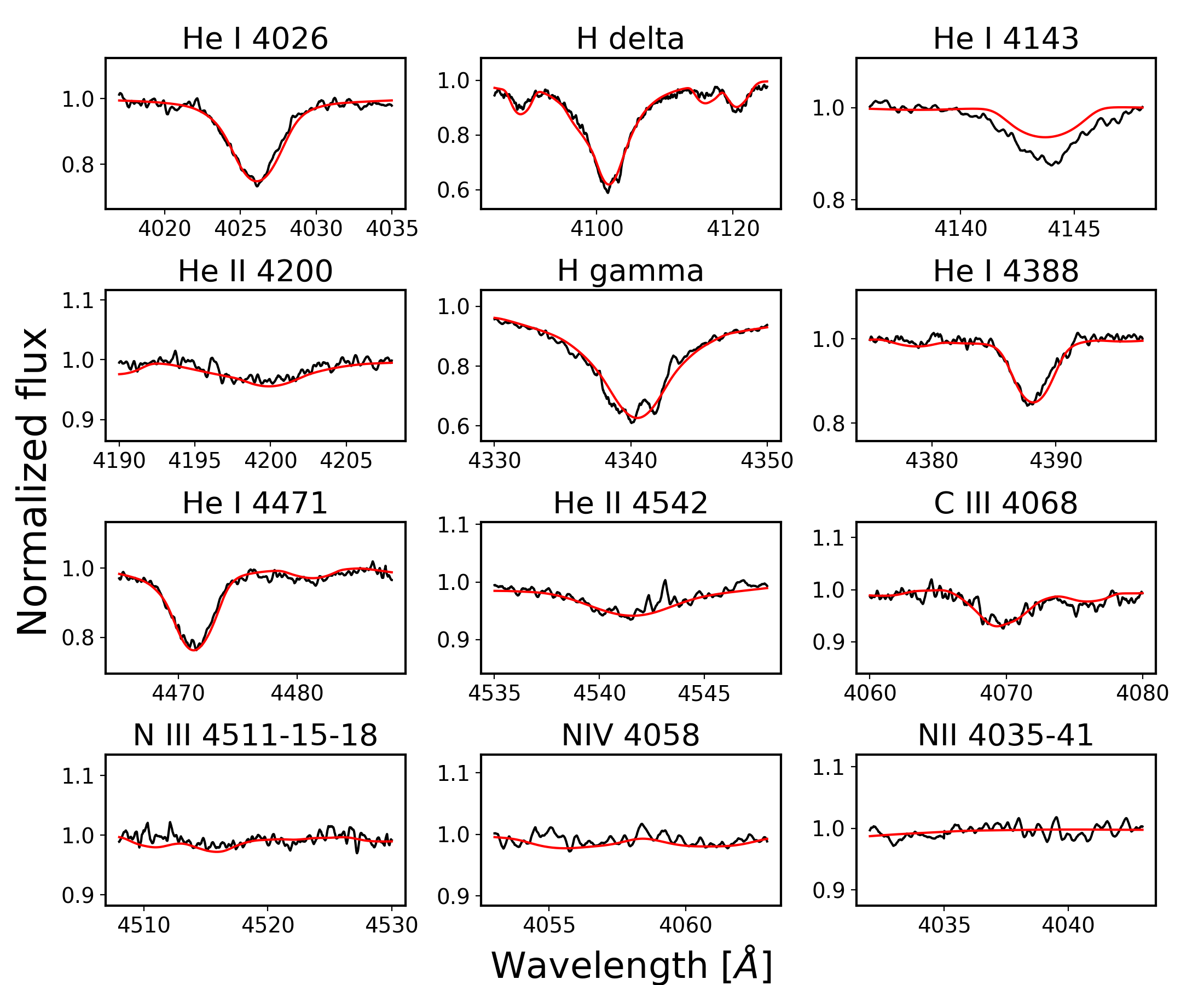}
    \includegraphics[width=7.cm, bb=5 0 453 346,clip]{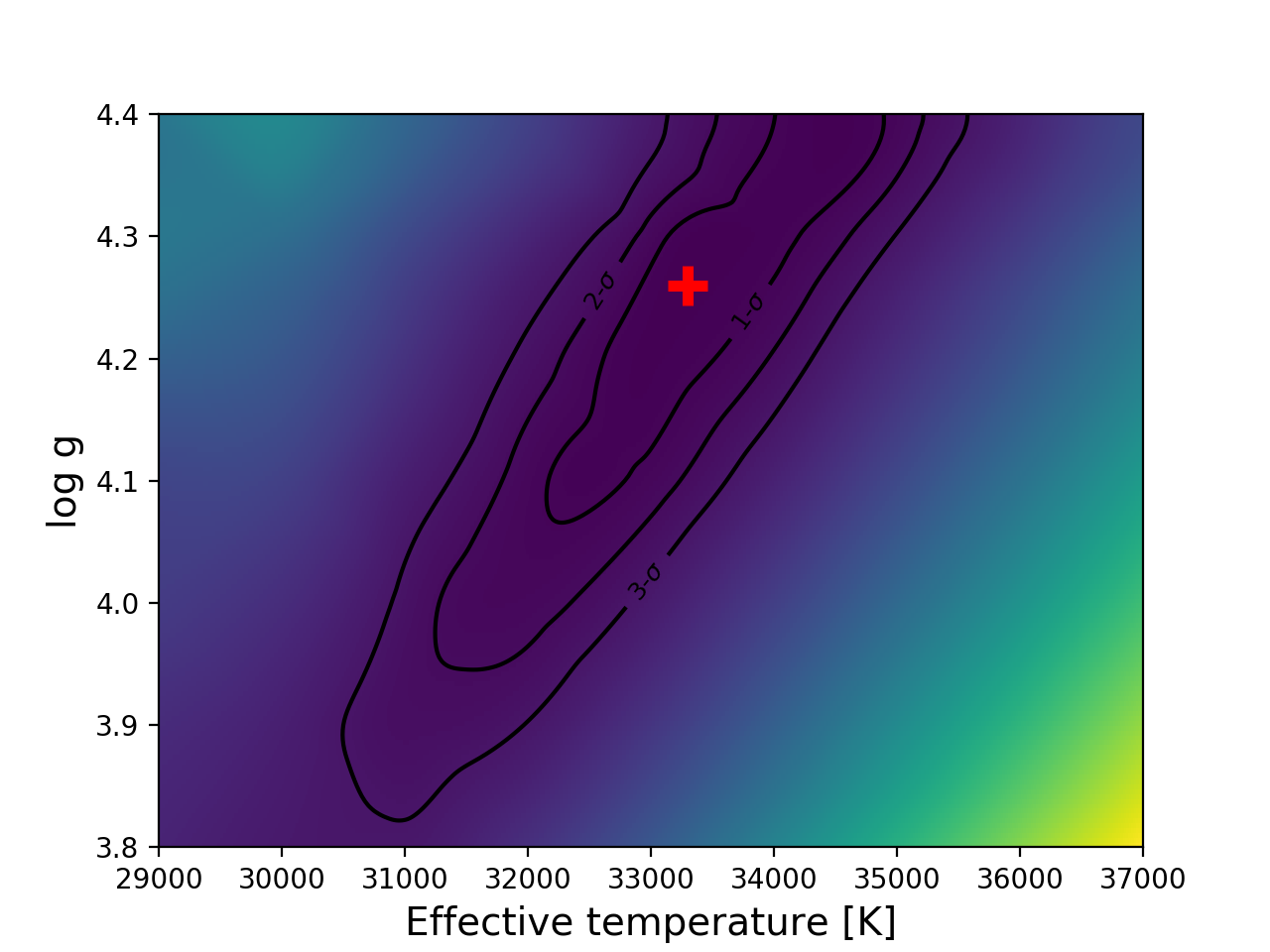}
    \includegraphics[width=7cm]{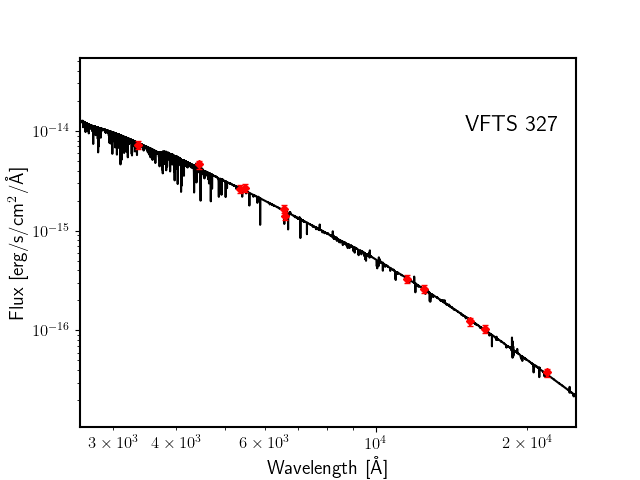}
    \includegraphics[width=6.5cm]{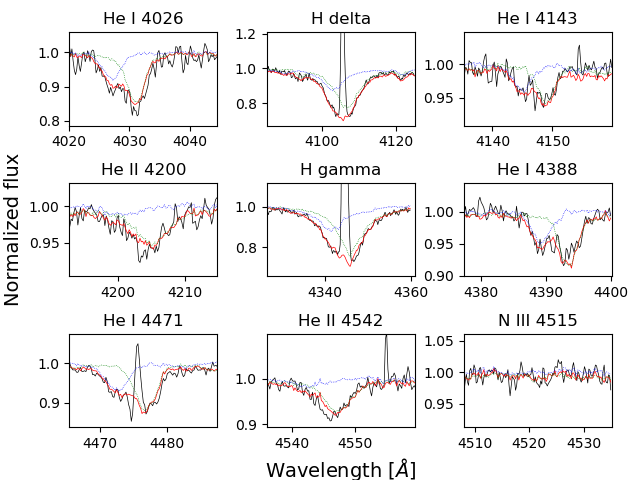}
    \includegraphics[width=7cm, bb=5 0 453 346,clip]{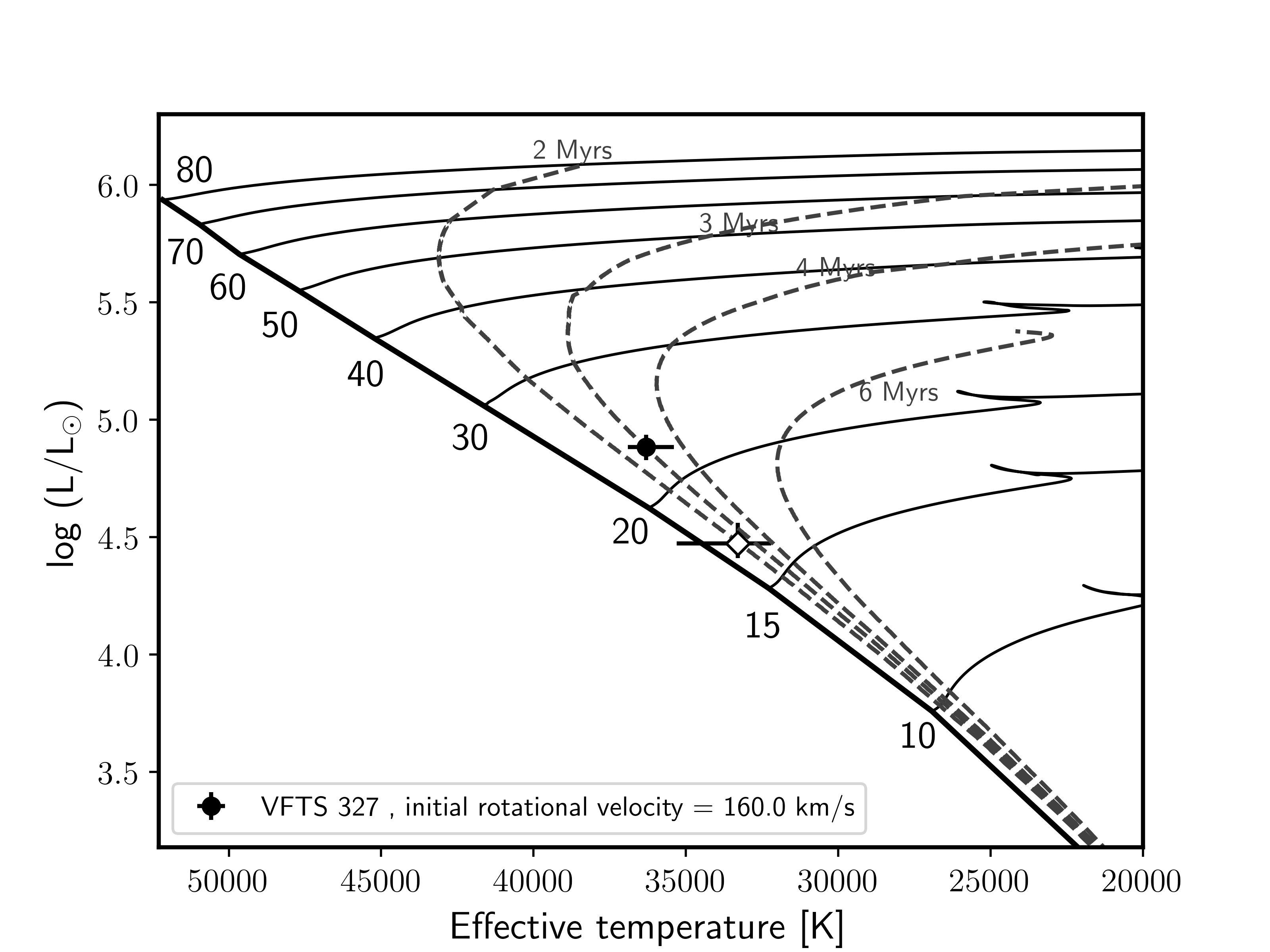}
    \includegraphics[width=7cm, bb=5 0 453 346,clip]{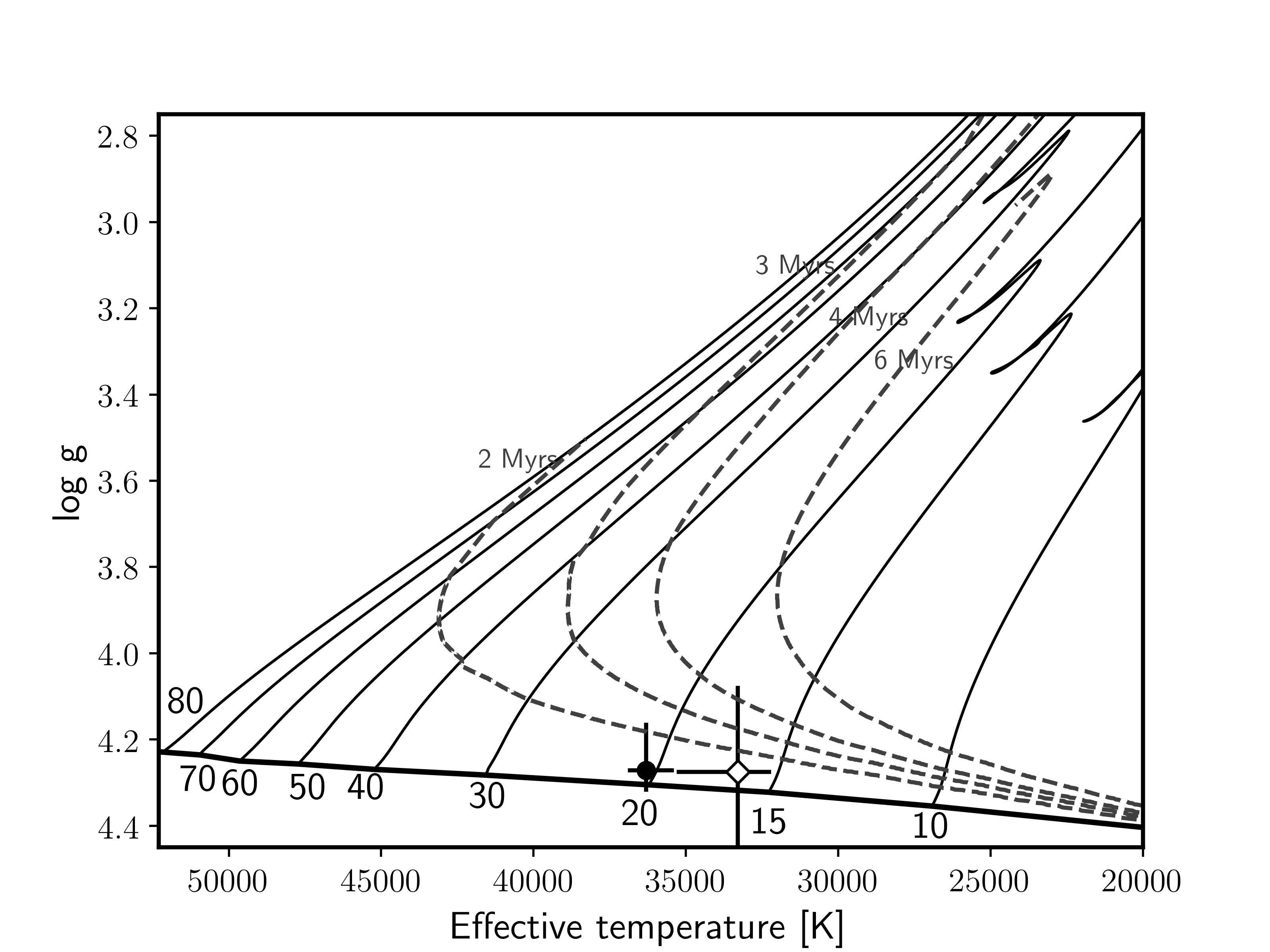}
    \caption{Same as Fig.\,\ref{fig:042} but for VFTS\,327.} \label{fig:327} 
  \end{figure*} 
 \clearpage

 \begin{figure*}[t!]
    \centering
    \includegraphics[width=6.cm]{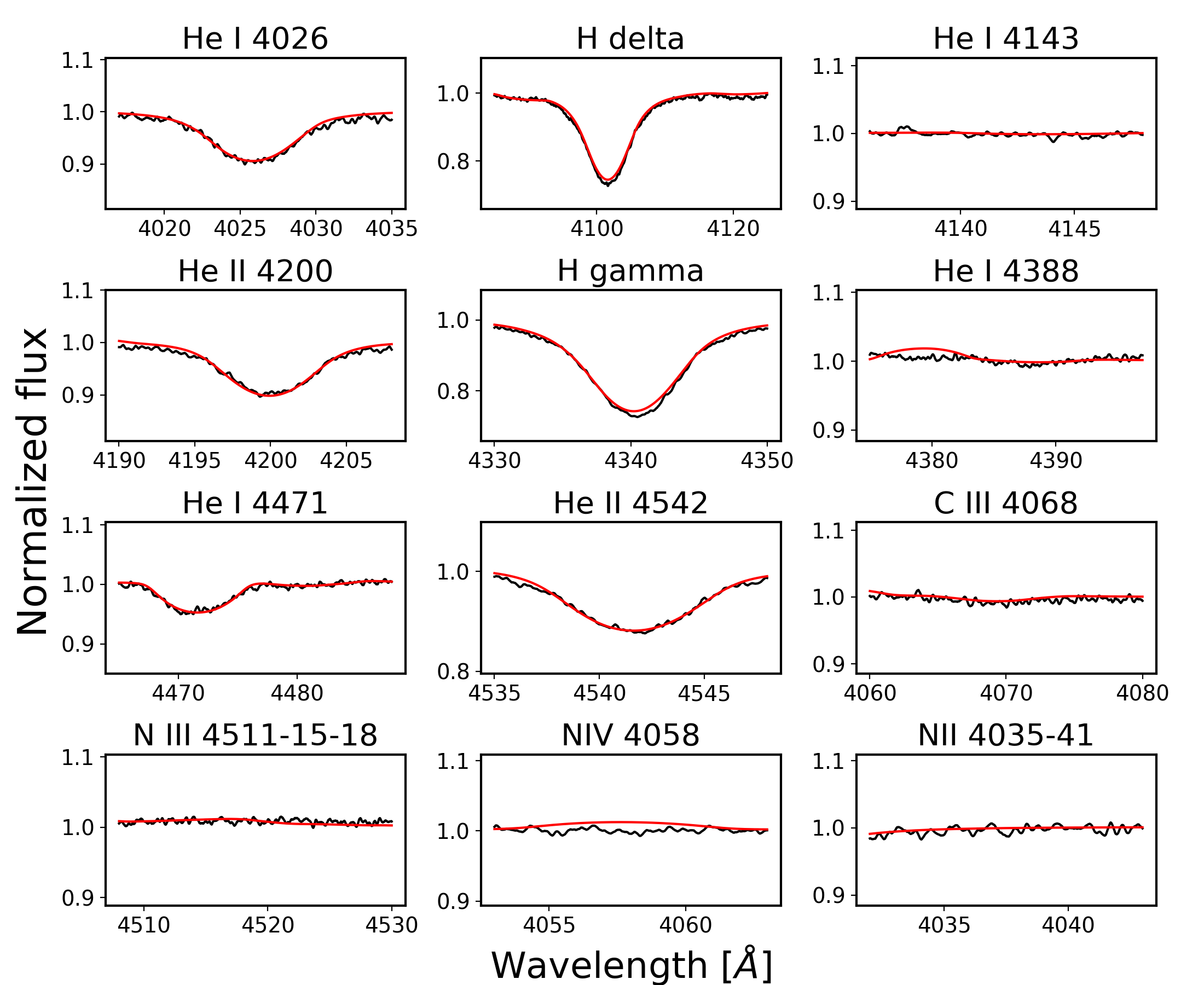}
    \includegraphics[width=7.cm, bb=5 0 453 346,clip]{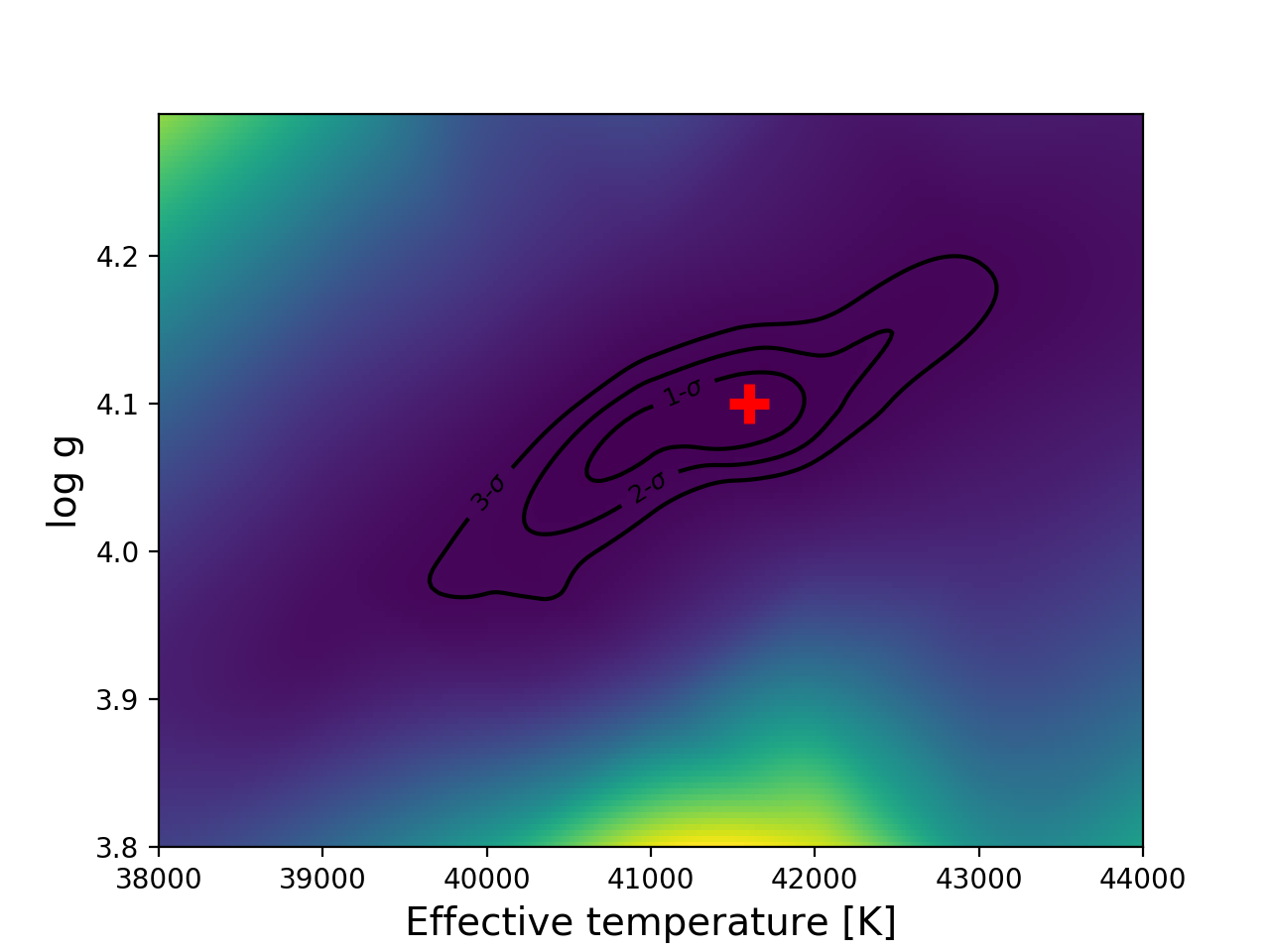}
    \includegraphics[width=6.cm]{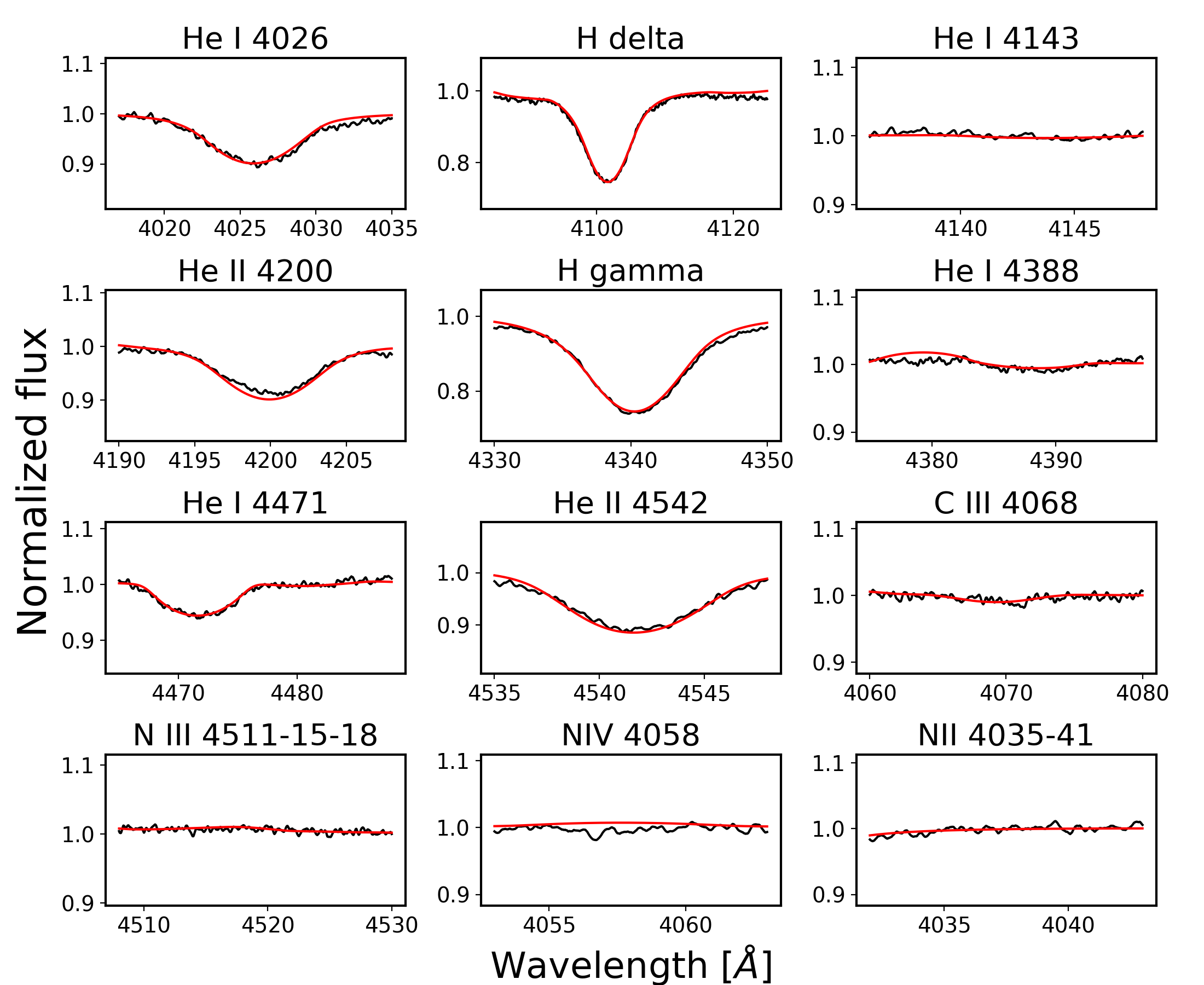}
    \includegraphics[width=7.cm, bb=5 0 453 346,clip]{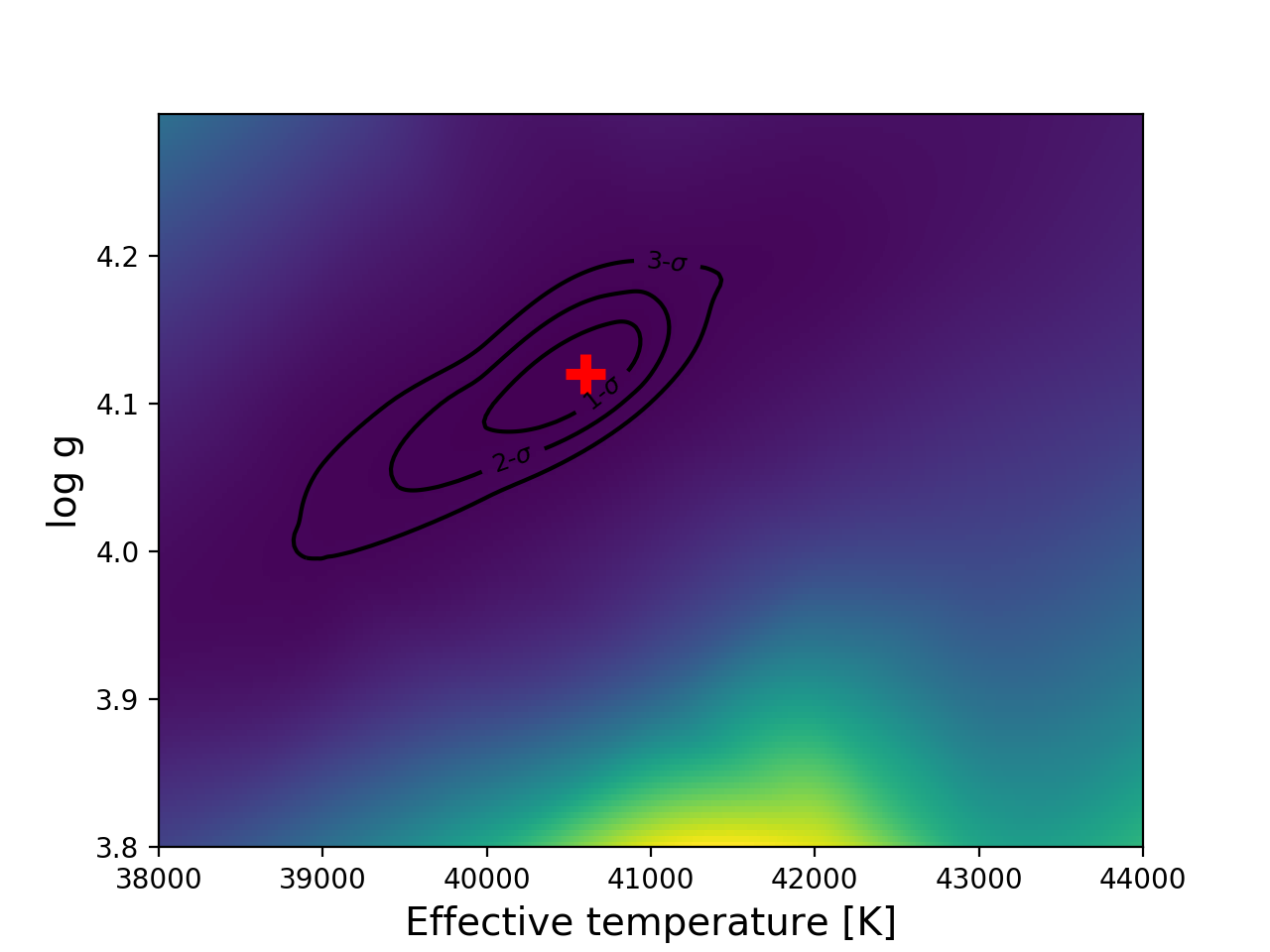}
    \includegraphics[width=7cm]{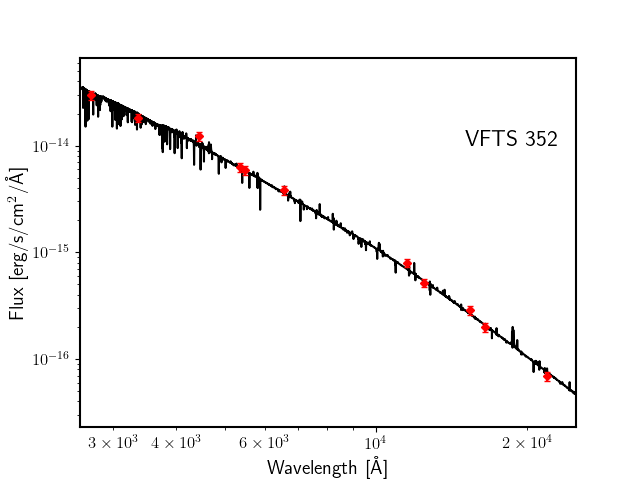}
    \includegraphics[width=6.5cm]{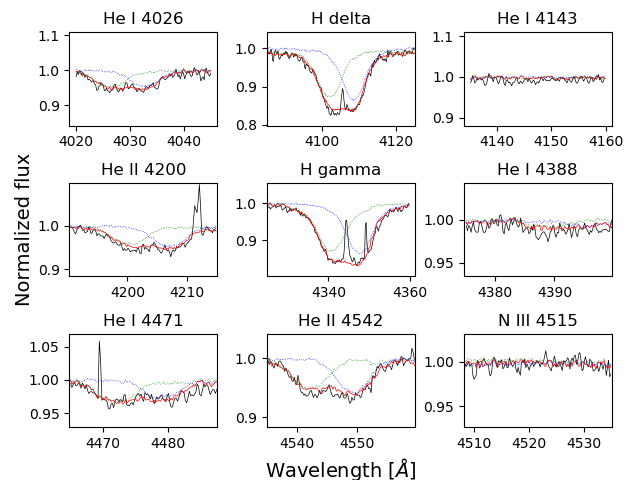}
    \includegraphics[width=7cm, bb=5 0 453 346,clip]{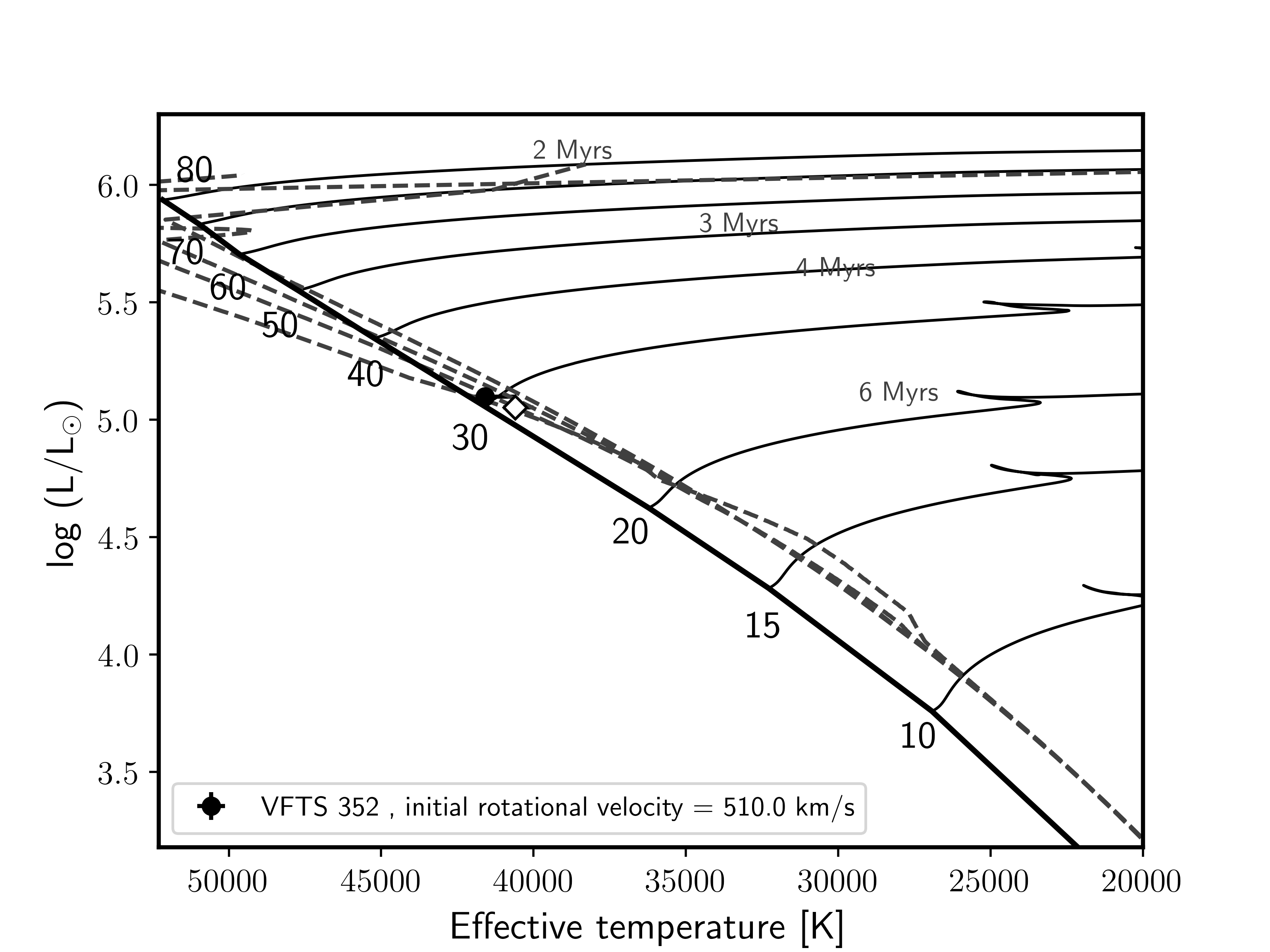}
    \includegraphics[width=7cm, bb=5 0 453 346,clip]{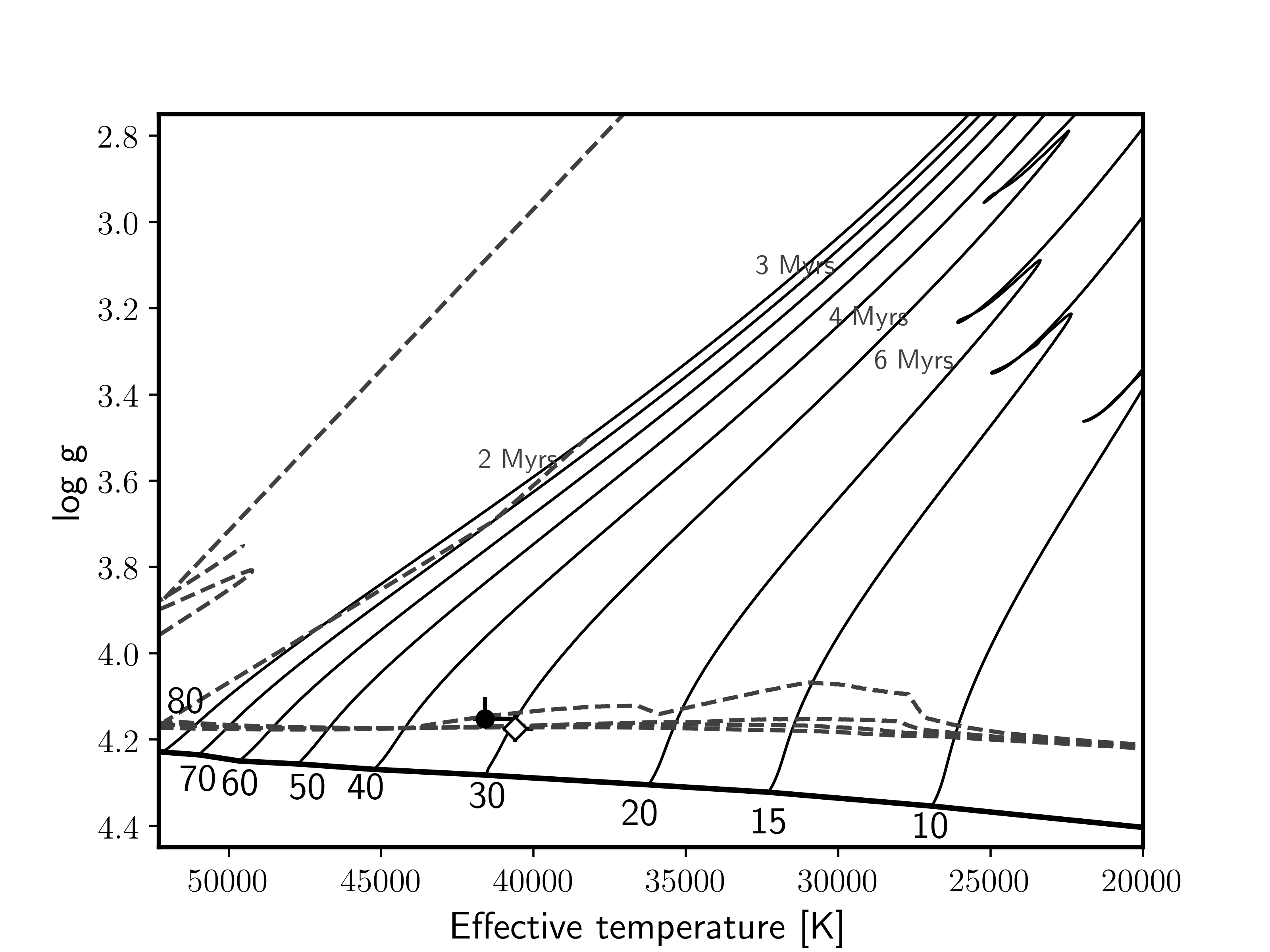}
    \caption{Same as Fig.\,\ref{fig:042} but for VFTS\,352.} \label{fig:352} 
  \end{figure*}   
  \clearpage

 \begin{figure*}[t!]
    \centering
    \includegraphics[width=6.cm]{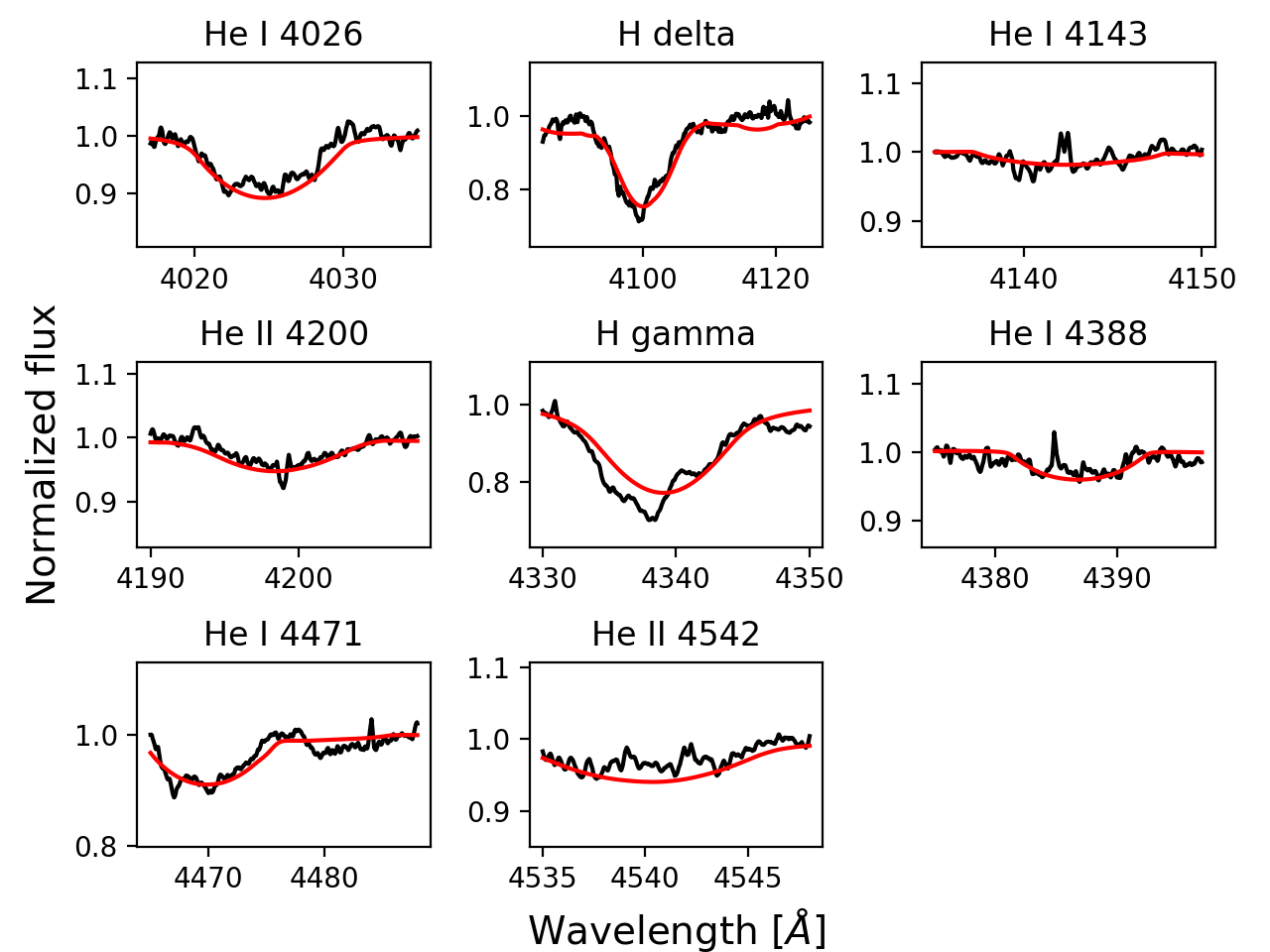}
    \includegraphics[width=7.cm, bb=5 0 453 346,clip]{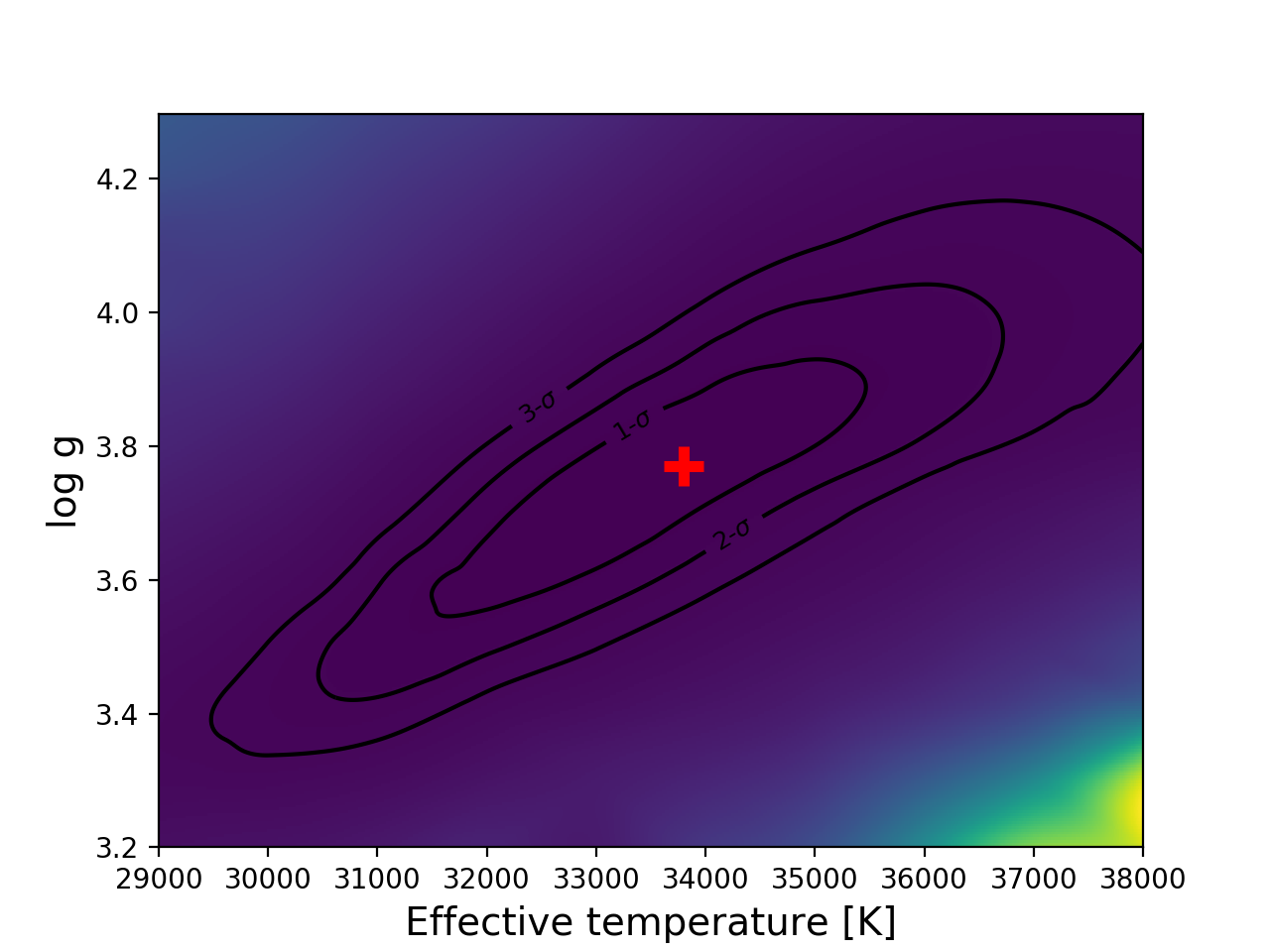}
    \includegraphics[width=6.cm]{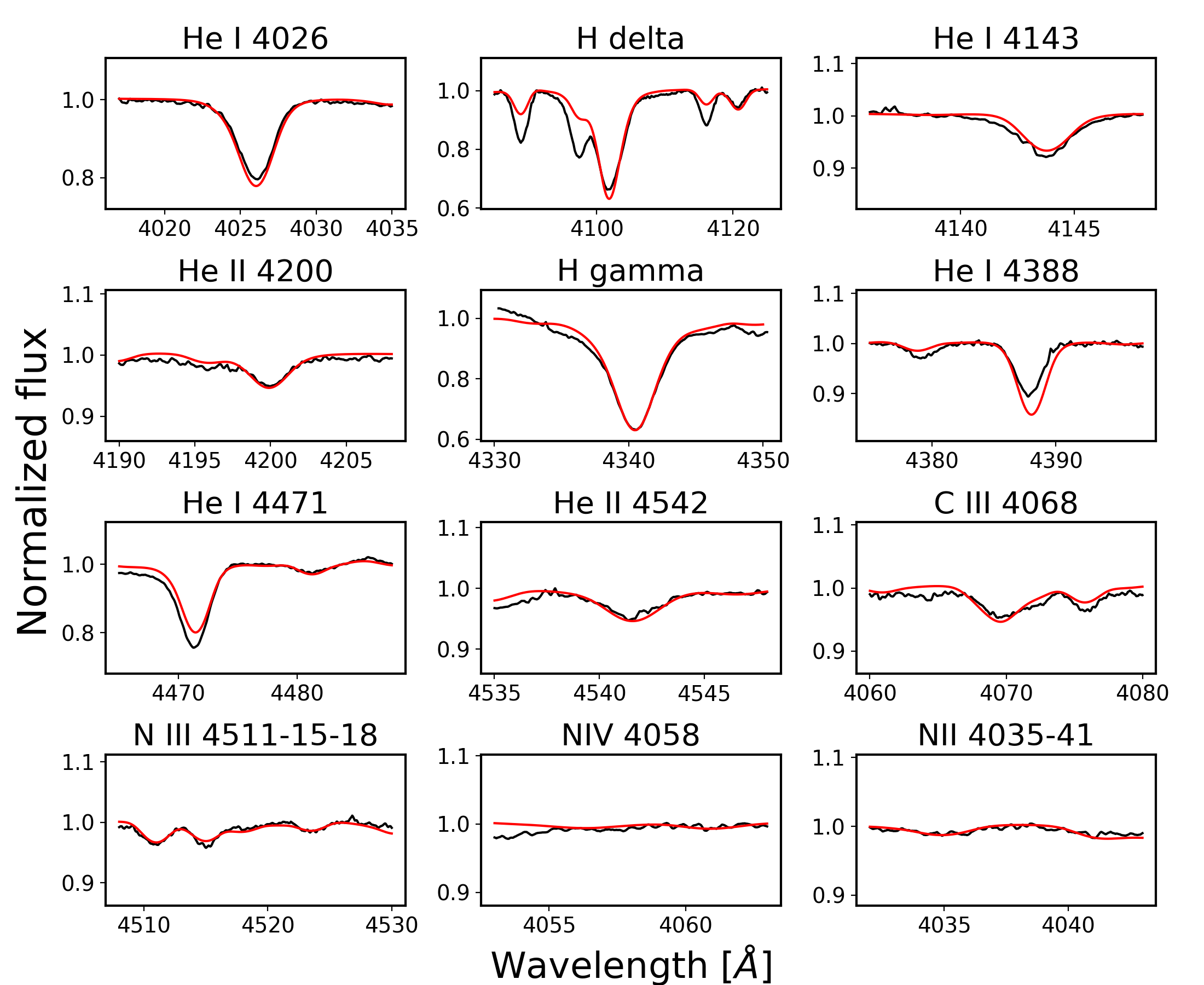}
    \includegraphics[width=7.cm, bb=5 0 453 346,clip]{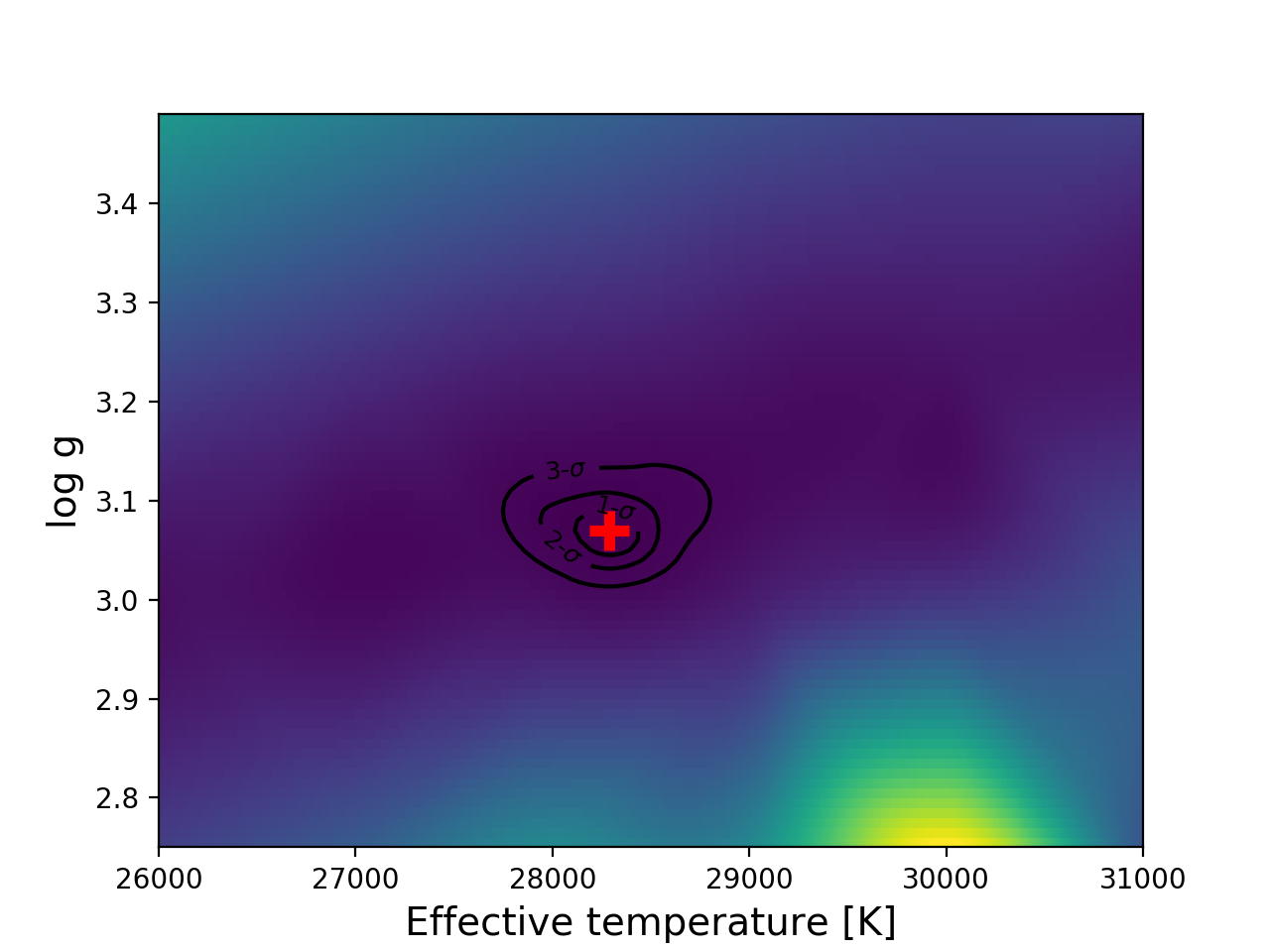}
    \includegraphics[width=7cm]{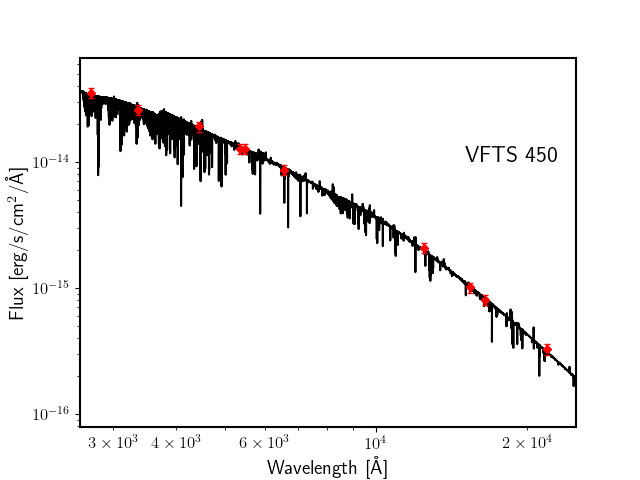}
    \includegraphics[width=6.5cm]{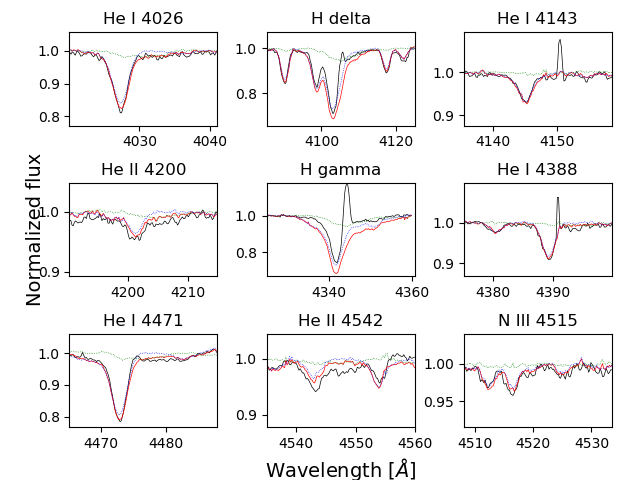}
    \includegraphics[width=7cm, bb=5 0 453 346,clip]{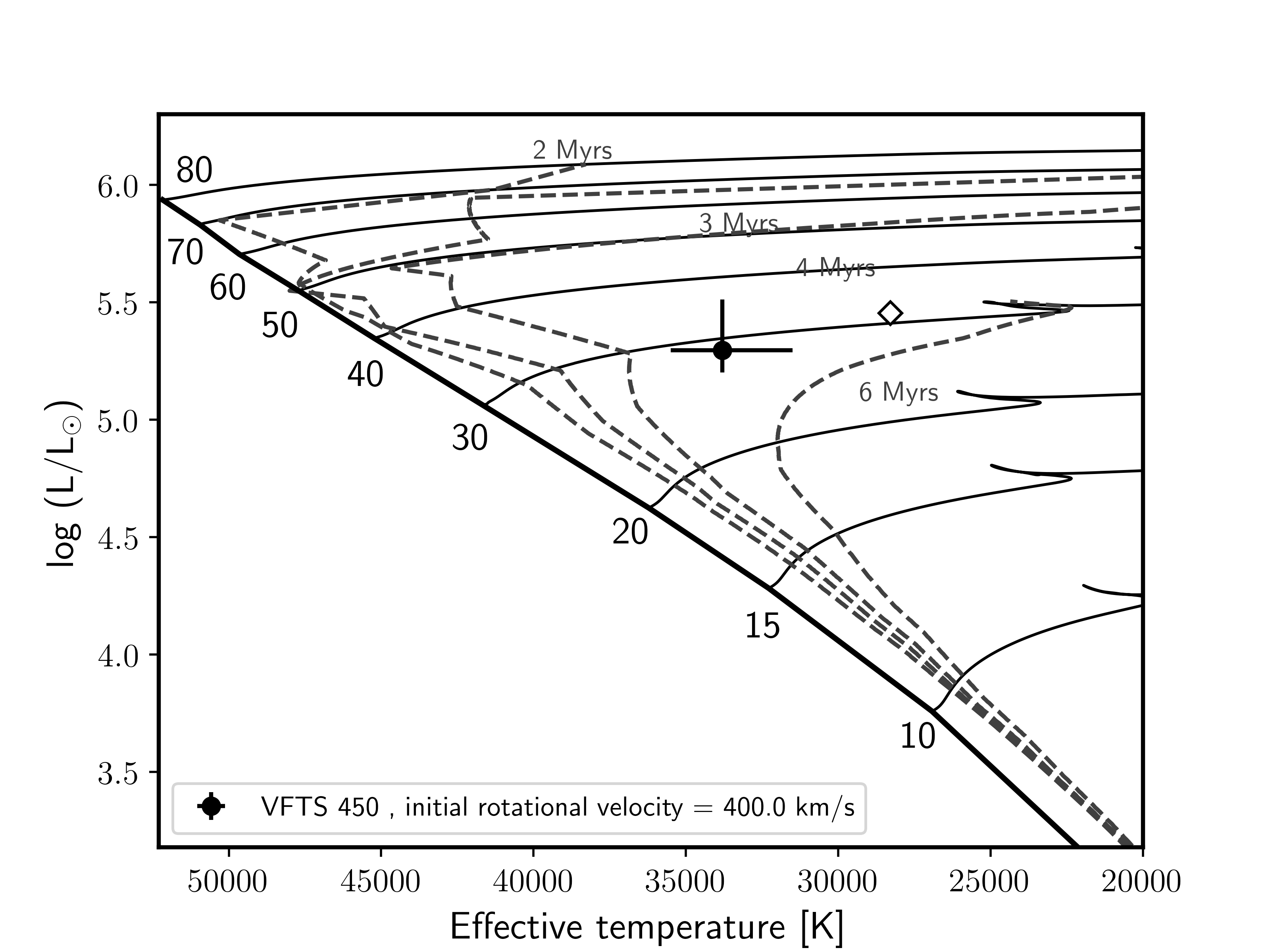}
    \includegraphics[width=7cm, bb=5 0 453 346,clip]{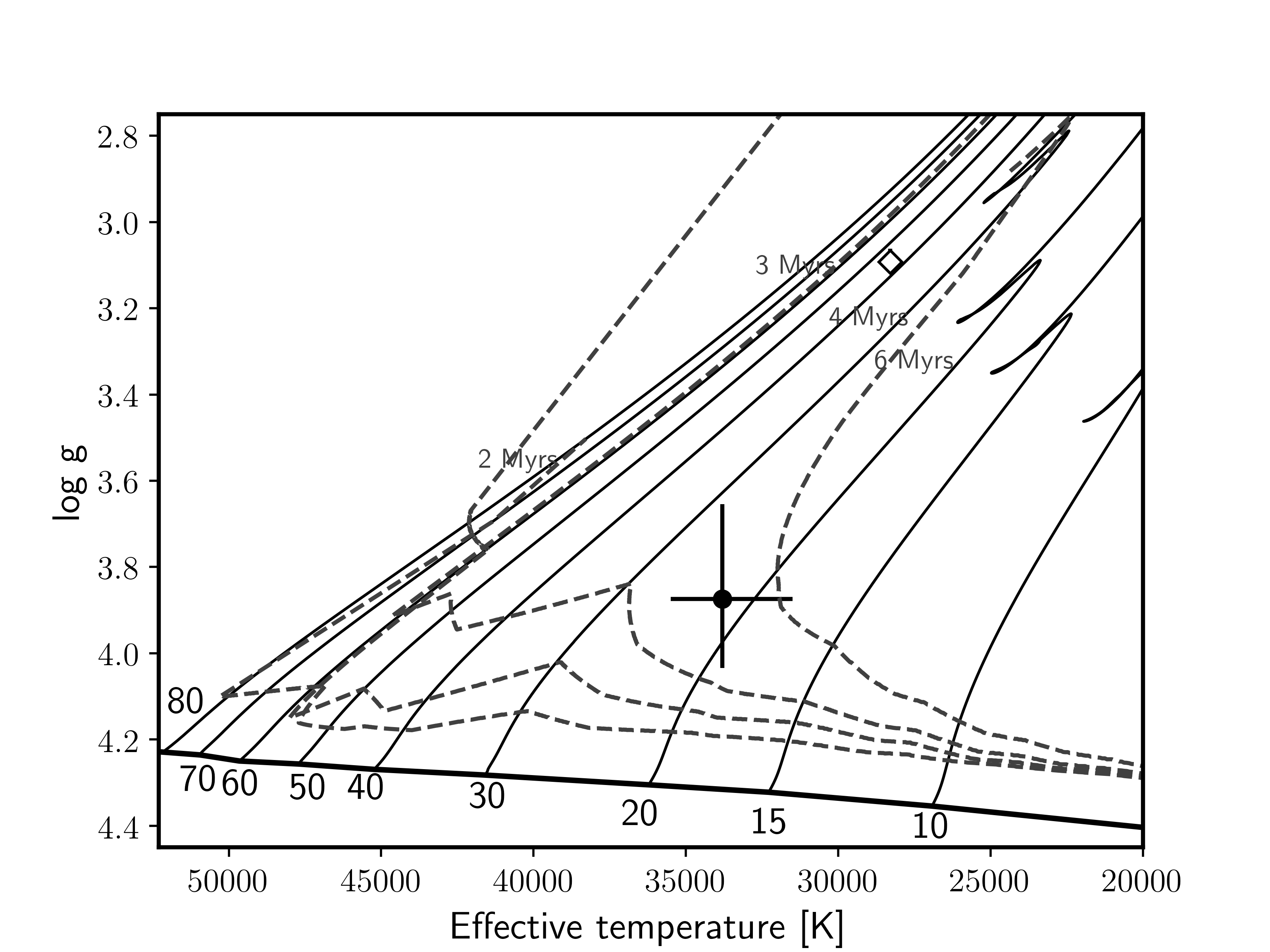}
    \caption{Same as Fig.\,\ref{fig:042} but for VFTS\,450.} \label{fig:450} 
  \end{figure*} 
 \clearpage

 \begin{figure*}[t!]
    \centering
    \includegraphics[width=6.cm]{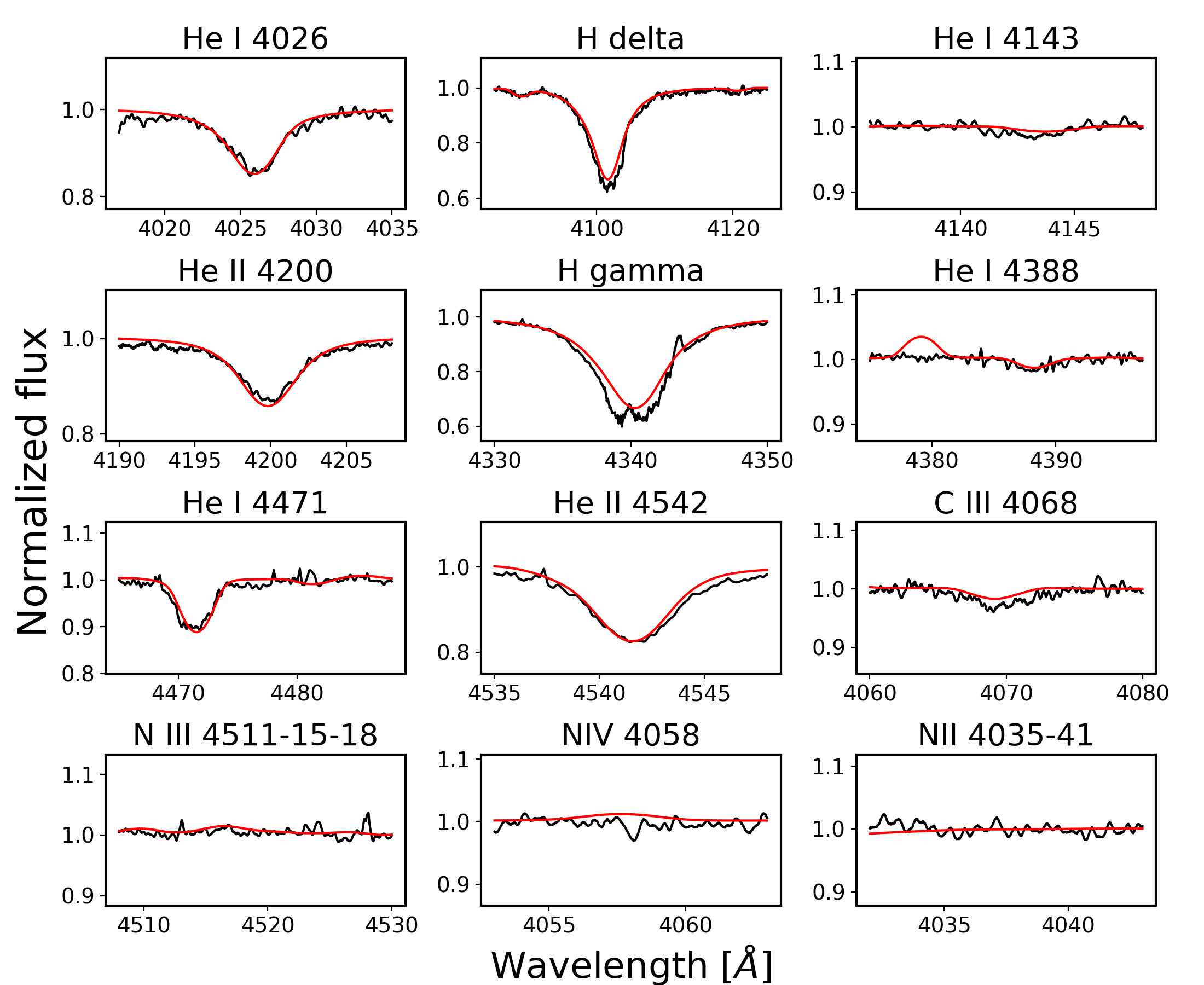}
    \includegraphics[width=7.cm, bb=5 0 453 346,clip]{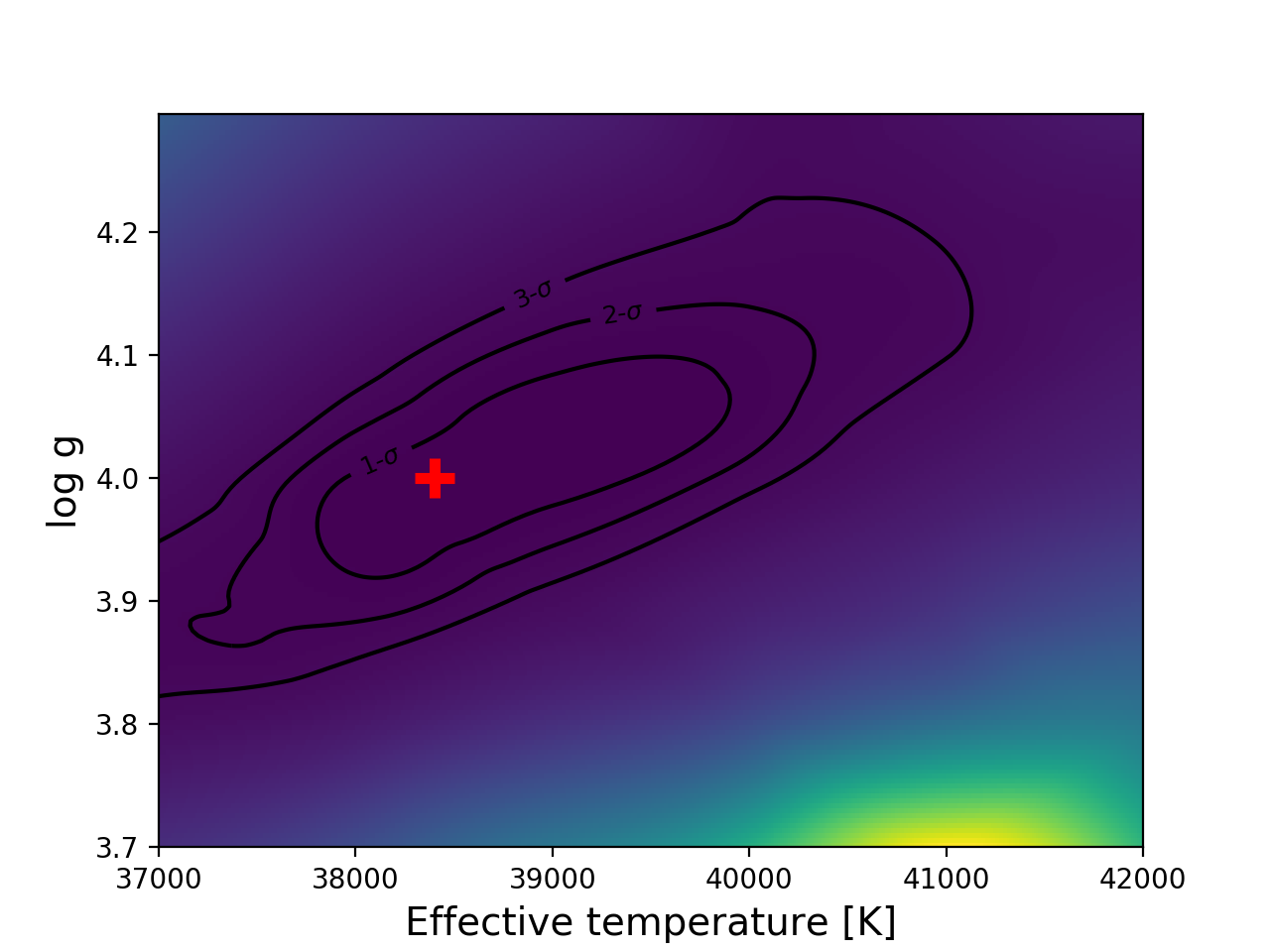}
    \includegraphics[width=6.cm]{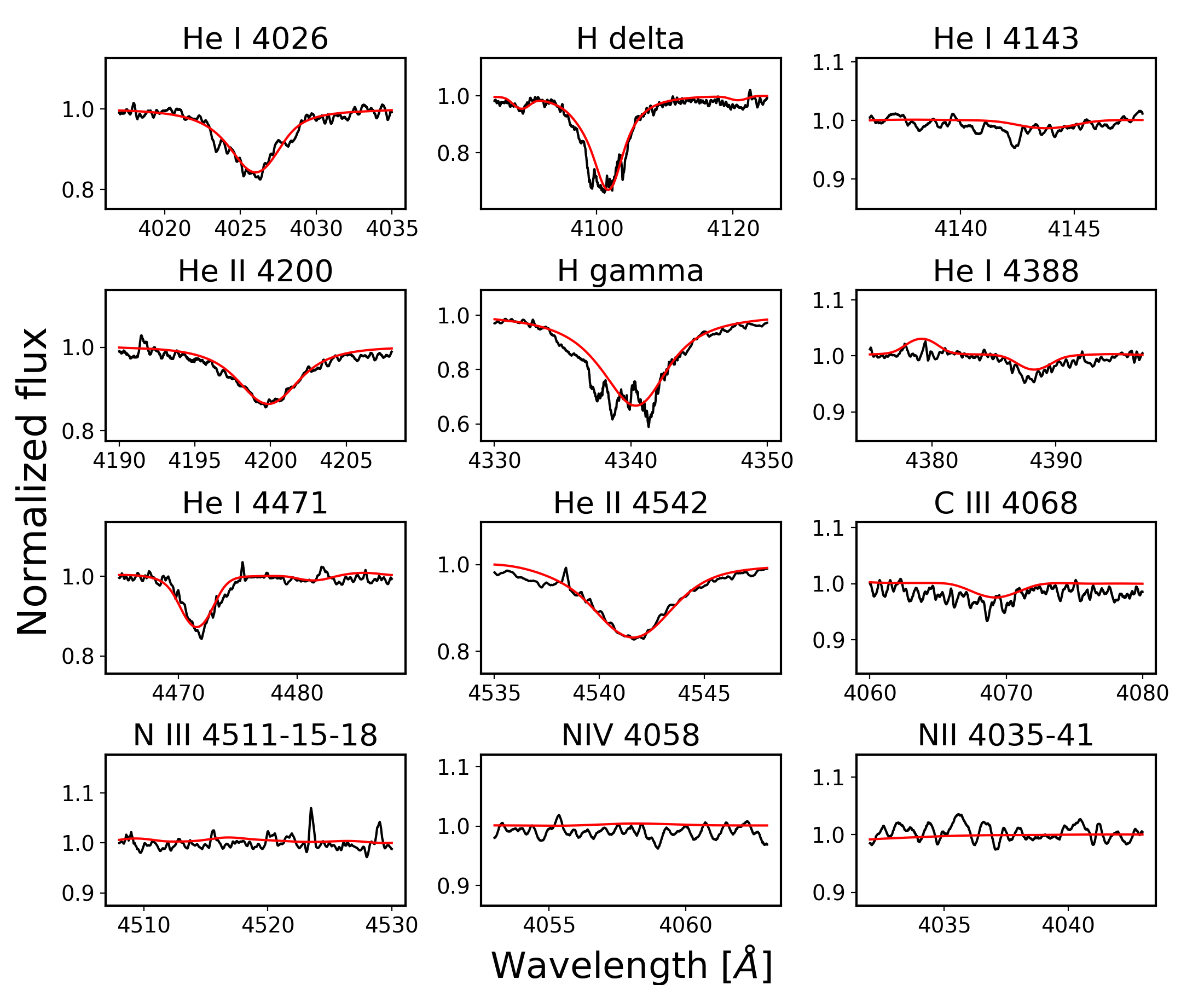}
    \includegraphics[width=7.cm, bb=5 0 453 346,clip]{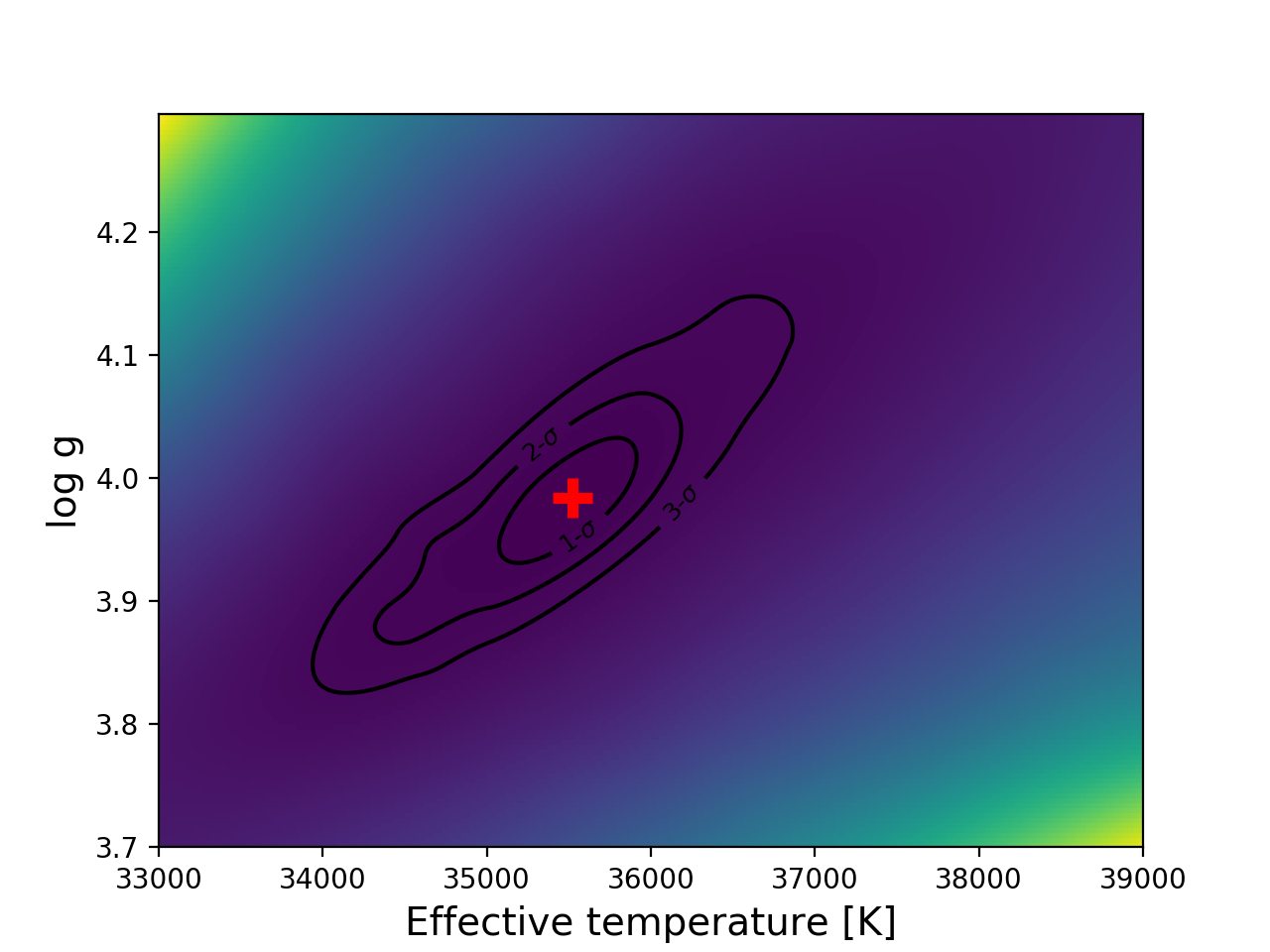}
    \includegraphics[width=7cm]{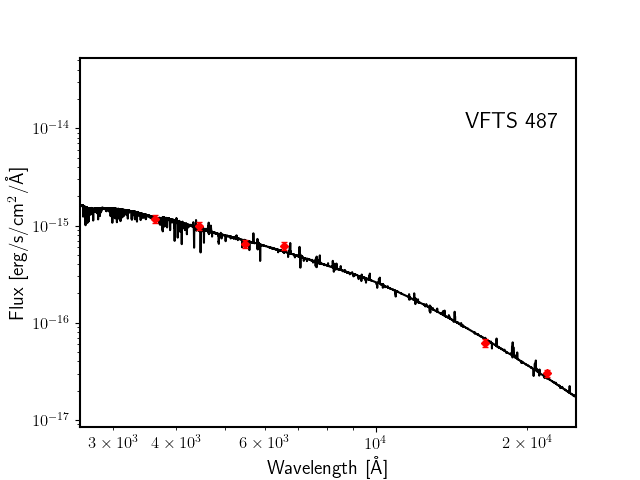}
    \includegraphics[width=6.5cm]{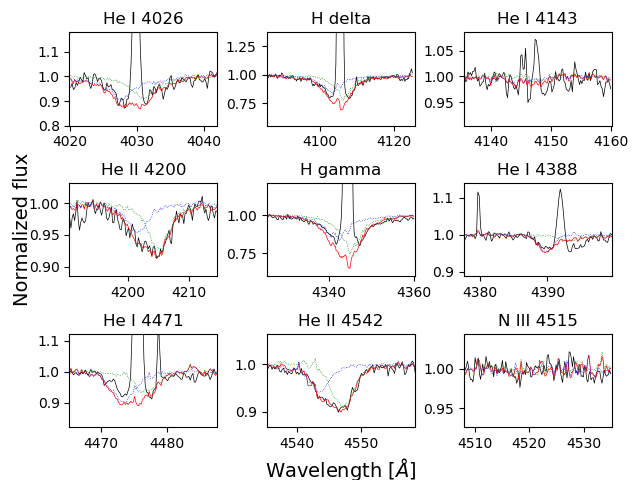}
    \includegraphics[width=7cm, bb=5 0 453 346,clip]{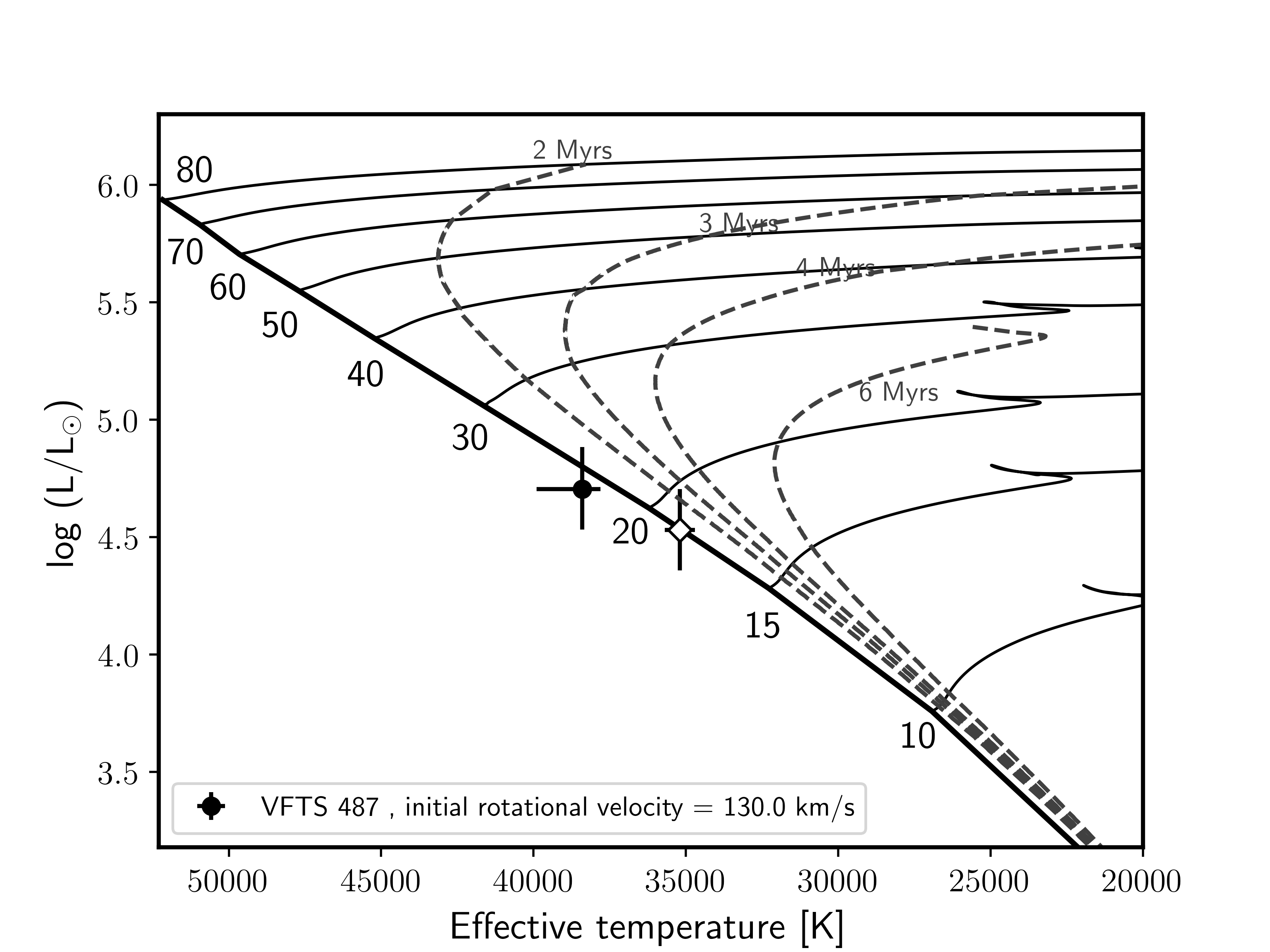}
    \includegraphics[width=7cm, bb=5 0 453 346,clip]{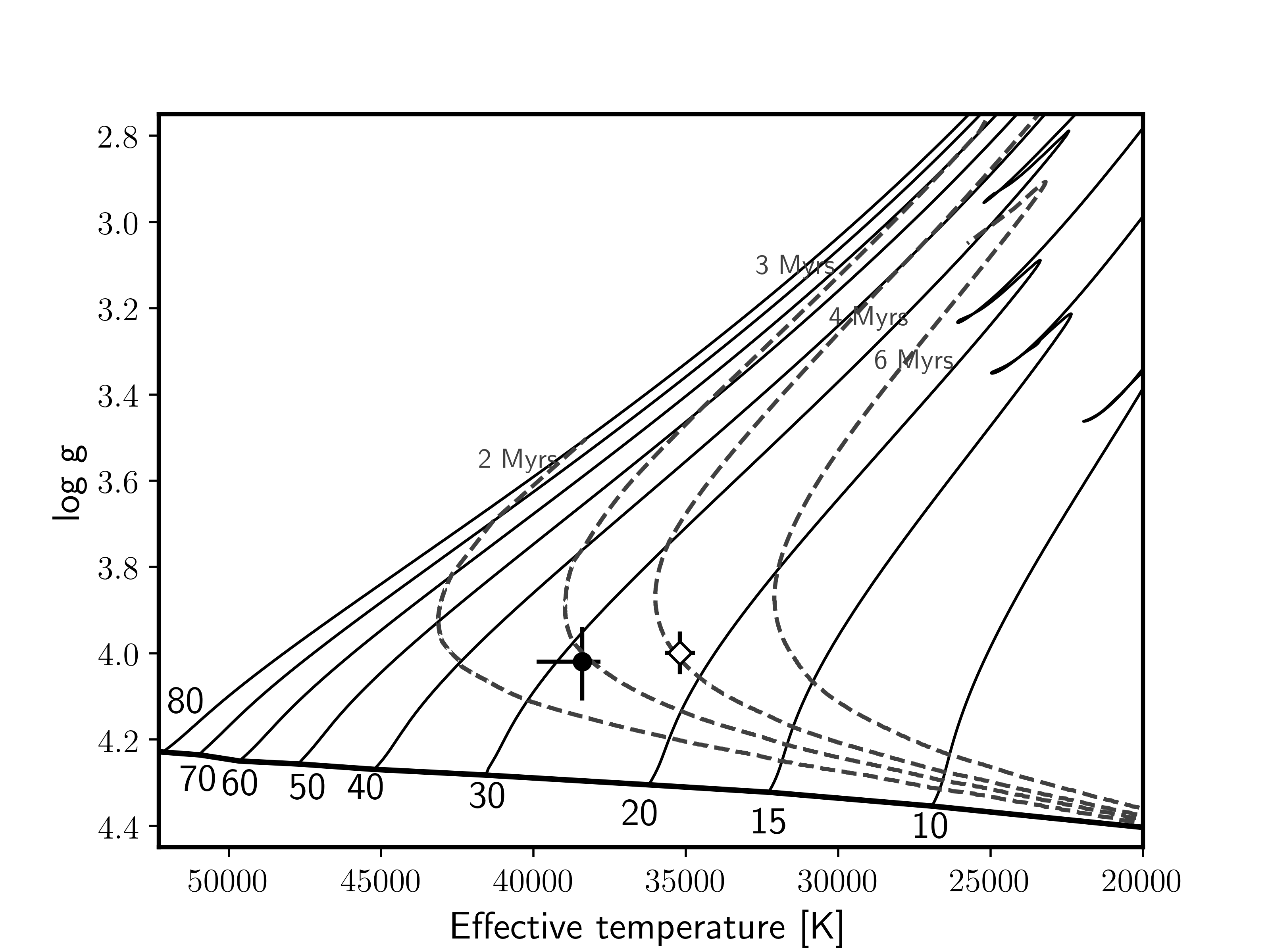}
    \caption{Same as Fig.\,\ref{fig:042} but for VFTS\,487.} \label{fig:487} 
  \end{figure*} 
   \clearpage
       
 \begin{figure*}[t!]
    \centering
    \includegraphics[width=6.cm]{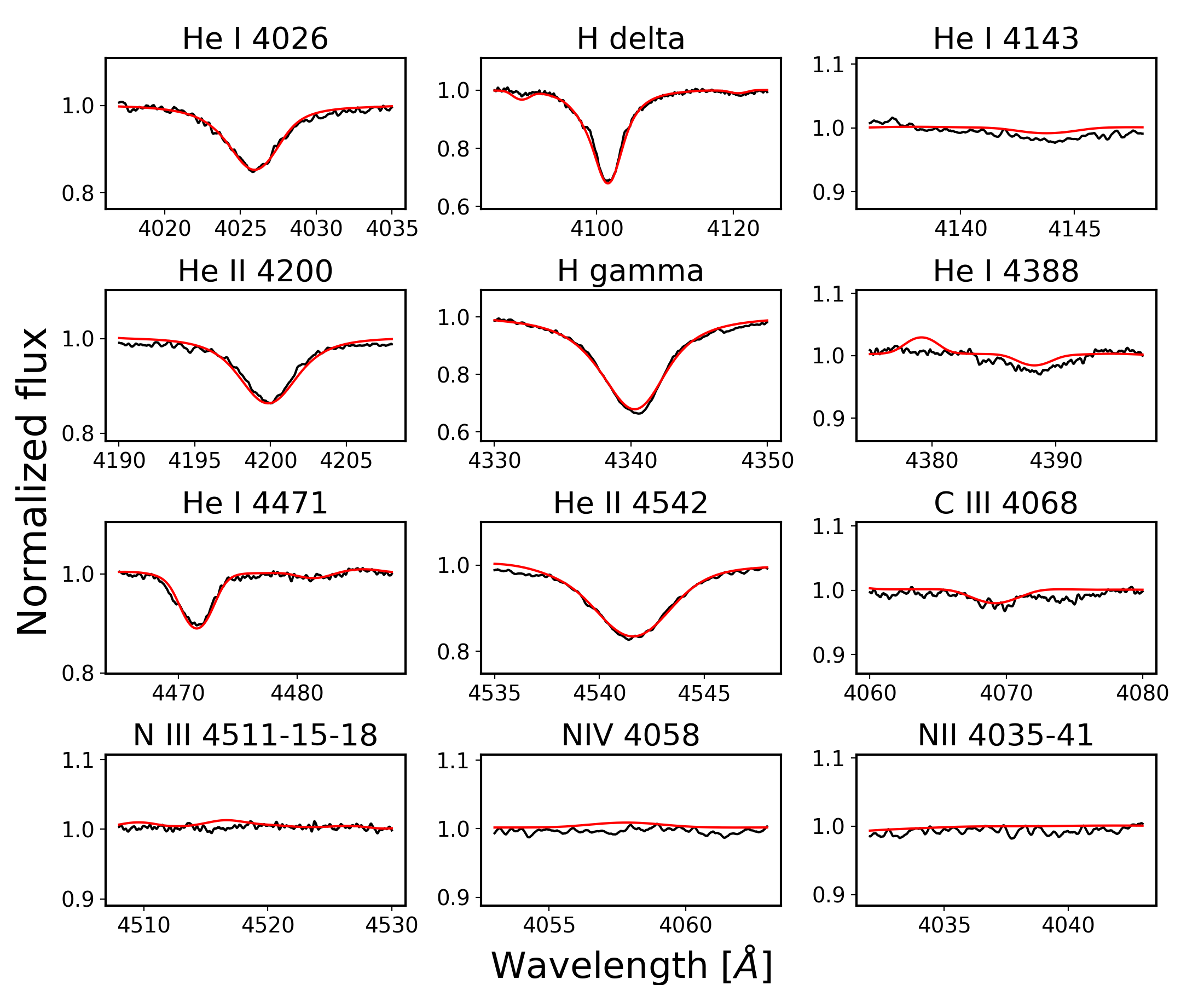}
    \includegraphics[width=7.cm, bb=5 0 453 346,clip]{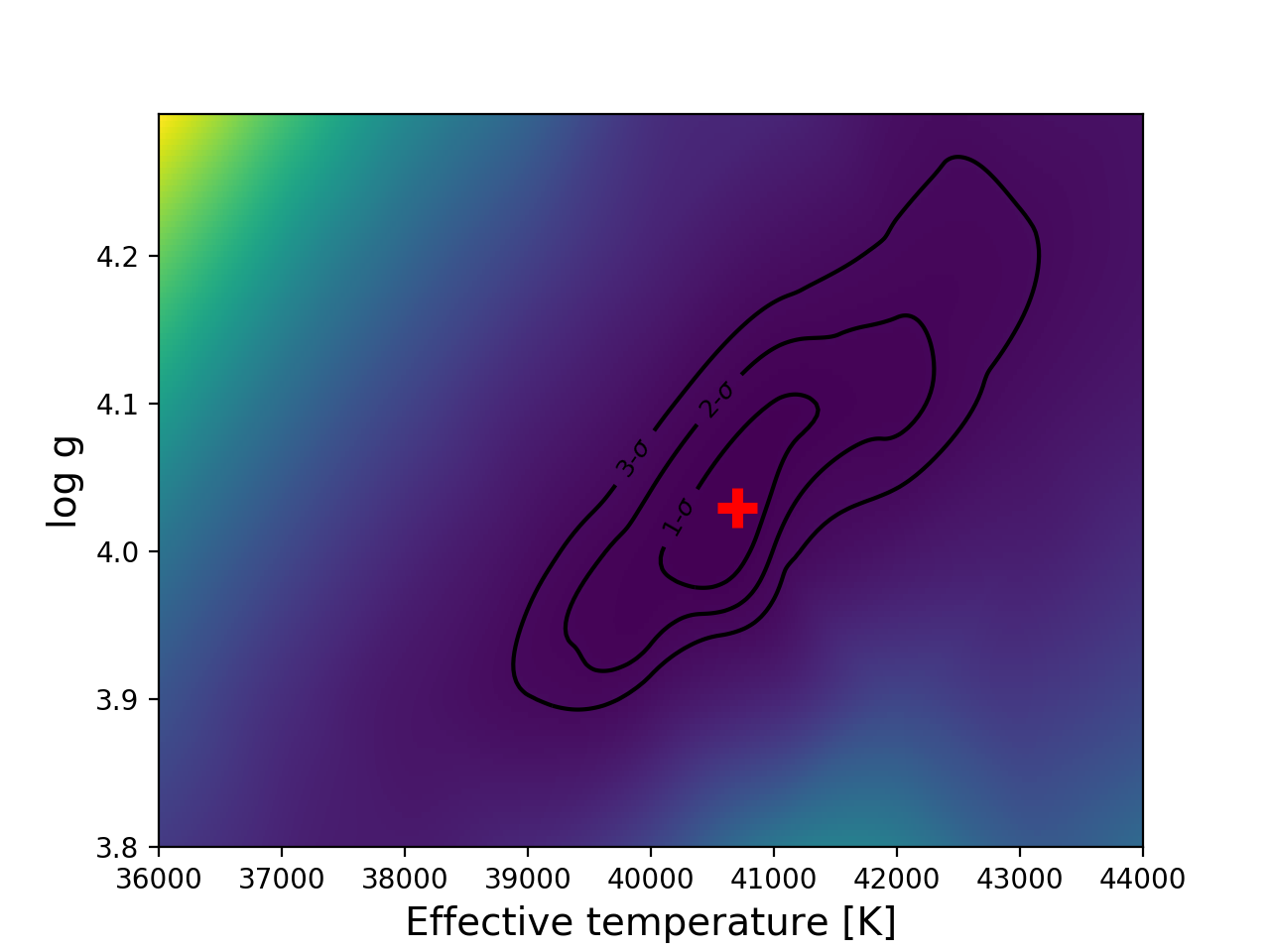}
    \includegraphics[width=6.cm]{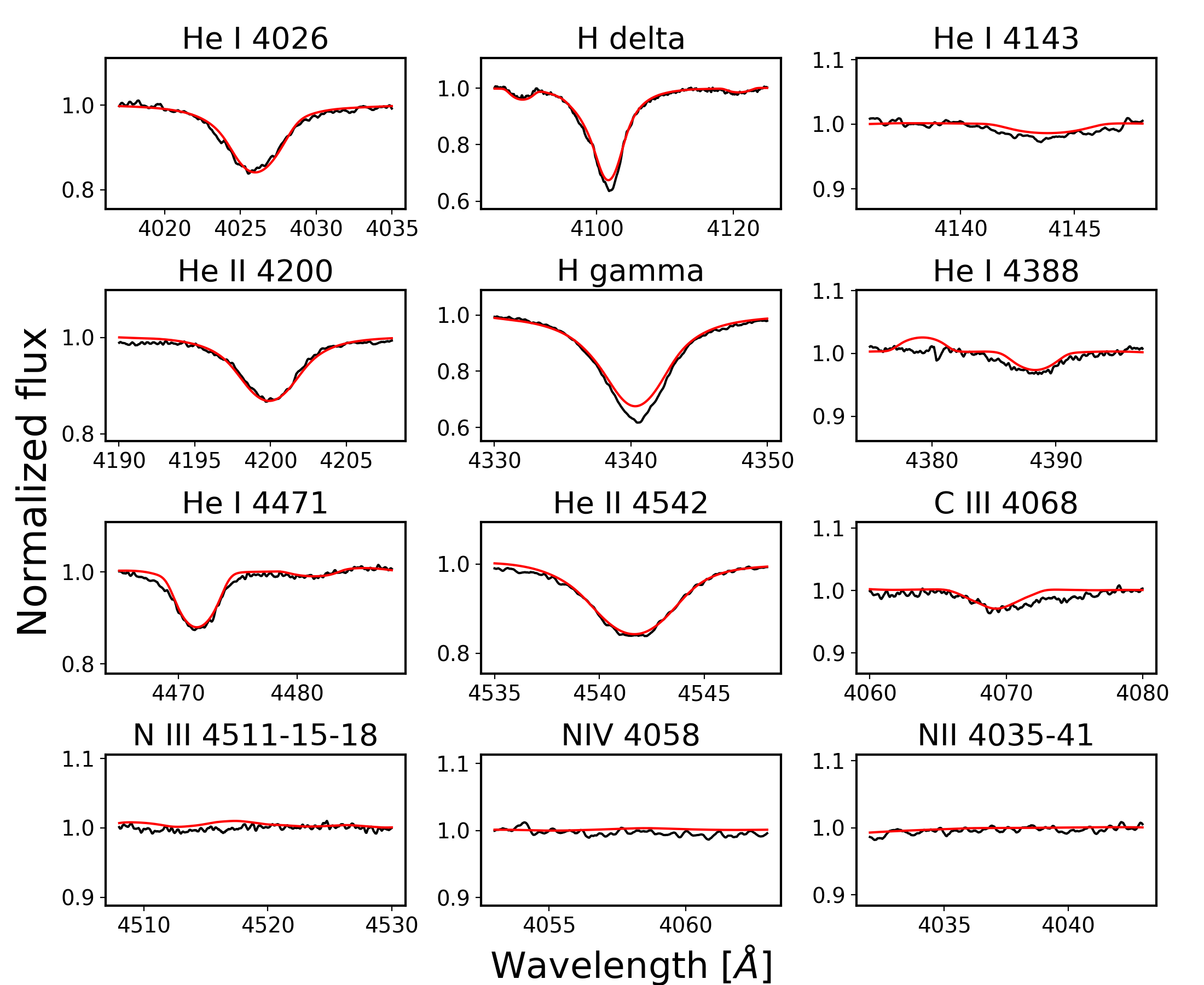}
    \includegraphics[width=7.cm, bb=5 0 453 346,clip]{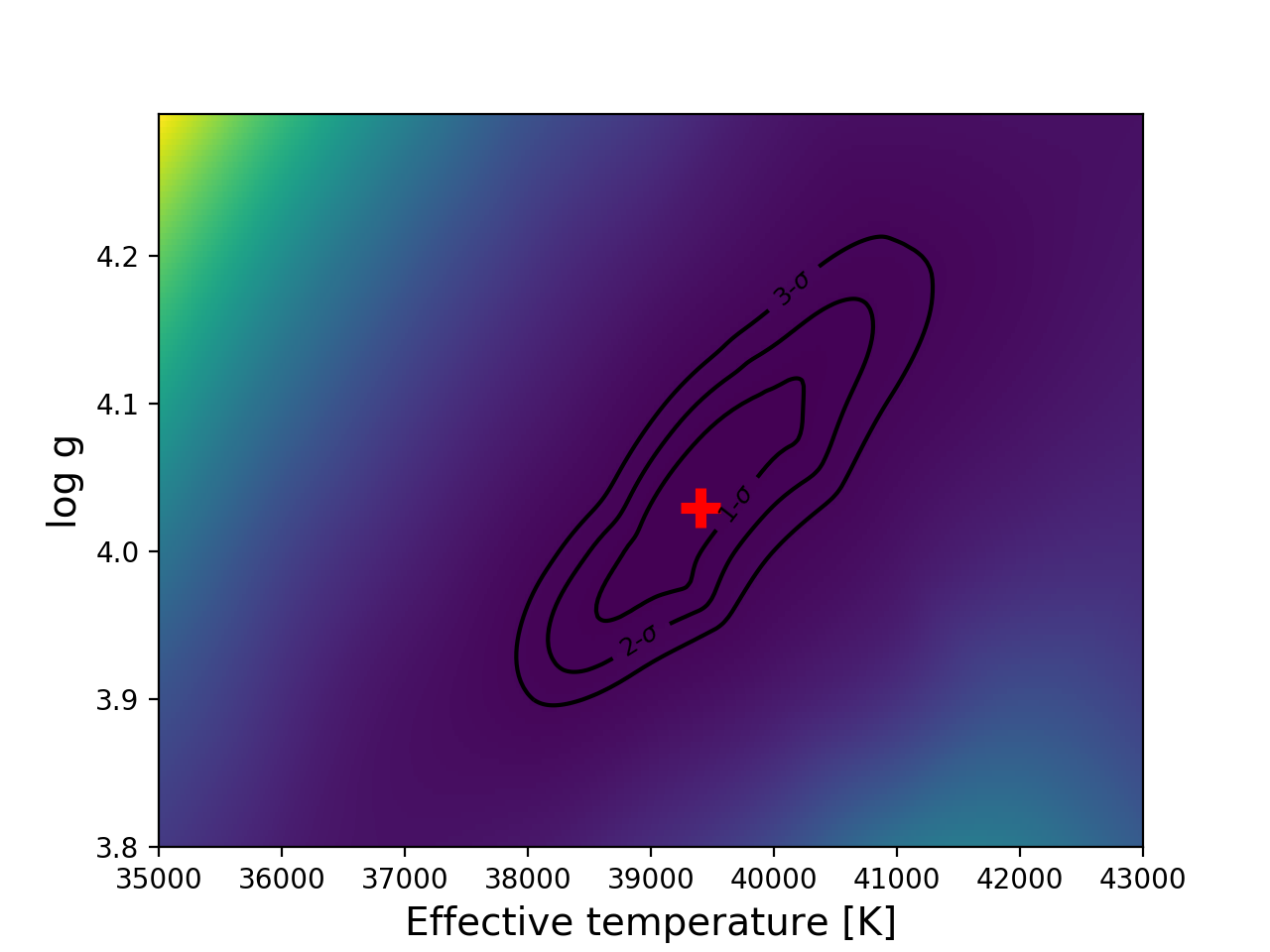}
    \includegraphics[width=7cm]{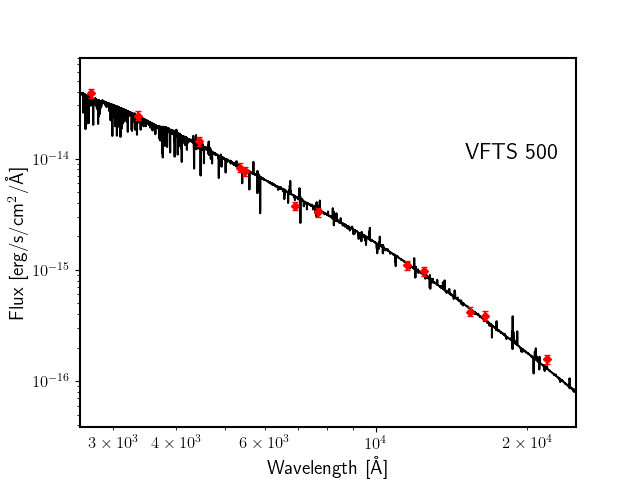}
    \includegraphics[width=6.5cm]{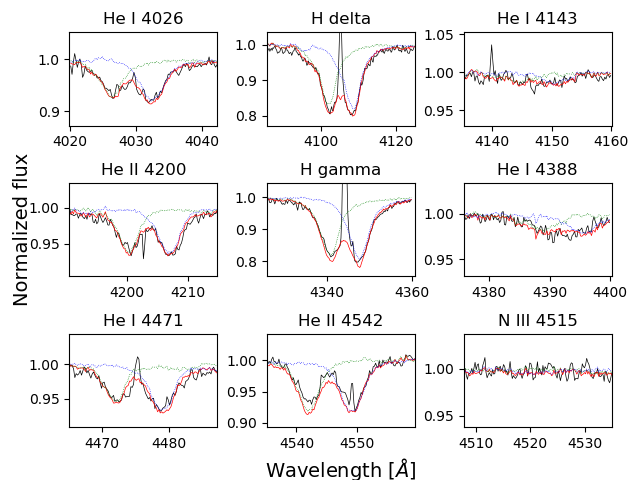}
    \includegraphics[width=7cm, bb=5 0 453 346,clip]{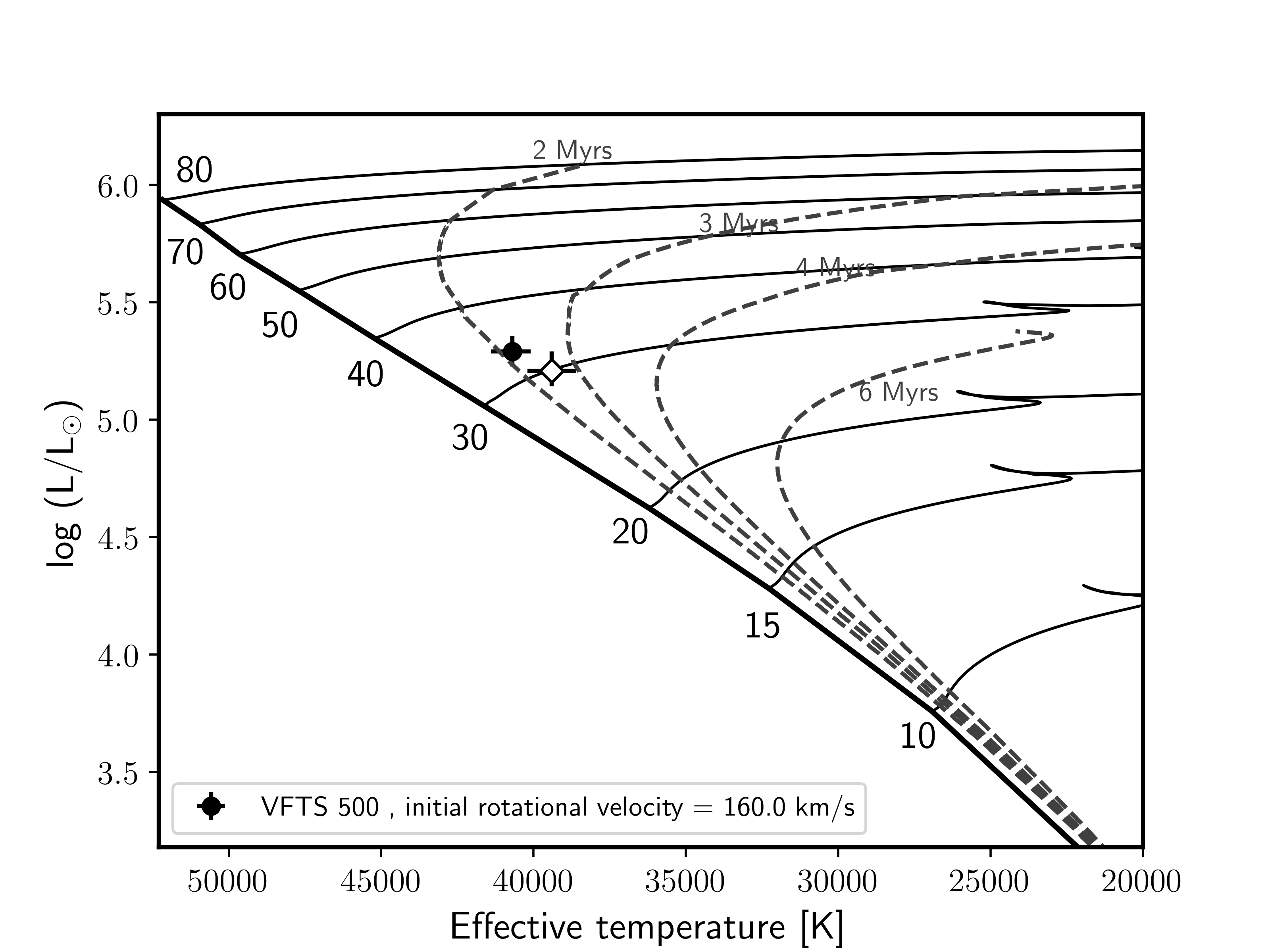}
    \includegraphics[width=7cm, bb=5 0 453 346,clip]{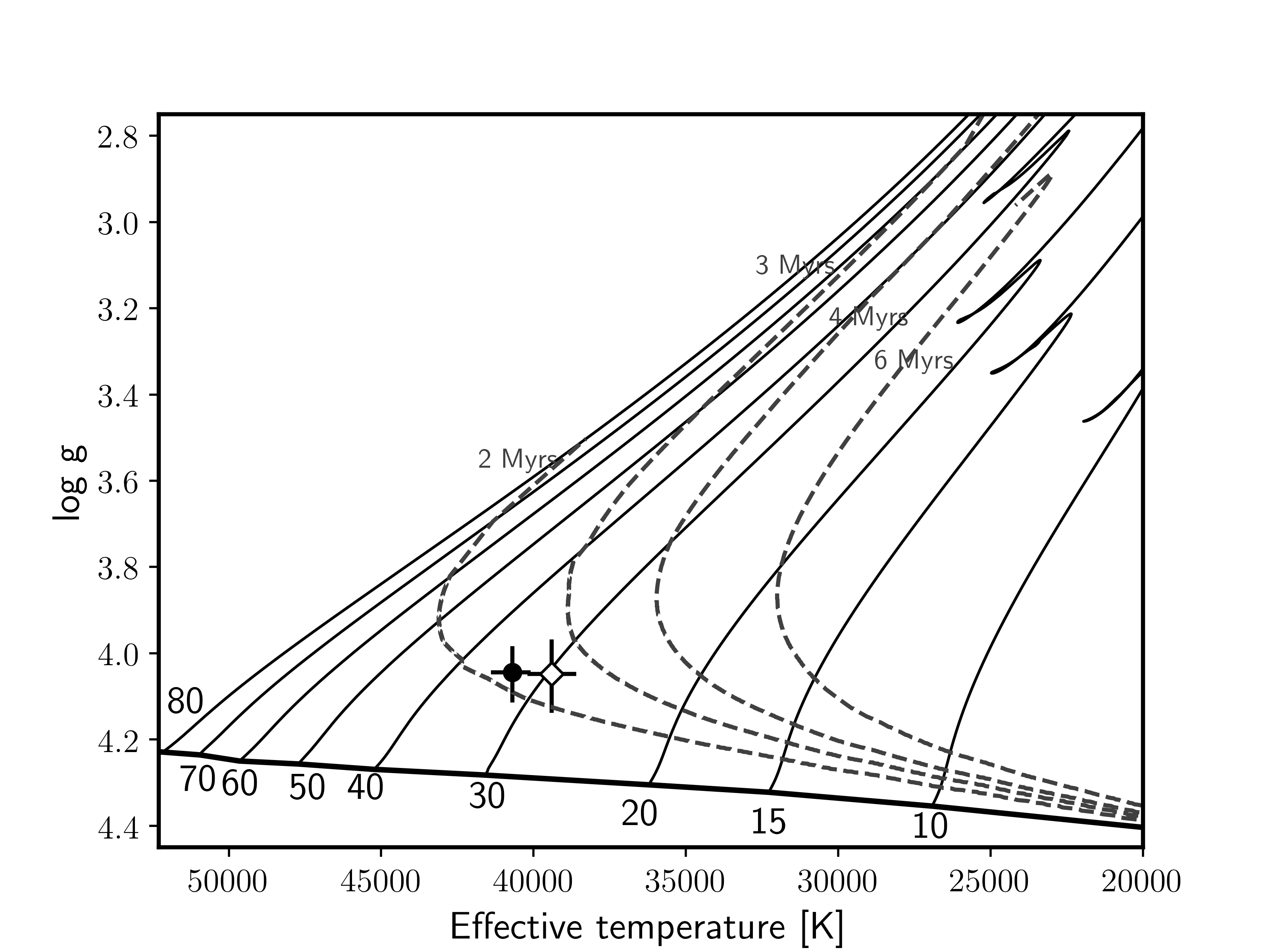}
    \caption{Same as Fig.\,\ref{fig:042} but for VFTS\,500.} \label{fig:500} 
  \end{figure*} 
 \clearpage
      
 \begin{figure*}[t!]
    \centering
    \includegraphics[width=6.cm]{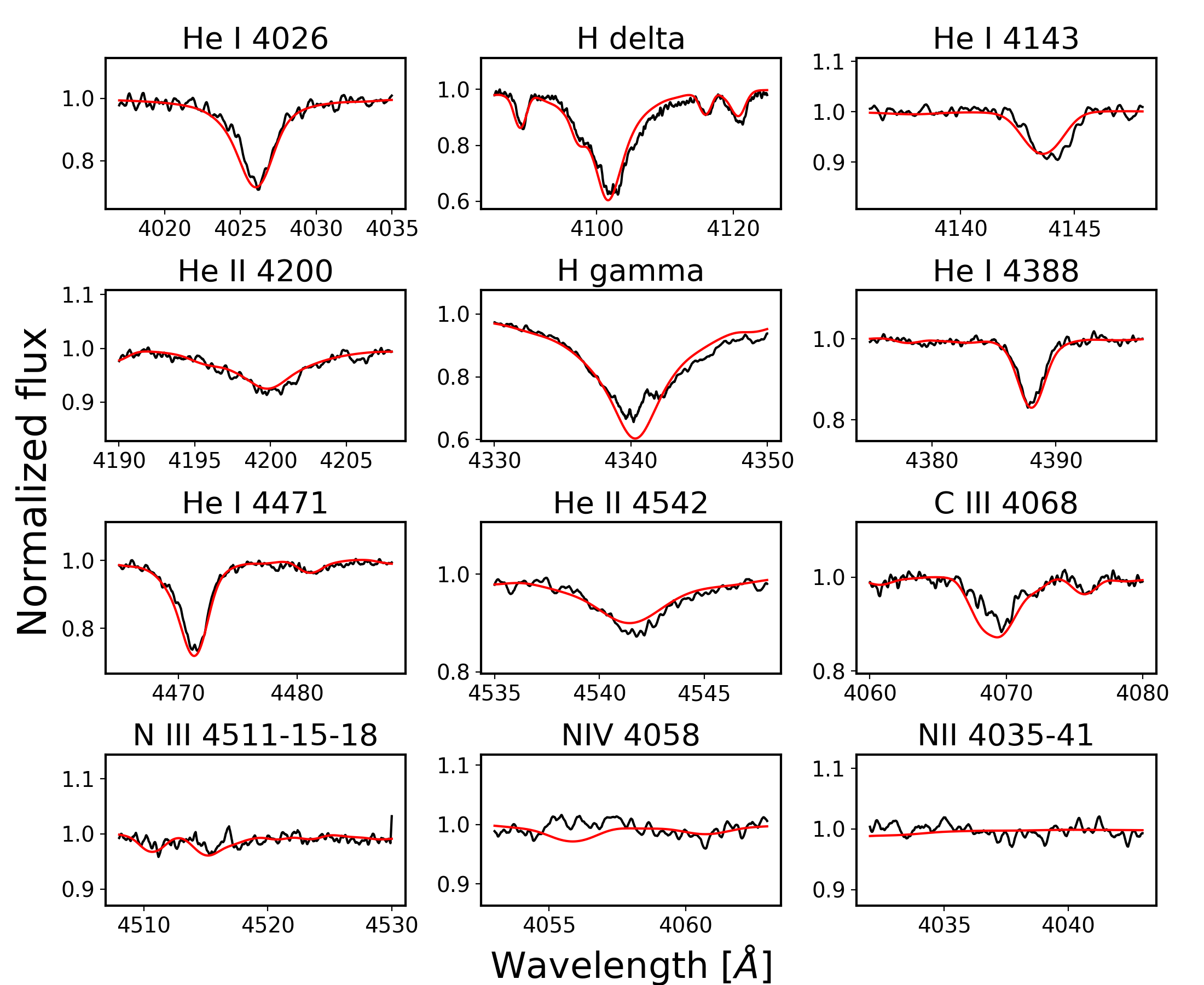}
    \includegraphics[width=7.cm, bb=5 0 453 346,clip]{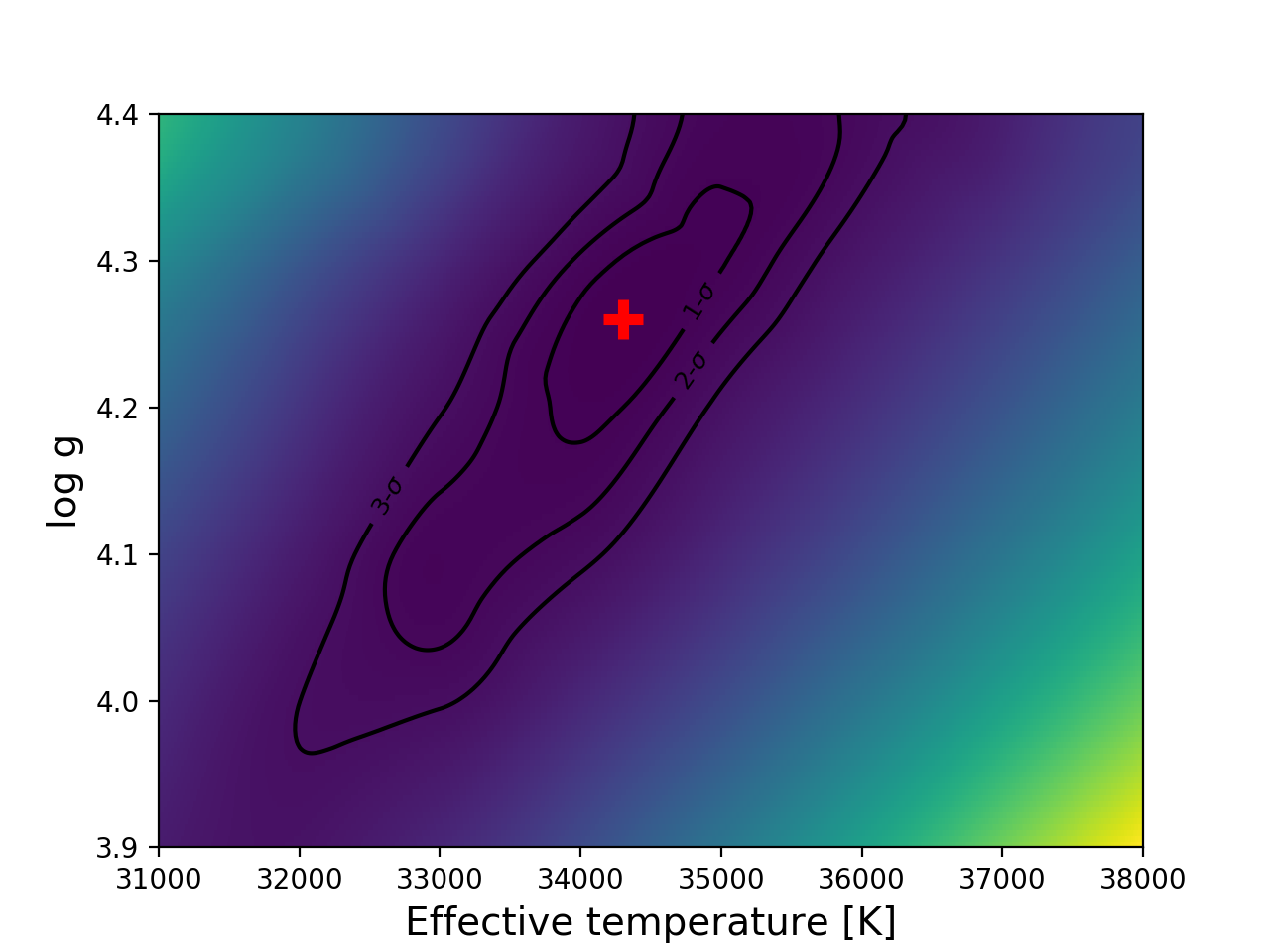}
    \includegraphics[width=6.cm]{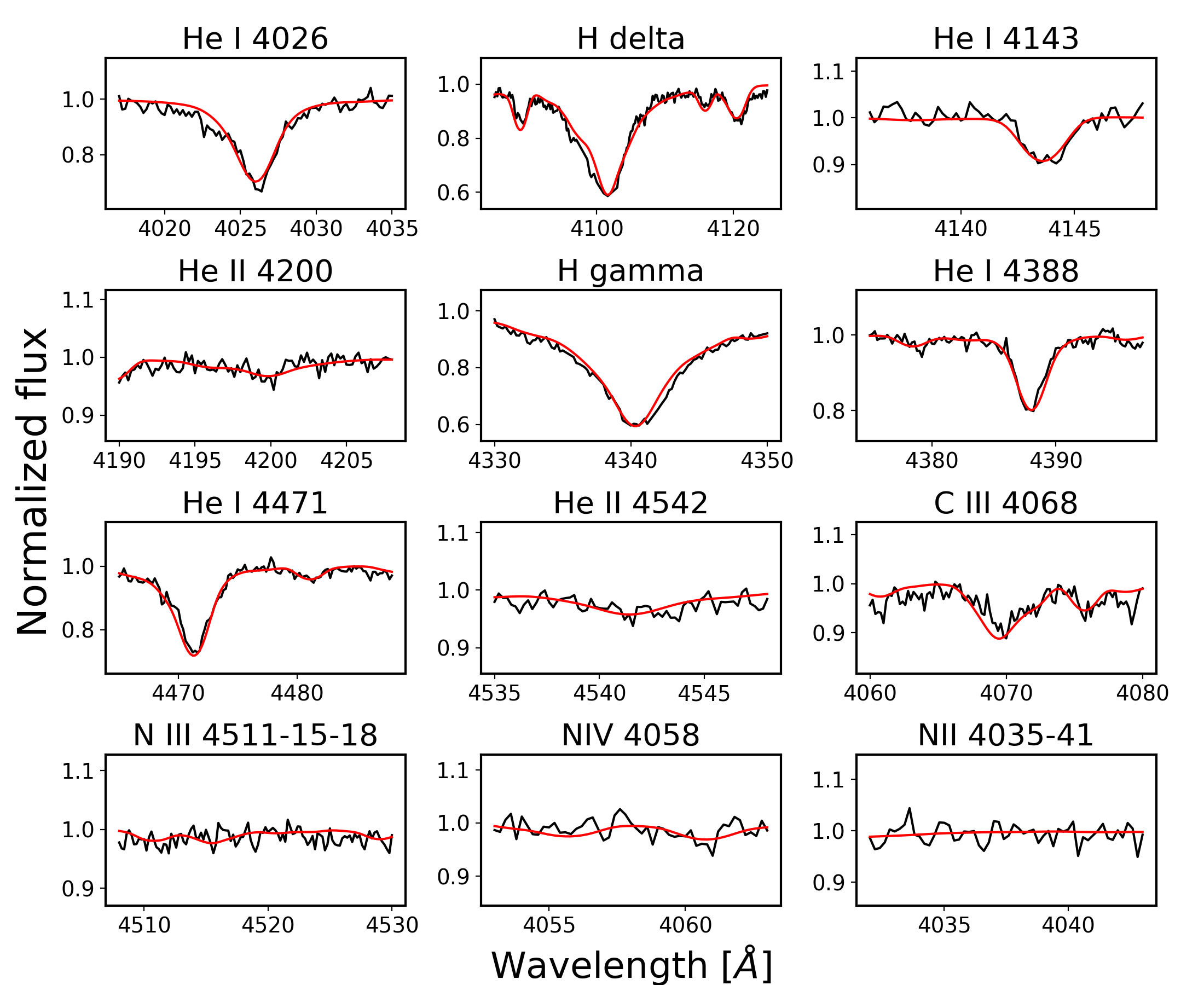}
    \includegraphics[width=7.cm, bb=5 0 453 346,clip]{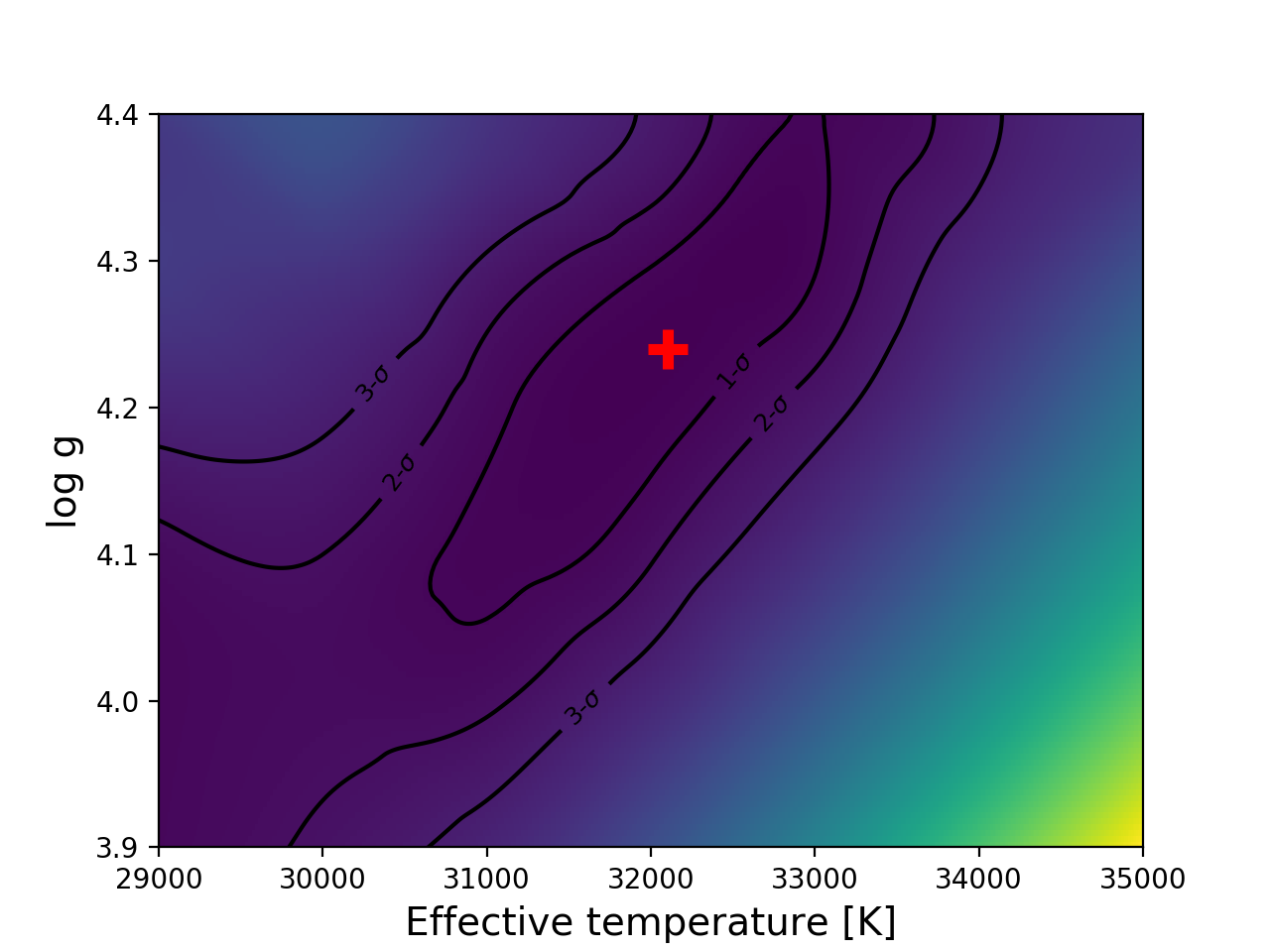}
    \includegraphics[width=7cm]{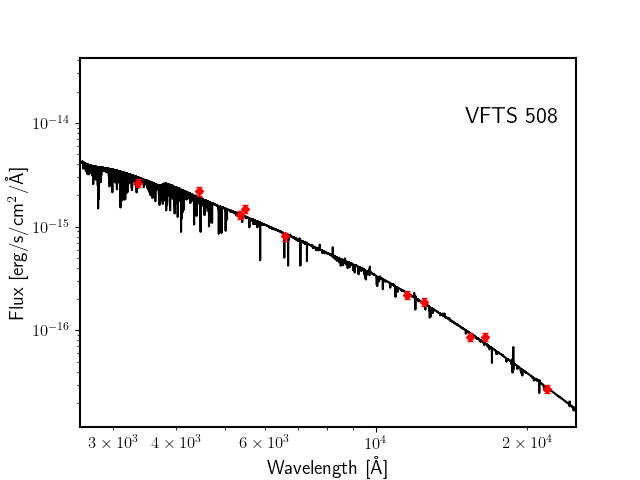}
    \includegraphics[width=6.5cm]{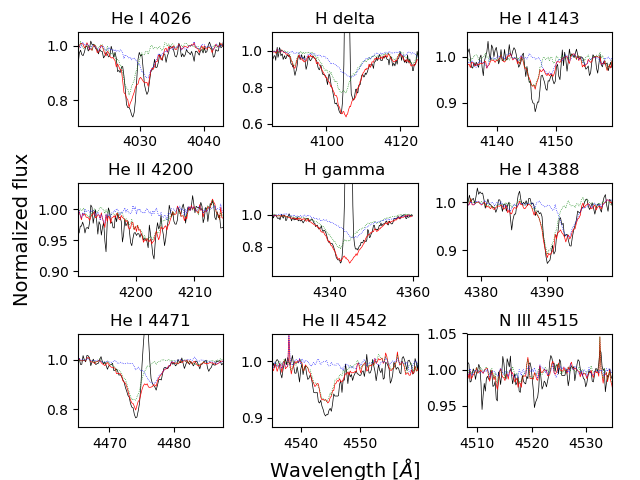}
    \includegraphics[width=7cm, bb=5 0 453 346,clip]{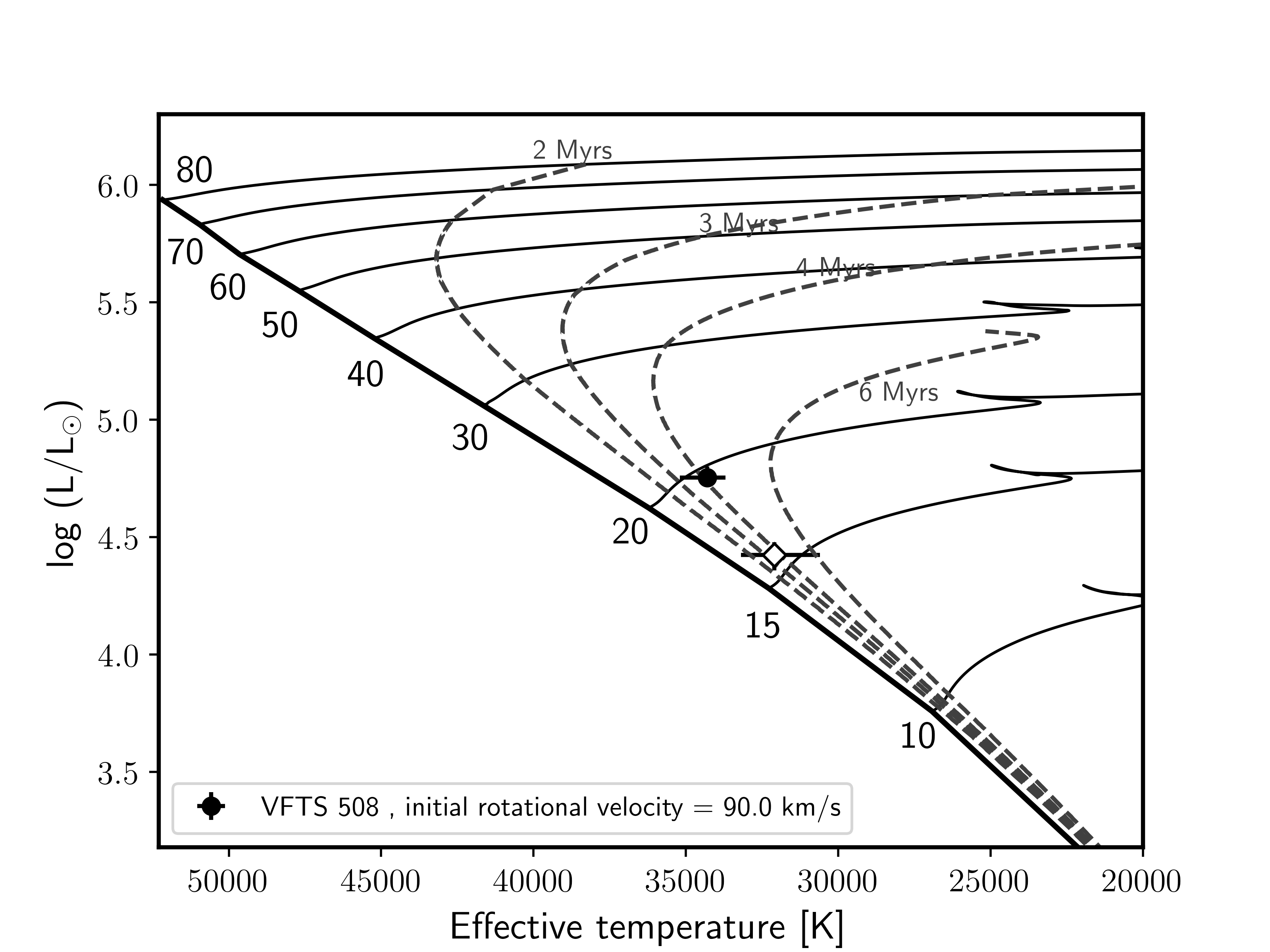}
    \includegraphics[width=7cm, bb=5 0 453 346,clip]{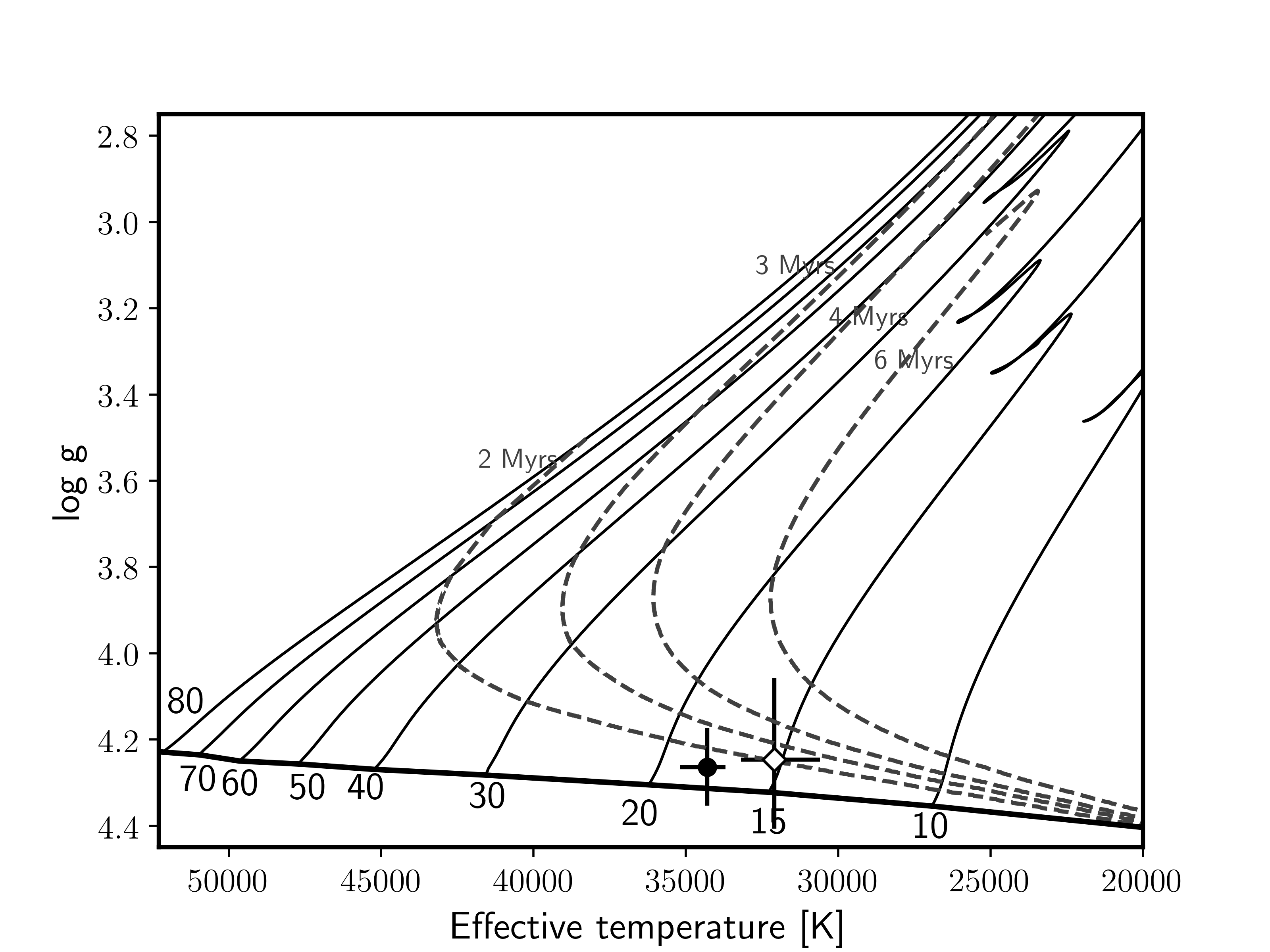}
    \caption{Same as Fig.\,\ref{fig:042} but for VFTS\,508.}\label{fig:508} 
  \end{figure*} 
\clearpage

 \begin{figure*}[t!]
    \centering
     \includegraphics[width=6.cm]{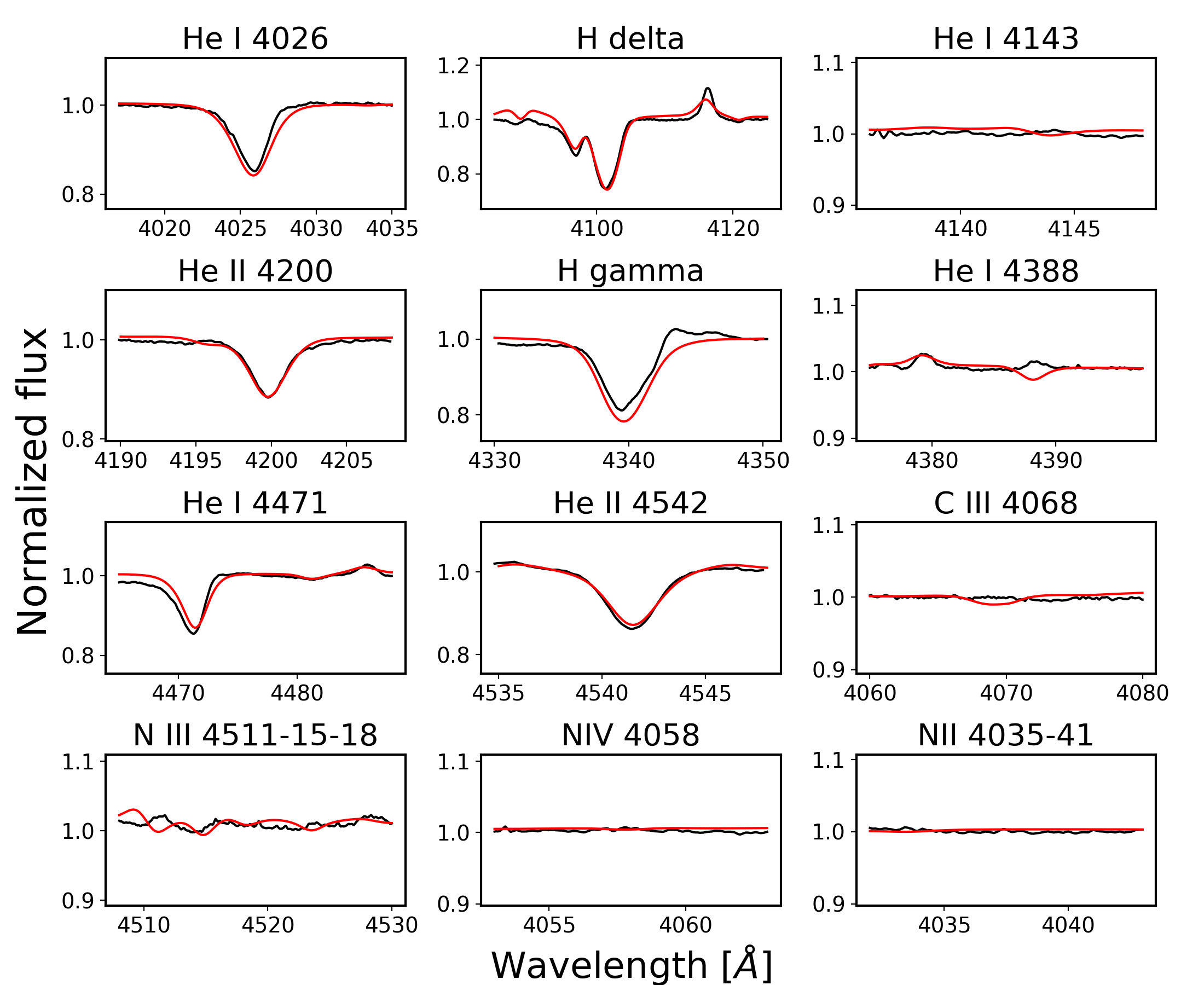}
    \includegraphics[width=6.cm]{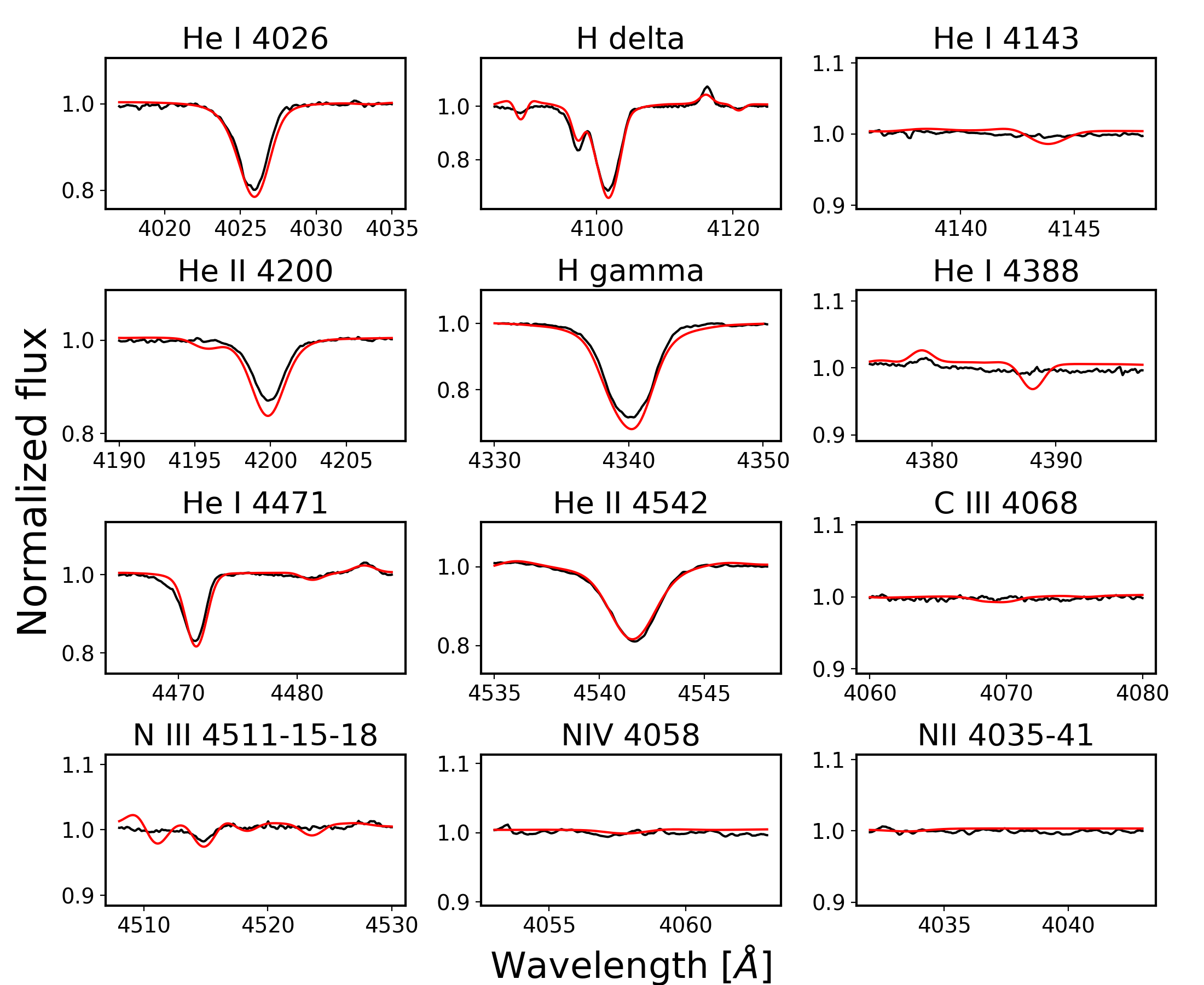}
    \includegraphics[width=7cm]{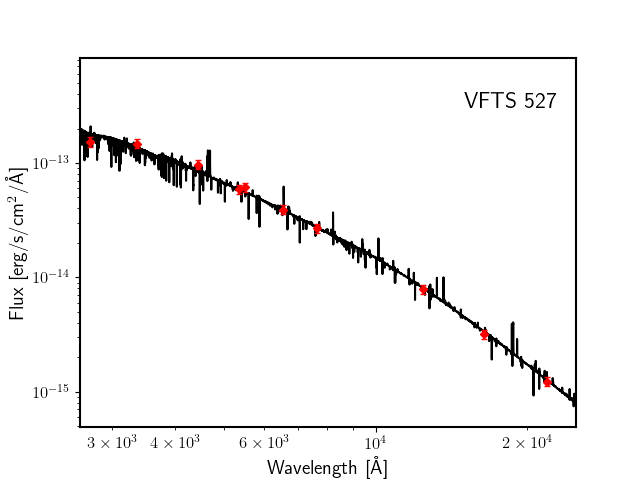}
    \includegraphics[width=6.5cm]{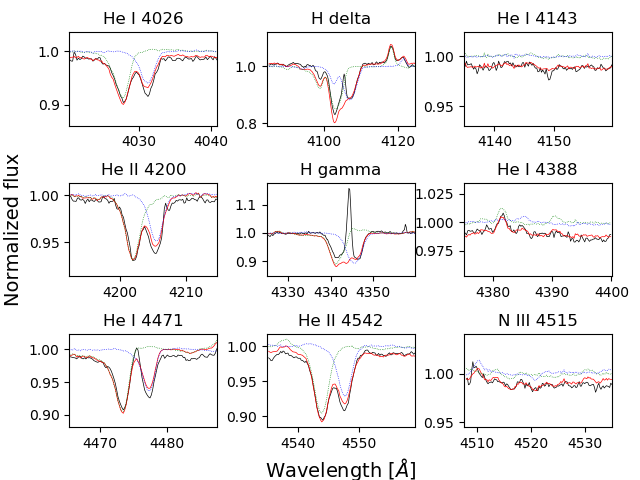}
    \includegraphics[width=7cm, bb=5 0 453 346,clip]{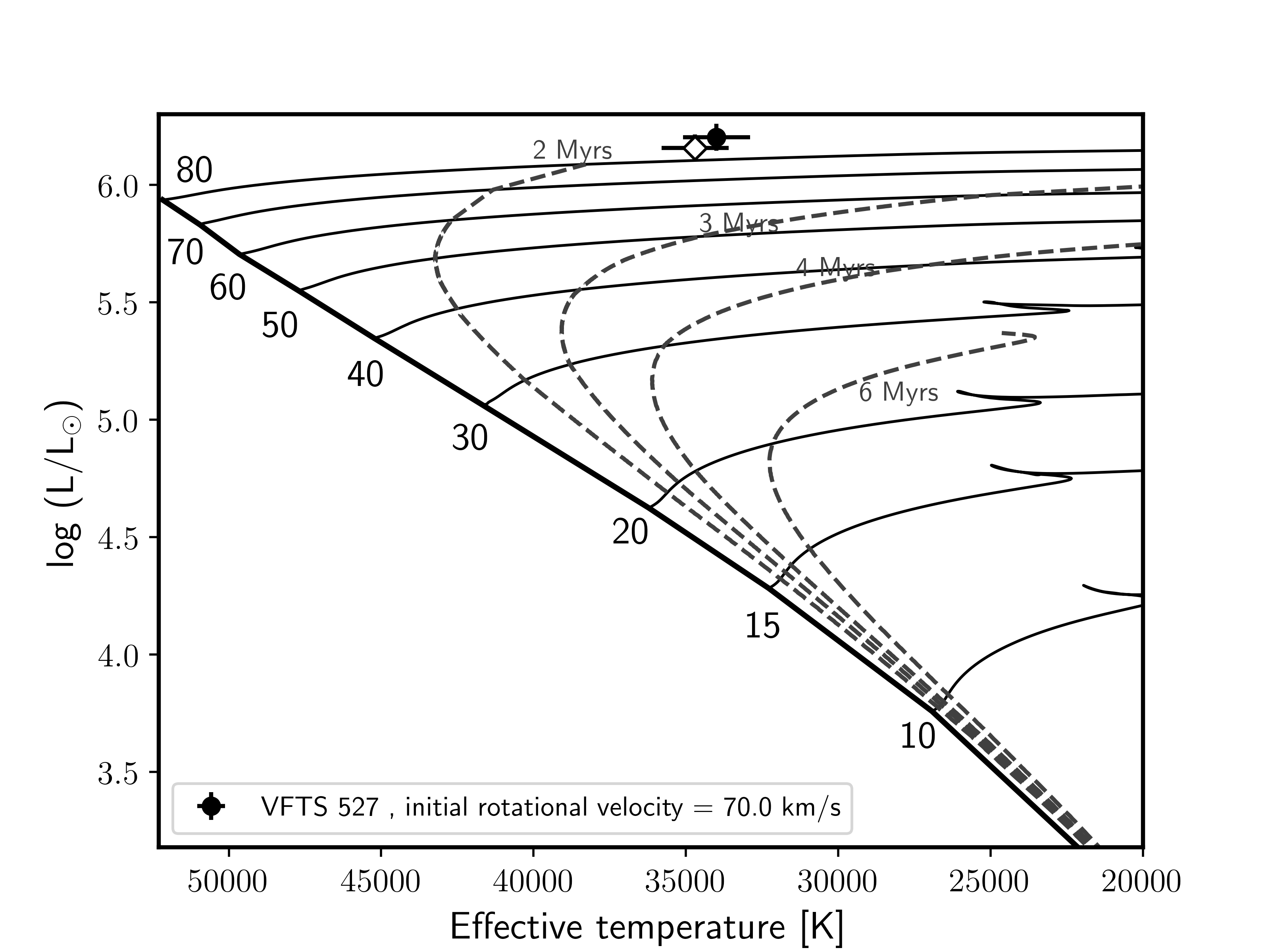}
    \includegraphics[width=7cm, bb=5 0 453 346,clip]{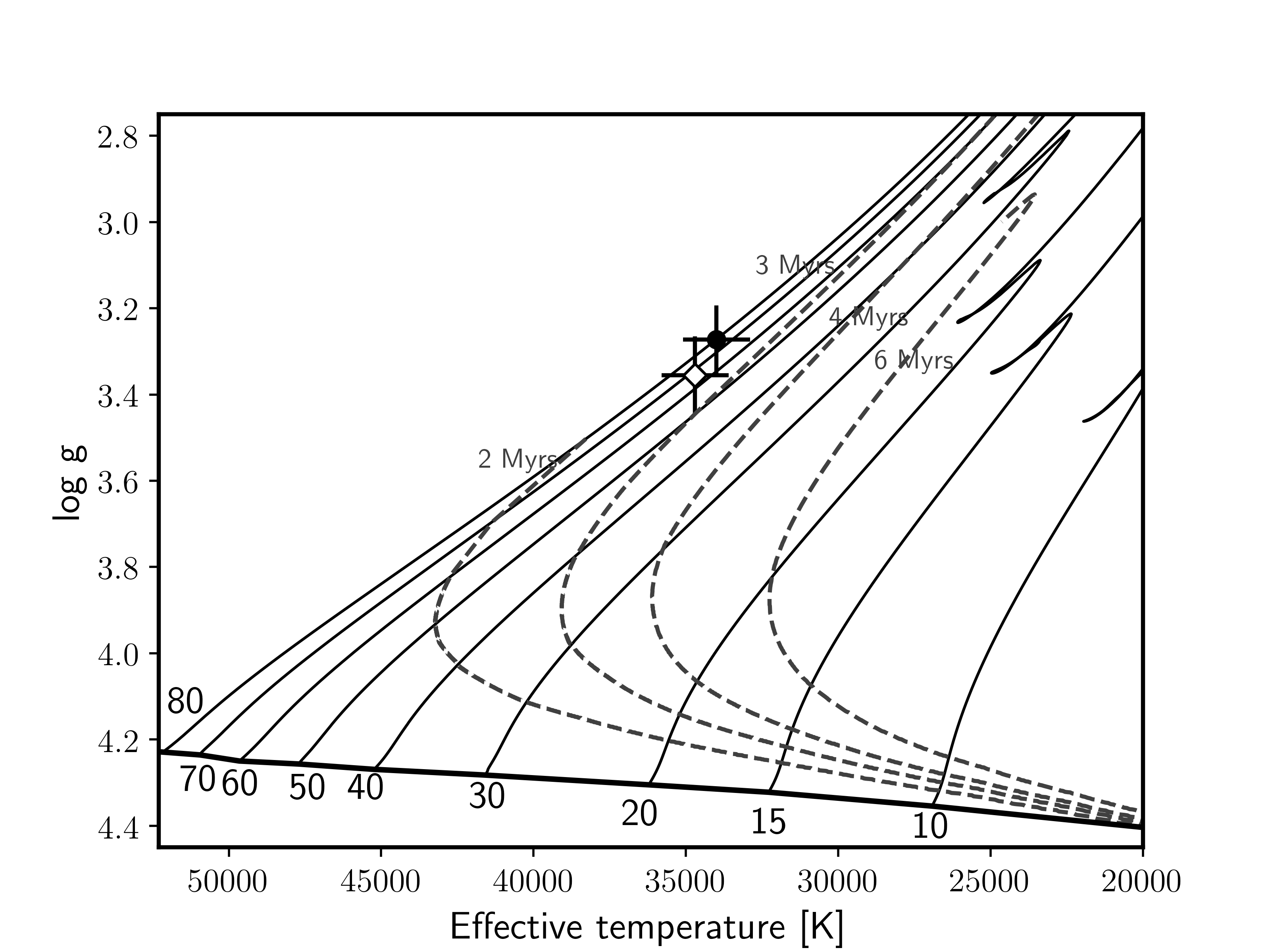}
    \caption{From top to bottom and left to right: (1) Best-fit model of VFTS\,527 primary (red line) compared with the disentangled spectrum (black line), (2)same as for (1) but for the secondary. (3) Spectral Energy Distribution of VFTS\,527. (4) Comparison between the disentangled spectra (scaled by the brightness factor of each component and shifted by their radial velocities) and one observed spectrum (5) Individual HRD (Left) and (6) $\logg - \teff$ diagram of the stars of VFTS\,527.} \label{fig:527} 
  \end{figure*} 
 \clearpage

 \begin{figure*}[t!]
    \centering
    \includegraphics[width=6.cm]{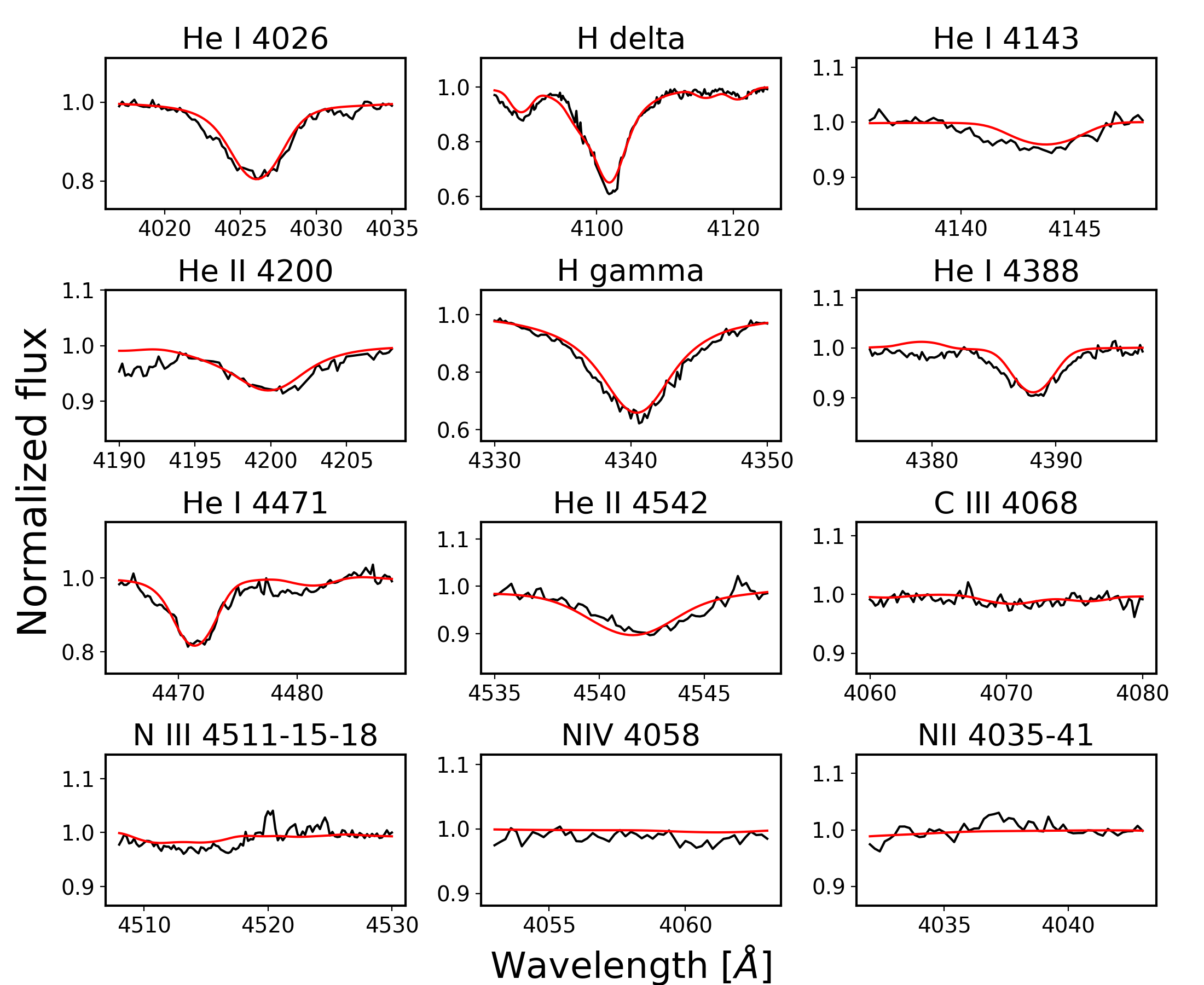}
    \includegraphics[width=7.cm, bb=5 0 453 346,clip]{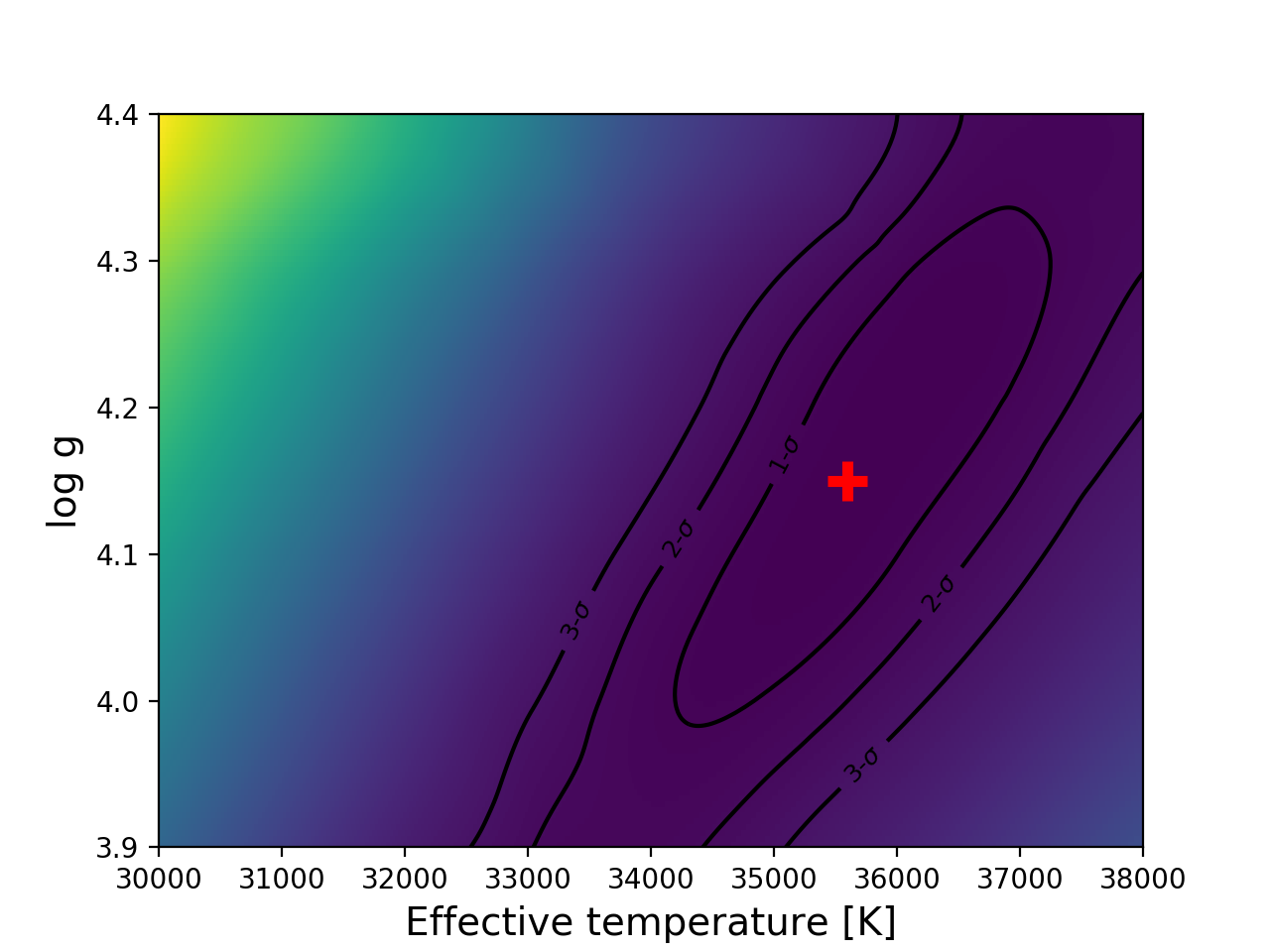}
    \includegraphics[width=6.cm]{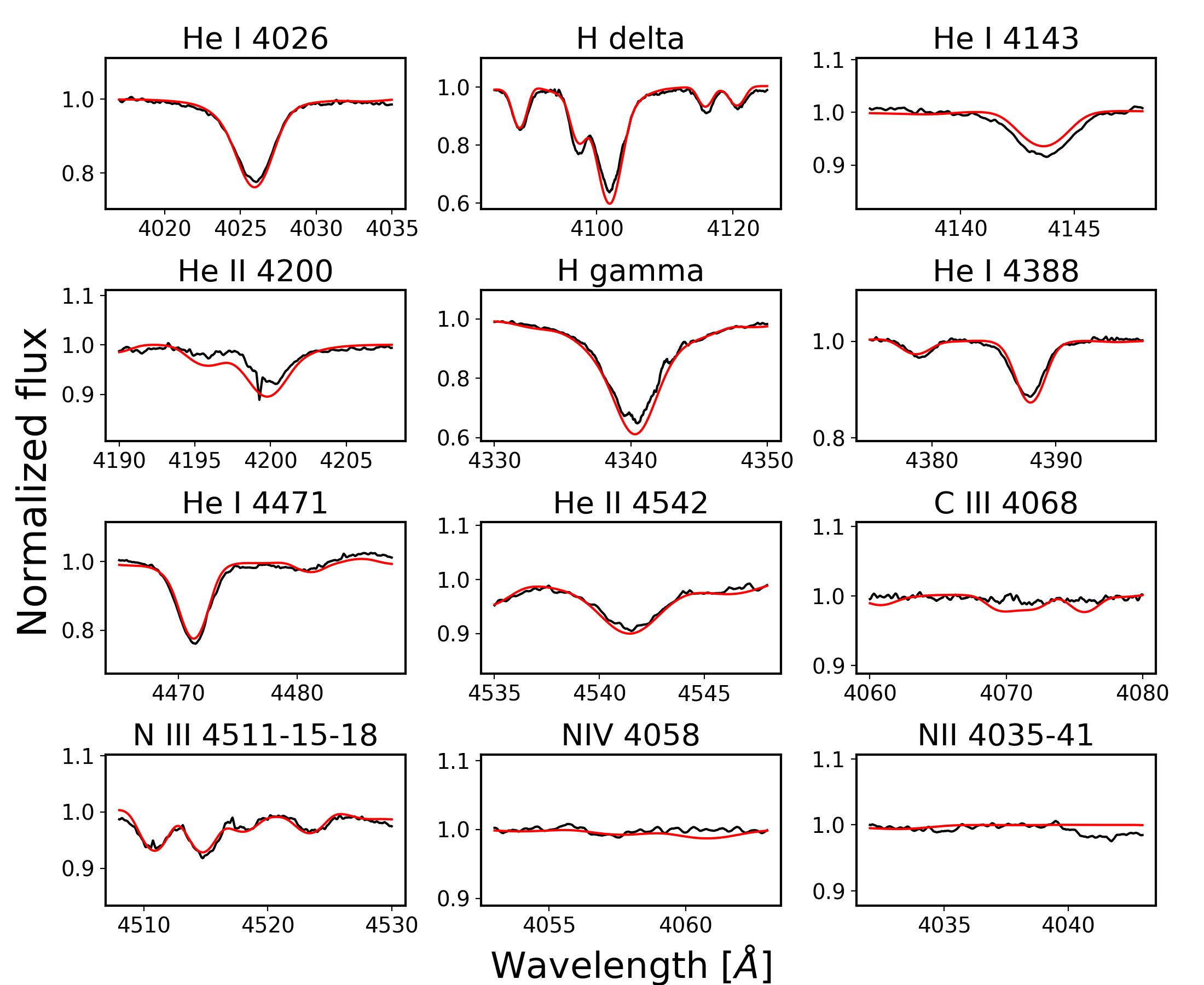}
    \includegraphics[width=7.cm, bb=5 0 453 346,clip]{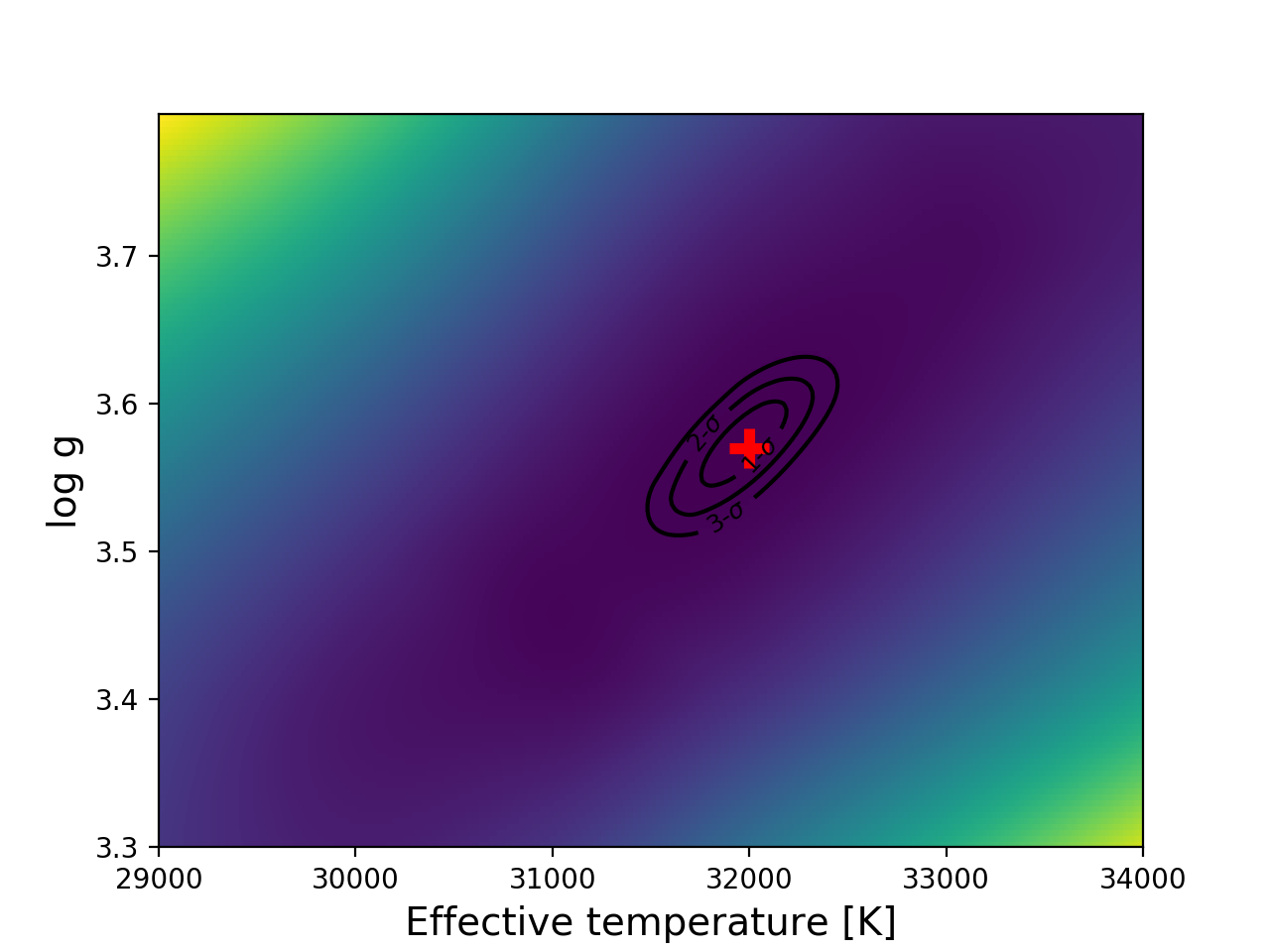}
    \includegraphics[width=7cm]{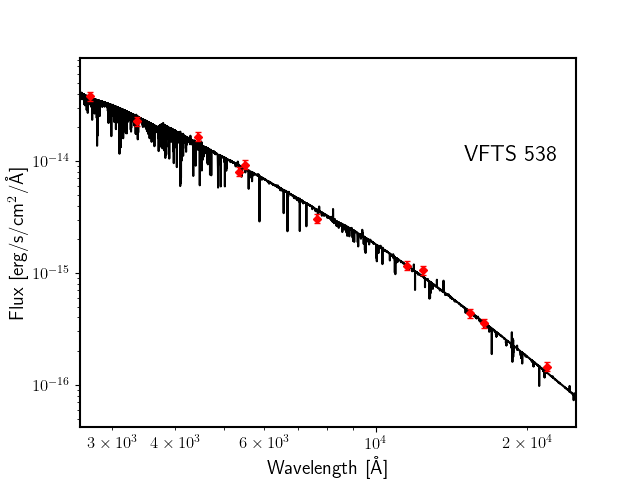}
    \includegraphics[width=6.5cm]{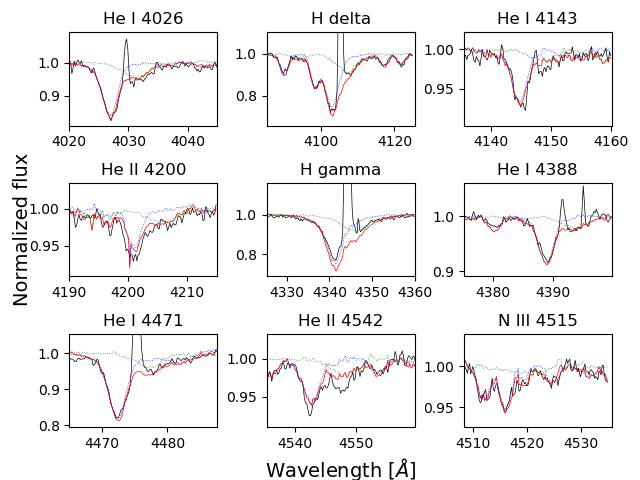}
    \includegraphics[width=7cm, bb=5 0 453 346,clip]{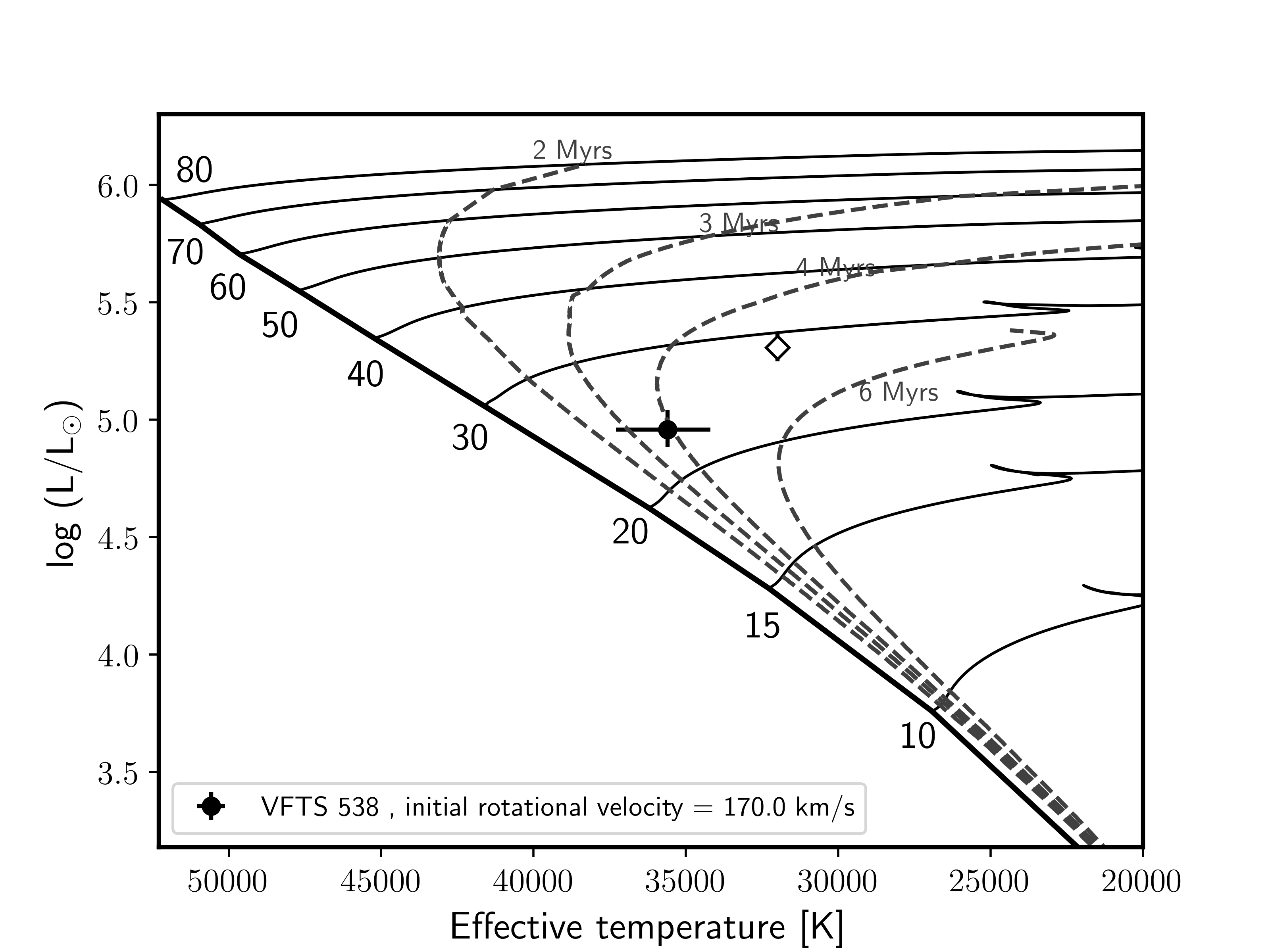}
    \includegraphics[width=7cm, bb=5 0 453 346,clip]{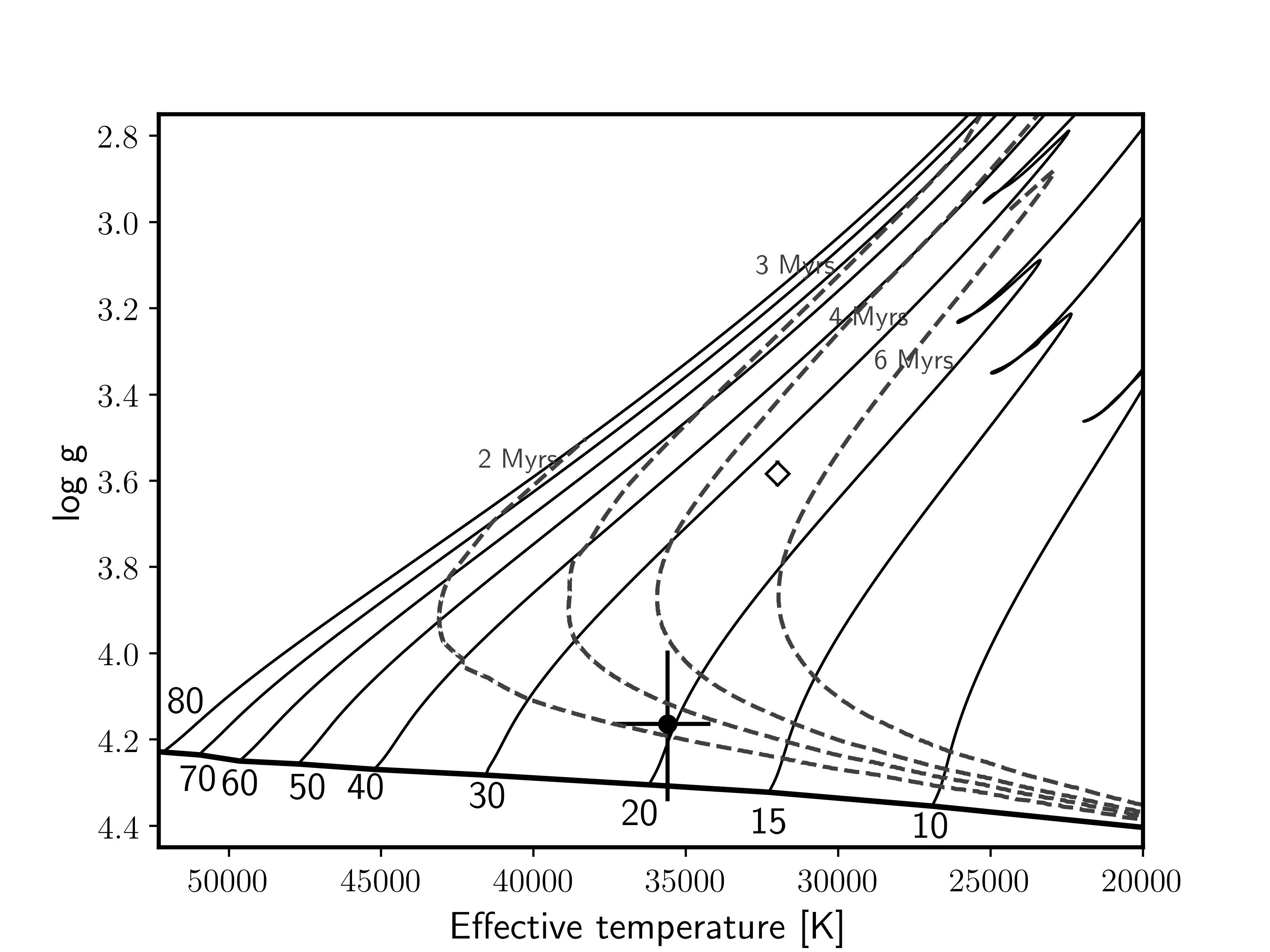}
    \caption{Same as Fig.\,\ref{fig:042} but for VFTS\,538.} \label{fig:538} 
  \end{figure*} 
 \clearpage
           
 \begin{figure*}[t!]
    \centering
    \includegraphics[width=6.cm]{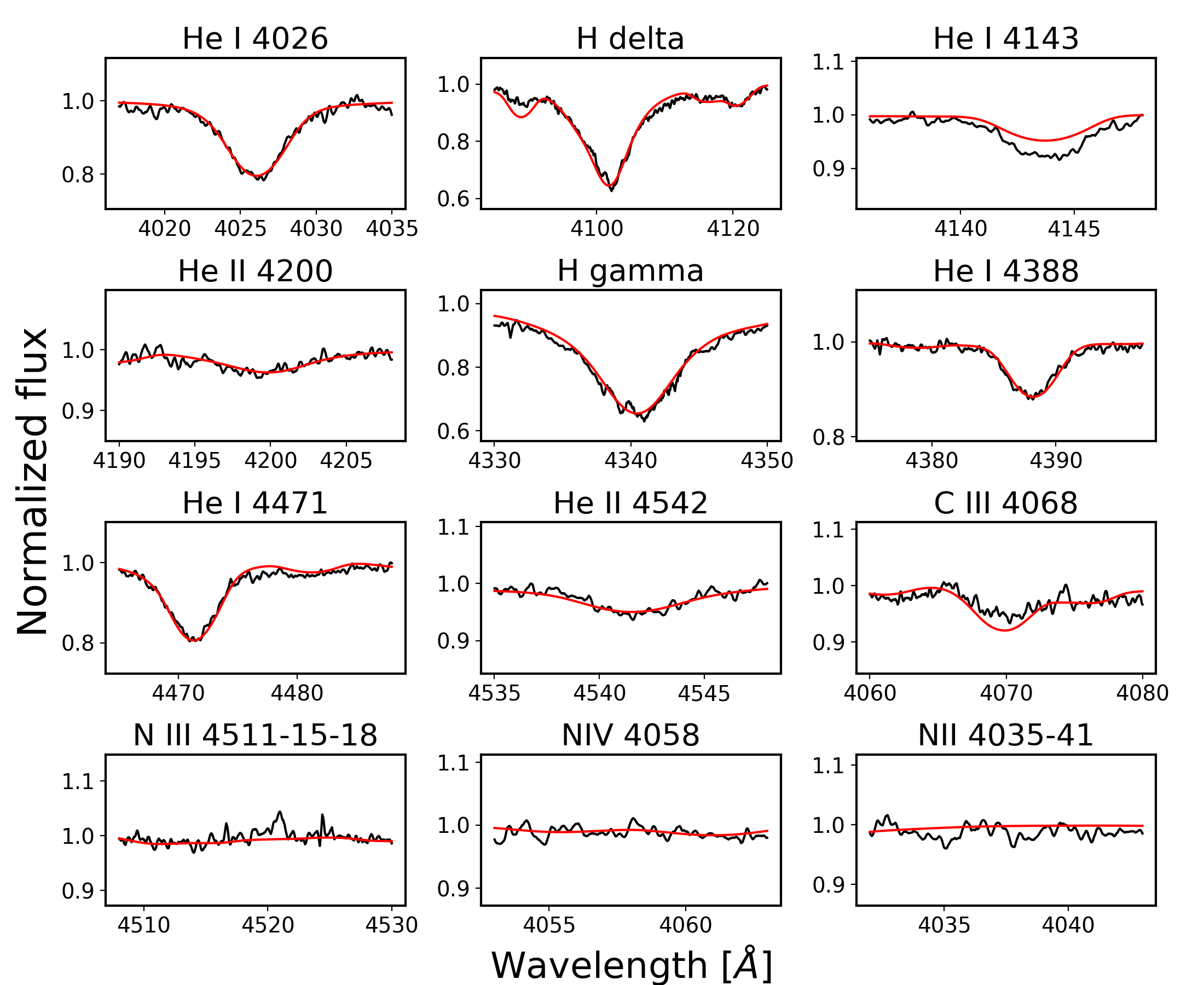}
    \includegraphics[width=7.cm, bb=5 0 453 346,clip]{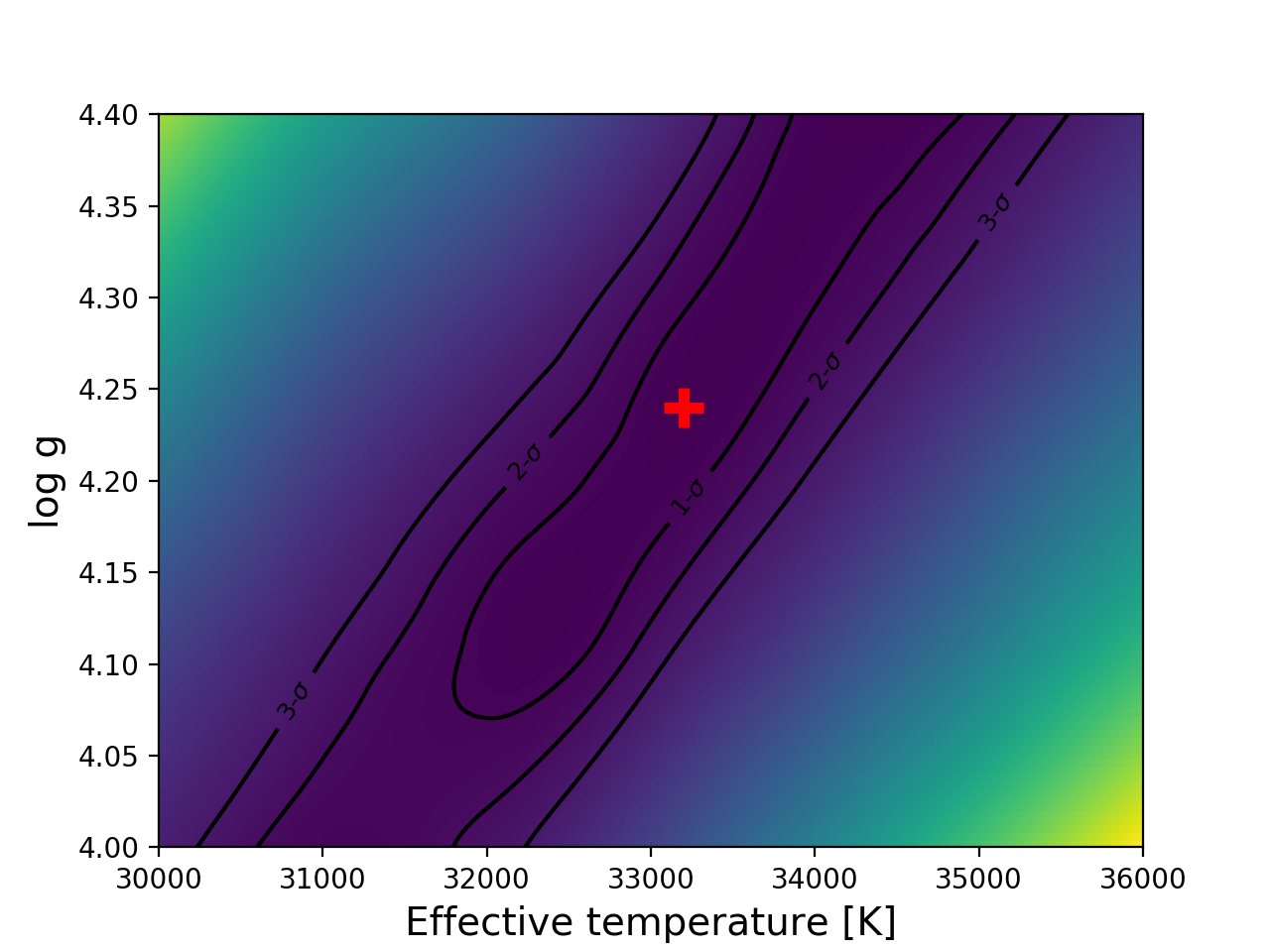}
    \includegraphics[width=6.cm]{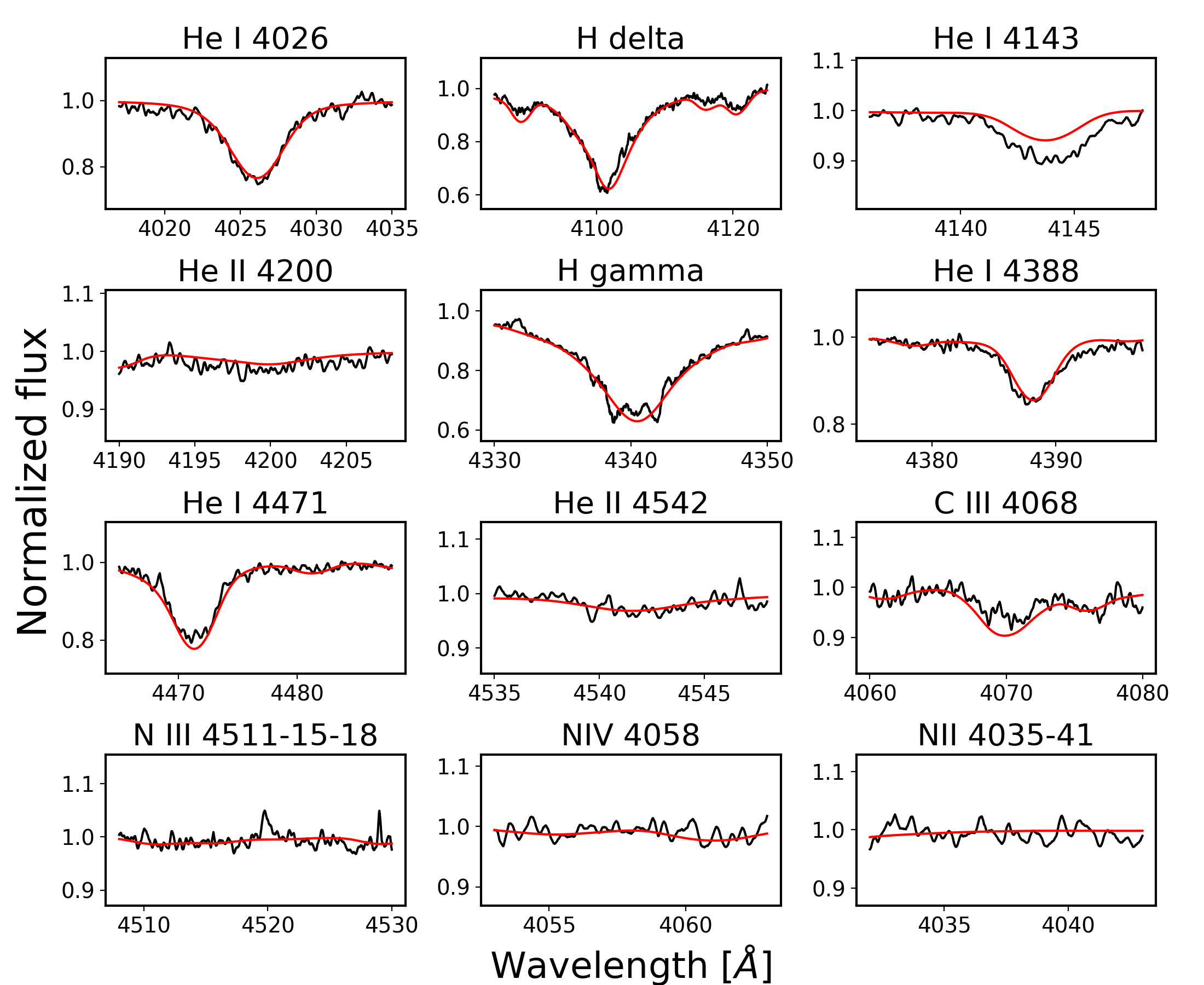}
    \includegraphics[width=7.cm, bb=5 0 453 346,clip]{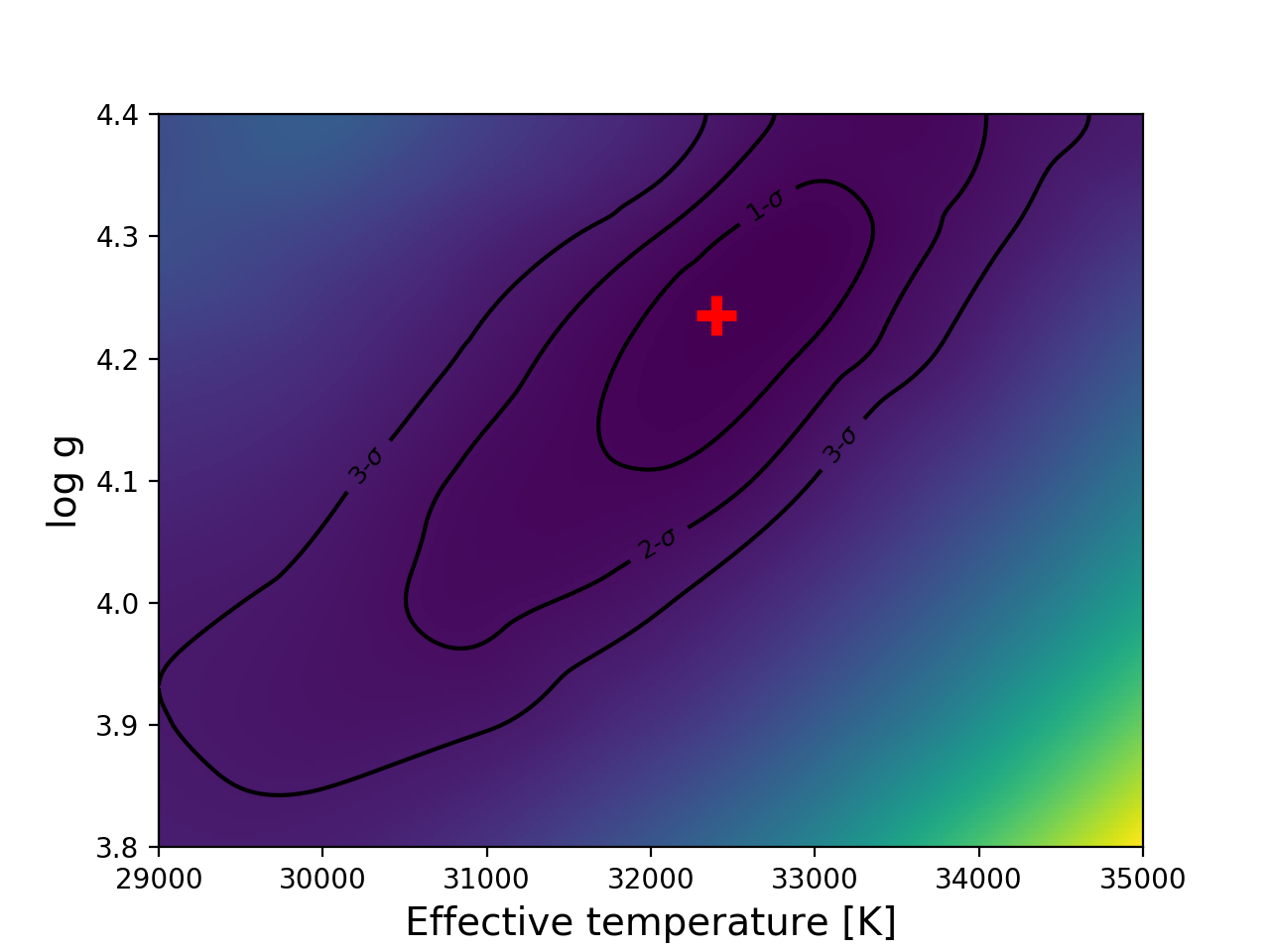}
    \includegraphics[width=7cm]{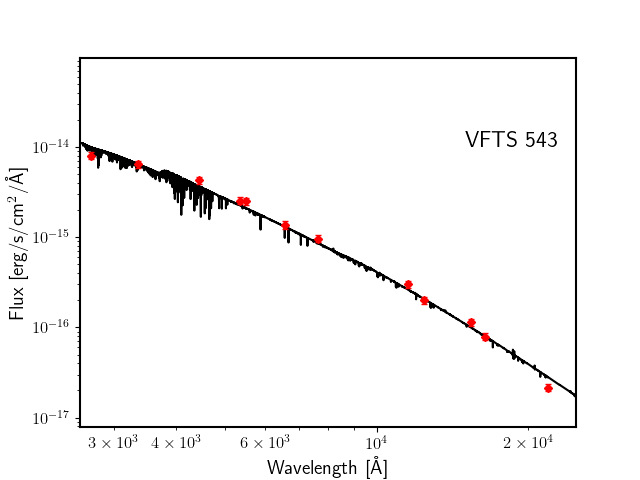}
    \includegraphics[width=6.5cm]{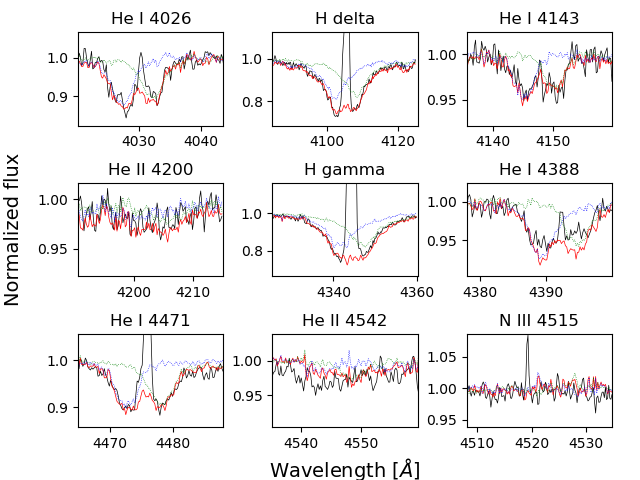}
    \includegraphics[width=7cm, bb=5 0 453 346,clip]{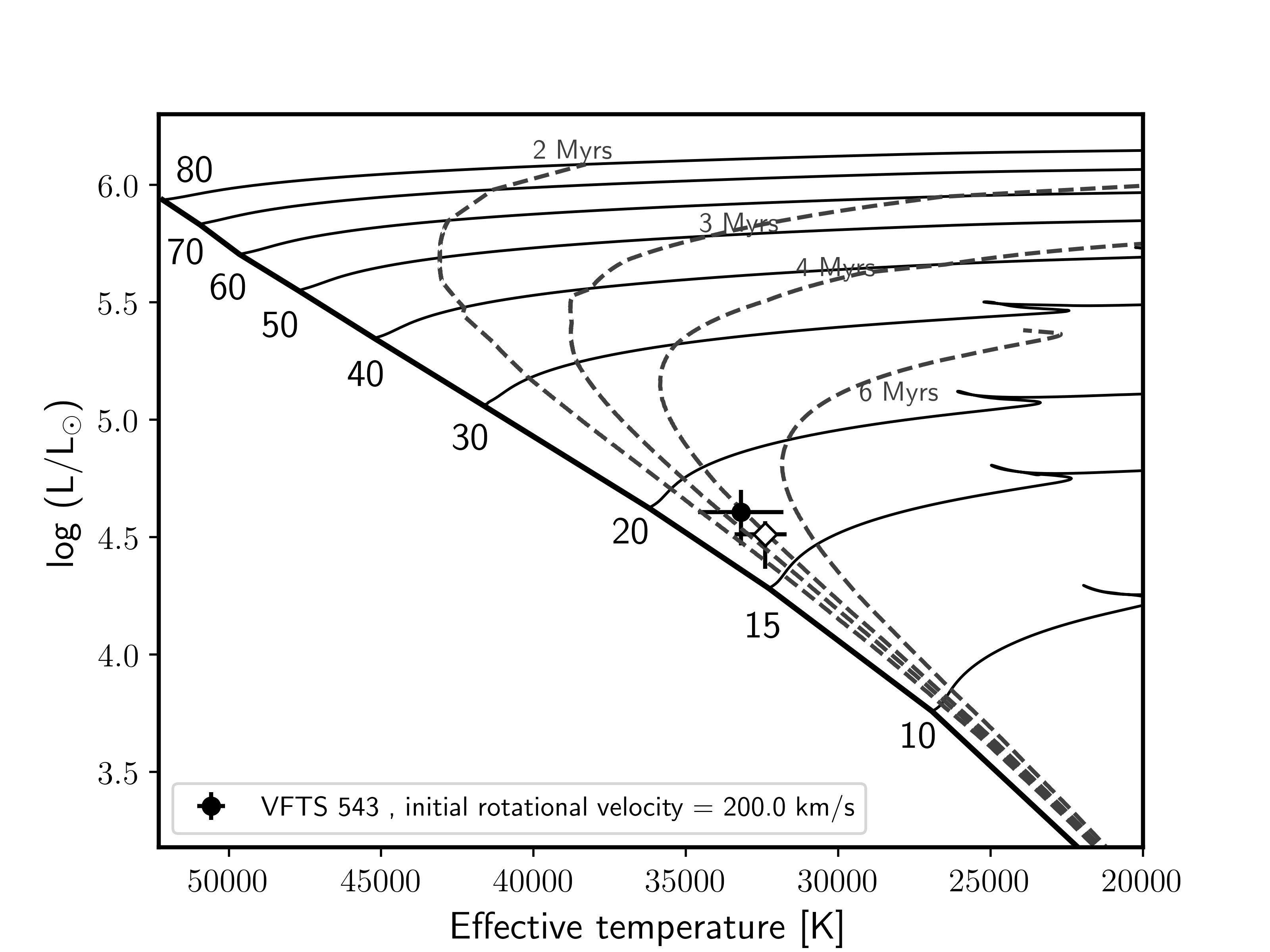}
    \includegraphics[width=7cm, bb=5 0 453 346,clip]{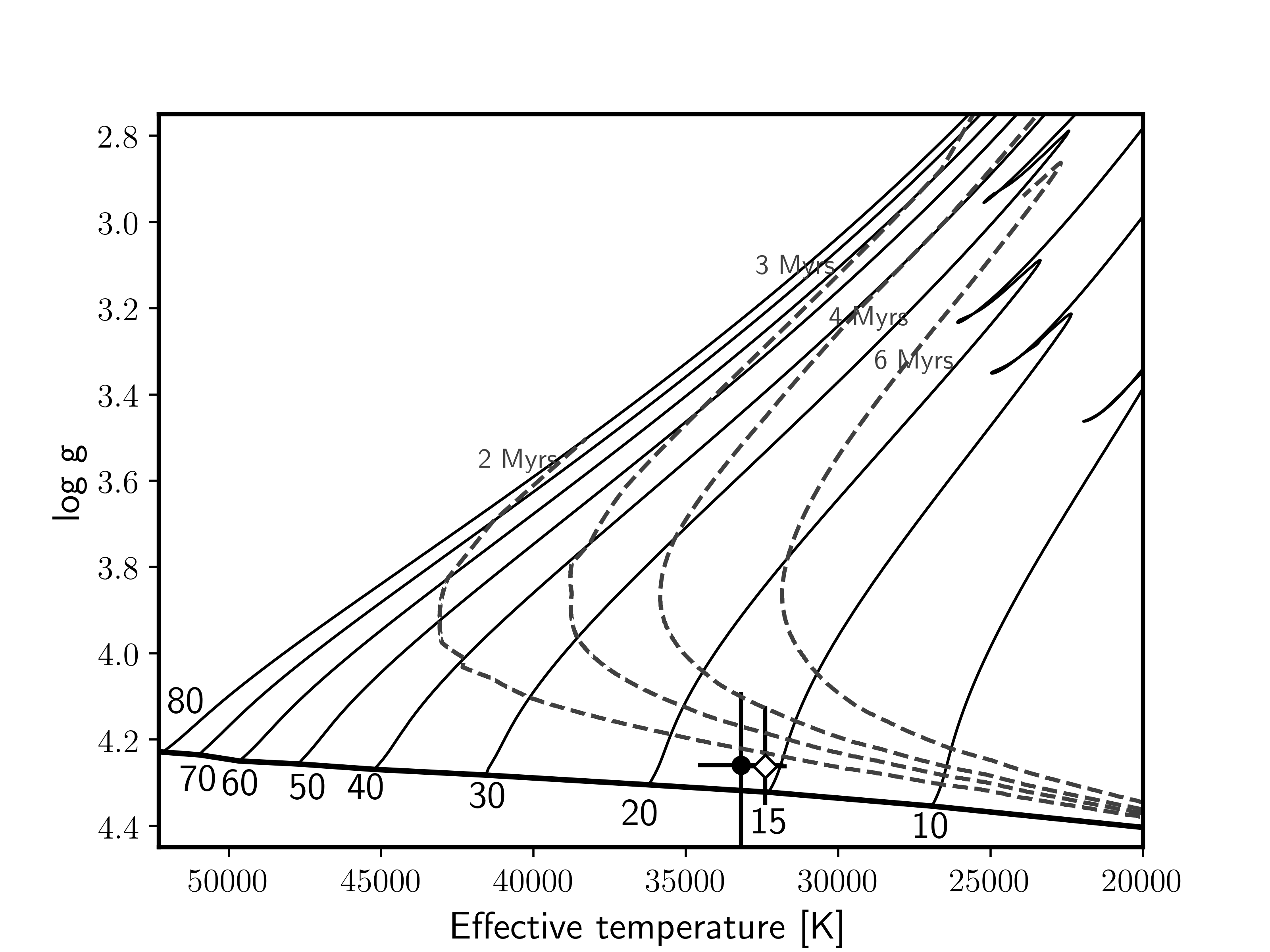}
    \caption{Same as Fig.\,\ref{fig:042} but for VFTS\,543.} \label{fig:543} 
  \end{figure*} 
 \clearpage
       
 \begin{figure*}[t!]
    \centering
    \includegraphics[width=6.cm]{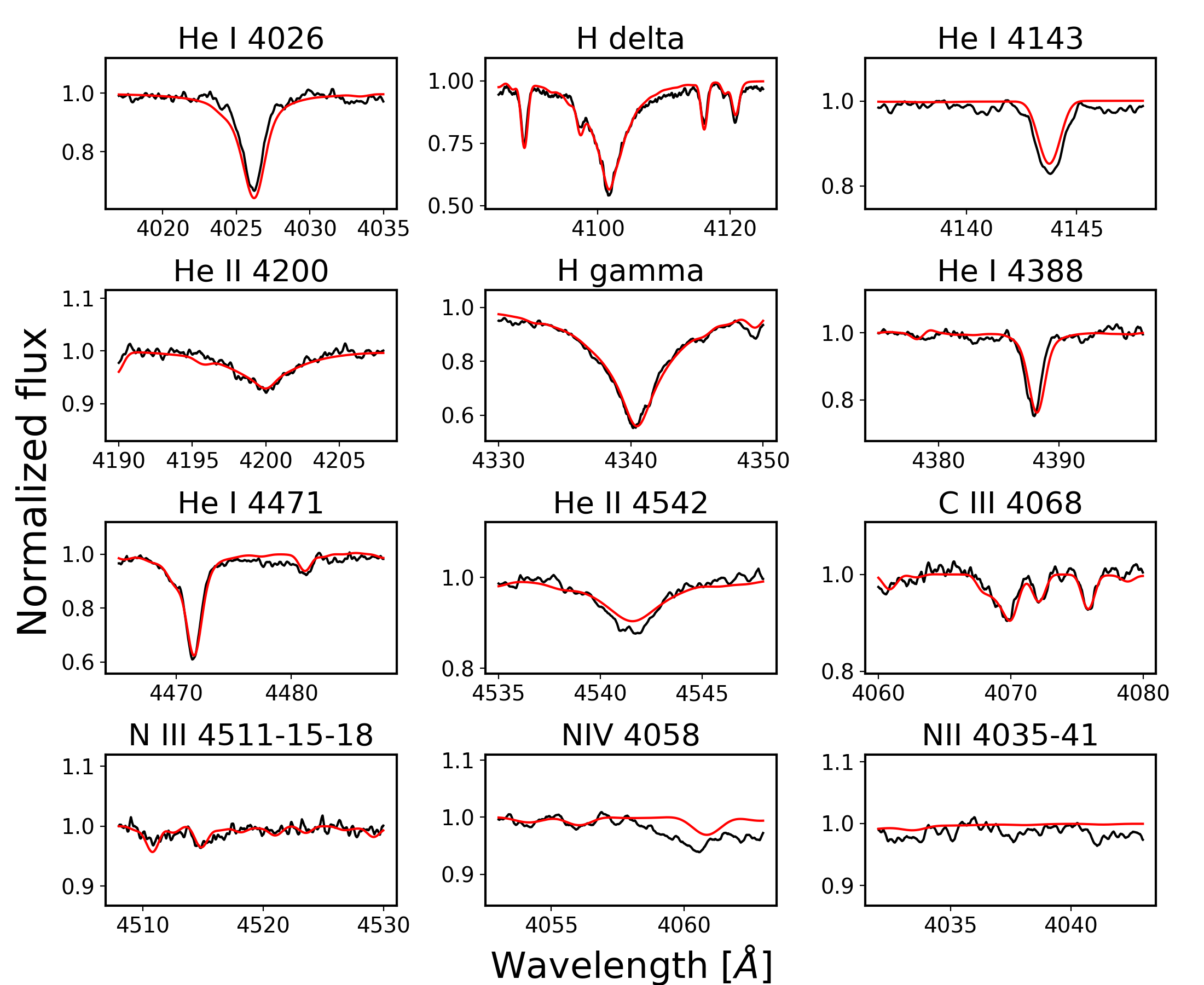}
    \includegraphics[width=7.cm, bb=5 0 453 346,clip]{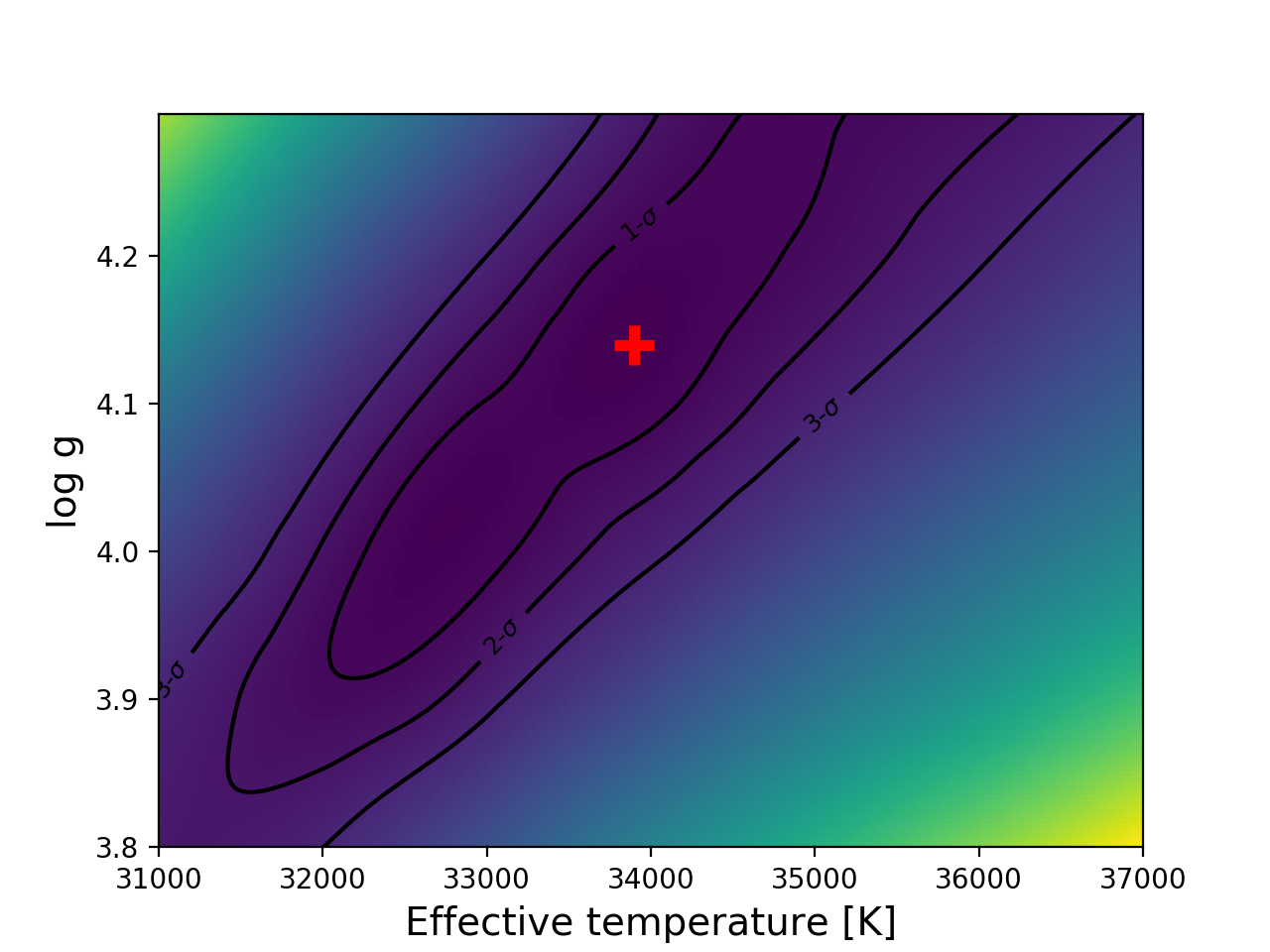}
    \includegraphics[width=6.cm]{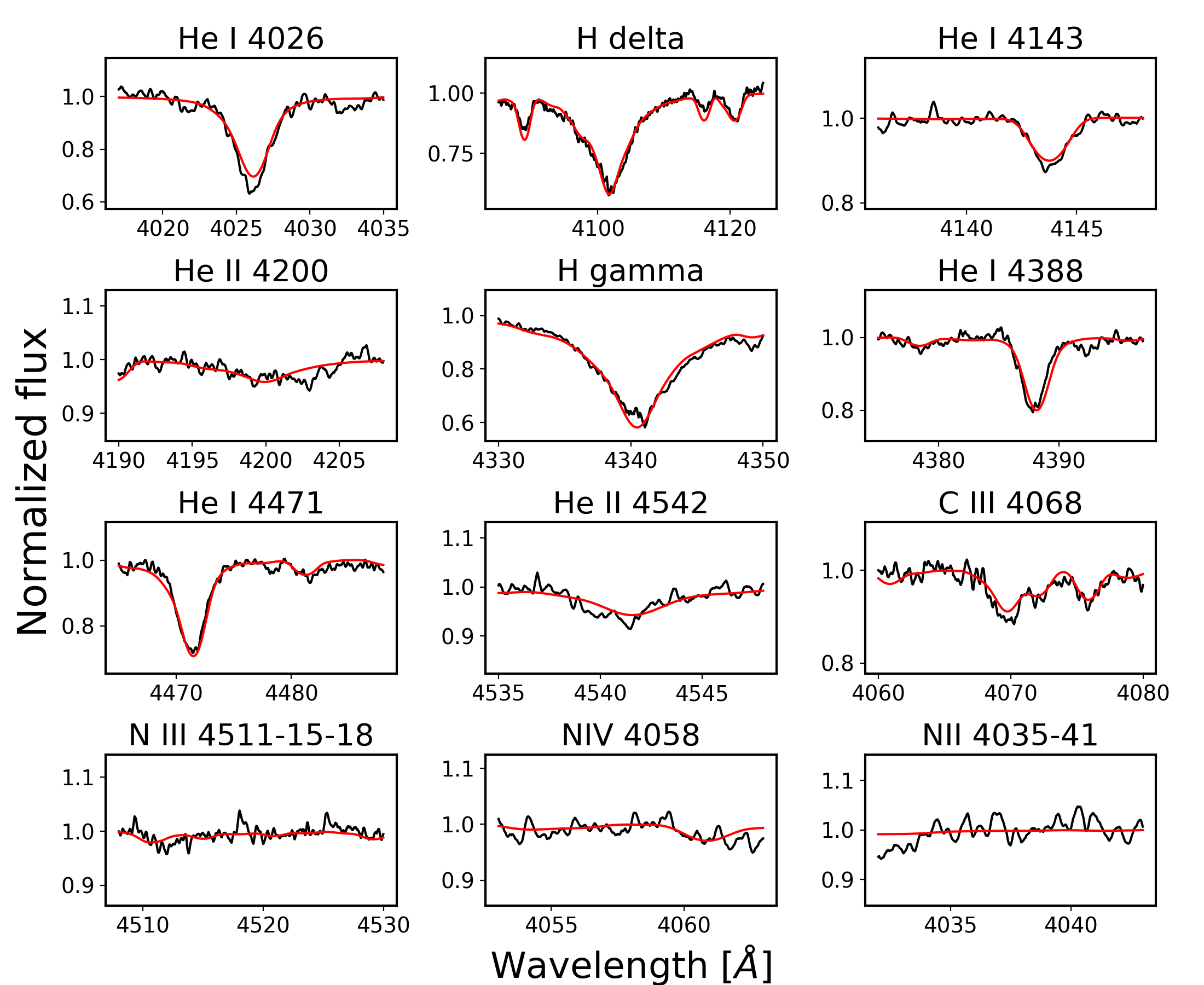}
    \includegraphics[width=7.cm, bb=5 0 453 346,clip]{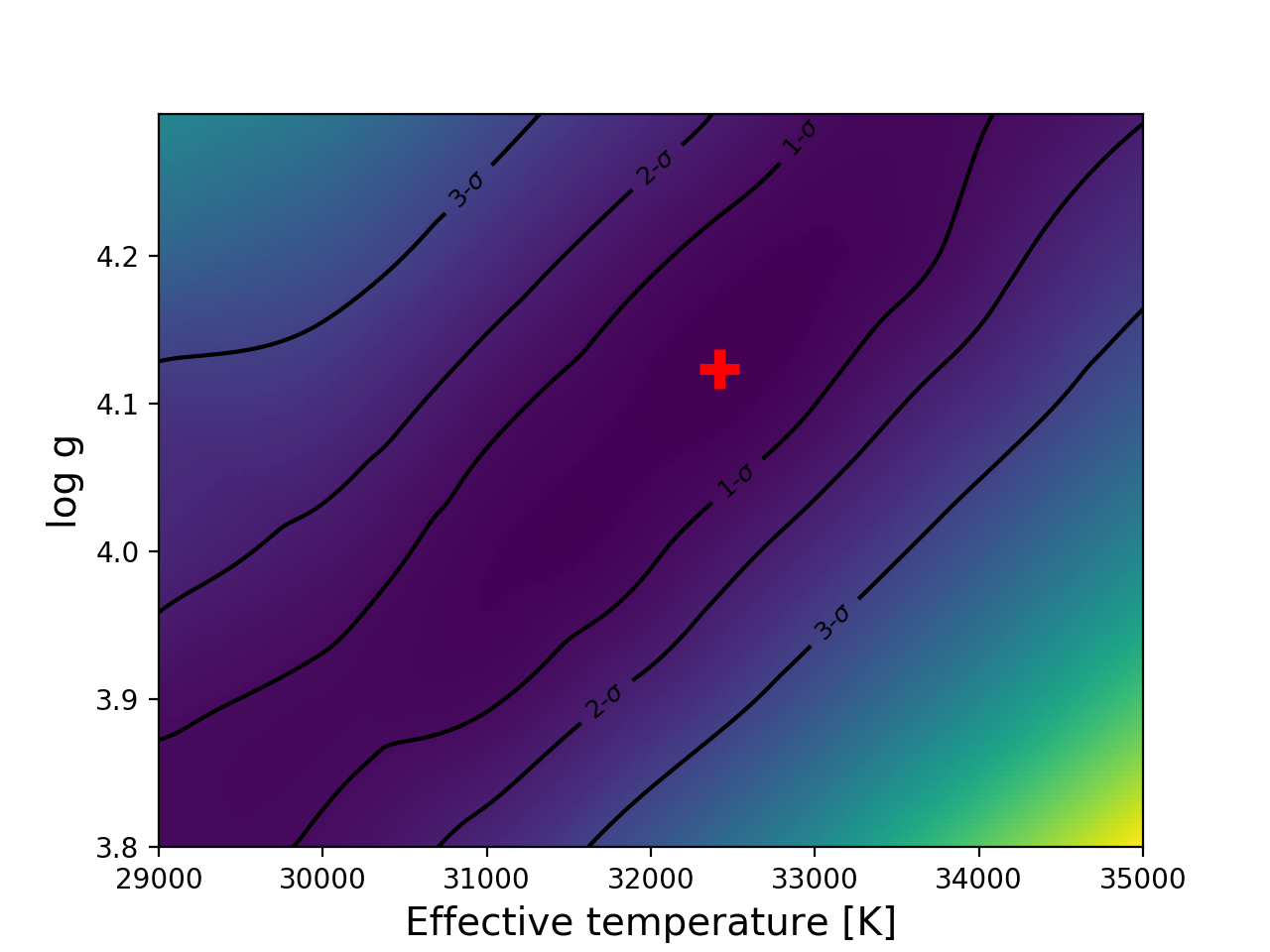}
    \includegraphics[width=7cm]{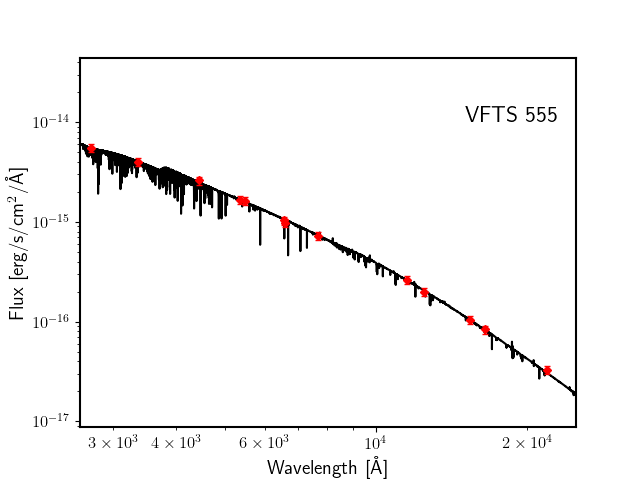}
    \includegraphics[width=6.5cm]{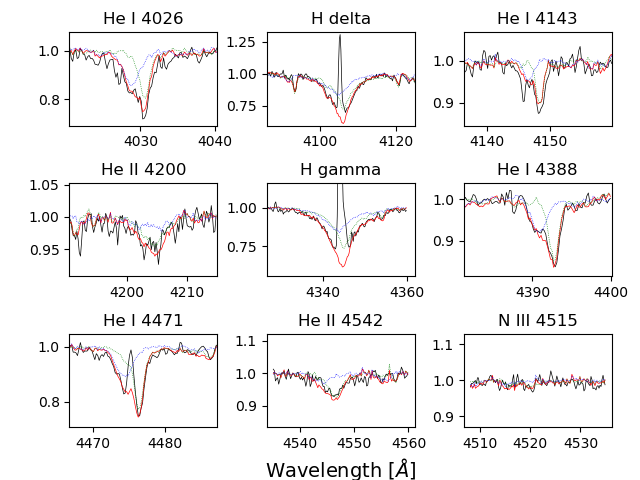}
    \includegraphics[width=7cm, bb=5 0 453 346,clip]{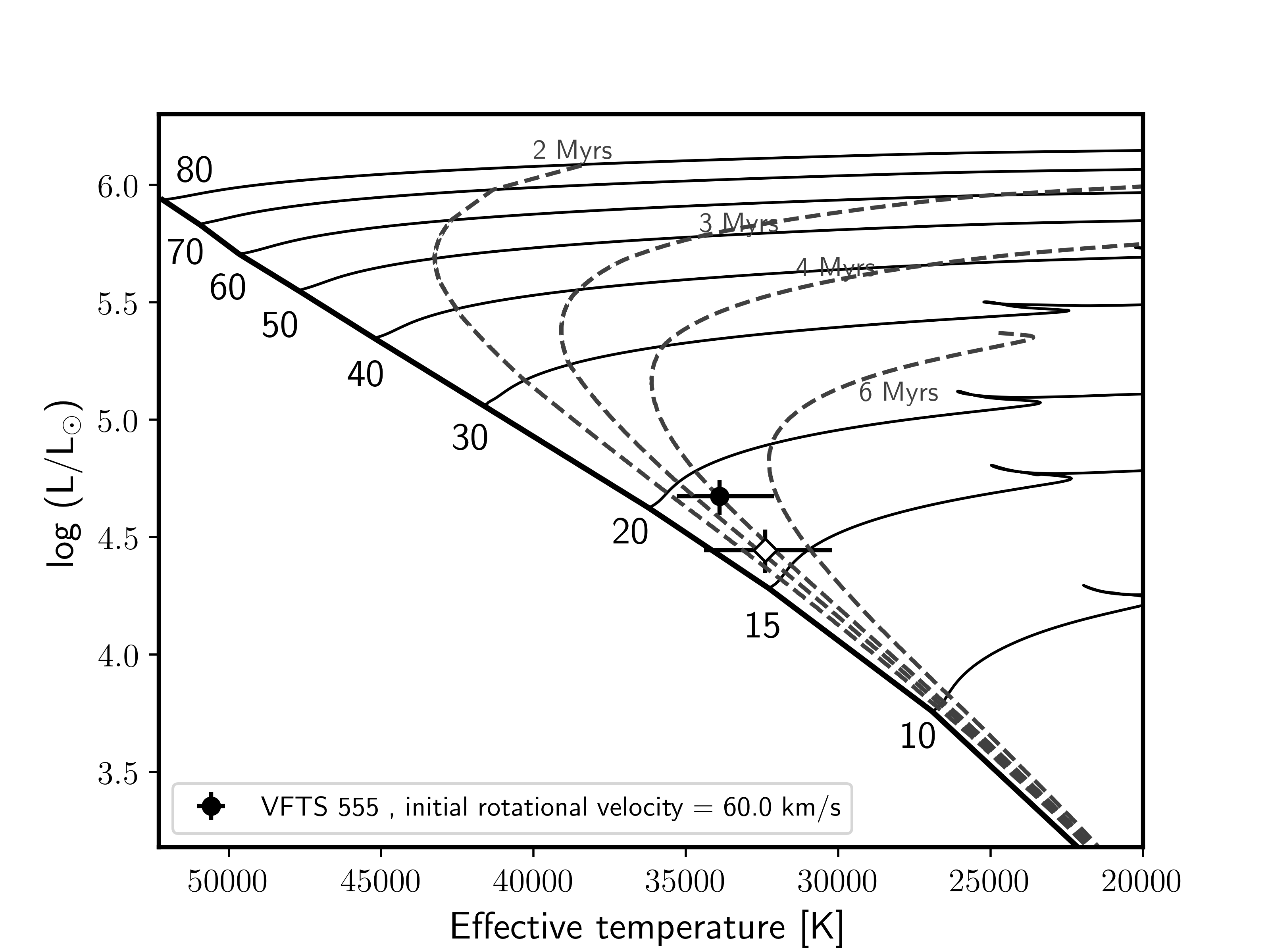}
    \includegraphics[width=7cm, bb=5 0 453 346,clip]{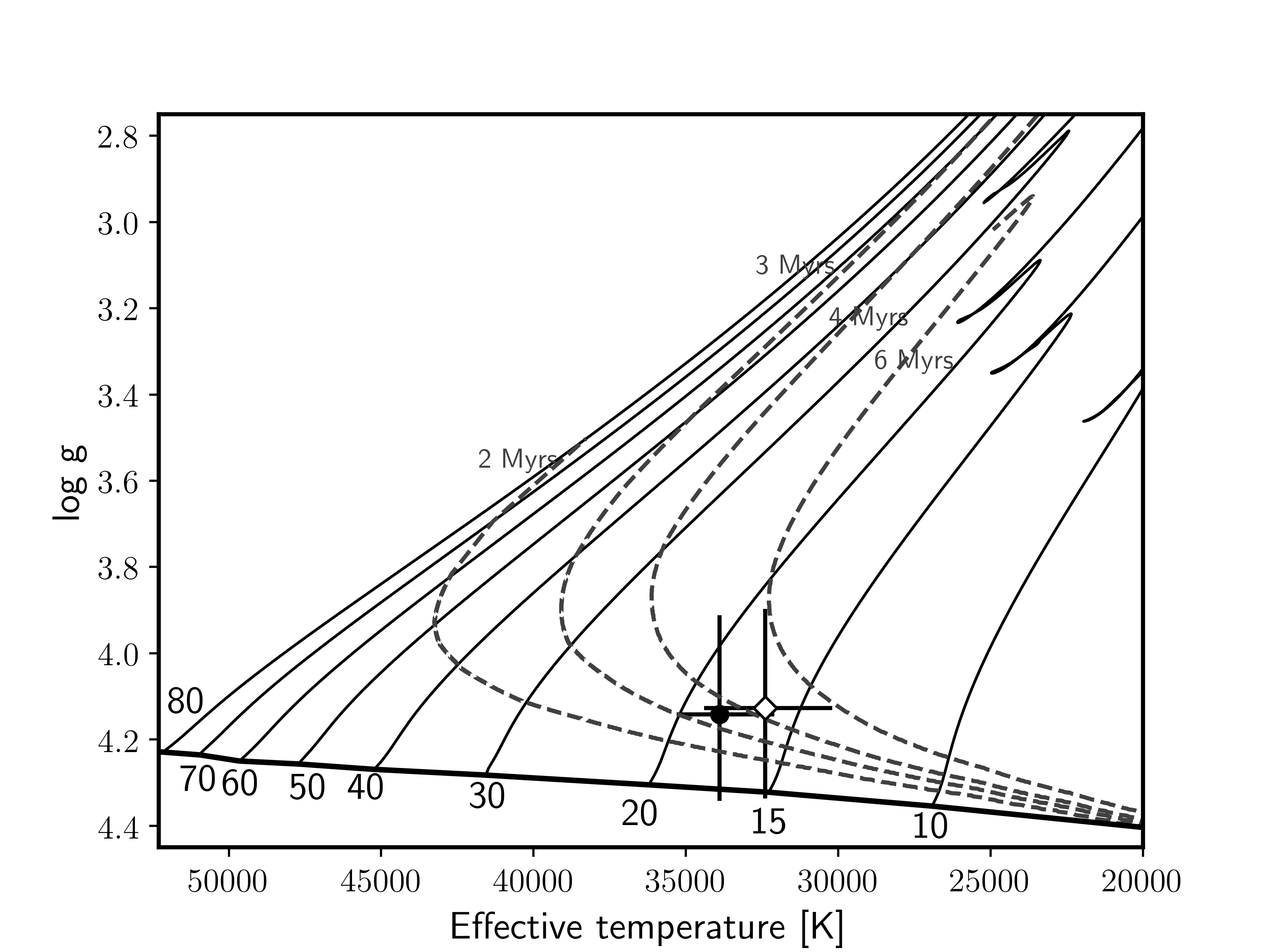}
    \caption{Same as Fig.\,\ref{fig:042} but for VFTS\,555.} \label{fig:555} 
  \end{figure*}  
  \clearpage
        
  \begin{figure*}[t!]
    \centering
    \includegraphics[width=6.cm]{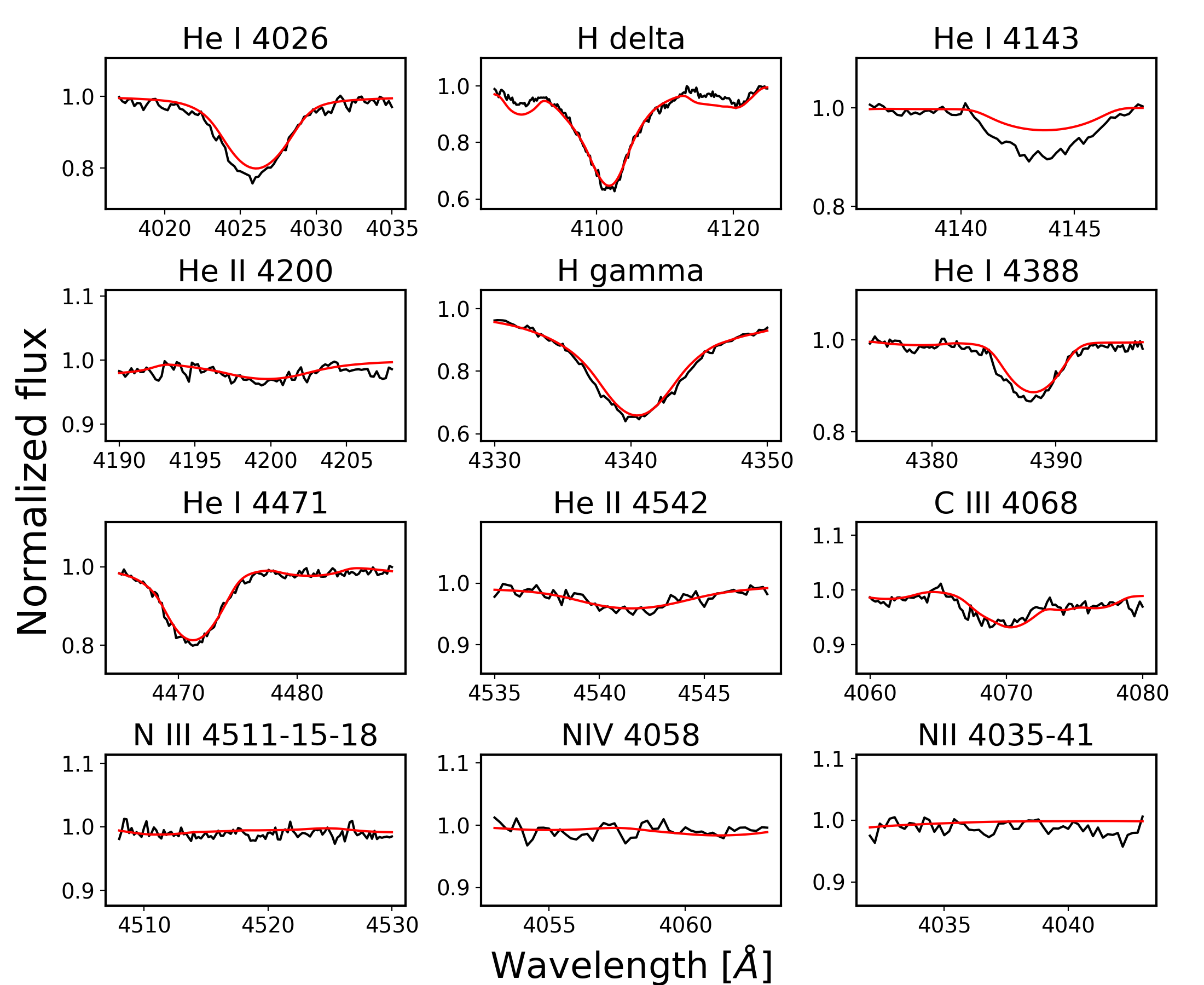}
    \includegraphics[width=7.cm, bb=5 0 453 346,clip]{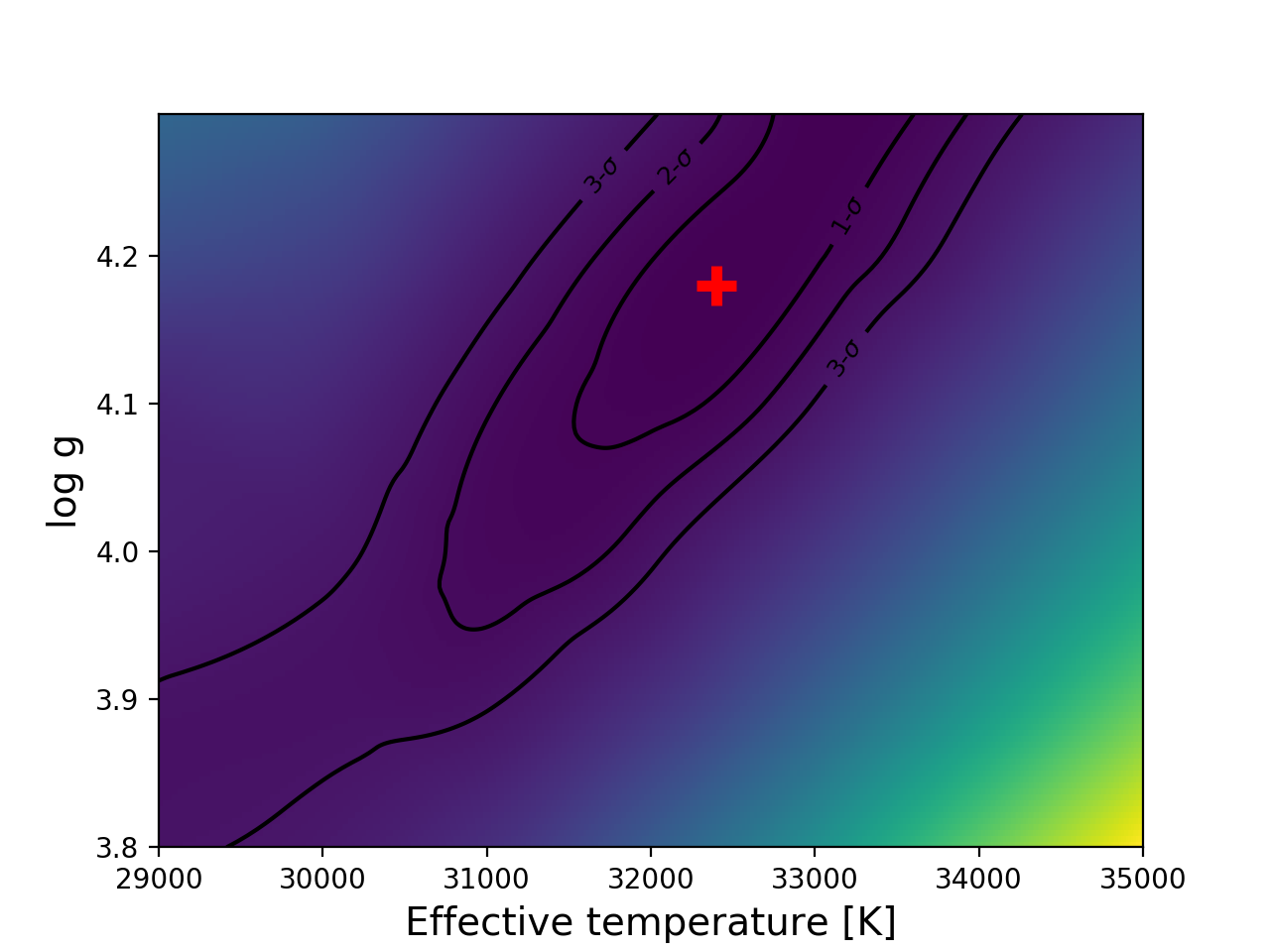}
    \includegraphics[width=6.cm]{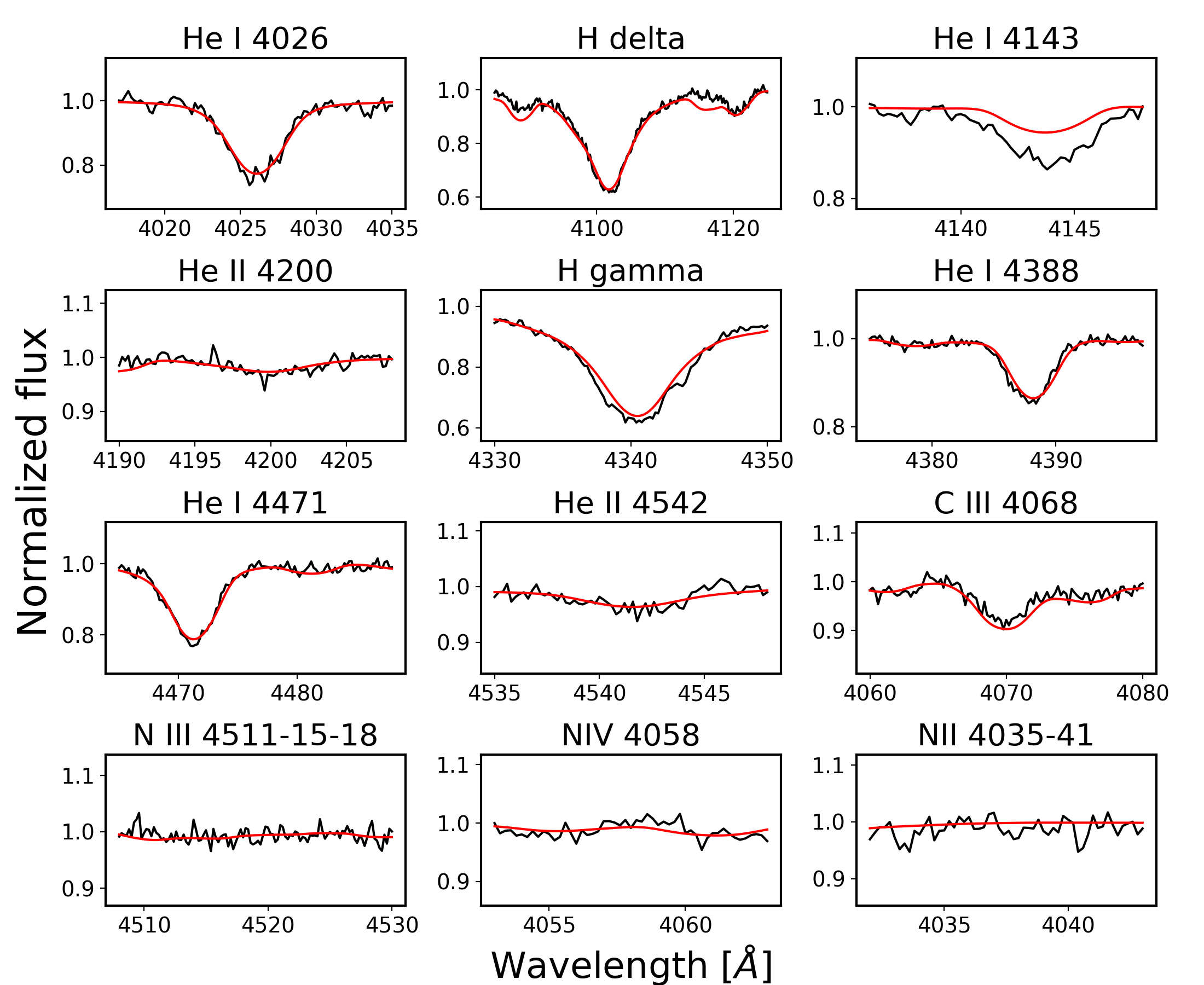}
    \includegraphics[width=7.cm, bb=5 0 453 346,clip]{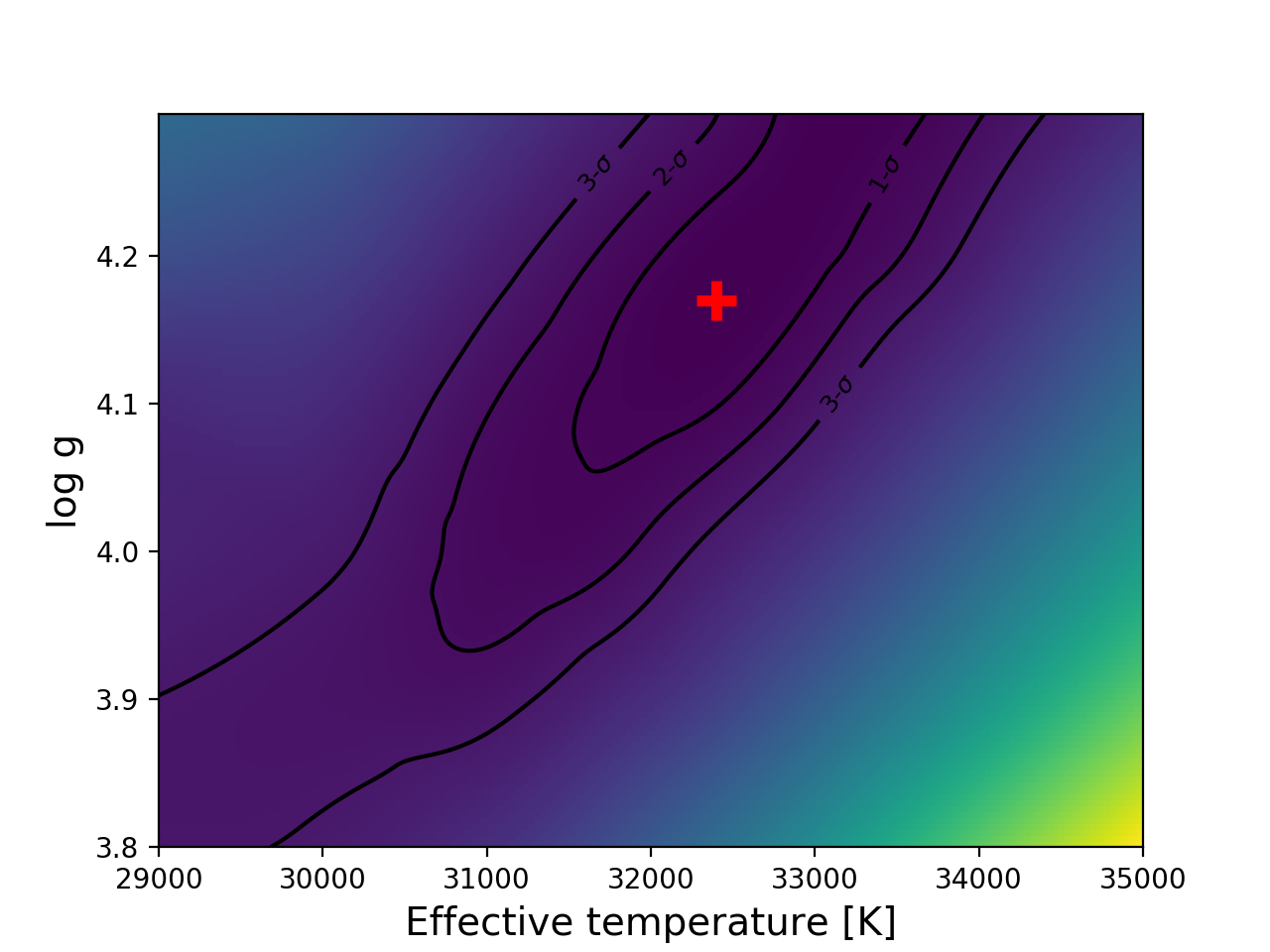}
    \includegraphics[width=7cm]{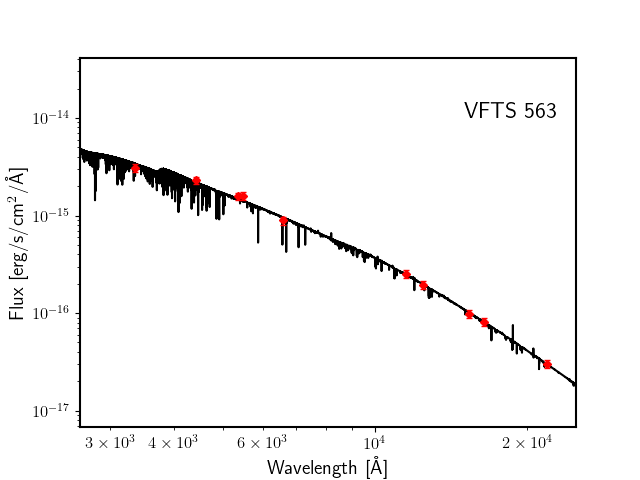}
    \includegraphics[width=6.5cm]{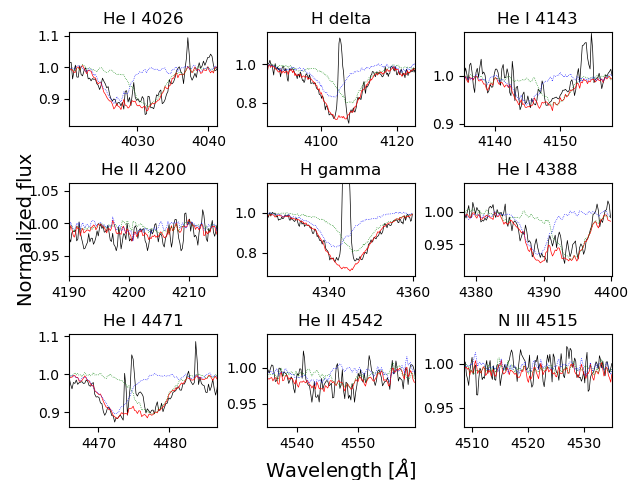}
    \includegraphics[width=7cm, bb=5 0 453 346,clip]{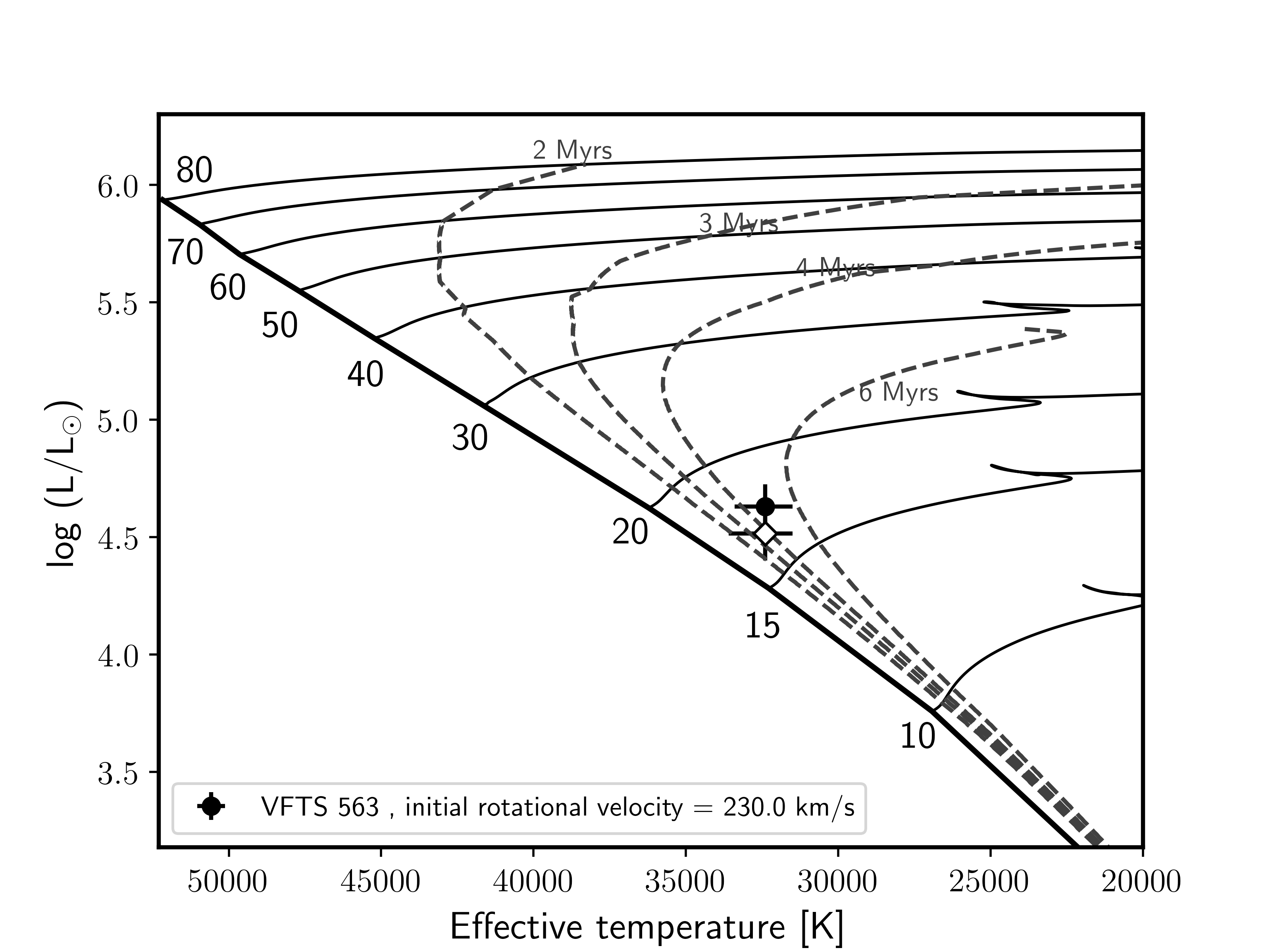}
    \includegraphics[width=7cm, bb=5 0 453 346,clip]{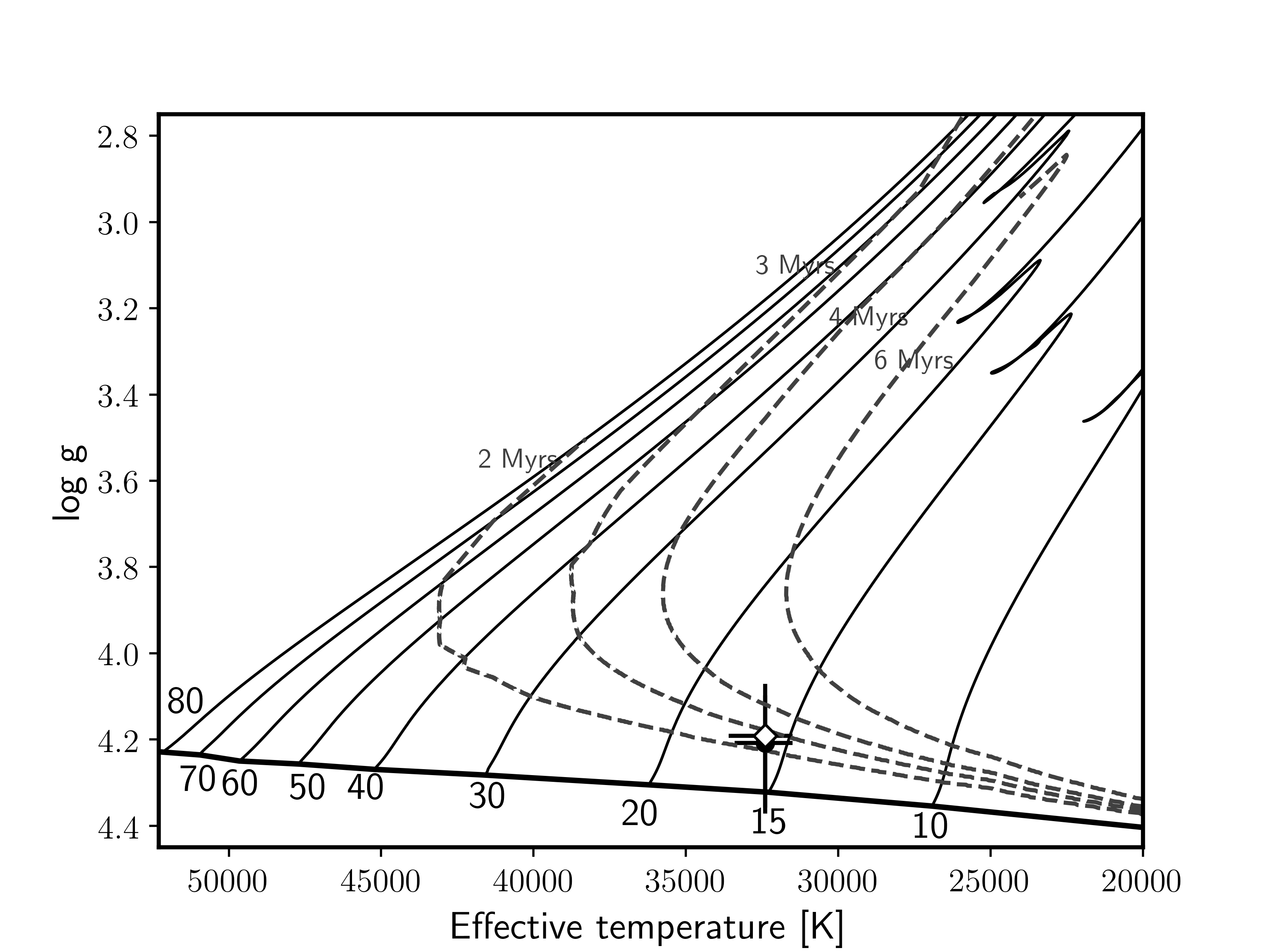}
    \caption{Same as Fig.\,\ref{fig:042} but for VFTS\,563.} \label{fig:563} 
  \end{figure*}  
   \clearpage
           
    \begin{figure*}[t!]
    \centering
    \includegraphics[width=6.cm]{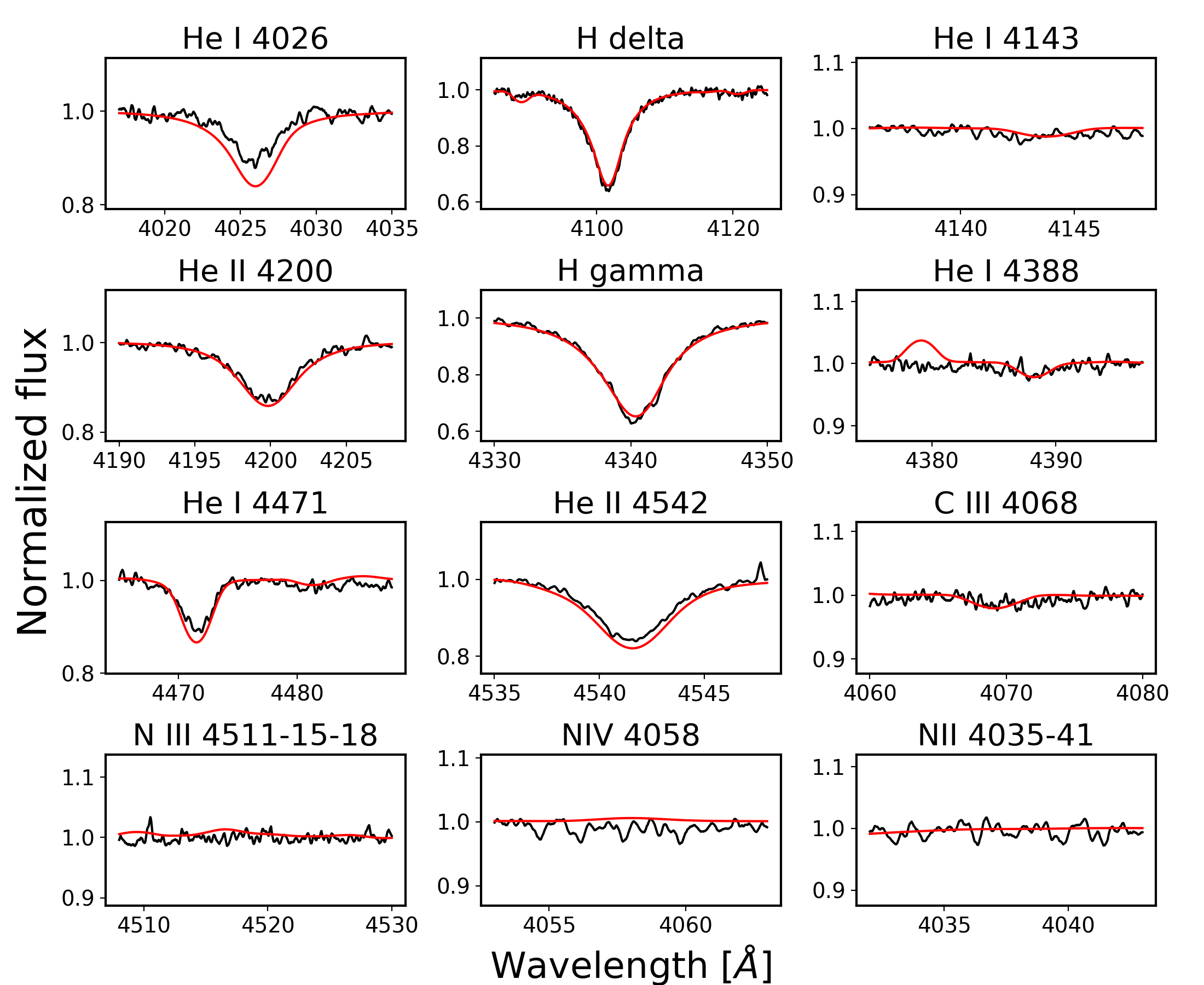}
    \includegraphics[width=7.cm, bb=5 0 453 346,clip]{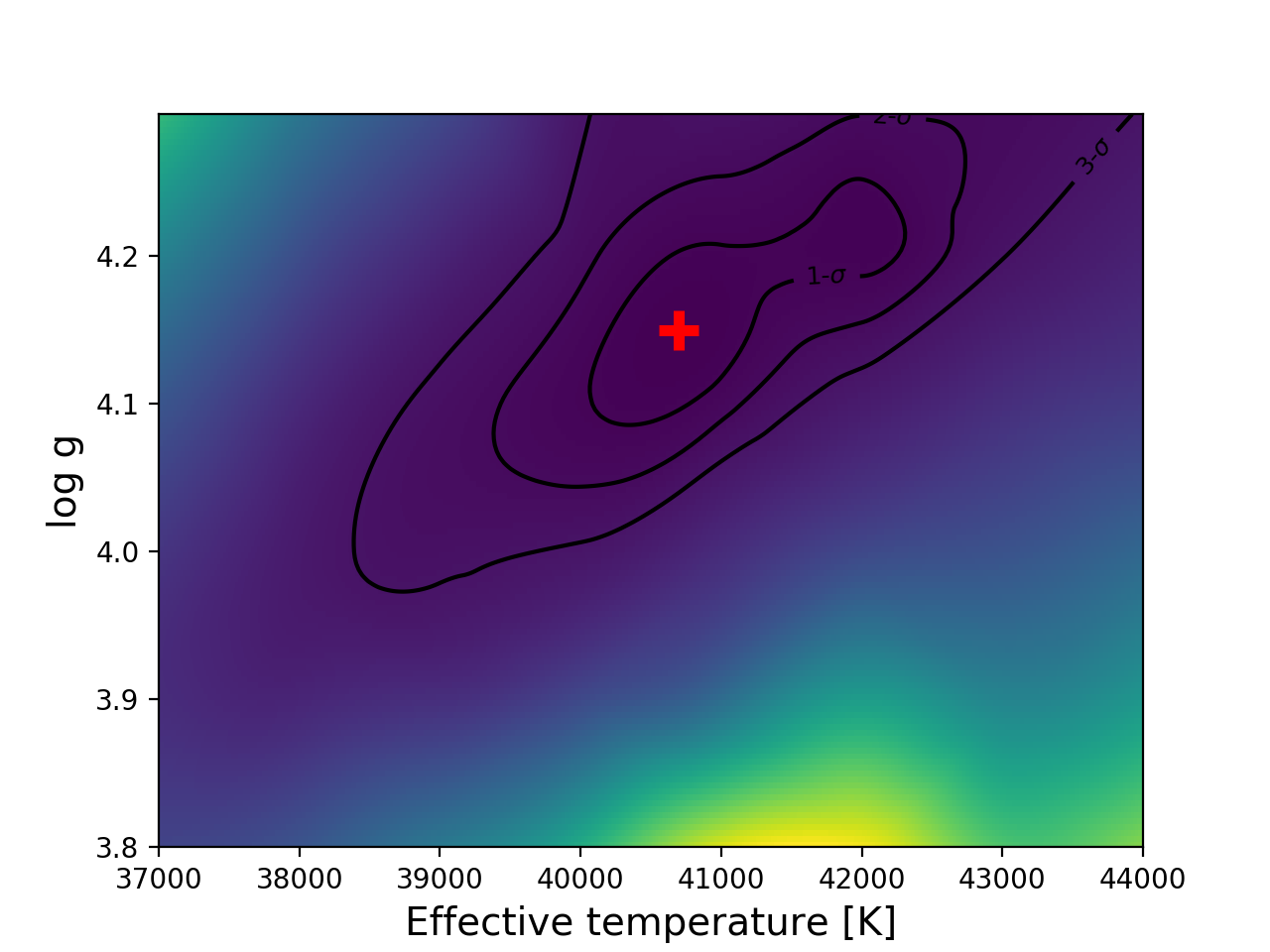}
    \includegraphics[width=6.cm]{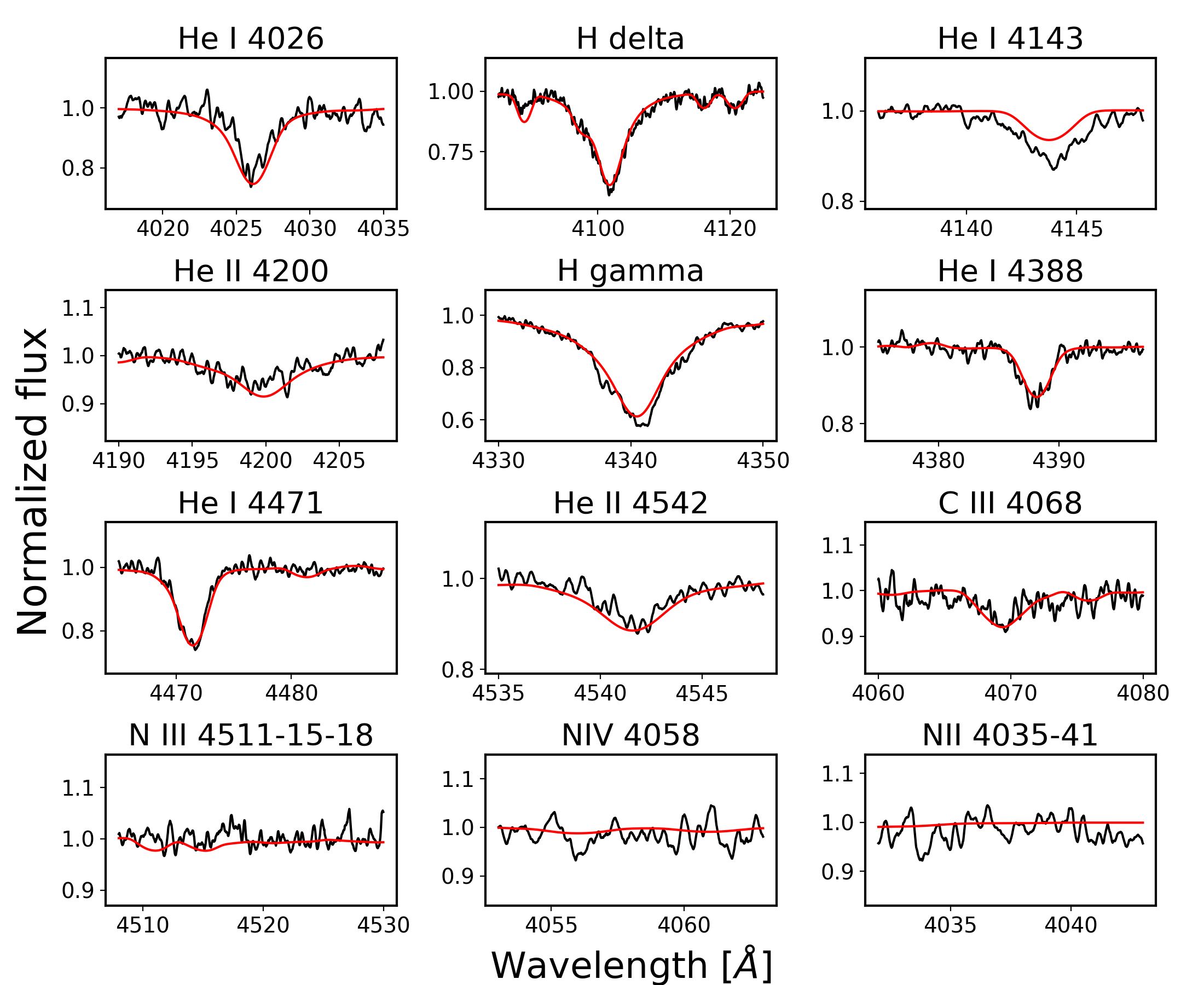}
    \includegraphics[width=7.cm, bb=5 0 453 346,clip]{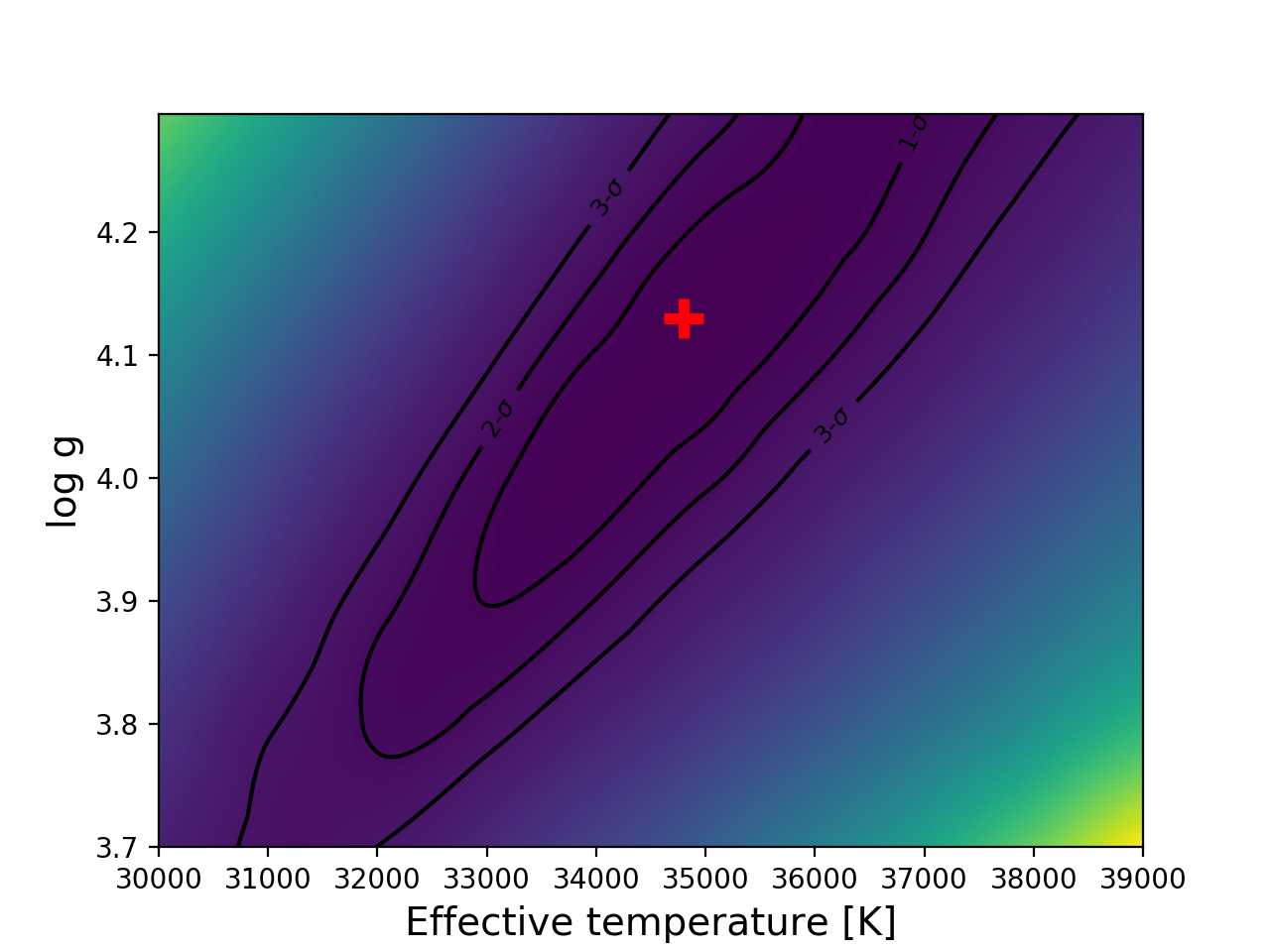}
    \includegraphics[width=7cm]{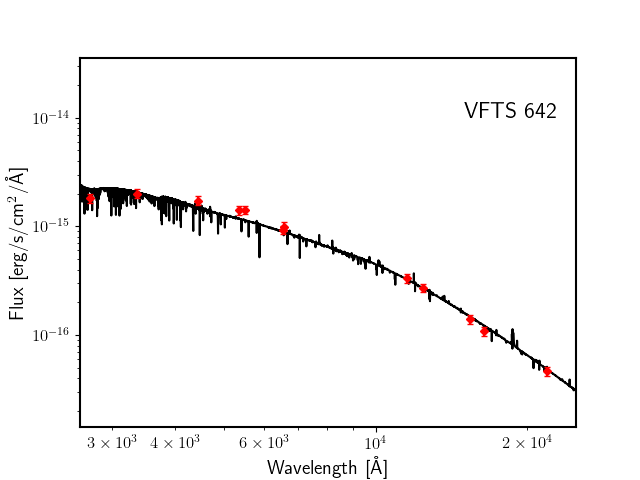}
    \includegraphics[width=6.5cm]{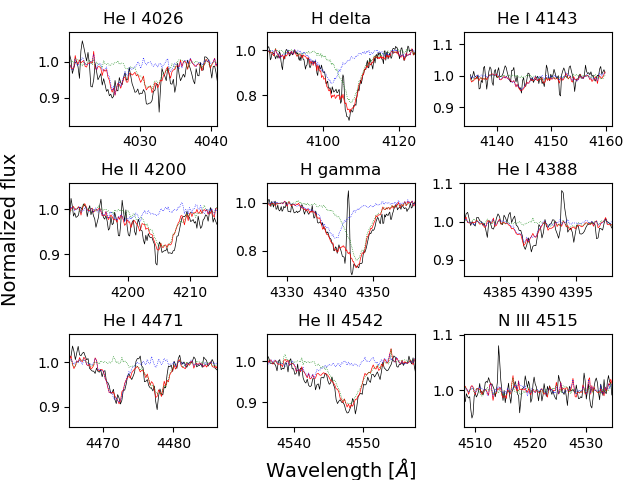}
    \includegraphics[width=7cm, bb=5 0 453 346,clip]{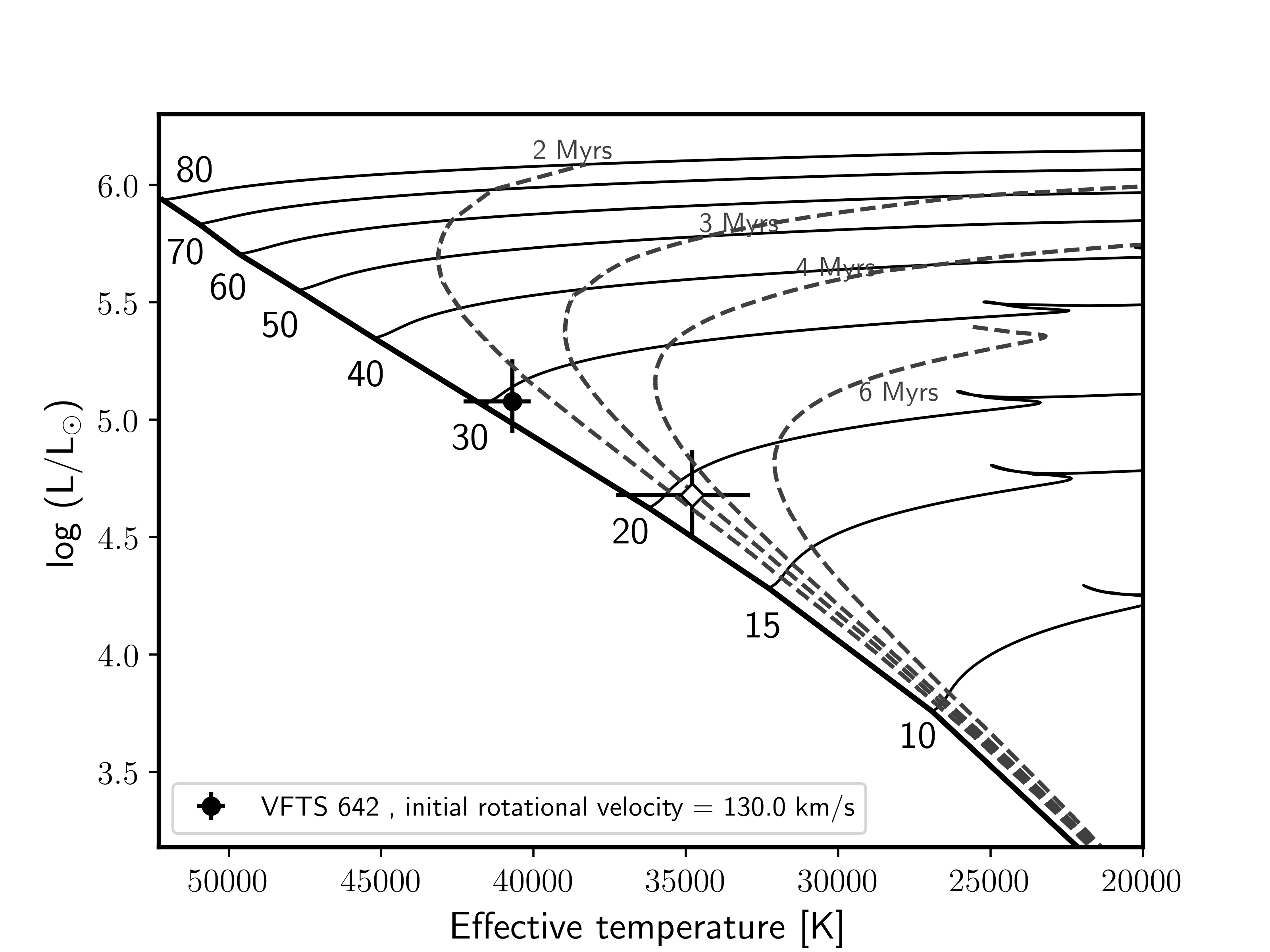}
    \includegraphics[width=7cm, bb=5 0 453 346,clip]{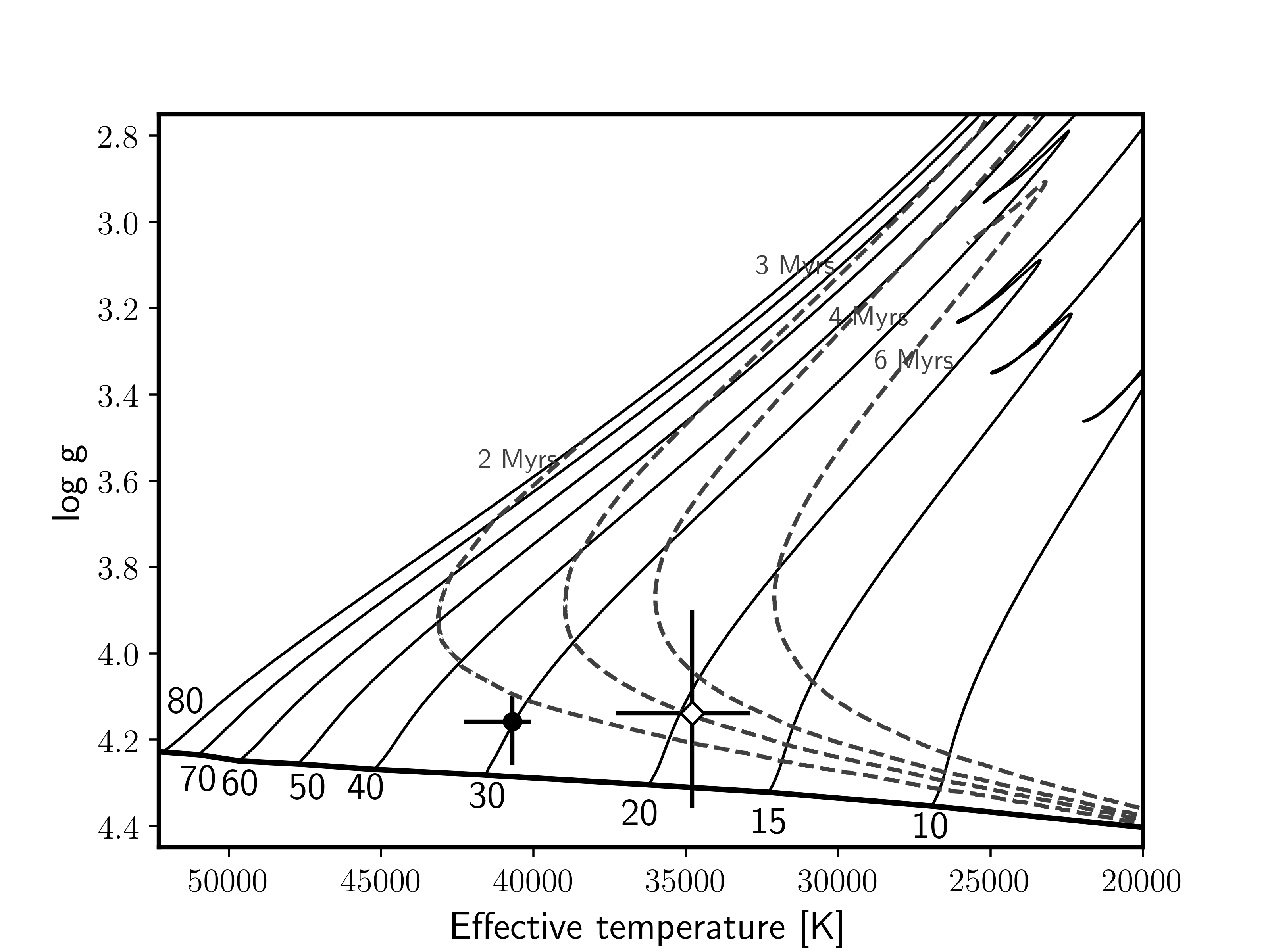}
    \caption{Same as Fig.\,\ref{fig:042} but for VFTS\,642.} \label{fig:642} 
  \end{figure*}
 \clearpage

    \begin{figure*}[t!]
    \centering
    \includegraphics[width=6.cm]{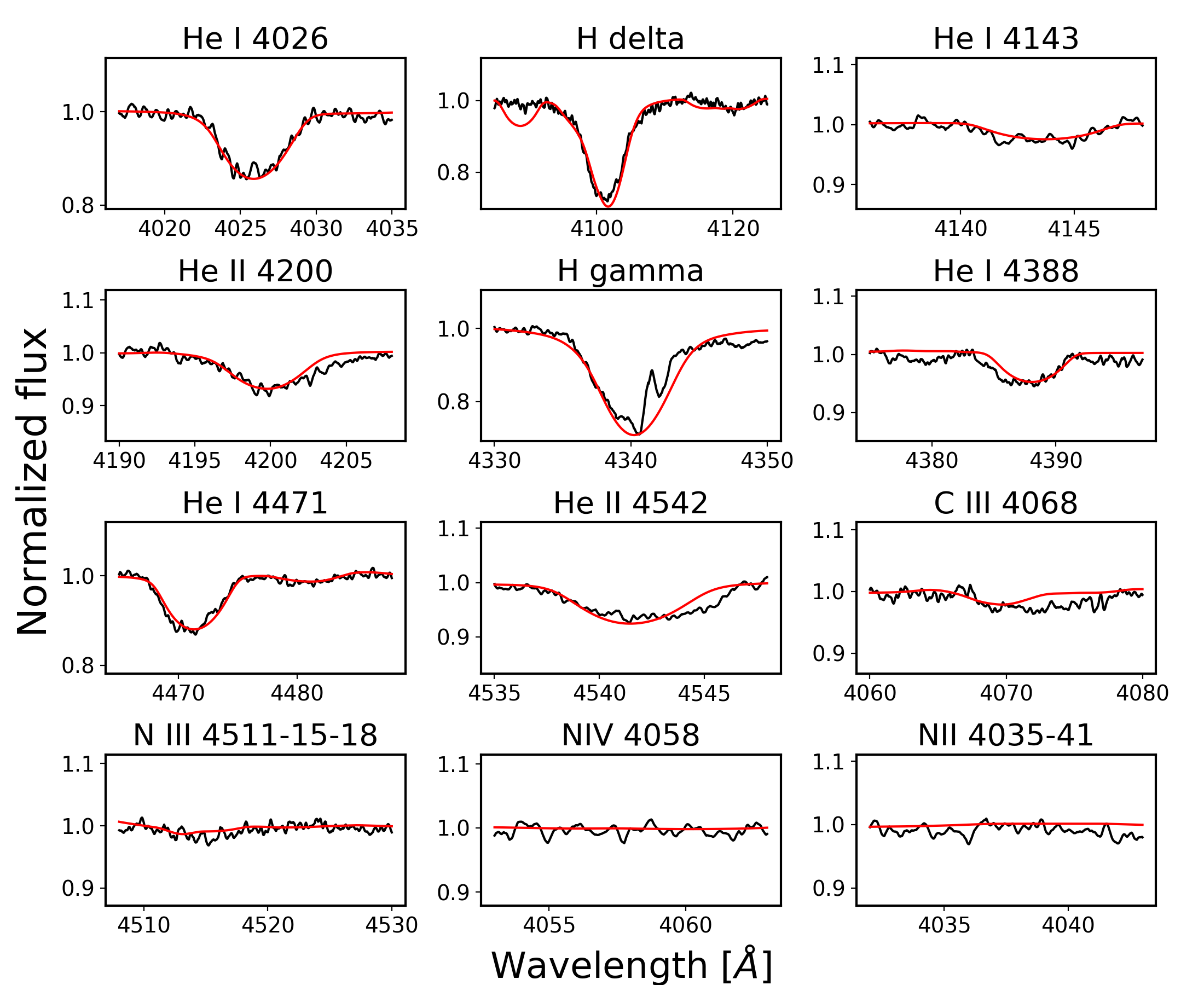}
    \includegraphics[width=7.cm, bb=5 0 453 346,clip]{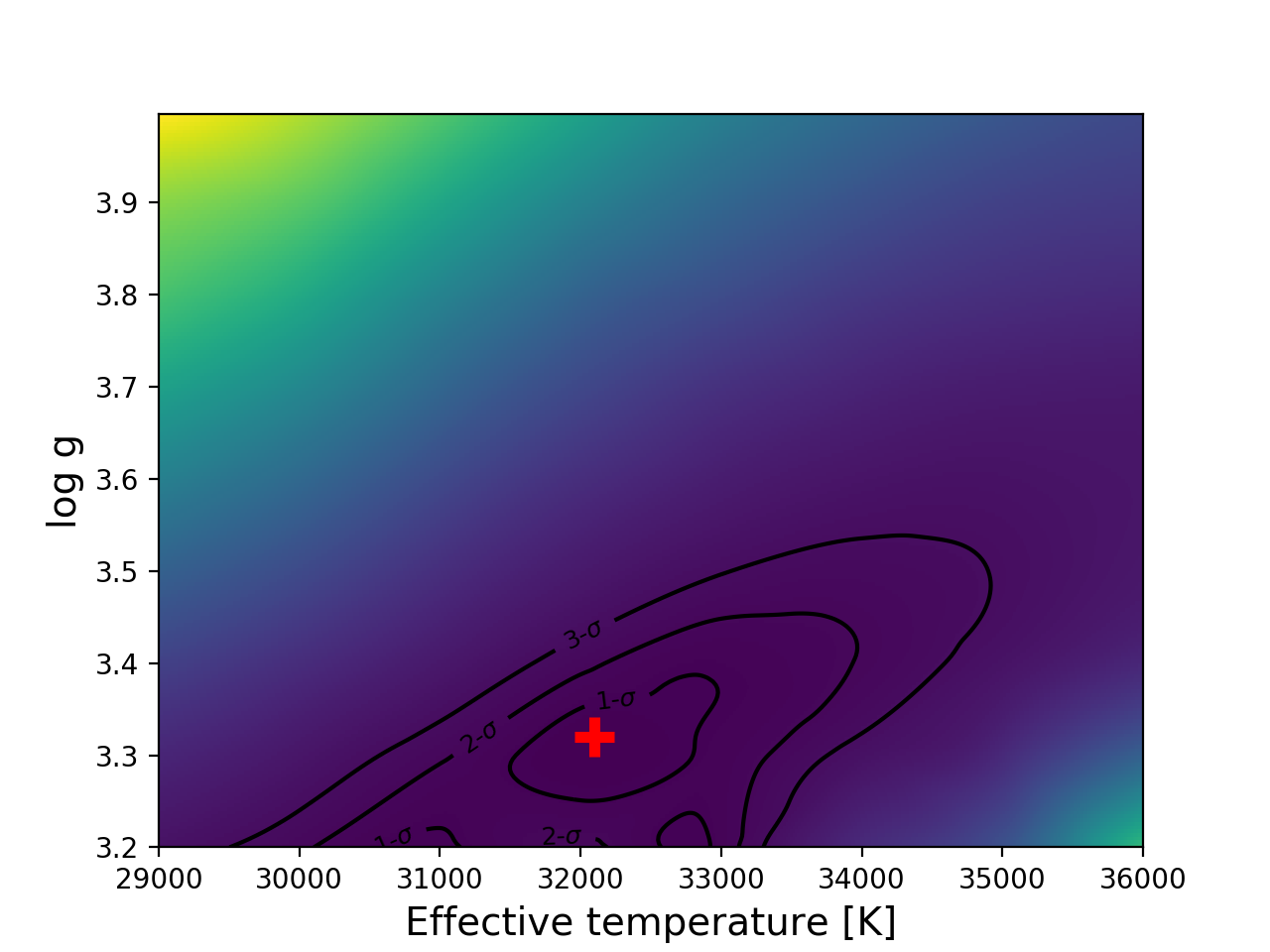}
    \includegraphics[width=6.cm]{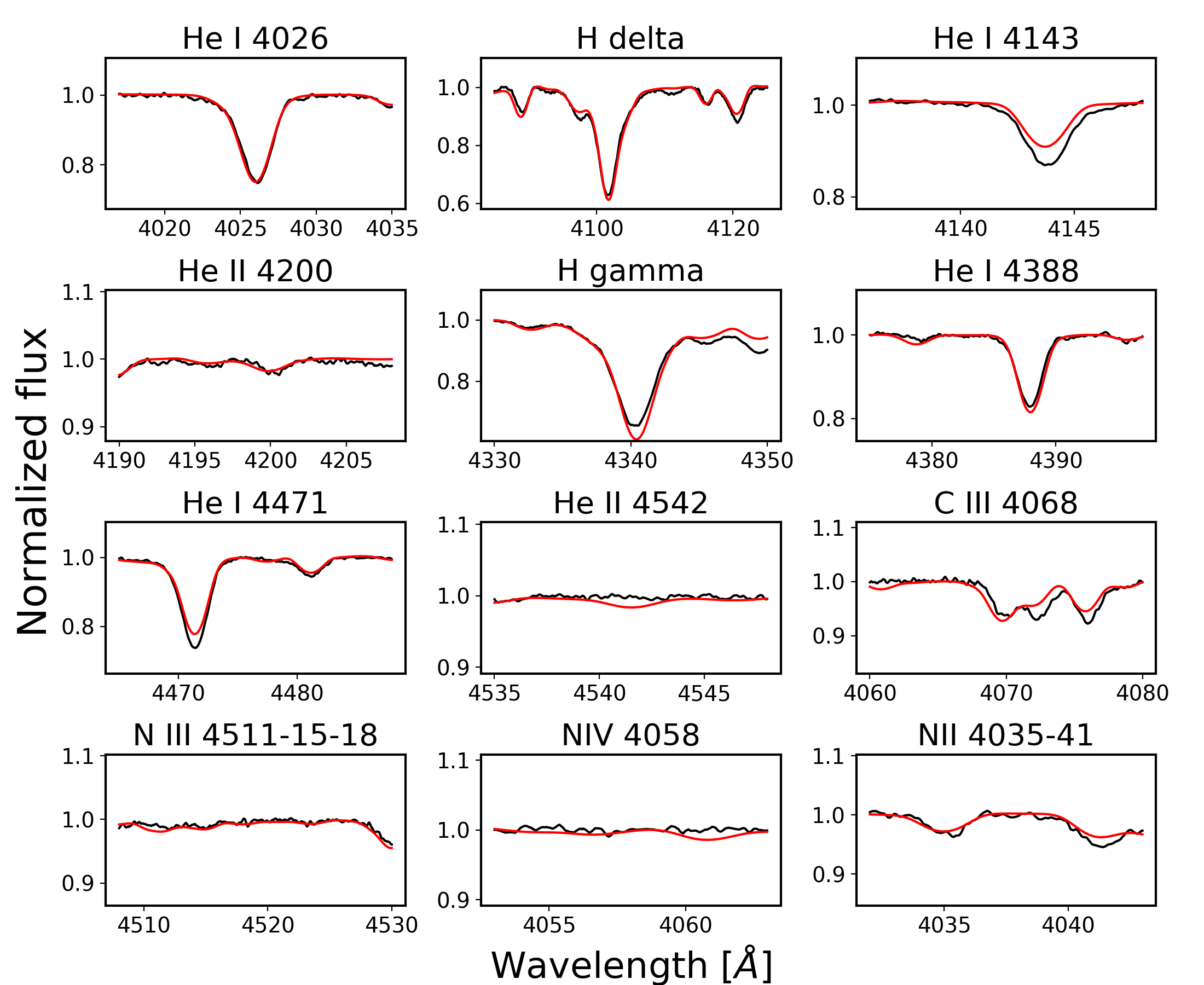}
    \includegraphics[width=7.cm, bb=5 0 453 346,clip]{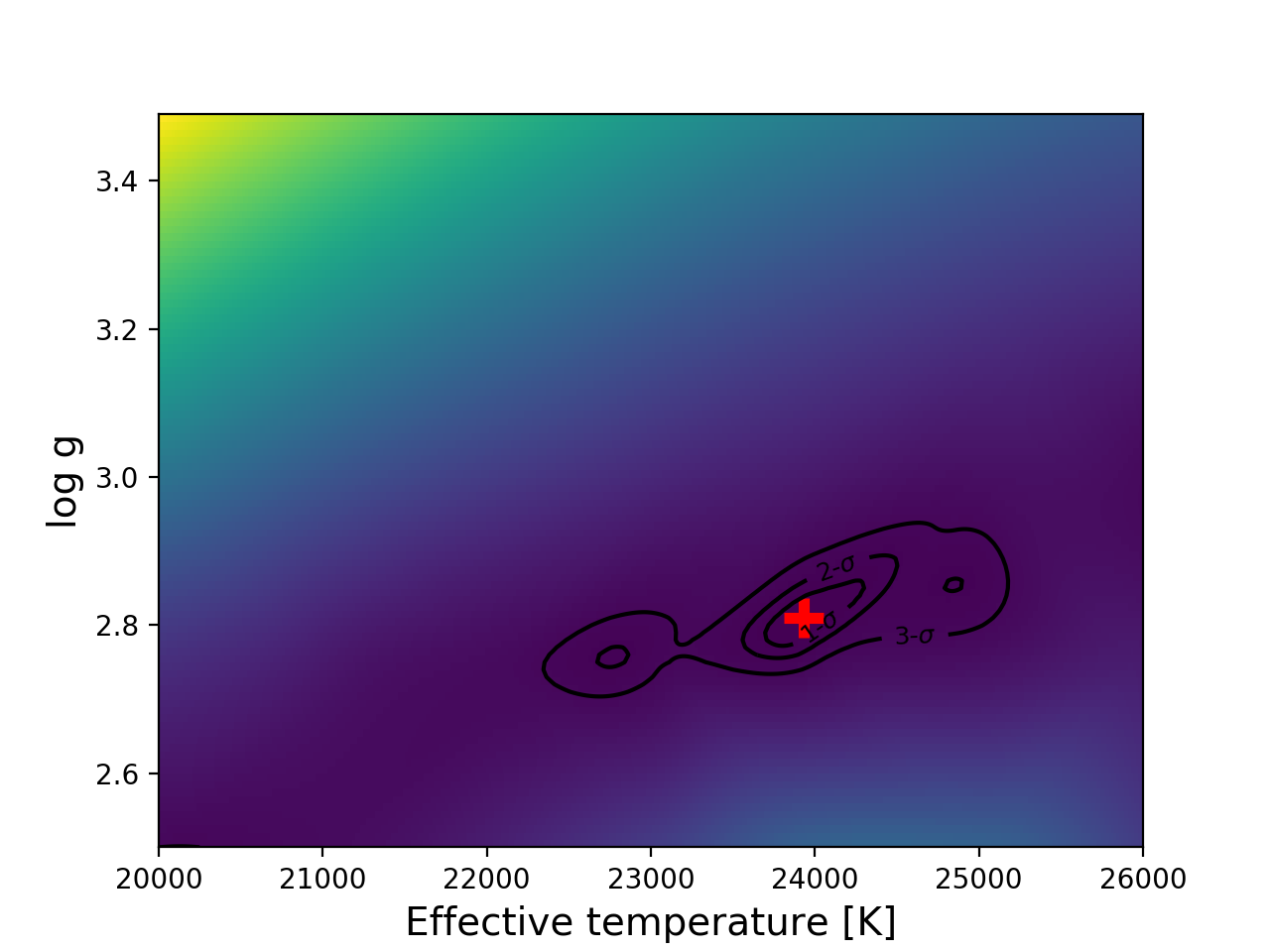}
    \includegraphics[width=7cm]{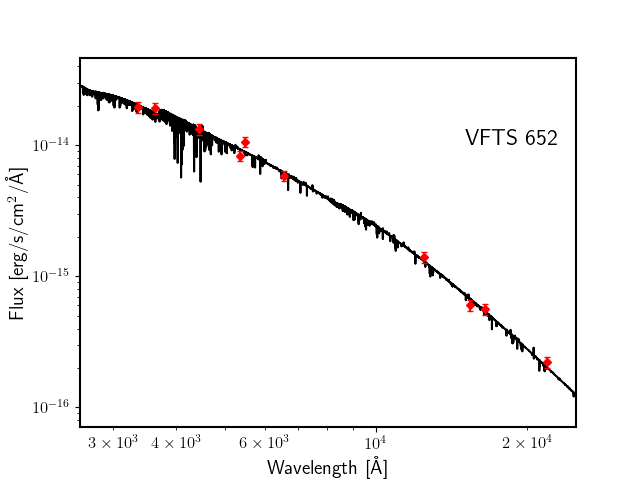}
    \includegraphics[width=6.5cm]{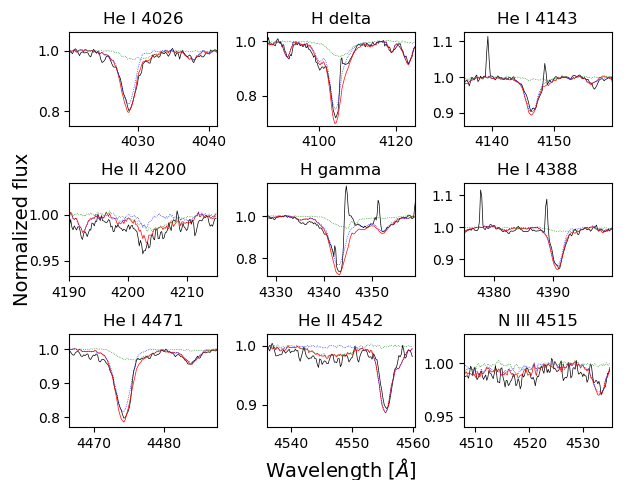}
    \includegraphics[width=7cm, bb=5 0 453 346,clip]{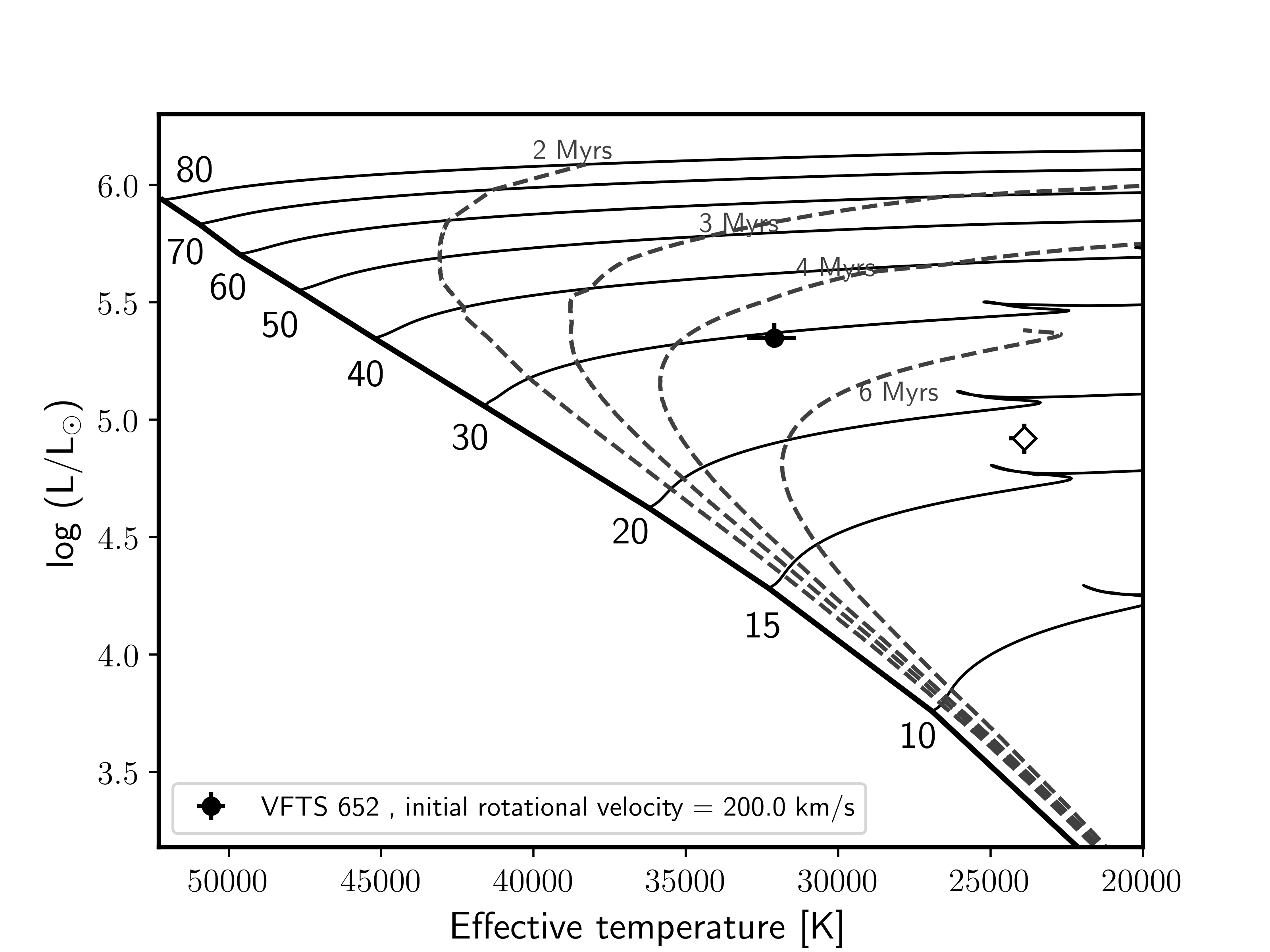}
    \includegraphics[width=7cm, bb=5 0 453 346,clip]{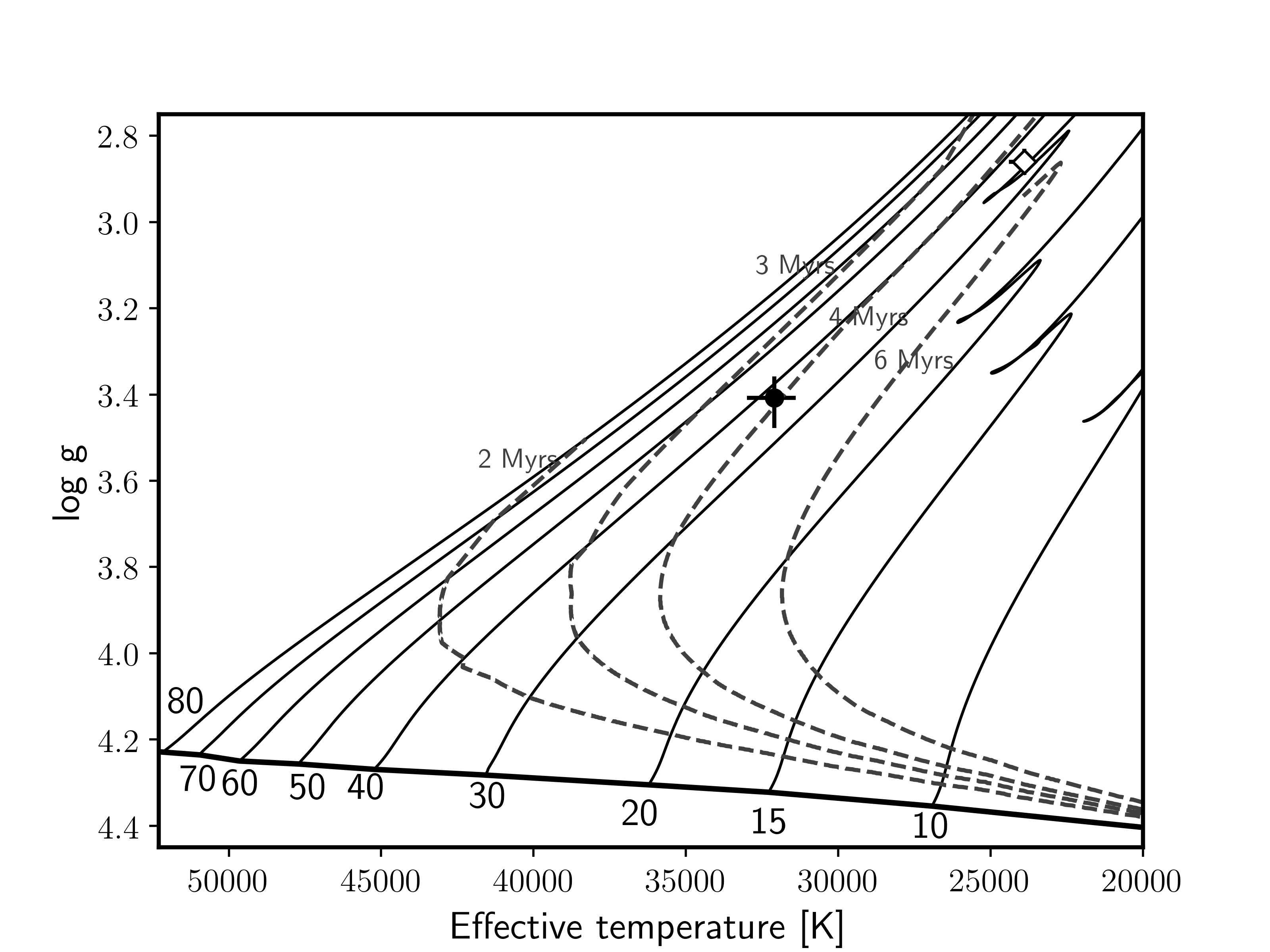}
    \caption{Same as Fig.\,\ref{fig:042} but for VFTS\,652.} \label{fig:652} 
  \end{figure*}
 \clearpage

    \begin{figure*}[t!]
    \centering
    \includegraphics[width=6.cm]{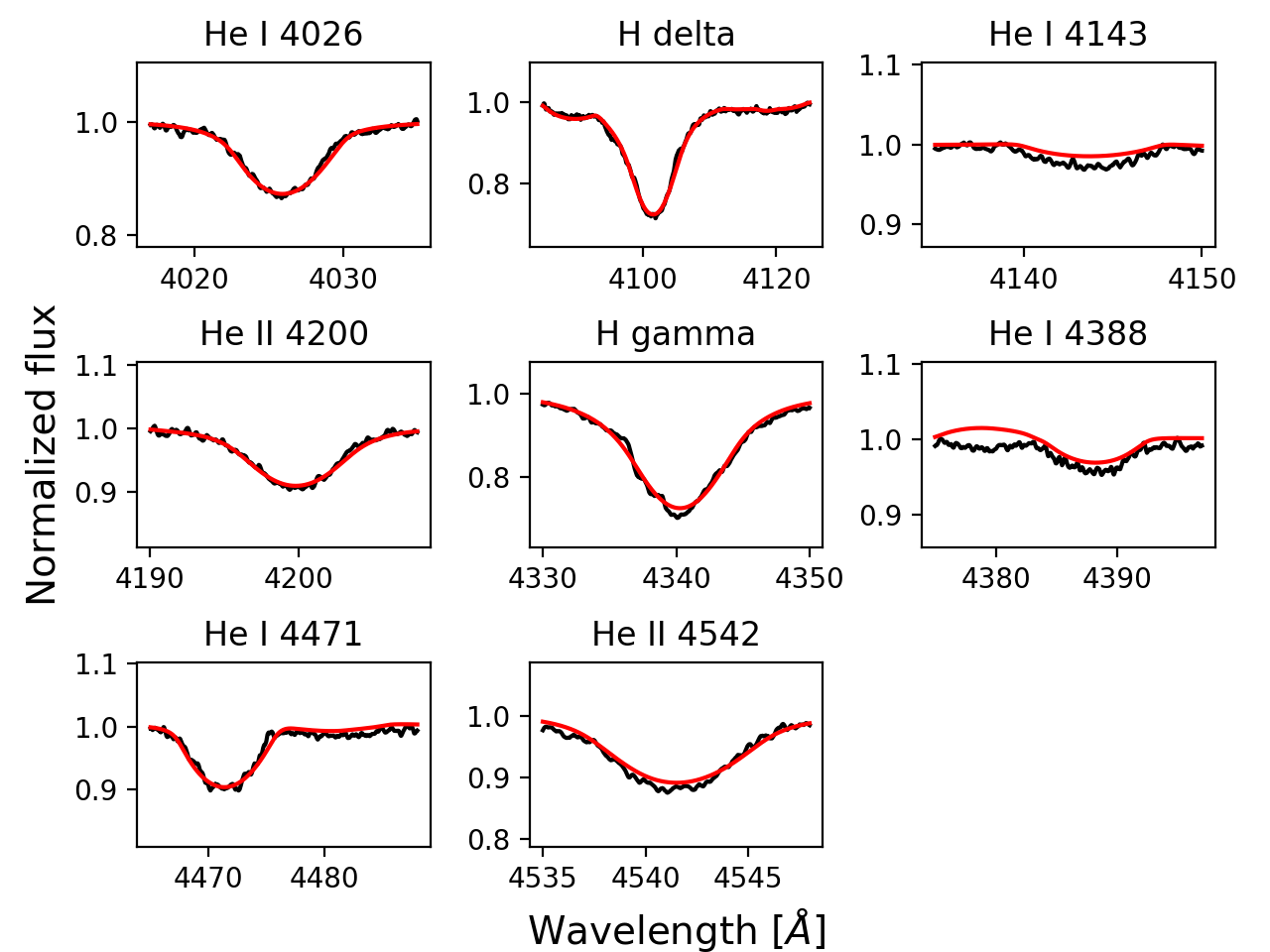}
    \includegraphics[width=7.cm, bb=5 0 453 346,clip]{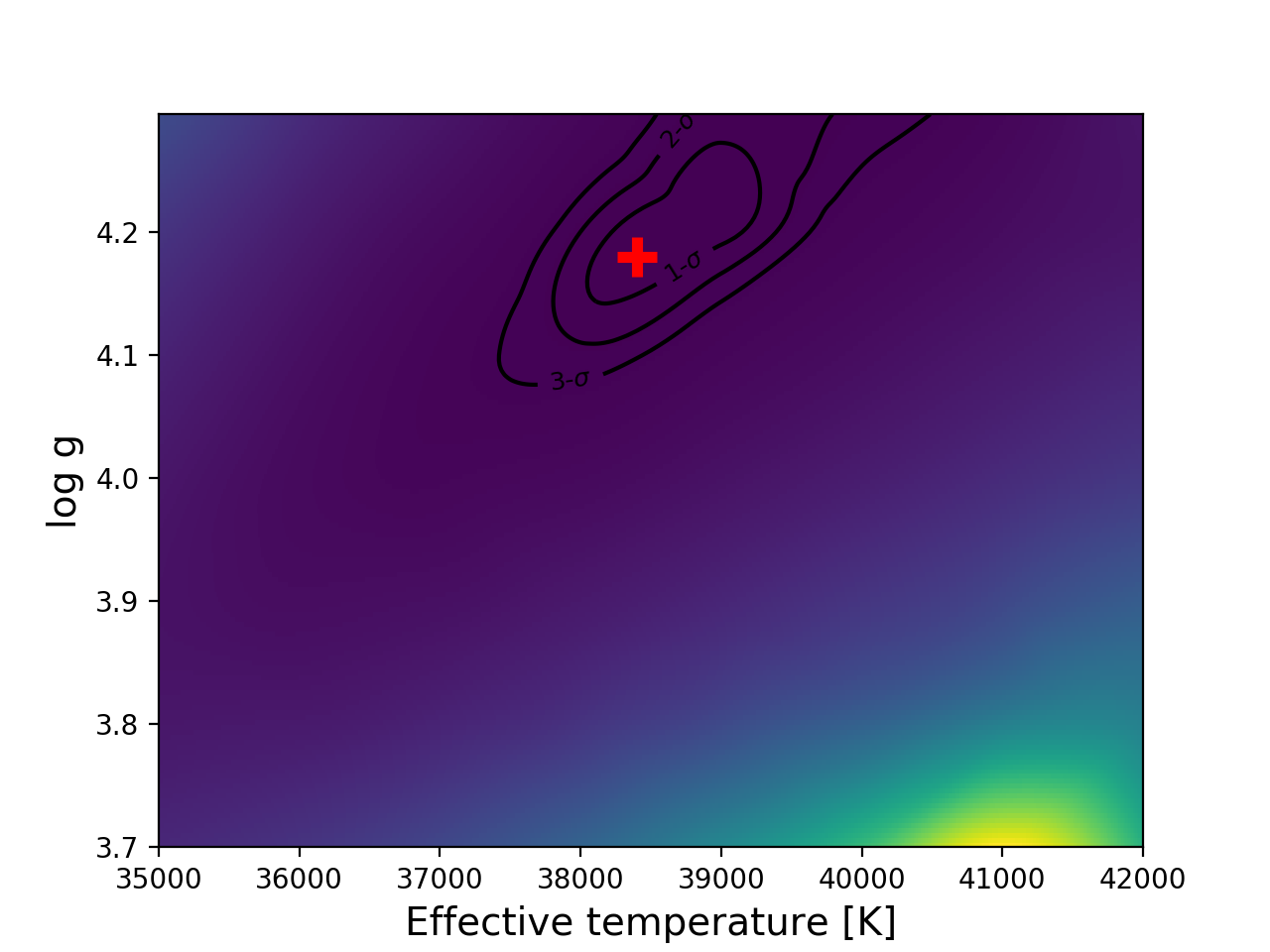}
    \includegraphics[width=6.cm]{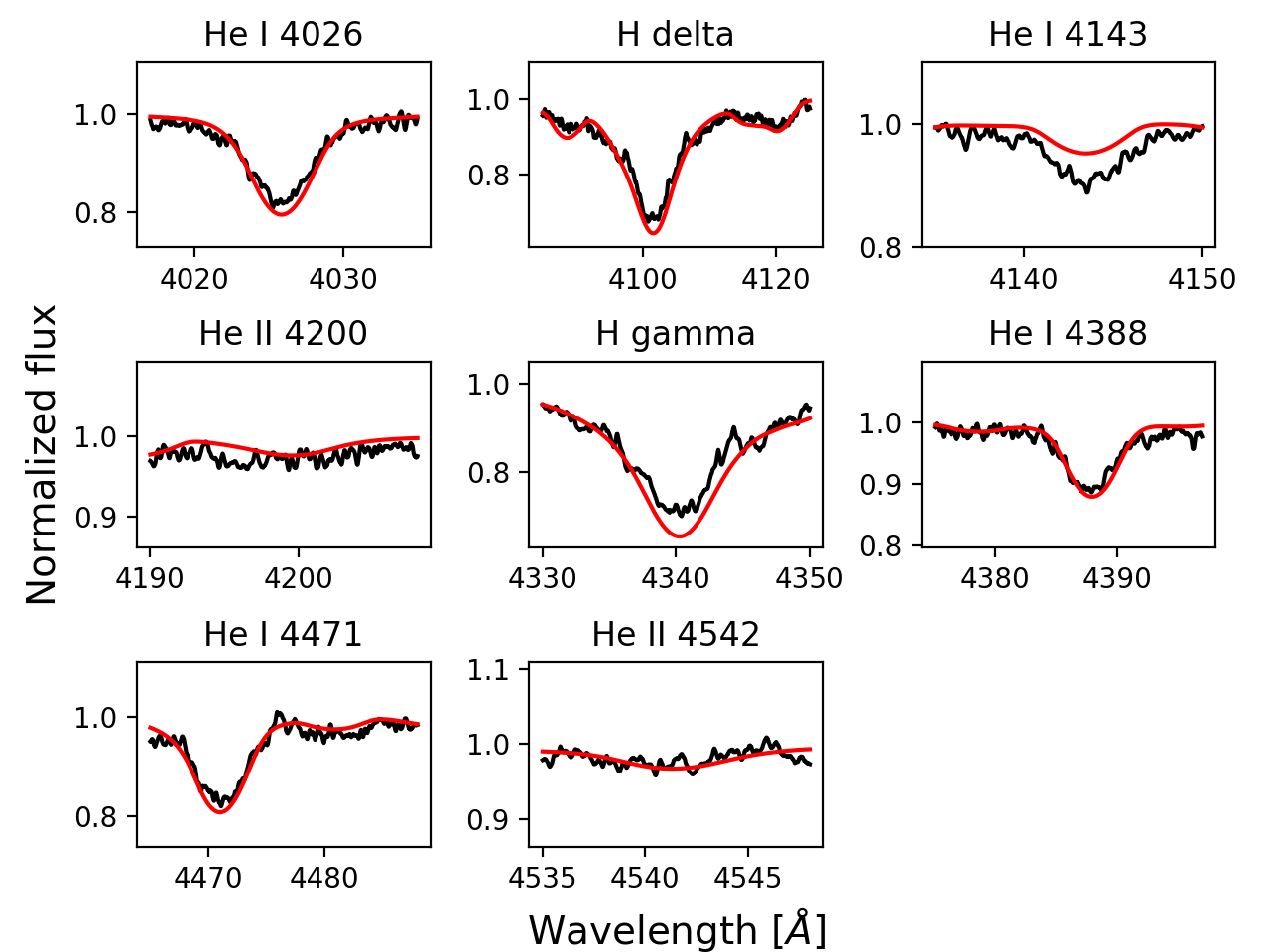}
    \includegraphics[width=7.cm, bb=5 0 453 346,clip]{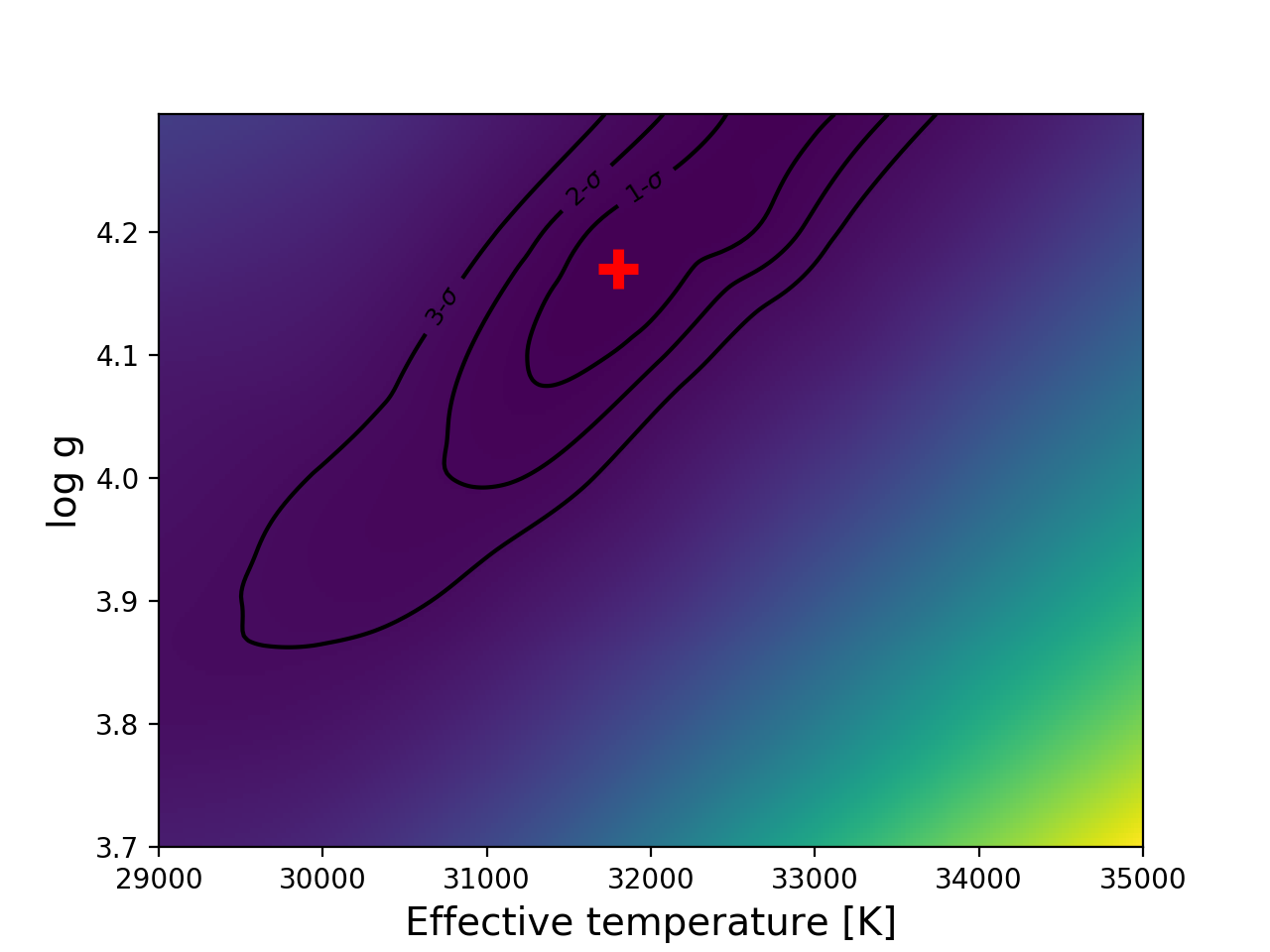}
    \includegraphics[width=7cm]{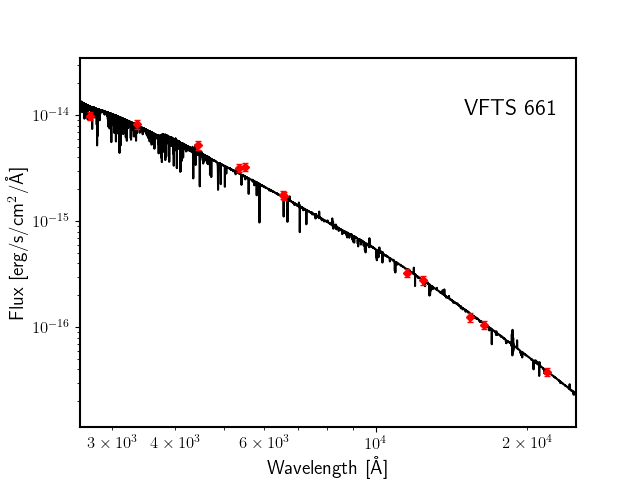}
    \includegraphics[width=6.5cm]{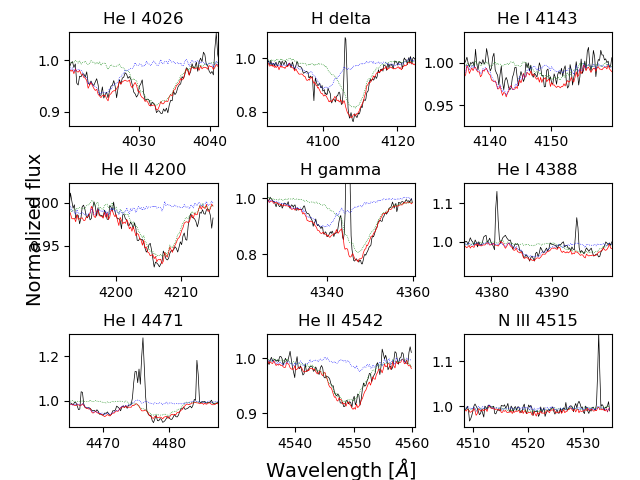}
    \includegraphics[width=7cm, bb=5 0 453 346,clip]{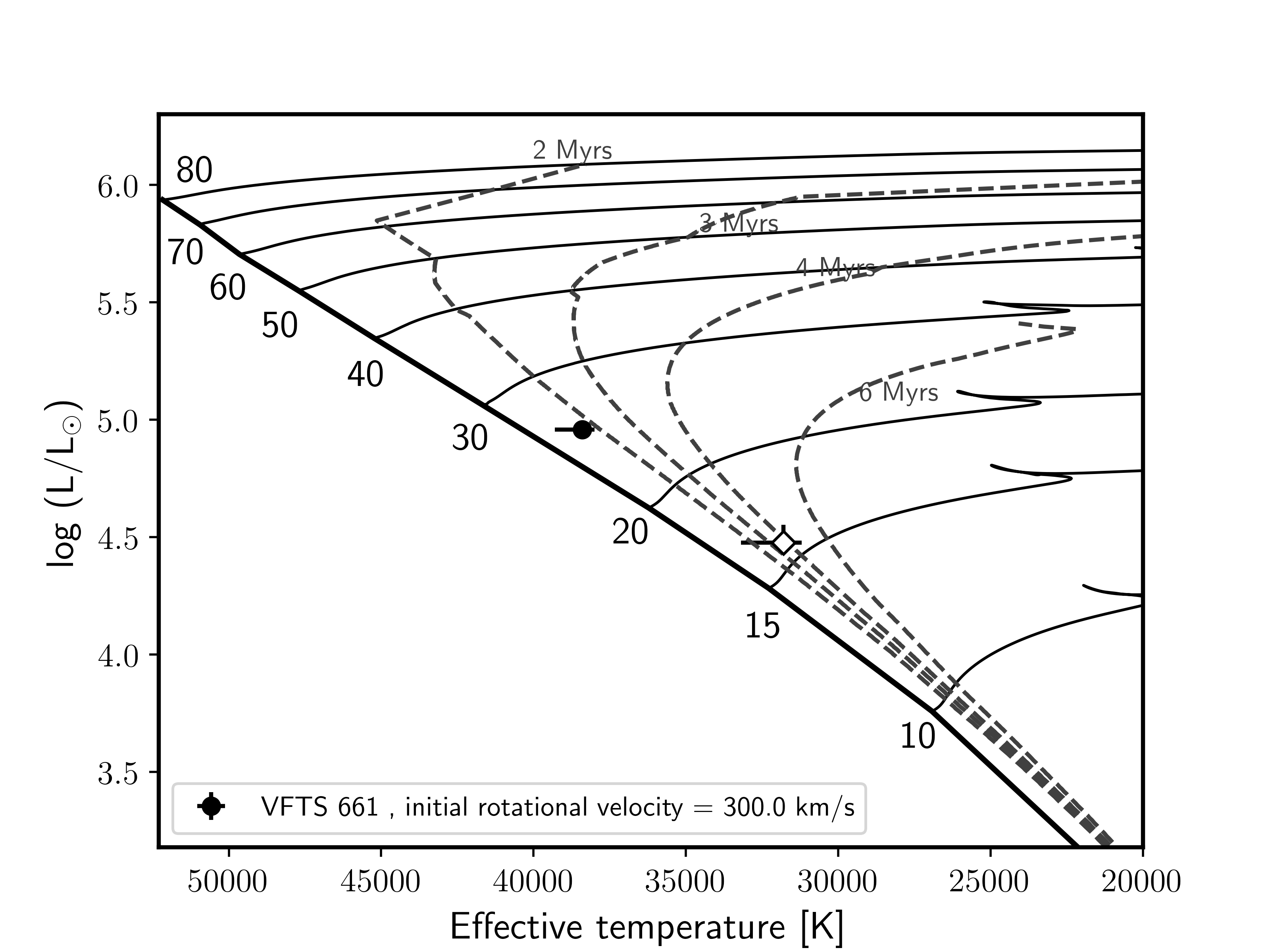}
    \includegraphics[width=7cm, bb=5 0 453 346,clip]{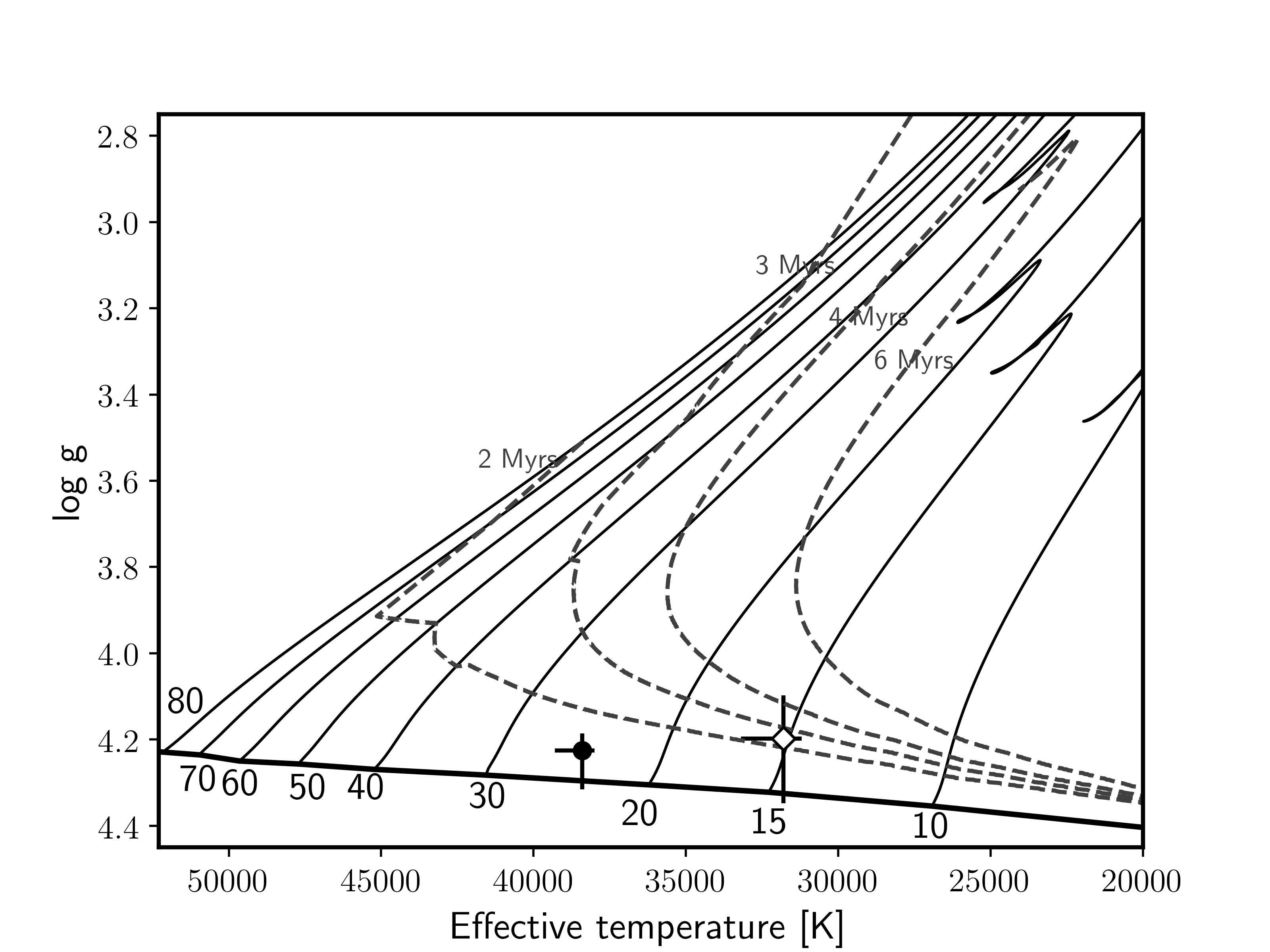}
    \caption{Same as Fig.\,\ref{fig:042} but for VFTS\,661.} \label{fig:661} 
  \end{figure*}
 \clearpage

    %\subsection{VFTS\,047 secondary - O8.5\,V}

    \begin{figure*}[t!]
    \centering
    \includegraphics[width=6.cm]{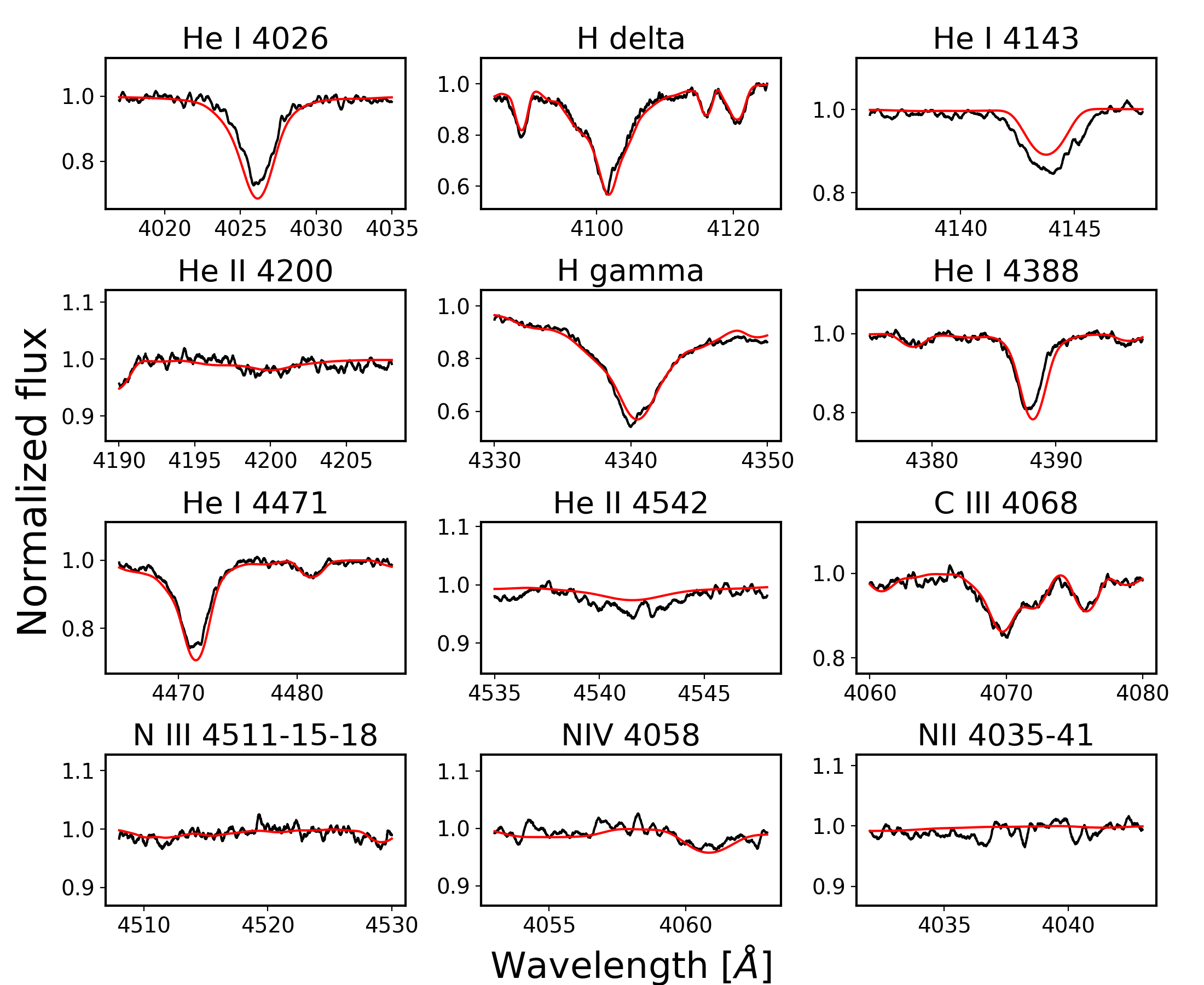}
    \includegraphics[width=7.cm, bb=5 0 453 346,clip]{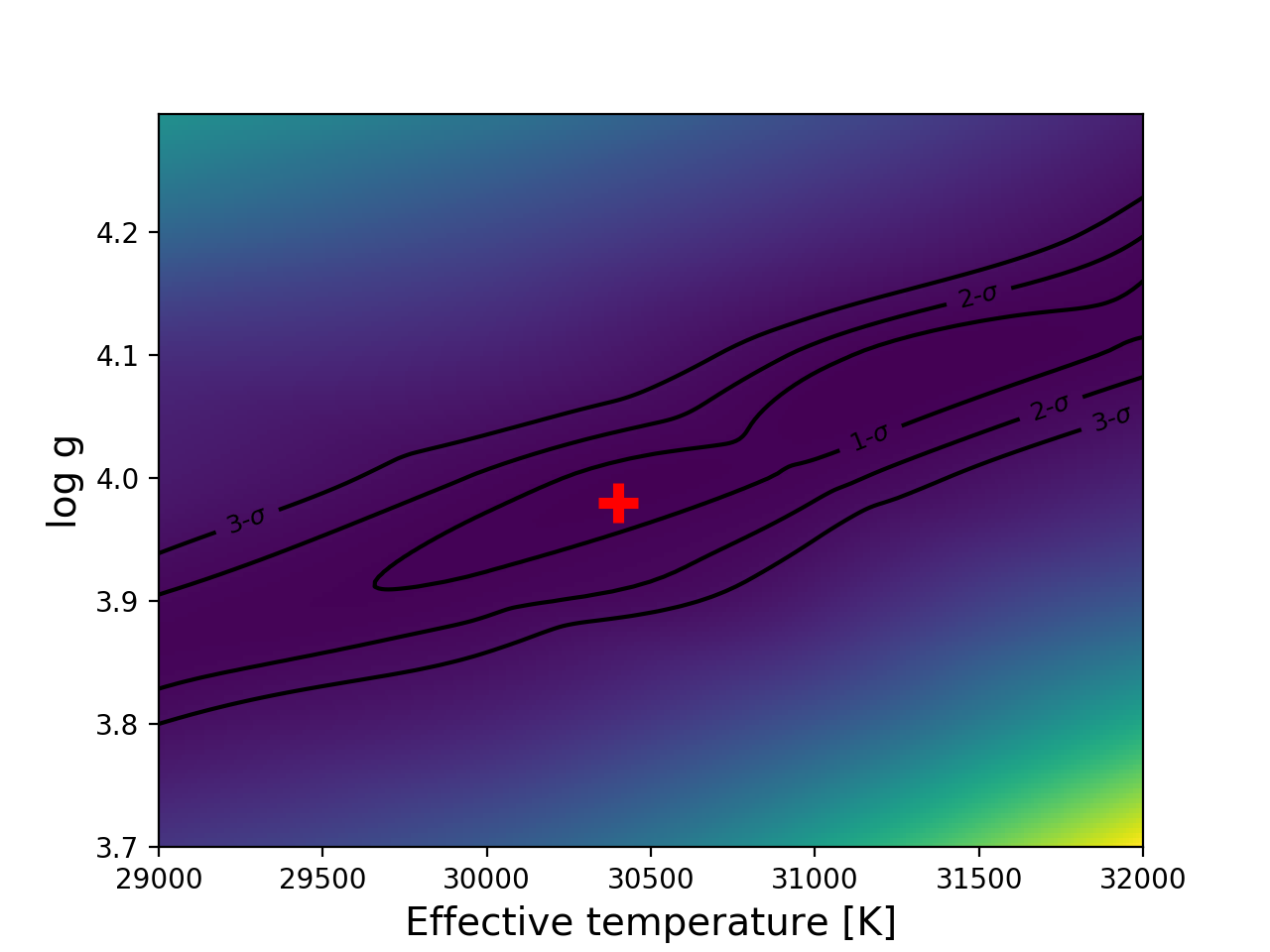}
    \includegraphics[width=6.cm]{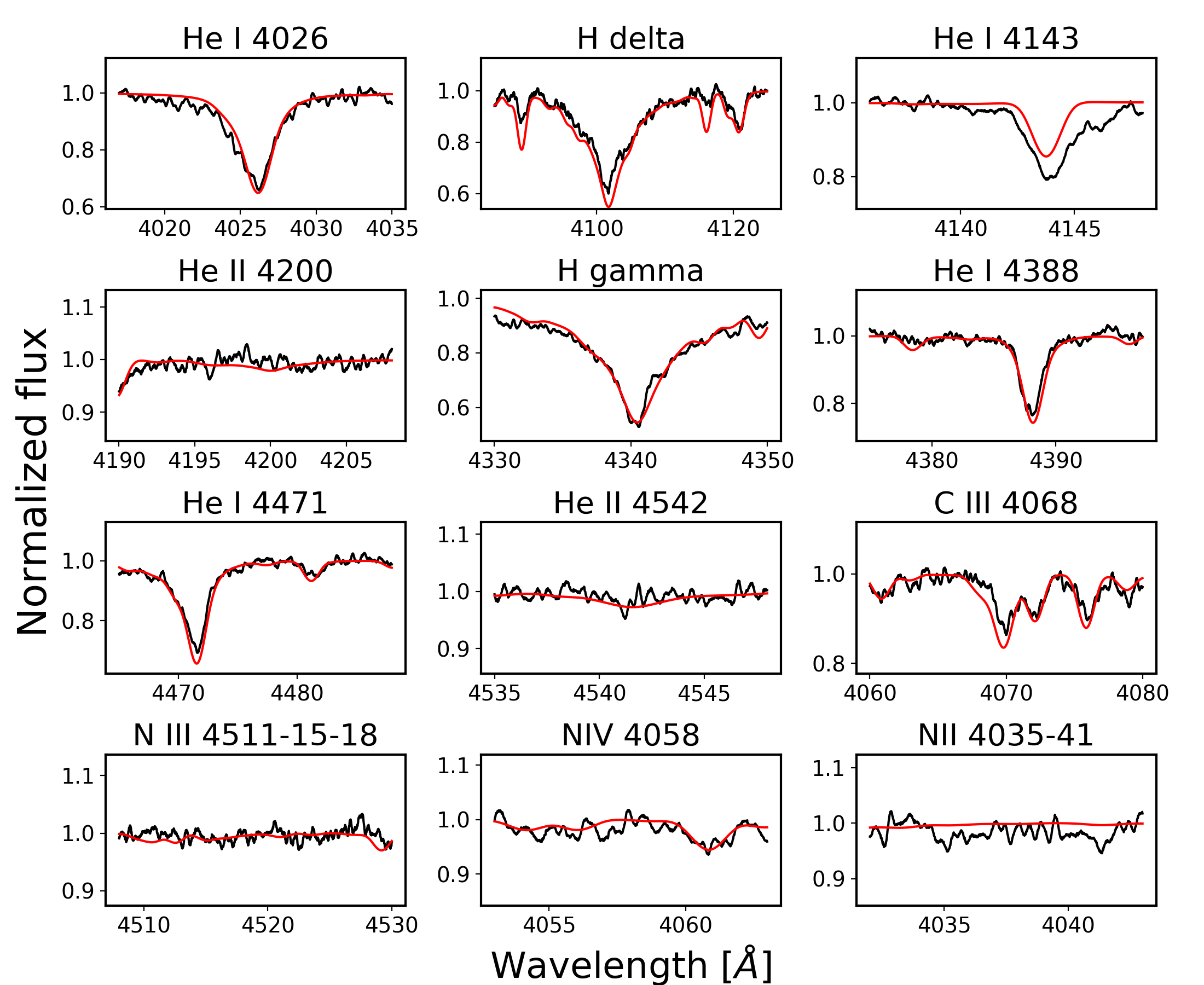}
    \includegraphics[width=7.cm, bb=5 0 453 346,clip]{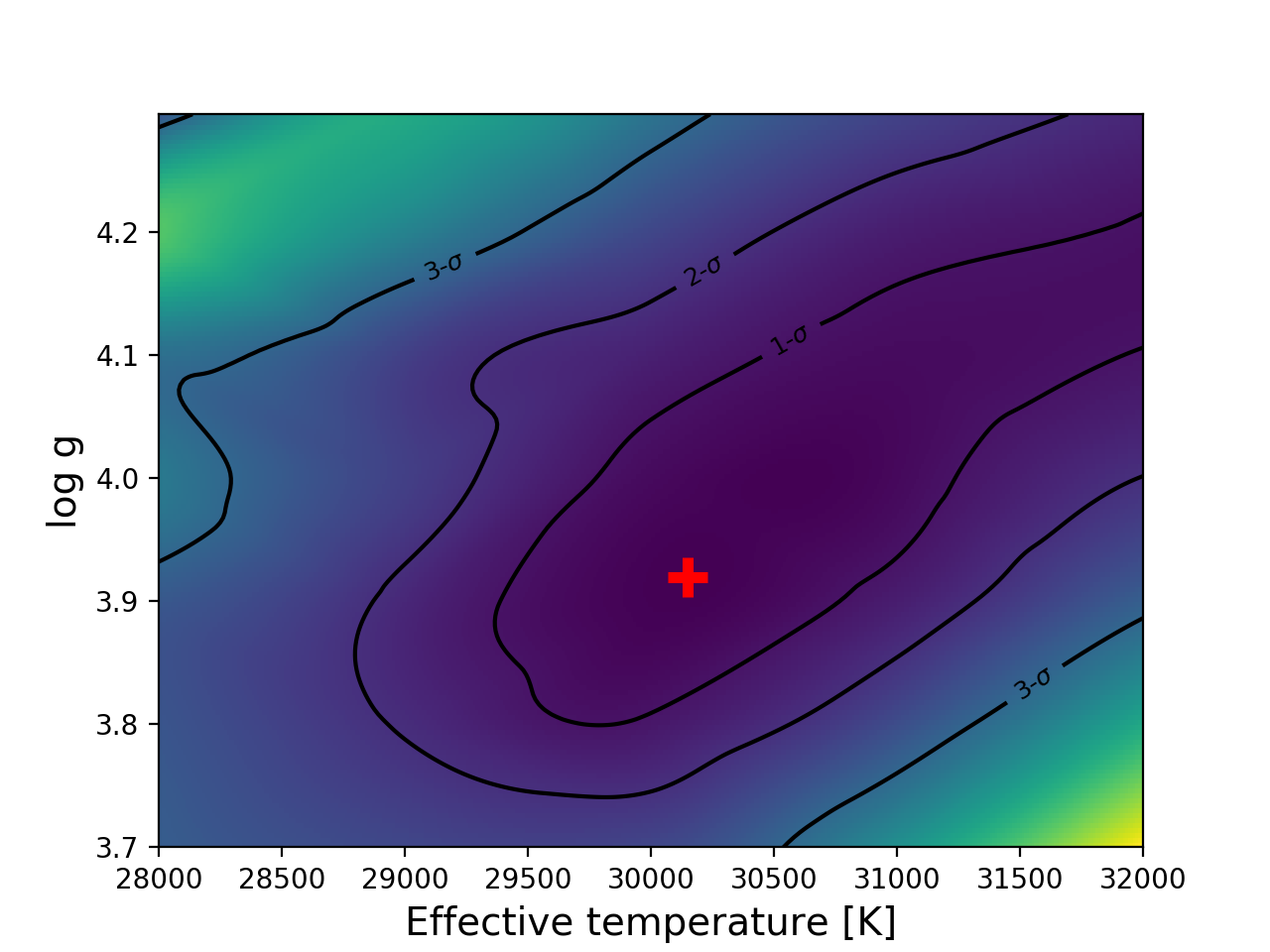}
    \includegraphics[width=7cm]{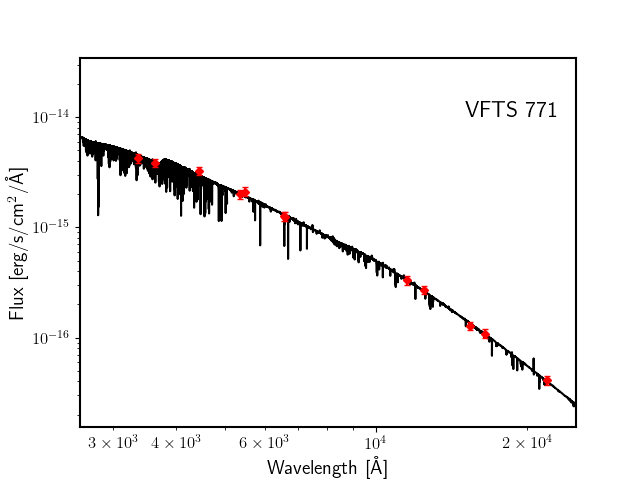}
    \includegraphics[width=6.5cm]{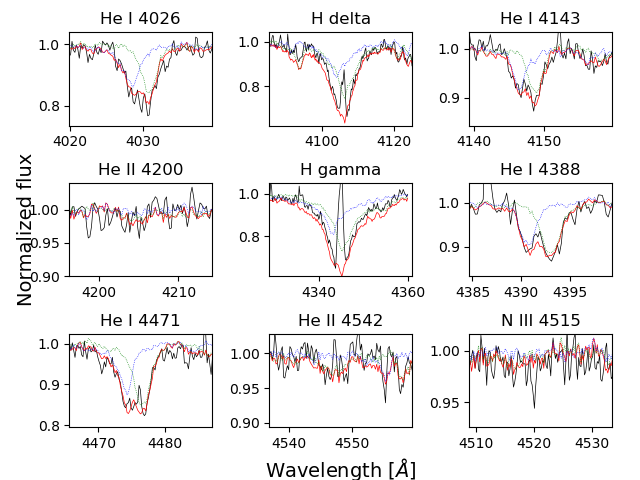}
    \includegraphics[width=7cm, bb=5 0 453 346,clip]{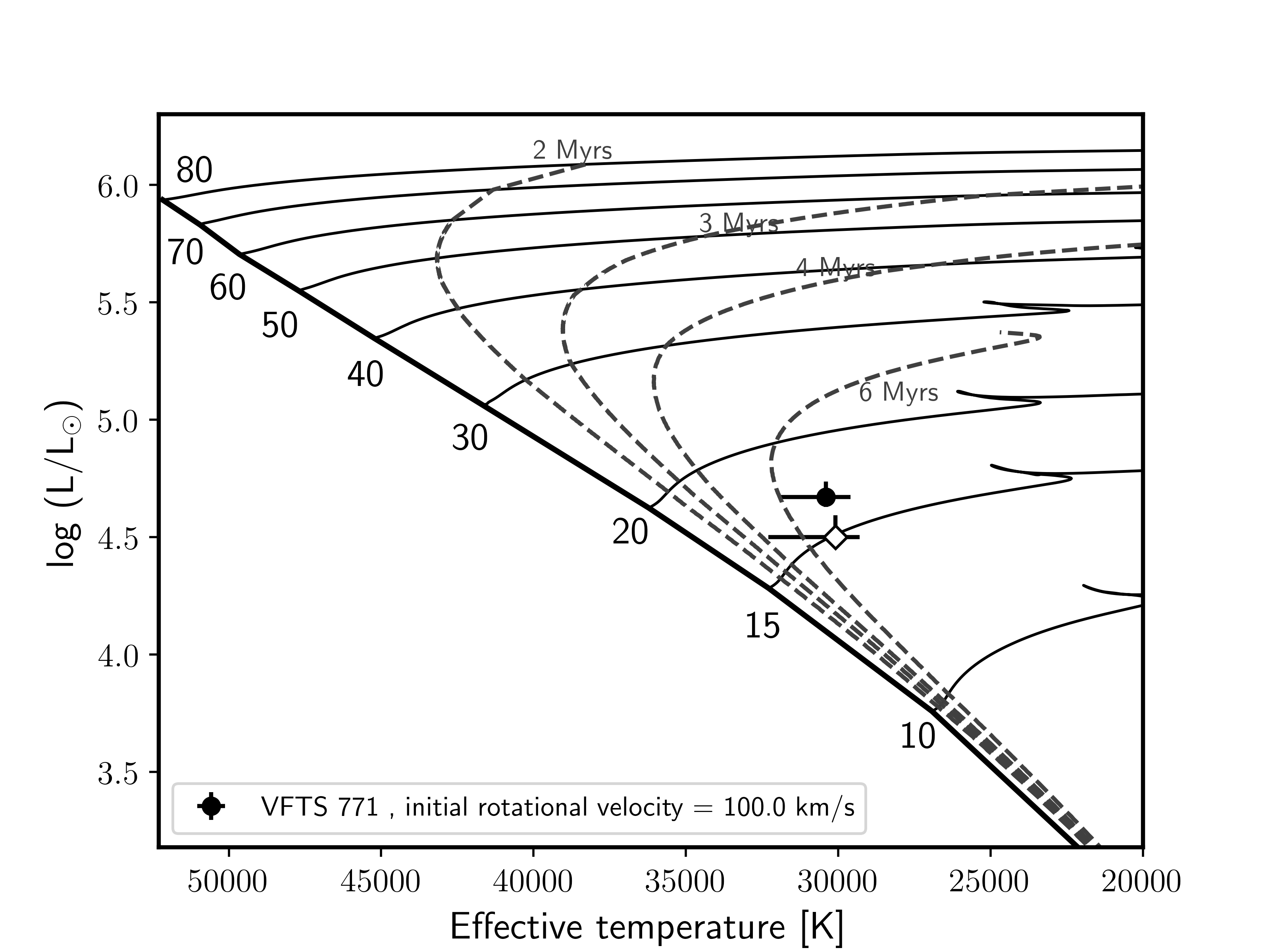}
    \includegraphics[width=7cm, bb=5 0 453 346,clip]{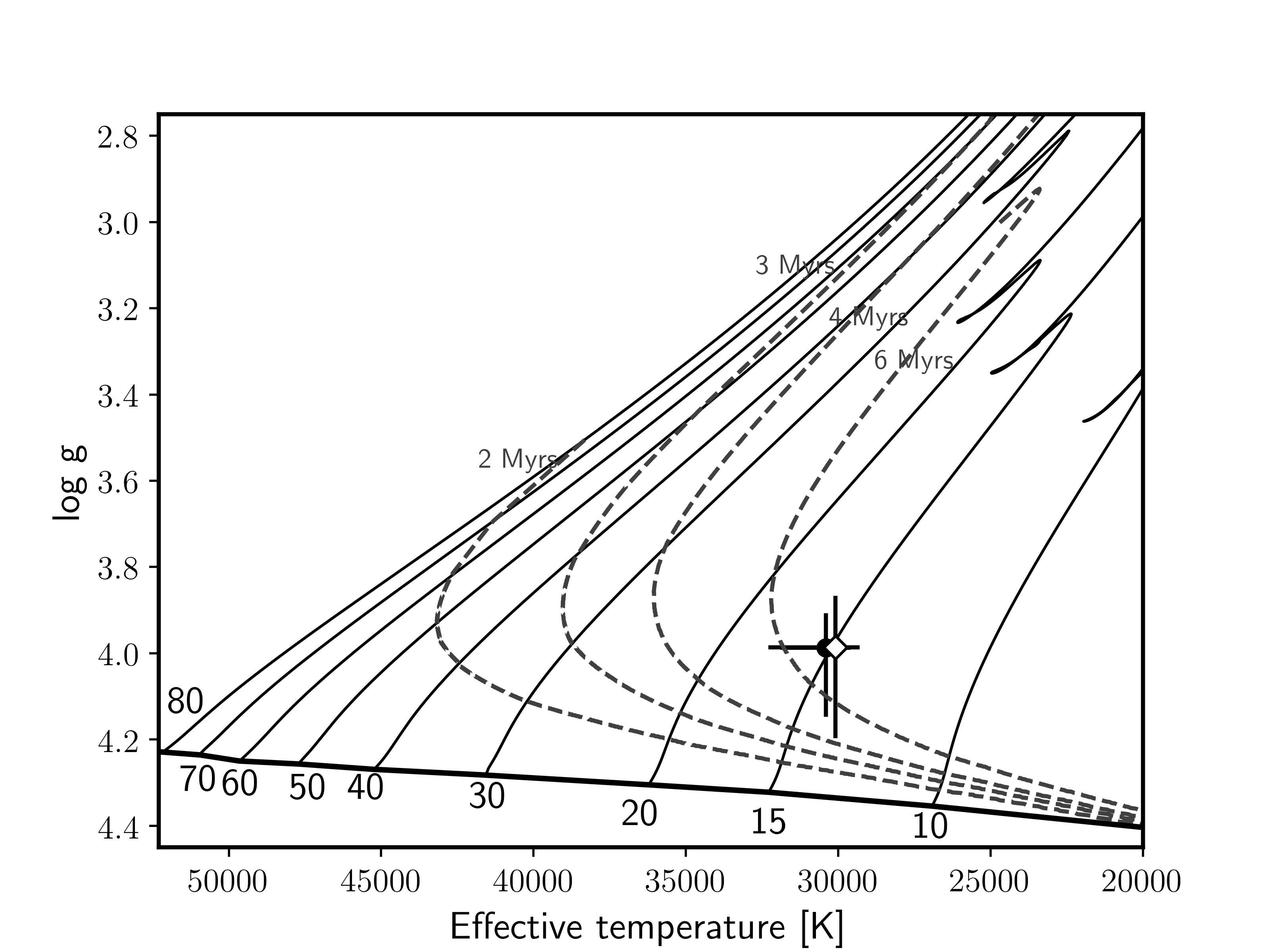}
    \caption{Same as Fig.\,\ref{fig:042} but for VFTS\,771.} \label{fig:771} 
  \end{figure*}
 \clearpage

    \begin{figure*}[t!]
    \centering
    \includegraphics[width=6.cm]{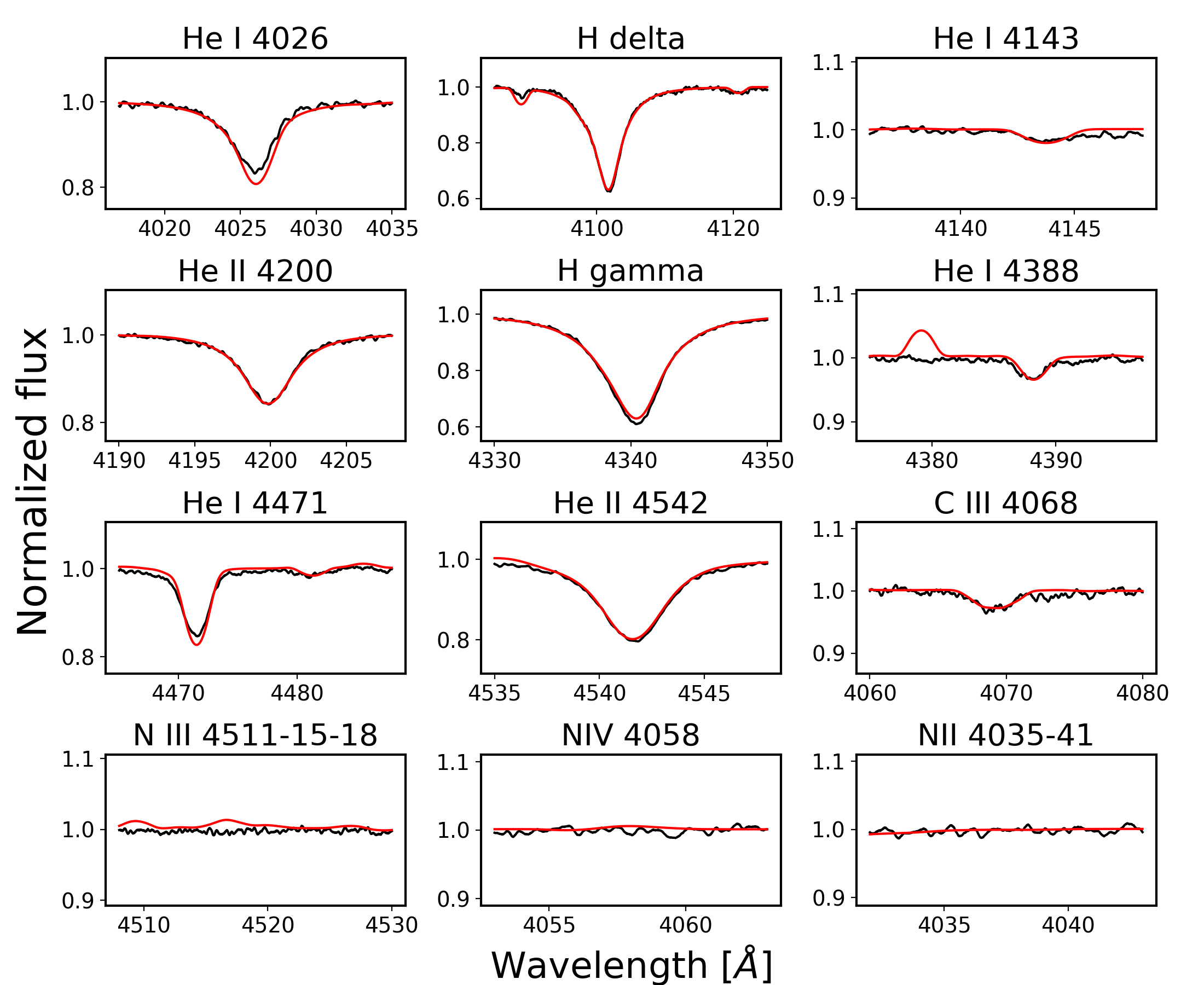}
    \includegraphics[width=7.cm, bb=5 0 453 346,clip]{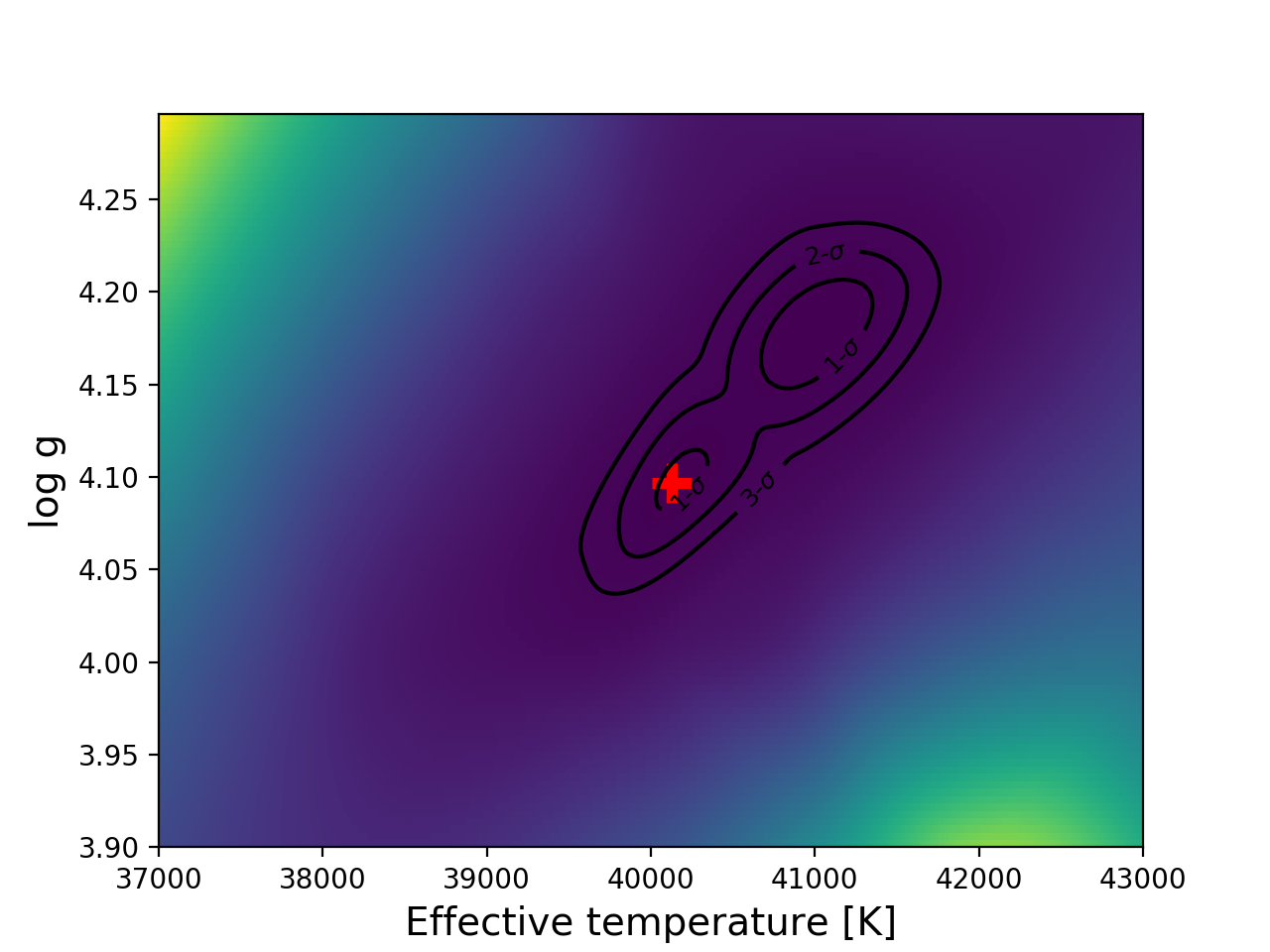}
    \includegraphics[width=6.cm]{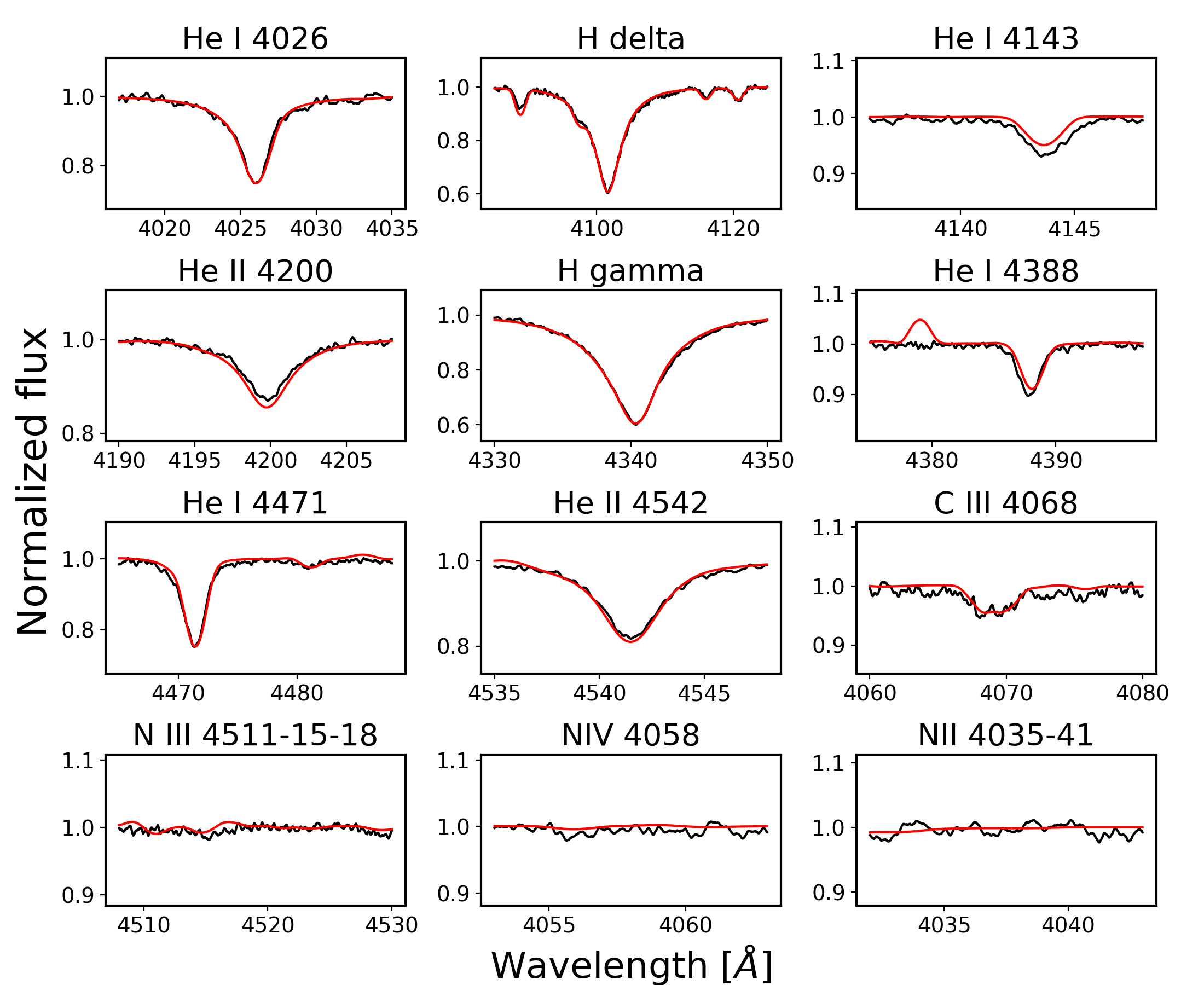}
    \includegraphics[width=7.cm, bb=5 0 453 346,clip]{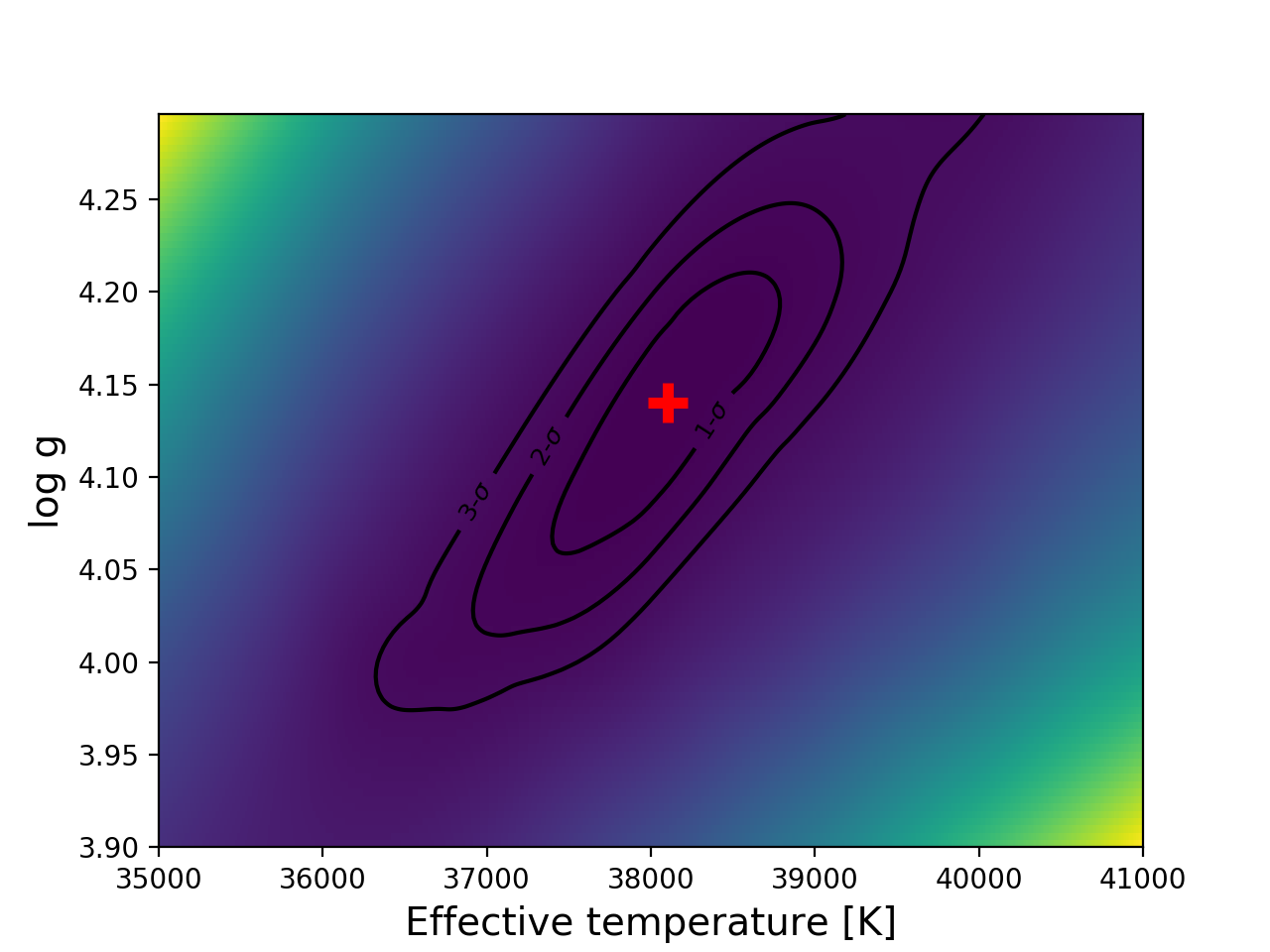}
    \includegraphics[width=7cm]{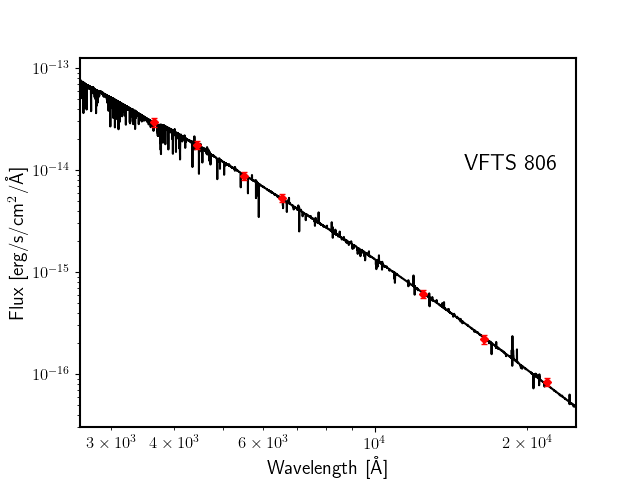}
    \includegraphics[width=6.5cm]{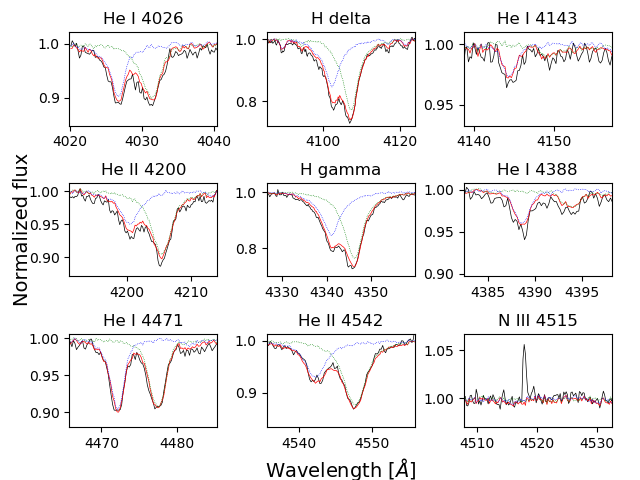}
    \includegraphics[width=7cm, bb=5 0 453 346,clip]{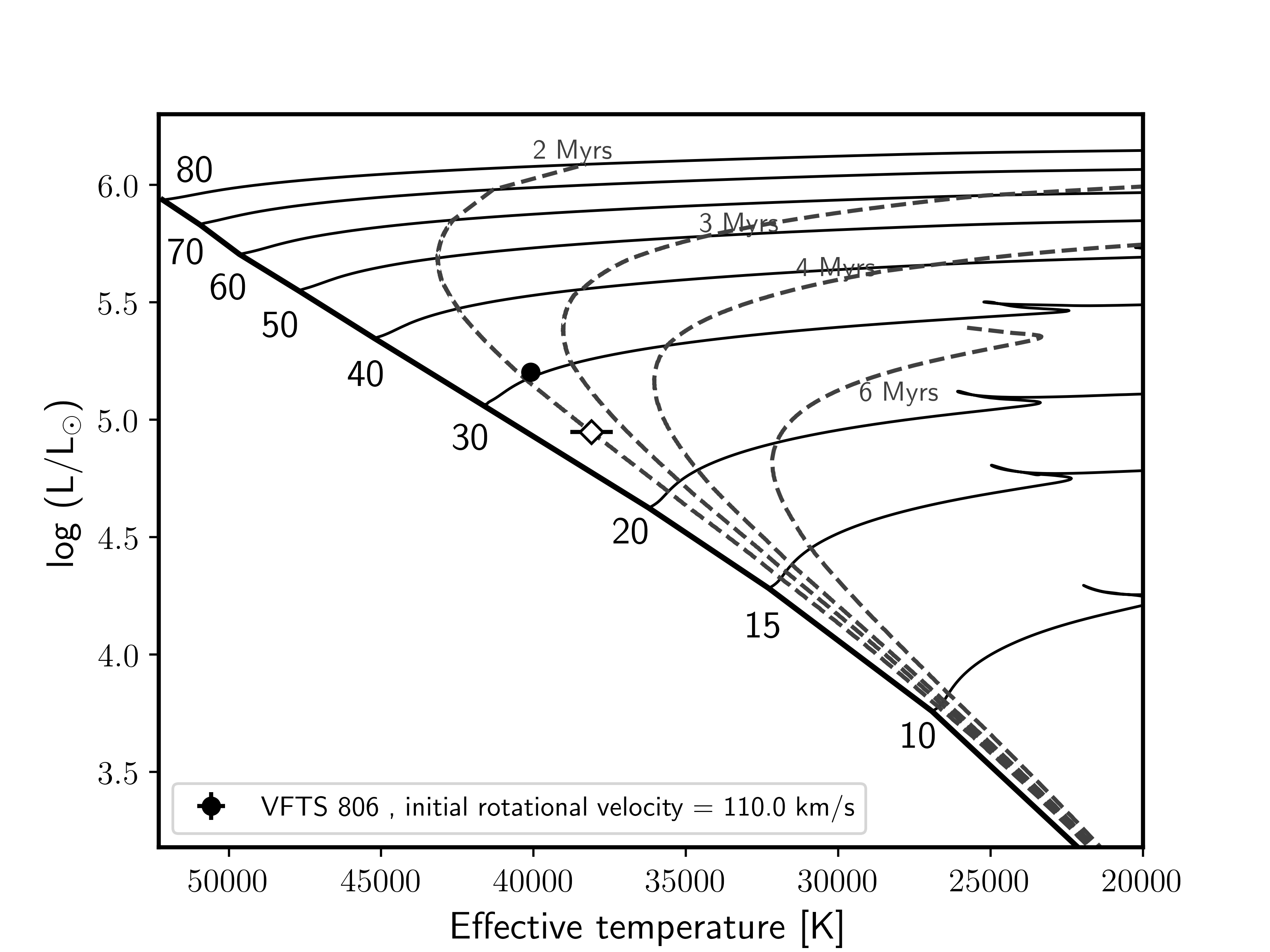}
    \includegraphics[width=7cm, bb=5 0 453 346,clip]{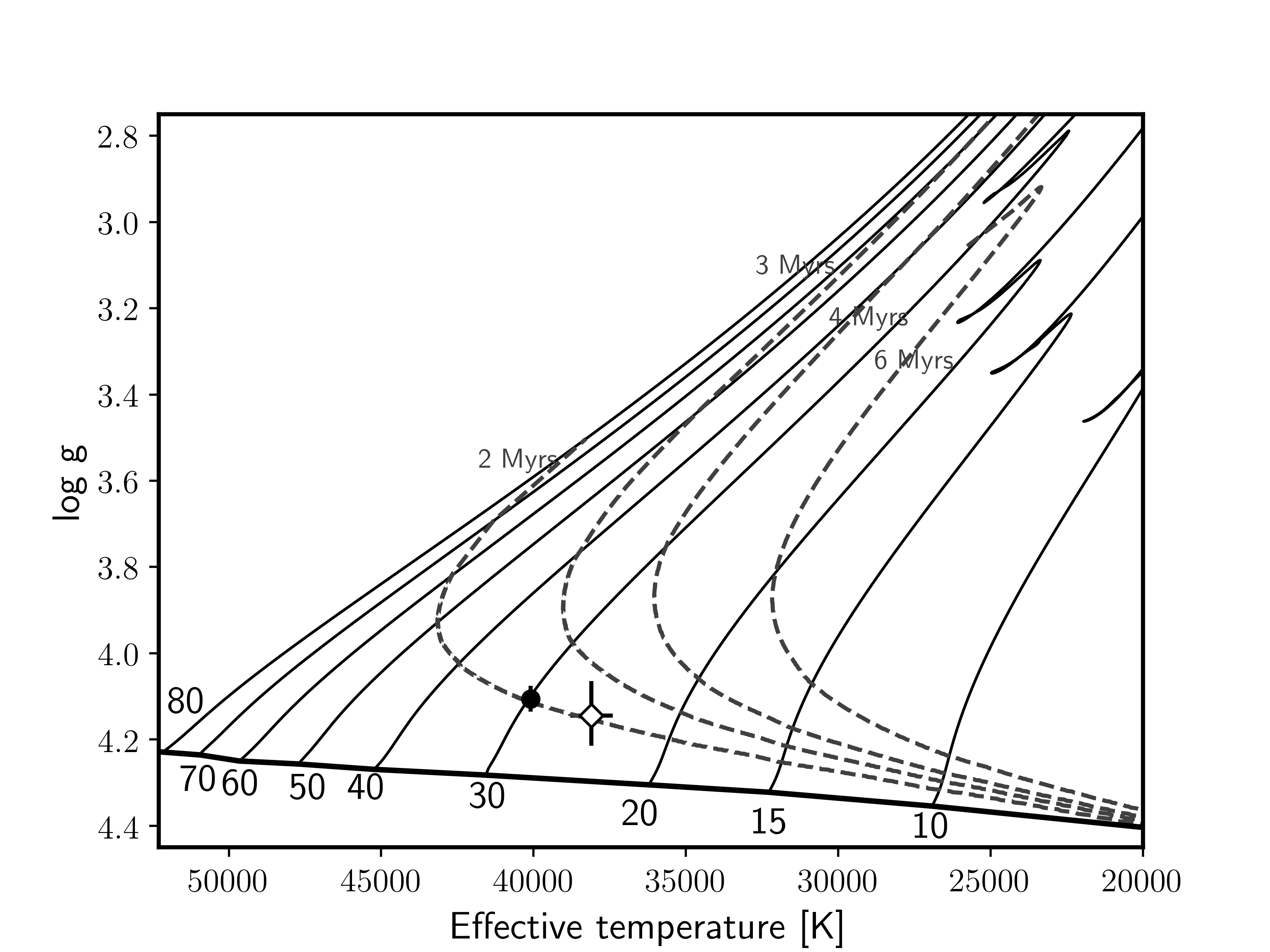}
    \caption{Same as Fig.\,\ref{fig:042} but for VFTS\,806.} \label{fig:806} 
  \end{figure*}
 \clearpage

\section{System magnitudes}

\begin{sidewaystable*}
\caption{\label{tab:magnitude} Photometry of the systems in our sample.}
\centering
%\addtolength{\tabcolsep}{-3.5pt}   
\begin{tabular}{lcccccccccccccccc}
\hline\hline
Star & F275W$^{a}$ & F336W$^{a}$ & F555W$^{a}$ & F658N$^{a}$ & F775W$^{a}$ & F110W$^{a}$ & F160W$^{a}$ & $U$ & $B^{b}$ &$V^{b}$ & $G^{c}$ & $R$ & $J^{b}$ & $H^{b}$ & $K^{b}$ & $A_V$ \\
\hline
\hline
VFTS042 & 12.87 & 13.29 & 14.68 & 14.57 & 14.68 & 14.77 & 14.82 &  -- & 14.54 & 14.66 & 14.67 &  -- & 14.75 & 14.77 & 14.82 & 0.92 \\[1pt] 

VFTS047 & 16.45 & 16.39 & 17.06 & 16.51 & 16.37 &  -- &  -- &  -- & 17.19 & 16.91 &  -- &  -- & 15.56 & 15.32 & 15.17 & 3.00 \\[1pt] 

VFTS055 & 14.73 & 14.73 & 15.68 &  -- & 15.30 &  -- &  -- &  -- & 15.79 & 15.56 & 15.58 &  -- & 14.84 & 14.78 & 14.62  & 2.24\\[1pt] 

VFTS061 & 13.95 & 14.10 & 15.45 &  -- & 15.17 &  -- &  -- &  -- & 15.49 & 15.35 & 15.39 & 15.30 & 14.91 & 14.79 & 14.48  & 1.70 \\[1pt] 

VFTS063 & 12.81 & 13.07 & 14.28 &  -- &  -- &  -- &  -- &  -- & 14.23 & 14.40 & 14.21 & 14.16 & 13.88 & 13.79 & 13.73 &  1.67 \\[1pt] 

VFTS066 & 14.40 & 14.50 & 15.61 & 15.39 & 15.43 & 15.27 & 15.21 &  -- & 15.64 & 15.54 &  -- &  -- & 15.20 & 15.14 & 15.09 & 1.66 \\[1pt] 

VFTS094 & 13.00 & 13.15 & 14.17 &  -- &  -- & 13.52 & 13.40 & 13.52 & 14.21 & 14.16 & 14.06 &  -- & 13.48 & 13.32 & 13.32 & 1.96  \\[1pt] 

VFTS114 & 14.82 & 14.99 & 16.04 & 15.73 &  -- & 15.42 & 15.25 &  -- & 16.14 & 15.97 &  -- &  -- & 15.33 & 15.21 & 15.09 & 2.04 \\[1pt] 

VFTS116 &  -- & 15.61 & 16.44 &  -- &  -- & 15.98 & 15.86 & 15.90 & 16.65 & 16.44 &  -- &  -- & 15.89 & 15.76 & 15.71  & 1.91\\[1pt] 

VFTS140 & 15.14 & 15.20 & 16.13 & 15.81 &  -- & 15.52 & 15.39 & 15.51 & 16.28 & 16.05 &  -- &  -- & 15.48 & 15.37 & 15.29  & 2.09\\[1pt] 

VFTS174 &  -- & 14.69 & 15.58 & 15.29 & 15.26 & 15.13 & 15.02 &  -- & 15.75 & 15.70 & 15.53 &  -- & 15.18 & 14.97 & 14.96  & 1.89 \\[1pt] 

VFTS176 & 13.52 & 13.61 & 15.02 & 14.76 &  -- & 15.02 & 14.85 &  -- & 14.81 & 15.00 & 14.87 &  -- & 14.63 & 14.65 & 14.47 &  1.38 \\[1pt] 

VFTS187 & 15.05 & 15.01 & 15.89 & 15.62 &  -- & 15.43 & 15.32 &  -- & 16.03 & 16.00 & 15.83 & 15.87 & 15.32 & 15.24 & 15.19  & 2.03 \\[1pt] 

VFTS197 & 12.20 & 12.52 & 13.87 &  -- &  -- & 13.83 & 13.91 & 12.90 & 13.59 & 13.70 & 13.83 & 13.92 & 13.82 & 13.85 & 13.78  & 1.12 \\[1pt] 

VFTS217 &  -- & 12.27 & 13.79 &  -- &  -- & 13.76 & 13.85 &  -- & 13.68 & 13.74 & 13.73 & 13.90 & 13.61 & 13.87 & 13.87 & 1.01 \\[1pt] 

VFTS327 &  -- & 14.13 & 15.41 & 15.25 &  -- & 15.23 & 15.18 &  -- & 15.34 & 15.33 & 15.34 &  -- & 15.22 & 15.14 & 15.12  & 1.38 \\[1pt] 

VFTS352 & 12.75 & 13.13 & 14.46 &  -- &  -- & 14.27 & 14.27 &  -- & 14.28 & 14.49 & 14.43 &  -- & 14.47 & 14.43 & 14.47  & 1.18 \\[1pt] 

VFTS450 & 12.57 & 12.75 & 13.69 &  -- &  -- &  -- & 12.90 &  -- & 13.80 & 13.66 & 13.54 &  -- & 12.96 & 12.91 & 12.79  & 1.90 \\[1pt] 

VFTS487 &  15.95$^d$ &  16.16$^d$ & 17.23$^d$ & 17.08$^d$ & 17.06$^d$ & 16.94$^d$ & 16.90$^d$ &  16.38 & 17.00 & 16.88 &  -- &  -- &  -- & 15.69 & 15.37  & 2.55 \\[1pt] 

VFTS500 & 12.47 & 12.82 & 14.16 &  -- & 14.00 & 13.92 & 13.85 &  -- & 14.11 & 14.19 &  -- & 14.28 & 13.78 & 13.70 & 13.57  & 1.49 \\[1pt] 

VFTS508 &  -- & 15.23 & 16.17 & 15.86 &  -- & 15.67 & 15.56 &  -- & 16.15 & 15.98 &  -- &  -- & 15.57 & 15.33 & 15.48  & 1.93 \\[1pt] 

VFTS527 & 10.97 & 10.86 & 12.02 &  -- & 11.72 &  -- &  -- &  -- & 12.04 & 11.94 & 11.90 &  -- & 11.51 & 11.41 & 11.35  & 1.72 \\[1pt] 

VFTS538 & 12.48 & 12.89 & 14.19 &  -- & 14.08 & 13.86 & 13.81 &  -- & 13.96 & 13.99 &  -- &  -- & 13.70 & 13.79 & 13.66  & 1.36 \\[1pt] 

VFTS543 & 14.17 & 14.26 & 15.44 & 15.28 & 15.33 & 15.32 & 15.27 &  -- & 15.43 & 15.41 &  -- &  -- & 15.50 & 15.43 & 15.74  & 1.27 \\[1pt] 

VFTS555 & 14.58 & 14.78 & 15.90 & 15.64 & 15.64 & 15.47 & 15.36 &  -- & 15.96 & 15.88 & 15.84 &  -- & 15.51 & 15.37 & 15.29  & 1.73 \\[1pt] 

VFTS563 &  -- & 15.06 & 15.96 & 15.75 &  -- & 15.51 & 15.43 &  -- & 16.10 & 15.91 &  -- &  -- & 15.54 & 15.40 & 15.37  & 1.85 \\[1pt] 

VFTS642 & 15.78 & 15.53 & 16.08 & 15.63 &  -- & 15.22 & 15.05 &  -- & 16.41 & 16.03 & 15.96 &  -- & 15.17 & 15.08 & 14.90  & 2.66 \\[1pt] 

VFTS652 &  -- & 13.06 & 14.15 &  -- &  -- &  -- & 13.46 & 13.35 & 14.20 & 13.84 & 13.96 &  -- & 13.39 & 13.30 & 13.21  & 1.76 \\[1pt] 

VFTS661 & 13.94 & 14.00 & 15.19 & 15.02 &  -- & 15.24 & 15.17 &  -- & 15.21 & 15.13 & 15.25 &  -- & 15.15 & 15.12 & 15.12  & 1.37\\[1pt] 

VFTS771 &  -- & 14.72 & 15.70 & 15.38 &  -- & 15.22 & 15.14 & 15.09 & 15.72 & 15.60 & 15.62 &  -- & 15.17 & 15.08 & 15.04  & 1.79 \\[1pt] 

VFTS806 &  -- &  -- &  -- &  -- &  -- &  -- &  -- & 12.88 & 13.89 & 14.06 & 14.06 &  -- & 14.29 & 14.32 & 14.26 &  0.71 \\[1pt]
\hline
\end{tabular}

\tablebib{
$a$:~\citet{sabbi16}; $b$: \citet{evans11,kato07}; $c$: \citet{gaia18}.
}
\tablefoot{Typical errors on the magnitudes are 0.01 and on $A_V$ 0.1. \\
$^d$: there is a large discrepancy for VFTS\,487 between \citet{sabbi16} and the $UBVGRJHK$ magnitudes. Given that the threshold to select the VFTS target was $V=17$ we use the latter magnitudes for our analysis.}
\end{sidewaystable*}

\end{appendix}
\end{document}